# UWB Array Design Using *Variable* $Z_0$ Technology and Central Force Optimization (ver. 2)

**Richard A. Formato**
Registered Patent Attorney & Consulting Engineer
Cataldo & Fisher, LLC, 400 TradeCenter, Suite 5900
Woburn, MA 01801 USA
rf2@ieee.org, rformato@cataldofisher.com

**Abstract —** This note applies *Variable* $Z_0$ technology to the design of an Ultra Wideband (UWB) Yagi-Uda array optimized using Central Force Optimization. *Variable* $Z_0$ is a novel and proprietary approach to antenna design and optimization that treats the feed system characteristic impedance, $Z_0$, as a design *variable* instead of a *fixed* design parameter as is traditionally done. *Variable* $Z_0$ is applicable to any antenna design or optimization methodology, and using it will generally produce better antenna designs across any user-specified set of performance objectives.

## 1. Introduction

*Variable* $Z_0$ [1] is a new antenna design and optimization methodology that produces better antennas than traditional methodology. *Variable* $Z_0$ is a proprietary approach that appears to have been heretofore overlooked. As used herein, *design* refers to the process of specifying a complete set of parameters defining an antenna meeting specific performance objectives, while *optimization* refers to specifying a complete set of parameters defining the antenna that *best* meets specific performance objectives. Following standard usage, $Z_0$ is the antenna feed system characteristic impedance. This new technology should be especially useful for increasing impedance bandwidth (IBW), but *Variable* $Z_0$ can be used to achieve any performance objectives, including those that do not consider IBW at all.

Traditional antenna design and optimization methods, techniques, processes, or procedures ("methodology") treat $Z_0$ as a *fixed* design *parameter* with a constant value specified at the start of the methodology. In traditional methodology $Z_0$ is not a ***variable*** quantity whose value is determined ***by*** the methodology. This distinction is fundamental and quite important, because traditional methodology excludes from the outset *all* designs that could provide better performance by using some other value of $Z_0$. *Variable* $Z_0$ improves on traditional methodology by treating $Z_0$ as another design *variable* in the set of variable antenna system parameters to be determined by the methodology. This note illustrates its use by applying it to the design of a numerically optimized ultra wideband (UWB) Yagi-Uda array with excellent results.

## 2. Typical performance objectives and $Z_0$ as a design *variable*

The term "bandwidth" refers generally to the range of frequencies over which some specific antenna performance measure is met. For example, "gain bandwidth" is the frequency range over which a mini-mum power gain is achieved; and so on with respect to other performance measures. IBW is defined as the frequency band or bands within which the antenna input impedance, $Z_{in} = R_{in} + jX_{in}$, $j = \sqrt{-1}$, is matched to the feed system characteristic impedance, $Z_0$, within specified limits. The required degree of matching can be specified in terms of the antenna's actual input impedance (resistance, $R_{in}$, and reactance, $X_{in}$) as a function of frequency, or, as is more often the case, in terms of a maximum voltage standing wave ratio (VSWR). IBW typically is specified as VSWR // $Z_0 \leq 2{:}1$ (// denoting "relative to"), which is equivalent to a return loss (scattering parameter $S_{11}$) approximately less than $-10\,\mathrm{dB}$. Other VSWR thresholds can be used instead, and frequently are. Zehforoosh *et al.* [2] describe the design of an UWB microstrip antenna and provide a good summary of $Z_0$'s significance as a design parameter in the context of IBW.

Following the traditional methodology, $Z_0$ in [2] is a fixed parameter, not a variable quantity as it is in *Variable* $Z_0$. While *Variable* $Z_0$ is useful in any antenna design, its widest range of applicability likely will be in improving IBW, which consequently is emphasized in the design example considered here. The current state-of-the-art, which fixes $Z_0$'s value, limits achievable IBW (as well as meeting other performance objectives) because better results are obtainable when $Z_0$ is considered to be a design *variable* whose value is determined *by* the design or optimization methodology, instead of being a user-supplied constant at the outset of that methodology.

## 3. Design example: UWB Yagi-Uda array

Two 6-element Yagi-Uda ("Yagi") arrays are described. The arrays comprise a driven element (DE) excited by the radio-frequency (RF) source flanked by a parasitic reflector (REF) on one side and four parasitic directors ($\mathrm{D}_k$, $k = 1,\ldots,4$) on the other. All elements are PEC (Perfectly Electrically Conducting). One Yagi is designed using *Variable* $Z_0$, which treats $Z_0$ as an unknown variable whose value is to be determined by an optimization algorithm. The second Yagi array is designed following traditional methodology in which $Z_0$ is assigned the fixed value, in this case $50\,\Omega$ resistive. Each antenna was optimized using Central Force Optimization (CFO), a deterministic metaheuristic that has performed well against recognized benchmark functions and problems in applied electromagnetics [3-10]. The Yagis' performance was computed by NEC-4D (Numerical Electromagnetics Code ver. 4 double precision) [11,12]. NEC is a widely used Method of Moments (MoM) wire structure modeling program developed at Lawrence Livermore National Laboratory (LLNL). A freeware version of NEC-2 is available online (source code and executables) [13].

The optimized antennas are shown in **Figure 1** (as visualized using 4nec2 [13], the red circle indicating the RF source). Interestingly, the *Variable* $Z_0$ design looks more "normal" than the fixed $Z_0$ design. Its REF element is longer than DE, and the directors are spaced closer together with increasing distance along the boom (+X-axis), which are typical Yagi characteristics. By comparison, the fixed $Z_0$ design has a short reflector and directors whose spacing increases with distance from DE, which is somewhat unusual.

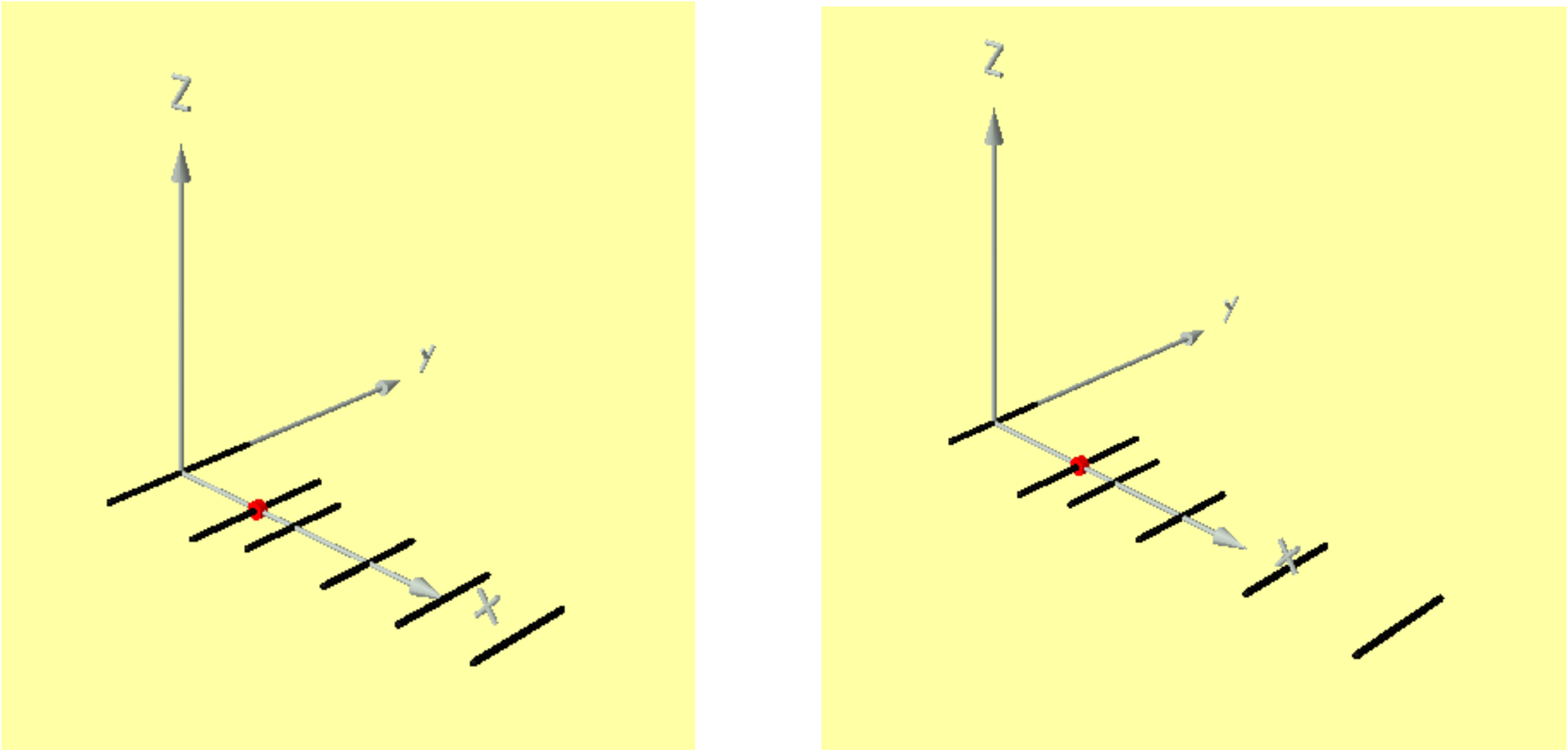

$Z_0 = 105.64\ \Omega$ $Z_0 = 50\ \Omega$

**Figure 1.** CFO-optimized *Variable* $Z_0$ (left) and fixed $Z_0$ (right) Yagis.

The following fitness (or "objective") function was employed (to be maximized by CFO):

$$F(Z_0, L_i, S_j) = c_1 \cdot G_{fwd}(f_L) - c_2 \cdot VSWR(f_L) + c_3 \cdot G_{fwd}(f_C) - c_4 \cdot VSWR(f_C) + c_5 \cdot G_{fwd}(f_U) - c_6 \cdot VSWR(f_U).$$

$L_i$, $i = 1,..,6$ and $S_j$, $j = 1,..,5$ are the element lengths and spacings, respectively. There are twelve decision variables in the optimization problem: six element lengths, five element spacings, and, very importantly, $Z_0$. *VSWR* is the voltage standing wave ratio *relative to* $Z_0$. The optimization problem thus is 12-dimensional (12-D). The constants $c_i$ are the empirically determined weighting coefficients shown in **Table 1**. $G_{fwd}(f)$ the Yagi's forward directivity (along the +X-axis; that is, $\theta = 90°, \phi = 0°$ in NEC's right-handed spherical polar coordinate system [11,12]). $f_L$, $f_C$, and $f_U$, respectively, the lower, center, and upper frequencies defining the symmet-rical band within which the Yagi is optimized. In this case, the band center was $f_C = 299.8$ MHz with $f_{L,U} = f_C \mp 50$ MHz, so that the wavelength at $f_C$ is $\lambda_C = 1$ meter [note that NEC requires coordinates input in meters, not wavelengths]. In post processing, the array performance was computed from 200-400 MHz every 0.1 MHz (computed data are tabulated in **Appendices II** & **III**).

Of course, the antenna designer is free to specify any desired fitness function, and its specific form will produce different antenna designs as a consequence of the different decision space landscape. Some objective functions may introduce $Z_0$ as a variable indirectly, as is done here through the VSWR; or it may be introduced explicitly as, for example, in the second and third objective functions used in [8].

**Table 1.** Yagi fitness coefficients.

| $c_1$ | $c_2$ | $c_3$ | $c_4$ | $c_5$ | $c_6$ |
|---|---|---|---|---|---|
| 0.2 | 4 | 1 | 8 | 1 | 0.8 |

A parameter-free CFO implementation was used as described in [5] with $N_d = 12$, $\frac{N_p}{N_d} = 4$ (no stepping), $N_t = 250$, $N_\gamma = 11$, $F_{rep}^{init} = 0.5$, $\Delta F_{rep} = 0.1$, $F_{rep}^{\min} = 0.05$, $\alpha = 1$, $\beta = 1$, $G = 2$, $\Delta t = 0.5$. Decision space adaptation was applied every 20th time step with an early termination criterion of fitness saturation for 25 consecutive steps (variation $\leq 10^{-6}$). Note that in [1] a 6-element Yagi was modeled as a 13-dimensional problem as a matter of programming convenience in dealing with how data arrays in CFO legacy code were dimensioned. In this case, the CFO program's data arrays were redimensioned for the Yagi problem to reflect the actual 12-D decision space dimensionality. **Appendix I** contains the complete source code listing for the program that was used to compute the data reported here.

The decision space geometry variables were $0.2 \leq L_i \leq 0.6\lambda_C$ and $0.1 \leq S_j \leq 0.5\lambda_C$. The *variable* feed system impedance was bounded by $25 \leq Z_0 \leq 250\ \Omega$ (purely resistive). $Z_0$ was held constant for the fixed $Z_0$ case by setting $50 \leq Z_0 \leq 50\ \Omega$ (of course, the designer is free to specify any desired range of values for $Z_0$, which is a useful attribute of *Variable* $Z_0$ because it allows complete flexibility in matching "standard" impedance values if this is an important consideration). All array elements had the same radius $0.00635\lambda_C$ (0.5-inch diameter elements at 299.8 MHz). Lengths were rounded to three decimal places in the NEC4 model (except radius) and $Z_0$ to two places. For the *Variable* $Z_0$ Yagi, CFO's best returned fitness was $-1.189...$ (CFO probe #48) with $\gamma = 0.9$ at step 210 (90,528 function evaluations). The corresponding values for the fixed $Z_0$ array are $-13.645...$ (probe #44) with $\gamma = 0.6$ at step 49 (39,072 function evaluations). The computed CFO probe coordinates appear in **Figure 2.**

Using *Variable* $Z_0$, the optimum feed system impedance computed by CFO was $Z_0 = 105.64\ \Omega$. Although Yagi arrays are generally considered "low" impedance antennas (typically $|Z_{in}| \approx 10-30\ \Omega$), this optimized feed system $Z_0$ is relatively high. This value, however, is more readily matched to the "standard" values of $50-75\ \Omega$, which is a fortuitous outcome. The optimized arrays' NEC input files appear in **Figure 3**, and their geometries are tabulated in **Table 2**. The spacings have been added to specify each element's position ($X-$coordinate, "boom dist") along the Yagi's boom ($+X-$axis). Because $\lambda_C = 1$ meter, the entries in **Table 2** are also in wavelengths at $f_C$, so that these results can be used to scale the CFO-optimized UWB Yagi to any other frequency.

**Figures 4-6**, respectively, show the evolution of CFO's best fitness, $D_{avg}$, and best probe number. $D_{avg}$ is the average distance between the probe returning the best fitness and all other probes, step-by-step. As CFO's probes converge on an optimum, $D_{avg}$ decreases, thus serving as a measure of how tightly the probes have clustered. These plots are characteristic of CFO's behavior both on antenna problems [6,9] and on standard benchmark functions [5,10]. CFO was selected because of the author's familiarity with the algorithm and its deterministic nature. But any number of other commonly employed algorithms, such as Particle Swarm (PSO) [14], Ant Colony (ACO) [15], Group Search Optimizer (GSO) [16], Differential Evolution (DE) [17-19], or Genetic Algorithm (GA) [20] could be used instead. This "product by process" approach applies to any methodology, deterministic ones like CFO; stochastic metaheuristics like PSO, ACO, GSO DE, or GA; analytic approaches such as extended Wu-King impedance loading [4]; or even "seat of the pants" design or optimization based on experience, intuition, or a "best guess." The specific design or optimization methodology is *entirely irrelevant* to the novelty of treating $Z_0$ as a design variable instead of a fixed parameter. Thus, *Variable* $Z_0$ can be used advantageously with *any* design or optimization methodology.

```
YAGI
Best Fitness = -1.18919692392253 returned by
Probe # 48 at Time Step 210

P48 coordinates:
1  .31883055
2  .14047133
3  .27812909
4  .25622996
5  .23761514
6  .56163747
7  .50718407
8  .36523335
9  .35859015
10  .35989697
11  .35959582
12  105.64385547
```

```
YAGI
Best Fitness = -13.6452736240978 returned by
Probe # 44 at Time Step 49

P44 coordinates:
1  .37022425
2  .13786736
3  .25716011
4  .36513622
5  .35287928
6  .35716011
7  .48852411
8  .35716011
9  .35266668
10  .33530139
11  .36045709
12  50
```

**Figure 2.** CFO results for *Variable* $Z_0$ (left) and fixed $Z_0$ (right) Yagis.

```
CM File: YAGI.NEC
CM YAGI ARRAY IN FREE SPACE
CM Band center frequency, Fc = 299.8 MHz
CM Freq step = 50 MHz +/- Fc
CM Run ID: 07282011_001156
CM Fitness function:
CM .2*Gfwd(L)-4*VSWR(L)+1*Gfwd(M)-8*VSWR(M)+
CM 1*Gfwd(U)-.8*VSWR(U)
CM where L,M,U are lower/mid/upper
frequencies
CM Zo=105.64 ohms
CM Note: All dimensions are in METERS.
CM File ID 07282011122236
CM Nd= 12, p= 48, j= 210
CE
GW1,9,0.,-.281,0.,0.,.281,0.,.00635
GW2,9,.319,-.254,0.,.319,.254,0.,.00635
GW3,9,.459,-.183,0.,.459,.183,0.,.00635
GW4,9,.737,-.179,0.,.737,.179,0.,.00635
GW5,9,.993,-.18,0.,.993,.18,0.,.00635
GW6,9,1.231,-.18,0.,1.231,.18,0.,.00635
GE
FR 0,3,0,0,249.8,50.
EX 0,2,5,1,1.,0.
RP 0,2,2,1001,0.,0.,90.,180.,100000.
EN
```

```
CM File: YAGI.NEC
CM YAGI ARRAY IN FREE SPACE
CM Band center frequency, Fc = 299.8 MHz
CM Freq step = 50 MHz +/- Fc
CM Run ID: 07282011_124011
CM Fitness function:
CM .2*Gfwd(L)-4*VSWR(L)+1*Gfwd(M)-8*VSWR(M)+
CM 1*Gfwd(U)-.8*VSWR(U)
CM where L,M,U are lower/mid/upper
frequencies
CM Zo=50 ohms
CM Note: All dimensions are in METERS.
CM File ID 07282011180440
CM Nd= 12, p= 44, j= 49
CE
GW1,9,0.,-.179,0.,0.,.179,0.,.00635
GW2,9,.37,-.244,0.,.37,.244,0.,.00635
GW3,9,.508,-.179,0.,.508,.179,0.,.00635
GW4,9,.765,-.176,0.,.765,.176,0.,.00635
GW5,9,1.13,-.168,0.,1.13,.168,0.,.00635
GW6,9,1.483,-.18,0.,1.483,.18,0.,.00635
GE
FR 0,3,0,0,249.8,50.
EX 0,2,5,1,1.,0.
RP 0,2,2,1001,0.,0.,90.,180.,100000.
EN
```

**Figure 3.** NEC input files for CFO-optimized *Variable* $Z_0$ (left) and fixed $Z_0$ (right) Yagis.

**Figure 7** plots the *Variable* $Z_0$ and fixed $Z_0$ array VSWR curves from 225 to 350 MHz, and **Table 3** summarizes the bandwidth data for three different VSWR thresholds (2:1, 2.5:1, and 3:1). In the table, $f_1$ and $f_2$ are the lower and upper frequency limits (MHz) corresponding to VSWRs below the specified threshold, and $\Delta f = f_2 - f_1$ is the bandwidth in MHz. The fractional bandwidth in percent (relative to the band center frequency) is computed as $BW(\%) = \frac{200\Delta f}{f_1 + f_2}$. The *Variable* $Z_0$ Yagi exhibits UWB performance (fractional bandwidth $\geq 25\%$) at all VSWR thresholds, whereas the fixed $Z_0$ antenna is not UWB at any threshold. *Variable* $Z_0$ methodology resulted in nearly doubling the Yagi's $VSWR \leq 2:1$ bandwidth from 13.43% at $Z_0 = 50\ \Omega$ to 26.34% at $Z_0 = 105.64\ \Omega$. This improvement is dramatic, and directly attributable to *Variable* $Z_0$ because CFO is a deterministic metaheuristic (the improvement thus cannot be a consequence of an optimizer's stochasticity). With respect to matching this array to a "standard" $50\ \Omega$ feed system impedance, a 2.113:1 impedance ratio broadband transformer or other suitable matching network is required, which easily is accomplished with state-of-the-art matching techniques. By injecting an additional degree of freedom in the traditional antenna design or optimization methodology, *Variable* $Z_0$ technology has produced a substantially better antenna design.

**Table 2.** Geometry of *Variable* $Z_0$ and fixed $Z_0$ CFO-optimized Yagi arrays.

| *element #* | *Variable* $Z_0$ ($Z_0 = 105.64\,\Omega$) | | Fixed $Z_0$ ($Z_0 = 50\,\Omega$) | |
|---|---|---|---|---|
| | *length* ($\lambda_C$) | *boom dist* ($\lambda_C$) | *length* ($\lambda_C$) | *boom dist* ($\lambda_C$) |
| 1 (REF) | 0.562 | 0 | 0.358 | 0 |
| 2 (DE) | 0.508 | 0.319 | 0.488 | 0.370 |
| 3 ($D_1$) | 0.366 | 0.459 | 0.358 | 0.508 |
| 4 ($D_2$) | 0.358 | 0.737 | 0.352 | 0.765 |
| 5 ($D_3$) | 0.360 | 0.993 | 0.336 | 1.130 |
| 6 ($D_4$) | 0.360 | 1.231 | 0.360 | 1.483 |

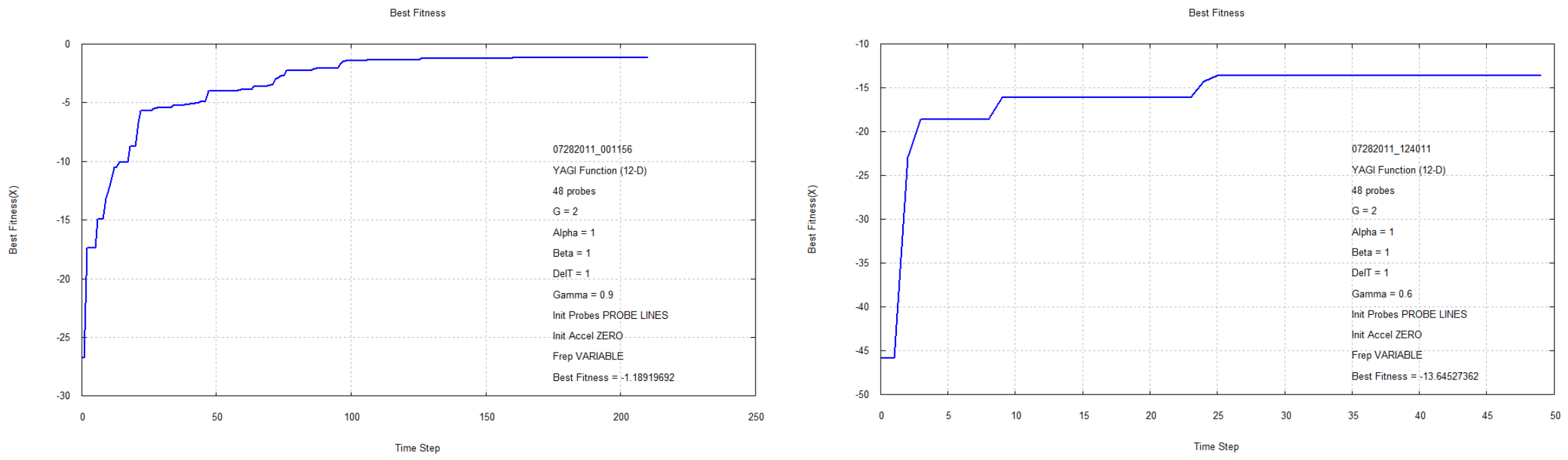

**Figure 4.** Evolution of CFO's best fitness, *Variable* $Z_0$ (left) and fixed $Z_0$ (right).

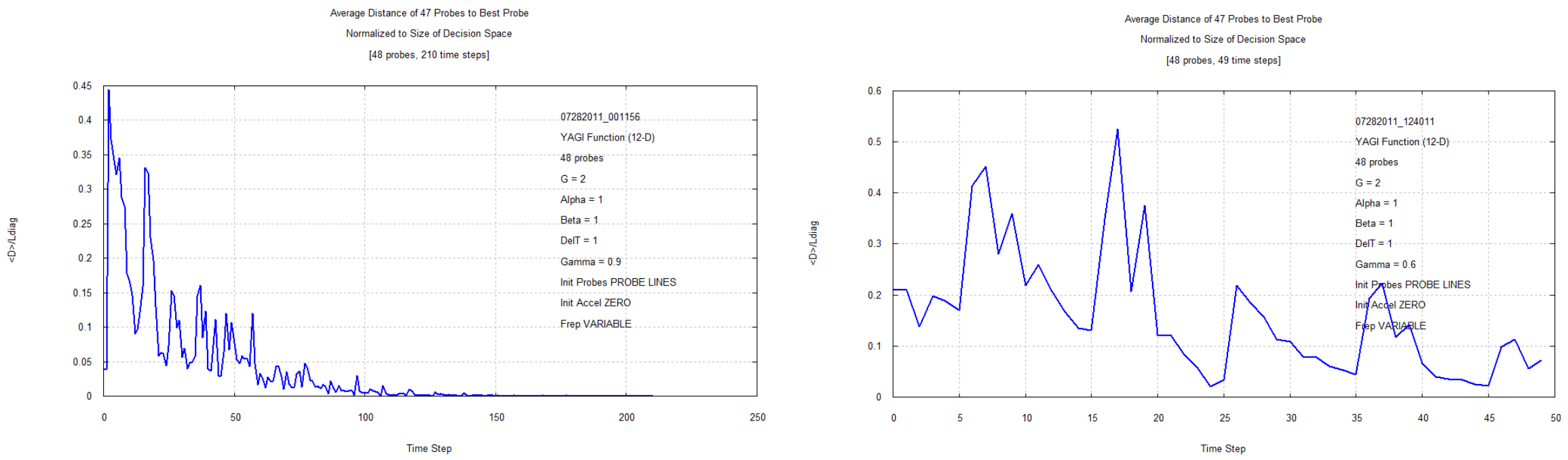

**Figure 5.** Evolution of CFO's $D_{avg}$, *Variable* $Z_0$ (left) and fixed $Z_0$ (right).

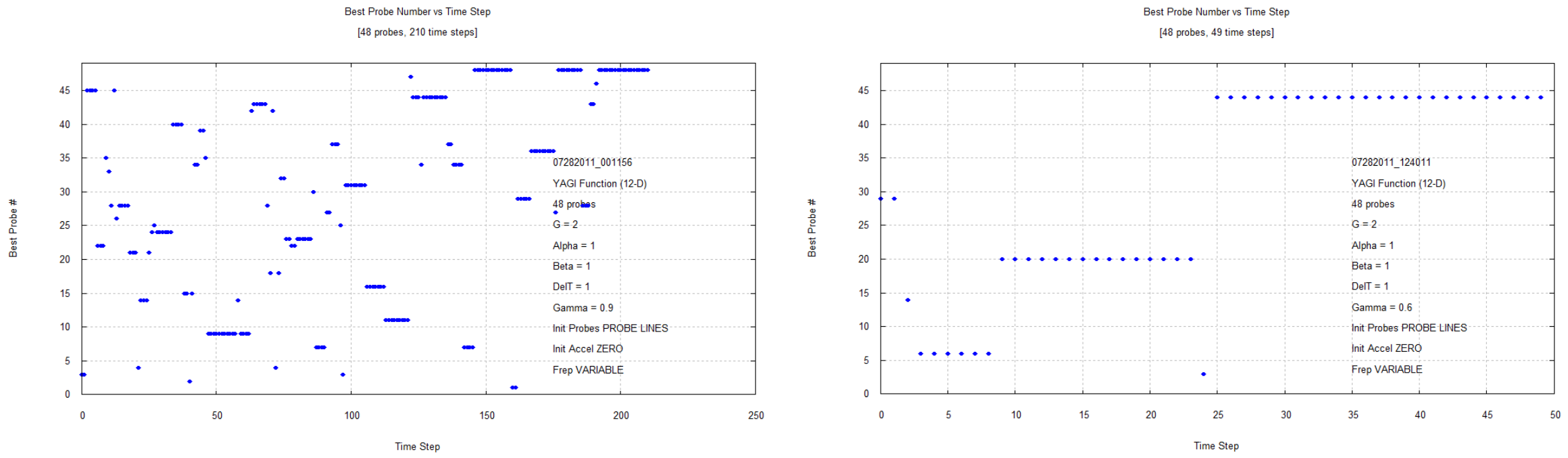

**Figure 6.** Evolution of CFO's best probe, *Variable* $Z_0$ (left) and fixed $Z_0$ (right).

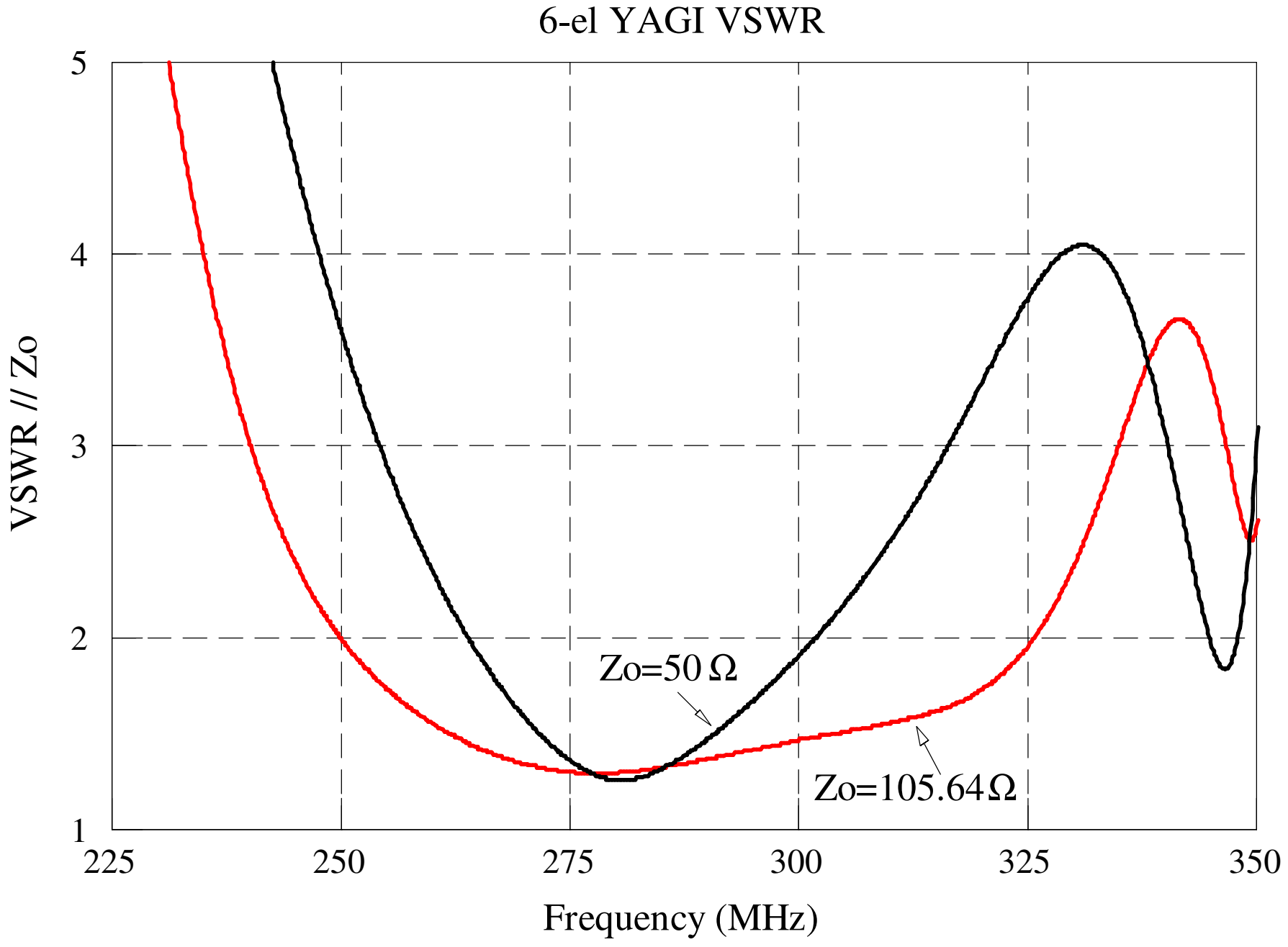

**Figure 7.** VSWR // $Z_0$

**Table 3.** Bandwidth as a function of VSWR threshold (frequencies in MHz).

| $VSWR // Z_0$ Threshold | *Variable* $Z_0 = 105.64\ \Omega$ | | | | *Fixed* $Z_0 = 50\ \Omega$ | | | |
|---|---|---|---|---|---|---|---|---|
| | $f_1$ | $f_2$ | $\Delta f$ | $BW(\%)$ | $f_1$ | $f_2$ | $\Delta f$ | $BW(\%)$ |
| 2.0 : 1 | 249.9 | 325.7 | 75.8 | 26.34 | 263.9 | 301.9 | 38.0 | 13.43 |
| 2.5 : 1 | 244.0 | 331.1 | 87.1 | 30.29 | 258.5 | 310.0 | 51.5 | 18.12 |
| 3.0 : 1 | 240.2 | 334.9 | 94.7 | 32.93 | 254.2 | 316.3 | 62.1 | 21.77 |

**Figure 8** plots the *Variable* $Z_0$ and fixed $Z_0$ Yagis' forward gain and front-to-back ratio (FBR) [note that in this case directivity power gain are equal because the PEC array elements result in 100% radiation efficiency]. The *Variable* $Z_0$ array has a generally flatter gain curve with better values. Both arrays exhibit a sharp drop in gain at the high end of the band, with the *Variable* $Z_0$ array's fall-off being very rapid. As to FBR, the *Variable* $Z_0$ array performs better across the entire band, with values generally above 10 dB beyond 250 MHz. The fixed $Z_0$ Yagi's performance generally is several dB worse at all frequencies.

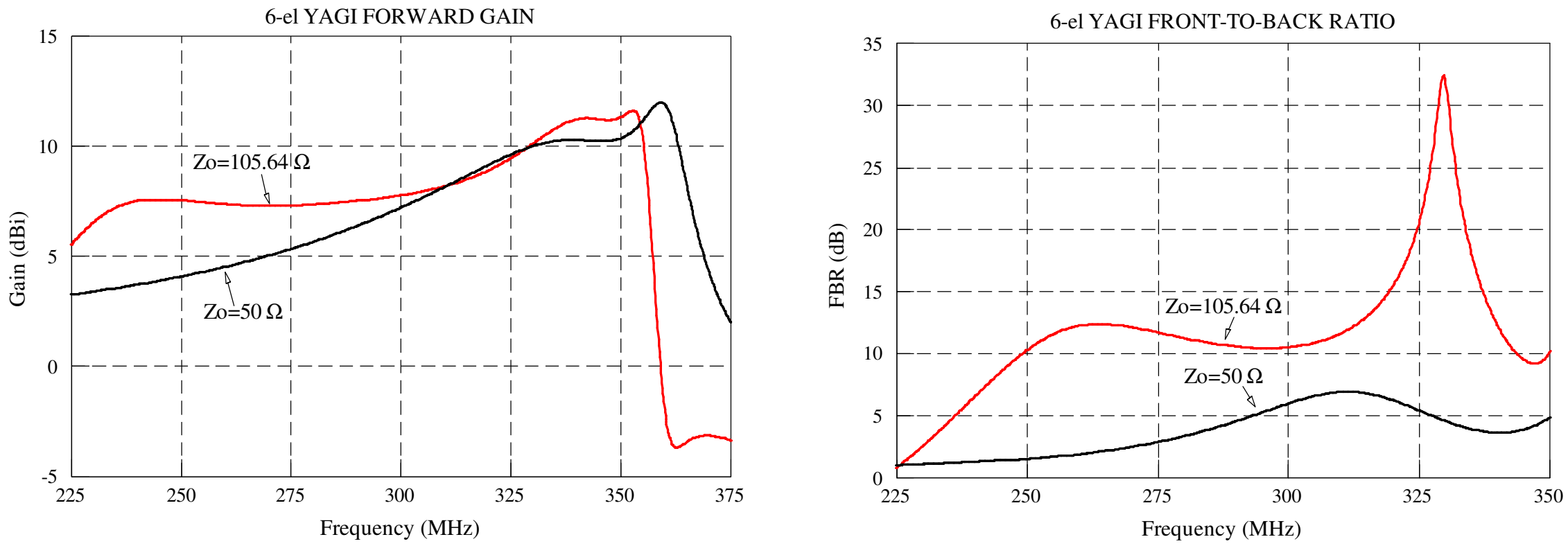

**Figure 8.** Forward gain (left) and FBR (right).

**Figure 9** shows the input resistance and reactance. For both the *Variable* $Z_0$ and fixed $Z_0$ designs, the variation of $R_{in}$ and $X_{in}$ is moderate across the band. For the *Variable* $Z_0$ array, for example, $R_{in}$ is between about 26 and 120 ohms at all frequencies from 225 to 350 MHz, while $X_{in}$ varies from about -90 to +140 ohms. The ranges are similar for the fixed $Z_0$ array. Each antenna exhibits a single resonance ($X_{in} = 0$), the *Variable* $Z_0$ array's near 270 MHz and the fixed $Z_0$'s near 280 MHz. The moderate variations in input impedance and absence of strong resonances are what account for each antenna's significant IBW, although it is evident that *Variable* $Z_0$ technology was able to substantially extend the Yagi's IBW into the UWB range, whereas the traditional methodology in which $Z_0$ is a fixed design parameter could not.

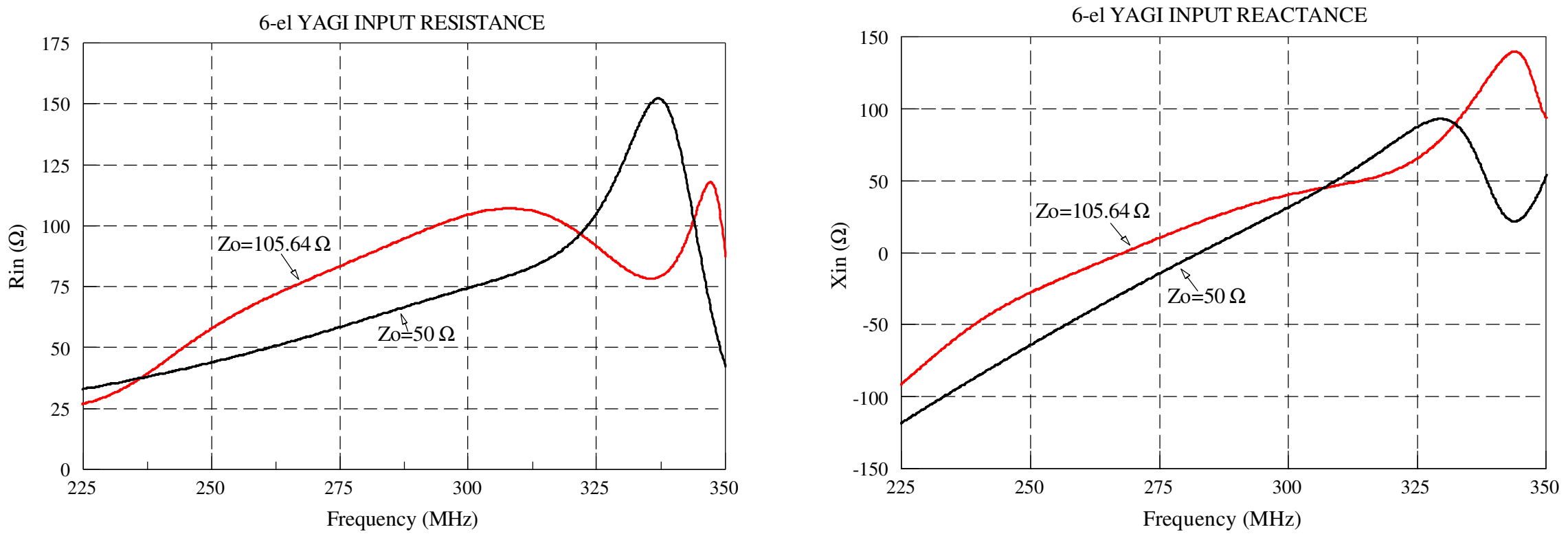

**Figure 9.** Yagi input resistance (left) and reactance (right).

## 4. Conclusion

This paper provides an example of applying *Variable* $Z_0$ technology to the design of an Ultra Wideband Yagi-Uda array. Even though Yagis are generally considered "narrowband" antennas, *Variable* $Z_0$ technology produced a UWB Yagi design with good gain and FBR over a fractional bandwidth greater than 26% (nearly twice the bandwidth of the CFO-optimized fixed $Z_0$ array). *Variable* $Z_0$ is a new approach to antenna design in which the feed system characteristic impedance, $Z_0$, is treated as a *variable* quantity whose value is determined *by* the design or optimization methodology. *Variable* $Z_0$ is a heretofore overlooked and fundamentally different antenna design methodology because it departs from the traditional methodology of treating $Z_0$ as a *fixed* design *parameter* whose value is specified at the outset and never changes. *Variable* $Z_0$ introduces into the antenna design space an additional degree of freedom that makes it easier to achieve specific performance goals, as the design example shows. *Variable* $Z_0$ should be especially useful for improving IBW, but it will be helpful in achieving any desired antenna performance objectives, even objectives that do not involve IBW directly.

***13 August 2011***
***Brewster, Massachusetts***

***Variable*** $Z_0$ ***is a registered Service Mark.***

* * * * *

## References

1. Formato, R. A., "A Novel Methodology for Antenna Design and Optimization: *Variable* $Z_0$ (ver. 2)," http://arXiv.org/abs/1107.1437, July 2011.

2. Zehforoosh, Y., Ghobadi, C., and Nourinia, J., "Antenna Design for Ultra Wideband Application Using a New Multilayer Structure," *PIERS Online*, vol. 2, no. 6, 2006, pp. 544-549.

3. Formato, R. A., "Central Force Optimization: A New Metaheuristic with Applications in Applied Electromagnetics," *Prog. Electromagnetics Research*, vol. PIER 77, pp. 425-491, 2007: http://www.jpier.org/pier/pier.php?paper=07082403 (doi:10.2528/PIER07082403).

4. Formato, R. A., "New Techniques for Increasing Antenna Bandwidth with Impedance Loading," *Progress in Electromagnetics Research B*, Vol. 29, 269-288, 2011.

5. Formato, R. A., “Parameter-Free Deterministic Global Search with Simplified Central Force Optimization,” in **Advanced Intelligent Computing Theories and Applications (ICIC2010)**, Lecture Notes in Computer Science (D.-S. Huang, Z. Zhao, V. Bevilacqua and J, C, Figueroa, Eds.), LNCS 6215, pp. 309–318, Springer-Verlag Berlin Heidelberg, 2010.

6. Formato, R. A., “Improved CFO Algorithm for Antenna Optimization,” *Prog. Electromagnetics Research B*, pp. 405-425, 2010. http://www.jpier.org/pierb/pier.php?paper=09112309 (doi:10.2528/PIERB09112309).

7. Formato, R. A., “Central Force Optimization with Variable Initial Probes and Adaptive Decision Space,” *Applied Mathematics and Computation*, Vol. 217, 2011, pp. 8866-8872.

8. Formato, R. A., “Issues in Antenna Optimization – A Monopole Case Study,” March, 2011. http://arXiv.org/abs/1103.5629.

9. Formato, R. A, “Central Force Optimization Applied to the PBM Suite of Antenna Benchmarks,” Feb., 2010. http://arXiv.org/abs/1003.0221.

10. Formato, R. A, “Comparative Results: Group Search Optimizer and Central Force Optimization,” Feb., 2010. http://arXiv.org/abs/1002.2798.

11. Burke, G. J., and Poggio, A. J., “Numerical Electromagnetics Code (NEC) – Method of Moments,” Parts I, II and III, January 1981, UCID-19934, Lawrence Livermore National Laboratory, Livermore, California, USA

12. Burke, G. J., “Numerical Electromagnetics Code – NEC-4, Method of Moments, Part I: User’s Manual and Part II: Program Description – Theory,” January 1992, UCRL-MA-109338, Lawrence Livermore National Laboratory, Livermore, California, USA (https://ipo.llnl.gov/technology/software/softwaretitles/nec.php).

13. (i) “4nec2” antenna modeling freeware by Arie Voors, available online at http://home.ict.nl/~arivoors/. (ii) Unofficial Numerical Electromagnetic Code (NEC) Archives, online at http://www.si-list.net/swindex.html

14. Kennedy, J., and Eberhart, R., “Particle Swarm Optimization,” *Proc. IEEE Conf. On Neural Networks*, 4 (Nov/Dec), 1942-1948, 1995.

15. Dorigo, M., Maniezzo,V., and Colorni, A., “Positive feedback as a search strategy,” Dipartimento di Elettronica, Politecnico di Milano, Italy, Tech. Rep. 91-016, 1991: http://iridia.ulb.ac.be/~mdorigo/pub_x_subj.html

16. He, S., Wu, Q. H., Saunders, J. R., “Group Search Optimizer: An Optimization Algorithm Inspired by Animal Searching Behavior,” IEEE *Trans. Evol. Computation*, vol. 13, no. 5, pp. 973-990, Oct. 2009.

17. Storn, R., and K. V. Price, “Differential evolution:| A simple and efficient adaptive scheme for global optimization over continuous spaces," Technical Report TR-95-012, ICSI (Univ. of California, Berkeley), 1995. http://www.icsi.berkeley.edu/~storn/litera.html.

18. Storn, R., and K. V. Price, “Minimizing the real functions of the ICEC 1996 contest by differential evolution," *Proc. of the 1996 IEEE International Conference on Evolutionary Computation*, 842-844, Nagoya, Japan, 1996.

19. A. Chowdhury, A. Ghosh, R. Giri, and S. Das, "Optimization of Antenna Configuration with a Fitness-adaptive Differential Evolution Algorithm," *Progress In Electromagnetics Research B*, Vol. 26, 291-319, 2010.

20. Jones, E. A., and Joines, W. T., “Design of Yagi-Uda Antennas Using Genetic Algorithms,” IEEE Trans. Antennas and Propagation, vol. 45, no. 9, September 1997, pp. 1386-1392.

# Appendix I. CFO Source Code Listing

```
'Program 'CFO_YAGI_07-27-2011.BAS'
'Compiled with Power Basic/Windows Compiler 9.04.0122 (www.PowerBasic.com).

'CFO-OPTIMIZED R-LOADED (1 RESISTOR) BOWTIE IN FREE SPACE AND
'YAGI-UDA ARRAY IN FREE SPACE WITH Zo AS AN OPTIMIZATION PARAMETER.

'CFO ARRAY DIMENSIONING CHANGED FROM PREVIOUS VERSION TO AVOID
'ASSIGNING A FIXED "SPACING" OF ZERO TO ELEMENT 'REF.'
'DIMENSIONALITY OF YAGI PROBLEM IS NOW 2*#elements WHEREAS
'IN THE PREVIOUS VERSION IT WAS 2*#elements+1.

'LAST MOD 07-28-2011 ~1240 HRS EDT

'****************** YAGI FITNESS FUNCTION ***********************
'THE USER CAN USE THE FOLLOWING FITNESS FUNCTION:
'        F=[A*FwdGain(DBi)-B*|Zo-Max(Rin)|-
'           C*Max|Xin|-D*(MaxVSWR-MinVSWR)]/(A+B+C+D)
'WHERE THE COEFFICIENTS A, B, C and D are USER-SPECIFIED IN THE
'FILE YagiCoeff.TXT.  THIS FILE MUST CONTAIN FOUR LINES WITH A,
'B, C, AND D, RESPECTIVELY, ON ONE LINE EACH.  IF THE FILE DDOES
'NOT EXIST, A DEFAULT FILE IS CREATED WITH A=20, B=2, C=3, D=4.
'OR, SOME OTHER FITNESS MAY BE HARDWIRED IN THE YAGI FUNCTION.
'CHECK THE SOURCE LISTING TO SEE WHICH IS BEING USED.

'****************** YAGI CFO SETUP PARAMETERS *******************
'THE TEXT FILE YAGI_CFO.CFG CONTAINS THREE CFO SETUP PARAMETERS:
'   Nt& (# time steps)
'   NumGammas%(# gamma values)
'   MinProbesPerDimension% (min # probes/dim on probe lines)
'   MaxProbesPerDimension% (max # probes/dim on probe lines)
'EACH ONE MUST BE ENTERED AS A NUMBER ON A SEPARATE LINE.  IF THE
'FILE YAGI_CFO.CFG DOES NOT EXIST, A DEFAULT VERSION IS CREATED.
'THIS APPLIES ONLY TO THE YAGI MODEL.  OTHER BUILT-IN BENCHMARK
'FUNCTIONS USE THE HARDWIRED INTERNAL VALUES FOR THESE PARAMETERS.

'****************** IMPORTANT MODELING NOTE *********************
'THE YAGI ELEMENT SEGMENT LENGTH IN NEC CAN BE VARIABLE OR FIXED.
'IF FIXED, THEN THE NUMBER OF SEGMENTS VARIES FROM ONE ELEMENT TO
'THE NEXT, BUT ALL SEGMENTS ARE PERFECTLY ALIGNED, WHICH, AS A
'GENERAL RULE IS RECOMMENDED IN THE NEC MODELING GUIDELINES.  BUT
'EXPERIENCE SUGGESTS THAT THIS IS NOT NECESSARILY THE BEST
'APPROACH UNDER ALL CIRCUMSTANCES, SO VARIABLE LENGTH SEGMENTS
'ARE ALLOWED AS WELL.  THE SETUP FILE 'YagiSeg.TXT' CONTAINS TWO
'LINES.  THE FIRST IS EITHER THE WORD 'VARIABLE' (NO QUOTES) OR
'THE WORD 'FIXED,' WHICH DETERMINES WHETHER VARIABLE LENGTH OR
'FIXED LENGTH SEGMENTS ARE USED.  THE NEXT LINE IS THE SEGMENT
'LENGTH IN WAVELENGTHS (RANGE 0.02-0.2).  THIS FILE WILL BE
'CREATED IF IT DOES NOT EXIST USING VARIABLE SEGMENT LENGTH AS
'THE DEFAULT WITH 9 SEGMENTS PER ARRAY ELEMENT.  NOTE THAT IF
'FIXED SEGMENT LENGTH IS USED, EACH ELEMENT'S PHYSICAL LENGTH
'IS ADJUSTED TO BE AN INTEGER NUMBER OF SEGMENTS IN ORDER TO
'ALIGN EVERY SEGMENT.
'****************************************************************

'****************** BOWTIE FITNESS FUNCTION *********************
'THE HARDWIRED BOWTIE FITNESS FUNCTION IS:
'F=[Min(Eff)+5*Min(Gmax)]/{|Zo-Max(Rin)|*[Max(VSWR)-Min(VSWR)]*
'                                       [Max(Xin)-Min(Xin)]}

'**************** BOWTIE CFO SETUP PARAMETERS ********************
'THE TEXT FILE BOWTIE_CFO.CFG CONTAINS THREE CFO SETUP PARAMETERS:
'   Nt& (# time steps)
'   NumGammas%(# gamma values)
'   MinProbesPerDimension% (min # probes/dim on probe lines)
'   MaxProbesPerDimension% (max # probes/dim on probe lines)
'EACH ONE MUST BE ENTERED AS A NUMBER ON A SEPARATE LINE.  IF THE
'FILE BOWTIE_CFO.CFG DOES NOT EXIST, A DEFAULT VERSION IS CREATED.
'THIS APPLIES ONLY TO THE BOWTIE MODEL.  OTHER BUILT-IN BENCHMARK
'FUNCTIONS USE THE HARDWIRED INTERNAL VALUES FOR THESE PARAMETERS.

'==============================================================
'NOTE: ALL PBM FUNCTIONS HAVE WIRE RADIUS SET TO 0.00001 LAMBDA
'==============================================================

'THIS PROGRAM IMPLEMENTS A SIMPLE VERSION OF "CENTRAL
'FORCE OPTIMIZATION."  IT IS DISTRIBUTED FREE OF CHARGE
'TO INCREASE AWARENESS OF CFO AND TO ENCOURAGE EXPERI-
'MENTATION WITH THE ALGORITHM.

'CFO IS A MULTIDIMENSIONAL SEARCH AND OPTIMIZATION
'ALGORITHM THAT LOCATES THE GLOBAL MAXIMA OF A FUNCTION.
'UNLIKE MOST OTHER ALGORITHMS, CFO IS COMPLETELY DETERMIN-
'ISTIC, SO THAT EVERY RUN WITH THE SAME SETUP PRODUCES
'THE SAME RESULTS.

'(c) 2006-2011 Richard A. Formato

'ALL RIGHTS RESERVED WORLDWIDE

'THIS PROGRAM IS FREEWARE.  IT MAY BE COPIED AND
'DISTRIBUTED WITHOUT LIMITATION AS LONG AS THIS
'COPYRIGHT NOTICE AND THE GNUPLOT AND REFERENCE
'INFORMATION BELOW ARE INCLUDED WITHOUT MODIFICATION,
'AND AS LONG AS NO FEE OR COMPENSATION IS CHARGED,
'INCLUDING "TIE-IN" OR "BUNDLING" FEES CHARGED FOR
'OTHER PRODUCTS.

'==================================================================

'THIS PROGRAM REQUIRES wgnuplot.exe TO DISPLAY PLOTS.
'Gnuplot is a copyrighted freeware plotting program
'available at http://www.gnuplot.info/index.html.

'IT ALSO REQUIRES A VERSION OF THE Numerical Electromagnetics
'Code (NEC) in order to run the PBM benchmarks, Bowtie and Yagi
'antenna models.  If this file is not present, a runtime error
'occurs.  Remove the code that checks for the NEC EXE is there
'is no interest in the these functions.

'==================================================================
'CFO REFERENCES (author: Richard A. Formato unless otherwise noted)
'------------------------------------------------------------------

'"A Novel Methodology for Antenna Design and Optimization: Variable Zo," July 2011, http://arXiv.org/abs/1107.1437.

'"Central Force Optimization with Variable Initial Probes and Adaptive Decision Space," Applied Mathematics and Computation,
'Vol. 217, 2011, pp. 8866-8872, 2011.

'"New Techniques for Increasing Antenna Bandwidth with Impedance Loading," Progress In Electromagnetics Research B, vol. 29, pp. 269-288, 2011,
'http://www.jpier.org/pierb/pier.php?paper=11021904.

'"Parameter-Free Deterministic Global Search with Simplified Central Force Optimization," in Advanced Intelligent Computing Theories and
'Applications (ICIC2010), Lecture Notes in Computer Science (D.-S. Huang, Z. Zhao, V. Bevilacqua and J, C, Figueroa,Eds.), LNCS 6215, pp. 309-318,
'Springer-Verlag Berlin Heidelberg, 2010 [book chapter].

'"Antenna Benchmark Performance and Array Synthesis using Central Force Optimization," IET (U.K.) Microwaves, Antennas & Propagation vol. 4, no. 5,
```

```
'pp. 583-592, 2010.  (doi: 10.1049/iet-map.2009.0147). [with G. M. Qubati and N. I. Dib].

'"Improved CFO Algorithm for Antenna Optimization," Prog. Electromagnetics Research B, pp. 405-425, 2010
'http://www.jpier.org/pierb/pier.php?paper=09112309  (doi:10.2528/PIERB09112309).

'"Central Force Optimization and NEOs - First Cousins?," Journal of Multiple-Valued Logic and Soft Computing, vol. 16, pp. 547-565, 2010.

'"Convergence Analysis and Performance of the Extended Artificial Physics Optimization Algorithm," Applied Mathematics and Computation (in press)
'[with L. Xie and J. Zeng].

'"Central Force Optimization with Variable Initial Probes and Adaptive Decision Space," Applied Mathematics and Computation, July 2011.

'"High-Performance Indoor VHF-UHF Antennas: Technology Update Report," National Association of Broadcasters (NAB), FASTROAD (Flexible Advanced
'Services for Television and Radio On All Devices), Technology Advocacy Program, 15 May 2010 [with M. W. Cross and E. Merulla]
'http://www.nabfastroad.org/NABHighperformanceIndoorTVantennaRpt.pdf

'"Central Force Optimisation: a New Gradient-Like Metaheuristic for Multidimensional Search and Optimisation," Int. J. Bio-Inspired Computation,
'vol. 1, no. 4, pp. 217-238, 2009.  (doi: 10.1504/IJBIC.2009.024721).

'"Central Force Optimization: A New Deterministic Gradient-Like Optimization Metaheuristic," OPSEARCH, Journal of the Operations Research Society of India,
'vol. 46, no. 1, pp. 25-51, 2009.  (doi: 10.1007/s12597-009-0003-4).

'"Central Force Optimization: A New Computational Framework for Multidimensional Search and Optimization," in Nature Inspired Cooperative Strategies for
'Optimization (NICSO 2007), Studies in Computational Intelligence 129 (N. Krasnogor, G. Nicosia, M. Pavone, and D. Pelta, Eds.), vol. 129, Springer-Verlag,
'Heidelberg, 2008 [book chapter].

'"Central Force Optimization: A New Metaheuristic with Applications in Applied Electromagnetics," Prog. Electromagnetics Research, vol. PIER 77,
'pp. 425-491, 2007: http://www.jpier.org/pier/pier.php?paper=07082403 (doi:10.2528/PIER07082403).

'"Are Near Earth Objects the Key to Optimization Theory?," Dec., 2009.  http://arxiv.org/abs/0912.1394.

'"Pseudorandomness in Central Force Optimization," Jan., 2010.  http://arxiv.org/abs/1001.0317.

'"Central Force Optimization Applied to the PBM Suite of Antenna Benchmarks," Feb., 2010, http://arxiv.org/abs/1003.0221.

'"On the Utility of Directional Information for Repositioning Errant Probes in Central Force Optimization," May, 2010, http://arxiv.org/abs/1005.5490.

'"Issues in Antenna Optimization - A Monopole Case Study," Mar., 2011.  http://arXiv.org/abs/1103.5629.

'=============================================================================================================================================================
========

#COMPILE EXE

#DIM ALL

%USEMACROS = 1

#INCLUDE "Win32API.inc"

DEFEXT A-Z

'------ EQUATES -----

%IDC_FRAME1     = 101
%IDC_FRAME2     = 102

%IDC_Function_Number1  = 121
%IDC_Function_Number2  = 122
%IDC_Function_Number3  = 123
%IDC_Function_Number4  = 124
%IDC_Function_Number5  = 125
%IDC_Function_Number6  = 126
%IDC_Function_Number7  = 127
%IDC_Function_Number8  = 128
%IDC_Function_Number9  = 129
%IDC_Function_Number10 = 130
%IDC_Function_Number11 = 131
%IDC_Function_Number12 = 132
%IDC_Function_Number13 = 133
%IDC_Function_Number14 = 134
%IDC_Function_Number15 = 135
%IDC_Function_Number16 = 136
%IDC_Function_Number17 = 137
%IDC_Function_Number18 = 138
%IDC_Function_Number19 = 139
%IDC_Function_Number20 = 140
%IDC_Function_Number21 = 141
%IDC_Function_Number22 = 142
%IDC_Function_Number23 = 143
%IDC_Function_Number24 = 144
%IDC_Function_Number25 = 145
%IDC_Function_Number26 = 146
%IDC_Function_Number27 = 147
%IDC_Function_Number28 = 148
%IDC_Function_Number29 = 149
%IDC_Function_Number30 = 150
%IDC_Function_Number31 = 151
%IDC_Function_Number32 = 152
%IDC_Function_Number33 = 153
%IDC_Function_Number34 = 154
%IDC_Function_Number35 = 155
%IDC_Function_Number36 = 156
%IDC_Function_Number37 = 157
%IDC_Function_Number38 = 158
%IDC_Function_Number39 = 159
%IDC_Function_Number40 = 160
%IDC_Function_Number41 = 161
%IDC_Function_Number42 = 162
%IDC_Function_Number43 = 163
%IDC_Function_Number44 = 164
%IDC_Function_Number45 = 165
%IDC_Function_Number46 = 166
%IDC_Function_Number47 = 167
%IDC_Function_Number48 = 168
%IDC_Function_Number49 = 169
%IDC_Function_Number50 = 170

'------------------------------- GLOBAL CONSTANTS & SYMBOLS ---------------------------

GLOBAL c1, c2, c3, c4, c5, c6 AS EXT 'YAGI fitness function coefficients

GLOBAL Fit1$, Fit2$

GLOBAL NumRadPattAngles%

GLOBAL YagiSegmentLength$, YagiCoefficients$

GLOBAL NumYagiElements% 'added 05/12/2011 instead of passing in function call

GLOBAL YAGIsecsPerRun, YagiSegmentLengthWvln, YagiFitnessCoefficients() AS EXT

GLOBAL BowtieSegmentLength$, BowtieCoefficients$

GLOBAL NumBowtieElements% 'added 05/12/2011 instead of passing in function call

GLOBAL BOWTIEsecsPerRun, BowtieSegmentLengthWvln, BowtieFitnessCoefficients() AS EXT

GLOBAL XiOffset() AS EXT 'offset array for Rosenbrock F6 function
```

```
GLOBAL XiMin(), XiMax(), DiagLength, StartingXiMin(), StartingXiMax() AS EXT 'decision space boundaries, length of diagonal
GLOBAL Aij() AS EXT 'array for Shekel's Foxholes function
GLOBAL EulerConst, Pi, Pi2, Pi4, TwoPi, FourPi, e, Root2 AS EXT 'mathematical constants
GLOBAL Alphabet$, Digits$, RunID$  'upper/lower case alphabet, digits 0-9 & Run ID
GLOBAL Quote$, SpecialCharacters$  'quotation mark & special symbols
GLOBAL Mu0, Eps0, c, eta0 AS EXT   'E&M constants
GLOBAL Rad2Deg, Deg2Rad, Feet2Meters, Meters2Feet, Inches2Meters, Meters2Inches AS EXT 'conversion factors
GLOBAL Miles2Meters, Meters2Miles, NautMi2Meters, Meters2NautMi AS EXT                 'conversion factors
GLOBAL ScreenWidth&, ScreenHeight& 'screen width & height
GLOBAL xOffset&, yOffset&          'offsets for probe plot windows
GLOBAL FunctionNumber%
GLOBAL AddNoiseToPBM2$
'------------------------------ TEST FUNCTION DECLARATIONS --------------------------------
DECLARE FUNCTION F1(R(),Nd%,p%,j&)              'F1 (n-D)
DECLARE FUNCTION F2(R(),Nd%,p%,j&)              'F2(n-D)
DECLARE FUNCTION F3(R(),Nd%,p%,j&)              'F3 (n-D)
DECLARE FUNCTION F4(R(),Nd%,p%,j&)              'F4 (n-D)
DECLARE FUNCTION F5(R(),Nd%,p%,j&)              'F5 (n-D)
DECLARE FUNCTION F6(R(),Nd%,p%,j&)              'F6 (n-D)
DECLARE FUNCTION F7(R(),Nd%,p%,j&)              'F7 (n-D)
DECLARE FUNCTION F8(R(),Nd%,p%,j&)              'F8 (n-D)
DECLARE FUNCTION F9(R(),Nd%,p%,j&)              'F9 (n-D)
DECLARE FUNCTION F10(R(),Nd%,p%,j&)             'F10 (n-D)
DECLARE FUNCTION F11(R(),Nd%,p%,j&)             'F11 (n-D)
DECLARE FUNCTION F12(R(),Nd%,p%,j&)             'F12 (n-D)
DECLARE FUNCTION u(Xi,a,k,m)                    'Auxiliary function for F12 & F13
DECLARE FUNCTION F13(R(),Nd%,p%,j&)             'F13 (n-D)
DECLARE FUNCTION F14(R(),Nd%,p%,j&)             'F14 (n-D)
DECLARE FUNCTION F15(R(),Nd%,p%,j&)             'F15 (n-D)
DECLARE FUNCTION F16(R(),Nd%,p%,j&)             'F16 (n-D)
DECLARE FUNCTION F17(R(),Nd%,p%,j&)             'F17 (n-D)
DECLARE FUNCTION F18(R(),Nd%,p%,j&)             'F18 (n-D)
DECLARE FUNCTION F19(R(),Nd%,p%,j&)             'F19 (n-D)
DECLARE FUNCTION F20(R(),Nd%,p%,j&)             'F20 (n-D)
DECLARE FUNCTION F21(R(),Nd%,p%,j&)             'F21 (n-D)
DECLARE FUNCTION F22(R(),Nd%,p%,j&)             'F22 (n-D)
DECLARE FUNCTION F23(R(),Nd%,p%,j&)             'F23 (n-D)
DECLARE FUNCTION F24(R(),Nd%,p%,j&)             'F24 (n-D)
DECLARE FUNCTION F25(R(),Nd%,p%,j&)             'F25 (n-D)
DECLARE FUNCTION F26(R(),Nd%,p%,j&)             'F26 (n-D)
DECLARE FUNCTION F27(R(),Nd%,p%,j&)             'F27 (n-D)
DECLARE FUNCTION ParrottF4(R(),Nd%,p%,j&)       'Parrott F4 (1-D)
DECLARE FUNCTION SGO(R(),Nd%,p%,j&)             'SGO Function (2-D)
DECLARE FUNCTION GoldsteinPrice(R(),Nd%,p%,j&)  'Goldstein-Price Function (2-D)
DECLARE FUNCTION StepFunction(R(),Nd%,p%,j&)    'Step Function (n-D)
DECLARE FUNCTION Schwefel226(R(),Nd%,p%,j&)     'Schwefel Prob. 2.26 (n-D)
DECLARE FUNCTION Colville(R(),Nd%,p%,j&)        'Colville Function (4-D)
DECLARE FUNCTION Griewank(R(),Nd%,p%,j&)        'Griewank (n-D)
DECLARE FUNCTION Himmelblau(R(),Nd%,p%,j&)      'Himmelblau (2-D)
DECLARE FUNCTION Rosenbrock(R(),Nd%,p%,j&)      'Rosenbrock (n-D)
DECLARE FUNCTION Sphere(R(),Nd%,p%,j&)          'Sphere (n-D)
DECLARE FUNCTION HimmelblauNLO(R(),Nd%,p%,j&)   'Himmelblau NLO (5-D)
DECLARE FUNCTION Tripod(R(),Nd%,p%,j&)          'Tripod (2-D)
DECLARE FUNCTION Sign(X)                        'Auxiliary function for Tripod
DECLARE FUNCTION RosenbrockF6(R(),Nd%,p%,j&)    'Rosenbrock F6 (10-D)
DECLARE FUNCTION CompressionSpring(R(),Nd%,p%,j&)'Compression Spring (3-D)
DECLARE FUNCTION GearTrain(R(),Nd%,p%,j&)       'Gear Train (4-D)
DECLARE FUNCTION PBM_1(R(),Nd%,p%,j&)           'PBM Benchmark #1
DECLARE FUNCTION PBM_2(R(),Nd%,p%,j&)           'PBM Benchmark #2
DECLARE FUNCTION PBM_3(R(),Nd%,p%,j&)           'PBM Benchmark #3
DECLARE FUNCTION PBM_4(R(),Nd%,p%,j&)           'PBM Benchmark #4
DECLARE FUNCTION PBM_5(R(),Nd%,p%,j&)           'PBM Benchmark #5
DECLARE FUNCTION BOWTIE(R(),Nd%,p%,j&)          'FREE-SPACE RLC-LOADED BOWTIE
DECLARE FUNCTION YAGI_ARRAY(R(),Nd%,p%,j&)      'FREE-SPACE YAGI ARRAY
```

```
'----------------------------------- SUB DECLARATIONS -------------------------------------

DECLARE SUB SieveOfEratosthenes(N&&,Primes&&(),NumPrimes&&)

DECLARE SUB IPD_Halton(Np%,Nd%,Nt&,R(),Gamma)

DECLARE SUB ReplaceCommentCard(NECfile$)

DECLARE SUB CopyBestMatrices(Np%,Nd%,Nt&,R(),M(),Rbest(),Mbest())

DECLARE SUB CheckNECFiles(NECfileError$)

DECLARE SUB GetTestFunctionNumber(FunctionName$)

DECLARE SUB FillArrayAij

DECLARE SUB Plot3DbestProbeTrajectories(NumTrajectories%,M(),R(),Np%,Nd%,LastStep&,FunctionName$)

DECLARE SUB Plot2DbestProbeTrajectories(NumTrajectories%,M(),R(),Np%,Nd%,LastStep&,FunctionName$)

DECLARE SUB Plot2DindividualProbeTrajectories(NumTrajectories%,M(),R(),Np%,Nd%,LastStep&,FunctionName$)

DECLARE SUB Show2Dprobes(R(),Np%,Nt&,j&,Frep,BestFitness,BestProbeNumber%,BestTimeStep&,FunctionName$,RepositionFactor$,Gamma)

DECLARE SUB Show3Dprobes(R(),Np%,Nd%,Nt&,j&,Frep,BestFitness,BestProbeNumber%,BestTimeStep&,FunctionName$,RepositionFactor$,Gamma)

DECLARE SUB StatusWindow(FunctionName$,StatusWindowHandle???)

DECLARE SUB
PlotResults(FunctionName$,Nd%,Np%,BestFitnessOverall,BestNpNd%,BestGamma,Neval&&,Rbest(),Mbest(),BestProbeNumberOverall%,BestTimeStepOverall&,LastStepBestR
un&,Alpha,Beta)

DECLARE SUB
DisplayRunParameters(FunctionName$,Nd%,Np%,Nt&,G,DeltaT,Alpha,Beta,Frep,R(),A(),M(),PlaceInitialProbes$,InitialAcceleration$,RepositionFactor$,RunCFO$,Shri
nkDS$,CheckForEarlyTermination$)

DECLARE SUB GetBestFitness(M(),Np%,StepNumber&,BestFitness,BestProbeNumber%,BestTimeStep&)

DECLARE SUB
Tabulate1DprobeCoordinates(Max1DprobesPlotted%,Nd%,Np%,LastStep&,G,DeltaT,Alpha,Beta,Frep,R(),M(),PlaceInitialProbes$,InitialAcceleration$,RepositionFactor
$,FunctionName$,Gamma)

DECLARE SUB
GetPlotAnnotation(PlotAnnotation$,Nd%,Np%,Nt&,G,DeltaT,Alpha,Beta,Frep,M(),PlaceInitialProbes$,InitialAcceleration$,RepositionFactor$,FunctionName$,Gamma)

DECLARE SUB
ChangeRunParameters(NumProbesPerDimension%,Np%,Nd%,Nt&,G,Alpha,Beta,DeltaT,Frep,PlaceInitialProbes$,InitialAcceleration$,RepositionFactor$,FunctionName$)

DECLARE SUB CLEANUP

DECLARE SUB
Plot1DprobePositions(Max1DprobesPlotted%,Nd%,Np%,LastStep&,G,DeltaT,Alpha,Beta,Frep,R(),M(),PlaceInitialProbes$,InitialAcceleration$,RepositionFactor$,Func
tionName$,Gamma)

DECLARE SUB DisplayMmatrix(Np%,Nt&,M())

DECLARE SUB DisplayMbestMatrix(Np%,Nt&,Mbest())

DECLARE SUB DisplayMmatrixThisTimeStep(Np%,j&,M())

DECLARE SUB DisplayAmatrix(Np%,Nd%,Nt&,A())

DECLARE SUB DisplayAmatrixThisTimeStep(Np%,Nd%,j&,A())

DECLARE SUB DisplayRmatrix(Np%,Nd%,Nt&,R())

DECLARE SUB DisplayRmatrixThisTimeStep(Np%,Nd%,j&,R(),Gamma)

DECLARE SUB DisplayXiMinMax(Nd%,XiMin(),XiMax())

DECLARE SUB DisplayRunParameters2(FunctionName$,Nd%,Np%,Nt&,G,DeltaT,Alpha,Beta,Frep,PlaceInitialProbes$,InitialAcceleration$,RepositionFactor$)

DECLARE SUB
PlotBestProbeVsTimeStep(Nd%,Np%,LastStep&,G,DeltaT,Alpha,Beta,Frep,M(),PlaceInitialProbes$,InitialAcceleration$,RepositionFactor$,FunctionName$,Gamma)

DECLARE SUB
PlotBestFitnessEvolution(Nd%,Np%,LastStep&,G,DeltaT,Alpha,Beta,Frep,M(),PlaceInitialProbes$,InitialAcceleration$,RepositionFactor$,FunctionName$,Gamma)

DECLARE SUB
PlotAverageDistance(Nd%,Np%,LastStep&,G,DeltaT,Alpha,Beta,Frep,M(),PlaceInitialProbes$,InitialAcceleration$,RepositionFactor$,FunctionName$,R(),DiagLength,
Gamma)

DECLARE SUB Plot2Dfunction(FunctionName$,R())

DECLARE SUB Plot1Dfunction(FunctionName$,R())

DECLARE SUB
GetFunctionRunParameters(FunctionName$,Nd%,Np%,Nt&,G,DeltaT,Alpha,Beta,Frep,R(),A(),M(),DiagLength,PlaceInitialProbes$,InitialAcceleration$,RepositionFacto
r$)

DECLARE SUB InitialProbeDistribution(Np%,Nd%,Nt&,R(),PlaceInitialProbes$,Gamma)

DECLARE SUB RetrieveErrantProbes(Np%,Nd%,j&,R(),Frep)

DECLARE SUB Retrieveerrantprobes2(Np%,Nd%,j&,R(),A(),Frep)

DECLARE SUB CFO(FunctionName$,Nd%,Nt&,R(),A(),M(),DiagLength,BestFitnessOverall,BestNpNd%,BestGamma,Neval&&,Rbest(),_
          Mbest(),BestProbeNumberOverall%,BestTimeStepOverall&,LastStepBestRun&,Alpha,Beta) 'Self-contained CFO routine -> NO USER-SPECIFIED
PARAMETERS

DECLARE SUB IPD(Np%,Nd%,Nt&,R(),Gamma)

DECLARE SUB ResetDecisionSpaceBoundaries(Nd%)

DECLARE SUB ThreeDplot(PlotFileName$,PlotTitle$,Annotation$,xCoord$,yCoord$,zCoord$, _
                    XaxisLabel$,YaxisLabel$,ZaxisLabel$,zMin$,zMax$,GnuPlotEXE$,A$)

DECLARE SUB ThreeDplot2(PlotFileName$,PlotTitle$,Annotation$,xCoord$,yCoord$,zCoord$,XaxisLabel$,_
                     YaxisLabel$,ZaxisLabel$,zMin$,zMax$,GnuPlotEXE$,A$,xStart$,xStop$,yStart$,yStop$)

DECLARE SUB ThreeDplot3(PlotFileName$,PlotTitle$,Annotation$,xCoord$,yCoord$,zCoord$, _
                     XaxisLabel$,YaxisLabel$,ZaxisLabel$,zMin$,zMax$,GnuPlotEXE$,xStart$,xStop$,yStart$,yStop$)

DECLARE SUB TwoDplot(PlotFileName$,PlotTitle$,xCoord$,yCoord$,XaxisLabel$,YaxisLabel$, _
                  LogXaxis$,LogYaxis$,xMin$,xMax$,yMin$,yMax$,xTics$,yTics$,GnuPlotEXE$,LineType$,Annotation$)

DECLARE SUB TwoDplot2Curves(PlotFileName1$,PlotFileName2$,PlotTitle$,Annotation$,xCoord$,yCoord$,XaxisLabel$,YaxisLabel$, _
                         LogXaxis$,LogYaxis$,xMin$,xMax$,yMin$,yMax$,xTics$,yTics$,GnuPlotEXE$,LineSize)

DECLARE SUB TwoDplot3curves(NumCurves%,PlotFileName1$,PlotFileName2$,PlotFileName3$,PlotTitle$,Annotation$,xCoord$,yCoord$,XaxisLabel$,YaxisLabel$, _
                         LogXaxis$,LogYaxis$,xMin$,xMax$,yMin$,yMax$,xTics$,yTics$,GnuPlotEXE$)

DECLARE SUB CreateGNUplotINIfile(PlotWindowULC_X%,PlotWindowULC_Y%,PlotWindowWidth%,PlotWindowHeight%)

DECLARE SUB Delay(NumSecs)

DECLARE SUB MathematicalConstants

DECLARE SUB AlphabetAndDigits
```

```
DECLARE SUB SpecialSymbols

DECLARE SUB EMconstants

DECLARE SUB ConversionFactors

DECLARE SUB ShowConstants

DECLARE SUB
GetNECdata(NECoutputFile$,NumFreqs%,NumRadPattAngles%,Zo,FrequencyMHZ(),RadEfficiencyPCT(),MaxGainDBI(),MinGainDBI(),RinOhms(),XinOhms(),VSWR(),ForwardGain
DBi(),FileStatus$,FileID$)

DECLARE SUB
GetYagiNECdata(NECoutputFile$,NumFreqs%,NumRadPattAngles%,Zo,FrequencyMHZ(),RadEfficiencyPCT(),MaxGainDBI(),MinGainDBI(),RinOhms(),XinOhms(),VSWR(),Forward
GainDBi(),RearGainDBi(),FileStatus$,FileID$)

DECLARE SUB ComplexMultiply(ReA,ImA,ReB,ImB,ReC,ImC)

DECLARE SUB ComplexDivide(ReA,ImA,ReB,ImB,ReC,ImC)

'------ FUNCTION DECLARATIONS -------

DECLARE FUNCTION DecimalToVanDerCorputBaseN(N&&,Nbase%)

DECLARE FUNCTION ProbeWeight2(Nd%,Np%,R(),M(),p%,j&)

DECLARE FUNCTION ProbeWeight(Nd%,R(),p%,j&) 'computes a 'weighting factor' based on probe's position (greater weight if closer to decision space boundary)

DECLARE FUNCTION SlopeRatio(M(),Np%,StepNumber&)

DECLARE CALLBACK FUNCTION DlgProc

DECLARE FUNCTION HasFITNESSsaturated$(Nsteps&,j&,Np%,Nd%,M(),R(),DiagLength)

DECLARE FUNCTION HasDAVGsaturated$(Nsteps&,j&,Np%,Nd%,M(),R(),DiagLength)

DECLARE FUNCTION OscillationInDavg$(j&,Np%,Nd%,M(),R(),DiagLength)

DECLARE FUNCTION DavgThisStep(j&,Np%,Nd%,M(),R(),DiagLength)

DECLARE FUNCTION NoSpaces$(X,NumDigits%)

DECLARE FUNCTION FormatFP$(X,Ndigits%)

DECLARE FUNCTION FormatInteger$(M%)

DECLARE FUNCTION TerminateNowForSaturation$(j&,Nd%,Np%,Nt&,G,DeltaT,Alpha,Beta,R(),A(),M())

DECLARE FUNCTION MagVector(V(),N%)

DECLARE FUNCTION UniformDeviate(u&&)

DECLARE FUNCTION RandomNum(a,b)

DECLARE FUNCTION GaussianDeviate(Mu,Sigma)

DECLARE FUNCTION UnitStep(X)

DECLARE FUNCTION Fibonacci&&(N%)

DECLARE FUNCTION ObjectiveFunction(R(),Nd%,p%,j&,FunctionName$)

DECLARE FUNCTION UnitStep(X)

DECLARE FUNCTION FP2String2$(X!)

DECLARE FUNCTION FP2String$(X)

DECLARE FUNCTION Int2String$(X%)

DECLARE FUNCTION StandingWaveRatio(Zo,ReZ,ImZ)

'====================================================================================================

'----- MAIN PROGRAM ------

FUNCTION PBMAIN () AS LONG

'    ------ CFO Parameters -----

    LOCAL Nd%, Np%, Nt&

    LOCAL G, DeltaT, Alpha, Beta, Frep AS EXT

    LOCAL PlaceInitialProbes$, InitialAcceleration$, RepositionFactor$

    LOCAL R(), A(), M(), Rbest(), Mbest() AS EXT    'position, acceleration & fitness matrices

    LOCAL FunctionName$           'name of objective function

'    ------------------------- Miscellaneous Setup Parameters -----------------------

    LOCAL N%, O%, P%, i%, YN&, Neval&&, NevalTotal&&, BestNpNd%, NumTrajectories%, Max1DprobesPlotted%, LastStepBestRun&, Pass%

    LOCAL A$, RunCFO$, CFOversion$, NECfileError$, RunStart$, RunStop$

    LOCAL BestGamma, BestFitnessThisRun, BestFitnessOverall, StartTime, StopTime, OptimumFitness AS EXT

    LOCAL BestProbeNum%, BestTimeStep&, BestProbeNumberOverall%, BestTimeStepOverall&, StatusWindowHandle???

'    -------------------- Global Constants ---------------------

    REDIM Aij(1 TO 2, 1 TO 25) '(GLOBAL array for Shekel's Foxholes function)

    CALL FillArrayAij

    CALL MathematicalConstants 'NOTE: Calling order is important!!

    CALL AlphabetAndDigits 'NOTE: GLOBAL VARIABLE RunID$ IS SET HERE

    CALL SpecialSymbols

    CALL EMconstants

    CALL ConversionFactors        ': CALL ShowConstants 'to verify constants have been set

    BOWTIEsecsPerRun = 3##

'    --------------------------- General Setup ----------------------------

    CFOversion$ = "CFO Ver. 06-25-2011"

    RANDOMIZE TIMER  'seed random number generator with program start time

    DESKTOP GET SIZE TO ScreenWidth&, ScreenHeight&  'get screen size (global variables)

    IF DIR$("wgnuplot.exe") = "" THEN
        MSGBOX("WARNING!  'wgnuplot.exe' not found.  Run terminated.") : EXIT FUNCTION
```

```
    END IF

    IF DIR$("Fitness")    <> "" THEN KILL "Fitness"
    IF DIR$("Davg")       <> "" THEN KILL "Davg"
    IF DIR$("Best Probe") <> "" THEN KILL "Best Probe"

'   ----------------------------------------------------------------------- CFO RUN PARAMETERS -------------------------------------------------------------------------

    CALL GetTestFunctionNumber(FunctionName$)' : exit function 'DEBUG

    CALL
GetFunctionRunParameters(FunctionName$,Nd%,Np%,Nt&,G,DeltaT,Alpha,Beta,Frep,R(),A(),M(),DiagLength,PlaceInitialProbes$,InitialAcceleration$,RepositionFactor$) 'NOTE: Parameters returned but not used in this version!!

    REDIM R(1 TO Np%, 1 TO Nd%, 0 TO Nt&), A(1 TO Np%, 1 TO Nd%, 0 TO Nt&), M(1 TO Np%, 0 TO Nt&) 'position, acceleration & fitness matrices

    REDIM Rbest(1 TO Np%, 1 TO Nd%, 0 TO Nt&), Mbest(1 TO Np%, 0 TO Nt&) 'overall best position & fitness matrices

'   -------- PLOT 1D and 2D FUNCTIONS ON-SCREEN FOR VISUALIZATION --------

    IF Nd% = 2 AND INSTR(FunctionName$,"PBM_") > 0 THEN

        CALL CheckNECFiles(NECfileError$)

        IF NECfileError$ = "YES" THEN
            EXIT FUNCTION
        ELSE
            MSGBOX("Begin computing plot of function "+FunctionName$+"?  May take a while - be patient...")
        END IF

    END IF

    SELECT CASE Nd%
        CASE 1 : CALL Plot1Dfunction(FunctionName$,R()) : REDIM R(1 TO Np%, 1 TO Nd%, 0 TO Nt&) 'erases coordinate data in R()used to plot function
        CASE 2 : CALL Plot2Dfunction(FunctionName$,R()) : REDIM R(1 TO Np%, 1 TO Nd%, 0 TO Nt&) 'ditto
    END SELECT

'   -----------------------------------------------------------------------------  RUN CFO ----------------------------------------------------------------------------------

    YN& = MSGBOX("RUN CFO ON FUNCTION " + FunctionName$ + "?"+CHR$(13)+CHR$(13)+"Get some coffee & sit back...",%MB_YESNO,"CONFIRM RUN") : IF YN& = %IDYES
THEN RunCFO$ = "YES"

    IF RunCFO$ = "YES" THEN

        StartTime = TIMER : RunStart$ = "Started at "+TIME$+", "+DATE$


'       ------------- BOWTIE -----------
        IF FunctionName$ = "BOWTIE" THEN
            N% = FREEFILE : OPEN "RunTime_BOWTIE.DAT" FOR OUTPUT AS #N% : PRINT #N%, RunStart$ : CLOSE #N%

            IF DIR$("BOWTIE_CFO.CFG") = "" THEN
                N% = FREEFILE
                OPEN "BOWTIE_CFO.CFG" FOR OUTPUT AS #N%
                PRINT #N%, "150" : PRINT #N%, "11" : PRINT #N%, "2" : PRINT #N%, "4" 'CFO parameters Nt&, NumGammas%, Min/MaxProbesPerDimension%,
respectively
                CLOSE #N%
            END IF

        END IF

'       ------------- YAGI -----------
        IF FunctionName$ = "YAGI" THEN
            N% = FREEFILE : OPEN "RunTime_YAGI.DAT" FOR OUTPUT AS #N% : PRINT #N%, RunStart$ : CLOSE #N%

            IF DIR$("YAGI_CFO.CFG") = "" THEN
                N% = FREEFILE
                OPEN "YAGI_CFO.CFG" FOR OUTPUT AS #N%
                PRINT #N%, "150" : PRINT #N%, "11" : PRINT #N%, "2" : PRINT #N%, "4" 'CFO parameters Nt&, NumGammas%, Min/MaxProbesPerDimension%,
respectively
                CLOSE #N%
            END IF

        END IF

        CALL
CFO(FunctionName$,Nd%,Nt&,R(),A(),M(),DiagLength,BestFitnessOverall,BestNpNd%,BestGamma,Neval&&,Rbest(),Mbest(),BestProbeNumberOverall%,BestTimeStepOverall
&,LastStepBestRun&,Alpha,Beta)

        StopTime = TIMER : RunStop$ = "Ended at "+TIME$+", "+DATE$

        Np% = BestNpNd%*Nd%

'       -------------- Add Best Run Results to CFO Setup Parameter File ---------------

        N% = FREEFILE
        OPEN "CFO_"+RunID$+".PAR" FOR APPEND AS #N%
            PRINT #N%,""
            PRINT #N%,"Best Run Data:"
            PRINT #N%,"--------------"
            PRINT #N%,"Fitness    = "+STR$(ROUND(BestFitnessOverall,8))
            PRINT #N%,"Np/Nd      = "+STR$(BestNpNd%)
            PRINT #N%,"Nd         = "+STR$(Nd%)
            PRINT #N%,"Np         = "+STR$(Np%)
            PRINT #N%,"Gamma      = "+STR$(BestGamma)
            PRINT #N%,"Probe #    = "+STR$(BestProbeNumberOverall%)
            PRINT #N%,"#Time Step = "+STR$(BestTimeStepOverall&)
            PRINT #N%,"#Last Step = "+STR$(LastStepBestRun&)
        CLOSE #N%

        CALL
PlotResults(FunctionName$,Nd%,Np%,BestFitnessOverall,BestNpNd%,BestGamma,Neval&&,Rbest(),Mbest(),BestProbeNumberOverall%,BestTimeStepOverall&,LastStepBestR
un&,Alpha,Beta)

'       ------------- BOWTIE -----------
        IF FunctionName$ = "BOWTIE" THEN

            N% = FREEFILE
            OPEN "RunTime_BOWTIE.DAT" FOR APPEND AS #N%
                PRINT #N%,RunStop$
                PRINT #N%,"Runtime = "+STR$(ROUND((StopTime-StartTime)/3600##,2))+" hrs"
                PRINT #N%,"   Note: runtime will be incorrect if start/stop times transition midnight..."
                PRINT #N%,FunctionName$+", Total Function Evaluations = "+STR$(Neval&&)
                PRINT #N%,"Avg Time/Run = " +STR$(ROUND((StopTime-StartTime)/Neval&&,6))
            CLOSE #N%

            IF DIR$("BestBOWTIE.NEC") <> "" THEN KILL "BestBOWTIE.NEC"
            IF DIR$("BestBOWTIE.OUT") <> "" THEN KILL "BestBOWTIE.OUT"
            OptimumFitness  = BOWTIE(Rbest(),Nd%,BestProbeNumberOverall%,BestTimeStepOverall&)
            CALL ReplaceCommentCard("BOWTIE.NEC")
            NAME "BOWTIE.NEC" AS "BestBOWTIE.NEC"
            CALL ReplaceCommentCard("BOWTIE.OUT")
            NAME "BOWTIE.OUT" AS "BestBOWTIE.OUT"

            O% = FREEFILE
            OPEN "BestBOWTIE.NEC" FOR APPEND AS #O%
                PRINT #O%,"RLC-LOADED BOWTIE"
```

```
                PRINT #O%,"" : PRINT #O%,"OPTIMUM FITNESS = "+STR$(ROUND(OptimumFitness,3))+" @ Fo = 299.8 MHz, "+DATE$+", "+TIME$
                PRINT #O%,"" : PRINT #O%,USING$("Best Gamma: ##.###     BestNp/Nd: ###     Nt: #####     Neval: #######     LastStep:
#####",BestGamma,BestNpNd,Nt&,Neval&&,LastStepBestRun&)
                PRINT #O%, ""
                PRINT #O%,RunStart$
                PRINT #O%,RunStop$
            CLOSE #O%

            P% = FREEFILE
            OPEN "BestBOWTIE.OUT" FOR APPEND AS #P%
                PRINT #P%,"RLC-LOADED BOWTIE"
                PRINT #P%,"" : PRINT #P%,"OPTIMUM FITNESS = "+STR$(ROUND(OptimumFitness,3))+" @ Fo = 299.8 MHz, "+DATE$+", "+TIME$
                PRINT #P%,"" : PRINT #P%,USING$("Best Gamma: ##.###     BestNp/Nd: ###     Nt: #####     Neval: #######     LastStep:
#####",BestGamma,BestNpNd,Nt&,Neval&&,LastStepBestRun&)
                PRINT #P%, ""
                PRINT #P%,RunStart$
                PRINT #P%,RunStop$
            CLOSE #P%

            SHELL "Read_NEC_Output_File_BOWTIE.exe"

        END IF 'FunctionName$ = "BOWTIE"

'       ---------- YAGI ARRAY --------
        IF FunctionName$ = "YAGI" THEN

            N% = FREEFILE
            OPEN "RunTime_YAGI.DAT" FOR APPEND AS #N%
                PRINT #N%,RunStop$
                PRINT #N%,"Runtime = "+STR$(ROUND((StopTime-StartTime)/3600##,2))+" hrs"
                PRINT #N%,"   Note: runtime will be incorrect if start/stop times transition midnight..."
                PRINT #N%,FunctionName$+", Total Function Evaluations = "+STR$(Neval&&)
                PRINT #N%,"Avg Time/Run = " +STR$(ROUND((StopTime-StartTime)/Neval&&,6))
            CLOSE #N%

            IF DIR$("BestYAGI.NEC") <> "" THEN KILL "BestYAGI.NEC"
            IF DIR$("BestYAGI.OUT") <> "" THEN KILL "BestYAGI.OUT"
            OptimumFitness  = YAGI_ARRAY(Rbest(),Nd%,BestProbeNumberOverall%,BestTimeStepOverall&)
            CALL ReplaceCommentCard("YAGI.NEC")
            NAME "YAGI.NEC" AS "BestYAGI.NEC"
            CALL ReplaceCommentCard("YAGI.OUT")
            NAME "YAGI.OUT" AS "BestYAGI.OUT"

            O% = FREEFILE
            OPEN "BestYAGI.NEC" FOR APPEND AS #O%
                PRINT #O%,"YAGI ARRAY"
                PRINT #O%,"" : PRINT #O%,"OPTIMUM FITNESS = "+STR$(ROUND(OptimumFitness,3))+" @ Fo = 299.8 MHz, "+DATE$+", "+TIME$
                PRINT #O%,"" : PRINT #O%,USING$("Best Gamma: ##.###     BestNp/Nd: ###     Nt: #####     Neval: #######     LastStep:
#####",BestGamma,BestNpNd,Nt&,Neval&&,LastStepBestRun&)
                PRINT #O%, ""
                PRINT #O%,RunStart$
                PRINT #O%,RunStop$
            CLOSE #O%

            P% = FREEFILE
            OPEN "BestYAGI.OUT" FOR APPEND AS #P%
                PRINT #P%,"YAGI ARRAY"
                PRINT #P%,"" : PRINT #P%,"OPTIMUM FITNESS = "+STR$(ROUND(OptimumFitness,3))+" @ Fo = 299.8 MHz, "+DATE$+", "+TIME$
                PRINT #P%,"" : PRINT #P%,USING$("Best Gamma: ##.###     BestNp/Nd: ###     Nt: #####     Neval: #######     LastStep:
#####",BestGamma,BestNpNd,Nt&,Neval&&,LastStepBestRun&)
                PRINT #P%, ""
                PRINT #P%,RunStart$
                PRINT #P%,RunStop$
            CLOSE #P%

            SHELL "Read_NEC_Output_File_YAGI_ARRAY.exe"

        END IF 'FunctionName$ = "YAGI"

        MSGBOX(FunctionName$+CHR$(13)+"Total Function Evaluations = "+STR$(Neval&&)+CHR$(13)+"Runtime = "+STR$(ROUND((StopTime-StartTime)/3600##,2))+_
                               " hrs, Avg Time/Run = " +STR$(ROUND((StopTime-StartTime)/Neval&&,6)))

        NAME "Fitness" AS "Fitness"+RunID$+".DAT" : NAME "Davg" AS "Davg"+RunID$+".DAT" : NAME "Best Probe" AS "BestProbe"+RunID$+".DAT" 'append Run ID and
'DAT' to data file names to identify file type

'       ------------ Cleanup Temp Files --------------
        IF DIR$("cmd2d.gp") <> "" THEN KILL "cmd2d.gp"
        IF DIR$("cmd3d.gp") <> "" THEN KILL "cmd3d.gp"
        IF DIR$("NECtemp")  <> "" THEN KILL "NECtemp"

    END IF 'RunCFO$ = "YES"

ExitPBMAIN:

END FUNCTION 'PBMAIN()

'======================================================================================  CFO SUBROUTINE
=========================================================================================

SUB
CFO(FunctionName$,Nd%,Nt&,R(),A(),M(),DiagLength,BestFitnessOverall,BestNpNd%,BestGamma,Neval&&,Rbest(),Mbest(),BestProbeNumberOverall%,BestTimeStepOverall
&,LastStepBestRun&,Alpha,Beta)

LOCAL p%, i%, j& 'Standard Indices: Probe #, Coordinate #, Time Step #

LOCAL Np% 'Number of Probes

LOCAL MinProbesPerDimension%, MaxProbesPerDimension% 'Min/Max # probes per dimension (even #'s)

LOCAL k%, L% 'Dummy summation indices

LOCAL M% 'file number

LOCAL NumProbesPerDimension%, GammaNumber%, NumGammas% 'Probes/dimension on each probe line; gamm point #; # gamma points

LOCAL SumSQ, Denom, Numerator, Gamma, Frep, DeltaFrep, FrepInit, MinFrep AS EXT

LOCAL BestProbeNumber%, BestTimeStep&, LastStep&, BestProbeNumberThisRun%, BestTimeStepThisRun&

LOCAL BestFitness, BestFitnessThisRun, eta AS EXT

LOCAL FitnessSaturation$

'<<<<<<<<<<<<<< NOTE: THIS IS A 'PARAMETER FREE' CFO IMPLEMENTATION.  SEE THE ICIC2010 REFERENCE ABOVE FOR DETAILS. >>>>>>>>>>>>>>

'---------------- Initial Parameter Values ------------------

NumGammas% = 3'11 'default value for functions other than BOWTIE

Alpha = 1## : Beta = 1## 'CFO EXPONENTS Alpha & Beta

Nt& = 1000 'set to a large value expecting early termination

IF FunctionName$ = "F7" THEN Nt& = 100 'to reduce runtime because this function contains random noise

Neval&& = 0

FrepInit  = 0.5##
```

```
DeltaFrep = 0.1##

MinFrep   = 0.05##

SELECT CASE Nd% 'set Np%/Nd% based on Nd% to avoid excessive runtimes

    CASE   1 TO   6 : MinProbesPerDimension% = 2 : MaxProbesPerDimension% = 14
    CASE   7 TO  10 : MinProbesPerDimension% = 2 : MaxProbesPerDimension% = 12
    CASE  11 TO  15 : MinProbesPerDimension% = 2 : MaxProbesPerDimension% = 10
    CASE  16 TO  20 : MinProbesPerDimension% = 2 : MaxProbesPerDimension% =  8
    CASE  21 TO  30 : MinProbesPerDimension% = 2 : MaxProbesPerDimension% =  6
    CASE ELSE       : MinProbesPerDimension% = 2 : MaxProbesPerDimension% =  4

END SELECT

'------------ BOWTIE -----------
IF FunctionName$ = "BOWTIE" THEN
    M% = FREEFILE
    OPEN "BOWTIE_CFO.CFG" FOR INPUT AS #M%
        INPUT #M%, Nt& : INPUT #M%, NumGammas% : INPUT #M%, MinProbesPerDimension% : INPUT #M%, MaxProbesPerDimension%
    CLOSE #M%
    IF Nt&                    < 5 OR Nt&                    > 500 THEN Nt& = 150
    IF NumGammas%             < 2 OR NumGammas%             > 31  THEN NumGammas% = 11
    IF MinProbesPerDimension% < 2 OR MinProbesPerDimension% > 10  THEN MinProbesPerDimension% = 2
    IF MaxProbesPerDimension% < 2 OR MaxProbesPerDimension% > 10  THEN MaxProbesPerDimension% = 6
    IF MinProbesPerDimension% > MinProbesPerDimension% THEN SWAP MinProbesPerDimension%,MinProbesPerDimension%
END IF

'--------- YAGI ARRAY --------
IF FunctionName$ = "YAGI" THEN
    M% = FREEFILE
    OPEN "YAGI_CFO.CFG" FOR INPUT AS #M%
        INPUT #M%, Nt& : INPUT #M%, NumGammas% : INPUT #M%, MinProbesPerDimension% : INPUT #M%, MaxProbesPerDimension%
    CLOSE #M%
    IF Nt&                    < 5 OR Nt&                    > 500 THEN Nt& = 150
    IF NumGammas%             < 2 OR NumGammas%             > 31  THEN NumGammas% = 11
    IF MinProbesPerDimension% < 2 OR MinProbesPerDimension% > 10  THEN MinProbesPerDimension% = 2
    IF MaxProbesPerDimension% < 2 OR MaxProbesPerDimension% > 10  THEN MaxProbesPerDimension% = 6
    IF MinProbesPerDimension% > MinProbesPerDimension% THEN SWAP MinProbesPerDimension%,MinProbesPerDimension%
END IF

'--------------

LastStep& = Nt&

'IF FunctionName$ = "BOWTIE" THEN
'    MaxProbesPerDimension% =  6 'to reduce runtime for LOADED BOWTIE
'    MSGBOX("Approximate runtime: "+STR$(ROUND(BOWTIEsecsPerRun*NumGammas*Nd%*MaxProbesPerDimension%*Nt&/40##,1))+" minutes")
'END IF

'   ------------------------- Save Setup Parameters for This Run -----------------------------

    M% = FREEFILE

    OPEN "CFO_"+RunID$+".PAR" FOR OUTPUT AS #M%
        PRINT #M%,"CFO RUN PARAMETERS, Run ID: "+RunID$
        PRINT #M%,"----------------------------"+STRING$(LEN(RunID$),"-")
        PRINT #M%,"Alpha           = "+STR$(Alpha)
        PRINT #M%,"Beta            = "+STR$(Beta)
        PRINT #M%,"Nt              = "+STR$(Nt&)
        PRINT #M%,"Initial Frep    = "+STR$(FrepInit)
        PRINT #M%,"Delta Frep      = "+STR$(DeltaFrep)
        PRINT #M%,"Min Frep        = "+STR$(MinFrep)
        PRINT #M%,"# Gammas        = "+STR$(NumGammas%)
        PRINT #M%,"Nt              = "+STR$(Nt&)
        PRINT #M%,"Min Probes/Dim  = "+STR$(MinProbesPerDimension%)
        PRINT #M%,"Max Probes/Dim  = "+STR$(MaxProbesPerDimension%)
    CLOSE #M%

'   ------------------------- Np/Nd LOOP ---------------------

BestFitnessOverall = -1E4200 'very large negative number...

FOR NumProbesPerDimension% = MinProbesPerDimension% TO MaxProbesPerDimension% STEP 2

'FOR NumProbesPerDimension% = 4 TO MaxProbesPerDimension% STEP 2 'original code

Np% = NumProbesPerDimension%*Nd%

'   ------------------------- GAMMA LOOP ---------------------

FOR GammaNumber% = 1 TO NumGammas%

Gamma = (GammaNumber%-1)/(NumGammas%-1)

REDIM R(1 TO Np%, 1 TO Nd%, 0 TO Nt&), A(1 TO Np%, 1 TO Nd%, 0 TO Nt&), M(1 TO Np%, 0 TO Nt&) 're-initializes Position Vector/Acceleration/Fitness matrices
to zero

'STEP (A1) ------------ Compute Initial Probe Distribution (Step 0)----------------

    CALL IPD(Np%,Nd%,Nt&,R(),Gamma) 'Probe Line IPD intersecting on diagonal at a point determined by Gamma

'STEP (A2) ------------ Compute Initial Fitness Matrix (Step 0) ------------------------

    FOR p% = 1 TO Np% : M(p%,0) = ObjectiveFunction(R(),Nd%,p%,0,FunctionName$) : INCR Neval&& : NEXT p%

'STEP (A3) ------------ Set Initial Probe Accelerations to ZERO (Step 0)----------------

    FOR p% = 1 TO Np% : FOR i% = 1 TO Nd% : A(p%,i%,0) = 0## : NEXT i% : NEXT p%

'STEP (A4) ------------ Initialize Frep ----------------

    Frep = FrepInit '0.5##

'   ===================================== LOOP ON TIME STEPS STARTING AT STEP #1 ==========================================

    BestFitnessThisRun = M(1,0)

    FOR j& = 1 TO Nt&

'STEP (B) ----------- Compute Probe Position Vectors for this Time Step --------

        FOR p% = 1 TO Np% : FOR i% = 1 TO Nd% : R(p%,i%,j&) = R(p%,i%,j&-1) + A(p%,i%,j&-1) : NEXT i% : NEXT p% 'note: factor of 1/2 combined with G=2 to
produce unity coefficient

'STEP (C) ----------- Retrieve Errant Probes ---------------

       CALL RetrieveErrantProbes(Np%,Nd%,j&,R(),Frep)

'         CALL RetrieveErrantprobes2(Np%,Nd%,j&,R(),A(),Frep) 'added 04-01-10

'STEP (D) ----------- Compute Fitness Matrix for Current Probe Distribution ---------

       FOR p% = 1 TO Np% : M(p%,j&) = ObjectiveFunction(R(),Nd%,p%,j&,FunctionName$) : INCR Neval&& : NEXT p%

'STEP (E) ----------- Compute Accelerations Based on Current Probe Distribution & Fitnesses ----------------

        FOR p% = 1 TO Np%
```

```
            FOR i% = 1 TO Nd%
                A(p%,i%,j&) = 0
                FOR k% = 1 TO Np%
                    IF k% <> p% THEN
                        SumSQ = 0## : FOR L% = 1 TO Nd%  : SumSQ = SumSQ + (R(k%,L%,j&)-R(p%,L%,j&))^2 : NEXT L% 'dummy index
                        IF SumSQ <> 0## THEN 'to avoid zero denominator (added 03-20-10) [NOTE THAT POWER BASIC DOES NOT FAULT ON DIVIDE-BY-ZERO...]
                            Denom = SQR(SumSQ) : Numerator = UnitStep((M(k%,j&)-M(p%,j&)))*(M(k%,j&)-M(p%,j&))
                            A(p%,i%,j&) = A(p%,i%,j&) + (R(k%,i%,j&)-R(p%,i%,j&))*Numerator^Alpha/Denom^Beta 'ORIGINAL VERSION WITH VARIABLE Alpha & Beta
                        END IF 'added 03-20-10
                    END IF
                NEXT k% 'dummy index
            NEXT i% 'coord (dimension) #
        NEXT p% 'probe #
'   --------- Get Best Fitness & Corresponding Probe # and Time Step ---------
    CALL GetBestFitness(M(),Np%,j&,BestFitness,BestProbeNumber%,BestTimeStep&)
    IF BestFitness >= BestFitnessThisRun THEN
        BestFitnessThisRun = BestFitness : BestProbeNumberThisRun% = BestProbeNumber% : BestTimeStepThisRun& = BestTimeStep&
    END IF
'   ----- Increment Frep -----
    Frep = Frep + DeltaFrep
    IF Frep > 1## THEN Frep = MinFrep 'keep Frep in range [0.05,1]
'   ---------- Starting at Step #20 Shrink Decision Space Around Best Probe Every 20th Step -----------
'   IF j& MOD 10 = 0 AND j& >= 20 THEN
    IF j& MOD 20 = 0 AND j& >= 20 THEN
        FOR i% = 1 TO Nd% : XiMin(i%) = XiMin(i%)+(R(BestProbeNumber%,i%,BestTimeStep&)-XiMin(i%))/2## : XiMax(i%) = XiMax(i%)-(XiMax(i%)-
R(BestProbeNumber%,i%,BestTimeStep&))/2## : NEXT i% 'shrink DS by 0.5
       CALL RetrieveErrantProbes(Np%,Nd%,j&,R(),Frep) 'TO RETRIEVE PROBES LYING OUTSIDE SHRUNKEN DS 'ADDED 02-07-2010
'        CALL RetrieveErrantprobes2(Np%,Nd%,j&,R(),A(),Frep) 'added 04-01-10
    END IF
'STEP (F) ----------- Check for Early Run Termination ---------
    IF HasFITNESSsaturated$(25,j&,Np%,Nd%,M(),R(),DiagLength) = "YES" THEN
        LastStep& = j&
        EXIT FOR
    END IF
    NEXT j& 'END TIME STEP LOOP
'---------------- Best Overall Fitness & Corresponding Parameters -------------------
IF BestFitnessThisRun >= BestFitnessOverall THEN
    BestFitnessOverall = BestFitnessThisRun : BestProbeNumberOverall% = BestProbeNumberThisRun% : BestTimeStepOverall& = BestTimeStepThisRun&
    BestNpNd% = NumProbesPerDimension%      : BestGamma = Gamma : LastStepBestRun& = LastStep&
    CALL CopyBestMatrices(Np%,Nd%,Nt&,R(),M(),Rbest(),Mbest())
END IF
'STEP (G) ----- Reset Decision Space Boundaries to Initial Values -----
CALL ResetDecisionSpaceBoundaries(Nd%)
NEXT GammaNumber% 'END GAMMA LOOP
NEXT NumProbesPerDimension% 'END Np/Nd LOOP
END SUB 'CFO()
'=====================================================================================================================================================
========
SUB IPD_Halton(Np%,Nd%,Nt&,R(),Gamma) 'added 06-29-2011
LOCAL DeltaXi, DelX1, DelX2, Di AS EXT
LOCAL NumProbesPerDimension%, p%, i%, k%, NumX1points%, NumX2points%, x1pointNum%, x2pointNum%
            IF Nd% > 1 THEN
                NumProbesPerDimension% = Np%\Nd% 'even #
            ELSE
                NumProbesPerDimension% = Np%
            END IF
            FOR i% = 1 TO Nd%
                FOR p% = 1 TO Np%
                    R(p%,i%,0) = XiMin(i%) + Gamma*(XiMax(i%)-XiMin(i%))
                NEXT Np%
            NEXT i%
            FOR i% = 1 TO Nd% 'place probes probe line-by-probe line (i% is dimension number)
                DeltaXi = (XiMax(i%)-XiMin(i%))/(NumProbesPerDimension%-1)
                FOR k% = 1 TO NumProbesPerDimension%
                    p% = k% + NumProbesPerDimension%*(i%-1) 'probe #
                    R(p%,i%,0) = XiMin(i%) + (k%-1)*DeltaXi
```

```
            NEXT k%

        NEXT i%

END SUB 'IPD_Halton()

'-------------------

FUNCTION DecimalToVanDerCorputBaseN(N&&,Nbase%)

'Returns the decimal value of the VDC base Nbase% 'radical inverse function' to create a Low Discrepancy sequence of decimal values.
'Maps integer argument into a quasirandom real number on the interval [0,1).

'Refs: (1) "Improved Particle Swarm Optimization with Low-Discrepancy Sequences,"
'           Pant, Thangaraj, Grosan & Abraham, 2008 IEEE Congress on Evolutionary Computation (CEC2008), p. 3011.

'      (2) "Quasirandom Ramblings," Computing Science column in American Scientist Magazine, July-Augist 2011, p. 282.
'          (www.american scientist.org)

'CHECK VALUES FOR BASE 2
'-----------------------
'Integer(N&&)  Base 2 Representation   Digits Reversed    VDC Decimal Value
'------------  ---------------------   ---------------    -----------------
'      0                    0                  0             0.0
'      2                   10                 01             0.25
'      3                   11                 11             0.75
'      4                  100                001             0.125
'      6                  110                011             0.375
'      7                  111                111             0.875
'     15                 1111               1111             0.9375
'    120              1111000            0001111             0.1171875
'    121              1111001            1001111             0.6171875
'    532           1000010100         0010100001             0.1572265625

'CHECK VALUES FOR BASE 3
'-----------------------
'Integer(N&&)  Base 3 Representation   Digits Reversed    VDC Decimal Value
'------------  ---------------------   ---------------    -----------------
'      0                    0                  0             0.0
'      2                    2                  2             0.666...
'      3                   10                 01             0.111...
'      7                   21                 12             0.555...
'    120                11110              01111             0.164609053497942
'    532               201201             102102             0.422496570544719

'CHECK VALUES FOR BASE 5
'-----------------------
'Integer(N&&)  Base 5 Representation   Digits Reversed    VDC Decimal Value
'------------  ---------------------   ---------------    -----------------
'      0                    0                  0             0.0
'      1                    1                  1             0.2
'      2                    2                  2             0.4
'      3                    3                  3             0.6
'      4                    4                  4             0.8
'      5                   10                 01             0.04
'      7                   12                 21             0.44
'     17                   32                 23             0.52
'    121                  441                144             0.392
'    532                 4112               2114             0.4544

LOCAL S AS EXT, R AS EXT, I&&, D&&

    S = 0##

    IF N&& > 2^63-2 THEN
        MSGBOX("WARNING! VDC arg > 2^63-2.  Zero returned.") : GOTO ExitVDC
    END IF

    I&& = N&& : R = 1##/Nbase%

    DO WHILE I&& <> 0 : D&& = I&& MOD Nbase% : S = S + D&&*R : I&& = (I&&-D&&)/Nbase% : R = R/Nbase% : LOOP

ExitVDC:

DecimalToVanDerCorputBaseN = S

END FUNCTION

'-----------

SUB SieveOfEratosthenes(N&&,Primes&&(),NumPrimes&&)

'Returns primes between 2 and N&& and their number.

'NOTE: There are 303 primes between 2 and 2000. Accuracy of this routine has been checked against table of the first 1000 primes (# primes up to N&&=7919).

'Ref: Caldwell, C., "The Prime Pages," Univ. Tennessee at Martin, http://primes.utm.edu

LOCAL i&&, iStart&&, p&&, j&&, TempArray&&()

    REDIM TempArray&&(1 TO N&&)

    TempArray&&(1) = 0 : FOR i&& = 2 TO N&& : TempArray&&(i&&) = 1 : NEXT i&&

    p&& = 2 : WHILE p&&^2 =< N&& : j&& = p&&^2 : WHILE j&& =< N&& : TempArray&&(j&&) = 0 : j&& = j&& + p&& : WEND : p&& = p&& + 1 : WEND 'inserts 0 or 1
depending on whether or not # is a prime

    NumPrimes&& = 0

    FOR i&& = 1 TO N&&  'converts the 1's into their corresponding prime numbers & retrieves # primes between 2 and N&&
        IF TempArray&&(i&&) = 1 THEN INCR NumPrimes&&
        TempArray&&(i&&) = i&&*TempArray&&(i&&)
    NEXT i&&

    REDIM Primes&&(1 TO NumPrimes&&)

    iStart&& = 1
    FOR j&& = 1 TO NumPrimes&&
        FOR i&& = iStart&& TO N&&
            IF TempArray&&(i&&) <> 0 THEN
                Primes&&(j&&) = TempArray&&(i&&) : iStart&& = i&&+1 : EXIT FOR
            END IF
        NEXT i&&
    NEXT j&&

END SUB 'SieveOfEratosthenes()

'-----------------------------

SUB IPD(Np%,Nd%,Nt&,R(),Gamma)

LOCAL DeltaXi, DelX1, DelX2, Di AS EXT

LOCAL NumProbesPerDimension%, p%, i%, k%, NumX1points%, NumX2points%, x1pointNum%, x2pointNum%

        IF Nd% > 1 THEN

            NumProbesPerDimension% = Np%\Nd% 'even #
```

```
        ELSE
            NumProbesPerDimension% = Np%
        END IF
        FOR i% = 1 TO Nd%
            FOR p% = 1 TO Np%
                R(p%,i%,0) = XiMin(i%) + Gamma*(XiMax(i%)-XiMin(i%))
            NEXT Np%
        NEXT i%
        FOR i% = 1 TO Nd% 'place probes probe line-by-probe line (i% is dimension number)
            DeltaXi = (XiMax(i%)-XiMin(i%))/(NumProbesPerDimension%-1)
            FOR k% = 1 TO NumProbesPerDimension%
                p% = k% + NumProbesPerDimension%*(i%-1) 'probe #
                R(p%,i%,0) = XiMin(i%) + (k%-1)*DeltaXi
            NEXT k%
        NEXT i%
END SUB 'IPD()
'----
FUNCTION HasFITNESSsaturated$(Nsteps&,j&,Np%,Nd%,M(),R(),DiagLength)
LOCAL A$, B$
LOCAL k&, p%
LOCAL BestFitness, SumOfBestFitnesses, BestFitnessStepJ, FitnessSatTOL AS EXT
    A$ = "NO" : B$ = "j="+STR$(j&)+CHR$(13)
    FitnessSatTOL = 0.000001## 'tolerance for FITNESS saturation
    IF j& < Nsteps& + 10 THEN GOTO ExitHasFITNESSsaturated 'execute at least 10 steps after averaging interval before performing this check
    SumOfBestFitnesses = 0##
    FOR k& = j&-Nsteps&+1 TO j& 'GET BEST FITNESSES STEP-BY-STEP FOR Nsteps& INCLUDING THIS STEP j& AND COMPUTE AVERAGE VALUE.
'       BestFitness = M(k&,1) 'ORIG CODE 03-23-2010: THIS IS A MISTAKE!
        BestFitness = -1E4200 'THIS LINE CORRECTED 03-23-2010 PER DISCUSSION WITH ROB GREEN.
                              'INITIALIZE BEST FITNESS AT k&-th TIME STEP TO AN EXTREMELY LARGE NEGATIVE NUMBER.
        FOR p% = 1 TO Np% 'PROBE-BY-PROBE GET MAXIMUM FITNESS
            IF M(p%,k&) >= BestFitness THEN BestFitness = M(p%,k&)
        NEXT p%
        IF k& = j& THEN BestFitnessStepJ = BestFitness  'IF AT THE END OF AVERAGING INTERVAL, SAVE BEST FITNESS FOR CURRENT TIME STEP j&
        SumOfBestFitnesses = SumOfBestFitnesses + BestFitness
        B$ = B$ + "k="+STR$(k&)+"  BestFit="+STR$(BestFitness)+"   SumFit="+STR$(SumOfBestFitnesses)+CHR$(13)
    NEXT k&
    IF ABS(SumOfBestFitnesses/Nsteps&-BestFitnessStepJ) =< FitnessSatTOL THEN A$ = "YES" 'saturation if (avg value - last value) are within TOL
ExitHasFITNESSsaturated:
    HasFITNESSsaturated$ = A$
END FUNCTION 'HasFITNESSsaturated$()
'-----------------------------------
SUB RetrieveErrantprobes2(Np%,Nd%,j&,R(),A(),Frep) 'added 04-01-10
LOCAL ErrantProbe$
LOCAL p%, i%, k%
LOCAL Xik, dMax, Eta(), EtaStar, SumSQ, MagAj1 AS EXT
REDIM Eta(1 TO Nd%, 1 TO 2) 'Eta(i%,k%)
    FOR p% = 1 TO Np%
        ErrantProbe$ = "NO" 'presume each probe is inside DS
        FOR i% = 1 TO Nd% 'check to see if probe p lies outside DS (any coordinate exceeding a bounday)
            IF (R(p%,i%,j&) > XiMax(i%) OR R(p%,i%,j&) < XiMin(i%)) AND A(p%,i%,j&-1) <> 0## THEN 'probe lies outside DS
                ErrantProbe$ = "YES" : EXIT FOR
            END IF
        NEXT i%
        IF ErrantProbe$ = "YES" THEN 'reposition probe p inside DS with acceleration direction preserved
            FOR i% = 1 TO Nd% 'compute array of Eta values
                FOR k% = 1 TO 2
                    SELECT CASE k%
                        CASE 1 : Xik = XiMin(i%)
                        CASE 2 : Xik = XiMax(i%)
                    END SELECT
                    Eta(i%,k%) = (Xik-R(p%,i%,j&-1))/A(p%,i%,j&-1)
                NEXT k%
            NEXT i%
            EtaStar = 1E4200 'very large positive number
            FOR i% = 1 TO Nd% 'get min Eta value > 0
                FOR k% = 1 TO 2
                    IF Eta(i%,k%) =< EtaStar AND Eta(i%,k%) >= 0## THEN EtaStar = Eta(i%,k%)
```

```
                NEXT k%
            NEXT i%
            SumSQ = 0## : FOR i% = 1 TO Nd% : SumSQ = SumSQ + A(p%,i%,j&-1)^2 : NEXT i% : MagAj1 = SQR(SumSQ) 'magnitude of acceleration vector A(probe #p) at step j&-1
            dMax = EtaStar*MagAj1 'distance to nearest boundary plane from position of probe p% at step j&-1
            FOR i% = 1 TO Nd% 'change probe p's i-th coordinate in proportion to the ratio of position vector lengths at steps j& and j&-1
                R(p%,i%,j&) = R(p%,i%,j&-1) + Frep*dMax*A(p%,i%,j&-1)/MagAj1 'preserves acceleration directional information by scaling dMax by Frep, which is arbitrary but precisely known
            NEXT i%
        END IF 'ErrantProbe$ = "YES"
    NEXT p%
END SUB 'Retrieveerrantprobes2()
'------------------------------
SUB RetrieveErrantProbes(Np%,Nd%,j&,R(),Frep)
LOCAL p%, i%
    FOR p% = 1 TO Np%
        FOR i% = 1 TO Nd%
            IF R(p%,i%,j&) < XiMin(i%) THEN R(p%,i%,j&) = MAX(XiMin(i%) + Frep*(R(p%,i%,j&-1)-XiMin(i%)),XiMin(i%)) 'CHANGED 02-07-10
            IF R(p%,i%,j&) > XiMax(i%) THEN R(p%,i%,j&) = MIN(XiMax(i%) - Frep*(XiMax(i%)-R(p%,i%,j&-1)),XiMax(i%))
        NEXT i%
    NEXT p%
END SUB 'RetrieveErrantProbes()
'------------------------------
SUB ResetDecisionSpaceBoundaries(Nd%)
    LOCAL i%
    FOR i% = 1 TO Nd% : XiMin(i%) = StartingXiMin(i%) : XiMax(i%) = StartingXiMax(i%) : NEXT i%
END SUB
'------
SUB CopyBestMatrices(Np%,Nd%,Nt&,R(),M(),Rbest(),Mbest())
LOCAL p%, i%, j&
REDIM Rbest(1 TO Np%, 1 TO Nd%, 0 TO Nt&), Mbest(1 TO Np%, 0 TO Nt&) 're-initializes Best Position Vector/Fitness matrices to zero
    FOR p% = 1 TO Np%
        FOR i% = 1 TO Nd%
            FOR j& = 0 TO Nt&
                Rbest(p%,i%,j&) = R(p%,i%,j&) : Mbest(p%,j&) = M(p%,j&)
            NEXT j&
        NEXT i%
    NEXT p%
END SUB
'------
'FORGET THIS IDEA !!!!
FUNCTION ProbeWeight(Nd%,R(),p%,j&) 'computes a 'weighting factor' based on probe's position (greater weight if closer to decision space boundary)
LOCAL MinDistCoordinate%, i% 'Dimension number.  Remember, XiMin(), XiMax()& DiagLength are GLOBAL.
LOCAL MinDistance, dStar, MaxWeight AS EXT
    MinDistance = DiagLength 'largest dimension of decision space
    FOR i% = 1 TO Nd% 'compute distance to closest boundary
        IF ABS(R(p%,i%,j&)-XiMin(i%)) =< MinDistance THEN
            MinDistance = ABS(R(p%,i%,j&)-XiMin(i%)) : MinDistCoordinate% = i%
        END IF
        IF ABS(XiMax(i%)-R(p%,i%,j&)) =< MinDistance THEN
            MinDistance = ABS(XiMax(i%)-R(p%,i%,j&)) : MinDistCoordinate% = i%
        END IF
    NEXT i%
    dStar = MinDistance/(XiMax(MinDistCoordinate%)-XiMin(MinDistCoordinate%)) 'normalized minimum distance, [0-1]
    MaxWeight = 2##
'    ProbeWeight = 1## + 2##*MaxWeight*abs(dStar-0.5##)
    ProbeWeight = 1## + 4##*MaxWeight*(dStar-0.5##)^2
END FUNCTION 'ProbeWeight
'------------------------
'FORGET THIS IDEA !!!!
FUNCTION ProbeWeight2(Nd%,Np%,R(),M(),p%,j&) 'computes a 'weighting factor' based on probe's position (greater weight if closer to decision space boundary)
LOCAL MinDistCoordinate%, ProbeNum%, BestProbeThisStep%, i% 'Dimension number.  Remember, XiMin(), XiMax()& DiagLength are GLOBAL.
LOCAL Distance, SumSQ, dStar, MaxWeight, BestFitnessThisStep AS EXT
    BestFitnessThisStep = M(1,j&)
    FOR ProbeNum% = 1 TO Np% 'get number of best probe this step
```

```
        IF M(ProbeNum%,j&) >= BestFitnessThisStep THEN

            BestFitnessThisStep = M(ProbeNum%,j&) : BestProbeThisStep% = ProbeNum%

        END IF

    NEXT ProbeNum%

    SumSQ = 0##

    FOR i% = 1 TO Nd% 'compute distance from probe #p% to thes best probe this step

        SumSQ = (R(p%,i%,j&) - R(BestProbeThisStep%,i%,j&))^2

    NEXT i%

    Distance = SQR(SumSQ)

    dStar = Distance/DiagLength 'range [0-1]

'   ----------------- Compute weight Factor ---------------------

    MaxWeight = 0##

'    ProbeWeight2 = 1## + 2##*MaxWeight*abs(dStar-0.5##)

    ProbeWeight2 = 1## + 4##*MaxWeight*(dStar-0.5##)^2

END FUNCTION 'ProbeWeight2

'--------------------------

FUNCTION SlopeRatio(M(),Np%,StepNumber&)

LOCAL p% 'probe #

LOCAL NumSteps%

LOCAL BestFitnessAtStepNumber, BestFitnessAtStepNumberMinus1, BestFitnessAtStepNumberMinus2, Z AS EXT

    Z = 1## 'assumes no slope change

    IF StepNumber& < 10 THEN GOTO ExitSlopeRatio 'need at least 10 steps for this test

    NumSteps% = 2

    BestFitnessAtStepNumber       = M(1,StepNumber&)             : FOR p% = 1 TO Np% : IF M(p%,StepNumber&)             >= BestFitnessAtStepNumber
THEN BestFitnessAtStepNumber = M(p%,StepNumber&)                  : NEXT p%

    BestFitnessAtStepNumberMinus1 = M(1,StepNumber&-NumSteps%)   : FOR p% = 1 TO Np% : IF M(p%,StepNumber&-NumSteps%)   >= BestFitnessAtStepNumberMinus1
THEN BestFitnessAtStepNumberMinus1 = M(p%,StepNumber&-NumSteps%)   : NEXT p%

    BestFitnessAtStepNumberMinus2 = M(1,StepNumber&-2*NumSteps%) : FOR p% = 1 TO Np% : IF M(p%,StepNumber&-2*NumSteps%) >= BestFitnessAtStepNumberMinus2
THEN BestFitnessAtStepNumberMinus2 = M(p%,StepNumber&-2*NumSteps%) : NEXT p%

    Z = (BestFitnessAtStepNumber-BestFitnessAtStepNumberMinus1)/(BestFitnessAtStepNumberMinus1-BestFitnessAtStepNumberMinus2)

ExitSlopeRatio:

    SlopeRatio = Z

END FUNCTION

'-----------

SUB
PlotResults(FunctionName$,Nd%,Np%,BestFitnessOverall,BestNpNd%,BestGamma,Neval&&,Rbest(),Mbest(),BestProbeNumberOverall%,BestTimeStepOverall&,LastStepBestR
un&,Alpha,Beta)

LOCAL LastStep&, BestFitnessProbeNumber%, BestFitnessTimeStep&, NumTrajectories%, Max1DprobesPlotted%, i%

LOCAL RepositionFactor$, PlaceInitialProbes$, InitialAcceleration$, A$, B$

LOCAL G, DeltaT, Frep AS EXT

'    G = 2## : DeltaT = 1## : Frep = 0.5## : RepositionFactor$ = "VARIABLE" : PlaceInitialProbes$ = "UNIFORM ON-AXIS " : InitialAcceleration$ = "FIXED"
'THESE ARE NOW HARDWIRED IN THE CFO EQUATIONS 'ORIG CODE 05/14/2011
    G = 2## : DeltaT = 1## : Frep = 0.5## : RepositionFactor$ = "VARIABLE" : PlaceInitialProbes$ = "PROBE LINES" : InitialAcceleration$ = "ZERO" 'MOD
05/14/2011

    B$ = "" : IF Nd% > 1 THEN B$ = "s"

    A$ = FunctionName$ + CHR$(13) +_
         "Best Fitness = " + REMOVE$(STR$(BestFitnessOverall),ANY" ")       + " returned by" + CHR$(13) +_
         "Probe # "        + REMOVE$(STR$(BestProbeNumberOverall%),ANY" ") +_
         " at Time Step "  + REMOVE$(STR$(BestTimeStepOverall&),ANY" ")     + CHR$(13) + CHR$(13) + "P" + REMOVE$(STR$(BestProbeNumberOverall%),ANY" ") + "
coordinate" + B$ + ":" + CHR$(13)

    FOR i% = 1 TO Nd% : A$ = A$ + STR$(i%)+"    "+REMOVE$(STR$(ROUND(Rbest(BestProbeNumberOverall%,i%,BestTimeStepOverall&),8)),ANY" ")+CHR$(13) : NEXT i%

    MSGBOX(A$)

'   ---------------------------------------------- PLOT EVOLUTION OF BEST FITNESS, AVG DISTANCE & BEST PROBE # ------------------------------------------
-------------

    CALL
PlotBestFitnessEvolution(Nd%,Np%,LastStepBestRun&,G,DeltaT,Alpha,Beta,Frep,Mbest(),PlaceInitialProbes$,InitialAcceleration$,RepositionFactor$,FunctionName$
,BestGamma)

'    MSGBOX("Enter for next plot...")

    CALL
PlotAverageDistance(Nd%,Np%,LastStepBestRun&,G,DeltaT,Alpha,Beta,Frep,Mbest(),PlaceInitialProbes$,InitialAcceleration$,RepositionFactor$,FunctionName$,Rbes
t(),DiagLength,BestGamma)

'    MSGBOX("Enter for next plot...")

    CALL
PlotBestProbeVsTimeStep(Nd%,Np%,LastStepBestRun&,G,DeltaT,Alpha,Beta,Frep,Mbest(),PlaceInitialProbes$,InitialAcceleration$,RepositionFactor$,FunctionName$,
BestGamma)

'   --------------------------------------------- PLOT TRAJECTORIES OF BEST PROBES FOR 2/3-D FUNCTIONS ---------------------------------------------------

    IF Nd% = 2 THEN

        NumTrajectories% = 10 : CALL Plot2DbestProbeTrajectories(NumTrajectories%,Mbest(),Rbest(),Np%,Nd%,LastStepBestRun&,FunctionName$)

        NumTrajectories% = 16 : CALL Plot2DindividualProbeTrajectories(NumTrajectories%,Mbest(),Rbest(),Np%,Nd%,LastStepBestRun&,FunctionName$)

    END IF

    IF Nd% = 3 THEN

        NumTrajectories% = 4 : CALL Plot3DbestProbeTrajectories(NumTrajectories%,Mbest(),Rbest(),Np%,Nd%,LastStepBestRun&,FunctionName$)

    END IF

'   ----------- For 1-D Objective Functions, Tabulate Probe Coordinates & if Np% =< Max1DprobesPlotted% Plot Evolution of Probe Positions ------------
```

```
    IF Nd% = 1 THEN

        Max1DprobesPlotted% = 15

        CALL
Tabulate1DprobeCoordinates(Max1DprobesPlotted%,Nd%,Np%,LastStepBestRun&,G,DeltaT,Alpha,Beta,Frep,Rbest(),Mbest(),PlaceInitialProbes$,InitialAcceleration$,R
epositionFactor$,FunctionName$,BestGamma)

        IF Np% =< Max1DprobesPlotted% THEN _
        CALL
Plot1DprobePositions(Max1DprobesPlotted%,Nd%,Np%,LastStepBestRun&,G,DeltaT,Alpha,Beta,Frep,Rbest(),Mbest(),PlaceInitialProbes$,InitialAcceleration$,Reposit
ionFactor$,FunctionName$,BestGamma)

        CALL CLEANUP 'delete probe coordinate plot files, if any

    END IF

END SUB 'PlotResults()

'==============================================================================================================================================================
========================================================

SUB CheckNECFiles(NECfileError$)

LOCAL N%

    NECfileError$ = "NO"

'   ---------------------------- NEC Files Required for PBM Antenna Benchmarks -----------------------------

    IF DIR$("n41_2k1.exe") = "" THEN

            MSGBOX("WARNING!  'n41_2k1.exe' not found.  Run terminated.") : NECfileError$ = "YES" : EXIT SUB

    END IF

'   --------------- These Files are Required for NEC-4.1, Some for NEC-2 -------------------

    N% = FREEFILE : OPEN "ENDERR.DAT"   FOR OUTPUT AS #N% : PRINT #N%, "NO"      : CLOSE #N%

    N% = FREEFILE : OPEN "FILE_MSG.DAT" FOR OUTPUT AS #N% : PRINT #N%, "NO"      : CLOSE #N%

    N% = FREEFILE : OPEN "NHSCALE.DAT"  FOR OUTPUT AS #N% : PRINT #N%, "0.00001" : CLOSE #N%

END SUB

'------

SUB GetBestFitness(M(),Np%,StepNumber&,BestFitness,BestProbeNumber%,BestTimeStep&)

LOCAL p%, i&, A$

    BestFitness = M(1,0)

    FOR i& = 0 TO StepNumber&

        FOR p% = 1 TO Np%

            IF M(p%,i&) >= BestFitness THEN

                BestFitness = M(p%,i&) : BestProbeNumber% = p% : BestTimeStep& = i&

            END IF

        NEXT p%

    NEXT i&

END SUB

'============================================================ FUNCTION DEFINITIONS ============================================================

FUNCTION ObjectiveFunction(R(),Nd%,p%,j&,FunctionName$) 'Objective function to be MAXIMIZED is defined here

    SELECT CASE FunctionName$

        CASE "ParrottF4"     : ObjectiveFunction = ParrottF4(R(),Nd%,p%,j&)      'Parrott F4 (1-D)

        CASE "SGO"           : ObjectiveFunction = SGO(R(),Nd%,p%,j&)            'SGO Function (2-D)

        CASE "GP"            : ObjectiveFunction = GoldsteinPrice(R(),Nd%,p%,j&) 'Goldstein-Price Function (2-D)

        CASE "STEP"          : ObjectiveFunction = StepFunction(R(),Nd%,p%,j&)   'Step Function (n-D)

        CASE "SCHWEFEL_226"  : ObjectiveFunction = Schwefel226(R(),Nd%,p%,j&)    'Schwefel Prob. 2.26 (n-D)

        CASE "COLVILLE"      : ObjectiveFunction = Colville(R(),Nd%,p%,j&)       'Colville Function (4-D)

        CASE "GRIEWANK"      : ObjectiveFunction = Griewank(R(),Nd%,p%,j&)       'Griewank Function (n-D)

        CASE "HIMMELBLAU"    : ObjectiveFunction = Himmelblau(R(),Nd%,p%,j&)     'Himmelblau Function (2-D)

        CASE "ROSENBROCK"    : ObjectiveFunction = Rosenbrock(R(),Nd%,p%,j&)     'Rosenbrock Function (n-D)

        CASE "SPHERE"        : ObjectiveFunction = Sphere(R(),Nd%,p%,j&)         'Sphere Function (n-D)

        CASE "HIMMELBLAUNLO" : ObjectiveFunction = HIMMELBLAUNLO(R(),Nd%,p%,j&)  'Himmelblau NLO (5-D)

        CASE "TRIPOD"        : ObjectiveFunction = Tripod(R(),Nd%,p%,j&)         'Tripod (2-D)

        CASE "ROSENBROCKF6"  : ObjectiveFunction = RosenbrockF6(R(),Nd%,p%,j&)   'RosebrockF6 (10-D)

        CASE "COMPRESSIONSPRING"  : ObjectiveFunction = CompressionSpring(R(),Nd%,p%,j&)   'Compression Spring (3-D)

        CASE "GEARTRAIN"     : ObjectiveFunction = GearTrain(R(),Nd%,p%,j&)      'Gear Train (4-D)

'       ------------------------- GSO Paper Benchmark Functions ---------------------------

        CASE  "F1"           : ObjectiveFunction = F1(R(),Nd%,p%,j&)             'F1  (n-D)
        CASE  "F2"           : ObjectiveFunction = F2(R(),Nd%,p%,j&)             'F2  (n-D)
        CASE  "F3"           : ObjectiveFunction = F3(R(),Nd%,p%,j&)             'F3  (n-D)
        CASE  "F4"           : ObjectiveFunction = F4(R(),Nd%,p%,j&)             'F4  (n-D)
        CASE  "F5"           : ObjectiveFunction = F5(R(),Nd%,p%,j&)             'F5  (n-D)
        CASE  "F6"           : ObjectiveFunction = F6(R(),Nd%,p%,j&)             'F6  (n-D)
        CASE  "F7"           : ObjectiveFunction = F7(R(),Nd%,p%,j&)             'F7  (n-D)
        CASE  "F8"           : ObjectiveFunction = F8(R(),Nd%,p%,j&)             'F8  (n-D)
        CASE  "F9"           : ObjectiveFunction = F9(R(),Nd%,p%,j&)             'F9  (n-D)
        CASE "F10"           : ObjectiveFunction = F10(R(),Nd%,p%,j&)            'F10 (n-D)
        CASE "F11"           : ObjectiveFunction = F11(R(),Nd%,p%,j&)            'F11 (n-D)
        CASE "F12"           : ObjectiveFunction = F12(R(),Nd%,p%,j&)            'F12 (n-D)
        CASE "F13"           : ObjectiveFunction = F13(R(),Nd%,p%,j&)            'F13 (n-D)
        CASE "F14"           : ObjectiveFunction = F14(R(),Nd%,p%,j&)            'F14 (2-D)
        CASE "F15"           : ObjectiveFunction = F15(R(),Nd%,p%,j&)            'F15 (4-D)
        CASE "F16"           : ObjectiveFunction = F16(R(),Nd%,p%,j&)            'F16 (2-D)
        CASE "F17"           : ObjectiveFunction = F17(R(),Nd%,p%,j&)            'F17 (2-D)
        CASE "F18"           : ObjectiveFunction = F18(R(),Nd%,p%,j&)            'F18 (2-D)
        CASE "F19"           : ObjectiveFunction = F19(R(),Nd%,p%,j&)            'F19 (3-D)
        CASE "F20"           : ObjectiveFunction = F20(R(),Nd%,p%,j&)            'F20 (6-D)
        CASE "F21"           : ObjectiveFunction = F21(R(),Nd%,p%,j&)            'F21 (4-D)
```

```
        CASE "F22"           : ObjectiveFunction = F22(R(),Nd%,p%,j&)            'F22 (4-D)
        CASE "F23"           : ObjectiveFunction = F23(R(),Nd%,p%,j&)            'F23 (4-D)

'       ------------------------------ PBM Antenna Benchmarks ------------------------------

        CASE "PBM_1"         : ObjectiveFunction = PBM_1(R(),Nd%,p%,j&)          'PBM_1 (2-D)
        CASE "PBM_2"         : ObjectiveFunction = PBM_2(R(),Nd%,p%,j&)          'PBM_2 (2-D)
        CASE "PBM_3"         : ObjectiveFunction = PBM_3(R(),Nd%,p%,j&)          'PBM_3 (2-D)
        CASE "PBM_4"         : ObjectiveFunction = PBM_4(R(),Nd%,p%,j&)          'PBM_4 (2-D)
        CASE "PBM_5"         : ObjectiveFunction = PBM_5(R(),Nd%,p%,j&)          'PBM_5 (2-D)

'       --------------------------- LOADED BOWTIE IN FREE SPACE -------------------------
        CASE "BOWTIE"          : ObjectiveFunction = BOWTIE(R(),Nd%,p%,j&)       'FREE SPACE LOADED BOWTIE

'       --------------------------- YAGI ARRAY IN FREE SPACE --------------------------
        CASE "YAGI"            : ObjectiveFunction = YAGI_ARRAY(R(),Nd%,p%,j&)   'FREE SPACE YAGI ARRAY

    END SELECT

END FUNCTION 'ObjectiveFunction()

'------

SUB GetFunctionRunParameters(FunctionName$,Nd%,Np%,Nt&,G,DeltaT,Alpha,Beta,Frep,R(),A(),M(),_
                             DiagLength,PlaceInitialProbes$,InitialAcceleration$,RepositionFactor$)

LOCAL i%, NumProbesPerDimension%, NN%, NumCollinearElements%

LOCAL A$

    SELECT CASE FunctionName$

        CASE "ParrottF4"

            Nd% = 1 : Np% = 3

            REDIM XiMin(1 TO Nd%), XiMax(1 TO Nd%) : XiMin(1) = 0## : XiMax(1) = 1##

            REDIM StartingXiMin(1 TO Nd%), StartingXiMax(1 TO Nd%) : FOR i% = 1 TO Nd% : StartingXiMin(i%) = XiMin(i%) : StartingXiMax(i%) = XiMax(i%) :
NEXT i%

        CASE "SGO"

            Nd% = 2 : Np% = 8

            REDIM XiMin(1 TO Nd%), XiMax(1 TO Nd%) : FOR i% = 1 TO Nd% : XiMin(i%) = -50## : XiMax(i%) = 50## : NEXT i%

            REDIM StartingXiMin(1 TO Nd%), StartingXiMax(1 TO Nd%) : FOR i% = 1 TO Nd% : StartingXiMin(i%) = XiMin(i%) : StartingXiMax(i%) = XiMax(i%) :
NEXT i%

        CASE "GP"

            Nd% = 2 : Np% = 8

            REDIM XiMin(1 TO Nd%), XiMax(1 TO Nd%) : FOR i% = 1 TO Nd% : XiMin(i%) = -100## : XiMax(i%) = 100## : NEXT i%

            REDIM StartingXiMin(1 TO Nd%), StartingXiMax(1 TO Nd%) : FOR i% = 1 TO Nd% : StartingXiMin(i%) = XiMin(i%) : StartingXiMax(i%) = XiMax(i%) :
NEXT i%

        CASE "STEP"

            Nd% = 2 : Np% = 8

            REDIM XiMin(1 TO Nd%), XiMax(1 TO Nd%) : FOR i% = 1 TO Nd% : XiMin(i%) = -100## : XiMax(i%) = 100## : NEXT i%

'           REDIM XiMin(1 TO Nd%), XiMax(1 TO Nd%) : XiMin(1) = 72## : XiMax(1) = 78## : XiMin(2) = 27## : XiMax(2) = 33## 'use this to plot STEP detail

            REDIM StartingXiMin(1 TO Nd%), StartingXiMax(1 TO Nd%) : FOR i% = 1 TO Nd% : StartingXiMin(i%) = XiMin(i%) : StartingXiMax(i%) = XiMax(i%) :
NEXT i%

        CASE "SCHWEFEL_226"

            Nd% = 30 : Np% = 120

            REDIM XiMin(1 TO Nd%), XiMax(1 TO Nd%) : FOR i% = 1 TO Nd% : XiMin(i%) = -500## : XiMax(i%) = 500## : NEXT i%

            REDIM StartingXiMin(1 TO Nd%), StartingXiMax(1 TO Nd%) : FOR i% = 1 TO Nd% : StartingXiMin(i%) = XiMin(i%) : StartingXiMax(i%) = XiMax(i%) :
NEXT i%

        CASE "COLVILLE"

            Nd% = 4 : Np% = 16

            REDIM XiMin(1 TO Nd%), XiMax(1 TO Nd%) : FOR i% = 1 TO Nd% : XiMin(i%) = -10## : XiMax(i%) = 10## : NEXT i%

            REDIM StartingXiMin(1 TO Nd%), StartingXiMax(1 TO Nd%) : FOR i% = 1 TO Nd% : StartingXiMin(i%) = XiMin(i%) : StartingXiMax(i%) = XiMax(i%) :
NEXT i%

        CASE "GRIEWANK"

            Nd% = 2 : Np% = 8

            REDIM XiMin(1 TO Nd%), XiMax(1 TO Nd%) : FOR i% = 1 TO Nd% : XiMin(i%) = -600## : XiMax(i%) = 600## : NEXT i%

            REDIM StartingXiMin(1 TO Nd%), StartingXiMax(1 TO Nd%) : FOR i% = 1 TO Nd% : StartingXiMin(i%) = XiMin(i%) : StartingXiMax(i%) = XiMax(i%) :
NEXT i%

       CASE "HIMMELBLAU"

            Nd% = 2 : Np% = 8

            REDIM XiMin(1 TO Nd%), XiMax(1 TO Nd%) : FOR i% = 1 TO Nd% : XiMin(i%) = -6## : XiMax(i%) = 6## : NEXT i%

            REDIM StartingXiMin(1 TO Nd%), StartingXiMax(1 TO Nd%) : FOR i% = 1 TO Nd% : StartingXiMin(i%) = XiMin(i%) : StartingXiMax(i%) = XiMax(i%) :
NEXT i%

        CASE "ROSENBROCK" '(n-D)

            Nd% = 2 : Np% = 8

            REDIM XiMin(1 TO Nd%), XiMax(1 TO Nd%) : FOR i% = 1 TO Nd% : XiMin(i%) = -2## : XiMax(i%) = 2## : NEXT i% :'XiMin(i%) = -6## : XiMax(i%) = 6##
: NEXT i%

            REDIM StartingXiMin(1 TO Nd%), StartingXiMax(1 TO Nd%) : FOR i% = 1 TO Nd% : StartingXiMin(i%) = XiMin(i%) : StartingXiMax(i%) = XiMax(i%) :
NEXT i%

        CASE "SPHERE" '(n-D)

            Nd% = 2 : Np% = 8

            REDIM XiMin(1 TO Nd%), XiMax(1 TO Nd%) : FOR i% = 1 TO Nd% : XiMin(i%) = -100## : XiMax(i%) = 100## : NEXT i%

            REDIM StartingXiMin(1 TO Nd%), StartingXiMax(1 TO Nd%) : FOR i% = 1 TO Nd% : StartingXiMin(i%) = XiMin(i%) : StartingXiMax(i%) = XiMax(i%) :
NEXT i%

        CASE "HIMMELBLAUNLO" '(5-D)

            Nd% = 5 : Np% = 20
```

```
            REDIM XiMin(1 TO Nd%), XiMax(1 TO Nd%)

            XiMin(1) = 78## : XiMax(1) = 102##
            XiMin(2) = 33## : XiMax(2) = 45##
            XiMin(3) = 27## : XiMax(3) = 45##
            XiMin(4) = 27## : XiMax(4) = 45##
            XiMin(5) = 27## : XiMax(5) = 45##

            REDIM StartingXiMin(1 TO Nd%), StartingXiMax(1 TO Nd%) : FOR i% = 1 TO Nd% : StartingXiMin(i%) = XiMin(i%) : StartingXiMax(i%) = XiMax(i%) :
NEXT i%

        CASE "TRIPOD" '(2-D)

            Nd% = 2 : Np% = 8

            REDIM XiMin(1 TO Nd%), XiMax(1 TO Nd%) : FOR i% = 1 TO Nd% : XiMin(i%) = -100## : XiMax(i%) = 100## : NEXT i%

            REDIM StartingXiMin(1 TO Nd%), StartingXiMax(1 TO Nd%) : FOR i% = 1 TO Nd% : StartingXiMin(i%) = XiMin(i%) : StartingXiMax(i%) = XiMax(i%) :
NEXT i%

        CASE "ROSENBROCKF6" '(10-D)

            Nd% = 10
            Np% = 40

            REDIM XiOffset(1 TO Nd%)

'            XiOffset(1) =  81.0232##
'            XiOffset(2) = -48.3950##
'            XiOffset(3) =  19.2316##
'            XiOffset(4) =  -2.5231##
'            XiOffset(5) =  70.4338##
'            XiOffset(6) =  47.1774##
'            XiOffset(7) =  -7.8358##
'            XiOffset(8) = -86.6693##
'            XiOffset(9) =  57.8532##
'            XiOffset(10) = 0##

'            XiOffset(1)  =  80##
'            XiOffset(2)  = -50##
'            XiOffset(3)  =  20##
'            XiOffset(4)  =  -3##
'            XiOffset(5)  =  70##
'            XiOffset(6)  =  47##
'            XiOffset(7)  =  -8##
'            XiOffset(8)  = -87##
'            XiOffset(9)  =  58##
'            XiOffset(10) =   0##

            XiOffset(1)  =   5##
            XiOffset(2)  = -25##
            XiOffset(3)  =   5##
            XiOffset(4)  = -15##
            XiOffset(5)  =   5##
            XiOffset(6)  = -25##
            XiOffset(7)  =  25##
            XiOffset(8)  =  -5##
            XiOffset(9)  =   5##
            XiOffset(10) = -15##

'            XiOffset(1)  =  0##
'            XiOffset(2)  =  81.0232##
'            XiOffset(3)  = -48.3950##
'            XiOffset(4)  =  19.2316##
'            XiOffset(5)  =  -2.5231##
'            XiOffset(6)  =  70.4338##
'            XiOffset(7)  =  47.1774##
'            XiOffset(8)  =  -7.8358##
'            XiOffset(9)  = -86.6693##
'            XiOffset(10) =  57.8532##

            REDIM XiMin(1 TO Nd%), XiMax(1 TO Nd%) : FOR i% = 1 TO Nd% : XiMin(i%) = -100## : XiMax(i%) = 100## : NEXT i%

            REDIM StartingXiMin(1 TO Nd%), StartingXiMax(1 TO Nd%) : FOR i% = 1 TO Nd% : StartingXiMin(i%) = XiMin(i%) : StartingXiMax(i%) = XiMax(i%) :
NEXT i%

         CASE "COMPRESSIONSPRING" '(3-D)

            Nd% = 3 : Np% = 12

            REDIM XiMin(1 TO Nd%), XiMax(1 TO Nd%)

            XiMin(1) = 1##     : XiMax(1) =  70## 'integer values only!!
            XiMin(2) = 0.6##   : XiMax(2) =   3##
            XiMin(3) = 0.207## : XiMax(3) = 0.5##

            REDIM StartingXiMin(1 TO Nd%), StartingXiMax(1 TO Nd%) : FOR i% = 1 TO Nd% : StartingXiMin(i%) = XiMin(i%) : StartingXiMax(i%) = XiMax(i%) :
NEXT i%

         CASE "GEARTRAIN" '(4-D)

            Nd% = 4 : Np% = 16

            REDIM XiMin(1 TO Nd%), XiMax(1 TO Nd%) : FOR i% = 1 TO Nd% : XiMin(i%) = 12# : XiMax(i%) = 60## : NEXT i%

            REDIM StartingXiMin(1 TO Nd%), StartingXiMax(1 TO Nd%) : FOR i% = 1 TO Nd% : StartingXiMin(i%) = XiMin(i%) : StartingXiMax(i%) = XiMax(i%) :
NEXT i%

        CASE "F1" '(n-D)

            Nd% = 30 : Np% = 60

            REDIM XiMin(1 TO Nd%), XiMax(1 TO Nd%) : FOR i% = 1 TO Nd% : XiMin(i%) = -100## : XiMax(i%) = 100## : NEXT i%

            REDIM StartingXiMin(1 TO Nd%), StartingXiMax(1 TO Nd%) : FOR i% = 1 TO Nd% : StartingXiMin(i%) = XiMin(i%) : StartingXiMax(i%) = XiMax(i%) :
NEXT i%

       CASE "F2" '(n-D)

            Nd% = 30 : Np% = 60

            REDIM XiMin(1 TO Nd%), XiMax(1 TO Nd%) : FOR i% = 1 TO Nd% : XiMin(i%) = -10## : XiMax(i%) = 10## : NEXT i%

            REDIM StartingXiMin(1 TO Nd%), StartingXiMax(1 TO Nd%) : FOR i% = 1 TO Nd% : StartingXiMin(i%) = XiMin(i%) : StartingXiMax(i%) = XiMax(i%) :
NEXT i%

      CASE "F3" '(n-D)

            Nd% = 30 : Np% = 60

            REDIM XiMin(1 TO Nd%), XiMax(1 TO Nd%) : FOR i% = 1 TO Nd% : XiMin(i%) = -100## : XiMax(i%) = 100## : NEXT i%

            REDIM StartingXiMin(1 TO Nd%), StartingXiMax(1 TO Nd%) : FOR i% = 1 TO Nd% : StartingXiMin(i%) = XiMin(i%) : StartingXiMax(i%) = XiMax(i%) :
NEXT i%

      CASE "F4" '(n-D)

            Nd% = 30 : Np% = 60
```

```
            REDIM XiMin(1 TO Nd%), XiMax(1 TO Nd%) : FOR i% = 1 TO Nd% : XiMin(i%) = -100## : XiMax(i%) = 100## : NEXT i%
            REDIM StartingXiMin(1 TO Nd%), StartingXiMax(1 TO Nd%) : FOR i% = 1 TO Nd% : StartingXiMin(i%) = XiMin(i%) : StartingXiMax(i%) = XiMax(i%) :
NEXT i%
      CASE "F5" '(n-D)
            Nd% = 30 : Np% = 60
            REDIM XiMin(1 TO Nd%), XiMax(1 TO Nd%) : FOR i% = 1 TO Nd% : XiMin(i%) = -30## : XiMax(i%) = 30## : NEXT i%
            REDIM StartingXiMin(1 TO Nd%), StartingXiMax(1 TO Nd%) : FOR i% = 1 TO Nd% : StartingXiMin(i%) = XiMin(i%) : StartingXiMax(i%) = XiMax(i%) :
NEXT i%
       CASE "F6" '(n-D) STEP
            Nd% = 30 : Np% = 60
            REDIM XiMin(1 TO Nd%), XiMax(1 TO Nd%) : FOR i% = 1 TO Nd% : XiMin(i%) = -100## : XiMax(i%) = 100## : NEXT i%
            REDIM StartingXiMin(1 TO Nd%), StartingXiMax(1 TO Nd%) : FOR i% = 1 TO Nd% : StartingXiMin(i%) = XiMin(i%) : StartingXiMax(i%) = XiMax(i%) :
NEXT i%
       CASE "F7" '(n-D)
            Nd% = 30 : Np% = 60
            REDIM XiMin(1 TO Nd%), XiMax(1 TO Nd%) : FOR i% = 1 TO Nd% : XiMin(i%) = -1.28## : XiMax(i%) = 1.28## : NEXT i%
            REDIM StartingXiMin(1 TO Nd%), StartingXiMax(1 TO Nd%) : FOR i% = 1 TO Nd% : StartingXiMin(i%) = XiMin(i%) : StartingXiMax(i%) = XiMax(i%) :
NEXT i%
       CASE "F8" '(n-D)
            Nd% = 30 : Np% = 60
            REDIM XiMin(1 TO Nd%), XiMax(1 TO Nd%) : FOR i% = 1 TO Nd% : XiMin(i%) = -500## : XiMax(i%) = 500## : NEXT i%
            REDIM StartingXiMin(1 TO Nd%), StartingXiMax(1 TO Nd%) : FOR i% = 1 TO Nd% : StartingXiMin(i%) = XiMin(i%) : StartingXiMax(i%) = XiMax(i%) :
NEXT i%
       CASE "F9" '(n-D)
            Nd% = 30 : Np% = 60
            REDIM XiMin(1 TO Nd%), XiMax(1 TO Nd%) : FOR i% = 1 TO Nd% : XiMin(i%) = -5.12## : XiMax(i%) = 5.12## : NEXT i%
            REDIM StartingXiMin(1 TO Nd%), StartingXiMax(1 TO Nd%) : FOR i% = 1 TO Nd% : StartingXiMin(i%) = XiMin(i%) : StartingXiMax(i%) = XiMax(i%) :
NEXT i%
       CASE "F10" '(n-D) Ackley's Function
            Nd% = 30 : Np% = 60
            REDIM XiMin(1 TO Nd%), XiMax(1 TO Nd%) : FOR i% = 1 TO Nd% : XiMin(i%) = -32## : XiMax(i%) = 32## : NEXT i%
            REDIM StartingXiMin(1 TO Nd%), StartingXiMax(1 TO Nd%) : FOR i% = 1 TO Nd% : StartingXiMin(i%) = XiMin(i%) : StartingXiMax(i%) = XiMax(i%) :
NEXT i%
       CASE "F11" '(n-D)
            Nd% = 30 : Np% = 60
            REDIM XiMin(1 TO Nd%), XiMax(1 TO Nd%) : FOR i% = 1 TO Nd% : XiMin(i%) = -600## : XiMax(i%) = 600## : NEXT i%
            REDIM StartingXiMin(1 TO Nd%), StartingXiMax(1 TO Nd%) : FOR i% = 1 TO Nd% : StartingXiMin(i%) = XiMin(i%) : StartingXiMax(i%) = XiMax(i%) :
NEXT i%
        CASE "F12" '(n-D) Penalized #1
            Nd% = 30 : Np% = 60
            REDIM XiMin(1 TO Nd%), XiMax(1 TO Nd%) : FOR i% = 1 TO Nd% : XiMin(i%) = -50## : XiMax(i%) = 50## : NEXT i%
'           REDIM XiMin(1 TO Nd%), XiMax(1 TO Nd%) : FOR i% = 1 TO Nd% : XiMin(i%) = -5## : XiMax(i%) = 5## : NEXT i% 'use this interval for second run to
improve performance
            REDIM StartingXiMin(1 TO Nd%), StartingXiMax(1 TO Nd%) : FOR i% = 1 TO Nd% : StartingXiMin(i%) = XiMin(i%) : StartingXiMax(i%) = XiMax(i%) :
NEXT i%
       CASE "F13" '(n-D) Penalized #2
            Nd% = 30 : Np% = 60
            REDIM XiMin(1 TO Nd%), XiMax(1 TO Nd%) : FOR i% = 1 TO Nd% : XiMin(i%) = -50## : XiMax(i%) = 50## : NEXT i%
            REDIM StartingXiMin(1 TO Nd%), StartingXiMax(1 TO Nd%) : FOR i% = 1 TO Nd% : StartingXiMin(i%) = XiMin(i%) : StartingXiMax(i%) = XiMax(i%) :
NEXT i%
       CASE "F14" '(2-D) Shekel's Foxholes
            Nd% = 2 : Np% = 8
            REDIM XiMin(1 TO Nd%), XiMax(1 TO Nd%) : FOR i% = 1 TO Nd% : XiMin(i%) = -65.536## : XiMax(i%) = 65.536## : NEXT i%
            REDIM StartingXiMin(1 TO Nd%), StartingXiMax(1 TO Nd%) : FOR i% = 1 TO Nd% : StartingXiMin(i%) = XiMin(i%) : StartingXiMax(i%) = XiMax(i%) :
NEXT i%
       CASE "F15" '(4-D) Kowalik's Function
            Nd% = 4
            Np% = 16
            REDIM XiMin(1 TO Nd%), XiMax(1 TO Nd%) : FOR i% = 1 TO Nd% : XiMin(i%) = -5## : XiMax(i%) = 5## : NEXT i%
            REDIM StartingXiMin(1 TO Nd%), StartingXiMax(1 TO Nd%) : FOR i% = 1 TO Nd% : StartingXiMin(i%) = XiMin(i%) : StartingXiMax(i%) = XiMax(i%) :
NEXT i%
       CASE "F16" '(2-D) Camel Back
            Nd% = 2 : Np% = 8
            REDIM XiMin(1 TO Nd%), XiMax(1 TO Nd%) : FOR i% = 1 TO Nd% : XiMin(i%) = -5## : XiMax(i%) = 5## : NEXT i%
            REDIM StartingXiMin(1 TO Nd%), StartingXiMax(1 TO Nd%) : FOR i% = 1 TO Nd% : StartingXiMin(i%) = XiMin(i%) : StartingXiMax(i%) = XiMax(i%) :
NEXT i%
       CASE "F17" '(2-D) Branin
            Nd% = 2 : Np% = 8
            REDIM XiMin(1 TO Nd%), XiMax(1 TO Nd%) : XiMin(1) = -5## : XiMax(1) = 10## : XiMin(2) = 0## : XiMax(2) = 15##
            REDIM StartingXiMin(1 TO Nd%), StartingXiMax(1 TO Nd%) : FOR i% = 1 TO Nd% : StartingXiMin(i%) = XiMin(i%) : StartingXiMax(i%) = XiMax(i%) :
NEXT i%
       CASE "F18" '(2-D) Goldstein-Price
            Nd% = 2 : Np% = 8
```

```
        REDIM XiMin(1 TO Nd%), XiMax(1 TO Nd%) : XiMin(1) = -2## : XiMax(1) = 2## : XiMin(2) = -2## : XiMax(2) = 2##

        REDIM StartingXiMin(1 TO Nd%), StartingXiMax(1 TO Nd%) : FOR i% = 1 TO Nd% : StartingXiMin(i%) = XiMin(i%) : StartingXiMax(i%) = XiMax(i%) :
NEXT i%

    CASE "F19" '(3-D) Hartman's Family #1

        Nd% = 3 : Np% = 12

        REDIM XiMin(1 TO Nd%), XiMax(1 TO Nd%) : FOR i% = 1 TO Nd% : XiMin(i%) = 0## : XiMax(i%) = 1## : NEXT i%

        REDIM StartingXiMin(1 TO Nd%), StartingXiMax(1 TO Nd%) : FOR i% = 1 TO Nd% : StartingXiMin(i%) = XiMin(i%) : StartingXiMax(i%) = XiMax(i%) :
NEXT i%

    CASE "F20" '(6-D) Hartman's Family #2

        Nd% = 6 : Np% = 24

        REDIM XiMin(1 TO Nd%), XiMax(1 TO Nd%) : FOR i% = 1 TO Nd% : XiMin(i%) = 0## : XiMax(i%) = 1## : NEXT i%

        REDIM StartingXiMin(1 TO Nd%), StartingXiMax(1 TO Nd%) : FOR i% = 1 TO Nd% : StartingXiMin(i%) = XiMin(i%) : StartingXiMax(i%) = XiMax(i%) :
NEXT i%

    CASE "F21" '(4-D) Shekel's Family m=5

        Nd% = 4 : Np% = 16

        REDIM XiMin(1 TO Nd%), XiMax(1 TO Nd%) : FOR i% = 1 TO Nd% : XiMin(i%) = 0## : XiMax(i%) = 10## : NEXT i%

        REDIM StartingXiMin(1 TO Nd%), StartingXiMax(1 TO Nd%) : FOR i% = 1 TO Nd% : StartingXiMin(i%) = XiMin(i%) : StartingXiMax(i%) = XiMax(i%) :
NEXT i%

    CASE "F22" '(4-D) Shekel's Family m=7

        Nd% = 4 : Np% = 16

        REDIM XiMin(1 TO Nd%), XiMax(1 TO Nd%) : FOR i% = 1 TO Nd% : XiMin(i%) = 0## : XiMax(i%) = 10## : NEXT i%

        REDIM StartingXiMin(1 TO Nd%), StartingXiMax(1 TO Nd%) : FOR i% = 1 TO Nd% : StartingXiMin(i%) = XiMin(i%) : StartingXiMax(i%) = XiMax(i%) :
NEXT i%

    CASE "F23" '(4-D) Shekel's Family m=10

        Nd% = 4 : Np% = 16

        REDIM XiMin(1 TO Nd%), XiMax(1 TO Nd%) : FOR i% = 1 TO Nd% : XiMin(i%) = 0## : XiMax(i%) = 10## : NEXT i%

        REDIM StartingXiMin(1 TO Nd%), StartingXiMax(1 TO Nd%) : FOR i% = 1 TO Nd% : StartingXiMin(i%) = XiMin(i%) : StartingXiMax(i%) = XiMax(i%) :
NEXT i%

    CASE "PBM_1" '2-D

        Nd%                    = 2
        NumProbesPerDimension% = 2 '4 '20
        Np%                    = NumProbesPerDimension%*Nd%

        Nt&      = 100
        G        = 2##
        Alpha    = 2##
        Beta     = 2##
        DeltaT   = 1##
        Frep     = 0.5##

        PlaceInitialProbes$  = "UNIFORM ON-AXIS"
        InitialAcceleration$ = "ZERO"
        RepositionFactor$    = "VARIABLE" '"FIXED"

        Np% = NumProbesPerDimension%*Nd%

        REDIM XiMin(1 TO Nd%), XiMax(1 TO Nd%)

        XiMin(1) = 0.5## : XiMax(1) = 3## 'dipole length, L, in Wavelengths
        XiMin(2) = 0##   : XiMax(2) = Pi2 'polar angle, Theta, in Radians

        REDIM StartingXiMin(1 TO Nd%), StartingXiMax(1 TO Nd%) : FOR i% = 1 TO Nd% : StartingXiMin(i%) = XiMin(i%) : StartingXiMax(i%) = XiMax(i%) :
NEXT i%

        NN% = FREEFILE : OPEN "INFILE.DAT" FOR OUTPUT AS #NN% : PRINT #NN%,"PBM1.NEC" : PRINT #NN%,"PBM1.OUT" : CLOSE #NN% 'NEC Input/Output Files

    CASE "PBM_2" '2-D

        AddNoiseToPBM2$ = "NO" '"YES" '"NO" '"YES"

        Nd%                    = 2
        NumProbesPerDimension% = 4 '20
        Np%                    = NumProbesPerDimension%*Nd%

        Nt&      = 100
        G        = 2##
        Alpha    = 2##
        Beta     = 2##
        DeltaT   = 1##
        Frep     = 0.5##

        PlaceInitialProbes$  = "UNIFORM ON-AXIS"
        InitialAcceleration$ = "ZERO"
        RepositionFactor$    = "VARIABLE" '"FIXED"

        Np% = NumProbesPerDimension%*Nd%

        REDIM XiMin(1 TO Nd%), XiMax(1 TO Nd%)

        XiMin(1) = 5## : XiMax(1) = 15## 'dipole separation, D, in Wavelengths
        XiMin(2) = 0## : XiMax(2) = Pi   'polar angle, Theta, in Radians

        REDIM StartingXiMin(1 TO Nd%), StartingXiMax(1 TO Nd%) : FOR i% = 1 TO Nd% : StartingXiMin(i%) = XiMin(i%) : StartingXiMax(i%) = XiMax(i%) :
NEXT i%

        NN% = FREEFILE : OPEN "INFILE.DAT" FOR OUTPUT AS #NN% : PRINT #NN%,"PBM2.NEC" : PRINT #NN%,"PBM2.OUT" : CLOSE #NN%

    CASE "PBM_3" '2-D

        Nd%                    = 2
        NumProbesPerDimension% = 4 '20
        Np%                    = NumProbesPerDimension%*Nd%

        Nt&      = 100
        G        = 2##
        Alpha    = 2##
        Beta     = 2##
        DeltaT   = 1##
        Frep     = 0.5##

        PlaceInitialProbes$  = "UNIFORM ON-AXIS"
        InitialAcceleration$ = "ZERO"
        RepositionFactor$    = "VARIABLE" '"FIXED"

        Np% = NumProbesPerDimension%*Nd%
```

```
            REDIM XiMin(1 TO Nd%), XiMax(1 TO Nd%)

            XiMin(1) = 0## : XiMax(1) = 4## 'Phase Parameter, Beta (0-4)
            XiMin(2) = 0## : XiMax(2) = Pi  'polar angle, Theta, in Radians

            REDIM StartingXiMin(1 TO Nd%), StartingXiMax(1 TO Nd%) : FOR i% = 1 TO Nd% : StartingXiMin(i%) = XiMin(i%) : StartingXiMax(i%) = XiMax(i%) :
NEXT i%

            NN% = FREEFILE : OPEN "INFILE.DAT" FOR OUTPUT AS #NN% : PRINT #NN%,"PBM3.NEC" : PRINT #NN%,"PBM3.OUT" : CLOSE #NN%

       CASE "PBM_4" '2-D

            Nd%                    = 2
            NumProbesPerDimension% = 4 '6 '2 '4 '20
            Np%                    = NumProbesPerDimension%*Nd%

            Nt&      = 100
            G        = 2##
            Alpha    = 2##
            Beta     = 2##
            DeltaT   = 1##
            Frep     = 0.5##

            PlaceInitialProbes$  = "UNIFORM ON-AXIS"
            InitialAcceleration$ = "ZERO"
            RepositionFactor$    = "VARIABLE" '"FIXED"

            Np% = NumProbesPerDimension%*Nd%

            REDIM XiMin(1 TO Nd%), XiMax(1 TO Nd%)

            XiMin(1) = 0.5##   : XiMax(1) = 1.5##  'ARM LENGTH (NOT Total Length), wavelengths (0.5-1.5)
            XiMin(2) = Pi/18## : XiMax(2) = Pi/2## 'Inner angle, Alpha, in Radians (Pi/18-Pi/2)

            REDIM StartingXiMin(1 TO Nd%), StartingXiMax(1 TO Nd%) : FOR i% = 1 TO Nd% : StartingXiMin(i%) = XiMin(i%) : StartingXiMax(i%) = XiMax(i%) :
NEXT i%

            NN% = FREEFILE : OPEN "INFILE.DAT" FOR OUTPUT AS #NN% : PRINT #NN%,"PBM4.NEC" : PRINT #NN%,"PBM4.OUT" : CLOSE #NN%

       CASE "PBM_5"

            NumCollinearElements%  = 6 '30 'EVEN or ODD: 6,7,10,13,16,24 used by PBM

            Nd%                    = NumCollinearElements% - 1
            NumProbesPerDimension% = 4 '20
            Np%                    = NumProbesPerDimension%*Nd%

            Nt&      = 100
            G        = 2##
            Alpha    = 2##
            Beta     = 2##
            DeltaT   = 1##
            Frep     = 0.5##

            PlaceInitialProbes$  = "UNIFORM ON-AXIS"
            InitialAcceleration$ = "ZERO"
            RepositionFactor$    = "VARIABLE" '"FIXED"

            Nd% = NumCollinearElements% - 1

            Np% = NumProbesPerDimension%*Nd%

            REDIM XiMin(1 TO Nd%), XiMax(1 TO Nd%) : FOR i% = 1 TO Nd% : XiMin(i%) = 0.5## : XiMax(i%) = 1.5## : NEXT i%

            REDIM StartingXiMin(1 TO Nd%), StartingXiMax(1 TO Nd%) : FOR i% = 1 TO Nd% : StartingXiMin(i%) = XiMin(i%) : StartingXiMax(i%) = XiMax(i%) :
NEXT i%

            NN% = FREEFILE : OPEN "INFILE.DAT" FOR OUTPUT AS #NN% : PRINT #NN%,"PBM5.NEC" : PRINT #NN%,"PBM5.OUT" : CLOSE #NN%

'      ================================================= END PBM BENCHMARKS =================================================

       CASE "BOWTIE" 'FREE SPACE LOADED BOWTIE

'           --------- Check for NEC Executable ----------

            IF DIR$("NEC41D_4K_053011.EXE") = "" THEN
                MSGBOX("WARNING! NEC4.1D executable NEC41D_4K_053011.EXE not found!"+CHR$(13)+CHR$(13)+_
                       "          RUN TERMINATED.")+CHR$(13)+CHR$(13)
                EXIT SUB
            END IF

            NumRadPattAngles% = 19' : NN% = FREEFILE : OPEN "NumPattAngles.DAT" FOR OUTPUT AS #NN% : PRINT #NN%, NumRadPattAngles% : CLOSE #NN% 'for use by
ReadNEC program

'           ------- Get Segmentation -------
'            IF DIR$("BowtieSeg.TXT") = "" THEN 'create BowtieSeg.TXT if it doesn't exist
'                    NN% = FREEFILE : OPEN "BowtieSeg.TXT" FOR OUTPUT AS NN% : PRINT #NN%,"Variable" : PRINT #NN%,"0.05" : CLOSE #NN%
'            END IF
'            NN% = FREEFILE : OPEN "BowtieSeg.TXT" FOR INPUT AS NN% : INPUT #NN%,BowtieSegmentLength$ : INPUT #NN%,BowtieSegmentLengthWvln : CLOSE #NN%
'            IF INSTR(UCASE$(BowtieSegmentLength$),"V") > 0 THEN BowtieSegmentLength$ = "VARIABLE"
'            IF INSTR(UCASE$(BowtieSegmentLength$),"F") > 0 THEN BowtieSegmentLength$ = "FIXED"
'            IF INSTR(UCASE$(BowtieSegmentLength$),"V") = 0 AND INSTR(UCASE$(BowtieSegmentLength$),"F") = 0 THEN BowtieSegmentLength$ = "VARIABLE"
'            IF BowtieSegmentLengthWvln < 0.02## OR BowtieSegmentLengthWvln > 0.2## THEN BowtieSegmentLengthWvln = 0.05## 'range 0.02-0.2 wavelengths

'           ----- Get Bowtie Fitness Function Coefficients -----  NOTE: OPTIONALLY USED IN FITNESS FUNCTION (SEE BOWTIE FUNCTION FOR ACTUAL CALCULATION)
            REDIM BowtieFitnessCoefficients(1 TO 3)
            IF DIR$("BowtieCoeff.TXT") = "" THEN 'create BowtieCoeff.TXT if it doesn't exist
                    NN% = FREEFILE : OPEN "BowtieCoeff.TXT" FOR OUTPUT AS NN% : PRINT #NN%,"40" : PRINT #NN%,"2" : PRINT #NN%,"3" : CLOSE #NN%
            END IF
            NN% = FREEFILE
            OPEN "BowtieCoeff.TXT" FOR INPUT AS NN%
                INPUT #NN%,BowtieFitnessCoefficients(1) : INPUT #NN%,BowtieFitnessCoefficients(2) : INPUT #NN%,BowtieFitnessCoefficients(3)
            CLOSE #NN%

            BowtieCoefficients$ = "A="+REMOVE$(STR$(BowtieFitnessCoefficients(1)),ANY" ")+_
                                  ", B="+REMOVE$(STR$(BowtieFitnessCoefficients(2)),ANY" ")+_
                                  ", C="+REMOVE$(STR$(BowtieFitnessCoefficients(3)),ANY" ")

'BOWTIE DECISION SPACE BOUNDARY ARRAY:

'    Array Element #                Design Variable
'    ---------------       -----------------------------------
'          1                Bowtie atm length (wavelengths)
'          2                Bowtie HALF angle (degrees)
'          3                Loading segment number #1
'          4                Loading resistance #1 (ohms)
'          5                Zo

            Nd% = 5 'FIXED DIMENSIONALITY BASED ON ABOVE DESIGN PARAMETERS

            REDIM XiMin(1 TO Nd%), XiMax(1 TO Nd%)

            XiMin(1) = 0.01## : XiMax(1) = 0.08## 'min/max arm length (METERS) (>~1/4-wave at 1 GHz)
            XiMin(2) = 10##   : XiMax(2) = 80##   'min/max HALF angle (degrees)
            XiMin(3) = 1      : XiMax(3) = 9      'min/max loading segment number #1 (NOTE FIXED SEGMENTATION AT 9 SEGS !!!)
            XiMin(4) = 1##    : XiMax(4) = 1000## 'min/max loading resistance, R1 (ohms)
            XiMin(5) = 50##   : XiMax(5) = 1000## 'Zo, ohms
```

```
            REDIM StartingXiMin(1 TO Nd%), StartingXiMax(1 TO Nd%)
            FOR i% = 1 TO Nd% : StartingXiMin(i%) = XiMin(i%) : StartingXiMax(i%) = XiMax(i%) : NEXT i%

            NN% = FREEFILE : OPEN "INFILE.DAT" FOR OUTPUT AS #NN% : PRINT #NN%,"BOWTIE.NEC" : PRINT #NN%,"BOWTIE.OUT" : CLOSE #NN%

'       ============================================================ END BOWTIE ====================================================

        CASE "YAGI" 'FREE SPACE YAGI ARRAY

'           --------- Check for NEC Executable ----------

            IF DIR$("NEC41D_4K_053011.EXE") = "" THEN
                MSGBOX("WARNING! NEC4.1D executable NEC41D_4K_053011.EXE not found!"+CHR$(13)+CHR$(13)+_
                       "          RUN TERMINATED.")+CHR$(13)+CHR$(13)
                EXIT SUB
            END IF

            NumRadPattAngles% = 2' : NN% = FREEFILE : OPEN "NumPattAngles.DAT" FOR OUTPUT AS #NN% : PRINT #NN%, NumRadPattAngles% : CLOSE #NN% 'for use by
ReadNEC program

'           ------- Get Segmentation -------
            IF DIR$("YagiSeg.TXT") = "" THEN 'create YagiSeg.TXT if it doesn't exist
                    NN% = FREEFILE : OPEN "YagiSeg.TXT" FOR OUTPUT AS NN% : PRINT #NN%,"Variable" : PRINT #NN%,"0.05" : CLOSE #NN%
            END IF
            NN% = FREEFILE : OPEN "YagiSeg.TXT" FOR INPUT AS NN% : INPUT #NN%,YagiSegmentLength$ : INPUT #NN%,YagiSegmentLengthWvln : CLOSE #NN%
            IF INSTR(UCASE$(YagiSegmentLength$),"V") > 0 THEN YagiSegmentLength$ = "VARIABLE"
            IF INSTR(UCASE$(YagiSegmentLength$),"F") > 0 THEN YagiSegmentLength$ = "FIXED"
            IF INSTR(UCASE$(YagiSegmentLength$),"V") = 0 AND INSTR(UCASE$(YagiSegmentLength$),"F") = 0 THEN YagiSegmentLength$ = "VARIABLE"
            IF YagiSegmentLengthWvln < 0.02## OR YagiSegmentLengthWvln > 0.2## THEN YagiSegmentLengthWvln = 0.05## 'range 0.02-0.2 wavelengths

'           ---------- Get Yagi Fitness Function Coefficients -----------
            REDIM YagiFitnessCoefficients(1 TO 4)
            IF DIR$("YagiCoeff.TXT") = "" THEN 'create YagiCoeff.TXT if it doesn't exist
'A=20, B=2, C=3, D=4.
                    NN% = FREEFILE : OPEN "YagiCoeff.TXT" FOR OUTPUT AS NN% : PRINT #NN%,"20" : PRINT #NN%,"2" : PRINT #NN%,"3" : PRINT #NN%,"4" : CLOSE
#NN%
            END IF
            NN% = FREEFILE
            OPEN "YagiCoeff.TXT" FOR INPUT AS NN%
                INPUT #NN%,YagiFitnessCoefficients(1) : INPUT #NN%,YagiFitnessCoefficients(2) : INPUT #NN%,YagiFitnessCoefficients(3) : INPUT
#NN%,YagiFitnessCoefficients(4)
            CLOSE #NN%

            YagiCoefficients$ = "A="+REMOVE$(STR$(YagiFitnessCoefficients(1)),ANY" ")+_
                                ", B="+REMOVE$(STR$(YagiFitnessCoefficients(2)),ANY" ")+_
                                ", C="+REMOVE$(STR$(YagiFitnessCoefficients(3)),ANY" ")+_
                                ", D="+REMOVE$(STR$(YagiFitnessCoefficients(4)),ANY" ")

'   ------------------- Alternate Fitness Coefficients ---------------

    c1 = 0.2## : c2 = 4# : c3 = 1## : c4 = 8## : c5 = 1## : c6 = 0.8##

    NN% = FREEFILE
    OPEN "Alt_Yagi_Coeff.DAT" FOR OUTPUT AS #NN%
        PRINT #NN%,"c1="+REMOVE$(STR$(c1),ANY" ")
        PRINT #NN%,"c2="+REMOVE$(STR$(c2),ANY" ")
        PRINT #NN%,"c3="+REMOVE$(STR$(c3),ANY" ")
        PRINT #NN%,"c4="+REMOVE$(STR$(c4),ANY" ")
        PRINT #NN%,"c5="+REMOVE$(STR$(c5),ANY" ")
        PRINT #NN%,"c6="+REMOVE$(STR$(c6),ANY" ")
    CLOSE #NN%

    Fit1$ = "CM "+REMOVE$(STR$(c1),ANY" ")+"*Gfwd(L)-"+REMOVE$(STR$(c2),ANY" ")+"*VSWR(L)+"+REMOVE$(STR$(c3),ANY" ")+"*Gfwd(M)-"+REMOVE$(STR$(c4),ANY"
")+"*VSWR(M)+"
    Fit2$ = "CM "+REMOVE$(STR$(c5),ANY" ")+"*Gfwd(U)-"+REMOVE$(STR$(c6),ANY" ")+"*VSWR(U)"

'           ----------------- # Array Elements -------------------
            A$ = INPUTBOX$("# Array Elements","YAGI ELEMENTS","6")

            NumYagiElements% = VAL(A$)

            Nd% =2*NumYagiElements% 'spacing/length are the two optimization parameters defining each element, AND Zo IS AS A DESIGN <<<VARIABLE>>>.
                                                                                                            '====================================
'YAGI DECISION SPACE BOUNDARY ARRAY:

'    Array Element #                             Design Variable
'    ---------------      ---------------------------------------------------------------------
'          1
'          TO               Yagi element spacing along boom from previous element, wavelengths
'   NumYagiElements-1

'   NumYagiElements
'          TO                               Yagi element length, wavelengths
'   2*NumYagiElements-1

'   2*NumYagiElements (Nd)                          <<<< Zo >>>>

            REDIM XiMin(1 TO Nd%), XiMax(1 TO Nd%)

            FOR i% = 1 TO NumYagiElements%-1 : XiMin(i%) = 0.1## : XiMax(i%) = 0.5## : NEXT i% 'min/max SPACING, wvln @ 299.8 MHz

            FOR i% = NumYagiElements% TO Nd%-1
                XiMin(i%) = 0.2## : XiMax(i%) = 0.6## 'min/max element TOTAL LENGTHS, wvln @ 299.8 MHz
            NEXT i%

            XiMin(Nd%) = 25## : XiMax(i%) = 250##    'min/max values of Zo (OHMS) 'USED FOR 07282011_001156 DESIGN

'            XiMin(Nd%) = 50## : XiMax(i%) = 50##    'min/max values of Zo (OHMS) 'TO FIX Zo AT 50 OHMS

            REDIM StartingXiMin(1 TO Nd%), StartingXiMax(1 TO Nd%)
            FOR i% = 1 TO Nd% : StartingXiMin(i%) = XiMin(i%) : StartingXiMax(i%) = XiMax(i%) : NEXT i%

            NN% = FREEFILE : OPEN "INFILE.DAT" FOR OUTPUT AS #NN% : PRINT #NN%,"YAGI.NEC" : PRINT #NN%,"YAGI.OUT" : CLOSE #NN%

'       ============================================================ END YAGI ====================================================

'   ========================================================================================
'   NOTE - DON'T FORGET TO ADD NEW TEST FUNCTIONS TO FUNCTION ObjectiveFunction() ABOVE !!
'   ========================================================================================

    END SELECT

    IF Nd% > 100 THEN Nt& = MIN(Nt&,200) 'to avoid array dimensioning problems

    DiagLength = 0## : FOR i% = 1 TO Nd% : DiagLength = DiagLength + (XiMax(i%)-XiMin(i%))^2 : NEXT i% : DiagLength = SQR(DiagLength) 'compute length of
decision space principal diagonal

END SUB 'GetFunctionRunParameters()

'--------------------------------

FUNCTION ParrottF4(R(),Nd%,p%,j&) 'Parrott F4 (1-D)

'MAXIMUM = 1 AT ~0.0796875... WITH ZERO OFFSET (SEEMS TO WORK BEST WITH JUST 3 PROBES, BUT NOT ALLOWED IN THIS VERSION...)

'References:

'Beasley, D., D. R. Bull, and R. R. Martin, "A Sequential Niche Technique for Multimodal
'Function Optimization," Evol. Comp. (MIT Press), vol. 1, no. 2, 1993, pp. 101-125
```

```
'(online at http://citeseer.ist.psu.edu/beasley93sequential.html).

'Parrott, D., and X. Li, "Locating and Tracking Multiple Dynamic Optima by a Particle Swarm
'Model Using Speciation," IEEE Trans. Evol. Computation, vol. 10, no. 4, Aug. 2006, pp. 440-458.


LOCAL Z, x, offset AS EXT

    offset = 0##

    x = R(p%,1,j&)

    Z = EXP(-2##*LOG(2##)*((x-0.08##-offset)/0.854##)^2)*(SIN(5##*Pi*((x-offset)^0.75##-0.05##)))^6 'WARNING! This is a NATURAL LOG, NOT Log10!!!

    ParrottF4 = Z

END FUNCTION 'ParrottF4()

'-----------------------------

FUNCTION SGO(R(),Nd%,p%,j&) 'SGO Function (2-D)

'MAXIMUM = ~130.8323226... @ ~(-2.8362075...,-2.8362075...) WITH ZERO OFFSET.

'Reference:

'Hsiao, Y., Chuang, C., Jiang, J., and Chien, C., "A Novel Optimization Algorithm: Space
'Gravitational Optimization," Proc. of 2005 IEEE International Conference on Systems, Man,
'and Cybernetics, 3, 2323-2328. (2005)

    LOCAL x1, x2, Z, t1, t2, SGOx1offset, SGOx2offset AS EXT

    SGOx1offset = 0## : SGOx2offset = 0##

'   SGOx1offset = 40## : SGOx2offset = 10##

    x1 = R(p%,1,j&) - SGOx1offset : x2 = R(p%,2,j&) - SGOx2offset

    t1 = x1^4 - 16##*x1^2 + 0.5##*x1 : t2 = x2^4 - 16##*x2^2 + 0.5##*x2

    Z = t1 + t2

    SGO = -Z

END FUNCTION 'SGO()

'------------------

FUNCTION GoldsteinPrice(R(),Nd%,p%,j&) 'Goldstein-Price Function (2-D)

'MAXIMUM = -3 @ (0,-1) WITH ZERO OFFSET.

'Reference:

'Cui, Z., Zeng, J., and Sun, G. (2006) 'A Fast Particle Swarm Optimization,' Int'l. J.
'Innovative Computing, Information and Control, vol. 2, no. 6, December, pp. 1365-1380.

    LOCAL Z, x1, x2, offset1, offset2, t1, t2 AS EXT

    offset1 = 0## : offset2 = 0##

'    offset1 = 20## : offset2 = -10##

    x1 = R(p%,1,j&)-offset1 : x2 = R(p%,2,j&)-offset2

    t1 = 1##+(x1+x2+1##)^2*(19##-14##*x1+3##*x1^2-14##*x2+6##*x1*x2+3##*x2^2)

    t2 = 30##+(2##*x1-3##*x2)^2*(18##-32##*x1+12##*x1^2+48##*x2-36##*x1*x2+27##*x2^2)

    Z  = t1*t2

    GoldsteinPrice = -Z

END FUNCTION 'GoldsteinPrice()

'-----------

FUNCTION StepFunction(R(),Nd%,p%,j&) 'Step Function (n-D)

'MAXIMUM VALUE = 0 @ [Offset]^n.

'Reference:

'Yao, X., Liu, Y., and Lin, G., "Evolutionary Programming Made Faster,"
'IEEE Trans. Evolutionary Computation, Vol. 3, No. 2, 82-102, Jul. 1999.


    LOCAL Offset, Z AS EXT

    LOCAL i%

    Z = 0## : Offset = 0## '75.123## '0##

    FOR i% = 1 TO Nd%

        IF Nd% = 2 AND i% = 1 THEN Offset = 75 '75##

        IF Nd% = 2 AND i% = 2 THEN Offset = 35 '30 '35##

        Z = Z + INT((R(p%,i%,j&)-Offset) + 0.5##)^2

    NEXT i%

    StepFunction = -Z

END FUNCTION 'StepFunction()

'-----------

FUNCTION Schwefel226(R(),Nd%,p%,j&) 'Schwefel Problem 2.26 (n-D)

'MAXIMUM = 12,569.5 @ [420.8687]^30 (30-D CASE).

'Reference:

'Yao, X., Liu, Y., and Lin, G., "Evolutionary Programming Made Faster,"
'IEEE Trans. Evolutionary Computation, Vol. 3, No. 2, 82-102, Jul. 1999.

    LOCAL Z, Xi AS EXT

    LOCAL i%

    Z = 0##

    FOR i% = 1 TO Nd%

        Xi = R(p%,i%,j&)

        Z = Z + Xi*SIN(SQR(ABS(Xi)))
```

```
    NEXT i%

    Schwefel226 = Z

END FUNCTION 'SCHWEFEL226()

'-----------

FUNCTION Colville(R(),Nd%,p%,j&) 'Colville Function (4-D)

'MAXIMUM = 0 @ (1,1,1,1) WITH ZERO OFFSET.

'Reference:

'Doo-Hyun, and Se-Young, O., "A New Mutation Rule for Evolutionary Programming Motivated from
'Backpropagation Learning," IEEE Trans. Evolutionary Computation, Vol. 4, No. 2, pp. 188-190,
'July 2000.

    LOCAL Z, x1, x2, x3, x4, offset AS EXT

    offset = 0## '7.123##

    x1 = R(p%,1,j&)-offset : x2 = R(p%,2,j&)-offset : x3 = R(p%,3,j&)-offset : x4 = R(p%,4,j&)-offset

    Z =  100##*(x2-x1^2)^2 + (1##-x1)^2  + _
          90##*(x4-x3^2)^2 + (1##-x3)^2  + _
        10.1##*((x2-1##)^2 + (x4-1##)^2) + _
        19.8##*(x2-1##)*(x4-1##)

    Colville = -Z

END FUNCTION 'Colville()

'-----------

FUNCTION Griewank(R(),Nd%,p%,j&) 'Griewank (n-D)

'Max of zero at (0,...,0)

'Reference: Yao, X., Liu, Y., and Lin, G., "Evolutionary Programming Made Faster,"
'IEEE Trans. Evolutionary Computation, Vol. 3, No. 2, 82-102, Jul. 1999.

    LOCAL Offset, Sum, Prod, Z, Xi AS EXT

    LOCAL i%

    Sum = 0## : Prod = 1##

    Offset = 75.123##

    FOR i% = 1 TO Nd%

        Xi = R(p%,i%,j&) - Offset

        Sum = Sum + Xi^2

        Prod = Prod*COS(Xi/SQR(i%))

    NEXT i%

    Z = Sum/4000## - Prod + 1##

    Griewank = -Z

END FUNCTION 'Griewank()

'-----------

FUNCTION Himmelblau(R(),Nd%,p%,j&) 'Himmelblau (2-D)

    LOCAL Z, x1, x2, offset AS EXT

    offset = 0##

    x1 = R(p%,1,j&)-offset : x2 = R(p%,2,j&)-offset

    Z = 200## - (x1^2 + x2 -11##)^2 - (x1+x2^2-7##)^2

    Himmelblau = Z

END FUNCTION 'Himmelblau()

'-----------

FUNCTION Rosenbrock(R(),Nd%,p%,j&) 'Rosenbrock (n-D)

'MAXIMUM = 0 @ [1,...,1]^n (n-D CASE).

'Reference: Yao, X., Liu, Y., and Lin, G., "Evolutionary Programming Made Faster,"
'IEEE Trans. Evolutionary Computation, Vol. 3, No. 2, 82-102, Jul. 1999.

    LOCAL Z, Xi, Xi1 AS EXT

    LOCAL i%

    Z = 0##

    FOR i% = 1 TO Nd%-1

        Xi  = R(p%,i%,j&) : Xi1 = R(p%,i%+1,j&)

        Z = Z + 100##*(Xi1-Xi^2)^2 + (Xi-1##)^2

    NEXT i%

    Rosenbrock = -Z

END FUNCTION 'ROSENBROCK()

'-----------

FUNCTION Sphere(R(),Nd%,p%,j&) 'Sphere (n-D)

'MAXIMUM = 0 @ [0,...,0]^n (n-D CASE).

'Reference: Yao, X., Liu, Y., and Lin, G., "Evolutionary Programming Made Faster,"
'IEEE Trans. Evolutionary Computation, Vol. 3, No. 2, 82-102, Jul. 1999.

    LOCAL Z, Xi, Xi1 AS EXT

    LOCAL i%

    Z = 0##

    FOR i% = 1 TO Nd%
```

```
        Xi  = R(p%,i%,j&)
        Z = Z + Xi^2
    NEXT i%
    Sphere = -Z
END FUNCTION 'SPHERE()
'-----------
FUNCTION HimmelblauNLO(R(),Nd%,p%,j&) 'Himmelblau non-linear optimization (5-D)
'MAXIMUM ~ 31025.5562644972 @ (78.0,33.0,27.0709971052,45.0,44.9692425501)
'Reference: "Constrained Optimization using CODEQ," Mahamed G.H. Omran & Ayed Salman,
'Chaos, Solitons and Tractals, 42(2009), 662-668
    LOCAL Z, x1, x2, x3, x4, x5, g1, g2, g3 AS EXT
    Z = 1E4200
    x1 = R(p%,1,j&) : x2 = R(p%,2,j&) : x3 = R(p%,3,j&) : x4 = R(p%,4,j&) : x5 = R(p%,5,j&)
    g1 = 85.334407## + 0.0056858##*x2*x5 + 0.00026##*x1*x4   - 0.0022053##*x3*x5
    g2 = 80.51249##  + 0.0071317##*x2*x5 + 0.0029955##*x1*x2 + 0.0021813##*x3*x3
    g3 = 9.300961##  + 0.0047026##*x3*x5 + 0.0012547##*x1*x3 + 0.0019085##*x3*x4
    IF g1## < 0 OR g1 > 92## OR g2 < 90## OR g2 > 110## OR g3 < 20## OR g3 > 25## THEN GOTO ExitHimmelblauNLO
    Z = 5.3578547##*x3*x3 + 0.8356891##*x1*x5 + 37.29329##*x1 - 40792.141##
ExitHimmelblauNLO:
    HimmelblauNLO = -Z
END FUNCTION 'HimmelblauNLO()
'-----------
FUNCTION Tripod(R(),Nd%,p%,j&) 'Tripod (2-D)
'MAXIMUM = 0 at (0,-50)
'Reference: "Appendix: A mini-benchmark," Maurice Clerc
    LOCAL Z, x1, x2, s1, s2, t1, t2, t3 AS EXT
    x1 = R(p%,1,j&) : x2 = R(p%,2,j&)
    s1 = Sign(x1) : s2 = Sign(x2)
    t1 = (1##-s2)*(ABS(x1)+ABS(x2+50##))
    t2 = 0.5##*(1##+s2)*(1##-s1)*(1##+ABS(x1+50##)+ABS(x2-50##))
    t3 = (1##+s1)*(2##+ABS(x1-50##)+ABS(x2-50##))
    Z = 0.5##*(t1 + t2 + t3)
    Tripod = -Z
END FUNCTION 'Tripod()
'---------------------
FUNCTION Sign(X)
LOCAL Z AS EXT
    Z = 1## : IF X =< 0## THEN Z = -1##
    Sign = Z
END FUNCTION
'-----------
FUNCTION RosenbrockF6(R(),Nd%,p%,j&) 'Rosenbrock F6 (10-D)
'WARNING !!  03-19-10  THIS FUNCTION CONTAINS ERRORS.  SEE CLERC'S EMAIL!
'MAXIMUM = 394 at (0,-50)
'Reference: "Appendix: A mini-benchmark," Maurice Clerc (NOTE: Uses his notation...)
    LOCAL Z, Xi, Xi1, Zi, Zi1, Sum AS EXT
    LOCAL i%
    Sum = 0##
    FOR i% = 2 TO Nd%
        Xi = R(p%,i%,j&) : Xi1 = R(p%,i%-1,j&)
        Zi = Xi - XiOffset(i%) + 1## : Zi1 = Xi1 - XiOffset(i%-1) + 1##
        Sum = Sum + 100##*(Zi1^2-Zi)^2  + (Zi1-1##)^2
    NEXT i%
    Z = 390## + Sum
    RosenbrockF6 = -Z
END FUNCTION 'RosenbrockF6()
'---------------------
FUNCTION CompressionSpring(R(),Nd%,p%,j&) 'Compression Spring (3-D)
'MAXIMUM = 394 at (0,-50)
'Reference: "Appendix: A mini-benchmark," Maurice Clerc (NOTE: Uses his notation...)
LOCAL Z, x1, x2, x3, g1, g2, g3, g4, g5, Cf, Fmax, S, Lf, Lmax, SigP, SigPM, Fp, K, SigW AS EXT
    Z = 1E4200
    x1 = ROUND(R(p%,1,j&),0) : x2 = R(p%,2,j&)  : x3 = ROUND(R(p%,3,j&),3)
    Cf = 1## + 0.75##*x3/(x2-x3)+0.615##*x3/x2
    Fmax = 1000## : S = 189000## : Lmax = 14## : SigPM = 6## : Fp = 300## : SigW = 1.25##
```

```
    K = 11.5##*1E6*x3^4/(8##*x1*x2^3)
    Lf = Fmax/K + 1.05##*(x1+2##)*x3
    SigP = Fp/K
    g1 = 8##*Cf*Fmax*x2/(Pi*x3^3) - S
    g2 = Lf - Lmax
    g3 = SigP - SigPM
    g4 = SigP -Fp/K 'WARNING!  03-19-10.  THIS IS SATISFIED EXACTLY (SEE CLERC'S EMAIL - TYPO IN HIS BENCHMARKS)
    g5 = SigW - (Fmax-Fp)/K
    IF g1 > 0## OR g2 > 0## OR g3 > 0## OR g4 > 0## OR g5 > 0## THEN GOTO ExitCompressionSpring
    Z = Pi^2*x2*x3^2*(x1+1##)/4##
ExitCompressionSpring:
    CompressionSpring = -Z
END FUNCTION 'CompressionSpring
'---------------------
FUNCTION GearTrain(R(),Nd%,p%,j&) 'GearTrain (4-D)
'MAXIMUM = 394 at (0,-50)
'Reference: "Appendix: A mini-benchmark," Maurice Clerc (NOTE: Uses his notation...)
LOCAL Z, x1, x2, x3, x4 AS EXT
    x1 = ROUND(R(p%,1,j&),0) : x2 = ROUND(R(p%,2,j&),0)
    x3 = ROUND(R(p%,3,j&),0) : x4 = ROUND(R(p%,4,j&),0)
    Z = (1##/6.931##-x1*x2/(x3*x4))^2
    GearTrain = -Z
END FUNCTION 'GearTrain
'---------------------
FUNCTION F1(R(),Nd%,p%,j&) 'F1 (n-D)
'MAXIMUM = ZERO (n-D CASE).
'Reference:
    LOCAL Z, Xi AS EXT
    LOCAL i%
    Z = 0##
    FOR i% = 1 TO Nd%
        Xi = R(p%,i%,j&)
        Z = Z + Xi^2
    NEXT i%
    F1 = -Z
END FUNCTION 'F1
'-----------
FUNCTION F2(R(),Nd%,p%,j&) 'F2 (n-D)
'MAXIMUM = ZERO (n-D CASE).
'Reference:
    LOCAL Sum, prod, Z, Xi AS EXT
    LOCAL i%
    Z = 0## : Sum = 0## : Prod = 1##
    FOR i% = 1 TO Nd%
        Xi = R(p%,i%,j&)
        Sum  = Sum+ ABS(Xi)
        Prod = Prod*ABS(Xi)
    NEXT i%
    Z = Sum + Prod
    F2 = -Z
END FUNCTION 'F2
'-----------
FUNCTION F3(R(),Nd%,p%,j&) 'F3 (n-D)
'MAXIMUM = ZERO (n-D CASE).
'Reference:
    LOCAL Z, Xk, Sum AS EXT
    LOCAL i%, k%
    Z = 0##
    FOR i% = 1 TO Nd%
        Sum = 0##
        FOR k% = 1 TO i%
            Xk = R(p%,k%,j&)
            Sum = Sum + Xk
        NEXT k%
        Z = Z + Sum^2
```

```
    NEXT i%
    F3 = -Z
END FUNCTION 'F3

'-----------
FUNCTION F4(R(),Nd%,p%,j&) 'F4 (n-D)
'MAXIMUM = ZERO (n-D CASE).
'Reference:
    LOCAL Z, Xi, MaxXi AS EXT
    LOCAL i%
    MaxXi = -1E4200
    FOR i% = 1 TO Nd%
        Xi = R(p%,i%,j&)
        IF ABS(Xi) >= MaxXi THEN MaxXi = ABS(Xi)
    NEXT i%
    F4 = -MaxXi
END FUNCTION 'F4
'-----------
FUNCTION F5(R(),Nd%,p%,j&) 'F5 (n-D)
'MAXIMUM = ZERO (n-D CASE).
'Reference:
    LOCAL Z, Xi, XiPlus1 AS EXT
    LOCAL i%
    Z = 0##
    FOR i% = 1 TO Nd%-1
        Xi      = R(p%,i%,j&)
        XiPlus1 = R(p%,i%+1,j&)
        Z = Z + (100##*(XiPlus1-Xi^2)^2+(Xi-1##))^2
    NEXT i%
    F5 = -Z
END FUNCTION 'F5
'-----------
FUNCTION F6(R(),Nd%,p%,j&) 'F6 (n-D STEP)
'MAXIMUM VALUE = 0 @ [Offset]^n.
'Reference:
'Yao, X., Liu, Y., and Lin, G., “Evolutionary Programming Made Faster,”
'IEEE Trans. Evolutionary Computation, Vol. 3, No. 2, 82-102, Jul. 1999.

    LOCAL Z AS EXT
    LOCAL i%
    Z = 0##
    FOR i% = 1 TO Nd%
        Z = Z + INT(R(p%,i%,j&) + 0.5##)^2
    NEXT i%
    F6 = -Z
END FUNCTION 'F6
'-----------
FUNCTION F7(R(),Nd%,p%,j&) 'F7
'MAXIMUM VALUE = 0 @ [Offset]^n.
'Reference:
'Yao, X., Liu, Y., and Lin, G., “Evolutionary Programming Made Faster,”
'IEEE Trans. Evolutionary Computation, Vol. 3, No. 2, 82-102, Jul. 1999.

    LOCAL Z, Xi AS EXT
    LOCAL i%
    Z = 0##
    FOR i% = 1 TO Nd%
        Xi = R(p%,i%,j&)
        Z = Z + i%*Xi^4
    NEXT i%
    F7 = -Z - RandomNum(0##,1##)
END FUNCTION 'F7
'-----------
FUNCTION F8(R(),Nd%,p%,j&) '(n-D) F8 [Schwefel Problem 2.26]
'MAXIMUM = 12,569.5 @ [420.8687]^30 (30-D CASE).
'Reference:
'Yao, X., Liu, Y., and Lin, G., “Evolutionary Programming Made Faster,”
```

```
'IEEE Trans. Evolutionary Computation, Vol. 3, No. 2, 82-102, Jul. 1999.
    LOCAL Z, Xi AS EXT
    LOCAL i%
    Z = 0##
    FOR i% = 1 TO Nd%
        Xi = R(p%,i%,j&)
        Z  = Z - Xi*SIN(SQR(ABS(Xi)))
    NEXT i%
    F8 = -Z
END FUNCTION 'F8
'-----------
FUNCTION F9(R(),Nd%,p%,j&) '(n-D) F9 [Rastrigin]
'MAXIMUM = ZERO (n-D CASE).
'Reference:
'Yao, X., Liu, Y., and Lin, G., "Evolutionary Programming Made Faster,"
'IEEE Trans. Evolutionary Computation, Vol. 3, No. 2, 82-102, Jul. 1999.
    LOCAL Z, Xi AS EXT
    LOCAL i%
    Z = 0##
    FOR i% = 1 TO Nd%
        Xi = R(p%,i%,j&)
        Z  = Z + (Xi^2 - 10##*COS(TwoPi*Xi) + 10##)^2
    NEXT i%
    F9 = -Z
END FUNCTION 'F9
'-----------
FUNCTION F10(R(),Nd%,p%,j&) '(n-D) F10 [Ackley's Function]
'MAXIMUM = ZERO (n-D CASE).
'Reference:
'Yao, X., Liu, Y., and Lin, G., "Evolutionary Programming Made Faster,"
'IEEE Trans. Evolutionary Computation, Vol. 3, No. 2, 82-102, Jul. 1999.
    LOCAL Z, Xi, Sum1, Sum2 AS EXT
    LOCAL i%
    Z = 0## : Sum1 = 0## : Sum2 = 0##
    FOR i% = 1 TO Nd%
        Xi   = R(p%,i%,j&)
        Sum1 = Sum1 + Xi^2
        Sum2 = Sum2 + COS(TwoPi*Xi)
    NEXT i%
    Z = -20##*EXP(-0.2##*SQR(Sum1/Nd%)) - EXP(Sum2/Nd%) + 20## + e
    F10 = -Z
END FUNCTION 'F10
'-----------
FUNCTION F11(R(),Nd%,p%,j&) '(n-D) F11
'MAXIMUM = ZERO (n-D CASE).
'Reference:
'Yao, X., Liu, Y., and Lin, G., "Evolutionary Programming Made Faster,"
'IEEE Trans. Evolutionary Computation, Vol. 3, No. 2, 82-102, Jul. 1999.
    LOCAL Z, Xi, Sum, Prod AS EXT
    LOCAL i%
    Z = 0## : Sum = 0## : Prod = 1##
    FOR i% = 1 TO Nd%
        Xi   = R(p%,i%,j&)
        Sum  = Sum + (Xi-100##)^2
        Prod = Prod*COS((Xi-100##)/SQR(i%))
    NEXT i%
    Z = Sum/4000## - Prod + 1##
    F11 = -Z
END FUNCTION 'F11
'-----
FUNCTION u(Xi,a,k,m)
LOCAL Z AS EXT
    Z = 0##
    SELECT CASE Xi
        CASE > a  : Z = k*(Xi-a)^m
        CASE < -a : Z = k*(-Xi-a)^m
```

```
    END SELECT
    u = Z
END FUNCTION
'-----------
FUNCTION F12(R(),Nd%,p%,j&) '(n-D) F12, Penalized #1
'Ref: Yao(1999).  Max=0 @ (-1,-1,...,-1), -50=<Xi=<50.
    LOCAL Offset, Sum1, Sum2, Z, X1, Y1, Xn, Yn, Xi, Yi, XiPlus1, YiPlus1 AS EXT
    LOCAL i%, m%, A$
    X1 = R(p%,1,j&)   : Y1 = 1## + (X1+1##)/4##
    Xn = R(p%,Nd%,j&) : Yn = 1## + (Xn+1##)/4##
    Sum1 = 0##
    FOR i% = 1 TO Nd%-1
        Xi      = R(p%,i%,j&)  : Yi      = 1## + (Xi+1##)/4##
        XiPlus1 = R(p%,i%+1,j&): YiPlus1 = 1## + (XiPlus1+1##)/4##
        Sum1 = Sum1 + (Yi-1##)^2*(1##+10##*(SIN(Pi*YiPlus1))^2)
    NEXT i%
    Sum1 = Sum1 + 10##*(SIN(Pi*Y1))^2 + (Yn-1##)^2
    Sum1 = Pi*Sum1/Nd%
    Sum2 = 0##
    FOR i% = 1 TO Nd%
        Xi = R(p%,i%,j&)
        Sum2 = Sum2 + u(Xi,10##,100##,4##)
    NEXT i%
    Z = Sum1 + Sum2
    F12 = -Z
END FUNCTION 'F12()
'------------------
FUNCTION F13(R(),Nd%,p%,j&) '(n-D) F13, Penalized #2
'Ref: Yao(1999).  Max=0 @ (1,1,...,1), -50=<Xi=<50.
    LOCAL Offset, Sum1, Sum2, Z, Xi, Xn, XiPlus1, X1 AS EXT
    LOCAL i%, m%, A$
    X1 = R(p%,1,j&) : Xn = R(p%,Nd%,j&)
    Sum1 = 0##
    FOR i% = 1 TO Nd%-1
        Xi  = R(p%,i%,j&) : XiPlus1 = R(p%,i%+1,j&)
        Sum1 = Sum1 + (Xi-1##)^2*(1##+(SIN(3##*Pi*XiPlus1))^2)
    NEXT i%
    Sum1 = Sum1 + (SIN(Pi*3##*X1))^2 +(Xn-1##)^2*(1##+(SIN(TwoPi*Xn))^2)
    Sum2 = 0##
    FOR i% = 1 TO Nd%
        Xi = R(p%,i%,j&)
        Sum2 = Sum2 + u(Xi,5##,100##,4##)
    NEXT i%
    Z = Sum1/10## + Sum2
    F13 = -Z
END FUNCTION 'F13()
'------------------
SUB FillArrayAij  'needed for function F14, Shekel's Foxholes

    Aij(1,1)=-32##  : Aij(1,2)=-16##  : Aij(1,3)=0##  : Aij(1,4)=16##  : Aij(1,5)=32##
    Aij(1,6)=-32##  : Aij(1,7)=-16##  : Aij(1,8)=0##  : Aij(1,9)=16##  : Aij(1,10)=32##
    Aij(1,11)=-32## : Aij(1,12)=-16## : Aij(1,13)=0## : Aij(1,14)=16## : Aij(1,15)=32##
    Aij(1,16)=-32## : Aij(1,17)=-16## : Aij(1,18)=0## : Aij(1,19)=16## : Aij(1,20)=32##
    Aij(1,21)=-32## : Aij(1,22)=-16## : Aij(1,23)=0## : Aij(1,24)=16## : Aij(1,25)=32##

    Aij(2,1)=-32##  : Aij(2,2)=-32## : Aij(2,3)=-32## : Aij(2,4)=-32## : Aij(2,5)=-32##
    Aij(2,6)=-16##  : Aij(2,7)=-16## : Aij(2,8)=-16## : Aij(2,9)=-16## : Aij(2,10)=-16##
    Aij(2,11)=0##   : Aij(2,12)=0##  : Aij(2,13)=0##  : Aij(2,14)=0##  : Aij(2,15)=0##
    Aij(2,16)=16##  : Aij(2,17)=16## : Aij(2,18)=16## : Aij(2,19)=16## : Aij(2,20)=16##
    Aij(2,21)=32##  : Aij(2,22)=32## : Aij(2,23)=32## : Aij(2,24)=32## : Aij(2,25)=32#

END SUB
'-----
FUNCTION F14(R(),Nd%,p%,j&) 'F14 (2-D) Shekel's Foxholes (INVERTED...)
    LOCAL Sum1, Sum2, Z, Xi AS EXT
    LOCAL i%, jj%
    Sum1 = 0##
    FOR jj% = 1 TO 25
        Sum2 = 0##
        FOR i% = 1 TO 2
            Xi = R(p%,i%,j&)
            Sum2 = Sum2 + (Xi-Aij(i%,jj%))^6
```

```
        NEXT i%
        Sum1 = Sum1 + 1##/(jj%+Sum2)
    NEXT j%
    Z = 1##/(0.002##+Sum1)
    F14 = -Z
END FUNCTION 'F14
'-----------
FUNCTION F16(R(),Nd%,p%,j&) 'F16 (2-D) 6-Hump Camel-Back
    LOCAL x1, x2, Z AS EXT
    x1 = R(p%,1,j&) : x2 = R(p%,2,j&)
    Z = 4##*x1^2 - 2.1##*x1^4 + x1^6/3## + x1*x2 - 4*x2^2 + 4*x2^4
    F16 = -Z
END FUNCTION 'F16
'-----------
FUNCTION F15(R(),Nd%,p%,j&) 'F15 (4-D) Kowalik's Function
'Global maximum = -0.0003075 @ (0.1928,0.1908,0.1231,0.1358)
    LOCAL x1, x2, x3, x4, Num, Denom, Z, Aj(), Bj() AS EXT
    LOCAL jj%
    REDIM Aj(1 TO 11), Bj(1 TO 11)
    Aj(1)  = 0.1957## : Bj(1)  = 1##/0.25##
    Aj(2)  = 0.1947## : Bj(2)  = 1##/0.50##
    Aj(3)  = 0.1735## : Bj(3)  = 1##/1.00##
    Aj(4)  = 0.1600## : Bj(4)  = 1##/2.00##
    Aj(5)  = 0.0844## : Bj(5)  = 1##/4.00##
    Aj(6)  = 0.0627## : Bj(6)  = 1##/6.00##
    Aj(7)  = 0.0456## : Bj(7)  = 1##/8.00##
    Aj(8)  = 0.0342## : Bj(8)  = 1##/10.0##
    Aj(9)  = 0.0323## : Bj(9)  = 1##/12.0##
    Aj(10) = 0.0235## : Bj(10) = 1##/14.0##
    Aj(11) = 0.0246## : Bj(11) = 1##/16.0##
    Z = 0##
    x1 = R(p%,1,j&) : x2 = R(p%,2,j&) : x3 = R(p%,3,j&) : x4 = R(p%,4,j&)
    FOR jj% = 1 TO 11
       Num   = x1*(Bj(jj%)^2+Bj(jj%)*x2)
       Denom = Bj(jj%)^2+Bj(jj%)*x3+x4
       Z = Z + (Aj(jj%)-Num/Denom)^2
    NEXT jj%
    F15 = -Z
END FUNCTION 'F15
'-----------
FUNCTION F17(R(),Nd%,p%,j&) 'F17, (2-D) Branin
'Global maximum = -0.398 @ (-3.142.12.275), (3.142,2.275), (9.425,2.425)
    LOCAL x1, x2, Z AS EXT
    x1 = R(p%,1,j&) : x2 = R(p%,2,j&)
    Z = (x2-5.1##*x1^2/(4##*Pi^2)+5##*x1/Pi-6##)^2 + 10##*(1##-1##/(8##*Pi))*COS(x1) + 10##
    F17 = -Z
END FUNCTION 'F17
'-----------
FUNCTION F18(R(),Nd%,p%,j&) 'Goldstein-Price 2-D Test Function
'Global maximum = -3 @ (0,-1)
    LOCAL Z, x1, x2, t1, t2 AS EXT
    x1 = R(p%,1,j&) : x2 = R(p%,2,j&)
    t1 = 1##+(x1+x2+1##)^2*(19##-14##*x1+3##*x1^2-14##*x2+6##*x1*x2+3##*x2^2)
    t2 = 30##+(2##*x1-3##*x2)^2*(18##-32##*x1+12##*x1^2+48##*x2-36##*x1*x2+27##*x2^2)
    Z  = t1*t2
    F18 = -Z
END FUNCTION 'F18()
'-----------
FUNCTION F19(R(),Nd%,p%,j&) 'F19 (3-D) Hartman's Family #1
'Global maximum = 3.86 @ (0.114,0.556,0.852)
    LOCAL Xi, Z, Sum, Aji(), Cj(), Pji() AS EXT
    LOCAL i%, jj%, m%
    REDIM Aji(1 TO 4, 1 TO 3), Cj(1 TO 4), Pji(1 TO 4, 1 TO 3)
    Aji(1,1) = 3.0## : Aji(1,2) = 10## : Aji(1,3) = 30## : Cj(1) = 1.0##
    Aji(2,1) = 0.1## : Aji(2,2) = 10## : Aji(2,3) = 35## : Cj(2) = 1.2##
    Aji(3,1) = 3.0## : Aji(3,2) = 10## : Aji(3,3) = 30## : Cj(3) = 3.0##
    Aji(4,1) = 0.1## : Aji(4,2) = 10## : Aji(4,3) = 35## : Cj(4) = 3.2##
    Pji(1,1) = 0.36890## : Pji(1,2) = 0.1170## : Pji(1,3) = 0.2673##
    Pji(2,1) = 0.46990## : Pji(2,2) = 0.4387## : Pji(2,3) = 0.7470##
    Pji(3,1) = 0.10910## : Pji(3,2) = 0.8732## : Pji(3,3) = 0.5547##
    Pji(4,1) = 0.03815## : Pji(4,2) = 0.5743## : Pji(4,3) = 0.8828##
    Z = 0##
```

```
    FOR jj% = 1 TO 4

        Sum = 0##

        FOR i% = 1 TO 3

            Xi = R(p%,i%,j&)

            Sum = Sum + Aji(jj%,i%)*(Xi-Pji(jj%,i%))^2

        NEXT i%

        Z = Z + Cj(jj%)*EXP(-Sum)

    NEXT jj%

    F19 = Z

END FUNCTION 'F19

'-----------

FUNCTION F20(R(),Nd%,p%,j&) 'F20 (6-D) Hartman's Family #2

'Global maximum = 3.32 @ (0.201,0.150,0.477,0.275,0.311,0.657)

    LOCAL Xi, Z, Sum, Aji(), Cj(), Pji() AS EXT

    LOCAL i%, jj%, m%

    REDIM Aji(1 TO 4, 1 TO 6), Cj(1 TO 4), Pji(1 TO 4, 1 TO 6)

    Aji(1,1) = 10.0## : Aji(1,2) = 3.00## : Aji(1,3) = 17.0## : Cj(1) = 1.0##
    Aji(2,1) = 0.05## : Aji(2,2) = 10.0## : Aji(2,3) = 17.0## : Cj(2) = 1.2##
    Aji(3,1) = 3.00## : Aji(3,2) = 3.50## : Aji(3,3) = 1.70## : Cj(3) = 3.0##
    Aji(4,1) = 17.0## : Aji(4,2) = 8.00## : Aji(4,3) = 0.05## : Cj(4) = 3.2##

    Aji(1,4) = 3.5## : Aji(1,5) = 1.7## : Aji(1,6) =  8##
    Aji(2,4) = 0.1## : Aji(2,5) =   8## : Aji(2,6) = 14##
    Aji(3,4) =  10## : Aji(3,5) =  17## : Aji(3,6) =  8##
    Aji(4,4) =  10## : Aji(4,5) = 0.1## : Aji(4,6) = 14##

    Pji(1,1) = 0.13120## : Pji(1,2) = 0.1696## : Pji(1,3) = 0.5569##
    Pji(2,1) = 0.23290## : Pji(2,2) = 0.4135## : Pji(2,3) = 0.8307##
    Pji(3,1) = 0.23480## : Pji(3,2) = 0.1415## : Pji(3,3) = 0.3522##
    Pji(4,1) = 0.40470## : Pji(4,2) = 0.8828## : Pji(4,3) = 0.8732##

    Pji(1,4) = 0.01240## : Pji(1,5) = 0.8283## : Pji(1,6) = 0.5886##
    Pji(2,4) = 0.37360## : Pji(2,5) = 0.1004## : Pji(2,6) = 0.9991##
    Pji(3,4) = 0.28830## : Pji(3,5) = 0.3047## : Pji(3,6) = 0.6650##
    Pji(4,4) = 0.57430## : Pji(4,5) = 0.1091## : Pji(4,6) = 0.0381##

    Z = 0##

    FOR jj% = 1 TO 4

        Sum = 0##

        FOR i% = 1 TO 6

            Xi = R(p%,i%,j&)

            Sum = Sum + Aji(jj%,i%)*(Xi-Pji(jj%,i%))^2

        NEXT i%

        Z = Z + Cj(jj%)*EXP(-Sum)

    NEXT jj%

    F20 = Z

END FUNCTION 'F20

'-----------

FUNCTION F21(R(),Nd%,p%,j&) 'F21 (4-D) Shekel's Family m=5

'Global maximum = 10

    LOCAL Xi, Z, Sum, Aji(), Cj() AS EXT

    LOCAL i%, jj%, m%

    m% = 5 : REDIM Aji(1 TO m%, 1 TO 4), Cj(1 TO m%)

    Aji(1,1)  = 4## : Aji(1,2)  =   4## : Aji(1,3)  = 4## : Aji(1,4)  =   4## : Cj(1)  = 0.1##
    Aji(2,1)  = 1## : Aji(2,2)  =   1## : Aji(2,3)  = 1## : Aji(2,4)  =   1## : Cj(2)  = 0.2##
    Aji(3,1)  = 8## : Aji(3,2)  =   8## : Aji(3,3)  = 8## : Aji(3,4)  =   8## : Cj(3)  = 0.2##
    Aji(4,1)  = 6## : Aji(4,2)  =   6## : Aji(4,3)  = 6## : Aji(4,4)  =   6## : Cj(4)  = 0.4##
    Aji(5,1)  = 3## : Aji(5,2)  =   7## : Aji(5,3)  = 3## : Aji(5,4)  =   7## : Cj(5)  = 0.4##

    Z = 0##

    FOR jj% = 1 TO m%  'NOTE:  Index jj% is used to avoid same variable name as j&

        Sum = 0##

        FOR i% = 1 TO 4 'Shekel's family is 4-D only

            Xi = R(p%,i%,j&)

            Sum = Sum + (Xi-Aji(jj%,i%))^2

        NEXT i%

        Z = Z + 1##/(Sum + Cj(jj%))

    NEXT jj%

    F21 = Z

END FUNCTION 'F21

'-----------

FUNCTION F22(R(),Nd%,p%,j&) 'F22 (4-D) Shekel's Family m=7

'Global maximum = 10

    LOCAL Xi, Z, Sum, Aji(), Cj() AS EXT

    LOCAL i%, jj%, m%

    m% = 7 : REDIM Aji(1 TO m%, 1 TO 4), Cj(1 TO m%)

    Aji(1,1)  = 4## : Aji(1,2)  =   4## : Aji(1,3)  = 4## : Aji(1,4)  =   4## : Cj(1)  = 0.1##
    Aji(2,1)  = 1## : Aji(2,2)  =   1## : Aji(2,3)  = 1## : Aji(2,4)  =   1## : Cj(2)  = 0.2##
    Aji(3,1)  = 8## : Aji(3,2)  =   8## : Aji(3,3)  = 8## : Aji(3,4)  =   8## : Cj(3)  = 0.2##
```

```
    Aji(4,1)  = 6## : Aji(4,2)  =   6## : Aji(4,3)  = 6## : Aji(4,4)  =   6## : Cj(4)  = 0.4##
    Aji(5,1)  = 3## : Aji(5,2)  =   7## : Aji(5,3)  = 3## : Aji(5,4)  =   7## : Cj(5)  = 0.4##
    Aji(6,1)  = 2## : Aji(6,2)  =   9## : Aji(6,3)  = 2## : Aji(6,4)  =   9## : Cj(6)  = 0.6##
    Aji(7,1)  = 5## : Aji(7,2)  =   5## : Aji(7,3)  = 3## : Aji(7,4)  =   3## : Cj(7)  = 0.3##

    Z = 0##

    FOR jj% = 1 TO m%  'NOTE:  Index jj% is used to avoid same variable name as j&

        Sum = 0##

        FOR i% = 1 TO 4 'Shekel's family is 4-D only

            Xi = R(p%,i%,j&)

            Sum = Sum + (Xi-Aji(jj%,i%))^2

        NEXT i%

        Z = Z + 1##/(Sum + Cj(jj%))

    NEXT jj%

    F22 = Z

END FUNCTION 'F22

'-----------

FUNCTION F23(R(),Nd%,p%,j&) 'F23 (4-D) Shekel's Family m=10

'Global maximum = 10

    LOCAL Xi, Z, Sum, Aji(), Cj() AS EXT

    LOCAL i%, jj%, m%

    m% = 10 : REDIM Aji(1 TO m%, 1 TO 4), Cj(1 TO m%)

    Aji(1,1)  = 4## : Aji(1,2)  =   4## : Aji(1,3)  = 4## : Aji(1,4)  =   4## : Cj(1)  = 0.1##
    Aji(2,1)  = 1## : Aji(2,2)  =   1## : Aji(2,3)  = 1## : Aji(2,4)  =   1## : Cj(2)  = 0.2##
    Aji(3,1)  = 8## : Aji(3,2)  =   8## : Aji(3,3)  = 8## : Aji(3,4)  =   8## : Cj(3)  = 0.2##
    Aji(4,1)  = 6## : Aji(4,2)  =   6## : Aji(4,3)  = 6## : Aji(4,4)  =   6## : Cj(4)  = 0.4##
    Aji(5,1)  = 3## : Aji(5,2)  =   7## : Aji(5,3)  = 3## : Aji(5,4)  =   7## : Cj(5)  = 0.4##
    Aji(6,1)  = 2## : Aji(6,2)  =   9## : Aji(6,3)  = 2## : Aji(6,4)  =   9## : Cj(6)  = 0.6##
    Aji(7,1)  = 5## : Aji(7,2)  =   5## : Aji(7,3)  = 3## : Aji(7,4)  =   3## : Cj(7)  = 0.3##
    Aji(8,1)  = 8## : Aji(8,2)  =   1## : Aji(8,3)  = 8## : Aji(8,4)  =   1## : Cj(8)  = 0.7##
    Aji(9,1)  = 6## : Aji(9,2)  =   2## : Aji(9,3)  = 6## : Aji(9,4)  =   2## : Cj(9)  = 0.5##
    Aji(10,1) = 7## : Aji(10,2) = 3.6## : Aji(10,3) = 7## : Aji(10,4) = 3.6## : Cj(10) = 0.5##

    Z = 0##

    FOR jj% = 1 TO m%  'NOTE:  Index jj% is used to avoid same variable name as j&

        Sum = 0##

        FOR i% = 1 TO 4 'Shekel's family is 4-D only

            Xi = R(p%,i%,j&)

            Sum = Sum + (Xi-Aji(jj%,i%))^2

        NEXT i%

        Z = Z + 1##/(Sum + Cj(jj%))

    NEXT jj%

    F23 = Z

END FUNCTION 'F23

'======================================================== END FUNCTION DEFINITIONS =============================================================

SUB Plot2DbestProbeTrajectories(NumTrajectories%,M(),R(),Np%,Nd%,LastStep&,FunctionName$)

LOCAL TrajectoryNumber%, ProbeNumber%, StepNumber&, N%, M%, ProcID???

LOCAL MaximumFitness, MinimumFitness AS EXT

LOCAL BestProbeThisStep%()

LOCAL BestFitnessThisStep(), TempFitness() AS EXT

LOCAL Annotation$, xCoord$, yCoord$, GnuPlotEXE$, PlotWithLines$

    Annotation$    = ""

    PlotWithLines$ = "YES" '"NO"

    NumTrajectories% = MIN(Np%,NumTrajectories%)

    GnuPlotEXE$ = "wgnuplot.exe"

'   ---------------- Get Min/Max Fitnesses -----------------

    MaximumFitness = M(1,0) : MinimumFitness = M(1,0)  'Note:  M(p%,j&)

    FOR StepNumber& = 0 TO LastStep&

        FOR ProbeNumber% = 1 TO Np%

            IF M(ProbeNumber%,StepNumber&) >= MaximumFitness THEN MaximumFitness = M(ProbeNumber%,StepNumber&)

            IF M(ProbeNumber%,StepNumber&) =< MinimumFitness THEN MinimumFitness = M(ProbeNumber%,StepNumber&)

        NEXT ProbeNumber%

    NEXT StepNumber%

'   ------------- Copy Fitness Array M() into TempFitness to Preserve M() ----------------

    REDIM TempFitness(1 TO Np%, 0 TO LastStep&)

    FOR StepNumber& = 0 TO LastStep&

        FOR ProbeNumber% = 1 TO Np%

            TempFitness(ProbeNumber%,StepNumber&) = M(ProbeNumber%,StepNumber&)

        NEXT ProbeNumber%

    NEXT StepNumber%

'   ------------ LOOP ON TRAJECTORIES -----------

    FOR TrajectoryNumber% = 1 TO NumTrajectories%

'       --------------- Get Trajectory Coordinate Data -----------------
```

```
        REDIM BestFitnessThisStep(0 TO LastStep&), BestProbeThisStep%(0 TO LastStep&)

        FOR StepNumber& = 0 TO LastStep&

            BestFitnessThisStep(StepNumber&) = TempFitness(1,StepNumber&)

            FOR ProbeNumber% = 1 TO Np%

                IF TempFitness(ProbeNumber%,StepNumber&) >= BestFitnessThisStep(StepNumber&) THEN

                    BestFitnessThisStep(StepNumber&) = TempFitness(ProbeNumber%,StepNumber&)

                    BestProbeThisStep%(StepNumber&)  = ProbeNumber%

                END IF

            NEXT ProbeNumber%

        NEXT StepNumber&

'   ----- Create Plot Data File -----

    N% = FREEFILE

    SELECT CASE TrajectoryNumber%

        CASE 1  : OPEN "t1"  FOR OUTPUT AS #N%
        CASE 2  : OPEN "t2"  FOR OUTPUT AS #N%
        CASE 3  : OPEN "t3"  FOR OUTPUT AS #N%
        CASE 4  : OPEN "t4"  FOR OUTPUT AS #N%
        CASE 5  : OPEN "t5"  FOR OUTPUT AS #N%
        CASE 6  : OPEN "t6"  FOR OUTPUT AS #N%
        CASE 7  : OPEN "t7"  FOR OUTPUT AS #N%
        CASE 8  : OPEN "t8"  FOR OUTPUT AS #N%
        CASE 9  : OPEN "t9"  FOR OUTPUT AS #N%
        CASE 10 : OPEN "t10" FOR OUTPUT AS #N%

    END SELECT

'   ------------ Write Plot File Data ------------

    FOR StepNumber& = 0 TO LastStep&

        PRINT #N%, USING$("######.########
######.########",R(BestProbeThisStep%(StepNumber&),1,StepNumber&),R(BestProbeThisStep%(StepNumber&),2,StepNumber&))

        TempFitness(BestProbeThisStep%(StepNumber&),StepNumber&) = MinimumFitness 'so that same max will not be found for next trajectory

    NEXT StepNumber%

    CLOSE #N%

    NEXT TrajectoryNumber%

'   ------------------------- Plot Trajectories --------------------------

    CALL CreateGNUplotINIfile(0.13##*ScreenWidth&,0.18##*ScreenHeight&,0.7##*ScreenHeight&,0.7##*ScreenHeight&)

    Annotation$ = ""

    N% = FREEFILE

    OPEN "cmd2d.gp" FOR OUTPUT AS #N%

        PRINT #N%, "set xrange ["+REMOVE$(STR$(XiMin(1)),ANY"" )+":"+REMOVE$(STR$(XiMax(1)),ANY" ")+"]"
        PRINT #N%, "set yrange ["+REMOVE$(STR$(XiMin(2)),ANY" ")+":"+REMOVE$(STR$(XiMax(2)),ANY" ")+"]"

        'PRINT #N%, "set label "      + Quote$ + Annotation$ + Quote$ + " at graph " + xCoord$ + "," + yCoord$
        PRINT #N%, "set grid xtics " + "10"
        PRINT #N%, "set grid ytics " + "10"
        PRINT #N%, "set grid mxtics"
        PRINT #N%, "set grid mytics"
        PRINT #N%, "show grid"
        PRINT #N%, "set title "  + Quote$ + "2D "+ FunctionName$+" TRAJECTORIES OF PROBES WITH BEST\nFITNESSES (ORDERED BY FITNESS)" + "\n" + RunID$ +
Quote$
        PRINT #N%, "set xlabel " + Quote$ + "x1\n\n"                                   + Quote$
        PRINT #N%, "set ylabel " + Quote$ + "\nx2"                                     + Quote$

        IF PlotWithLines$ = "YES" THEN

            SELECT CASE NumTrajectories%

                CASE 1  : PRINT #N%, "plot "+Quote$+"t1"+Quote$+" w l lw 3"
                CASE 2  : PRINT #N%, "plot "+Quote$+"t1"+Quote$+" w l lw 3,"+Quote$+"t2"+Quote$+" w l"
                CASE 3  : PRINT #N%, "plot "+Quote$+"t1"+Quote$+" w l lw 3,"+Quote$+"t2"+Quote$+" w l,"+Quote$+"t3"+Quote$+" w l"
                CASE 4  : PRINT #N%, "plot "+Quote$+"t1"+Quote$+" w l lw 3,"+Quote$+"t2"+Quote$+" w l,"+Quote$+"t3"+Quote$+" w l,"+Quote$+"t4"+Quote$+" w
l"
                CASE 5  : PRINT #N%, "plot "+Quote$+"t1"+Quote$+" w l lw 3,"+Quote$+"t2"+Quote$+" w l,"+Quote$+"t3"+Quote$+" w l,"+Quote$+"t4"+Quote$+" w
l,"+Quote$+"t5"+Quote$+" w l"
                CASE 6  : PRINT #N%, "plot "+Quote$+"t1"+Quote$+" w l lw 3,"+Quote$+"t2"+Quote$+" w l,"+Quote$+"t3"+Quote$+" w l,"+Quote$+"t4"+Quote$+" w
l,"+Quote$+"t5"+Quote$+" w l,"+Quote$+"t6"+Quote$+" w l"
                CASE 7  : PRINT #N%, "plot "+Quote$+"t1"+Quote$+" w l lw 3,"+Quote$+"t2"+Quote$+" w l,"+Quote$+"t3"+Quote$+" w l,"+Quote$+"t4"+Quote$+" w
l,"+Quote$+"t5"+Quote$+" w l,"+Quote$+"t6"+Quote$+" w l,"+_
                                      Quote$+"t7"+Quote$+" w l"
                CASE 8  : PRINT #N%, "plot "+Quote$+"t1"+Quote$+" w l lw 3,"+Quote$+"t2"+Quote$+" w l,"+Quote$+"t3"+Quote$+" w l,"+Quote$+"t4"+Quote$+" w
l,"+Quote$+"t5"+Quote$+" w l,"+Quote$+"t6"+Quote$+" w l,"+_
                                      Quote$+"t7"+Quote$+" w l,"     +Quote$+"t8"+Quote$+" w l"
                CASE 9  : PRINT #N%, "plot "+Quote$+"t1"+Quote$+" w l lw 3,"+Quote$+"t2"+Quote$+" w l,"+Quote$+"t3"+Quote$+" w l,"+Quote$+"t4"+Quote$+" w
l,"+Quote$+"t5"+Quote$+" w l,"+Quote$+"t6"+Quote$+" w l,"+_
                                      Quote$+"t7"+Quote$+" w l,"     +Quote$+"t8"+Quote$+" w l,"+Quote$+"t9"+Quote$+" w l"
                CASE 10 : PRINT #N%, "plot "+Quote$+"t1"+Quote$+" w l lw 3,"+Quote$+"t2"+Quote$+" w l,"+Quote$+"t3"+Quote$+" w l,"+Quote$+"t4"+Quote$+" w
l,"+Quote$+"t5"+Quote$+" w l,"+Quote$+"t6"+Quote$+" w l,"+_
                                      Quote$+"t7"+Quote$+" w l,"     +Quote$+"t8"+Quote$+" w l,"+Quote$+"t9"+Quote$+" w l,"+Quote$+"t10"+Quote$+" w
l"
            END SELECT

        ELSE

            SELECT CASE NumTrajectories%

                CASE 1  : PRINT #N%, "plot "+Quote$+"t1"+Quote$+" lw 2"
                CASE 2  : PRINT #N%, "plot "+Quote$+"t1"+Quote$+" lw 2,"+Quote$+"t2"+Quote$
                CASE 3  : PRINT #N%, "plot "+Quote$+"t1"+Quote$+" lw 2,"+Quote$+"t2"+Quote$+" ,"+Quote$+"t3"+Quote$
                CASE 4  : PRINT #N%, "plot "+Quote$+"t1"+Quote$+" lw 2,"+Quote$+"t2"+Quote$+" ,"+Quote$+"t3"+Quote$+" ,"+Quote$+"t4"+Quote$
                CASE 5  : PRINT #N%, "plot "+Quote$+"t1"+Quote$+" lw 2,"+Quote$+"t2"+Quote$+" ,"+Quote$+"t3"+Quote$+" ,"+Quote$+"t4"+Quote$+"
,"+Quote$+"t5"+Quote$
                CASE 6  : PRINT #N%, "plot "+Quote$+"t1"+Quote$+" lw 2,"+Quote$+"t2"+Quote$+" ,"+Quote$+"t3"+Quote$+" ,"+Quote$+"t4"+Quote$+"
,"+Quote$+"t5"+Quote$+" ,"+Quote$+"t6"+Quote$
                CASE 7  : PRINT #N%, "plot "+Quote$+"t1"+Quote$+" lw 2,"+Quote$+"t2"+Quote$+" ,"+Quote$+"t3"+Quote$+" ,"+Quote$+"t4"+Quote$+"
,"+Quote$+"t5"+Quote$+" ,"+Quote$+"t6"+Quote$+" ,"+_
                                      Quote$+"t7"+Quote$
                CASE 8  : PRINT #N%, "plot "+Quote$+"t1"+Quote$+" lw 2,"+Quote$+"t2"+Quote$+" ,"+Quote$+"t3"+Quote$+" ,"+Quote$+"t4"+Quote$+"
,"+Quote$+"t5"+Quote$+" ,"+Quote$+"t6"+Quote$+" ,"+_
                                      Quote$+"t7"+Quote$+" ,"    +Quote$+"t8"+Quote$
                CASE 9  : PRINT #N%, "plot "+Quote$+"t1"+Quote$+" lw 2,"+Quote$+"t2"+Quote$+" ,"+Quote$+"t3"+Quote$+" ,"+Quote$+"t4"+Quote$+"
,"+Quote$+"t5"+Quote$+" ,"+Quote$+"t6"+Quote$+" ,"+_
                                      Quote$+"t7"+Quote$+" ,"    +Quote$+"t8"+Quote$+" ,"+Quote$+"t9"+Quote$
```

```
                CASE 10 : PRINT #N%, "plot "+Quote$+"t1"+Quote$+" lw 2,"+Quote$+"t2"+Quote$+" ,"+Quote$+"t3"+Quote$+" ,"+Quote$+"t4"+Quote$+"
,"+Quote$+"t5"+Quote$+" ,"+Quote$+"t6"+Quote$+" ,"+_
                                             Quote$+"t7"+Quote$+" ,"    +Quote$+"t8"+Quote$+" ,"+Quote$+"t9"+Quote$+" ,"+Quote$+"t10"+Quote$
            END SELECT

        END IF

    CLOSE #N%

    ProcID??? = SHELL(GnuPlotEXE$+" cmd2d.gp -") : CALL Delay(1##)

END SUB 'Plot2DbestProbeTrajectories()

'----

SUB Plot2DindividualProbeTrajectories(NumTrajectories%,M(),R(),Np%,Nd%,LastStep&,FunctionName$)

LOCAL ProbeNumber%, StepNumber&, N%, ProcID???

LOCAL Annotation$, xCoord$, yCoord$, GnuPlotEXE$, PlotWithLines$

    NumTrajectories% = MIN(Np%,NumTrajectories%)

    Annotation$    = ""

    PlotWithLines$ = "YES" '"NO"

    GnuPlotEXE$ = "wgnuplot.exe"

'   -------------- LOOP ON PROBES ---------------

    FOR ProbeNumber% = 1 TO MIN(NumTrajectories%,Np%)

'   ----- Create Plot Data File -----

    N% = FREEFILE

    SELECT CASE ProbeNumber%

        CASE 1  : OPEN "p1"  FOR OUTPUT AS #N%
        CASE 2  : OPEN "p2"  FOR OUTPUT AS #N%
        CASE 3  : OPEN "p3"  FOR OUTPUT AS #N%
        CASE 4  : OPEN "p4"  FOR OUTPUT AS #N%
        CASE 5  : OPEN "p5"  FOR OUTPUT AS #N%
        CASE 6  : OPEN "p6"  FOR OUTPUT AS #N%
        CASE 7  : OPEN "p7"  FOR OUTPUT AS #N%
        CASE 8  : OPEN "p8"  FOR OUTPUT AS #N%
        CASE 9  : OPEN "p9"  FOR OUTPUT AS #N%
        CASE 10 : OPEN "p10" FOR OUTPUT AS #N%
        CASE 11 : OPEN "p11" FOR OUTPUT AS #N%
        CASE 12 : OPEN "p12" FOR OUTPUT AS #N%
        CASE 13 : OPEN "p13" FOR OUTPUT AS #N%
        CASE 14 : OPEN "p14" FOR OUTPUT AS #N%
        CASE 15 : OPEN "p15" FOR OUTPUT AS #N%
        CASE 16 : OPEN "p16" FOR OUTPUT AS #N%

    END SELECT

'   ------------ Write Plot File Data ------------

    FOR StepNumber& = 0 TO LastStep&

        PRINT #N%, USING$("######.######## ######.########",R(ProbeNumber%,1,StepNumber&),R(ProbeNumber%,2,StepNumber&))

    NEXT StepNumber%

    CLOSE #N%

    NEXT ProbeNumber%

'   -------------------------------------------- Plot Trajectories ---------------------------------------------

'usage:  CALL CreateGNUplotINIfile(PlotWindowULC_X%,PlotWindowULC_Y%,PlotWindowWidth%,PlotWindowHeight%)

    CALL CreateGNUplotINIfile(0.17##*ScreenWidth&,0.22##*ScreenHeight&,0.7##*ScreenHeight&,0.7##*ScreenHeight&)

    Annotation$ = ""

    N% = FREEFILE

    OPEN "cmd2d.gp" FOR OUTPUT AS #N%

        PRINT #N%, "set xrange ["+REMOVE$(STR$(XiMin(1)),ANY"" )+":"+REMOVE$(STR$(XiMax(1)),ANY" ")+"]"
        PRINT #N%, "set yrange ["+REMOVE$(STR$(XiMin(2)),ANY" ")+":"+REMOVE$(STR$(XiMax(2)),ANY" ")+"]"

        PRINT #N%, "set grid xtics " + "10"
        PRINT #N%, "set grid ytics " + "10"
        PRINT #N%, "set grid mxtics"
        PRINT #N%, "set grid mytics"
        PRINT #N%, "show grid"
        PRINT #N%, "set title "  + Quote$ + "2D "+ FunctionName$+" INDIVIDUAL PROBE TRAJECTORIES\n(ORDERED BY PROBE #)" + "\n" + RunID$ + Quote$
        PRINT #N%, "set xlabel " + Quote$ + "x1\n\n"                                   + Quote$
        PRINT #N%, "set ylabel " + Quote$ + "\nx2"                                     + Quote$

        IF PlotWithLines$ = "YES" THEN

            SELECT CASE NumTrajectories%

                CASE 1  : PRINT #N%, "plot "+Quote$+"p1"  +Quote$+" w l lw 1"
                CASE 2  : PRINT #N%, "plot "+Quote$+"p1"  +Quote$+" w l lw 1,"+Quote$+"p2"+Quote$+" w l"
                CASE 3  : PRINT #N%, "plot "+Quote$+"p1"  +Quote$+" w l lw 1,"+Quote$+"p2"+Quote$+" w l,"+Quote$+"p3"+Quote$+" w l"
                CASE 4  : PRINT #N%, "plot "+Quote$+"p1"  +Quote$+" w l lw 1,"+Quote$+"p2"+Quote$+" w l,"+Quote$+"p3"+Quote$+" w l,"+Quote$+"p4"+Quote$+" w
l"
                CASE 5  : PRINT #N%, "plot "+Quote$+"p1"  +Quote$+" w l lw 1,"+Quote$+"p2"+Quote$+" w l,"+Quote$+"p3"+Quote$+" w l,"+Quote$+"p4"+Quote$+" w
l,"+Quote$+"p5"+Quote$+" w l"
                CASE 6  : PRINT #N%, "plot "+Quote$+"p1"  +Quote$+" w l lw 1,"+Quote$+"p2"+Quote$+" w l,"+Quote$+"p3"+Quote$+" w l,"+Quote$+"p4"+Quote$+" w
l,"+Quote$+"p5"+Quote$+" w l,"+Quote$+"p6"+Quote$+" w l"
                CASE 7  : PRINT #N%, "plot "+Quote$+"p1"  +Quote$+" w l lw 1,"+Quote$+"p2"+Quote$+" w l,"+Quote$+"p3"+Quote$+" w l,"+Quote$+"p4"+Quote$+" w
l,"+Quote$+"p5"+Quote$+" w l,"+Quote$+"p6"+Quote$+" w l,"+_
                                             Quote$+"p7"  +Quote$+" w l"
                CASE 8  : PRINT #N%, "plot "+Quote$+"p1"  +Quote$+" w l lw 1,"+Quote$+"p2"+Quote$+" w l,"+Quote$+"p3"+Quote$+" w l,"+Quote$+"p4"+Quote$+" w
l,"+Quote$+"p5"+Quote$+" w l,"+Quote$+"p6"+Quote$+" w l,"+_
                                             Quote$+"p7"  +Quote$+" w l,"     +Quote$+"p8"+Quote$+" w l"
                CASE 9  : PRINT #N%, "plot "+Quote$+"p1"  +Quote$+" w l lw 1,"+Quote$+"p2"+Quote$+" w l,"+Quote$+"p3"+Quote$+" w l,"+Quote$+"p4"+Quote$+" w
l,"+Quote$+"p5"+Quote$+" w l,"+Quote$+"p6"+Quote$+" w l,"+_
                                             Quote$+"p7"  +Quote$+" w l,"     +Quote$+"p8"+Quote$+" w l,"+Quote$+"p9"+Quote$+" w l"

                CASE 10 : PRINT #N%, "plot "+Quote$+"p1"  +Quote$+" w l lw 1,"+Quote$+"p2" +Quote$+" w l,"+Quote$+"p3" +Quote$+" w l,"+Quote$+"p4"
+Quote$+" w l,"+Quote$+"p5"+Quote$+" w l,"+Quote$+"p6"+Quote$+" w l,"+_
                                             Quote$+"p7"  +Quote$+" w l,"     +Quote$+"p8" +Quote$+" w l,"+Quote$+"p9" +Quote$+" w
l,"+Quote$+"p10"+Quote$+" w l"

                CASE 11 : PRINT #N%, "plot "+Quote$+"p1"  +Quote$+" w l lw 1,"+Quote$+"p2" +Quote$+" w l,"+Quote$+"p3" +Quote$+" w l,"+Quote$+"p4"
+Quote$+" w l,"+Quote$+"p5" +Quote$+" w l,"+Quote$+"p6" +Quote$+" w l,"+_
                                             Quote$+"p7"  +Quote$+" w l,"     +Quote$+"p8" +Quote$+" w l,"+Quote$+"p9" +Quote$+" w
l,"+Quote$+"p10"+Quote$+" w l,"+Quote$+"p11"+Quote$+" w l"

                CASE 12 : PRINT #N%, "plot "+Quote$+"p1"  +Quote$+" w l lw 1,"+Quote$+"p2" +Quote$+" w l,"+Quote$+"p3" +Quote$+" w l,"+Quote$+"p4"
+Quote$+" w l,"+Quote$+"p5" +Quote$+" w l,"+Quote$+"p6" +Quote$+" w l,"+_
```

```
                                             Quote$+"p7"  +Quote$+" w l,"     +Quote$+"p8" +Quote$+" w l,"+Quote$+"p9" +Quote$+" w
l,"+Quote$+"p10"+Quote$+" w l,"+Quote$+"p11"+Quote$+" w l,"+Quote$+"p12"+Quote$+" w l"

                CASE 13 : PRINT #N%, "plot "+Quote$+"p1"  +Quote$+" w l lw 1,"+Quote$+"p2" +Quote$+" w l,"+Quote$+"p3" +Quote$+" w l,"+Quote$+"p4"
+Quote$+" w l,"+Quote$+"p5" +Quote$+" w l,"+Quote$+"p6" +Quote$+" w l,"+_
                                             Quote$+"p7"  +Quote$+" w l,"     +Quote$+"p8" +Quote$+" w l,"+Quote$+"p9" +Quote$+" w
l,"+Quote$+"p10"+Quote$+" w l,"+Quote$+"p11"+Quote$+" w l,"+Quote$+"p12"+Quote$+" w l,"+_
                                             Quote$+"p13 "+Quote$+" w l"

                CASE 14 : PRINT #N%, "plot "+Quote$+"p1"  +Quote$+" w l lw 1,"+Quote$+"p2" +Quote$+" w l,"+Quote$+"p3" +Quote$+" w l,"+Quote$+"p4"
+Quote$+" w l,"+Quote$+"p5" +Quote$+" w l,"+Quote$+"p6" +Quote$+" w l,"+_
                                             Quote$+"p7"  +Quote$+" w l,"     +Quote$+"p8" +Quote$+" w l,"+Quote$+"p9" +Quote$+" w
l,"+Quote$+"p10"+Quote$+" w l,"+Quote$+"p11"+Quote$+" w l,"+Quote$+"p12"+Quote$+" w l,"+_
                                             Quote$+"p13" +Quote$+" w l,"     +Quote$+"p14"+Quote$+" w l"

                CASE 15 : PRINT #N%, "plot "+Quote$+"p1"  +Quote$+" w l lw 1,"+Quote$+"p2" +Quote$+" w l,"+Quote$+"p3" +Quote$+" w l,"+Quote$+"p4"
+Quote$+" w l,"+Quote$+"p5" +Quote$+" w l,"+Quote$+"p6" +Quote$+" w l,"+_
                                             Quote$+"p7"  +Quote$+" w l,"     +Quote$+"p8" +Quote$+" w l,"+Quote$+"p9" +Quote$+" w
l,"+Quote$+"p10"+Quote$+" w l,"+Quote$+"p11"+Quote$+" w l,"+Quote$+"p12"+Quote$+" w l,"+_
                                             Quote$+"p13" +Quote$+" w l,"     +Quote$+"p14"+Quote$+" w l,"+Quote$+"p15"+Quote$+" w l"

                CASE 16 : PRINT #N%, "plot "+Quote$+"p1"  +Quote$+" w l lw 1,"+Quote$+"p2" +Quote$+" w l,"+Quote$+"p3" +Quote$+" w l,"+Quote$+"p4"
+Quote$+" w l,"+Quote$+"p5" +Quote$+" w l,"+Quote$+"p6" +Quote$+" w l,"+_
                                             Quote$+"p7"  +Quote$+" w l,"     +Quote$+"p8" +Quote$+" w l,"+Quote$+"p9" +Quote$+" w
l,"+Quote$+"p10"+Quote$+" w l,"+Quote$+"p11"+Quote$+" w l,"+Quote$+"p12"+Quote$+" w l,"+_
                                             Quote$+"p13" +Quote$+" w l,"     +Quote$+"p14"+Quote$+" w l,"+Quote$+"p15"+Quote$+" w
l,"+Quote$+"p16"+Quote$+" w l"
            END SELECT

        ELSE

            SELECT CASE NumTrajectories%

                CASE 1  : PRINT #N%, "plot "+Quote$+"p1"+Quote$+" lw 1"
                CASE 2  : PRINT #N%, "plot "+Quote$+"p1"+Quote$+" lw 1,"+Quote$+"p2"+Quote$
                CASE 3  : PRINT #N%, "plot "+Quote$+"p1"+Quote$+" lw 1,"+Quote$+"p2"+Quote$+" ,"+Quote$+"p3"+Quote$
                CASE 4  : PRINT #N%, "plot "+Quote$+"p1"+Quote$+" lw 1,"+Quote$+"p2"+Quote$+" ,"+Quote$+"p3"+Quote$+" ,"+Quote$+"p4"+Quote$
                CASE 5  : PRINT #N%, "plot "+Quote$+"p1"+Quote$+" lw 1,"+Quote$+"p2"+Quote$+" ,"+Quote$+"p3"+Quote$+" ,"+Quote$+"p4"+Quote$+"
,"+Quote$+"p5"+Quote$
                CASE 6  : PRINT #N%, "plot "+Quote$+"p1"+Quote$+" lw 1,"+Quote$+"p2"+Quote$+" ,"+Quote$+"p3"+Quote$+" ,"+Quote$+"p4"+Quote$+"
,"+Quote$+"p5"+Quote$+" ,"+Quote$+"p6"+Quote$
                CASE 7  : PRINT #N%, "plot "+Quote$+"p1"+Quote$+" lw 1,"+Quote$+"p2"+Quote$+" ,"+Quote$+"p3"+Quote$+" ,"+Quote$+"p4"+Quote$+"
,"+Quote$+"p5"+Quote$+" ,"+Quote$+"p6"+Quote$+" ,"+_
                                             Quote$+"p7"+Quote$
                CASE 8  : PRINT #N%, "plot "+Quote$+"p1"+Quote$+" lw 1,"+Quote$+"p2"+Quote$+" ,"+Quote$+"p3"+Quote$+" ,"+Quote$+"p4"+Quote$+"
,"+Quote$+"p5"+Quote$+" ,"+Quote$+"p6"+Quote$+" ,"+_
                                             Quote$+"p7"+Quote$+" ,"    +Quote$+"p8"+Quote$
                CASE 9  : PRINT #N%, "plot "+Quote$+"p1"+Quote$+" lw 1,"+Quote$+"p2"+Quote$+" ,"+Quote$+"p3"+Quote$+" ,"+Quote$+"p4"+Quote$+"
,"+Quote$+"p5"+Quote$+" ,"+Quote$+"p6"+Quote$+" ,"+_
                                             Quote$+"p7"+Quote$+" ,"    +Quote$+"p8"+Quote$+" ,"+Quote$+"p9"+Quote$

                CASE 10 : PRINT #N%, "plot "+Quote$+"p1"+Quote$+" lw 1," +Quote$+"p2" +Quote$+" ,"+Quote$+"p3"+Quote$+" ,"+Quote$+"p4"  +Quote$+"
,"+Quote$+"p5"+Quote$+" ,"+Quote$+"p6"+Quote$+" ,"+_
                                             Quote$+"p7"+Quote$+" ,"     +Quote$+"p8" +Quote$+" ,"+Quote$+"p9"+Quote$+" ,"+Quote$+"p10" +Quote$

                CASE 11 : PRINT #N%, "plot "+Quote$+"p1"+Quote$+" lw 1," +Quote$+"p2" +Quote$+" ,"+Quote$+"p3"+Quote$+" ,"+Quote$+"p4"  +Quote$+"
,"+Quote$+"p5" +Quote$+" ,"+Quote$+"p6"+Quote$+" ,"+_
                                             Quote$+"p7"+Quote$+" ,"     +Quote$+"p8" +Quote$+" ,"+Quote$+"p9"+Quote$+" ,"+Quote$+"p10" +Quote$+"
,"+Quote$+"p11"+Quote$

                CASE 12 : PRINT #N%, "plot "+Quote$+"p1"+Quote$+" lw 1," +Quote$+"p2" +Quote$+" ,"+Quote$+"p3"+Quote$+" ,"+Quote$+"p4"  +Quote$+"
,"+Quote$+"p5" +Quote$+" ,"+Quote$+"p6" +Quote$+" ,"+_
                                             Quote$+"p7"+Quote$+" ,"     +Quote$+"p8" +Quote$+" ,"+Quote$+"p9"+Quote$+" ,"+Quote$+"p10" +Quote$+"
,"+Quote$+"p11"+Quote$+" ,"+Quote$+"p12"+Quote$

                CASE 13 : PRINT #N%, "plot "+Quote$+"p1" +Quote$+" lw 1,"+Quote$+"p2" +Quote$+" ,"+Quote$+"p3"+Quote$+" ,"+Quote$+"p4"  +Quote$+"
,"+Quote$+"p5" +Quote$+" ," +Quote$+"p6" +Quote$+" ,"+_
                                             Quote$+"p7" +Quote$+" ,"    +Quote$+"p8" +Quote$+" ,"+Quote$+"p9"+Quote$+" ,"+Quote$+"p10" +Quote$+"
,"+Quote$+"p11"+Quote$+" ," +Quote$+"p12"+Quote$+" ,"+_
                                             Quote$+"p13"+Quote$

                CASE 14 : PRINT #N%, "plot "+Quote$+"p1" +Quote$+" lw 1,"+Quote$+"p2" +Quote$+" ,"+Quote$+"p3"+Quote$+" ,"+Quote$+"p4"  +Quote$+"
,"+Quote$+"p5" +Quote$+" ," +Quote$+"p6" +Quote$+" ,"+_
                                             Quote$+"p7" +Quote$+" ,"    +Quote$+"p8" +Quote$+" ,"+Quote$+"p9"+Quote$+" ,"+Quote$+"p10" +Quote$+"
,"+Quote$+"p11"+Quote$+" ," +Quote$+"p12"+Quote$+" ,"+_
                                             Quote$+"p13"+Quote$+" ,"    +Quote$+"p14"+Quote$

                CASE 15 : PRINT #N%, "plot "+Quote$+"p1" +Quote$+" lw 1,"+Quote$+"p2" +Quote$+" ,"+Quote$+"p3" +Quote$+" ,"+Quote$+"p4" +Quote$+"
,"+Quote$+"p5" +Quote$+" ," +Quote$+"p6" +Quote$+" ,"+_
                                             Quote$+"p7" +Quote$+" ,"    +Quote$+"p8" +Quote$+" ,"+Quote$+"p9" +Quote$+" ,"+Quote$+"p10"+Quote$+"
,"+Quote$+"p11"+Quote$+" ," +Quote$+"p12"+Quote$+" ,"+_
                                             Quote$+"p13"+Quote$+" ,"    +Quote$+"p14"+Quote$+" ,"+Quote$+"p15"+Quote$

                CASE 16 : PRINT #N%, "plot "+Quote$+"p1" +Quote$+" lw 1,"+Quote$+"p2" +Quote$+" ,"+Quote$+"p3" +Quote$+" ,"+Quote$+"p4" +Quote$+"
,"+Quote$+"p5" +Quote$+" ," +Quote$+"p6" +Quote$+" ,"+_
                                             Quote$+"p7" +Quote$+" ,"    +Quote$+"p8" +Quote$+" ,"+Quote$+"p9" +Quote$+" ,"+Quote$+"p10"+Quote$+"
,"+Quote$+"p11"+Quote$+" ," +Quote$+"p12"+Quote$+" ,"+_
                                             Quote$+"p13"+Quote$+" ,"    +Quote$+"p14"+Quote$+" ,"+Quote$+"p15"+Quote$+" ,"+Quote$+"p16"+Quote$
            END SELECT

        END IF

    CLOSE #N%

    ProcID??? = SHELL(GnuPlotEXE$+" cmd2d.gp -") : CALL Delay(1##)

END SUB 'Plot2DindividualProbeTrajectories()

'----

SUB Plot3DbestProbeTrajectories(NumTrajectories%,M(),R(),Np%,Nd%,LastStep&,FunctionName$) 'XYZZY

LOCAL TrajectoryNumber%, ProbeNumber%, StepNumber&, N%, M%, ProcID???

LOCAL MaximumFitness, MinimumFitness AS EXT

LOCAL BestProbeThisStep%()

LOCAL BestFitnessThisStep(), TempFitness() AS EXT

LOCAL Annotation$, xCoord$, yCoord$, zCoord$, GnuPlotEXE$, PlotWithLines$

    Annotation$      = ""

    PlotWithLines$   = "NO" '"YES" '"NO"

    NumTrajectories% = MIN(Np%,NumTrajectories%)

    GnuPlotEXE$ = "wgnuplot.exe"

'   ---------------- Get Min/Max Fitnesses -----------------

    MaximumFitness = M(1,0) : MinimumFitness = M(1,0)  'Note:  M(p%,j&)

    FOR StepNumber& = 0 TO LastStep&

        FOR ProbeNumber% = 1 TO Np%

            IF M(ProbeNumber%,StepNumber&) >= MaximumFitness THEN MaximumFitness = M(ProbeNumber%,StepNumber&)
```

```
            IF M(ProbeNumber%,StepNumber&) =< MinimumFitness THEN MinimumFitness = M(ProbeNumber%,StepNumber&)

        NEXT ProbeNumber%

    NEXT StepNumber%

'   ------------- Copy Fitness Array M() into TempFitness to Preserve M() ----------------

    REDIM TempFitness(1 TO Np%, 0 TO LastStep&)

    FOR StepNumber& = 0 TO LastStep&

        FOR ProbeNumber% = 1 TO Np%

            TempFitness(ProbeNumber%,StepNumber&) = M(ProbeNumber%,StepNumber&)

        NEXT ProbeNumber%

    NEXT StepNumber%

'   ------------ LOOP ON TRAJECTORIES -----------

    FOR TrajectoryNumber% = 1 TO NumTrajectories%

'       --------------- Get Trajectory Coordinate Data -----------------

        REDIM BestFitnessThisStep(0 TO LastStep&), BestProbeThisStep%(0 TO LastStep&)

        FOR StepNumber& = 0 TO LastStep&

            BestFitnessThisStep(StepNumber&) = TempFitness(1,StepNumber&)

            FOR ProbeNumber% = 1 TO Np%

                IF TempFitness(ProbeNumber%,StepNumber&) >= BestFitnessThisStep(StepNumber&) THEN

                    BestFitnessThisStep(StepNumber&) = TempFitness(ProbeNumber%,StepNumber&)

                    BestProbeThisStep%(StepNumber&)  = ProbeNumber%

                END IF

            NEXT ProbeNumber%

        NEXT StepNumber&

'   ----- Create Plot Data File -----

    N% = FREEFILE

    SELECT CASE TrajectoryNumber%

        CASE 1  : OPEN "t1"  FOR OUTPUT AS #N%
        CASE 2  : OPEN "t2"  FOR OUTPUT AS #N%
        CASE 3  : OPEN "t3"  FOR OUTPUT AS #N%
        CASE 4  : OPEN "t4"  FOR OUTPUT AS #N%
        CASE 5  : OPEN "t5"  FOR OUTPUT AS #N%
        CASE 6  : OPEN "t6"  FOR OUTPUT AS #N%
        CASE 7  : OPEN "t7"  FOR OUTPUT AS #N%
        CASE 8  : OPEN "t8"  FOR OUTPUT AS #N%
        CASE 9  : OPEN "t9"  FOR OUTPUT AS #N%
        CASE 10 : OPEN "t10" FOR OUTPUT AS #N%

    END SELECT

'   ------------ Write Plot File Data ------------

    FOR StepNumber& = 0 TO LastStep&

        PRINT #N%, USING$("######.######## ######.########
######.########",R(BestProbeThisStep%(StepNumber&),1,StepNumber&),R(BestProbeThisStep%(StepNumber&),2,StepNumber&),R(BestProbeThisStep%(StepNumber&),3,Step
Number&))+CHR$(13)

        TempFitness(BestProbeThisStep%(StepNumber&),StepNumber&) = MinimumFitness 'so that same max will not be found for next trajectory

    NEXT StepNumber%

    CLOSE #N%

    NEXT TrajectoryNumber%

'   ------------------------- Plot Trajectories --------------------------

   'CALL CreateGNUplotINIfile(0.1##*ScreenWidth&,0.25##*ScreenHeight&,0.6##*ScreenHeight&,0.6##*ScreenHeight&)

    Annotation$ = ""

    N% = FREEFILE

    OPEN "cmd3d.gp" FOR OUTPUT AS #N%

    PRINT #N%, "set pm3d"
    PRINT #N%, "show pm3d"
    PRINT #N%, "set hidden3d"
    PRINT #N%, "set view 45, 45, 1, 1"

    PRINT #N%, "unset colorbox"

    PRINT #N%, "set xrange [" + REMOVE$(STR$(XiMin(1)),ANY"" ) + ":" + REMOVE$(STR$(XiMax(1)),ANY"" ) + "]"
    PRINT #N%, "set yrange [" + REMOVE$(STR$(XiMin(2)),ANY"" ) + ":" + REMOVE$(STR$(XiMax(2)),ANY"" ) + "]"
    PRINT #N%, "set zrange [" + REMOVE$(STR$(XiMin(3)),ANY"" ) + ":" + REMOVE$(STR$(XiMax(3)),ANY"" ) + "]"

    PRINT #N%, "set grid xtics ytics ztics"
    PRINT #N%, "show grid"
    PRINT #N%, "set title "  + Quote$ + "3D " + FunctionName$ + " PROBE TRAJECTORIES" + "\n" + RunID$ + Quote$
    PRINT #N%, "set xlabel " + Quote$ + "x1"                                       + Quote$
    PRINT #N%, "set ylabel " + Quote$ + "x2"                                       + Quote$
    PRINT #N%, "set zlabel " + Quote$ + "x3"                                       + Quote$

    IF PlotWithLines$ = "YES" THEN

        SELECT CASE NumTrajectories%

            CASE 1  : PRINT #N%, "splot "+Quote$+"t1"+Quote$+" w l lw 3"
            CASE 2  : PRINT #N%, "splot "+Quote$+"t1"+Quote$+" w l lw 3,"+Quote$+"t2"+Quote$+" w l"
            CASE 3  : PRINT #N%, "splot "+Quote$+"t1"+Quote$+" w l lw 3,"+Quote$+"t2"+Quote$+" w l,"+Quote$+"t3"+Quote$+" w l"
            CASE 4  : PRINT #N%, "splot "+Quote$+"t1"+Quote$+" w l lw 3,"+Quote$+"t2"+Quote$+" w l,"+Quote$+"t3"+Quote$+" w l,"+Quote$+"t4"+Quote$+" w l"
            CASE 5  : PRINT #N%, "splot "+Quote$+"t1"+Quote$+" w l lw 3,"+Quote$+"t2"+Quote$+" w l,"+Quote$+"t3"+Quote$+" w l,"+Quote$+"t4"+Quote$+" w
l,"+Quote$+"t5"+Quote$+" w l"
            CASE 6  : PRINT #N%, "splot "+Quote$+"t1"+Quote$+" w l lw 3,"+Quote$+"t2"+Quote$+" w l,"+Quote$+"t3"+Quote$+" w l,"+Quote$+"t4"+Quote$+" w
l,"+Quote$+"t5"+Quote$+" w l,"+Quote$+"t6"+Quote$+" w l"
            CASE 7  : PRINT #N%, "splot "+Quote$+"t1"+Quote$+" w l lw 3,"+Quote$+"t2"+Quote$+" w l,"+Quote$+"t3"+Quote$+" w l,"+Quote$+"t4"+Quote$+" w
l,"+Quote$+"t5"+Quote$+" w l,"+Quote$+"t6"+Quote$+" w l,"+_
                                    Quote$+"t7"+Quote$+" w l"
            CASE 8  : PRINT #N%, "splot "+Quote$+"t1"+Quote$+" w l lw 3,"+Quote$+"t2"+Quote$+" w l,"+Quote$+"t3"+Quote$+" w l,"+Quote$+"t4"+Quote$+" w
l,"+Quote$+"t5"+Quote$+" w l,"+Quote$+"t6"+Quote$+" w l,"+_
                                    Quote$+"t7"+Quote$+" w l,"     +Quote$+"t8"+Quote$+" w l"
```

```
            CASE 9  : PRINT #N%, "splot "+Quote$+"t1"+Quote$+" w l lw 3,"+Quote$+"t2"+Quote$+" w l,"+Quote$+"t3"+Quote$+" w l,"+Quote$+"t4"+Quote$+" w
l,"+Quote$+"t5"+Quote$+" w l,"+Quote$+"t6"+Quote$+" w l,"+_
                                          Quote$+"t7"+Quote$+" w l,"     +Quote$+"t8"+Quote$+" w l,"+Quote$+"t9"+Quote$+" w l"
            CASE 10 : PRINT #N%, "splot "+Quote$+"t1"+Quote$+" w l lw 3,"+Quote$+"t2"+Quote$+" w l,"+Quote$+"t3"+Quote$+" w l,"+Quote$+"t4"+Quote$+" w
l,"+Quote$+"t5"+Quote$+" w l,"+Quote$+"t6"+Quote$+" w l,"+_
                                          Quote$+"t7"+Quote$+" w l,"     +Quote$+"t8"+Quote$+" w l,"+Quote$+"t9"+Quote$+" w l,"+Quote$+"t10"+Quote$+" w l"
        END SELECT

    ELSE

        SELECT CASE NumTrajectories%

            CASE 1  : PRINT #N%, "splot "+Quote$+"t1"+Quote$+" lw 2"
            CASE 2  : PRINT #N%, "splot "+Quote$+"t1"+Quote$+" lw 2,"+Quote$+"t2"+Quote$
            CASE 3  : PRINT #N%, "splot "+Quote$+"t1"+Quote$+" lw 2,"+Quote$+"t2"+Quote$+" ,"+Quote$+"t3"+Quote$
            CASE 4  : PRINT #N%, "splot "+Quote$+"t1"+Quote$+" lw 2,"+Quote$+"t2"+Quote$+" ,"+Quote$+"t3"+Quote$+" ,"+Quote$+"t4"+Quote$
            CASE 5  : PRINT #N%, "splot "+Quote$+"t1"+Quote$+" lw 2,"+Quote$+"t2"+Quote$+" ,"+Quote$+"t3"+Quote$+" ,"+Quote$+"t4"+Quote$+"
,"+Quote$+"t5"+Quote$
            CASE 6  : PRINT #N%, "splot "+Quote$+"t1"+Quote$+" lw 2,"+Quote$+"t2"+Quote$+" ,"+Quote$+"t3"+Quote$+" ,"+Quote$+"t4"+Quote$+"
,"+Quote$+"t5"+Quote$+" ,"+Quote$+"t6"+Quote$
            CASE 7  : PRINT #N%, "splot "+Quote$+"t1"+Quote$+" lw 2,"+Quote$+"t2"+Quote$+" ,"+Quote$+"t3"+Quote$+" ,"+Quote$+"t4"+Quote$+"
,"+Quote$+"t5"+Quote$+" ,"+Quote$+"t6"+Quote$+" ,"+_
                                          Quote$+"t7"+Quote$
            CASE 8  : PRINT #N%, "splot "+Quote$+"t1"+Quote$+" lw 2,"+Quote$+"t2"+Quote$+" ,"+Quote$+"t3"+Quote$+" ,"+Quote$+"t4"+Quote$+"
,"+Quote$+"t5"+Quote$+" ,"+Quote$+"t6"+Quote$+" ,"+_
                                          Quote$+"t7"+Quote$+" ,"    +Quote$+"t8"+Quote$
            CASE 9  : PRINT #N%, "splot "+Quote$+"t1"+Quote$+" lw 2,"+Quote$+"t2"+Quote$+" ,"+Quote$+"t3"+Quote$+" ,"+Quote$+"t4"+Quote$+"
,"+Quote$+"t5"+Quote$+" ,"+Quote$+"t6"+Quote$+" ,"+_
                                          Quote$+"t7"+Quote$+" ,"    +Quote$+"t8"+Quote$+" ,"+Quote$+"t9"+Quote$
            CASE 10 : PRINT #N%, "splot "+Quote$+"t1"+Quote$+" lw 2,"+Quote$+"t2"+Quote$+" ,"+Quote$+"t3"+Quote$+" ,"+Quote$+"t4"+Quote$+"
,"+Quote$+"t5"+Quote$+" ,"+Quote$+"t6"+Quote$+" ,"+_
                                          Quote$+"t7"+Quote$+" ,"    +Quote$+"t8"+Quote$+" ,"+Quote$+"t9"+Quote$+" ,"+Quote$+"t10"+Quote$
        END SELECT

    END IF

    CLOSE #N%

    ProcID??? = SHELL(GnuPlotEXE$+" cmd3d.gp -") : CALL Delay(1##)

END SUB 'Plot3DbestProbeTrajectories()

'-----------

FUNCTION HasDAVGsaturated$(Nsteps&,j&,Np%,Nd%,M(),R(),DiagLength)

LOCAL A$

LOCAL k&

LOCAL SumOfDavg, DavgStepJ AS EXT

LOCAL DavgSatTOL AS EXT

    A$ = "NO"

    DavgSatTOL = 0.0005## 'tolerance for DAVG saturation

    IF j& < Nsteps& + 10 THEN GOTO ExitHasDAVGsaturated 'execute at least 10 steps after averaging interval before performing this check

    DavgStepJ = DavgThisStep(j&,Np%,Nd%,M(),R(),DiagLength)

    SumOfDavg = 0##

    FOR k& = j&-Nsteps&+1 TO j& 'check this step and previous (Nsteps&-1) steps

        SumOfDavg = SumOfDavg + DavgThisStep(k&,Np%,Nd%,M(),R(),DiagLength)

    NEXT k&

    IF ABS(SumOfDavg/Nsteps&-DavgStepJ) =< DavgSatTOL THEN A$ = "YES" 'saturation if (avg value - last value) are within TOL

ExitHasDAVGsaturated:

    HasDAVGsaturated$ = A$

END FUNCTION 'HasDAVGsaturated$()

'-----------

FUNCTION OscillationInDavg$(j&,Np%,Nd%,M(),R(),DiagLength)

LOCAL A$

LOCAL k&, NumSlopeChanges%

    A$ = "NO"

    NumSlopeChanges% = 0

    IF j& < 15 THEN GOTO ExitDavgOscillation 'wait at least 15 steps

    FOR k& = j&-10 TO j&-1 'check previous ten steps

        IF (DavgThisStep(k&,Np%,Nd%,M(),R(),DiagLength)-DavgThisStep(k&-1,Np%,Nd%,M(),R(),DiagLength))* _
           (DavgThisStep(k&+1,Np%,Nd%,M(),R(),DiagLength)-DavgThisStep(k&,Np%,Nd%,M(),R(),DiagLength)) < 0## THEN INCR NumSlopeChanges%

    NEXT j&

    IF NumSlopeChanges% >= 3 THEN A$ = "YES"

ExitDavgOscillation:

    OscillationInDavg$ = A$

END FUNCTION 'OscillationInDavg()

'------

FUNCTION DavgThisStep(j&,Np%,Nd%,M(),R(),DiagLength)

LOCAL BestFitness, TotalDistanceAllProbes, SumSQ AS EXT

LOCAL p%, k&, N%, i%, BestProbeNumber%, BestTimeStep&

'   ----------- Best Probe #, etc. -----------

    FOR k& = 0 TO j&

        BestFitness = M(1,k&)

        FOR p% = 1 TO Np%

            IF M(p%,k&) >= BestFitness THEN

                BestFitness = M(p%,k&) : BestProbeNumber% = p% : BestTimeStep& = k&

            END IF
```

```
        NEXT p% 'probe #
    NEXT k& 'time step
'   --------- Average Distance to Best Probe -----------
    TotalDistanceAllProbes = 0##
    FOR p% = 1 TO Np%
        SumSQ = 0##
        FOR i% = 1 TO Nd%
            SumSQ = SumSQ + (R(BestProbeNumber%,i%,BestTimeStep&)-R(p%,i%,j&))^2 'do not exclude p%=BestProbeNumber%(j&) from sum because it adds zero
         NEXT i%
        TotalDistanceAllProbes = TotalDistanceAllProbes + SQR(SumSQ)
    NEXT p%
    DavgThisStep = TotalDistanceAllProbes/(DiagLength*(Np%-1)) 'but exclude best prove from average
END FUNCTION 'DavgThisStep()
'-----------
SUB
PlotBestFitnessEvolution(Nd%,Np%,LastStep&,G,DeltaT,Alpha,Beta,Frep,Mbest(),PlaceInitialProbes$,InitialAcceleration$,RepositionFactor$,FunctionName$,Gamma)
LOCAL BestFitness(), GlobalBestFitness AS EXT
LOCAL PlotAnnotation$, PlotTitle$
LOCAL p%, j&, N%
    REDIM BestFitness(0 TO LastStep&)
    CALL
GetPlotAnnotation(PlotAnnotation$,Nd%,Np%,LastStep&,G,DeltaT,Alpha,Beta,Frep,Mbest(),PlaceInitialProbes$,InitialAcceleration$,RepositionFactor$,FunctionNam
e$,Gamma)
    GlobalBestFitness = Mbest(1,0)
    FOR j& = 0 TO LastStep&
'       BestFitness(j&) = Mbest(1,j&) 'orig code 03-23-2010
        BestFitness(j&) = -1E4200 'added 03-23-2010
        FOR p% = 1 TO Np%
            IF Mbest(p%,j&) >= BestFitness(j&)   THEN BestFitness(j&)   = Mbest(p%,j&)
            IF Mbest(p%,j&) >= GlobalBestFitness THEN GlobalBestFitness = Mbest(p%,j&)
        NEXT p% 'probe #
    NEXT j& 'time step
    N% = FREEFILE
    OPEN "Fitness" FOR OUTPUT AS #N%
'    print #N%,"From Sub PlotBestFitness..."
        FOR j& = 0 TO LastStep&
'            PRINT #N%, USING$("###### ##.######^^^^^^",j&,BestFitness(j&))
            PRINT #N%, USING$("###### #######.#########",j&,BestFitness(j&))
        NEXT j&
    CLOSE #N%
    PlotAnnotation$ = PlotAnnotation$ + "Best Fitness = " + REMOVE$(STR$(ROUND(GlobalBestFitness,8)),ANY" ")
    PlotTitle$ = "Best Fitness vs Time Step\n" + "[" + REMOVE$(STR$(Np%),ANY" ") + " probes, "+REMOVE$(STR$(LastStep&),ANY" ")+" time steps]"
    CALL CreateGNUplotINIfile(0.1##*ScreenWidth&,0.1##*ScreenHeight&,0.6##*ScreenWidth&,0.6##*ScreenHeight&)
    CALL TwoDplot("Fitness","Best Fitness","0.7","0.7","Time Step\n\n.",".\n\nBest Fitness(X)", _
                  "","","","","","","","","wgnuplot.exe"," with lines linewidth 2",PlotAnnotation$)
END SUB 'PlotBestFitnessEvolution()
'------
SUB
PlotAverageDistance(Nd%,Np%,LastStep&,G,DeltaT,Alpha,Beta,Frep,Mbest(),PlaceInitialProbes$,InitialAcceleration$,RepositionFactor$,FunctionName$,R(),DiagLen
gth,Gamma)
LOCAL Davg(), BestFitness(), TotalDistanceAllProbes, SumSQ AS EXT
LOCAL PlotAnnotation$, PlotTitle$
LOCAL p%, j&, N%, i%, BestProbeNumber%(), BestTimeStep&()
    REDIM Davg(0 TO LastStep&), BestFitness(0 TO LastStep&), BestProbeNumber%(0 TO LastStep&), BestTimeStep&(0 TO LastStep&)
    CALL
GetPlotAnnotation(PlotAnnotation$,Nd%,Np%,LastStep&,G,DeltaT,Alpha,Beta,Frep,Mbest(),PlaceInitialProbes$,InitialAcceleration$,RepositionFactor$,FunctionNam
e$,Gamma)
'   ----------- Best Probe #, etc. -----------
    FOR j& = 0 TO LastStep&
        BestFitness(j&) = Mbest(1,j&)
        FOR p% = 1 TO Np%
            IF Mbest(p%,j&) >= BestFitness(j&) THEN
                BestFitness(j&) = Mbest(p%,j&) : BestProbeNumber%(j&) = p% : BestTimeStep&(j&) = j& 'only probe number is used at this time, but other data
are computed for possible future use.
            END IF
        NEXT p% 'probe #
    NEXT j& 'time step
    N% = FREEFILE
'   --------- Average Distance to Best Probe -----------
    FOR j& = 0 TO LastStep&
```

```
        TotalDistanceAllProbes = 0##

        FOR p% = 1 TO Np%

            SumSQ = 0##

            FOR i% = 1 TO Nd%

                SumSQ = SumSQ + (R(BestProbeNumber%(j&),i%,j&)-R(p%,i%,j&))^2 'do not exclude p%=BestProbeNumber%(j&) from sum because it adds zero

            NEXT i%

            TotalDistanceAllProbes = TotalDistanceAllProbes + SQR(SumSQ)

        NEXT p%

        Davg(j&) = TotalDistanceAllProbes/(DiagLength*(Np%-1)) 'but exclude best prove from average

    NEXT j&

'   ------------ Create Plot Data File -----------

    OPEN "Davg" FOR OUTPUT AS #N%

        FOR j& = 0 TO LastStep&

            PRINT #N%, USING$("###### #######.#######",j&,Davg(j&))

        NEXT j&

    CLOSE #N%

    PlotTitle$ = "Average Distance of " + REMOVE$(STR$(Np%-1),ANY" ") + " Probes to Best Probe\nNormalized to Size of Decision Space\n" + _
                 "[" + REMOVE$(STR$(Np%),ANY" ") + " probes, " + REMOVE$(STR$(LastStep&),ANY" ") + " time steps]"

    CALL CreateGNUplotINIfile(0.2##*ScreenWidth&,0.2##*ScreenHeight&,0.6##*ScreenWidth&,0.6##*ScreenHeight&)

    CALL TwoDplot("Davg",PlotTitle$,"0.7","0.9","Time Step\n\n.",".\n\n<D>/Ldiag", _
                  "","","","","","","","","wgnuplot.exe"," with lines linewidth 2",PlotAnnotation$)

END SUB 'PlotAverageDistance()

'------

SUB
GetPlotAnnotation(PlotAnnotation$,Nd%,Np%,LastStep&,G,DeltaT,Alpha,Beta,Frep,Mbest(),PlaceInitialProbes$,InitialAcceleration$,RepositionFactor$,FunctionNam
e$,Gamma)

LOCAL A$

    A$ = "" : IF PlaceInitialProbes$ = "UNIFORM ON-AXIS" AND Nd% > 1 THEN A$ = " ("+REMOVE$(STR$(Np%/Nd%),ANY" ") + "/axis)"

    PlotAnnotation$ = RunID$ + "\n" + _
                      FunctionName$ + " Function" + " ("+ FormatInteger$(Nd%) + "-D) \n"    +_
                      FormatInteger$(Np%) + " probes"       + A$ + "\n" +_
                      "G = " + FormatFP$(G,2)               + "\n" +_
                      "Alpha = "      + FormatFP$(Alpha,1)   + "\n" +_
                      "Beta = "       + FormatFP$(Beta,1)    + "\n" +_
                      "DelT = "       + FormatFP$(DeltaT,1)  + "\n" +_
                      "Gamma = "      + FormatFP$(Gamma,3)   + "\n" +_
                      "Init Probes " + PlaceInitialProbes$  + "\n" +_
                      "Init Accel "  + InitialAcceleration$ + "\n" +_
                      "Frep "        + RepositionFactor$ + "\n"

END SUB

'------

SUB
PlotBestProbeVsTimeStep(Nd%,Np%,LastStep&,G,DeltaT,Alpha,Beta,Frep,Mbest(),PlaceInitialProbes$,InitialAcceleration$,RepositionFactor$,FunctionName$,Gamma)

LOCAL BestFitness AS EXT

LOCAL PlotAnnotation$, PlotTitle$

LOCAL p%, j&, N%, BestProbeNumber%()

    REDIM BestProbeNumber%(0 TO LastStep&)

    CALL
GetPlotAnnotation(PlotAnnotation$,Nd%,Np%,LastStep&,G,DeltaT,Alpha,Beta,Frep,Mbest(),PlaceInitialProbes$,InitialAcceleration$,RepositionFactor$,FunctionNam
e$,Gamma)

    FOR j& = 0 TO LastStep&

        Bestfitness = Mbest(1,j&)

        FOR p% = 1 TO Np%

            IF Mbest(p%,j&) >= BestFitness THEN

                BestFitness = Mbest(p%,j&) : BestProbeNumber%(j&) = p%

            END IF

        NEXT p% 'probe #

    NEXT j& 'time step

    N% = FREEFILE

    OPEN "Best Probe" FOR OUTPUT AS #N%

        FOR j& = 0 TO LastStep&

            PRINT #N%, USING$("###### #####",j&,BestProbeNumber%(j&))

        NEXT j&


    CLOSE #N%

    PlotTitle$ = "Best Probe Number vs Time Step\n" + "[" +REMOVE$(STR$(Np%),ANY" ") + " probes, " + REMOVE$(STR$(LastStep&),ANY" ") + " time steps]"

    CALL CreateGNUplotINIfile(0.15#*ScreenWidth&,0.15##*ScreenHeight&,0.6##*ScreenWidth&,0.6##*ScreenHeight&)

'USAGE: CALL
TwoDplot(PlotFileName$,PlotTitle$,xCoord$,yCoord$,XaxisLabel$,YaxisLabel$,LogXaxis$,LogYaxis$,xMin$,xMax$,yMin$,yMax$,xTics$,yTics$,GnuPlotEXE$,LineType$,A
nnotation$)

    CALL TwoDplot("Best Probe",PlotTitle$,"0.7","0.7","Time Step\n\n.",".\n\nBest Probe #","","","","","0",NoSpaces$(Np%+1,0),"","","wgnuplot.exe"," pt 8
ps .5 lw 1",PlotAnnotation$) 'pt, pointtype; ps, pointsize; lw, linewidth

END SUB 'PlotBestProbeVsTimeStep()

'------

FUNCTION FormatInteger$(M%) : FormatInteger$ = REMOVE$(STR$(M%),ANY" ") : END FUNCTION
```

```
'------
FUNCTION FormatFP$(X,Ndigits%)
LOCAL A$
    IF X = 0## THEN
        A$ = "0." : GOTO ExitFormatFP
    END IF
    A$ = REMOVE$(STR$(ROUND(ABS(X),Ndigits%)),ANY" ")
    IF ABS(X) < 1## THEN
        IF X > 0## THEN
            A$ = "0" + A$
        ELSE
            A$ = "-0" + A$
        END IF
    ELSE
        IF X < 0## THEN A$ = "-" + A$
    END IF
ExitFormatFP:
    FormatFP$ = A$
END FUNCTION
'-----------
SUB InitialProbeDistribution(Np%,Nd%,Nt&,R(),PlaceInitialProbes$,Gamma)
LOCAL DeltaXi, DelX1, DelX2, Di AS EXT
LOCAL NumProbesPerDimension%, p%, i%, k%, NumX1points%, NumX2points%, x1pointNum%, x2pointNum%, A$
    SELECT CASE PlaceInitialProbes$
        CASE "UNIFORM ON-AXIS"
            IF Nd% > 1 THEN
                NumProbesPerDimension% = Np%\Nd% 'even #
            ELSE
                NumProbesPerDimension% = Np%
            END IF
            FOR i% = 1 TO Nd%
                FOR p% = 1 TO Np%
                    R(p%,i%,0) = XiMin(i%) + Gamma*(XiMax(i%)-XiMin(i%))
                NEXT Np%
            NEXT i%
            FOR i% = 1 TO Nd% 'place probes probe line-by-probe line (i% is dimension number)
                DeltaXi = (XiMax(i%)-XiMin(i%))/(NumProbesPerDimension%-1)
                FOR k% = 1 TO NumProbesPerDimension%
                    p% = k% + NumProbesPerDimension%*(i%-1) 'probe #
                    R(p%,i%,0) = XiMin(i%) + (k%-1)*DeltaXi
                NEXT k%
            NEXT i%
        CASE "UNIFORM ON-DIAGONAL"
            FOR p% = 1 TO Np%
                FOR i% = 1 TO Nd%
                    DeltaXi = (XiMax(i%)-XiMin(i%))/(Np%-1)
                    R(p%,i%,0) = XiMin(i%) + (p%-1)*DeltaXi
                NEXT i%
            NEXT p%
        CASE "2D GRID"
            NumProbesPerDimension% = SQR(Np%) : NumX1points% = NumProbesPerDimension% : NumX2points% = NumX1points% 'broken down for possible future use
            DelX1 = (XiMax(1)-XiMin(1))/(NumX1points%-1)
            DelX2 = (XiMax(2)-XiMin(2))/(NumX2points%-1)
            FOR x1pointNum% = 1 TO NumX1points%
                FOR x2pointNum% = 1 TO NumX2points%
                    p% = NumX1points%*(x1pointNum%-1)+x2pointNum% 'probe #
                    R(p%,1,0) = XiMin(1) + DelX1*(x1pointNum%-1) 'x1 coord
                    R(p%,2,0) = XiMin(2) + DelX2*(x2pointNum%-1) 'x2 coord
                NEXT x2pointNum%
            NEXT x1pointNum%
        CASE "RANDOM"
            FOR p% = 1 TO Np%
                FOR i% = 1 TO Nd%
                    R(p%,i%,0) = XiMin(i%) + RandomNum(0##,1##)*(XiMax(i%)-XiMin(i%))
                NEXT i%
```

```
            NEXT p%

    END SELECT

END SUB 'InitialProbeDistribution()

'------

SUB
ChangeRunParameters(NumProbesPerDimension%,Np%,Nd%,Nt&,G,Alpha,Beta,DeltaT,Frep,PlaceInitialProbes$,InitialAcceleration$,RepositionFactor$,FunctionName$)
'THIS PROCEDURE NOT USED

LOCAL A$, DefaultValue$

    A$ = INPUTBOX$("# dimensions?","Change # Dimensions ("+FunctionName$+")",NoSpaces$(Nd%+0,0)) : Nd%    = VAL(A$) : IF Nd% < 1 OR Nd% > 500 THEN Nd% = 2

    IF Nd% > 1 THEN NumProbesPerDimension% = 2*((NumProbesPerDimension%+1)\2) 'require an even # probes on each probe line to avoid overlapping at origin
(in symmetrical spaces at least...)

    IF Nd% = 1 THEN NumProbesPerDimension% = MAX(NumProbesPerDimension%,3)    'at least 3 probes on x-axis for 1-D functions

    Np% = NumProbesPerDimension%*Nd%

    A$ = INPUTBOX$("# time steps?","Change # Steps ("+FunctionName$+")",NoSpaces$(Nt&+0,0)) : Nt&    = VAL(A$) : IF Nt& < 3                        THEN Nt&
= 50

    A$ = INPUTBOX$("Grav Const G?","Change G ("+FunctionName$+")",NoSpaces$(G,2))           : G      = VAL(A$) : IF G < -100##    OR G > 100##    THEN G
= 2##

    A$ = INPUTBOX$("Alpha?","Change Alpha ("+FunctionName$+")",NoSpaces$(Alpha,2))          : Alpha  = VAL(A$) : IF Alpha < -50## OR Alpha > 50## THEN
Alpha = 2##

    A$ = INPUTBOX$("Beta?","Change Beta ("+FunctionName$+")",NoSpaces$(Beta,2))             : Beta   = VAL(A$) : IF Beta  < -50## OR Beta  > 50## THEN Beta
= 2##'

    A$ = INPUTBOX$("Delta T","Change Delta-T ("+FunctionName$+")",NoSpaces$(DeltaT,2))      : DeltaT = VAL(A$) : IF DeltaT =< 0##                 THEN
DeltaT = 1##

    A$ = INPUTBOX$("Frep [0-1]?","Change Frep ("+FunctionName$+")",NoSpaces$(Frep,3))       : Frep   = VAL(A$) : IF Frep < 0##    OR Frep > 1##   THEN Frep
= 0.5##

'   ------------ Initial Probe Distribution ------------

    SELECT CASE PlaceInitialProbes$
        CASE "UNIFORM ON-AXIS"     : DefaultValue$ = "1"
        CASE "UNIFORM ON-DIAGONAL" : DefaultValue$ = "2"
        CASE "2D GRID"             : DefaultValue$ = "3"
        CASE "RANDOM"              : DefaultValue$ = "4"
    END SELECT

    A$ = INPUTBOX$("Initial Probes?"+CHR$(13)+"1 - UNIFORM ON-AXIS"+CHR$(13)+"2 - UNIFORM ON-DIAGONAL"+CHR$(13)+"3 - 2D GRID"+CHR$(13)+"4 -
RANDOM","Initial Probe Distribution ("+FunctionName$+")",DefaultValue$)

    IF VAL(A$) < 1 OR VAL(A$) > 4 THEN A$ = "1"

    SELECT CASE VAL(A$)
        CASE 1 : PlaceInitialProbes$ = "UNIFORM ON-AXIS"
        CASE 2 : PlaceInitialProbes$ = "UNIFORM ON-DIAGONAL"
        CASE 3 : PlaceInitialProbes$ = "2D GRID"
        CASE 4 : PlaceInitialProbes$ = "RANDOM"
    END SELECT

    IF Nd% = 1  AND PlaceInitialProbes$ = "UNIFORM ON-DIAGONAL" THEN PlaceInitialProbes$ = "UNIFORM ON-AXIS" 'cannot do diagonal in 1-D space

    IF Nd% <> 2 AND PlaceInitialProbes$ = "2D GRID" THEN PlaceInitialProbes$ = "UNIFORM ON-AXIS" '2D grid is available only in 2 dimensions!

'   ------------- Initial Acceleration -----------------

    SELECT CASE InitialAcceleration$
        CASE "ZERO"   : DefaultValue$ = "1"
        CASE "FIXED"  : DefaultValue$ = "2"
        CASE "RANDOM" : DefaultValue$ = "3"
    END SELECT

    A$ = INPUTBOX$("Initial Acceleration?"+CHR$(13)+"1 - ZERO"+CHR$(13)+"2 - FIXED"+CHR$(13)+"3 - RANDOM","Initial Acceleration
("+FunctionName$+")",DefaultValue$)

   IF VAL(A$) < 1 OR VAL(A$) > 3 THEN A$ = "1"

    SELECT CASE VAL(A$)
        CASE 1 : InitialAcceleration$ = "ZERO"
        CASE 2 : InitialAcceleration$ = "FIXED"
        CASE 3 : InitialAcceleration$ = "RANDOM"
    END SELECT

'   ----------- Reposition Factor ---------------

    SELECT CASE RepositionFactor$
        CASE "FIXED"    : DefaultValue$ = "1"
        CASE "VARIABLE" : DefaultValue$ = "2"
        CASE "RANDOM"   : DefaultValue$ = "3"
    END SELECT

    A$ = INPUTBOX$("Reposition Factor?"+CHR$(13)+"1 - FIXED"+CHR$(13)+"2 - VARIABLE"+CHR$(13)+"3 - RANDOM","Retrieve Probes
("+FunctionName$+")",DefaultValue$)

    IF VAL(A$) < 1 OR VAL(A$) > 3 THEN A$ = "1"

    SELECT CASE VAL(A$)
        CASE 1 : RepositionFactor$ = "FIXED"
        CASE 2 : RepositionFactor$ = "VARIABLE"
        CASE 3 : RepositionFactor$ = "RANDOM"
    END SELECT

END SUB 'ChangeRunParameters()

'------

FUNCTION NoSpaces$(X,NumDigits%) :  NoSpaces$ = REMOVE$(STR$(X,NumDigits%),ANY" ") : END FUNCTION

'-----------

FUNCTION TerminateNowForSaturation$(j&,Nd%,Np%,Nt&,G,DeltaT,Alpha,Beta,R(),A(),M())

LOCAL A$, i&, p%, NumStepsForAveraging&

LOCAL BestFitness, AvgFitness, FitnessTOL AS EXT 'terminate if avg fitness does not change over NumStepsForAveraging& time steps

    FitnessTOL = 0.00001## : NumStepsForAveraging& = 10

    A$ = "NO"

    IF j& >= NumStepsForAveraging+10 THEN 'wait until step 10 to start checking for fitness saturation

        AvgFitness = 0##

        FOR i& = j&-NumStepsForAveraging&+1 TO j& 'avg fitness over current step & previous NumStepsForAveraging&-1 steps
```

```
            BestFitness = M(1,i&)
            FOR p% = 1 TO Np%
                IF M(p%,i&) >= BestFitness THEN BestFitness = M(p%,i&)
            NEXT p%
            AvgFitness = AvgFitness + BestFitness
        NEXT i&
        AvgFitness = AvgFitness/NumStepsForAveraging&
        IF ABS(AvgFitness-BestFitness) < FitnessTOL THEN A$ = "YES" 'compare avg fitness to best fitness at this step
    END IF
    TerminateNowForSaturation$ = A$
END FUNCTION 'TerminateNowForSaturation$()
'-----------
FUNCTION MagVector(V(),N%) 'returns magnitude of Nx1 column vector V
LOCAL SumSq AS EXT
LOCAL i%
    SumSQ = 0## : FOR i% = 1 TO N% : SumSQ = SumSQ + V(i%)^2 : NEXT i% : MagVector = SQR(SumSQ)
END FUNCTION 'MagVector()
'---
FUNCTION UnitStep(X)
LOCAL Z AS EXT
    IF X < 0## THEN
        Z = 0##
    ELSE
        Z = 1##
    END IF
    UnitStep = Z
END FUNCTION 'UnitStep()
'---
SUB Plot1Dfunction(FunctionName$,R()) 'plots 1D function on-screen
LOCAL NumPoints%, i%, N%
LOCAL DeltaX, X AS EXT
    NumPoints% = 32001
    DeltaX = (XiMax(1)-XiMin(1))/(NumPoints%-1)
    N% = FREEFILE
    SELECT CASE FunctionName$
        CASE "ParrottF4" 'PARROTT F4 FUNCTION
            OPEN "ParrottF4" FOR OUTPUT AS #N%
                FOR i% = 1 TO NumPoints%
                    R(1,1,0) = XiMin(1) + (i%-1)*DeltaX
                    PRINT #N%, USING$("#.######  #.######",R(1,1,0),ParrottF4(R(),1,1,0))
                NEXT i%
            CLOSE #N%
            CALL CreateGNUplotINIfile(0.2##*ScreenWidth&,0.2##*ScreenHeight&,0.6##*ScreenWidth&,0.6##*ScreenHeight&)
            CALL TwoDplot("ParrottF4","Parrott F4 Function","0.7","0.7","X\n\n.",".\n\nParrott F4(X)","","","0","1","0","1","","","wgnuplot.exe"," with
lines linewidth 2","")
    END SELECT
END SUB
'------
SUB CLEANUP 'probe coordinate plot files
    IF DIR$("P1")  <> "" THEN KILL "P1"
    IF DIR$("P2")  <> "" THEN KILL "P2"
    IF DIR$("P3")  <> "" THEN KILL "P3"
    IF DIR$("P4")  <> "" THEN KILL "P4"
    IF DIR$("P5")  <> "" THEN KILL "P5"
    IF DIR$("P6")  <> "" THEN KILL "P6"
    IF DIR$("P7")  <> "" THEN KILL "P7"
    IF DIR$("P8")  <> "" THEN KILL "P8"
    IF DIR$("P9")  <> "" THEN KILL "P9"
    IF DIR$("P10") <> "" THEN KILL "P10"
    IF DIR$("P11") <> "" THEN KILL "P11"
    IF DIR$("P12") <> "" THEN KILL "P12"
    IF DIR$("P13") <> "" THEN KILL "P13"
    IF DIR$("P14") <> "" THEN KILL "P14"
    IF DIR$("P15") <> "" THEN KILL "P15"
END SUB
'------
SUB Plot2Dfunction(FunctionName$,R())
LOCAL A$
LOCAL NumPoints%, i%, k%, N%
LOCAL DelX1, DelX2, Z AS EXT
    SELECT CASE FunctionName$
        CASE "BOWTIE" : Numpoints% = 5
```

```
        CASE "YAGI"    : Numpoints% = 5
        CASE "PBM_1","PBM_2","PBM_3","PBM_4","PBM_5" : NumPoints% = 25
        CASE ELSE : NumPoints% = 100
    END SELECT
    N% = FREEFILE : OPEN "TwoDplot.DAT" FOR OUTPUT AS #N%
    DelX1 = (XiMax(1)-XiMin(1))/(NumPoints%-1) : DelX2 = (XiMax(2)-XiMin(2))/(NumPoints%-1)
    FOR i% = 1 TO NumPoints%
        R(1,1,0) = XiMin(1) + (i%-1)*DelX1 'x1 value
        FOR k% = 1 TO NumPoints%
            R(1,2,0) = XiMin(2) + (k%-1)*DelX2 'x2 value
            Z = ObjectiveFunction(R(),2,1,0,FunctionName$)
            PRINT #N%, USING$("######.###### ######.###### #######.######^^^^",R(1,1,0),R(1,2,0),Z)
        NEXT k%
        PRINT #N%, ""
    NEXT i%
    CLOSE #N%
    CALL CreateGNUplotINIfile(0.1##*ScreenWidth&,0.1##*ScreenHeight&,0.6##*ScreenWidth&,0.6##*ScreenHeight&)
    A$ = "" : IF INSTR(FunctionName$,"PBM_") > 0 THEN A$ = "Coarse "
    CALL ThreeDplot2("TwoDplot.DAT",A$+"Plot of "+FunctionName$+" Function","","0.6","0.6","1.2", _
                    "x1","x2","z=F(x1,x2)","","","wgnuplot.exe","","","","","")
END SUB
'------
    SUB TwoDplot3curves(NumCurves%,PlotFileName1$,PlotFileName2$,PlotFileName3$,PlotTitle$,Annotation$,xCoord$,yCoord$,XaxisLabel$,YaxisLabel$, _
                        LogXaxis$,LogYaxis$,xMin$,xMax$,yMin$,yMax$,xTics$,yTics$,GnuPlotEXE$)
        LOCAL N%
        LOCAL LineSize$
        LineSize$ = "2"
        N% = FREEFILE
        OPEN "cmd2d.gp" FOR OUTPUT AS #N%
            IF LogXaxis$ = "YES" AND LogYaxis$ = "NO"  THEN PRINT #N%, "set logscale x"
            IF LogXaxis$ = "NO"  AND LogYaxis$ = "YES" THEN PRINT #N%, "set logscale y"
            IF LogXaxis$ = "YES" AND LogYaxis$ = "YES" THEN PRINT #N%, "set logscale xy"
            IF xMin$ <> "" AND xMax$ <> "" THEN  PRINT #N%, "set xrange ["+xMin$+":"+xMax$+"]"
            IF yMin$ <> "" AND yMax$ <> "" THEN  PRINT #N%, "set yrange ["+yMin$+":"+yMax$+"]"
            PRINT #N%, "set label "+Quote$+AnnoTation$+Quote$+" at graph "+xCoord$+","+yCoord$
            PRINT #N%, "set grid xtics"
            PRINT #N%, "set grid ytics"
            PRINT #N%, "set xtics "+xTics$
            PRINT #N%, "set ytics "+yTics$
            PRINT #N%, "set grid mxtics"
            PRINT #N%, "set grid mytics"
            PRINT #N%, "set title "+Quote$+PlotTitle$+Quote$
            PRINT #N%, "set xlabel "+Quote$+XaxisLabel$+Quote$
            PRINT #N%, "set ylabel "+Quote$+YaxisLabel$+Quote$
            SELECT CASE NumCurves%
            CASE 1
            PRINT #N%, "plot " + Quote$ + PlotFileName1$ + Quote$ + " with lines linewidth " + LineSize$
            CASE 2
            PRINT #N%, "plot " + Quote$ + PlotFileName1$ + Quote$ + " with lines linewidth " + LineSize$+", " + _
                                 Quote$ + PlotFileName2$ + Quote$ + " with lines linewidth " + LineSize$
            CASE 3
            PRINT #N%, "plot " + Quote$ + PlotFileName1$ + Quote$ + " with lines linewidth " + LineSize$+", " + _
                                 Quote$ + PlotFileName2$ + Quote$ + " with lines linewidth " + LineSize$+", " + _
                                 Quote$ + PlotFileName3$ + Quote$ + " with lines linewidth " + LineSize$
            END SELECT
        CLOSE #N%
        SHELL(GnuPlotEXE$+" cmd2d.gp -")
        CALL Delay(1##)
    END SUB 'TwoDplot3Curves()
'---
FUNCTION Fibonacci&&(N%) 'RETURNS Nth FIBONACCI NUMBER
LOCAL i%, Fn&&, Fn1&&, Fn2&&
LOCAL A$
    IF N% > 91 OR N% < 0 THEN
        MSGBOX("ERROR!  Fibonacci argument"+STR$(N%)+" > 91.  Out of range or < 0...") : EXIT FUNCTION
    END IF
    SELECT CASE N%
        CASE 0: Fn&& = 1
        CASE ELSE
            Fn&& = 0 : Fn2&& = 1 : i% = 0
            FOR i% = 1 TO N%
                Fn&& = Fn1&& + Fn2&&
                Fn1&& = Fn2&&
                Fn2&& = Fn&&
            NEXT i% 'LOOP
```

```
    END SELECT

    Fibonacci&& = Fn&&

END FUNCTION 'Fibonacci&&()

'-----------

FUNCTION RandomNum(a,b) 'Returns random number X, a=< X < b.

    RandomNum = a + (b-a)*RND

END FUNCTION 'RandomNum()

'-----------

FUNCTION GaussianDeviate(Mu,Sigma) 'returns NORMAL (Gaussian) random deviate with mean Mu and standard deviation Sigma (variance = Sigma^2)

'Refs: (1) Press, W.H., Flannery, B.P., Teukolsky, S.A., and Vetterling, W.T., "Numerical Recipes: The Art of Scientific Computing,"
'          §7.2, Cambridge University Press, Cambridge, UK, 1986.
'      (2) Shinzato, T., "Box Muller Method," 2007, http://www.sp.dis.titech.ac.jp/~shinzato/boxmuller.pdf

LOCAL s, t, Z AS EXT

    s = RND : t = RND

    Z = Mu + Sigma*SQR(-2##*LOG(s))*COS(TwoPi*t)

    GaussianDeviate = Z

END FUNCTION 'GaussianDeviate()

'-----------

    SUB ContourPlot(PlotFileName$,PlotTitle$,Annotation$,xCoord$,yCoord$,zCoord$, _
                XaxisLabel$,YaxisLabel$,ZaxisLabel$,zMin$,zMax$,GnuPlotEXE$,A$)
        LOCAL N%

        N% = FREEFILE

        OPEN "cmd3d.gp" FOR OUTPUT AS #N%

            PRINT #N%, "show surface"
            PRINT #N%, "set hidden3d"
            IF zMin$ <> "" AND zMax$ <> "" THEN  PRINT #N%, "set zrange ["+zMin$+":"+zMax$+"]"
            PRINT #N%, "set label "+Quote$+AnnoTation$+Quote$+" at graph "+xCoord$+","+yCoord$+","+zCoord$
            PRINT #N%, "show label"
            PRINT #N%, "set grid xtics ytics ztics"
            PRINT #N%, "show grid"
            PRINT #N%, "set title "+Quote$+PlotTitle$+Quote$
            PRINT #N%, "set xlabel "+Quote$+XaxisLabel$+Quote$
            PRINT #N%, "set ylabel "+Quote$+YaxisLabel$+Quote$
            PRINT #N%, "set zlabel "+Quote$+ZaxisLabel$+Quote$
            PRINT #N%, "splot "+Quote$+PlotFileName$+Quote$+A$  '" notitle with linespoints" 'A$'" notitle with lines"
        CLOSE #N%

        SHELL(GnuPlotEXE$+" cmd3d.gp -")

    END SUB 'ContourPlot()

'---

    SUB ThreeDplot(PlotFileName$,PlotTitle$,Annotation$,xCoord$,yCoord$,zCoord$, _
               XaxisLabel$,YaxisLabel$,ZaxisLabel$,zMin$,zMax$,GnuPlotEXE$,A$)

        LOCAL N%, ProcessID???

        N% = FREEFILE

        OPEN "cmd3d.gp" FOR OUTPUT AS #N%

            PRINT #N%, "set pm3d"
            PRINT #N%, "show pm3d"
            IF zMin$ <> "" AND zMax$ <> "" THEN  PRINT #N%, "set zrange ["+zMin$+":"+zMax$+"]"
            PRINT #N%, "set label "+Quote$+AnnoTation$+Quote$+" at graph "+xCoord$+","+yCoord$+","+zCoord$
            PRINT #N%, "show label"
            PRINT #N%, "set grid xtics ytics ztics"
            PRINT #N%, "show grid"
            PRINT #N%, "set title "+Quote$+PlotTitle$+Quote$
            PRINT #N%, "set xlabel "+Quote$+XaxisLabel$+Quote$
            PRINT #N%, "set ylabel "+Quote$+YaxisLabel$+Quote$
            PRINT #N%, "set zlabel "+Quote$+ZaxisLabel$+Quote$
            PRINT #N%, "splot "+Quote$+PlotFileName$+Quote$+A$+" notitle"' with lines"
        CLOSE #N%

        SHELL(GnuPlotEXE$+" cmd3d.gp -") : CALL Delay(1##)

    END SUB 'ThreeDplot()

'---

    SUB ThreeDplot2(PlotFileName$,PlotTitle$,Annotation$,xCoord$,yCoord$,zCoord$, _
                XaxisLabel$,YaxisLabel$,ZaxisLabel$,zMin$,zMax$,GnuPlotEXE$,A$,xStart$,xStop$,yStart$,yStop$)

        LOCAL N%

        N% = FREEFILE

        OPEN "cmd3d.gp" FOR OUTPUT AS #N%

            PRINT #N%, "set pm3d"
            PRINT #N%, "show pm3d"
            PRINT #N%, "set hidden3d"
            PRINT #N%, "set view 45, 45, 1, 1"

            IF zMin$ <> "" AND zMax$ <> "" THEN  PRINT #N%, "set zrange ["+zMin$+":"+zMax$+"]"

            PRINT #N%, "set xrange [" + xStart$ + ":" + xStop$ + "]"
            PRINT #N%, "set yrange [" + yStart$ + ":" + yStop$ + "]"

            PRINT #N%, "set label "   + Quote$  + AnnoTation$ + Quote$+" at graph "+xCoord$+","+yCoord$+","+zCoord$
            PRINT #N%, "show label"
            PRINT #N%, "set grid xtics ytics ztics"
            PRINT #N%, "show grid"
            PRINT #N%, "set title "  + Quote$+PlotTitle$    + Quote$
            PRINT #N%, "set xlabel " + Quote$+XaxisLabel$   + Quote$
            PRINT #N%, "set ylabel " + Quote$+YaxisLabel$   + Quote$
            PRINT #N%, "set zlabel " + Quote$+ZaxisLabel$   + Quote$
            PRINT #N%, "splot "      + Quote$+PlotFileName$ + Quote$ + A$ + " notitle with lines"
        CLOSE #N%

        SHELL(GnuPlotEXE$+" cmd3d.gp -")

    END SUB 'ThreeDplot2()

'---

     SUB TwoDplot2Curves(PlotFileName1$,PlotFileName2$,PlotTitle$,Annotation$,xCoord$,yCoord$,XaxisLabel$,YaxisLabel$, _
                  LogXaxis$,LogYaxis$,xMin$,xMax$,yMin$,yMax$,xTics$,yTics$,GnuPlotEXE$,LineSize)
```

```
        LOCAL N%, ProcessID???

        N% = FREEFILE

        OPEN "cmd2d.gp" FOR OUTPUT AS #N%
            'print #N%, "set output "+Quote$+"test.plt"+Quote$ 'tried this 3/11/06, didn't work...

            IF LogXaxis$ = "YES" AND LogYaxis$ = "NO"  THEN PRINT #N%, "set logscale x"
            IF LogXaxis$ = "NO"  AND LogYaxis$ = "YES" THEN PRINT #N%, "set logscale y"
            IF LogXaxis$ = "YES" AND LogYaxis$ = "YES" THEN PRINT #N%, "set logscale xy"

            IF xMin$ <> "" AND xMax$ <> "" THEN  PRINT #N%, "set xrange ["+xMin$+":"+xMax$+"]"

            IF yMin$ <> "" AND yMax$ <> "" THEN  PRINT #N%, "set yrange ["+yMin$+":"+yMax$+"]"

            PRINT #N%, "set label "+Quote$+AnnoTation$+Quote$+" at graph "+xCoord$+","+yCoord$
            PRINT #N%, "set grid xtics"
            PRINT #N%, "set grid ytics"
            PRINT #N%, "set xtics "+xTics$
            PRINT #N%, "set ytics "+yTics$
            PRINT #N%, "set grid mxtics"
            PRINT #N%, "set grid mytics"
            PRINT #N%, "set title "+Quote$+PlotTitle$+Quote$
            PRINT #N%, "set xlabel "+Quote$+XaxisLabel$+Quote$
            PRINT #N%, "set ylabel "+Quote$+YaxisLabel$+Quote$

            PRINT #N%, "plot "+Quote$+PlotFileName1$+Quote$+" with lines linewidth "+REMOVE$(STR$(LineSize),ANY" ")+","+ _
                               Quote$+PlotFileName2$+Quote$+" with points pointsize 0.05"'+REMOVE$(STR$(LineSize),ANY" ")

        CLOSE #N%

        ProcessID??? = SHELL(GnuPlotEXE$+" cmd2d.gp -") : CALL Delay(1##)

    END SUB 'TwoDplot2Curves()

'---

    SUB Probe2Dplots(ProbePlotsFileList$,PlotTitle$,Annotation$,xCoord$,yCoord$,XaxisLabel$,YaxisLabel$, _
                     LogXaxis$,LogYaxis$,xMin$,xMax$,yMin$,yMax$,xTics$,yTics$,GnuPlotEXE$)

        LOCAL N%, ProcessID???

        N% = FREEFILE

        OPEN "cmd2d.gp" FOR OUTPUT AS #N%

            IF LogXaxis$ = "YES" AND LogYaxis$ = "NO"  THEN PRINT #N%, "set logscale x"
            IF LogXaxis$ = "NO"  AND LogYaxis$ = "YES" THEN PRINT #N%, "set logscale y"
            IF LogXaxis$ = "YES" AND LogYaxis$ = "YES" THEN PRINT #N%, "set logscale xy"

            IF xMin$ <> "" AND xMax$ <> "" THEN  PRINT #N%, "set xrange ["+xMin$+":"+xMax$+"]"

            IF yMin$ <> "" AND yMax$ <> "" THEN  PRINT #N%, "set yrange ["+yMin$+":"+yMax$+"]"

            PRINT #N%, "set label "+Quote$+AnnoTation$+Quote$+" at graph "+xCoord$+","+yCoord$
            PRINT #N%, "set grid xtics"
            PRINT #N%, "set grid ytics"
            PRINT #N%, "set xtics "+xTics$
            PRINT #N%, "set ytics "+yTics$
            PRINT #N%, "set grid mxtics"
            PRINT #N%, "set grid mytics"
            PRINT #N%, "set title "+Quote$+PlotTitle$+Quote$
            PRINT #N%, "set xlabel "+Quote$+XaxisLabel$+Quote$
            PRINT #N%, "set ylabel "+Quote$+YaxisLabel$+Quote$

            PRINT #N%, ProbePlotsFileList$

        CLOSE #N%

        ProcessID??? = SHELL(GnuPlotEXE$+" cmd2d.gp -") : CALL Delay(1##)

    END SUB 'Probe2Dplots()

'---

SUB Show2Dprobes(R(),Np%,Nt&,j&,Frep,BestFitness,BestProbeNumber%,BestTimeStep&,FunctionName$,RepositionFactor$,Gamma)

    LOCAL N%, p%

    LOCAL A$, PlotFileName$, PlotTitle$, Symbols$

    LOCAL xMin$, xMax$, yMin$, yMax$

    LOCAL s1, s2, s3, s4 AS EXT

    PlotFileName$ = "Probes("+REMOVE$(STR$(j&),ANY" ")+")"

    IF j& > 0 THEN 'PLOT PROBES AT THIS TIME STEP

        PlotTitle$ = "\nLOCATIONS OF "+REMOVE$(STR$(Np%),ANY" ") + " PROBES AT TIME STEP" + STR$(j&) + " / " + REMOVE$(STR$(Nt&),ANY" ") + "\n" + _
                     "Fitness = "+REMOVE$(STR$(ROUND(BestFitness,3)),ANY" ") + ", Probe #" + REMOVE$(STR$(BestProbeNumber%),ANY" ") + " at Step #" +
REMOVE$(STR$(BestTimeStep&),ANY" ") + _
                     "  [Frep = "+REMOVE$(STR$(Frep,4),ANY" ") + " " + RepositionFactor$ + "]\n"

    ELSE 'PLOT INITIAL PROBE DISTRIBUTION

        PlotTitle$ = "\nLOCATIONS OF "+REMOVE$(STR$(Np%),ANY" ") + " INITIAL PROBES FOR " + FunctionName$ + " FUNCTION\n[gamma =
"+STR$(ROUND(Gamma,3))+"]\n"

    END IF

    N% = FREEFILE : OPEN PlotFileName$ FOR OUTPUT AS #N%

        FOR p% = 1 TO Np% : PRINT #N%, USING$("######.#####    ######.#####",R(p%,1,j&),R(p%,2,j&)) : NEXT p%

    CLOSE #N%

    s1 = 1.1## : s2 = 1.1## : s3 = 1.1## : s4 = 1.1## 'expand plots axes by 10%

    IF XiMin(1) > 0## THEN s1 = 0.9##
    IF XiMax(1) < 0## THEN s2 = 0.9##
    IF XiMin(2) > 0## THEN s3 = 0.9##
    IF XiMax(2) < 0## THEN s4 = 0.9##

    xMin$ = REMOVE$(STR$(s1*XiMin(1),2),ANY" ")
    xMax$ = REMOVE$(STR$(s2*XiMax(1),2),ANY" ")
    yMin$ = REMOVE$(STR$(s3*XiMin(2),2),ANY" ")
    yMax$ = REMOVE$(STR$(s4*XiMax(2),2),ANY" ")

    CALL TwoDplot(PlotFileName$,PlotTitle$,"0.6","0.7","x1\n\n","\nx2","NO","NO",xMin$,xMax$,yMin$,yMax$,"5","5","wgnuplot.exe"," pointsize 1 linewidth
2","")

    KILL PlotFileName$ 'erase plot data file after probes have been displayed

END SUB 'Show2Dprobes()

'---
```

```
SUB Show3Dprobes(R(),Np%,Nd%,Nt&,j&,Frep,BestFitness,BestProbeNumber%,BestTimeStep&,FunctionName$,RepositionFactor$,Gamma)

    LOCAL N%, p%, PlotWindowULC_X%, PlotWindowULC_Y%, PlotWindowWidth%, PlotWindowHeight%, PlotWindowOffset%

    LOCAL A$, PlotFileName$, PlotTitle$, Symbols$, Annotation$

    LOCAL xMin$, xMax$, yMin$, yMax$, zMin$, zMax$

    LOCAL s1, s2, s3, s4, s5, s6 AS EXT

    PlotFileName$ = "Probes("+REMOVE$(STR$(j&),ANY" ")+")"

    IF j& > 0 THEN 'PLOT PROBES AT THIS TIME STEP

        PlotTitle$ = "\nLOCATIONS OF "+REMOVE$(STR$(Np%),ANY" ") + " PROBES AT TIME STEP" + STR$(j&) + " / " + REMOVE$(STR$(Nt&),ANY" ") + "\n" + _
                     "Fitness = "+REMOVE$(STR$(ROUND(BestFitness,3)),ANY" ") + ", Probe #" + REMOVE$(STR$(BestProbeNumber%),ANY" ") + " at Step #" +
REMOVE$(STR$(BestTimeStep&),ANY" ") + _
                     "  [Frep = "+REMOVE$(STR$(Frep,4),ANY" ") + " " + RepositionFactor$ + "]\n"

    ELSE 'PLOT INITIAL PROBE DISTRIBUTION

        A$ = "" : IF Gamma > 0## AND Gamma < 1## THEN A$ = "0"

        PlotTitle$ = "\n"+REMOVE$(STR$(Np%),ANY" ") + "-PROBE IPD FOR FUNCTION " + FunctionName$ + ", GAMMA = "+A$+REMOVE$(STR$(ROUND(Gamma,3)),ANY" ")

    END IF

'   ---------------- Probe Coordinates -----------------

    N% = FREEFILE : OPEN PlotFileName$ FOR OUTPUT AS #N%

        PRINT #N%, USING$("######.#####    ######.#####    ######.#####",R(1,1,j&),R(1,2,j&),R(1,3,j&)) 'This line repeats Probe #1's coordinates.  It's
necessary
                                                                                                 'to deal with a plotting artifact in Gnuplot!
        FOR p% = 1 TO Np% : PRINT #N%, USING$("######.#####    ######.#####    ######.#####",R(p%,1,j&),R(p%,2,j&),R(p%,3,j&)) : NEXT p%

    CLOSE #N%

'   ------------ Principal Diagonal -------------

    N% = FREEFILE : OPEN "diag" FOR OUTPUT AS #N%

        PRINT #N%, USING$("######.#####    ######.#####    ######.#####",XiMin(1),XiMin(2),XiMin(3))
        PRINT #N%, ""
        PRINT #N%, USING$("######.#####    ######.#####    ######.#####",XiMax(1),XiMax(2),XiMax(3))

    CLOSE #N%

'   ------------------- Probe Line #1 ------------------

    N% = FREEFILE : OPEN "probeline1" FOR OUTPUT AS #N%

        PRINT #N%, USING$("######.#####    ######.#####    ######.#####",R(1,1,j&),R(1,2,j&),R(1,3,j&))
        PRINT #N%, ""
        PRINT #N%, USING$("######.#####    ######.#####    ######.#####",R(Np%/Nd%,1,j&),R(Np%/Nd%,2,j&),R(Np%/Nd%,3,j&))

    CLOSE #N%

'   ------------------- Probe Line #2 ------------------

    N% = FREEFILE : OPEN "probeline2" FOR OUTPUT AS #N%

        PRINT #N%, USING$("######.#####    ######.#####    ######.#####",R(1+Np%/Nd%,1,j&),R(1+Np%/Nd%,2,j&),R(1+Np%/Nd%,3,j&))
        PRINT #N%, ""
        PRINT #N%, USING$("######.#####    ######.#####    ######.#####",R(2*Np%/Nd%,1,j&),R(2*Np%/Nd%,2,j&),R(2*Np%/Nd%,3,j&))

    CLOSE #N%

'   ------------------- Probe Line #3 ------------------

    N% = FREEFILE : OPEN "probeline3" FOR OUTPUT AS #N%

        PRINT #N%, USING$("######.#####    ######.#####    ######.#####",R(1+2*Np%/Nd%,1,j&),R(1+2*Np%/Nd%,2,j&),R(1+2*Np%/Nd%,3,j&))
        PRINT #N%, ""
        PRINT #N%, USING$("######.#####    ######.#####    ######.#####",R(3*Np%/Nd%,1,j&),R(3*Np%/Nd%,2,j&),R(3*Np%/Nd%,3,j&))

    CLOSE #N%

'   -------- RE-PLOT PROBE #1 BECAUSE OF SOME ARTIFACT THAT DROPS IT FROM PROBE LINE #1 ?????? ------------------

    N% = FREEFILE : OPEN "probe1" FOR OUTPUT AS #N%

        PRINT #N%, USING$("######.#####    ######.#####    ######.#####",R(1,1,j&),R(1,2,j&),R(1,3,j&))
        PRINT #N%, USING$("######.#####    ######.#####    ######.#####",R(1,1,j&),R(1,2,j&),R(1,3,j&))

    CLOSE #N%

    s1 = 1.1## : s2 = s1 : s3 = s1 : s4 = s1 : s5 = s1 : s6 = s1 'expand plots axes by 10%

    IF XiMin(1) > 0## THEN s1 = 0.9##
    IF XiMax(1) < 0## THEN s2 = 0.9##
    IF XiMin(2) > 0## THEN s3 = 0.9##
    IF XiMax(2) < 0## THEN s4 = 0.9##
    IF XiMin(3) > 0## THEN s5 = 0.9##
    IF XiMax(3) < 0## THEN s6 = 0.9##

    xMin$ = REMOVE$(STR$(s1*XiMin(1),2),ANY" ")
    xMax$ = REMOVE$(STR$(s2*XiMax(1),2),ANY" ")
    yMin$ = REMOVE$(STR$(s3*XiMin(2),2),ANY" ")
    yMax$ = REMOVE$(STR$(s4*XiMax(2),2),ANY" ")
    zMin$ = REMOVE$(STR$(s5*XiMin(2),2),ANY" ")
    zMax$ = REMOVE$(STR$(s6*XiMax(2),2),ANY" ")

'USAGE: CALL ThreeDplot3(PlotFileName$,PlotTitle$,Annotation$,xCoord$,yCoord$,zCoord$, _
'                        XaxisLabel$,YaxisLabel$,ZaxisLabel$,zMin$,zMax$,GnuPlotEXE$,xStart$,xStop$,yStart$,yStop$)

    PlotWindowULC_X% = 50 : PlotWindowULC_Y% = 50 : PlotWindowWidth% = 1000 : PlotWindowHeight% = 800

    PlotWindowOffset% = 100*Gamma

    CALL CreateGNUplotINIfile(PlotWindowULC_X%+PlotWindowOffset%,PlotWindowULC_Y%+PlotWindowOffset%,PlotWindowWidth%,PlotWindowHeight%)

    CALL ThreeDplot3(PlotFileName$,PlotTitle$,Annotation$,"0.6","0.7","0.8", _
                     "x1","x2","x3",zMin$,zMax$,"wgnuplot.exe",xMin$,xMax$,yMin$,yMax$)

    'KILL PlotFileName$ 'erase plot data file after probes have been displayed

END SUB 'Show3Dprobes()

'---

    SUB ThreeDplot3(PlotFileName$,PlotTitle$,Annotation$,xCoord$,yCoord$,zCoord$, _
                   XaxisLabel$,YaxisLabel$,ZaxisLabel$,zMin$,zMax$,GnuPlotEXE$,xStart$,xStop$,yStart$,yStop$)

        LOCAL N%, ProcID???

        N% = FREEFILE

        OPEN "cmd3d.gp" FOR OUTPUT AS #N%
```

```
            PRINT #N%, "set pm3d"
            PRINT #N%, "show pm3d"
            PRINT #N%, "set hidden3d"
'           PRINT #N%, "set view 45, 45, 1, 1"

            PRINT #N%, "set view 45, 60, 1, 1"

            IF zMin$ <> "" AND zMax$ <> "" THEN  PRINT #N%, "set zrange ["+zMin$+":"+zMax$+"]"

            PRINT #N%, "set xrange [" + xStart$ + ":" + xStop$ + "]"
            PRINT #N%, "set yrange [" + yStart$ + ":" + yStop$ + "]"

            PRINT #N%, "set label "   + Quote$  + AnnoTation$ + Quote$+" at graph "+xCoord$+","+yCoord$+","+zCoord$
            PRINT #N%, "show label"
            PRINT #N%, "set grid xtics ytics ztics"
            PRINT #N%, "show grid"
            PRINT #N%, "set title "  + Quote$+PlotTitle$     + Quote$
            PRINT #N%, "set xlabel " + Quote$+XaxisLabel$    + Quote$
            PRINT #N%, "set ylabel " + Quote$+YaxisLabel$    + Quote$
            PRINT #N%, "set zlabel " + Quote$+ZaxisLabel$    + Quote$
            PRINT #N%, "unset colorbox"
'           print #N%, "set style fill"

            PRINT #N%, "splot "      + Quote$+PlotFileName$  + Quote$ + " notitle lw 1 pt 8," _
                                     + Quote$ + "diag"       + Quote$ + " notitle w l," _
                                     + Quote$ + "probeline1" + Quote$ + " notitle w l," _
                                     + Quote$ + "probeline2" + Quote$ + " notitle w l," _
                                     + Quote$ + "probeline3" + Quote$ + " notitle w l"
        CLOSE #N%

        ProcID??? = SHELL(GnuPlotEXE$+" cmd3d.gp -")

        CALL Delay(1##)

    END SUB 'ThreeDplot3()

'----

    SUB TwoDplot(PlotFileName$,PlotTitle$,xCoord$,yCoord$,XaxisLabel$,YaxisLabel$, _
             LogXaxis$,LogYaxis$,xMin$,xMax$,yMin$,yMax$,xTics$,yTics$,GnuPlotEXE$,LineType$,Annotation$)

        LOCAL N%, ProcessID???

        N% = FREEFILE

        OPEN "cmd2d.gp" FOR OUTPUT AS #N%

            IF LogXaxis$ = "YES" AND LogYaxis$ = "NO"  THEN PRINT #N%, "set logscale x"
            IF LogXaxis$ = "NO"  AND LogYaxis$ = "YES" THEN PRINT #N%, "set logscale y"
            IF LogXaxis$ = "YES" AND LogYaxis$ = "YES" THEN PRINT #N%, "set logscale xy"

            IF xMin$ <> "" AND xMax$ <> "" THEN  PRINT #N%, "set xrange ["+xMin$+":"+xMax$+"]"
            IF yMin$ <> "" AND yMax$ <> "" THEN  PRINT #N%, "set yrange ["+yMin$+":"+yMax$+"]"

            PRINT #N%, "set label "      + Quote$ + Annotation$ + Quote$ + " at graph " + xCoord$ + "," + yCoord$
            PRINT #N%, "set grid xtics " + XTics$
            PRINT #N%, "set grid ytics " + yTics$
            PRINT #N%, "set grid mxtics"
            PRINT #N%, "set grid mytics"
            PRINT #N%, "show grid"
            PRINT #N%, "set title "  + Quote$+PlotTitle$+Quote$
            PRINT #N%, "set xlabel " + Quote$+XaxisLabel$+Quote$
            PRINT #N%, "set ylabel " + Quote$+YaxisLabel$+Quote$

            PRINT #N%, "plot "+Quote$+PlotFileName$+Quote$+" notitle"+LineType$

        CLOSE #N%

        ProcessID??? = SHELL(GnuPlotEXE$+" cmd2d.gp -") : CALL Delay(1##)

    END SUB 'TwoDplot()

'-----

    SUB CreateGNUplotINIfile(PlotWindowULC_X%,PlotWindowULC_Y%,PlotWindowWidth%,PlotWindowHeight%)

    LOCAL N%, WinPath$, A$, B$, WindowsDirectory$

    WinPath$ = UCASE$(ENVIRON$("Path"))'DIR$("C:\WINDOWS",23)

    DO
        B$ = A$

        A$ = EXTRACT$(WinPath$,";")

        WinPath$ = REMOVE$(WinPath$,A$+";")

        IF RIGHT$(A$,7) = "WINDOWS" OR A$ = B$ THEN EXIT LOOP

        IF RIGHT$(A$,5) = "WINNT"   OR A$ = B$ THEN EXIT LOOP

    LOOP

    WindowsDirectory$ = A$

    N% = FREEFILE

'   ----------- WGNUPLOT.INPUT FILE -----------
    OPEN WindowsDirectory$+"\wgnuplot.ini" FOR OUTPUT AS #N%

        PRINT #N%,"[WGNUPLOT]"
        PRINT #N%,"TextOrigin=0 0"
        PRINT #N%,"TextSize=640 150"
        PRINT #N%,"TextFont=Terminal,9"
        PRINT #N%,"GraphOrigin="+REMOVE$(STR$(PlotWindowULC_X%),ANY" ")+" "+REMOVE$(STR$(PlotWindowULC_Y%),ANY" ")
        PRINT #N%,"GraphSize="  +REMOVE$(STR$(PlotWindowWidth%),ANY" ")+" "+REMOVE$(STR$(PlotWindowHeight%),ANY" ")
        PRINT #N%,"GraphFont=Arial,10"
        PRINT #N%,"GraphColor=1"
        PRINT #N%,"GraphToTop=1"
        PRINT #N%,"GraphBackground=255 255 255"
        PRINT #N%,"Border=0 0 0 0 0"
        PRINT #N%,"Axis=192 192 192 2 2"
        PRINT #N%,"Line1=0 0 255 0 0"
        PRINT #N%,"Line2=0 255 0 0 1"
        PRINT #N%,"Line3=255 0 0 0 2"
        PRINT #N%,"Line4=255 0 255 0 3"
        PRINT #N%,"Line5=0 0 128 0 4"

    CLOSE #N%

    END SUB 'CreateGNUplotINIfile()

'------

    SUB Delay(NumSecs)

        LOCAL StartTime, StopTime AS EXT
```

```
        StartTime = TIMER

        DO UNTIL (StopTime-StartTime) >= NumSecs

            StopTime = TIMER

        LOOP

    END SUB 'Delay()

'-----

SUB MathematicalConstants
    EulerConst    = 0.577215664901532860606512##
    Pi            = 3.141592653589793238462643##
    Pi2           = Pi/2##
    Pi4           = Pi/4##
    TwoPi         = 2##*Pi
    FourPi        = 4##*Pi
    e             = 2.718281828459045235360287##
    Root2         = 1.4142135623730950488##
END SUB

'-----

SUB AlphabetAndDigits
    Alphabet$   = "ABCDEFGHIJKLMNOPQRSTUVWXYZabcdefghijklmnopqrstuvwxyz"
    Digits$     = "0123456789"
    RunID$      = REMOVE$(DATE$+"_"+TIME$,ANY Alphabet$+" :-/")
END SUB

'------

SUB SpecialSymbols
    Quote$             = CHR$(34) 'Quotation mark "
    SpecialCharacters$ = "'(),#:;/_"
END SUB

'-----

SUB EMconstants
    Mu0  = 4E-7##*Pi     'hy/meter
    Eps0 = 8.854##*1E-12 'fd/meter
    c    = 2.998E8##     'velocity of light, 1##/SQR(Mu0*Eps0) 'meters/sec
    eta0 = SQR(Mu0/Eps0) 'impedance of free space, ohms
END SUB

'------

SUB ConversionFactors
    Rad2Deg       = 180##/Pi
    Deg2Rad       = 1##/Rad2Deg
    Feet2Meters   = 0.3048##
    Meters2Feet   = 1##/Feet2Meters
    Inches2Meters = 0.0254##
    Meters2Inches = 1##/Inches2Meters
    Miles2Meters  = 1609.344##
    Meters2Miles  = 1##/Miles2Meters
    NautMi2Meters = 1852##
    Meters2NautMi = 1##/NautMi2Meters
END SUB

'------

SUB ShowConstants 'puts up msgbox showing all constants

LOCAL A$

A$ = _
"Mathematical Constants:"+CHR$(13)+_
"Euler const="+STR$(EulerConst)+CHR$(13)+_
"Pi="+STR$(Pi)+CHR$(13)+_
"Pi/2="+STR$(Pi2)+CHR$(13)+_
"Pi/4="+STR$(Pi4)+CHR$(13)+_
"2Pi="+STR$(TwoPi)+CHR$(13)+_
"4Pi="+STR$(FourPi)+CHR$(13)+_
"e="+STR$(e)+CHR$(13)+_
"Sqr2="+STR$(Root2)+CHR$(13)+CHR$(13)+_
"Alphabet, Digits & Special Characters:"+CHR$(13)+_
"Alphabet="+Alphabet$+CHR$(13)+_
"Digits="+Digits$+CHR$(13)+_
"quote="+Quote$+CHR$(13)+_
"Spec chars="+SpecialCharacters$+CHR$(13)+CHR$(13)+_
"E&M Constants:"+CHR$(13)+_
"Mu0="+STR$(Mu0)+CHR$(13)+_
"Eps0="+STR$(Eps0)+CHR$(13)+_
"c="+STR$(c)+CHR$(13)+_
"Eta0="+STR$(eta0)+CHR$(13)+CHR$(13)+_
"Conversion Factors:"+CHR$(13)+_
"Rad2Deg="+STR$(Rad2Deg)+CHR$(13)+_
"Deg2Rad="+STR$(Deg2Rad)+CHR$(13)+_
"Ft2meters="+STR$(Feet2Meters)+CHR$(13)+_
"Meters2Ft="+STR$(Meters2Feet)+CHR$(13)+_
"Inches2Meters="+STR$(Inches2Meters)+CHR$(13)+_
"Meters2Inches="+STR$(Meters2Inches)+CHR$(13)+_
"Miles2Meters="+STR$(Miles2Meters)+CHR$(13)+_
"Meters2Miles="+STR$(Meters2Miles)+CHR$(13)+_
"NautMi2Meters="+STR$(NautMi2Meters)+CHR$(13)+_
"Meters2NautMi="+STR$(Meters2NautMi)+CHR$(13)+CHR$(13)

MSGBOX(A$)

END SUB

'------

SUB DisplayRmatrix(Np%,Nd%,Nt&,R())

LOCAL p%, i%, j&, A$

    A$ = "Position Vector Matrix R()"+CHR$(13)

    FOR p% = 1 TO Np%

        FOR i% = 1 TO Nd%

            FOR j& = 0 TO Nt&

                A$ = A$ + "R("+STR$(p%)+", "+STR$(i%)+", "+STR$(j&)+ ") ="+STR$(R(p%,i%,j&)) + CHR$(13)

            NEXT j&

        NEXT i%

    NEXT p%

    MSGBOX(A$)

END SUB
```

```
'------
SUB DisplayRmatrixThisTimeStep(Np%,Nd%,j&,R(),Gamma)
LOCAL p%, i%, A$, B$
    A$ = "Position Vector Matrix R() at step "+STR$(j&)+", Gamma ="+STR$(Gamma)+":"+CHR$(13)+CHR$(13)
    FOR p% = 1 TO Np%
        A$ = A$ + "Probe#"+REMOVE$(STR$(p%),ANY" ")+": "
        B$ = ""
        FOR i% = 1 TO Nd%
             B$ = B$ + "  " + USING$("####.##",R(p%,i%,j&))
        NEXT i%
        A$ = A$ + B$ + CHR$(13)
    NEXT p%
    MSGBOX(A$)
END SUB
'------
SUB DisplayAmatrix(Np%,Nd%,Nt&,A())
LOCAL p%, i%, j&, A$
    A$ = "Acceleration Vector Matrix A()"+CHR$(13)
    FOR p% = 1 TO Np%
        FOR i% = 1 TO Nd%
            FOR j& = 0 TO Nt&
                A$ = A$ + "A("+STR$(p%)+", "+STR$(i%)+", "+STR$(j&)+ ") ="+STR$(A(p%,i%,j&)) + CHR$(13)
            NEXT j&
        NEXT i%
    NEXT p%
    MSGBOX(A$)
END SUB
'------
SUB DisplayAmatrixThisTimeStep(Np%,Nd%,j&,A())
LOCAL p%, i%, A$
    A$ = "Acceleration matrix A() at step "+STR$(j&)+":"+CHR$(13)
    FOR p% = 1 TO Np%
        FOR i% = 1 TO Nd%
            A$ = A$ + "A("+STR$(p%)+", "+STR$(i%)+", "+STR$(j&)+ ") ="+STR$(A(p%,i%,j&)) + CHR$(13)
        NEXT i%
    NEXT p%
    MSGBOX(A$)
END SUB
'------
SUB DisplayMmatrix(Np%,Nt&,M())
LOCAL p%, j&, A$
    A$ = "Fitness Matrix M()"+CHR$(13)
    FOR p% = 1 TO Np%
       FOR j& = 0 TO Nt&
          A$ = A$ + "M("+STR$(p%)+", "+STR$(j&)+ ") ="+STR$(M(p%,j&)) + CHR$(13)
       NEXT j&
    NEXT p%
    MSGBOX(A$)
END SUB
'------
SUB DisplayMbestMatrix(Np%,Nt&,Mbest())
LOCAL p%, j&, A$
    A$ = "Np= "+STR$(Np%)+"  Nt="+STR$(Nt&)+CHR$(13)+"Fitness Matrix Mbest()"+CHR$(13)
    FOR p% = 1 TO Np%
       FOR j& = 0 TO Nt&
          A$ = A$ + "Mbest("+STR$(p%)+", "+STR$(j&)+ ") ="+STR$(Mbest(p%,j&)) + CHR$(13)
       NEXT j&
    NEXT p%
    MSGBOX(A$)
END SUB
'------
SUB DisplayMmatrixThisTimeStep(Np%,j&,M())
LOCAL p%, A$
    A$ = "Fitness matrix M() at step "+STR$(j&)+":"+CHR$(13)
```

```
    FOR p% = 1 TO Np%

          A$ = A$ + "M("+STR$(p%)+", "+STR$(j&)+ ") ="+STR$(M(p%,j&)) + CHR$(13)

    NEXT p%

    MSGBOX(A$)

END SUB

'------

SUB DisplayXiMinMax(Nd%,XiMin(),XiMax())

LOCAL i%, A$

    A$ = ""

    FOR i% = 1 TO Nd%

        A$ = A$ + "XiMin("+STR$(i%)+" ) = "+STR$(XiMin(i%))+"   XiMax("+STR$(i%)+" ) = "+STR$(XiMax(i%)) + CHR$(13)

    NEXT i%

    MSGBOX(A$)

END SUB

'------

SUB DisplayRunParameters2(FunctionName$,Nd%,Np%,Nt&,G,DeltaT,Alpha,Beta,Frep,PlaceInitialProbes$,InitialAcceleration$,RepositionFactor$)

LOCAL A$

    A$ = "Function = "+ FunctionName$+CHR$(13)+_
         "Nd = "+STR$(Nd%)+CHR$(13)+_
         "Np = "+STR$(Np%)+CHR$(13)+_
         "Nt = "+STR$(Nt&)+CHR$(13)+_
         "G  = "+STR$(G)+CHR$(13)+_
         "DeltaT = "+STR$(DeltaT)+CHR$(13)+_
         "Alpha = "+STR$(Alpha)+CHR$(13)+_
         "Beta  = "+STR$(Beta)+CHR$(13)+_
         "Frep  = "+STR$(Frep)+CHR$(13)+_
         "Init Probes: "+PlaceInitialProbes$+CHR$(13)+_
         "Init Accel:  "+InitialAcceleration$+CHR$(13)+_
         "Retrive Method: "+RepositionFactor$+CHR$(13)

    MSGBOX(A$)

END SUB

'------

SUB
Tabulate1DprobeCoordinates(Max1DprobesPlotted%,Nd%,Np%,LastStep&,G,DeltaT,Alpha,Beta,Frep,R(),M(),PlaceInitialProbes$,InitialAcceleration$,RepositionFactor
$,FunctionName$,Gamma)

LOCAL N%, ProbeNum%, FileHeader$, A$, B$, C$, D$, E$, F$, H$, StepNum&, FieldNumber% 'kludgy, yes, but it accomplishes its purpose...

        CALL
GetPlotAnnotation(FileHeader$,Nd%,Np%,LastStep&,G,DeltaT,Alpha,Beta,Frep,M(),PlaceInitialProbes$,InitialAcceleration$,RepositionFactor$,FunctionName$,Gamma
)

        REPLACE "\n" WITH ", " IN FileHeader$

        FileHeader$ = LEFT$(FileHeader$,LEN(FileHeader$)-2)

        FileHeader$ = "PROBE COORDINATES" + CHR$(13) +_
                      "-----------------" + CHR$(13) + FileHeader$

        N% = FREEFILE : OPEN "ProbeCoordinates.DAT" FOR OUTPUT AS #N%

            A$ = "   Step #     " : B$ = "   ------     " : C$ = ""

            FOR ProbeNum% = 1 TO Np% 'create out data file header

                SELECT CASE ProbeNum%
                    CASE   1 TO   9 : E$ = ""   : F$ = "            " : H$ = "            "
                    CASE  10 TO  99 : E$ = "-"  : F$ = "           "  : H$ = "           "
                    CASE 100 TO 999 : E$ = "--" : F$ = "          "   : H$ = "          "
                END SELECT

                A$ = A$ + "P" + NoSpaces$(ProbeNum%+0,0) + F$ 'note: adding zero to ProbeNum% necessary to convert to floating point...

                B$ = B$ + E$ + "--" + H$

                C$ = C$ + "######.###    "

'                C$ = C$ + "##.#######"

            NEXT ProbeNum%

            PRINT #N%, FileHeader$ + CHR$(13) : PRINT #N%, A$ : PRINT #N%, B$

            FOR StepNum& = 0 TO LastStep&

                D$ = USING$("######   ",StepNum&)

                FOR ProbeNum% = 1 TO Np% : D$ = D$ + USING$(C$,R(ProbeNum%,1,StepNum&)) : NEXT ProbeNum%

                PRINT #N%, D$

            NEXT StepNum&

        CLOSE #N%

END SUB 'Tabulate1DprobeCoordinates()

'------

SUB
Plot1DprobePositions(Max1DprobesPlotted%,Nd%,Np%,LastStep&,G,DeltaT,Alpha,Beta,Frep,R(),M(),PlaceInitialProbes$,InitialAcceleration$,RepositionFactor$,Func
tionName$,Gamma)
   'plots on-screen 1D function probe positions vs time step if Np =< 10

LOCAL ProcessID???, N%, n1%, n2%, n3%, n4%, n5%, n6%, n7%, n8%, n9%, n10%, n11%, n12%, n13%, n14%, n15%, ProbeNum%, StepNum&, A$

LOCAL PlotAnnotation$

    IF Np% > Max1DprobesPlotted% THEN EXIT SUB

    CALL CLEANUP 'delete old "Px" plot files, if any

    ProbeNum% = 0

    DO 'create output data files, probe-by-probe
        INCR ProbeNum% : n1%  = FREEFILE : OPEN "P"+REMOVE$(STR$(ProbeNum%),ANY" ") FOR OUTPUT AS #n1%  : IF ProbeNum% = Np% THEN EXIT LOOP
        INCR ProbeNum% : n2%  = FREEFILE : OPEN "P"+REMOVE$(STR$(ProbeNum%),ANY" ") FOR OUTPUT AS #n2%  : IF ProbeNum% = Np% THEN EXIT LOOP
        INCR ProbeNum% : n3%  = FREEFILE : OPEN "P"+REMOVE$(STR$(ProbeNum%),ANY" ") FOR OUTPUT AS #n3%  : IF ProbeNum% = Np% THEN EXIT LOOP
```

```
        INCR ProbeNum% : n4%  = FREEFILE : OPEN "P"+REMOVE$(STR$(ProbeNum%),ANY" ") FOR OUTPUT AS #n4%  : IF ProbeNum% = Np% THEN EXIT LOOP
        INCR ProbeNum% : n5%  = FREEFILE : OPEN "P"+REMOVE$(STR$(ProbeNum%),ANY" ") FOR OUTPUT AS #n5%  : IF ProbeNum% = Np% THEN EXIT LOOP
        INCR ProbeNum% : n6%  = FREEFILE : OPEN "P"+REMOVE$(STR$(ProbeNum%),ANY" ") FOR OUTPUT AS #n6%  : IF ProbeNum% = Np% THEN EXIT LOOP
        INCR ProbeNum% : n7%  = FREEFILE : OPEN "P"+REMOVE$(STR$(ProbeNum%),ANY" ") FOR OUTPUT AS #n7%  : IF ProbeNum% = Np% THEN EXIT LOOP
        INCR ProbeNum% : n8%  = FREEFILE : OPEN "P"+REMOVE$(STR$(ProbeNum%),ANY" ") FOR OUTPUT AS #n8%  : IF ProbeNum% = Np% THEN EXIT LOOP
        INCR ProbeNum% : n9%  = FREEFILE : OPEN "P"+REMOVE$(STR$(ProbeNum%),ANY" ") FOR OUTPUT AS #n9%  : IF ProbeNum% = Np% THEN EXIT LOOP
        INCR ProbeNum% : n10% = FREEFILE : OPEN "P"+REMOVE$(STR$(ProbeNum%),ANY" ") FOR OUTPUT AS #n10% : IF ProbeNum% = Np% THEN EXIT LOOP
        INCR ProbeNum% : n11% = FREEFILE : OPEN "P"+REMOVE$(STR$(ProbeNum%),ANY" ") FOR OUTPUT AS #n11% : IF ProbeNum% = Np% THEN EXIT LOOP
        INCR ProbeNum% : n12% = FREEFILE : OPEN "P"+REMOVE$(STR$(ProbeNum%),ANY" ") FOR OUTPUT AS #n12% : IF ProbeNum% = Np% THEN EXIT LOOP
        INCR ProbeNum% : n13% = FREEFILE : OPEN "P"+REMOVE$(STR$(ProbeNum%),ANY" ") FOR OUTPUT AS #n13% : IF ProbeNum% = Np% THEN EXIT LOOP
        INCR ProbeNum% : n14% = FREEFILE : OPEN "P"+REMOVE$(STR$(ProbeNum%),ANY" ") FOR OUTPUT AS #n14  : IF ProbeNum% = Np% THEN EXIT LOOP
        INCR ProbeNum% : n15% = FREEFILE : OPEN "P"+REMOVE$(STR$(ProbeNum%),ANY" ") FOR OUTPUT AS #n15% : IF ProbeNum% = Np% THEN EXIT LOOP
    LOOP

    ProbeNum% = 0

    DO 'output probe positions as a function of time step
        INCR ProbeNum% : FOR StepNum& = 0 TO LastStep& : PRINT #n1%,  USING$("######  ######.########",StepNum&,R(ProbeNum%,1,StepNum&)) : NEXT StepNum& :
IF ProbeNum% = Np% THEN EXIT LOOP
        INCR ProbeNum% : FOR StepNum& = 0 TO LastStep& : PRINT #n2%,  USING$("######  ######.########",StepNum&,R(ProbeNum%,1,StepNum&)) : NEXT StepNum& :
IF ProbeNum% = Np% THEN EXIT LOOP
        INCR ProbeNum% : FOR StepNum& = 0 TO LastStep& : PRINT #n3%,  USING$("######  ######.########",StepNum&,R(ProbeNum%,1,StepNum&)) : NEXT StepNum& :
IF ProbeNum% = Np% THEN EXIT LOOP
        INCR ProbeNum% : FOR StepNum& = 0 TO LastStep& : PRINT #n4%,  USING$("######  ######.########",StepNum&,R(ProbeNum%,1,StepNum&)) : NEXT StepNum& :
IF ProbeNum% = Np% THEN EXIT LOOP
        INCR ProbeNum% : FOR StepNum& = 0 TO LastStep& : PRINT #n5%,  USING$("######  ######.########",StepNum&,R(ProbeNum%,1,StepNum&)) : NEXT StepNum& :
IF ProbeNum% = Np% THEN EXIT LOOP
        INCR ProbeNum% : FOR StepNum& = 0 TO LastStep& : PRINT #n6%,  USING$("######  ######.########",StepNum&,R(ProbeNum%,1,StepNum&)) : NEXT StepNum& :
IF ProbeNum% = Np% THEN EXIT LOOP
        INCR ProbeNum% : FOR StepNum& = 0 TO LastStep& : PRINT #n7%,  USING$("######  ######.########",StepNum&,R(ProbeNum%,1,StepNum&)) : NEXT StepNum& :
IF ProbeNum% = Np% THEN EXIT LOOP
        INCR ProbeNum% : FOR StepNum& = 0 TO LastStep& : PRINT #n8%,  USING$("######  ######.########",StepNum&,R(ProbeNum%,1,StepNum&)) : NEXT StepNum& :
IF ProbeNum% = Np% THEN EXIT LOOP
        INCR ProbeNum% : FOR StepNum& = 0 TO LastStep& : PRINT #n9%,  USING$("######  ######.########",StepNum&,R(ProbeNum%,1,StepNum&)) : NEXT StepNum& :
IF ProbeNum% = Np% THEN EXIT LOOP
        INCR ProbeNum% : FOR StepNum& = 0 TO LastStep& : PRINT #n10%, USING$("######  ######.########",StepNum&,R(ProbeNum%,1,StepNum&)) : NEXT StepNum& :
IF ProbeNum% = Np% THEN EXIT LOOP
        INCR ProbeNum% : FOR StepNum& = 0 TO LastStep& : PRINT #n11,  USING$("######  ######.########",StepNum&,R(ProbeNum%,1,StepNum&)) : NEXT StepNum& :
IF ProbeNum% = Np% THEN EXIT LOOP
        INCR ProbeNum% : FOR StepNum& = 0 TO LastStep& : PRINT #n12%, USING$("######  ######.########",StepNum&,R(ProbeNum%,1,StepNum&)) : NEXT StepNum& :
IF ProbeNum% = Np% THEN EXIT LOOP
        INCR ProbeNum% : FOR StepNum& = 0 TO LastStep& : PRINT #n13%, USING$("######  ######.########",StepNum&,R(ProbeNum%,1,StepNum&)) : NEXT StepNum& :
IF ProbeNum% = Np% THEN EXIT LOOP
        INCR ProbeNum% : FOR StepNum& = 0 TO LastStep& : PRINT #n14%, USING$("######  ######.########",StepNum&,R(ProbeNum%,1,StepNum&)) : NEXT StepNum& :
IF ProbeNum% = Np% THEN EXIT LOOP
        INCR ProbeNum% : FOR StepNum& = 0 TO LastStep& : PRINT #n15%, USING$("######  ######.########",StepNum&,R(ProbeNum%,1,StepNum&)) : NEXT StepNum& :
IF ProbeNum% = Np% THEN EXIT LOOP
    LOOP

    ProbeNum% = 0

    DO 'close output data files
        INCR ProbeNum% : CLOSE #n1%  : IF ProbeNum% = Np% THEN EXIT LOOP
        INCR ProbeNum% : CLOSE #n2%  : IF ProbeNum% = Np% THEN EXIT LOOP
        INCR ProbeNum% : CLOSE #n3%  : IF ProbeNum% = Np% THEN EXIT LOOP
        INCR ProbeNum% : CLOSE #n4%  : IF ProbeNum% = Np% THEN EXIT LOOP
        INCR ProbeNum% : CLOSE #n5%  : IF ProbeNum% = Np% THEN EXIT LOOP
        INCR ProbeNum% : CLOSE #n6%  : IF ProbeNum% = Np% THEN EXIT LOOP
        INCR ProbeNum% : CLOSE #n7%  : IF ProbeNum% = Np% THEN EXIT LOOP
        INCR ProbeNum% : CLOSE #n8%  : IF ProbeNum% = Np% THEN EXIT LOOP
        INCR ProbeNum% : CLOSE #n9%  : IF ProbeNum% = Np% THEN EXIT LOOP
        INCR ProbeNum% : CLOSE #n10% : IF ProbeNum% = Np% THEN EXIT LOOP
        INCR ProbeNum% : CLOSE #n11% : IF ProbeNum% = Np% THEN EXIT LOOP
        INCR ProbeNum% : CLOSE #n12% : IF ProbeNum% = Np% THEN EXIT LOOP
        INCR ProbeNum% : CLOSE #n13% : IF ProbeNum% = Np% THEN EXIT LOOP
        INCR ProbeNum% : CLOSE #n14% : IF ProbeNum% = Np% THEN EXIT LOOP
        INCR ProbeNum% : CLOSE #n15% : IF ProbeNum% = Np% THEN EXIT LOOP
    LOOP

    ProbeNum% = 0 : A$ = ""

    DO 'create file string for plot command file
        INCR ProbeNum% : A$ = A$ + Quote$ + "P"+REMOVE$(STR$(ProbeNum%),ANY" ") + Quote$ + " w l lw 2, " : IF ProbeNum% = Np% THEN EXIT LOOP
        INCR ProbeNum% : A$ = A$ + Quote$ + "P"+REMOVE$(STR$(ProbeNum%),ANY" ") + Quote$ + " w l lw 2, " : IF ProbeNum% = Np% THEN EXIT LOOP
        INCR ProbeNum% : A$ = A$ + Quote$ + "P"+REMOVE$(STR$(ProbeNum%),ANY" ") + Quote$ + " w l lw 2, " : IF ProbeNum% = Np% THEN EXIT LOOP
        INCR ProbeNum% : A$ = A$ + Quote$ + "P"+REMOVE$(STR$(ProbeNum%),ANY" ") + Quote$ + " w l lw 2, " : IF ProbeNum% = Np% THEN EXIT LOOP
        INCR ProbeNum% : A$ = A$ + Quote$ + "P"+REMOVE$(STR$(ProbeNum%),ANY" ") + Quote$ + " w l lw 2, " : IF ProbeNum% = Np% THEN EXIT LOOP
        INCR ProbeNum% : A$ = A$ + Quote$ + "P"+REMOVE$(STR$(ProbeNum%),ANY" ") + Quote$ + " w l lw 2, " : IF ProbeNum% = Np% THEN EXIT LOOP
        INCR ProbeNum% : A$ = A$ + Quote$ + "P"+REMOVE$(STR$(ProbeNum%),ANY" ") + Quote$ + " w l lw 2, " : IF ProbeNum% = Np% THEN EXIT LOOP
        INCR ProbeNum% : A$ = A$ + Quote$ + "P"+REMOVE$(STR$(ProbeNum%),ANY" ") + Quote$ + " w l lw 2, " : IF ProbeNum% = Np% THEN EXIT LOOP
        INCR ProbeNum% : A$ = A$ + Quote$ + "P"+REMOVE$(STR$(ProbeNum%),ANY" ") + Quote$ + " w l lw 2, " : IF ProbeNum% = Np% THEN EXIT LOOP
        INCR ProbeNum% : A$ = A$ + Quote$ + "P"+REMOVE$(STR$(ProbeNum%),ANY" ") + Quote$ + " w l lw 2, " : IF ProbeNum% = Np% THEN EXIT LOOP
        INCR ProbeNum% : A$ = A$ + Quote$ + "P"+REMOVE$(STR$(ProbeNum%),ANY" ") + Quote$ + " w l lw 2, " : IF ProbeNum% = Np% THEN EXIT LOOP
        INCR ProbeNum% : A$ = A$ + Quote$ + "P"+REMOVE$(STR$(ProbeNum%),ANY" ") + Quote$ + " w l lw 2, " : IF ProbeNum% = Np% THEN EXIT LOOP
        INCR ProbeNum% : A$ = A$ + Quote$ + "P"+REMOVE$(STR$(ProbeNum%),ANY" ") + Quote$ + " w l lw 2, " : IF ProbeNum% = Np% THEN EXIT LOOP
        INCR ProbeNum% : A$ = A$ + Quote$ + "P"+REMOVE$(STR$(ProbeNum%),ANY" ") + Quote$ + " w l lw 2, " : IF ProbeNum% = Np% THEN EXIT LOOP
        INCR ProbeNum% : A$ = A$ + Quote$ + "P"+REMOVE$(STR$(ProbeNum%),ANY" ") + Quote$ + " w l lw 2, " : IF ProbeNum% = Np% THEN EXIT LOOP
    LOOP

    A$ = LEFT$(A$,LEN(A$)-2)

    CALL
GetPlotAnnotation(PlotAnnotation$,Nd%,Np%,LastStep&,G,DeltaT,Alpha,Beta,Frep,M(),PlaceInitialProbes$,InitialAcceleration$,RepositionFactor$,FunctionName$,G
amma)

    N% = FREEFILE

    OPEN "cmd2d.gp" FOR OUTPUT AS #N%

        PRINT #N%, "set label "      + Quote$ + PlotAnnotation$ + Quote$ + " at graph 0.5,0.95"
        PRINT #N%, "set grid xtics"
        PRINT #N%, "set grid ytics"
        PRINT #N%, "set title "  + Quote$ + "Evolution of "    + FunctionName$ + " Probe Positions"+ "\n" + RunID$ + Quote$
        PRINT #N%, "set xlabel " + Quote$ + "Time Step"       + Quote$
        PRINT #N%, "set ylabel " + Quote$ + "Probe Coordinate" + Quote$
        PRINT #N%, "plot "       + A$

    CLOSE #N%

    CALL CreateGNUplotINIfile(0.2##*ScreenWidth&,0.2##*ScreenHeight&,0.6##*ScreenWidth&,0.6##*ScreenHeight&) 'USAGE: CALL
CreateGNUplotINIfile(PlotWindowULC_X%,PlotWindowULC_Y%,PlotWindowWidth%,PlotWindowHeight%)

    ProcessID??? = SHELL("wgnuplot.exe"+" cmd2d.gp -") : CALL Delay(5##) 'before SUB Cleanup is called

END SUB

'------

SUB
DisplayRunParameters(FunctionName$,Nd%,Np%,Nt&,G,DeltaT,Alpha,Beta,Frep,R(),A(),M(),PlaceInitialProbes$,InitialAcceleration$,RepositionFactor$,RunCFO$,Shri
nkDS$,CheckForEarlyTermination$)

LOCAL A$, B$, YN&

    B$ = "" : IF PlaceInitialProbes$ = "UNIFORM ON-AXIS" AND Nd% > 1 THEN B$ = "  ["+REMOVE$(STR$(Np%/Nd%),ANY" ") + "/axis]"

    RunCFO$ = "NO"

 A$ = "RUN CFO WITH THE" + CHR$(13) +_
```

```
         "FOLLOWING PARAMETERS?"                              + CHR$(13) + CHR$(13) +_
         "Function "        + FunctionName$                   + " (" + REMOVE$(STR$(Nd%),ANY" ") + "-D)" + CHR$(13) +_
         "# time steps = "  + REMOVE$(STR$(Nt&),ANY" ")       + CHR$(13) + _
         "Grav Const G = "  + REMOVE$(STR$(G,2),ANY" ")       + CHR$(13) + _
         "Delta-T = "       + REMOVE$(STR$(DeltaT,3),ANY" ")  + CHR$(13) + _
         "Exp Alpha = "     + REMOVE$(STR$(Alpha,3),ANY" ")   + CHR$(13) + _
         "Exp Beta = "      + REMOVE$(STR$(Beta,3),ANY" ")    + CHR$(13) + _
         "Frep = "          + REMOVE$(STR$(Frep,4),ANY" ")    + " ("+RepositionFactor$ + ")" + CHR$(13) + _
         "Initial Probes: " + PlaceInitialProbes$             + CHR$(13) + _
         "Initial Accel: "  + InitialAcceleration$            + CHR$(13) + _
         "Check for Early Termination? " + CheckForEarlyTermination$ + CHR$(13) + _
         "Shrink Decision Space? "       + ShrinkDS$ + CHR$(13) +CHR$(13)

'   lResult& = MSGBOX(txt$ [, [style&], title$])

    A$ = "RUN CFO ON FUNCTION " + FunctionName$ + "?"

    YN& = MSGBOX(A$,%MB_YESNO,"CONFIRM RUN")

    IF YN& = %IDYES THEN RunCFO$ = "YES"

END SUB

'------

SUB StatusWindow(FunctionName$,StatusWindowHandle???)

    GRAPHIC WINDOW "Run Progress, "+FunctionName$,0.08##*ScreenWidth&,0.08##*ScreenHeight&,0.25##*ScreenWidth&,0.17##*ScreenHeight& TO
StatusWindowHandle???

    GRAPHIC ATTACH StatusWindowHandle???,0,REDRAW

    GRAPHIC FONT "Lucida Console",8,0 '"Courier New",8,0 'Fixed width fonts

    GRAPHIC SET PIXEL (35,15) : GRAPHIC PRINT "  Initializing...      " : GRAPHIC REDRAW

END SUB

'------

SUB GetTestFunctionNumber(FunctionName$)

  LOCAL hDlg AS DWORD

  LOCAL N%, M%

  LOCAL FrameWidth&, FrameHeight&, BoxWidth&, BoxHeight&

' BoxWidth& = 276 : BoxHeight& = 300 : FrameWidth& = 82 : FrameHeight& = BoxHeight&-5

  BoxWidth& = 276 : BoxHeight& = 300 : FrameWidth& = 90 : FrameHeight& = BoxHeight&-5

  DIALOG NEW 0, "CENTRAL FORCE OPTIMIZATION TEST FUNCTIONS",,, BoxWidth&, BoxHeight&, %WS_CAPTION OR %WS_SYSMENU, 0 TO hDlg

'------------------------------------------------------------------

  CONTROL ADD FRAME,  hDlg, %IDC_FRAME1,  "Test Functions",      5,   2, FrameWidth&, FrameHeight&
  CONTROL ADD FRAME,  hDlg, %IDC_FRAME2,  "GSO Test Functions",  105, 2, FrameWidth&, 255

  CONTROL ADD OPTION, hDlg, %IDC_Function_Number1,  "Parrott F4",      10,  14, 70, 10, %WS_GROUP OR %WS_TABSTOP
  CONTROL ADD OPTION, hDlg, %IDC_Function_Number2,  "SGO",             10,  24, 70, 10
  CONTROL ADD OPTION, hDlg, %IDC_Function_Number3,  "Goldstein-Price",10,  34, 70, 10
  CONTROL ADD OPTION, hDlg, %IDC_Function_Number4,  "Step",            10,  44, 70, 10
  CONTROL ADD OPTION, hDlg, %IDC_Function_Number5,  "Schwefel 2.26",   10,  54, 70, 10
  CONTROL ADD OPTION, hDlg, %IDC_Function_Number6,  "Colville",        10,  64, 70, 10
  CONTROL ADD OPTION, hDlg, %IDC_Function_Number7,  "Griewank",        10,  74, 70, 10

  CONTROL ADD OPTION, hDlg, %IDC_Function_Number31, "PBM #1",          10,  84, 70, 10
  CONTROL ADD OPTION, hDlg, %IDC_Function_Number32, "PBM #2",          10,  94, 70, 10
  CONTROL ADD OPTION, hDlg, %IDC_Function_Number33, "PBM #3",          10, 104, 70, 10
  CONTROL ADD OPTION, hDlg, %IDC_Function_Number34, "PBM #4",          10, 114, 70, 10
  CONTROL ADD OPTION, hDlg, %IDC_Function_Number35, "PBM #5",          10, 124, 70, 10
  CONTROL ADD OPTION, hDlg, %IDC_Function_Number36, "Himmelblau",      10, 134, 70, 10
  CONTROL ADD OPTION, hDlg, %IDC_Function_Number37, "Rosenbrock",      10, 144, 70, 10
  CONTROL ADD OPTION, hDlg, %IDC_Function_Number38, "Sphere",          10, 154, 70, 10
  CONTROL ADD OPTION, hDlg, %IDC_Function_Number39, "HimmelblauNLO",   10, 164, 70, 10
  CONTROL ADD OPTION, hDlg, %IDC_Function_Number40, "Tripod",          10, 174, 70, 10
  CONTROL ADD OPTION, hDlg, %IDC_Function_Number41, "Rosenbrock F6",   10, 184, 70, 10
  CONTROL ADD OPTION, hDlg, %IDC_Function_Number42, "Comp Spring",     10, 194, 70, 10
  CONTROL ADD OPTION, hDlg, %IDC_Function_Number43, "Gear Train",      10, 204, 70, 10
  CONTROL ADD OPTION, hDlg, %IDC_Function_Number44, "Loaded Bowtie",   10, 214, 70, 10
  CONTROL ADD OPTION, hDlg, %IDC_Function_Number45, "Yagi Array",      10, 224, 70, 10
  CONTROL ADD OPTION, hDlg, %IDC_Function_Number46, "Reserved",        10, 234, 70, 10
  CONTROL ADD OPTION, hDlg, %IDC_Function_Number47, "Reserved",        10, 244, 70, 10
  CONTROL ADD OPTION, hDlg, %IDC_Function_Number48, "Reserved",        10, 254, 70, 10
  CONTROL ADD OPTION, hDlg, %IDC_Function_Number49, "Reserved",        10, 264, 70, 10
  CONTROL ADD OPTION, hDlg, %IDC_Function_Number50, "Reserved",        10, 274, 70, 10

' --------------------- Test Functions from GSO Paper ---------------------
  CONTROL ADD OPTION, hDlg, %IDC_Function_Number8,  "f1" , 120,  14, 40, 10
  CONTROL ADD OPTION, hDlg, %IDC_Function_Number9,  "f2" , 120,  24, 40, 10
  CONTROL ADD OPTION, hDlg, %IDC_Function_Number10, "f3" , 120,  34, 40, 10
  CONTROL ADD OPTION, hDlg, %IDC_Function_Number11, "f4" , 120,  44, 40, 10
  CONTROL ADD OPTION, hDlg, %IDC_Function_Number12, "f5" , 120,  54, 40, 10
  CONTROL ADD OPTION, hDlg, %IDC_Function_Number13, "f6" , 120,  64, 40, 10
  CONTROL ADD OPTION, hDlg, %IDC_Function_Number14, "f7" , 120,  74, 40, 10
  CONTROL ADD OPTION, hDlg, %IDC_Function_Number15, "f8" , 120,  84, 40, 10
  CONTROL ADD OPTION, hDlg, %IDC_Function_Number16, "f9" , 120,  94, 40, 10
  CONTROL ADD OPTION, hDlg, %IDC_Function_Number17, "f10", 120, 104, 40, 10
  CONTROL ADD OPTION, hDlg, %IDC_Function_Number18, "f11", 120, 114, 40, 10
  CONTROL ADD OPTION, hDlg, %IDC_Function_Number19, "f12", 120, 124, 40, 10
  CONTROL ADD OPTION, hDlg, %IDC_Function_Number20, "f13", 120, 134, 40, 10
  CONTROL ADD OPTION, hDlg, %IDC_Function_Number21, "f14", 120, 144, 40, 10
  CONTROL ADD OPTION, hDlg, %IDC_Function_Number22, "f15", 120, 154, 40, 10
  CONTROL ADD OPTION, hDlg, %IDC_Function_Number23, "f16", 120, 164, 40, 10
  CONTROL ADD OPTION, hDlg, %IDC_Function_Number24, "f17", 120, 174, 40, 10
  CONTROL ADD OPTION, hDlg, %IDC_Function_Number25, "f18", 120, 184, 40, 10
  CONTROL ADD OPTION, hDlg, %IDC_Function_Number26, "f19", 120, 194, 40, 10
  CONTROL ADD OPTION, hDlg, %IDC_Function_Number27, "f20", 120, 204, 40, 10
  CONTROL ADD OPTION, hDlg, %IDC_Function_Number28, "f21", 120, 214, 40, 10
  CONTROL ADD OPTION, hDlg, %IDC_Function_Number29, "f22", 120, 224, 40, 10
  CONTROL ADD OPTION, hDlg, %IDC_Function_Number30, "f23", 120, 234, 40, 10

  CONTROL SET OPTION  hDlg, %IDC_Function_Number1, %IDC_Function_Number1, %IDC_Function_Number3 'default to Parrott F4

'-----------------------------------------------------------------------

  CONTROL ADD BUTTON, hDlg, %IDOK, "&OK", 200, 0.45##*BoxHeight&, 50, 14

'-----------------------------------------------------------------------

  DIALOG SHOW MODAL hDlg CALL DlgProc

  CALL Delay(1##)

  IF FunctionNumber% < 1 OR FunctionNumber% > 45 THEN

    FunctionNumber% = 1 : MSGBOX("Error in function number...")
```

```
  END IF

    SELECT CASE FunctionNumber%

        CASE 1 : FunctionName$ = "ParrottF4"
        CASE 2 : FunctionName$ = "SGO"
        CASE 3 : FunctionName$ = "GP"
        CASE 4 : FunctionName$ = "STEP"
        CASE 5 : FunctionName$ = "SCHWEFEL_226"
        CASE 6 : FunctionName$ = "COLVILLE"
        CASE 7 : FunctionName$ = "GRIEWANK"
        CASE 8 : FunctionName$ = "F1"
        CASE 9 : FunctionName$ = "F2"
        CASE 10: FunctionName$ = "F3"
        CASE 11: FunctionName$ = "F4"
        CASE 12: FunctionName$ = "F5"
        CASE 13: FunctionName$ = "F6"
        CASE 14: FunctionName$ = "F7"
        CASE 15: FunctionName$ = "F8"
        CASE 16: FunctionName$ = "F9"
        CASE 17: FunctionName$ = "F10"
        CASE 18: FunctionName$ = "F11"
        CASE 19: FunctionName$ = "F12"
        CASE 20: FunctionName$ = "F13"
        CASE 21: FunctionName$ = "F14"
        CASE 22: FunctionName$ = "F15"
        CASE 23: FunctionName$ = "F16"
        CASE 24: FunctionName$ = "F17"
        CASE 25: FunctionName$ = "F18"
        CASE 26: FunctionName$ = "F19"
        CASE 27: FunctionName$ = "F20"
        CASE 28: FunctionName$ = "F21"
        CASE 29: FunctionName$ = "F22"
        CASE 30: FunctionName$ = "F23"
        CASE 31: FunctionName$ = "PBM_1"
        CASE 32: FunctionName$ = "PBM_2"
        CASE 33: FunctionName$ = "PBM_3"
        CASE 34: FunctionName$ = "PBM_4"
        CASE 35: FunctionName$ = "PBM_5"
        CASE 36: FunctionName$ = "HIMMELBLAU"
        CASE 37: FunctionName$ = "ROSENBROCK"
        CASE 38: FunctionName$ = "SPHERE"
        CASE 39: FunctionName$ = "HIMMELBLAUNLO"
        CASE 40: FunctionName$ = "TRIPOD"
        CASE 41: FunctionName$ = "ROSENBROCKF6"
        CASE 42: FunctionName$ = "COMPRESSIONSPRING"
        CASE 43: FunctionName$ = "GEARTRAIN"
        CASE 44: FunctionName$ = "BOWTIE"
        CASE 45: FunctionName$ = "YAGI"

    END SELECT

END SUB

'-----------

CALLBACK FUNCTION DlgProc() AS LONG

  '------------------------------------------------------------------
  ' Callback procedure for the main dialog
  '-----------------------------------------------------------------
  LOCAL c, lRes AS LONG, sText AS STRING

  SELECT CASE AS LONG CBMSG

  CASE %WM_INITDIALOG' %WM_INITDIALOG is sent right before the dialog is shown.

  CASE %WM_COMMAND              ' <- a control is calling

     SELECT CASE AS LONG CBCTL  ' <- look at control's id

     CASE %IDOK                 ' <- OK button or Enter key was pressed

         IF CBCTLMSG = %BN_CLICKED THEN
             '-----------------------------------------
             ' Loop through the Function_Number controls
             ' to see which one is selected
             '-----------------------------------------
             FOR c = %IDC_Function_Number1 TO %IDC_Function_Number50

                 CONTROL GET CHECK CBHNDL, c TO lRes

                 IF lRes THEN EXIT FOR

             NEXT 'c holds the id for selected test function.

             FunctionNumber% = c-120

             DIALOG END CBHNDL

         END IF

     END SELECT

  END SELECT

END FUNCTION

'---------------------------- PBM ANTENNA BENCHMARK FUNCTIONS --------------------------

'Reference for benchmarks PBM_1 through PBM_5:

'Pantoja, M F., Bretones, A. R., Martin, R. G., "Benchmark Antenna Problems for Evolutionary
'Optimization Algorithms," IEEE Trans. Antennas & Propagation, vol. 55, no. 4, April 2007,
'pp. 1111-1121

FUNCTION PBM_1(R(),Nd%,p%,j&) 'PBM Benchmark #1: Max D for Variable-Length CF Dipole

    LOCAL Z, LengthWaves, ThetaRadians AS EXT

    LOCAL N%, Nsegs%, FeedSegNum%

    LOCAL NumSegs$, FeedSeg$, HalfLength$, Radius$, ThetaDeg$, Lyne$, GainDB$

    LengthWaves  = R(p%,1,j&)

    ThetaRadians = R(p%,2,j&)

    ThetaDeg$ = REMOVE$(STR$(ROUND(ThetaRadians*Rad2Deg,2)),ANY" ")

    IF TALLY(ThetaDeg$,".") = 0 THEN ThetaDeg$ = ThetaDeg$+"."

    Nsegs% = 2*(INT(100*LengthWaves)\2)+1 '100 segs per wavelength, must be an odd #, VOLTAGE SOURCE

    FeedSegNum% = Nsegs%\2 + 1 'center segment number, VOLTAGE SOURCE

    NumSegs$    = REMOVE$(STR$(Nsegs%),ANY" ")

    FeedSeg$    = REMOVE$(STR$(FeedSegNum%),ANY" ")
```

```
    HalfLength$ = REMOVE$(STR$(ROUND(LengthWaves/2##,6)),ANY" ")

    IF TALLY(HalfLength$,".") = 0 THEN HalfLength$ = HalfLength$+"."

    Radius$     = "0.00001" 'REMOVE$(STR$(ROUND(LengthWaves/1000##,6)),ANY" ")

    N% = FREEFILE

    OPEN "PBM1.NEC" FOR OUTPUT AS #N%

        PRINT #N%,"CM File: PBM1.NEC"
        PRINT #N%,"CM Run ID "+DATE$+" "+TIME$
        PRINT #N%,"CM Nd="+STR$(Nd%)+", p="STR$(p%)+", j="+STR$(j&)
        PRINT #N%,"CM R(p,1,j)="+STR$(R(p%,1,j&))+", R(p,2,j)="+STR$(R(p%,2,j&))
        PRINT #N%,"CE"
        PRINT #N%,"GW 1,"+NumSegs$+",0.,0.,-"+HalfLength$+",0.,0.,"+HalfLength$+","+Radius$
        PRINT #N%,"GE"
        PRINT #N%,"EX 0,1,"+FeedSeg$+",0,1.,0." 'VOLTAGE SOURCE
        PRINT #N%,"FR 0,1,0,0,299.79564,0."
        PRINT #N%,"RP 0,1,1,1001,"+ThetaDeg$+",0.,0.,0.,1000." 'gain at 1000 wavelengths range
        PRINT #N%,"XQ"
        PRINT #N%,"EN"

    CLOSE #N%

'      - - ANGLES - -          - POWER GAINS -        - - - POLARIZATION - - -    - - - E(THETA) - - -    - - - E(PHI) - - -
'  THETA     PHI        VERT.   HOR.    TOTAL      AXIAL     TILT   SENSE     MAGNITUDE    PHASE      MAGNITUDE    PHASE
' DEGREES  DEGREES       DB      DB      DB        RATIO     DEG.              VOLTS/M    DEGREES      VOLTS/M    DEGREES
'   90.00     0.00       3.91 -999.99    3.91    0.00000     0.00  LINEAR    1.29504E-04     5.37    0.00000E+00    -5.24
'123456789x123456789x123456789x123456789x123456789x123456789x123456789x123456789x123456789x123456789x123456789x123456789x
'        10        20        30        40        50        60        70        80        90       100       110       120

    SHELL "n41_2k1.exe",0

    N% = FREEFILE

    OPEN "PBM1.OUT" FOR INPUT AS #N%

        WHILE NOT EOF(N%)

            LINE INPUT #N%, Lyne$

            IF INSTR(Lyne$,"DEGREES  DEGREES") > 0 THEN EXIT LOOP

        WEND 'position at next data line

        LINE INPUT #N%, Lyne$

    CLOSE #N%

    GainDB$ = REMOVE$(MID$(Lyne$,37,8),ANY" ")

    PBM_1 = 10^(VAL(GainDB$)/10##) 'Directivity

END FUNCTION 'PBM_1()

'----

FUNCTION PBM_2(R(),Nd%,p%,j&) 'PBM Benchmark #2: Max D for Variable-Separation Array of CF Dipoles

    LOCAL Z, DipoleSeparationWaves, ThetaRadians AS EXT

    LOCAL N%, i%

    LOCAL NumSegs$, FeedSeg$, Radius$, ThetaDeg$, Lyne$, GainDB$, Xcoord$, WireNum$

    DipoleSeparationWaves = R(p%,1,j&)

    ThetaRadians          = R(p%,2,j&)

    ThetaDeg$ = REMOVE$(STR$(ROUND(ThetaRadians*Rad2Deg,2)),ANY" ")

    IF TALLY(ThetaDeg$,".") = 0 THEN ThetaDeg$ = ThetaDeg$+"."

    NumSegs$ = "49"

    FeedSeg$ = "25"

    Radius$  = "0.00001"

    N% = FREEFILE

    OPEN "PBM2.NEC" FOR OUTPUT AS #N%

        PRINT #N%,"CM File: PBM2.NEC"
        PRINT #N%,"CM Run ID "+DATE$+" "+TIME$
        PRINT #N%,"CM Nd="+STR$(Nd%)+", p="STR$(p%)+", j="+STR$(j&)
        PRINT #N%,"CM R(p,1,j)="+STR$(R(p%,1,j&))+", R(p,2,j)="+STR$(R(p%,2,j&))
        PRINT #N%,"CE"

        FOR i% = -9 TO 9 STEP 2
            WireNum$ = REMOVE$(STR$((i%+11)\2),ANY" ")
            Xcoord$  = REMOVE$(STR$(i%*DipoleSeparationWaves/2##),ANY" ")
            PRINT #N%,"GW "+WireNum$+","+NumSegs$+","+Xcoord$+",0.,-0.25,"+Xcoord$+",0.,0.25,"+Radius$
        NEXT i%

        PRINT #N%,"GE"

        FOR i% = 1 TO 10
            PRINT #N%,"EX 0,"+REMOVE$(STR$(i%),ANY" ")+","+FeedSeg$+",0,1.,0." 'VOLTAGE SOURCE
        NEXT i%
        PRINT #N%,"FR 0,1,0,0,299.79564,0."
        PRINT #N%,"RP 0,1,1,1001,"+ThetaDeg$+",90.,0.,0.,1000." 'gain at 1000 wavelengths range
        PRINT #N%,"XQ"
        PRINT #N%,"EN"

    CLOSE #N%

'      - - ANGLES - -          - POWER GAINS -        - - - POLARIZATION - - -    - - - E(THETA) - - -    - - - E(PHI) - - -
'  THETA     PHI        VERT.   HOR.    TOTAL      AXIAL     TILT   SENSE     MAGNITUDE    PHASE      MAGNITUDE    PHASE
' DEGREES  DEGREES       DB      DB      DB        RATIO     DEG.              VOLTS/M    DEGREES      VOLTS/M    DEGREES
'   90.00     0.00       3.91 -999.99    3.91    0.00000     0.00  LINEAR    1.29504E-04     5.37    0.00000E+00    -5.24
'123456789x123456789x123456789x123456789x123456789x123456789x123456789x123456789x123456789x123456789x123456789x123456789x
'        10        20        30        40        50        60        70        80        90       100       110       120

    SHELL "n41_2k1.exe",0

    N% = FREEFILE

    OPEN "PBM2.OUT" FOR INPUT AS #N%

        WHILE NOT EOF(N%)

            LINE INPUT #N%, Lyne$

            IF INSTR(Lyne$,"DEGREES  DEGREES") > 0 THEN EXIT LOOP

        WEND 'position at next data line
```

```
        LINE INPUT #N%, Lyne$

    CLOSE #N%

    GainDB$ = REMOVE$(MID$(Lyne$,37,8),ANY" ")

    IF AddNoiseToPBM2$ = "YES" THEN

        Z = 10^(VAL(GainDB$)/10##) + GaussianDeviate(0##,0.4472##) 'Directivity with Gaussian noise (zero mean, 0.2 variance)

    ELSE

        Z = 10^(VAL(GainDB$)/10##) 'Directivity without noise

    END IF

    PBM_2 = Z

END FUNCTION 'PBM_2()

'----

FUNCTION PBM_3(R(),Nd%,p%,j&) 'PBM Benchmark #3: Max D for Circular Dipole Array

    LOCAL Beta, ThetaRadians, Alpha, ReV, ImV AS EXT

    LOCAL N%, i%

    LOCAL NumSegs$, FeedSeg$, Radius$, ThetaDeg$, Lyne$, GainDB$, Xcoord$, Ycoord$, WireNum$, ReEX$, ImEX$

    Beta         = R(p%,1,j&)

    ThetaRadians = R(p%,2,j&)

    ThetaDeg$ = REMOVE$(STR$(ROUND(ThetaRadians*Rad2Deg,2)),ANY" ")

    IF TALLY(ThetaDeg$,".") = 0 THEN ThetaDeg$ = ThetaDeg$+"."

    NumSegs$ = "49"

    FeedSeg$ = "25"

    Radius$  = "0.00001"

    N% = FREEFILE

    OPEN "PBM3.NEC" FOR OUTPUT AS #N%

        PRINT #N%,"CM File: PBM3.NEC"
        PRINT #N%,"CM Run ID "+DATE$+" "+TIME$
        PRINT #N%,"CM Nd="+STR$(Nd%)+", p="STR$(p%)+", j="+STR$(j&)
        PRINT #N%,"CM R(p,1,j)="+STR$(R(p%,1,j&))+", R(p,2,j)="+STR$(R(p%,2,j&))
        PRINT #N%,"CE"

        FOR i% = 1 TO 8
            WireNum$ = REMOVE$(STR$(i%),ANY" ")

            SELECT CASE i%
                CASE 1 : Xcoord$ = "1"        : Ycoord$ = "0"
                CASE 2 : Xcoord$ = "0.70711"  : Ycoord$ = "0.70711"
                CASE 3 : Xcoord$ = "0"        : Ycoord$ = "1"
                CASE 4 : Xcoord$ = "-0.70711" : Ycoord$ = "0.70711"
                CASE 5 : Xcoord$ = "-1"       : Ycoord$ = "0"
                CASE 6 : Xcoord$ = "-0.70711" : Ycoord$ = "-0.70711"
                CASE 7 : Xcoord$ = "0"        : Ycoord$ = "-1"
                CASE 8 : Xcoord$ = "0.70711"  : Ycoord$ = "-0.70711"
            END SELECT

            PRINT #N%,"GW "+WireNum$+","+NumSegs$+","+Xcoord$+","+Ycoord$+",-0.25,"+Xcoord$+","Ycoord$+",0.25,"+Radius$
        NEXT i%

        PRINT #N%,"GE"

        FOR i% = 1 TO 8
            Alpha = -COS(TwoPi*Beta*(i%-1))

            ReV   = COS(Alpha)
            ImV   = SIN(Alpha)

            ReEX$ = REMOVE$(STR$(ROUND(ReV,6)),ANY" ")
            ImEX$ = REMOVE$(STR$(ROUND(ImV,6)),ANY" ")

            IF TALLY(ReEX$,".") = 0 THEN ReEX$ = ReEX$+"."
            IF TALLY(ImEX$,".") = 0 THEN ImEX$ = ImEX$+"."

            PRINT #N%,"EX 0,"+REMOVE$(STR$(i%),ANY" ")+","+FeedSeg$+",0,"+ReEX$+","+ImEX$ 'VOLTAGE SOURCE
        NEXT i%

        PRINT #N%,"FR 0,1,0,0,299.79564,0."
        PRINT #N%,"RP 0,1,1,1001,"+ThetaDeg$+",0.,0.,0.,1000." 'gain at 1000 wavelengths range
        PRINT #N%,"XQ"
        PRINT #N%,"EN"

    CLOSE #N%

'      - - ANGLES - -          - POWER GAINS -        - - - POLARIZATION - - -    - - - E(THETA) - - -    - - - E(PHI) - - -
'  THETA      PHI        VERT.   HOR.    TOTAL       AXIAL     TILT   SENSE     MAGNITUDE    PHASE      MAGNITUDE    PHASE
' DEGREES  DEGREES       DB      DB      DB          RATIO     DEG.              VOLTS/M    DEGREES      VOLTS/M    DEGREES
'   90.00     0.00       3.91 -999.99    3.91     0.00000      0.00  LINEAR    1.29504E-04     5.37    0.00000E+00    -5.24
'123456789x123456789x123456789x123456789x123456789x123456789x123456789x123456789x123456789x123456789x123456789x123456789x
'        10        20        30        40        50        60        70        80        90       100       110       120

    SHELL "n41_2k1.exe",0

    N% = FREEFILE

    OPEN "PBM3.OUT" FOR INPUT AS #N%

        WHILE NOT EOF(N%)

            LINE INPUT #N%, Lyne$

            IF INSTR(Lyne$,"DEGREES  DEGREES") > 0 THEN EXIT LOOP

        WEND 'position at next data line

        LINE INPUT #N%, Lyne$

    CLOSE #N%

    GainDB$ = REMOVE$(MID$(Lyne$,37,8),ANY" ")

    PBM_3 = 10^(VAL(GainDB$)/10##) 'Directivity

END FUNCTION 'PBM_3()

'----
```

```
FUNCTION PBM_4(R(),Nd%,p%,j&) 'PBM Benchmark #4: Max D for Vee Dipole

    LOCAL TotalLengthWaves, AlphaRadians, ArmLength, Xlength, Zlength, Lfeed AS EXT

    LOCAL N%, i%, Nsegs%, FeedZcoord$

    LOCAL NumSegs$, Lyne$, GainDB$, Xcoord$, Zcoord$

    TotalLengthWaves = 2##*R(p%,1,j&)

    AlphaRadians     = R(p%,2,j&)

    Lfeed            = 0.01##

    FeedZcoord$      = REMOVE$(STR$(Lfeed),ANY" ")

    ArmLength = (TotalLengthWaves-2##*Lfeed)/2##

    Xlength  = ROUND(ArmLength*COS(AlphaRadians),6)

    Xcoord$  = REMOVE$(STR$(Xlength),ANY" ") : IF TALLY(Xcoord$,".") = 0 THEN Xcoord$ = Xcoord$+"."

    Zlength  = ROUND(ArmLength*SIN(AlphaRadians),6)

    Zcoord$  = REMOVE$(STR$(Zlength+Lfeed),ANY" ") : IF TALLY(Zcoord$,".") = 0 THEN Zcoord$ = Zcoord$+"."

    Nsegs%   = 2*(INT(TotalLengthWaves*100)\2) 'even number, total # segs

    NumSegs$ = REMOVE$(STR$(Nsegs%\2),ANY" ") '# segs per arm

    N% = FREEFILE

    OPEN "PBM4.NEC" FOR OUTPUT AS #N%

        PRINT #N%,"CM File: PBM4.NEC"
        PRINT #N%,"CM Run ID "+DATE$+" "+TIME$
        PRINT #N%,"CM Nd="+STR$(Nd%)+", p="STR$(p%)+", j="+STR$(j&)
        PRINT #N%,"CM R(p,1,j)="+STR$(R(p%,1,j&))+", R(p,2,j)="+STR$(R(p%,2,j&))
        PRINT #N%,"CE"

        PRINT #N%,"GW 1,5,0.,0.,-"+FeedZcoord$+",0.,0.,"+FeedZcoord$+",0.00001" 'feed wire, 1 segment, 0.01 wvln

        PRINT #N%,"GW 2,"+NumSegs$+",0.,0.,"+FeedZcoord$+","+Xcoord$+",0.,"+Zcoord$+",0.00001" 'upper arm

        PRINT #N%,"GW 3,"+NumSegs$+",0.,0.,-"+FeedZcoord$+","+Xcoord$+",0.,-"+Zcoord$+",0.00001" 'lower arm

        PRINT #N%,"GE"

        PRINT #N%,"EX 0,1,3,0,1.,0." 'VOLTAGE SOURCE

        PRINT #N%,"FR 0,1,0,0,299.79564,0."
        PRINT #N%,"RP 0,1,1,1001,90.,0.,0.,0.,1000." 'ENDFIRE gain at 1000 wavelengths range
        PRINT #N%,"XQ"
        PRINT #N%,"EN"

    CLOSE #N%

'      - - ANGLES - -          - POWER GAINS -        - - - POLARIZATION - - -    - - - E(THETA) - - -    - - - E(PHI) - - -
'  THETA     PHI        VERT.   HOR.    TOTAL      AXIAL     TILT   SENSE     MAGNITUDE    PHASE      MAGNITUDE    PHASE
' DEGREES  DEGREES      DB      DB       DB        RATIO     DEG.              VOLTS/M    DEGREES      VOLTS/M    DEGREES
'   90.00     0.00       3.91 -999.99    3.91    0.00000     0.00  LINEAR    1.29504E-04     5.37    0.00000E+00    -5.24
'123456789x123456789x123456789x123456789x123456789x123456789x123456789x123456789x123456789x123456789x123456789x123456789x
'        10        20        30        40        50        60        70        80        90       100       110       120

    SHELL "n41_2k1.exe",0

    N% = FREEFILE

    OPEN "PBM4.OUT" FOR INPUT AS #N%

        WHILE NOT EOF(N%)

            LINE INPUT #N%, Lyne$

            IF INSTR(Lyne$,"DEGREES  DEGREES") > 0 THEN EXIT LOOP

        WEND 'position at next data line

        LINE INPUT #N%, Lyne$

    CLOSE #N%

    GainDB$ = REMOVE$(MID$(Lyne$,37,8),ANY" ")

    PBM_4 = 10^(VAL(GainDB$)/10##) 'Directivity

END FUNCTION 'PBM_4()

'----

FUNCTION PBM_5(R(),Nd%,p%,j&) 'PBM Benchmark #5: N-element collinear array (Nd=N-1)

    LOCAL TotalLengthWaves, Di(), Ystart, Y1, Y2, SumDi AS EXT

    LOCAL N%, i%, q%

    LOCAL Lyne$, GainDB$

    REDIM Di(1 TO Nd%)

    FOR i% = 1 TO Nd%

        Di(i%) = R(p%,i%,j&) 'dipole separation, wavelengths

    NEXT i%

    TotalLengthWaves = 0##

    FOR i% = 1 TO Nd%

        TotalLengthwaves = TotalLengthWaves + Di(i%)

    NEXT i%

    TotalLengthWaves = TotalLengthWaves + 0.5## 'add half-wavelength of 1 meter at 299.8 MHz

    Ystart = -TotalLengthWaves/2##

    N% = FREEFILE

    OPEN "PBM5.NEC" FOR OUTPUT AS #N%

        PRINT #N%,"CM File: PBM5.NEC"
        PRINT #N%,"CM Run ID "+DATE$+" "+TIME$
        PRINT #N%,"CM Nd="+STR$(Nd%)+", p="STR$(p%)+", j="+STR$(j&)
        PRINT #N%,"CM R(p,1,j)="+STR$(R(p%,1,j&))+", R(p,2,j)="+STR$(R(p%,2,j&))
        PRINT #N%,"CE"

        FOR i% = 1 TO Nd%+1
```

```
            SumDi = 0##
            FOR q% = 1 TO i%-1
                SumDi = SumDi + Di(q%)
            NEXT q%
            Y1 = ROUND(Ystart + SumDi,6)
            Y2 = ROUND(Y1+0.5##,6) 'add one-half wavelength for other end of dipole
            PRINT #N%,"GW "+REMOVE$(STR$(i%),ANY" ")+",49,0.,"+REMOVE$(STR$(Y1),ANY" ")+",0.,0.,"+REMOVE$(STR$(Y2),ANY" ")+",0.,0.00001"
        NEXT i%
        PRINT #N%,"GE"
        FOR i% = 1 TO Nd%+1
            PRINT #N%,"EX 0,"+REMOVE$(STR$(i%),ANY" ")+",25,0,1.,0." 'VOLTAGE SOURCES
        NEXT i%
        PRINT #N%,"FR 0,1,0,0,299.79564,0."
        PRINT #N%,"RP 0,1,1,1001,90.,0.,0.,0.,1000." 'gain at 1000 wavelengths range
        PRINT #N%,"XQ"
        PRINT #N%,"EN"
    CLOSE #N%
'      - - ANGLES - -          - POWER GAINS -        - - - POLARIZATION - - -    - - - E(THETA) - - -    - - - E(PHI) - - -
'  THETA     PHI        VERT.   HOR.    TOTAL       AXIAL     TILT   SENSE     MAGNITUDE    PHASE      MAGNITUDE    PHASE
' DEGREES  DEGREES       DB      DB      DB         RATIO     DEG.              VOLTS/M    DEGREES      VOLTS/M    DEGREES
'   90.00     0.00       3.91 -999.99    3.91     0.00000     0.00  LINEAR    1.29504E-04     5.37    0.00000E+00    -5.24
'123456789x123456789x123456789x123456789x123456789x123456789x123456789x123456789x123456789x123456789x123456789x123456789x
'         10        20        30        40        50        60        70        80        90       100       110       120
    SHELL "n41_2k1.exe",0
    N% = FREEFILE
    OPEN "PBM5.OUT" FOR INPUT AS #N%
        WHILE NOT EOF(N%)
            LINE INPUT #N%, Lyne$
            IF INSTR(Lyne$,"DEGREES  DEGREES") > 0 THEN EXIT LOOP
        WEND 'position at next data line
        LINE INPUT #N%, Lyne$
    CLOSE #N%
    GainDB$ = REMOVE$(MID$(Lyne$,37,8),ANY" ")
    PBM_5 = 10^(VAL(GainDB$)/10##) 'Directivity
END FUNCTION 'PBM_5()
'----
FUNCTION BOWTIE(R(),Nd%,p%,j&) 'FREE SPACE BOWTIE
    LOCAL N%, i%, Nsegs%, NumFreqs%, ElemNum%, NumRadPattAngles%, ExcitedSegment%
    LOCAL Lyne$, FR1$, FileStatus$, FileID$
    LOCAL A, B, C AS EXT 'coefficients for Bowtie Fitness function
    LOCAL a_meters, SegLength, x1, x2, y1, y2, z1, z2 AS EXT
    LOCAL Zo, FrequencyMHZ(), RadEfficiencyPCT(), MaxGainDBI(), MinGainDBI(), RinOhms(), XinOhms(), VSWR(), ForwardGainDBi() AS EXT
    LOCAL Fitness, MinimumRadiationEfficiency, MaximumRadiationEfficiency, MinimumMaxGain, MaximumMaxGain, MinVSWR, MaxVSWR, MinRin, MaxRin, MinXin,
MaxXin, MaxFwdGain, MinFwdGain AS EXT
    LOCAL StartFreqMHZ!, StopFreqMHZ!, FreqStepMHZ!
    LOCAL ArmLengthMeters, AngleRadians, LoadResistance1Ohms, LoadResistance2Ohms AS EXT
    LOCAL LoadedSegNum1%, LoadedSegNum2%
    REDIM FrequencyMHZ(1 TO 1), RadEfficiencyPCT(1 TO 1), MaxGainDBI(1 TO 1), MinGainDBI(1 TO 1), RinOhms(1 TO 1), XinOhms(1 TO 1), VSWR(1 TO 1),
ForwardGainDBi(1 TO 1)
'DECISION SPACE BOUNDARY ARRAY:
'DECISION SPACE BOUNDARY ARRAY:
'    Array Element #                 Design Variable
'    ---------------      ------------------------------------
'          1                Bowtie atm length (meters)
'          2                Bowtie HALF angle (degrees)
'          3                Loading segment number #1
'          4                Loading resistance #1 (ohms)
'          5                Zo
'   IMPORTANT NOTE: SEGMENTATION IS FIXED AT 9 !!!!  BE SURE Nsegs% = 9 BELOW, OR CHANGE IT HERE AND IN SUB GetFunctionRunParameters().
'   ------------------------------------------------------------------------------------------***-------------------------------------
    Nsegs% = 9
    ArmLengthMeters     = ROUND(R(p%,1,j&),3)
    AngleRadians        = ROUND(R(p%,2,j&)*Deg2Rad,3) 'remember, this is the HALF angle
    LoadedSegNum1%      = MAX(INT(R(p%,3,j&)),1)
    LoadResistance1Ohms = ROUND(R(p%,4,j&),2)
    Zo                  = ROUND(R(p%,5,j&),1)
    a_meters = 0.0005## 'meters
'    Zo = 300## 'feed system characteristic impedance, OHMS, for computing VSWR
    NumRadPattAngles% = 19
'    IF BowtieSegmentLength$ = "FIXED" THEN SegLength = BowtieSegmentLengthWvln 'wavelengths, FIXED [SEE HEADER NOTE ON SEGMENTATION]
    StartFreqMHZ! = 800! : StopFreqMHZ! = 12000! : FreqStepMHZ! = 100! : NumFreqs% = 1 + (StopFreqMHZ!-StartFreqMHZ!)/FreqStepMHZ!
    FR1$ = "FR 0,"+Int2String$(NumFreqs%)+",0,0,"+FP2String2$(StartFreqMHZ!)+","+FP2String2$(FreqStepMHZ!)
'    PROCESS SET PRIORITY %NORMAL_PRIORITY_CLASS 'NORMAL PRIORITY TO AVOID PROBLEMS WRITING/READING NEC FILES
'    PROCESS SET PRIORITY %REALTIME_PRIORITY_CLASS 'PREEMPTS ALL OTHER PROCESSES -> CAN CAUSE PROBLEMS...
    N% = FREEFILE
    OPEN "BOWTIE.NEC" FOR OUTPUT AS #N%
```

```
        FileID$  = REMOVE$(DATE$+TIME$,ANY Alphabet$+" -:/")

        PRINT #N%,"CM File: BOWTIE.NEC"
        PRINT #N%,"CM R-LOADED BOWTIE IN FREE SPACE WITH"
        PRINT #N%,"CM Zo AS AN OPTIMIZATION PARAMETER."
        PRINT #N%,"CM Antenna in Y-Z plane."
        PRINT #N%,"CM Run ID: "+RunID$
        PRINT #N%,"CM Fitness function:
        PRINT #N%,"CM [Min(Eff)+5*Min(Gmax)]/[|Zo-MaxRin|*(MaxVSWR-MinVSWR)*(MaxXin-MinXin)]
        PRINT #N%,"CM Arm Length = "+REMOVE$(STR$(ArmLengthMeters),ANY" ")+" meters"
        PRINT #N%,"CM Bowtie HALF Angle = "+REMOVE$(STR$(ROUND(R(p%,2,j&),2)),ANY" ")+" degrees"
        PRINT #N%,"CM Zo = "+REMOVE$(STR$(Zo),ANY" ")+" ohms"
        PRINT #N%,"CM Rload = "+REMOVE$(STR$(ROUND(LoadResistance1Ohms,2)),ANY" ")+ " ohms"
        PRINT #N%,"CM Loaded Seg # = "+REMOVE$(STR$(LoadedSegNum1%),ANY" ")+"/"+REMOVE$(STR$(Nsegs%),ANY" ")
        PRINT #N%,"CM File ID "+FileID$
        PRINT #N%,"CM Nd ="+STR$(Nd%)+", p ="STR$(p%)+", j ="+STR$(j&)
        PRINT #N%,"CE"

        x1 = 0## : x2 = x1 : y1 = -0.01## : y2 = -y1 : z1 = 0## : z2 = z1 'feed wire coords (wvln)
        PRINT
#N%,"GW1,3,"+FP2String$(x1)+","+FP2String$(y1)+","+FP2String$(z1)+","+FP2String$(x2)+","FP2String$(y2)+","+FP2String$(z2)+","+FP2String$(a_meters) 'feed
wire, 3 segs

        y1 = 0.01## : y2 = ROUND(y1+ArmLengthMeters*COS(AngleRadians),3) : z1 = 0## : z2 = ROUND(ArmLengthMeters*SIN(AngleRadians),3) 'upper right arm
        PRINT
#N%,"GW2,"+Int2String$(Nsegs%)+","+FP2String$(x1)+","+FP2String$(y1)+","+FP2String$(z1)+","+FP2String$(x2)+","FP2String$(y2)+","+FP2String$(z2)+","+FP2Stri
ng$(a_meters) 'feed wire, 3 segs

        y1 = 0.01## : y2 = ROUND(y1+ArmLengthMeters*COS(AngleRadians),3) : z1 = 0## : z2 = -ROUND(ArmLengthMeters*SIN(AngleRadians),3) 'lower right arm
        PRINT
#N%,"GW3,"+Int2String$(Nsegs%)+","+FP2String$(x1)+","+FP2String$(y1)+","+FP2String$(z1)+","+FP2String$(x2)+","FP2String$(y2)+","+FP2String$(z2)+","+FP2Stri
ng$(a_meters) 'feed wire, 3 segs

        y1 = ROUND(0.01##+ArmLengthMeters*COS(AngleRadians),3) : y2 = y1 : z1 = ROUND(ArmLengthMeters*SIN(AngleRadians),3) : z2 = -z1 'right arm vertical
connecting wire
        PRINT
#N%,"GW4,"+Int2String$(Nsegs%)+","+FP2String$(x1)+","+FP2String$(y1)+","+FP2String$(z1)+","+FP2String$(x2)+","FP2String$(y2)+","+FP2String$(z2)+","+FP2Stri
ng$(a_meters) 'feed wire, 3 segs

        y1 = -0.01## : y2 = ROUND(y1-ArmLengthMeters*COS(AngleRadians),3) : z1 = 0## : z2 = ROUND(ArmLengthMeters*SIN(AngleRadians),3) 'upper left arm
        PRINT
#N%,"GW5,"+Int2String$(Nsegs%)+","+FP2String$(x1)+","+FP2String$(y1)+","+FP2String$(z1)+","+FP2String$(x2)+","FP2String$(y2)+","+FP2String$(z2)+","+FP2Stri
ng$(a_meters) 'feed wire, 3 segs

        y1 = -0.01## : y2 = ROUND(y1-ArmLengthMeters*COS(AngleRadians),3) : z1 = 0## : z2 = -ROUND(ArmLengthMeters*SIN(AngleRadians),3) 'lower left arm
        PRINT
#N%,"GW6,"+Int2String$(Nsegs%)+","+FP2String$(x1)+","+FP2String$(y1)+","+FP2String$(z1)+","+FP2String$(x2)+","FP2String$(y2)+","+FP2String$(z2)+","+FP2Stri
ng$(a_meters) 'feed wire, 3 segs

        y1 = -ROUND(0.01##+ArmLengthMeters*COS(AngleRadians),3) : y2 = y1 : z1 = ROUND(ArmLengthMeters*SIN(AngleRadians),3) : z2 = -z1 'left arm vertical
connecting wire
        PRINT
#N%,"GW7,"+Int2String$(Nsegs%)+","+FP2String$(x1)+","+FP2String$(y1)+","+FP2String$(z1)+","+FP2String$(x2)+","FP2String$(y2)+","+FP2String$(z2)+","+FP2Stri
ng$(a_meters) 'feed wire, 3 segs

        PRINT #N%,"GE"

'       ----------------------------------- LOAD RESISTOR #1 --------------------------------------
        IF LoadResistance1Ohms <> 0## THEN 'load bowtie only if loading resistance is not zero
            PRINT #N%,"LD0,2,"+Int2String$(LoadedSegNum1%)+","+Int2String$(LoadedSegNum1%)+","_
                                  +FP2String2$(ROUND(LoadResistance1Ohms,2))+",0.,0."'load upper right arm
            PRINT #N%,"LD0,3,"+Int2String$(LoadedSegNum1%)+","+Int2String$(LoadedSegNum1%)+","_
                                  +FP2String2$(ROUND(LoadResistance1Ohms,2))+",0.,0." 'load lower right arm
            PRINT #N%,"LD0,5,"+Int2String$(LoadedSegNum1%)+","+Int2String$(LoadedSegNum1%)+","_
                                  +FP2String2$(ROUND(LoadResistance1Ohms,2))+",0.,0." 'load upper left arm
            PRINT #N%,"LD0,6,"+Int2String$(LoadedSegNum1%)+","+Int2String$(LoadedSegNum1%)+","_
                                  +FP2String2$(ROUND(LoadResistance1Ohms,2))+",0.,0." 'load lower left arm
        END IF

'       ----------------------------------- LOAD RESISTOR #2 --------------------------------------
'        IF LoadResistance2Ohms <> 0## THEN 'load bowtie only if loading resistance is not zero
'        PRINT #N%,"LD0,2,"+Int2String$(LoadedSegNum2%)+","+Int2String$(LoadedSegNum2%)+","_
 '                             +FP2String2$(ROUND(LoadResistance2Ohms,2))+",0.,0."'load upper right arm
  '      PRINT #N%,"LD0,3,"+Int2String$(LoadedSegNum2%)+","+Int2String$(LoadedSegNum2%)+","_
   '                          +FP2String2$(ROUND(LoadResistance2Ohms,2))+",0.,0." 'load lower right arm
    '    PRINT #N%,"LD0,5,"+Int2String$(LoadedSegNum2%)+","+Int2String$(LoadedSegNum2%)+","_
     '                        +FP2String2$(ROUND(LoadResistance2Ohms,2))+",0.,0." 'load upper left arm
      '  PRINT #N%,"LD0,6,"+Int2String$(LoadedSegNum2%)+","+Int2String$(LoadedSegNum2%)+","_
       '                     +FP2String2$(ROUND(LoadResistance2Ohms,2))+",0.,0." 'load lower left arm
        'END IF

        PRINT #N%, FR1$ ' frequency card

        PRINT #N%, "EX 0,1,2,1,1.,0." 'excite center segment in DRIVEN ELEMENT #2 with 1+j0 volts

        PRINT #N%, "RP 0,"+Int2String$(NumRadPattAngles%)+",1,1001,0.,0.,"+FP2String2$(ROUND(90##/(NumRadPattAngles%-1),2))+",0.,100000." 'vertical
radiation pattern at Phi=0 (+X-axis)

        PRINT #N%, "EN"

    CLOSE #N%

'    SHELL "NEC2D_200_02-22-2011.EXE",0

    SHELL "NEC41D_4K_053011.EXE",0

    CALL
GetNECdata("BOWTIE.OUT",NumFreqs%,NumRadPattAngles%,Zo,FrequencyMHZ(),RadEfficiencyPCT(),MaxGainDBI(),MinGainDBI(),RinOhms(),XinOhms(),VSWR(),ForwardGainDB
i(),FileStatus$,FileID$)

    Fitness = -98765## 'default value

    IF FileStatus$ = "OK" THEN

        MinimumRadiationEfficiency = RadEfficiencyPCT(1) : MinimumMaxGain = MaxGainDBI(1) : MinVSWR = VSWR(1) : MaxVSWR = VSWR(1)

        MinRin = RinOhms(1) : MaxRin = RinOhms(1) :  MinXin = XinOhms(1) : MaxXin = XinOhms(1)

        MinFwdGain = ForwardGainDBi(1) : MaxFwdGain = ForwardGainDBi(1)

        FOR i% = 1 TO NumFreqs%
            IF RadEfficiencyPCT(i%) =< MinimumRadiationEfficiency THEN MinimumRadiationEfficiency = RadEfficiencyPCT(i%)
            IF MinimumRadiationEfficiency < 0## THEN EXIT FOR 'bad run -> use default fitness & exit
            IF RadEfficiencyPCT(i%) >= MaximumRadiationEfficiency THEN MaximumRadiationEfficiency = RadEfficiencyPCT(i%)
            IF MaxGainDBI(i%)       =< MinimumMaxGain             THEN MinimumMaxGain             = MaxGainDBI(i%)
            IF MaxGainDBI(i%)       >= MaximumMaxGain             THEN MaximumMaxGain             = MaxGainDBI(i%)
            IF VSWR(i%)             =< MinVSWR                    THEN MinVSWR                    = VSWR(i%)
            IF VSWR(i%)             >= MaxVSWR                    THEN MaxVSWR                    = VSWR(i%)
            IF RinOhms(i%)          =< MinRin                     THEN MinRin                     = RinOhms(i%)
            IF RinOhms(i%)          >= MaxRin                     THEN MaxRin                     = RinOhms(i%)
            IF XinOhms(i%)          =< MinXin                     THEN MinXin                     = XinOhms(i%)
            IF XinOhms(i%)          >= MaxXin                     THEN MaxXin                     = XinOhms(i%)
            IF ForwardGainDBi(i%)   >= MaxFwdGain                 THEN MaxFwdGain                 = ForwardGainDBi(i%)
            IF ForwardGainDBi(i%)   =< MinFwdGain                 THEN MinFwdGain                 = ForwardGainDBi(i%)
        NEXT i%

'        A = BowtieFitnessCoefficients(1) : B = BowtieFitnessCoefficients(2) : C = BowtieFitnessCoefficients(3) 'use this notation for consistency

'NOTE: Forward gain is in direction on +X-axis.
```

```
'        Fitness = (A*ForwardGainDBi-B*ABS(Zo-MaxRin)-C*(MAX(ABS(MaxXin),ABS(MinXin))))/(A+B+C) 'NOTE: THIS IS ONE OF AN INFINITY OF FITNESS FUNCTIONS.
                                                                                           'CHANGING THE FUNCTION CHANGES THE DECISION SPACE LANDSCAPE
                                                                                           'RESULTING IN COMPLETELY DIFFERENT ANTENNA DESIGNS.

        IF MinimumRadiationEfficiency >= 0## THEN Fitness = (MinimumRadiationEfficiency+5##*MinimumMaxGain)/(ABS(Zo-MaxRin)*(MaxVSWR-MinVSWR)*(MaxXin-
MinXin)) 'compute only if run is OK and min eff >= 0

    END IF

    N% = FREEFILE
    OPEN "BOWTIE.NEC" FOR APPEND AS #N%
        PRINT #N%,""
        PRINT #N%,"CM R-LOADED BOWTIE IN FREE SPACE WITH"
        PRINT #N%,"CM Zo AS AN OPTIMIZATION PARAMETER."
        PRINT #N%,"Run ID: "+RunID$
        PRINT #N%,USING$ ("Fitness            =#.####^^^^^^",Fitness)
        PRINT #N%,"Fwd Gain (dBi) Min/Max    ="+STR$(MinFwdGain)+"/"+STR$(MinFwdGain)
        PRINT #N%,"VSWR Min/Max      ="+STR$(ROUND(MinVSWR,2))+"/"+STR$(ROUND(MaxVSWR,2))+"//"+REMOVE$(STR$(Zo),ANY" ")
        PRINT #N%,"Rin Min/Max       ="+STR$(ROUND(MinRin,2))+"/"+STR$(ROUND(MaxRin,2))
        PRINT #N%,"Xin Min/Max       ="+STR$(ROUND(MinXin,2))+"/"+STR$(ROUND(MaxXin,2))
        PRINT #N%,"Gmax(dBi) Min/Max ="+STR$(ROUND(MinimumMaxGain,2))+"/"+STR$(ROUND(MaximumMaxGain,2))
        PRINT #N%,"Eff(%) Min/Max    ="+STR$(ROUND(MinimumRadiationEfficiency,2))+"/"+STR$(ROUND(MaximumRadiationEfficiency,2))
        PRINT #N%,""
    CLOSE #N%

    N% = FREEFILE
    OPEN "BOWTIE.OUT" FOR APPEND AS #N%
        PRINT #N%,""
        PRINT #N%,"CM R-LOADED BOWTIE IN FREE SPACE WITH"
        PRINT #N%,"CM Zo AS AN OPTIMIZATION PARAMETER."
        PRINT #N%,"Run ID: "+RunID$
        PRINT #N%,USING$ ("Fitness            =#.####^^^^^^",Fitness)
        PRINT #N%,"Fwd Gain (dBi) Min/Max    ="+STR$(MinFwdGain)+"/"+STR$(MinFwdGain)
        PRINT #N%,"VSWR Min/Max      ="+STR$(ROUND(MinVSWR,2))+"/"+STR$(ROUND(MaxVSWR,2))+"//"+REMOVE$(STR$(Zo),ANY" ")
        PRINT #N%,"Rin Min/Max       ="+STR$(ROUND(MinRin,2))+"/"+STR$(ROUND(MaxRin,2))
        PRINT #N%,"Xin Min/Max       ="+STR$(ROUND(MinXin,2))+"/"+STR$(ROUND(MaxXin,2))
        PRINT #N%,"Gmax(dBi) Min/Max ="+STR$(ROUND(MinimumMaxGain,2))+"/"+STR$(ROUND(MaximumMaxGain,2))
        PRINT #N%,"Eff(%) Min/Max    ="+STR$(ROUND(MinimumRadiationEfficiency,2))+"/"+STR$(ROUND(MaximumRadiationEfficiency,2))
        PRINT #N%,""
    CLOSE #N%

'      - - ANGLES - -          - POWER GAINS -        - - - POLARIZATION - - -    - - - E(THETA) - - -    - - - E(PHI) - - -
'  THETA     PHI        VERT.   HOR.    TOTAL       AXIAL     TILT   SENSE     MAGNITUDE    PHASE      MAGNITUDE    PHASE
' DEGREES  DEGREES       DB      DB      DB         RATIO     DEG.              VOLTS/M    DEGREES      VOLTS/M    DEGREES
'   90.00     0.00       3.91 -999.99    3.91    0.00000     0.00  LINEAR    1.29504E-04      5.37    0.00000E+00    -5.24
'123456789x123456789x123456789x123456789x123456789x123456789x123456789x123456789x123456789x123456789x123456789x123456789x
'        10        20        30        40        50        60        70        80        90       100       110       120

    BOWTIE = Fitness

END FUNCTION 'BOWTIE()

'--------------===-------

SUB ReplaceCommentCard(NECfile$)

LOCAL N%, M%

LOCAL Lyne$

    N% = FREEFILE : OPEN NECfile$ FOR INPUT AS #N%

    M% = FREEFILE : OPEN "NECtemp" FOR OUTPUT AS #M%

    WHILE NOT EOF(N%)
        LINE INPUT #N%, Lyne$
        IF INSTR(Lyne$,"BOWTIE.NEC") > 0 THEN REPLACE "BOWTIE.NEC" WITH "BESTBOWTIE.NEC" IN Lyne$
        PRINT #M%, Lyne$
    WEND

    CLOSE #M% : CLOSE #N%

    KILL NecFile$ : NAME "NECtemp" AS NECfile$

END SUB

'------

    FUNCTION FP2String2$(X!)
    LOCAL A$
        A$=LTRIM$(RTRIM$(STR$(X!)))
        IF TALLY(A$,".") = 0! THEN A$ = A$ + "."
        FP2String2$ = A$
    END FUNCTION

'---

    FUNCTION FP2String$(X)
    LOCAL A$
        A$=LTRIM$(RTRIM$(STR$(X)))
        IF TALLY(A$,".") = 0## THEN A$ = A$ + "."
        FP2String$ = A$
    END FUNCTION

'---

    FUNCTION Int2String$(X%)
    LOCAL A$
       A$=LTRIM$(RTRIM$(STR$(X%)))
       Int2String$ = A$
    END FUNCTION

'---

    SUB
GetNECdata(NECoutputFile$,NumFreqs%,NumRadPattAngles%,Zo,FrequencyMHZ(),RadEfficiencyPCT(),MaxGainDBI(),MinGainDBI(),RinOhms(),XinOhms(),VSWR(),ForwardGain
DBi(),FileStatus$,FileID$)

    LOCAL N%, idx%, AngleNum%

    LOCAL Lyne$, Dum$

    LOCAL GmaxDBI, GminDBI, FwdGainDBi AS EXT

    REDIM FrequencyMHZ(1 TO NumFreqs%),RadEfficiencyPCT(1 TO NumFreqs%),MaxGainDBI(1 TO NumFreqs%),MinGainDBI(1 TO NumFreqs%), _
          RinOhms(1 TO NumFreqs%),XinOhms(1 TO NumFreqs%),VSWR(1 TO NumFreqs%), ForwardGainDBi(1 TO NumFreqs%)

    FileStatus$ = "NOK"

    OPEN NECoutputFile$ FOR INPUT AS #N%

        WHILE NOT EOF(N%)

            LINE INPUT #N%, Lyne$

            IF INSTR(Lyne$,"RUN TIME") > 0 THEN FileStatus$ = "OK"

        WEND
```

```
    CLOSE #N%

    IF FileStatus$ <> "OK" THEN EXIT SUB

    OPEN NECoutputFile$ FOR INPUT AS #N%

        idx% = 1

        WHILE NOT EOF(N%)

            LINE INPUT #N%, Lyne$

            IF INSTR(Lyne$,"File ID") > 0 THEN 'CHECK THAT NEC OUTPUT FILE WAS COMPUTED FROM CURRENT NEC INPUT FILE BY COMPARIN FILE ID's (TO AVOID
CACHE/BUFFER PROBLEMS CREATD BY OS)
                IF FileID$ <> REMOVE$(Lyne$,ANY Alphabet$+" ") THEN MSGBOX("WARNING! NEC I/O File ID's Don't Match!"+CHR$(13)+"Lyne$   =
"+Lyne$+CHR$(13)+"FileID$ = "+FileID$)
            END IF

            IF INSTR(Lyne$,"FREQUENCY=") > 0 THEN
                Lyne$ = REMOVE$(Lyne$,"MHZ") : Lyne$ = REMOVE$(Lyne$,"FREQUENCY= ") : FrequencyMHZ(idx%) = VAL(Lyne$)
'MSGBOX("idx="+STR$(idx%)+"  F="+STR$(FrequencyMHZ(idx%)))
            END IF

            IF INSTR(Lyne$,"INPUT PARAMETERS") > 0 THEN
                LINE INPUT #N%, Dum$ : LINE INPUT #N%, Dum$ : LINE INPUT #N%, Dum$ 'skip three lines
                LINE INPUT #N%, Lyne$ 'input next line with impedance data
                RinOhms(idx%) = VAL(MID$(Lyne$,61,12)) : XinOhms(idx%) = VAL(MID$(Lyne$,73,12)) : VSWR(idx%) =
StandingWaveRatio(Zo,RinOhms(idx%),XinOhms(idx%))
            END IF
            IF INSTR(Lyne$,"EFFICIENCY") > 0 THEN RadEfficiencyPCT(idx%) = VAL(REMOVE$(Lyne$,ANY Alphabet$+" ="))
            IF INSTR(Lyne$,"E(THETA)") > 0 THEN
                LINE INPUT #N%, Dum$ : LINE INPUT #N%, Dum$ 'skip two lines
                GmaxDBI = -9999## : GminDBI = -GmaxDBI
                FOR AngleNum% = 1 TO NumRadPattAngles%
                    LINE INPUT #N%, Lyne$ 'input next TEN lines with pattern data
                    IF VAL(MID$(Lyne$,38,7)) >= GmaxDBI THEN GmaxDBI = VAL(MID$(Lyne$,38,7)) 'get max gain
                    IF (VAL(MID$(Lyne$,38,7)) =< GminDBI AND VAL(MID$(Lyne$,38,7)) >= -999.99##) THEN GminDBI = VAL(MID$(Lyne$,38,7)) 'get min gain
                    IF AngleNum% = NumRadPattAngles% THEN FwdGainDBi = VAL(MID$(Lyne$,38,7)) 'forward gain
                NEXT AngleNum%
                MaxGainDBI(idx%) = GmaxDBI
                MinGainDBI(idx%) = GminDBI
                ForwardGainDBi(idx%) = FwdGainDBi
                INCR idx%
            END IF

'msgbox("idx="+STR$(idx%))

         WEND

    CLOSE #N%

'   TAG   SEG.    VOLTAGE (VOLTS)         CURRENT (AMPS)         IMPEDANCE (OHMS)        ADMITTANCE (MHOS)      POWER
'   NO.   NO.    REAL         IMAG.       REAL         IMAG.       REAL         IMAG.       REAL         IMAG.     (WATTS)
'     1     1 1.00000E+00 0.00000E+00 7.17910E-06 8.93193E-04 8.99811E+00-1.11951E+03 7.17910E-06 8.93193E-04 3.58955E-06
'123456789x123456789x123456789x123456789x123456789x123456789x123456789x123456789x123456789x123456789x123456789x123456789x123456789x
'         10        20        30        40        50        60        70        80        90       100       110       120       130

'  - - ANGLES - -           - POWER GAINS -       - - - POLARIZATION - - -    - - - E(THETA) - - -    - - - E(PHI) - - -
'  THETA     PHI        VERT.   HOR.    TOTAL       AXIAL     TILT   SENSE     MAGNITUDE    PHASE      MAGNITUDE    PHASE
' DEGREES  DEGREES       DB      DB      DB         RATIO     DEG.              VOLTS/M    DEGREES      VOLTS/M    DEGREES
'    0.00     0.00    -999.99 -999.99 -999.99    0.00000      0.00            0.00000E+00  -240.17    0.00000E+00  -240.17
'   10.00     0.00     -18.97 -999.99  -18.97    0.00000      0.00  LINEAR    2.69380E-08   -61.44    0.00000E+00  -240.17
'123456789x123456789x123456789x123456789x123456789x123456789x123456789x123456789x123456789x123456789x123456789x123456789x123456789x
'         10        20        30        40        50        60        70        80        90       100       110       120       130

    END SUB 'GetNECdata()

'------------------------

    FUNCTION StandingWaveRatio(Zo,ReZ,ImZ)

    LOCAL ReRho, ImRho, MagRho, SWR AS EXT

        SWR = 9999##

        CALL ComplexDivide(ReZ-Zo,ImZ,ReZ+Zo,ImZ,ReRho,ImRho)

        MagRho = SQR(ReRho*ReRho+ImRho*ImRho)  'reflection coefficient

        IF MagRho <> 1## THEN SWR=(1##+MagRho)/(1##-MagRho)

        StandingWaveRatio = SWR

    END FUNCTION 'StandingWaveRatio()

'-----

    SUB ComplexMultiply(ReA,ImA,ReB,ImB,ReC,ImC)

'   Returns real and imaginary parts of product C=A*B

        ReC = ReA*ReB-ImA*ImB
        ImC = ImA*ReB+ReA*ImB
    END SUB

'-----

    SUB ComplexDivide(ReA,ImA,ReB,ImB,ReC,ImC)

'   Returns real and imaginary parts of quotient C=A/B

        LOCAL Denom AS EXT

        Denom = ReB*ReB+ImB*ImB
        ReC = (ReA*ReB+ImA*ImB)/Denom
        ImC = (ImA*ReB-ReA*ImB)/Denom
    END SUB

'----

FUNCTION YAGI_ARRAY(R(),Nd%,p%,j&) 'FREE SPACE YAGI

    LOCAL N%, i%, Nsegs%, NumFreqs%, ElemNum%, NumRadPattAngles%, ExcitedSegment%

    LOCAL Lyne$, FR1$, FileStatus$, FileID$

    LOCAL A, B, C, D AS EXT 'coefficients for Yagi Fitness function

    LOCAL a_wvln, SegLength, x1, x2, y1, y2, BoomDistance, Spacing, Length AS EXT

    LOCAL Zo, FrequencyMHZ(), RadEfficiencyPCT(), MaxGainDBI(), MinGainDBI(), RinOhms(), XinOhms(), VSWR(), ForwardGainDBi(), RearGainDBi() AS EXT

    LOCAL Fitness, MinimumRadiationEfficiency, MaximumRadiationEfficiency, MinimumMaxGain, MaximumMaxGain, MinVSWR, MaxVSWR, MinRin, MaxRin, MinXin,
MaxXin, MinFwdGain, MaxFwdGain, MinRearGain, MaxRearGain AS EXT

    LOCAL StartFreqMHZ!, StopFreqMHZ!, FreqStepMHZ!, FcMHz!
```

```
    REDIM FrequencyMHZ(1 TO 1), RadEfficiencyPCT(1 TO 1), MaxGainDBI(1 TO 1), MinGainDBI(1 TO 1), RinOhms(1 TO 1), XinOhms(1 TO 1), VSWR(1 TO 1),
ForwardGainDBi(1 TO 1), RearGainDBi(1 TO 1)

'INFO: YAGI DECISION SPACE BOUNDARY ARRAY:

'    Array Element #                               Design Variable
'    ---------------      ----------------------------------------------------------------------
'          1
'          TO               Yagi element spacing along boom from previous element, wavelengths
'   NumYagiElements-1

'   NumYagiElements
'          TO                               Yagi element length, wavelengths
'   2*NumYagiElements-1

'   2*NumYagiElements (Nd)                          <<<< Zo >>>>

'   ============================================================================================

    a_wvln = 0.00635## 'wire radius WAVELENGTHS at 299.8 MHz (0.5" DIAM ELEMENTS @ 299.8 MHz)

    Zo = ROUND(R(p%,Nd%,j&),2) 'feed system characteristic impedance, OHMS, as a design VARIABLE for computing VSWR

    NumRadPattAngles% = 2

    IF YagiSegmentLength$ = "FIXED" THEN SegLength = YagiSegmentLengthWvln 'wavelengths @ 299.8 MHz, FIXED [SEE HEADER NOTE ON SEGMENTATION]

    FcMHz! = 299.8! : FreqStepMHZ! = 50!

    StartFreqMHZ! = FcMHz!-FreqStepMHZ! : StopFreqMHZ! = FcMHz! + FreqStepMHZ!

    NumFreqs% = 1 + (StopFreqMHZ!-StartFreqMHZ!)/FreqStepMHZ! 'Center Frequency, Fc = 299.8 MHz (wvln = 1.000 meters)

    NumFreqs% = 2*(NumFreqs%\2)+1 'make it an odd # so three-point averaging can be done

    FR1$ = "FR 0,"+Int2String$(NumFreqs%)+",0,0,"+FP2String2$(StartFreqMHZ!)+","+FP2String2$(FreqStepMHZ!)

'    PROCESS SET PRIORITY %NORMAL_PRIORITY_CLASS 'NORMAL PRIORITY TO AVOID PROBLEMS WRITING/READING NEC FILES

'    PROCESS SET PRIORITY %REALTIME_PRIORITY_CLASS 'PREEMPTS ALL OTHER PROCESSES -> CAN CAUSE PROBLEMS...

    N% = FREEFILE

    OPEN "YAGI.NEC" FOR OUTPUT AS #N%

        FileID$  = REMOVE$(DATE$+TIME$,ANY Alphabet$+" -:/")

        PRINT #N%,"CM File: YAGI.NEC"
        PRINT #N%,"CM YAGI ARRAY IN FREE SPACE"
        PRINT #N%,"CM Band center frequency, Fc ="+STR$(FcMHz!)+" MHz"
        PRINT #N%,"CM Freq step ="+STR$(FreqStepMHZ!)+" MHz +/- Fc"
        PRINT #N%,"CM Run ID: "+RunID$
        PRINT #N%,"CM Fitness function:
        PRINT #N%, Fit1$ : PRINT #N%, Fit2$
        PRINT #N%,"CM where L,M,U are lower/mid/upper frequencies"
'         PRINT #N%,"CM MaxGfwd/[|Zo-MaxRin|*(MaxVSWR-MinVSWR)*(MaxXin-MinXin)]
'         PRINT #N%,"CM (A*MaxFwdGain-B*|Zo-MaxRin|-C*(MAX(|MaxXin|,|MinXin|))"
'         PRINT #N%,"CM                -D*(MaxVSWR-MinVSWR))/(A+B+C+D)"
'         PrINT #N%,"CM where "+YagiCoefficients$
        PRINT #N%,"CM Zo="+REMOVE$(STR$(Zo),ANY" ")+" ohms"
        PRINT #N%,"CM Note: All dimensions are in METERS."
        PRINT #N%,"CM File ID "+FileID$
        PRINT #N%,"CM Nd="+STR$(Nd%)+", p="STR$(p%)+", j="+STR$(j&)
        PRINT #N%,"CE"

        FOR ElemNum% = 1 TO NumYagiElements%

            BoomDistance = 0## 'element #1 (REF) on Y-axis
            IF ElemNum% > 1 THEN
                FOR i% = 1 TO ElemNum%-1 'add up element spacings to get position along boom
                    Spacing = ROUND(R(p%,i%,j&),3) 'element spacing from previous element, WAVELENGTHS @ 299.8 MHz
                    BoomDistance = BoomDistance + Spacing
                NEXT i%
            END IF

            Length  = R(p%,ElemNum%+NumYagiElements%-1,j&) 'element length, WAVELENGTHS

            Nsegs% = 9 'use same # segs each element unless seg LENGTH is FIXED, in which case use a different # segs in each element as follows

            IF YagiSegmentLength$ = "FIXED" THEN 'adjust # segments in each element and change element length to be an integer # of segments
                NSegs% = INT(Length/SegLength)
                NSegs% = 2*(Nsegs%\2)+1    'must be odd # to preserve symmetry about excitation at center
                Length = NSegs%*SegLength 'round element length to an integer multiple of segment length to guaranty segment alignment as recommended in
NEC Manual
            END IF

            IF ElemNum% = 2 THEN ExcitedSegment% = Nsegs%\2+1

            x1 = ROUND(BoomDistance,3) : x2 =  x1 'Yagi elements are parallel to Y-axis arrayed along +X-axis.  Round spacings & length to nearest 0.001
wavelength.

            y1 = ROUND(-Length/2##,3)  : y2 = -y1

            PRINT
#N%,"GW"+Int2String$(ElemNum%)+","+Int2String$(NSegs%)+","+FP2String$(x1)+","+FP2String$(y1)+",0.,"+FP2String$(x2)+","FP2String$(y2)+",0.,"+FP2String$(a_wv
ln)

        NEXT ElemNum%

        PRINT #N%,"GE"

        PRINT #N%, FR1$ ' frequency card

        PRINT #N%, "EX 0,2,"+Int2String$(ExcitedSegment%)+",1,1.,0." 'excite center segment in DRIVEN ELEMENT #2 with 1+j0 volts

        PRINT #N%, "RP 0,"+Int2String$(NumRadPattAngles%)+",2,1001,0.,0.,"+FP2String2$(ROUND(90##/(NumRadPattAngles%-1),2))+",180.,100000."

        PRINT #N%, "EN"

    CLOSE #N%

    SHELL "NEC41D_4K_053011.EXE",0

    CALL
GetYagiNECdata("YAGI.OUT",NumFreqs%,NumRadPattAngles%,Zo,FrequencyMHZ(),RadEfficiencyPCT(),MaxGainDBI(),MinGainDBI(),RinOhms(),XinOhms(),VSWR(),ForwardGain
DBi(),RearGainDBi(),FileStatus$,FileID$)

    Fitness = -98765## 'default value

    IF FileStatus$ = "OK" THEN

        MinimumRadiationEfficiency = RadEfficiencyPCT(1)
        MaximumRadiationEfficiency = RadEfficiencyPCT(1)
        MinimumMaxGain = MaxGainDBI(1)      : MaximumMaxGain = MaxGainDBI(1)
        MinVSWR        = VSWR(1)            : MaxVSWR        = VSWR(1)
        MinRin         = RinOhms(1)         : MaxRin         = RinOhms(1)
        MinXin         = XinOhms(1)         : MaxXin         = XinOhms(1)
        MinFwdGain     = ForwardGainDBi(1) : MaxFwdGain     = ForwardGainDBi(1)
        MinRearGain    = RearGainDBi(1)    : MaxRearGain    = RearGainDBi(1)
```

```
        FOR i% = 1 TO NumFreqs%
            IF RadEfficiencyPCT(i%) =< MinimumRadiationEfficiency THEN MinimumRadiationEfficiency = RadEfficiencyPCT(i%)
            IF RadEfficiencyPCT(i%) >= MaximumRadiationEfficiency THEN MaximumRadiationEfficiency = RadEfficiencyPCT(i%)
            IF MaxGainDBI(i%)       =< MinimumMaxGain             THEN MinimumMaxGain             = MaxGainDBI(i%)
            IF MaxGainDBI(i%)       >= MaximumMaxGain             THEN MaximumMaxGain             = MaxGainDBI(i%)
            IF VSWR(i%)             =< MinVSWR                    THEN MinVSWR                    = VSWR(i%)
            IF VSWR(i%)             >= MaxVSWR                    THEN MaxVSWR                    = VSWR(i%)
            IF RinOhms(i%)          =< MinRin                     THEN MinRin                     = RinOhms(i%)
            IF RinOhms(i%)          >= MaxRin                     THEN MaxRin                     = RinOhms(i%)
            IF XinOhms(i%)          =< MinXin                     THEN MinXin                     = XinOhms(i%)
            IF XinOhms(i%)          >= MaxXin                     THEN MaxXin                     = XinOhms(i%)
            IF ForwardGainDBi(i%)   >= MaxFwdGain                 THEN MaxFwdGain                 = ForwardGainDBi(i%)
            IF ForwardGainDBi(i%)   =< MinFwdGain                 THEN MinFwdGain                 = ForwardGainDBi(i%)
            IF RearGainDBi(i%)      >= MaxRearGain                THEN MaxRearGain                = RearGainDBi(i%)
            IF RearGainDBi(i%)      =< MinRearGain                THEN MinRearGain                = RearGainDBi(i%)

'MSGBOX("freq#"+STR$(i%)+"      MaxGfwd="+STR$(MaxFwdGain)+",  MinGfwd="+STR$(MinFwdGain)+"      MaxGrear"+STR$(MaxRearGain)+"
MinGrear="+STR$(MinRearGain))

        NEXT i%

        A = YagiFitnessCoefficients(1) : B = YagiFitnessCoefficients(2) : C = YagiFitnessCoefficients(3) : D = YagiFitnessCoefficients(4) 'use this
notation for consistency

'        Fitness = (A*MaxFwdGain-B*ABS(Zo-MaxRin)-C*(MAX(ABS(MaxXin),ABS(MinXin)))-D*(MaxVSWR-MinVSWR))/(A+B+C+D)  'Yagi radiation efficiency = 100%
because it isn't loaded.
        'NOTE: THIS IS ONE OF AN INFINITY OF FITNESS FUNCTIONS. CHANGING THE FUNCTION CHANGES THE DECISION SPACE LANDSCAPE RESULTING IN COMPLETELY
DIFFERENT ANTENNA DESIGNS.

        Fitness = c1*ForwardGainDBi(1)-c2*VSWR(1)+c3*ForwardGainDBi(NumFreqs%\2+1)-c4*VSWR(NumFreqs%\2+1)+c5*ForwardGainDBi(NumFreqs%)-c6*VSWR(NumFreqs%)

    END IF

    N% = FREEFILE
    OPEN "YAGI.NEC" FOR APPEND AS #N%
        PRINT #N%,""
        PRINT #N%,"YAGI ARRAY IN FREE SPACE"
        PRINT #N%,"CM Band center frequency, Fc = 299.8 MHz"
        PRINT #N%,"Run ID: "+RunID$
        PRINT #N%,USING$ ("Fitness            =###.#######",Fitness)
        PRINT #N%,"Max Fwd Gain (dBi) ="+STR$(MaxFwdGain)
        PRINT #N%,"VSWR Min/Max       ="+STR$(ROUND(MinVSWR,2))+"/"+STR$(ROUND(MaxVSWR,2))+"//"+REMOVE$(STR$(Zo),ANY" ")
        PRINT #N%,"Rin Min/Max        ="+STR$(ROUND(MinRin,2))+"/"+STR$(ROUND(MaxRin,2))
        PRINT #N%,"Xin Min/Max        ="+STR$(ROUND(MinXin,2))+"/"+STR$(ROUND(MaxXin,2))
        PRINT #N%,"Gmax(dBi) Min/Max  ="+STR$(ROUND(MinimumMaxGain,2))+"/"+STR$(ROUND(MaximumMaxGain,2))
        PRINT #N%,"Eff(%) Min/Max     ="+STR$(ROUND(MinimumRadiationEfficiency,2))+"/"+STR$(ROUND(MaximumRadiationEfficiency,2))
        PRINT #N%,""
    CLOSE #N%

    N% = FREEFILE
    OPEN "YAGI.OUT" FOR APPEND AS #N%
        PRINT #N%,""
        PRINT #N%,"YAGI ARRAY IN FREE SPACE"
        PRINT #N%,"CM Band center frequency, Fc = 299.8 MHz"
        PRINT #N%,"Run ID: "+RunID$
        PRINT #N%,USING$ ("Fitness            =###.#######",Fitness)
        PRINT #N%,"Max Fwd Gain (dBi) ="+STR$(MaxFwdGain)
        PRINT #N%,"VSWR Min/Max       ="+STR$(ROUND(MinVSWR,2))+"/"+STR$(ROUND(MaxVSWR,2))+"//"+REMOVE$(STR$(Zo),ANY" ")
        PRINT #N%,"Rin Min/Max        ="+STR$(ROUND(MinRin,2))+"/"+STR$(ROUND(MaxRin,2))
        PRINT #N%,"Xin Min/Max        ="+STR$(ROUND(MinXin,2))+"/"+STR$(ROUND(MaxXin,2))
        PRINT #N%,"Gmax(dBi) Min/Max  ="+STR$(ROUND(MinimumMaxGain,2))+"/"+STR$(ROUND(MaximumMaxGain,2))
        PRINT #N%,"Eff(%) Min/Max     ="+STR$(ROUND(MinimumRadiationEfficiency,2))+"/"+STR$(ROUND(MaximumRadiationEfficiency,2))
        PRINT #N%,""
    CLOSE #N%

'       - - ANGLES - -           - POWER GAINS -         - - - POLARIZATION - - -     - - - E(THETA) - - -     - - - E(PHI) - - -
'  THETA      PHI        VERT.   HOR.    TOTAL       AXIAL      TILT    SENSE      MAGNITUDE     PHASE       MAGNITUDE     PHASE
' DEGREES  DEGREES       DB      DB      DB          RATIO      DEG.               VOLTS/M     DEGREES       VOLTS/M     DEGREES
'   90.00     0.00       3.91 -999.99    3.91      0.00000      0.00  LINEAR     1.29504E-04      5.37     0.00000E+00     -5.24
'123456789x123456789x123456789x123456789x123456789x123456789x123456789x123456789x123456789x123456789x123456789x123456789x
'         10        20        30        40        50        60        70        80        90        100       110       120

    YAGI_ARRAY = Fitness

END FUNCTION 'YAGI_ARRAY()

'---

    SUB
GetYagiNECdata(NECoutputFile$,NumFreqs%,NumRadPattAngles%,Zo,FrequencyMHZ(),RadEfficiencyPCT(),MaxGainDBI(),MinGainDBI(),RinOhms(),XinOhms(),VSWR(),Forward
GainDBi(),RearGainDBi(),FileStatus$,FileID$)

    LOCAL N%, idx%, AngleNum%

    LOCAL Lyne$, Dum$

    LOCAL GmaxDBI, GminDBI, FwdGainDBi, RearGain AS EXT

    REDIM FrequencyMHZ(1 TO NumFreqs%),RadEfficiencyPCT(1 TO NumFreqs%),MaxGainDBI(1 TO NumFreqs%),MinGainDBI(1 TO NumFreqs%), _
          RinOhms(1 TO NumFreqs%),XinOhms(1 TO NumFreqs%),VSWR(1 TO NumFreqs%), ForwardGainDBi(1 TO NumFreqs%), RearGainDBi(1 TO NumFreqs%)

    FileStatus$ = "NOK"

    OPEN NECoutputFile$ FOR INPUT AS #N%

        WHILE NOT EOF(N%)

            LINE INPUT #N%, Lyne$

            IF INSTR(Lyne$,"RUN TIME") > 0 THEN FileStatus$ = "OK"

        WEND

    CLOSE #N%

    IF FileStatus$ <> "OK" THEN EXIT SUB

    OPEN NECoutputFile$ FOR INPUT AS #N%

        idx% = 1

        WHILE NOT EOF(N%)

            LINE INPUT #N%, Lyne$

            IF INSTR(Lyne$,"File ID") > 0 THEN 'CHECK THAT NEC OUTPUT FILE WAS COMPUTED FROM CURRENT NEC INPUT FILE BY COMPARIN FILE ID's (TO AVOID
CACHE/BUFFER PROBLEMS CREATD BY OS)
                IF FileID$ <> REMOVE$(Lyne$,ANY Alphabet$+" ") THEN MSGBOX("WARNING! NEC I/O File ID's Don't Match!"+CHR$(13)+"Lyne$   =
"+Lyne$+CHR$(13)+"FileID$ = "+FileID$)
            END IF

            IF INSTR(Lyne$,"FREQUENCY=") > 0 THEN
                Lyne$ = REMOVE$(Lyne$,"MHZ") : Lyne$ = REMOVE$(Lyne$,"FREQUENCY= ") : FrequencyMHZ(idx%) = VAL(Lyne$)
'MSGBOX("idx="+STR$(idx%)+"  F="+STR$(FrequencyMHZ(idx%)))
            END IF

            IF INSTR(Lyne$,"INPUT PARAMETERS") > 0 THEN
                LINE INPUT #N%, Dum$ : LINE INPUT #N%, Dum$ : LINE INPUT #N%, Dum$ 'skip three lines
```

```
                LINE INPUT #N%, Lyne$ 'input next line with impedance data
                RinOhms(idx%) = VAL(MID$(Lyne$,61,12)) : XinOhms(idx%) = VAL(MID$(Lyne$,73,12)) : VSWR(idx%) =
StandingWaveRatio(Zo,RinOhms(idx%),XinOhms(idx%))
            END IF
            IF INSTR(Lyne$,"EFFICIENCY") > 0 THEN RadEfficiencyPCT(idx%) = VAL(REMOVE$(Lyne$,ANY Alphabet$+" ="))
            IF INSTR(Lyne$,"E(THETA)") > 0 THEN
                LINE INPUT #N%, Dum$ : LINE INPUT #N%, Dum$ 'skip two lines
                GmaxDBI = -9999## : GminDBI = -GmaxDBI

'               ---------- FORWARD PATTERN -----------
                FOR AngleNum% = 1 TO NumRadPattAngles%
                    LINE INPUT #N%, Lyne$ 'input next NumRadPattAngles% lines with pattern data
                    IF VAL(MID$(Lyne$,38,7)) >= GmaxDBI THEN GmaxDBI = VAL(MID$(Lyne$,38,7)) 'get max gain
                    IF (VAL(MID$(Lyne$,38,7)) =< GminDBI AND VAL(MID$(Lyne$,38,7)) >= -999.99##) THEN GminDBI = VAL(MID$(Lyne$,38,7)) 'get min gain
                    IF AngleNum% = NumRadPattAngles% THEN FwdGainDBi = VAL(MID$(Lyne$,38,7)) 'forward gain
                NEXT AngleNum%
                MaxGainDBI(idx%)     = GmaxDBI
                MinGainDBI(idx%)     = GminDBI
                ForwardGainDBi(idx%) = FwdGainDBi

'               ------------- REAR PATTERN -----------
                FOR AngleNum% = 1 TO NumRadPattAngles%
                    LINE INPUT #N%, Lyne$ 'input next NumRadPattAngles% lines with pattern data
                    IF AngleNum%  = NumRadPattAngles% THEN RearGain = VAL(MID$(Lyne$,38,7)) 'rear gain
                NEXT AngleNum%
                RearGainDBi(idx%) = RearGain

                INCR idx%
            END IF

'msgbox("idx="+STR$(idx%))

         WEND

    CLOSE #N%

'   TAG   SEG.    VOLTAGE (VOLTS)         CURRENT (AMPS)         IMPEDANCE (OHMS)        ADMITTANCE (MHOS)      POWER
'   NO.   NO.    REAL        IMAG.       REAL        IMAG.       REAL        IMAG.       REAL        IMAG.     (WATTS)
'     1     1 1.00000E+00 0.00000E+00 7.17910E-06 8.93193E-04 8.99811E+00-1.11951E+03 7.17910E-06 8.93193E-04 3.58955E-06
'123456789x123456789x123456789x123456789x123456789x123456789x123456789x123456789x123456789x123456789x123456789x123456789x123456789x
'        10        20        30        40        50        60        70        80        90       100       110       120       130

'  - - ANGLES - -           - POWER GAINS -       - - - POLARIZATION - - -    - - - E(THETA) - - -    - - - E(PHI) - - -
'  THETA     PHI        VERT.   HOR.    TOTAL       AXIAL     TILT   SENSE     MAGNITUDE    PHASE      MAGNITUDE    PHASE
' DEGREES  DEGREES       DB      DB      DB        RATIO     DEG.              VOLTS/M    DEGREES      VOLTS/M    DEGREES
'    0.00     0.00    -999.99 -999.99 -999.99    0.00000     0.00            0.00000E+00  -240.17    0.00000E+00  -240.17
'   10.00     0.00     -18.97 -999.99  -18.97    0.00000     0.00  LINEAR    2.69380E-08   -61.44    0.00000E+00  -240.17
'123456789x123456789x123456789x123456789x123456789x123456789x123456789x123456789x123456789x123456789x123456789x123456789x123456789x
'        10        20        30        40        50        60        70        80        90       100       110       120       130

    END SUB 'GetYagiNECdata()

'----------------------------
'*****************************************************  END PROGRAM 'CFO_YAGI_07-27-2011.BAS'  *****************************************************
```

# Appendix II. *Variable* $Z_0$ Yagi Performance

```
CM File: YAGI.NEC
CM YAGI ARRAY IN FREE SPACE
CM Band center frequency, Fc = 299.8 MHz
CM Freq step = 50 MHz +/- Fc
CM Run ID: 07282011_001156
CM Fitness function:
CM .2*Gfwd(L)-4*VSWR(L)+1*Gfwd(M)-8*VSWR(M)+
CM 1*Gfwd(U)-.8*VSWR(U)
CM where L,M,U are lower/mid/upper frequencies
CM Zo=105.64 ohms
CM Note: All dimensions are in METERS.
CM File ID 07282011122236
CM Nd= 12, p= 48, j= 210

FREE SPACE YAGI: SUMMARY NEC DATA
=================================
```

| F(MHz) | Rad Eff (%) | Fwd Gain (dBi) | Rear Gain (dBi) | FB Ratio (dB) | Max Gain (dBi) | Min Gain (dBi) | Rin (ohms) | Xin (ohms) | VSWR//105.64 | Avg Pwr Gain** |
|---|---|---|---|---|---|---|---|---|---|---|
| 200.00 | 100.00 | 1.84 | 4.14 | -2.30 | 1.84 | 0.02 | 22.37 | -160.77 | 15.81 | 0.996 |
| 200.10 | 100.00 | 1.85 | 4.15 | -2.30 | 1.85 | 0.01 | 22.38 | -160.50 | 15.76 | 0.996 |
| 200.20 | 100.00 | 1.85 | 4.15 | -2.30 | 1.85 | 0.00 | 22.40 | -160.24 | 15.72 | 0.996 |
| 200.30 | 100.00 | 1.85 | 4.16 | -2.31 | 1.85 | -0.01 | 22.41 | -159.98 | 15.67 | 0.996 |
| 200.40 | 100.00 | 1.86 | 4.17 | -2.31 | 1.86 | -0.02 | 22.42 | -159.71 | 15.63 | 0.996 |
| 200.50 | 100.00 | 1.86 | 4.17 | -2.31 | 1.86 | -0.02 | 22.44 | -159.45 | 15.58 | 0.996 |
| 200.60 | 100.00 | 1.87 | 4.18 | -2.31 | 1.87 | -0.03 | 22.45 | -159.18 | 15.54 | 0.996 |
| 200.70 | 100.00 | 1.87 | 4.18 | -2.31 | 1.87 | -0.04 | 22.46 | -158.92 | 15.49 | 0.996 |
| 200.80 | 100.00 | 1.87 | 4.19 | -2.32 | 1.87 | -0.05 | 22.48 | -158.65 | 15.45 | 0.996 |
| 200.90 | 100.00 | 1.88 | 4.19 | -2.31 | 1.88 | -0.06 | 22.49 | -158.39 | 15.40 | 0.996 |
| 201.00 | 100.00 | 1.88 | 4.20 | -2.32 | 1.88 | -0.07 | 22.50 | -158.12 | 15.36 | 0.996 |
| 201.10 | 100.00 | 1.89 | 4.20 | -2.31 | 1.89 | -0.07 | 22.52 | -157.86 | 15.32 | 0.996 |
| 201.20 | 100.00 | 1.89 | 4.21 | -2.32 | 1.89 | -0.08 | 22.53 | -157.60 | 15.27 | 0.996 |
| 201.30 | 100.00 | 1.90 | 4.22 | -2.32 | 1.90 | -0.09 | 22.54 | -157.33 | 15.23 | 0.996 |
| 201.40 | 100.00 | 1.90 | 4.22 | -2.32 | 1.90 | -0.10 | 22.55 | -157.07 | 15.19 | 0.996 |
| 201.50 | 100.00 | 1.91 | 4.23 | -2.32 | 1.91 | -0.11 | 22.57 | -156.80 | 15.14 | 0.996 |
| 201.60 | 100.00 | 1.91 | 4.23 | -2.32 | 1.91 | -0.12 | 22.58 | -156.54 | 15.10 | 0.996 |
| 201.70 | 100.00 | 1.92 | 4.24 | -2.32 | 1.92 | -0.12 | 22.59 | -156.27 | 15.06 | 0.996 |
| 201.80 | 100.00 | 1.92 | 4.24 | -2.32 | 1.92 | -0.13 | 22.60 | -156.01 | 15.01 | 0.996 |
| 201.90 | 100.00 | 1.93 | 4.25 | -2.32 | 1.93 | -0.14 | 22.62 | -155.74 | 14.97 | 0.996 |
| 202.00 | 100.00 | 1.93 | 4.25 | -2.32 | 1.93 | -0.15 | 22.63 | -155.48 | 14.93 | 0.996 |
| 202.10 | 100.00 | 1.94 | 4.26 | -2.32 | 1.94 | -0.16 | 22.64 | -155.22 | 14.88 | 0.996 |
| 202.20 | 100.00 | 1.94 | 4.27 | -2.33 | 1.94 | -0.17 | 22.65 | -154.95 | 14.84 | 0.996 |
| 202.30 | 100.00 | 1.95 | 4.27 | -2.32 | 1.95 | -0.18 | 22.67 | -154.69 | 14.80 | 0.996 |
| 202.40 | 100.00 | 1.95 | 4.28 | -2.33 | 1.95 | -0.19 | 22.68 | -154.42 | 14.76 | 0.996 |
| 202.50 | 100.00 | 1.96 | 4.28 | -2.32 | 1.96 | -0.20 | 22.69 | -154.16 | 14.72 | 0.996 |
| 202.60 | 100.00 | 1.96 | 4.29 | -2.33 | 1.96 | -0.20 | 22.70 | -153.89 | 14.67 | 0.996 |
| 202.70 | 100.00 | 1.97 | 4.29 | -2.32 | 1.97 | -0.21 | 22.72 | -153.63 | 14.63 | 0.996 |
| 202.80 | 100.00 | 1.97 | 4.30 | -2.33 | 1.97 | -0.22 | 22.73 | -153.36 | 14.59 | 0.996 |
| 202.90 | 100.00 | 1.98 | 4.30 | -2.32 | 1.98 | -0.23 | 22.74 | -153.10 | 14.55 | 0.996 |
| 203.00 | 100.00 | 1.99 | 4.31 | -2.32 | 1.99 | -0.24 | 22.75 | -152.84 | 14.51 | 0.996 |
| 203.10 | 100.00 | 1.99 | 4.32 | -2.33 | 1.99 | -0.25 | 22.76 | -152.57 | 14.47 | 0.996 |
| 203.20 | 100.00 | 2.00 | 4.32 | -2.32 | 2.00 | -0.26 | 22.78 | -152.31 | 14.43 | 0.996 |
| 203.30 | 100.00 | 2.00 | 4.33 | -2.33 | 2.00 | -0.27 | 22.79 | -152.04 | 14.38 | 0.996 |
| 203.40 | 100.00 | 2.01 | 4.33 | -2.32 | 2.01 | -0.28 | 22.80 | -151.78 | 14.34 | 0.996 |
| 203.50 | 100.00 | 2.02 | 4.34 | -2.32 | 2.02 | -0.29 | 22.81 | -151.51 | 14.30 | 0.996 |
| 203.60 | 100.00 | 2.02 | 4.34 | -2.32 | 2.02 | -0.30 | 22.82 | -151.25 | 14.26 | 0.996 |
| 203.70 | 100.00 | 2.03 | 4.35 | -2.32 | 2.03 | -0.31 | 22.84 | -150.98 | 14.22 | 0.996 |
| 203.80 | 100.00 | 2.03 | 4.36 | -2.33 | 2.03 | -0.32 | 22.85 | -150.72 | 14.18 | 0.996 |
| 203.90 | 100.00 | 2.04 | 4.36 | -2.32 | 2.04 | -0.33 | 22.86 | -150.46 | 14.14 | 0.996 |
| 204.00 | 100.00 | 2.05 | 4.37 | -2.32 | 2.05 | -0.34 | 22.87 | -150.19 | 14.10 | 0.996 |
| 204.10 | 100.00 | 2.05 | 4.37 | -2.32 | 2.05 | -0.35 | 22.88 | -149.93 | 14.06 | 0.996 |
| 204.20 | 100.00 | 2.06 | 4.38 | -2.32 | 2.06 | -0.36 | 22.89 | -149.66 | 14.02 | 0.996 |
| 204.30 | 100.00 | 2.07 | 4.38 | -2.31 | 2.07 | -0.37 | 22.91 | -149.40 | 13.98 | 0.996 |
| 204.40 | 100.00 | 2.07 | 4.39 | -2.32 | 2.07 | -0.38 | 22.92 | -149.13 | 13.94 | 0.996 |
| 204.50 | 100.00 | 2.08 | 4.39 | -2.31 | 2.08 | -0.39 | 22.93 | -148.87 | 13.90 | 0.996 |
| 204.60 | 100.00 | 2.09 | 4.40 | -2.31 | 2.09 | -0.40 | 22.94 | -148.60 | 13.86 | 0.996 |
| 204.70 | 100.00 | 2.10 | 4.41 | -2.31 | 2.10 | -0.41 | 22.95 | -148.34 | 13.82 | 0.996 |
| 204.80 | 100.00 | 2.10 | 4.41 | -2.31 | 2.10 | -0.42 | 22.96 | -148.07 | 13.78 | 0.996 |
| 204.90 | 100.00 | 2.11 | 4.42 | -2.31 | 2.11 | -0.43 | 22.98 | -147.81 | 13.74 | 0.996 |
| 205.00 | 100.00 | 2.12 | 4.42 | -2.30 | 2.12 | -0.44 | 22.99 | -147.54 | 13.70 | 0.996 |
| 205.10 | 100.00 | 2.13 | 4.43 | -2.30 | 2.13 | -0.45 | 23.00 | -147.28 | 13.67 | 0.996 |
| 205.20 | 100.00 | 2.13 | 4.43 | -2.30 | 2.13 | -0.46 | 23.01 | -147.01 | 13.63 | 0.996 |
| 205.30 | 100.00 | 2.14 | 4.44 | -2.30 | 2.14 | -0.47 | 23.02 | -146.74 | 13.59 | 0.996 |
| 205.40 | 100.00 | 2.15 | 4.45 | -2.30 | 2.15 | -0.48 | 23.03 | -146.48 | 13.55 | 0.996 |
| 205.50 | 100.00 | 2.16 | 4.45 | -2.29 | 2.16 | -0.49 | 23.04 | -146.22 | 13.51 | 0.996 |
| 205.60 | 100.00 | 2.16 | 4.46 | -2.30 | 2.16 | -0.50 | 23.05 | -145.95 | 13.47 | 0.996 |
| 205.70 | 100.00 | 2.17 | 4.46 | -2.29 | 2.17 | -0.51 | 23.07 | -145.68 | 13.43 | 0.996 |
| 205.80 | 100.00 | 2.18 | 4.47 | -2.29 | 2.18 | -0.52 | 23.08 | -145.42 | 13.40 | 0.996 |
| 205.90 | 100.00 | 2.19 | 4.47 | -2.28 | 2.19 | -0.53 | 23.09 | -145.15 | 13.36 | 0.996 |
| 206.00 | 100.00 | 2.20 | 4.48 | -2.28 | 2.20 | -0.54 | 23.10 | -144.89 | 13.32 | 0.996 |
| 206.10 | 100.00 | 2.20 | 4.48 | -2.28 | 2.20 | -0.55 | 23.11 | -144.62 | 13.28 | 0.996 |
| 206.20 | 100.00 | 2.21 | 4.49 | -2.28 | 2.21 | -0.56 | 23.12 | -144.36 | 13.24 | 0.996 |
| 206.30 | 100.00 | 2.22 | 4.50 | -2.28 | 2.22 | -0.57 | 23.13 | -144.09 | 13.21 | 0.996 |
| 206.40 | 100.00 | 2.23 | 4.50 | -2.27 | 2.23 | -0.58 | 23.14 | -143.82 | 13.17 | 0.996 |
| 206.50 | 100.00 | 2.24 | 4.51 | -2.27 | 2.24 | -0.60 | 23.15 | -143.56 | 13.13 | 0.996 |
| 206.60 | 100.00 | 2.25 | 4.51 | -2.26 | 2.25 | -0.61 | 23.16 | -143.29 | 13.09 | 0.996 |
| 206.70 | 100.00 | 2.26 | 4.52 | -2.26 | 2.26 | -0.62 | 23.18 | -143.02 | 13.06 | 0.996 |
| 206.80 | 100.00 | 2.27 | 4.52 | -2.25 | 2.27 | -0.63 | 23.19 | -142.76 | 13.02 | 0.996 |
| 206.90 | 100.00 | 2.28 | 4.53 | -2.25 | 2.28 | -0.64 | 23.20 | -142.49 | 12.98 | 0.996 |
| 207.00 | 100.00 | 2.28 | 4.54 | -2.26 | 2.28 | -0.66 | 23.21 | -142.23 | 12.94 | 0.996 |
| 207.10 | 100.00 | 2.29 | 4.54 | -2.25 | 2.29 | -0.67 | 23.22 | -141.96 | 12.91 | 0.996 |
| 207.20 | 100.00 | 2.30 | 4.55 | -2.25 | 2.30 | -0.68 | 23.23 | -141.69 | 12.87 | 0.996 |
| 207.30 | 100.00 | 2.31 | 4.55 | -2.24 | 2.31 | -0.69 | 23.24 | -141.43 | 12.83 | 0.996 |
| 207.40 | 100.00 | 2.32 | 4.56 | -2.24 | 2.32 | -0.71 | 23.25 | -141.16 | 12.80 | 0.996 |
| 207.50 | 100.00 | 2.33 | 4.56 | -2.23 | 2.33 | -0.72 | 23.26 | -140.89 | 12.76 | 0.996 |
| 207.60 | 100.00 | 2.34 | 4.57 | -2.23 | 2.34 | -0.73 | 23.27 | -140.62 | 12.72 | 0.996 |
| 207.70 | 100.00 | 2.35 | 4.57 | -2.22 | 2.35 | -0.74 | 23.28 | -140.36 | 12.69 | 0.996 |
| 207.80 | 100.00 | 2.36 | 4.58 | -2.22 | 2.36 | -0.76 | 23.29 | -140.09 | 12.65 | 0.996 |
| 207.90 | 100.00 | 2.37 | 4.58 | -2.21 | 2.37 | -0.77 | 23.31 | -139.82 | 12.62 | 0.996 |
| 208.00 | 100.00 | 2.38 | 4.59 | -2.21 | 2.38 | -0.78 | 23.32 | -139.56 | 12.58 | 0.996 |
| 208.10 | 100.00 | 2.39 | 4.60 | -2.21 | 2.39 | -0.80 | 23.33 | -139.29 | 12.54 | 0.996 |
| 208.20 | 100.00 | 2.40 | 4.60 | -2.20 | 2.40 | -0.81 | 23.34 | -139.02 | 12.51 | 0.996 |
| 208.30 | 100.00 | 2.41 | 4.61 | -2.20 | 2.41 | -0.82 | 23.35 | -138.75 | 12.47 | 0.996 |
| 208.40 | 100.00 | 2.42 | 4.61 | -2.19 | 2.42 | -0.84 | 23.36 | -138.48 | 12.44 | 0.996 |
| 208.50 | 100.00 | 2.44 | 4.62 | -2.18 | 2.44 | -0.85 | 23.37 | -138.22 | 12.40 | 0.996 |
| 208.60 | 100.00 | 2.45 | 4.62 | -2.17 | 2.45 | -0.86 | 23.38 | -137.95 | 12.36 | 0.996 |
| 208.70 | 100.00 | 2.46 | 4.63 | -2.17 | 2.46 | -0.88 | 23.39 | -137.68 | 12.33 | 0.996 |
| 208.80 | 100.00 | 2.47 | 4.63 | -2.16 | 2.47 | -0.89 | 23.40 | -137.41 | 12.29 | 0.996 |
| 208.90 | 100.00 | 2.48 | 4.64 | -2.16 | 2.48 | -0.90 | 23.41 | -137.14 | 12.26 | 0.996 |
| 209.00 | 100.00 | 2.49 | 4.64 | -2.15 | 2.49 | -0.92 | 23.42 | -136.88 | 12.22 | 0.996 |
| 209.10 | 100.00 | 2.50 | 4.65 | -2.15 | 2.50 | -0.93 | 23.43 | -136.61 | 12.19 | 0.996 |
| 209.20 | 100.00 | 2.51 | 4.66 | -2.15 | 2.51 | -0.95 | 23.44 | -136.34 | 12.15 | 0.996 |
| 209.30 | 100.00 | 2.53 | 4.66 | -2.13 | 2.53 | -0.96 | 23.45 | -136.07 | 12.12 | 0.996 |
| 209.40 | 100.00 | 2.54 | 4.67 | -2.13 | 2.54 | -0.98 | 23.47 | -135.80 | 12.08 | 0.996 |
| 209.50 | 100.00 | 2.55 | 4.67 | -2.12 | 2.55 | -0.99 | 23.48 | -135.53 | 12.05 | 0.996 |
| 209.60 | 100.00 | 2.56 | 4.68 | -2.12 | 2.56 | -1.00 | 23.49 | -135.26 | 12.01 | 0.996 |
| 209.70 | 100.00 | 2.57 | 4.68 | -2.11 | 2.57 | -1.02 | 23.50 | -134.99 | 11.98 | 0.996 |
| 209.80 | 100.00 | 2.59 | 4.69 | -2.10 | 2.59 | -1.03 | 23.51 | -134.72 | 11.94 | 0.996 |
| 209.90 | 100.00 | 2.60 | 4.69 | -2.09 | 2.60 | -1.05 | 23.52 | -134.45 | 11.91 | 0.996 |
| 210.00 | 100.00 | 2.61 | 4.70 | -2.09 | 2.61 | -1.06 | 23.53 | -134.18 | 11.87 | 0.996 |
| 210.10 | 100.00 | 2.62 | 4.70 | -2.08 | 2.62 | -1.08 | 23.54 | -133.91 | 11.84 | 0.996 |
| 210.20 | 100.00 | 2.64 | 4.71 | -2.07 | 2.64 | -1.09 | 23.55 | -133.64 | 11.80 | 0.996 |
| 210.30 | 100.00 | 2.65 | 4.71 | -2.06 | 2.65 | -1.11 | 23.56 | -133.37 | 11.77 | 0.996 |
| 210.40 | 100.00 | 2.66 | 4.72 | -2.06 | 2.66 | -1.12 | 23.57 | -133.10 | 11.73 | 0.996 |
| 210.50 | 100.00 | 2.67 | 4.72 | -2.05 | 2.67 | -1.14 | 23.58 | -132.83 | 11.70 | 0.996 |
| 210.60 | 100.00 | 2.69 | 4.73 | -2.04 | 2.69 | -1.15 | 23.59 | -132.56 | 11.66 | 0.996 |
| 210.70 | 100.00 | 2.70 | 4.73 | -2.03 | 2.70 | -1.17 | 23.61 | -132.29 | 11.63 | 0.996 |
| 210.80 | 100.00 | 2.71 | 4.74 | -2.03 | 2.71 | -1.18 | 23.62 | -132.02 | 11.60 | 0.996 |
| 210.90 | 100.00 | 2.73 | 4.74 | -2.01 | 2.73 | -1.20 | 23.63 | -131.75 | 11.56 | 0.996 |
| 211.00 | 100.00 | 2.74 | 4.75 | -2.01 | 2.74 | -1.21 | 23.64 | -131.47 | 11.53 | 0.996 |
| 211.10 | 100.00 | 2.75 | 4.75 | -2.00 | 2.75 | -1.23 | 23.65 | -131.20 | 11.49 | 0.996 |
| 211.20 | 100.00 | 2.77 | 4.76 | -1.99 | 2.77 | -1.25 | 23.66 | -130.93 | 11.46 | 0.996 |
| 211.30 | 100.00 | 2.78 | 4.76 | -1.98 | 2.78 | -1.26 | 23.67 | -130.66 | 11.43 | 0.996 |
| 211.40 | 100.00 | 2.80 | 4.77 | -1.97 | 2.80 | -1.28 | 23.68 | -130.39 | 11.39 | 0.996 |
| 211.50 | 100.00 | 2.81 | 4.77 | -1.96 | 2.81 | -1.29 | 23.69 | -130.11 | 11.36 | 0.996 |
| 211.60 | 100.00 | 2.82 | 4.78 | -1.96 | 2.82 | -1.31 | 23.71 | -129.84 | 11.32 | 0.996 |
| 211.70 | 100.00 | 2.84 | 4.78 | -1.94 | 2.84 | -1.33 | 23.72 | -129.57 | 11.29 | 0.996 |
| 211.80 | 100.00 | 2.85 | 4.79 | -1.94 | 2.85 | -1.34 | 23.73 | -129.30 | 11.26 | 0.996 |
| 211.90 | 100.00 | 2.87 | 4.79 | -1.92 | 2.87 | -1.36 | 23.74 | -129.02 | 11.22 | 0.996 |
| 212.00 | 100.00 | 2.88 | 4.80 | -1.92 | 2.88 | -1.38 | 23.75 | -128.75 | 11.19 | 0.996 |
| 212.10 | 100.00 | 2.90 | 4.80 | -1.90 | 2.90 | -1.39 | 23.76 | -128.48 | 11.16 | 0.996 |
| 212.20 | 100.00 | 2.91 | 4.80 | -1.89 | 2.91 | -1.41 | 23.77 | -128.20 | 11.12 | 0.996 |
| 212.30 | 100.00 | 2.93 | 4.81 | -1.88 | 2.93 | -1.43 | 23.79 | -127.93 | 11.09 | 0.996 |
| 212.40 | 100.00 | 2.94 | 4.81 | -1.87 | 2.94 | -1.44 | 23.80 | -127.66 | 11.06 | 0.996 |
| 212.50 | 100.00 | 2.96 | 4.82 | -1.86 | 2.96 | -1.46 | 23.81 | -127.38 | 11.02 | 0.996 |
| 212.60 | 100.00 | 2.97 | 4.82 | -1.85 | 2.97 | -1.48 | 23.82 | -127.11 | 10.99 | 0.996 |
| 212.70 | 100.00 | 2.99 | 4.83 | -1.84 | 2.99 | -1.49 | 23.83 | -126.83 | 10.96 | 0.996 |
| 212.80 | 100.00 | 3.01 | 4.83 | -1.82 | 3.01 | -1.51 | 23.84 | -126.56 | 10.92 | 0.996 |
| 212.90 | 100.00 | 3.02 | 4.84 | -1.82 | 3.02 | -1.53 | 23.86 | -126.28 | 10.89 | 0.996 |
| 213.00 | 100.00 | 3.04 | 4.84 | -1.80 | 3.04 | -1.55 | 23.87 | -126.01 | 10.86 | 0.996 |
| 213.10 | 100.00 | 3.05 | 4.84 | -1.79 | 3.05 | -1.56 | 23.88 | -125.73 | 10.82 | 0.996 |
| 213.20 | 100.00 | 3.07 | 4.85 | -1.78 | 3.07 | -1.58 | 23.89 | -125.46 | 10.79 | 0.996 |
| 213.30 | 100.00 | 3.09 | 4.85 | -1.76 | 3.09 | -1.60 | 23.90 | -125.18 | 10.76 | 0.996 |
| 213.40 | 100.00 | 3.10 | 4.86 | -1.76 | 3.10 | -1.62 | 23.92 | -124.90 | 10.72 | 0.996 |
| 213.50 | 100.00 | 3.12 | 4.86 | -1.74 | 3.12 | -1.63 | 23.93 | -124.63 | 10.69 | 0.996 |
| 213.60 | 100.00 | 3.14 | 4.86 | -1.72 | 3.14 | -1.65 | 23.94 | -124.35 | 10.66 | 0.996 |
| 213.70 | 100.00 | 3.15 | 4.87 | -1.72 | 3.15 | -1.67 | 23.96 | -124.07 | 10.63 | 0.996 |
| 213.80 | 100.00 | 3.17 | 4.87 | -1.70 | 3.17 | -1.69 | 23.97 | -123.80 | 10.59 | 0.996 |
| 213.90 | 100.00 | 3.19 | 4.88 | -1.69 | 3.19 | -1.71 | 23.98 | -123.52 | 10.56 | 0.996 |
| 214.00 | 100.00 | 3.20 | 4.88 | -1.68 | 3.20 | -1.72 | 23.99 | -123.24 | 10.53 | 0.996 |
| 214.10 | 100.00 | 3.22 | 4.88 | -1.66 | 3.22 | -1.74 | 24.01 | -122.96 | 10.49 | 0.996 |
| 214.20 | 100.00 | 3.24 | 4.89 | -1.65 | 3.24 | -1.76 | 24.02 | -122.69 | 10.46 | 0.996 |
| 214.30 | 100.00 | 3.25 | 4.89 | -1.64 | 3.25 | -1.78 | 24.03 | -122.41 | 10.43 | 0.996 |
| 214.40 | 100.00 | 3.27 | 4.89 | -1.62 | 3.27 | -1.80 | 24.05 | -122.13 | 10.40 | 0.996 |
| 214.50 | 100.00 | 3.29 | 4.90 | -1.61 | 3.29 | -1.82 | 24.06 | -121.85 | 10.36 | 0.996 |
| 214.60 | 100.00 | 3.31 | 4.90 | -1.59 | 3.31 | -1.83 | 24.07 | -121.57 | 10.33 | 0.996 |
| 214.70 | 100.00 | 3.33 | 4.90 | -1.57 | 3.33 | -1.85 | 24.09 | -121.29 | 10.30 | 0.996 |
| 214.80 | 100.00 | 3.34 | 4.91 | -1.57 | 3.34 | -1.87 | 24.10 | -121.01 | 10.27 | 0.996 |
| 214.90 | 100.00 | 3.36 | 4.91 | -1.55 | 3.36 | -1.89 | 24.12 | -120.74 | 10.23 | 0.996 |
| 215.00 | 100.00 | 3.38 | 4.91 | -1.53 | 3.38 | -1.91 | 24.13 | -120.46 | 10.20 | 0.996 |
| 215.10 | 100.00 | 3.40 | 4.92 | -1.52 | 3.40 | -1.93 | 24.14 | -120.18 | 10.17 | 0.996 |
| 215.20 | 100.00 | 3.42 | 4.92 | -1.50 | 3.42 | -1.95 | 24.16 | -119.89 | 10.14 | 0.996 |
| 215.30 | 100.00 | 3.44 | 4.92 | -1.48 | 3.44 | -1.97 | 24.17 | -119.61 | 10.10 | 0.996 |
| 215.40 | 100.00 | 3.45 | 4.93 | -1.48 | 3.45 | -1.99 | 24.19 | -119.33 | 10.07 | 0.996 |
| 215.50 | 100.00 | 3.47 | 4.93 | -1.46 | 3.47 | -2.01 | 24.20 | -119.05 | 10.04 | 0.996 |
| 215.60 | 100.00 | 3.49 | 4.93 | -1.44 | 3.49 | -2.03 | 24.22 | -118.77 | 10.01 | 0.996 |
| 215.70 | 100.00 | 3.51 | 4.94 | -1.43 | 3.51 | -2.04 | 24.23 | -118.49 | 9.97 | 0.996 |
| 215.80 | 100.00 | 3.53 | 4.94 | -1.41 | 3.53 | -2.07 | 24.25 | -118.21 | 9.94 | 0.996 |
| 215.90 | 100.00 | 3.55 | 4.94 | -1.39 | 3.55 | -2.09 | 24.26 | -117.92 | 9.91 | 0.996 |
| 216.00 | 100.00 | 3.57 | 4.94 | -1.37 | 3.57 | -2.11 | 24.28 | -117.64 | 9.88 | 0.996 |
| 216.10 | 100.00 | 3.59 | 4.95 | -1.36 | 3.59 | -2.14 | 24.29 | -117.36 | 9.84 | 0.996 |
| 216.20 | 100.00 | 3.61 | 4.95 | -1.34 | 3.61 | -2.16 | 24.31 | -117.08 | 9.81 | 0.996 |
| 216.30 | 100.00 | 3.63 | 4.95 | -1.32 | 3.63 | -2.18 | 24.33 | -116.79 | 9.78 | 0.996 |
| 216.40 | 100.00 | 3.65 | 4.95 | -1.30 | 3.65 | -2.21 | 24.34 | -116.51 | 9.75 | 0.996 |
| 216.50 | 100.00 | 3.67 | 4.96 | -1.29 | 3.67 | -2.23 | 24.36 | -116.23 | 9.71 | 0.996 |
| 216.60 | 100.00 | 3.69 | 4.96 | -1.27 | 3.69 | -2.25 | 24.38 | -115.94 | 9.68 | 0.996 |
| 216.70 | 100.00 | 3.71 | 4.96 | -1.25 | 3.71 | -2.28 | 24.39 | -115.66 | 9.65 | 0.996 |
| 216.80 | 100.00 | 3.73 | 4.96 | -1.23 | 3.73 | -2.30 | 24.41 | -115.37 | 9.62 | 0.996 |
| 216.90 | 100.00 | 3.75 | 4.96 | -1.21 | 3.75 | -2.33 | 24.43 | -115.09 | 9.58 | 0.996 |
| 217.00 | 100.00 | 3.77 | 4.97 | -1.20 | 3.77 | -2.35 | 24.44 | -114.80 | 9.55 | 0.996 |
| 217.10 | 100.00 | 3.79 | 4.97 | -1.18 | 3.79 | -2.38 | 24.46 | -114.52 | 9.52 | 0.996 |
| 217.20 | 100.00 | 3.81 | 4.97 | -1.16 | 3.81 | -2.40 | 24.48 | -114.23 | 9.49 | 0.996 |
| 217.30 | 100.00 | 3.83 | 4.97 | -1.14 | 3.83 | -2.42 | 24.50 | -113.95 | 9.46 | 0.996 |
| 217.40 | 100.00 | 3.85 | 4.97 | -1.12 | 3.85 | -2.45 | 24.52 | -113.66 | 9.42 | 0.996 |
| 217.50 | 100.00 | 3.87 | 4.97 | -1.10 | 3.87 | -2.47 | 24.54 | -113.38 | 9.39 | 0.996 |
| 217.60 | 100.00 | 3.89 | 4.98 | -1.09 | 3.89 | -2.50 | 24.55 | -113.09 | 9.36 | 0.996 |
| 217.70 | 100.00 | 3.91 | 4.98 | -1.07 | 3.91 | -2.52 | 24.57 | -112.80 | 9.33 | 0.996 |
| 217.80 | 100.00 | 3.93 | 4.98 | -1.05 | 3.93 | -2.55 | 24.59 | -112.51 | 9.29 | 0.996 |
| 217.90 | 100.00 | 3.95 | 4.98 | -1.03 | 3.95 | -2.58 | 24.61 | -112.23 | 9.26 | 0.996 |
| 218.00 | 100.00 | 3.97 | 4.98 | -1.01 | 3.97 | -2.60 | 24.63 | -111.94 | 9.23 | 0.996 |
| 218.10 | 100.00 | 3.99 | 4.98 | -0.99 | 3.99 | -2.63 | 24.65 | -111.65 | 9.20 | 0.996 |
| 218.20 | 100.00 | 4.02 | 4.98 | -0.96 | 4.02 | -2.65 | 24.67 | -111.36 | 9.16 | 0.996 |
| 218.30 | 100.00 | 4.04 | 4.98 | -0.94 | 4.04 | -2.68 | 24.69 | -111.07 | 9.13 | 0.996 |
| 218.40 | 100.00 | 4.06 | 4.98 | -0.92 | 4.06 | -2.70 | 24.71 | -110.78 | 9.10 | 0.996 |
| 218.50 | 100.00 | 4.08 | 4.98 | -0.90 | 4.08 | -2.73 | 24.74 | -110.50 | 9.07 | 0.996 |
| 218.60 | 100.00 | 4.10 | 4.98 | -0.88 | 4.10 | -2.76 | 24.76 | -110.20 | 9.03 | 0.996 |
| 218.70 | 100.00 | 4.12 | 4.98 | -0.86 | 4.12 | -2.78 | 24.78 | -109.92 | 9.00 | 0.996 |
| 218.80 | 100.00 | 4.15 | 4.98 | -0.83 | 4.15 | -2.81 | 24.80 | -109.62 | 8.97 | 0.996 |
| 218.90 | 100.00 | 4.17 | 4.98 | -0.81 | 4.17 | -2.83 | 24.82 | -109.34 | 8.94 | 0.996 |
| 219.00 | 100.00 | 4.19 | 4.98 | -0.79 | 4.19 | -2.86 | 24.84 | -109.04 | 8.91 | 0.996 |
| 219.10 | 100.00 | 4.21 | 4.98 | -0.77 | 4.21 | -2.89 | 24.87 | -108.75 | 8.87 | 0.996 |
| 219.20 | 100.00 | 4.23 | 4.98 | -0.75 | 4.23 | -2.91 | 24.89 | -108.46 | 8.84 | 0.996 |
| 219.30 | 100.00 | 4.25 | 4.98 | -0.73 | 4.25 | -2.94 | 24.91 | -108.17 | 8.81 | 0.996 |
| 219.40 | 100.00 | 4.28 | 4.98 | -0.70 | 4.28 | -2.97 | 24.94 | -107.88 | 8.78 | 0.996 |
| 219.50 | 100.00 | 4.30 | 4.98 | -0.68 | 4.30 | -2.99 | 24.96 | -107.58 | 8.74 | 0.996 |
| 219.60 | 100.00 | 4.32 | 4.98 | -0.66 | 4.32 | -3.02 | 24.99 | -107.29 | 8.71 | 0.996 |
| 219.70 | 100.00 | 4.34 | 4.98 | -0.64 | 4.34 | -3.05 | 25.01 | -107.00 | 8.68 | 0.996 |
| 219.80 | 100.00 | 4.37 | 4.98 | -0.61 | 4.37 | -3.07 | 25.04 | -106.71 | 8.65 | 0.996 |
| 219.90 | 100.00 | 4.39 | 4.98 | -0.59 | 4.39 | -3.10 | 25.06 | -106.41 | 8.61 | 0.996 |
| 220.00 | 100.00 | 4.41 | 4.98 | -0.57 | 4.41 | -3.13 | 25.09 | -106.12 | 8.58 | 0.996 |
| 220.10 | 100.00 | 4.43 | 4.98 | -0.55 | 4.43 | -3.15 | 25.11 | -105.83 | 8.55 | 0.996 |
| 220.20 | 100.00 | 4.46 | 4.98 | -0.52 | 4.46 | -3.18 | 25.14 | -105.53 | 8.52 | 0.996 |
| 220.30 | 100.00 | 4.48 | 4.97 | -0.49 | 4.48 | -3.21 | 25.17 | -105.24 | 8.48 | 0.996 |
| 220.40 | 100.00 | 4.50 | 4.97 | -0.47 | 4.50 | -3.23 | 25.19 | -104.94 | 8.45 | 0.996 |
| 220.50 | 100.00 | 4.52 | 4.97 | -0.45 | 4.52 | -3.26 | 25.22 | -104.65 | 8.42 | 0.996 |
| 220.60 | 100.00 | 4.55 | 4.97 | -0.42 | 4.55 | -3.29 | 25.25 | -104.35 | 8.39 | 0.996 |
| 220.70 | 100.00 | 4.57 | 4.97 | -0.40 | 4.57 | -3.32 | 25.28 | -104.06 | 8.35 | 0.996 |
| 220.80 | 100.00 | 4.59 | 4.96 | -0.37 | 4.59 | -3.34 | 25.31 | -103.76 | 8.32 | 0.996 |
| 220.90 | 100.00 | 4.61 | 4.96 | -0.35 | 4.61 | -3.37 | 25.34 | -103.46 | 8.29 | 0.996 |
| 221.00 | 100.00 | 4.64 | 4.96 | -0.32 | 4.64 | -3.40 | 25.37 | -103.17 | 8.26 | 0.996 |
| 221.10 | 100.00 | 4.66 | 4.96 | -0.30 | 4.66 | -3.42 | 25.40 | -102.87 | 8.22 | 0.996 |
| 221.20 | 100.00 | 4.68 | 4.95 | -0.27 | 4.68 | -3.45 | 25.43 | -102.57 | 8.19 | 0.996 |
| 221.30 | 100.00 | 4.71 | 4.95 | -0.24 | 4.71 | -3.48 | 25.46 | -102.27 | 8.16 | 0.996 |
| 221.40 | 100.00 | 4.73 | 4.95 | -0.22 | 4.73 | -3.50 | 25.49 | -101.98 | 8.12 | 0.996 |
| 221.50 | 100.00 | 4.75 | 4.94 | -0.19 | 4.75 | -3.53 | 25.52 | -101.68 | 8.09 | 0.996 |
| 221.60 | 100.00 | 4.77 | 4.94 | -0.17 | 4.77 | -3.56 | 25.55 | -101.38 | 8.06 | 0.996 |
| 221.70 | 100.00 | 4.80 | 4.94 | -0.14 | 4.80 | -3.58 | 25.59 | -101.08 | 8.03 | 0.996 |
| 221.80 | 100.00 | 4.82 | 4.93 | -0.11 | 4.82 | -3.61 | 25.62 | -100.78 | 7.99 | 0.996 |
| 221.90 | 100.00 | 4.84 | 4.93 | -0.09 | 4.84 | -3.64 | 25.65 | -100.48 | 7.96 | 0.996 |
| 222.00 | 100.00 | 4.87 | 4.93 | -0.06 | 4.87 | -3.68 | 25.69 | -100.18 | 7.93 | 0.996 |
| 222.10 | 100.00 | 4.89 | 4.92 | -0.03 | 4.89 | -3.71 | 25.72 | -99.88 | 7.90 | 0.996 |
| 222.20 | 100.00 | 4.91 | 4.92 | -0.01 | 4.91 | -3.75 | 25.76 | -99.58 | 7.86 | 0.996 |
| 222.30 | 100.00 | 4.94 | 4.91 | 0.03 | 4.94 | -3.78 | 25.79 | -99.28 | 7.83 | 0.996 |
| 222.40 | 100.00 | 4.96 | 4.91 | 0.05 | 4.96 | -3.82 | 25.83 | -98.98 | 7.80 | 0.996 |
| 222.50 | 100.00 | 4.98 | 4.90 | 0.08 | 4.98 | -3.85 | 25.86 | -98.68 | 7.76 | 0.996 |
| 222.60 | 100.00 | 5.01 | 4.90 | 0.11 | 5.01 | -3.88 | 25.90 | -98.38 | 7.73 | 0.996 |
| 222.70 | 100.00 | 5.03 | 4.89 | 0.14 | 5.03 | -3.92 | 25.94 | -98.08 | 7.70 | 0.996 |
| 222.80 | 100.00 | 5.05 | 4.89 | 0.16 | 5.05 | -3.95 | 25.98 | -97.78 | 7.67 | 0.996 |
| 222.90 | 100.00 | 5.07 | 4.88 | 0.19 | 5.07 | -3.99 | 26.01 | -97.48 | 7.63 | 0.996 |
| 223.00 | 100.00 | 5.10 | 4.88 | 0.22 | 5.10 | -4.02 | 26.05 | -97.17 | 7.60 | 0.996 |
| 223.10 | 100.00 | 5.12 | 4.87 | 0.25 | 5.12 | -4.05 | 26.09 | -96.87 | 7.57 | 0.996 |
| 223.20 | 100.00 | 5.14 | 4.87 | 0.27 | 5.14 | -4.09 | 26.13 | -96.57 | 7.53 | 0.996 |
| 223.30 | 100.00 | 5.17 | 4.86 | 0.31 | 5.17 | -4.12 | 26.17 | -96.27 | 7.50 | 0.996 |
| 223.40 | 100.00 | 5.19 | 4.85 | 0.34 | 5.19 | -4.16 | 26.21 | -95.96 | 7.47 | 0.996 |
| 223.50 | 100.00 | 5.21 | 4.85 | 0.36 | 5.21 | -4.19 | 26.26 | -95.66 | 7.44 | 0.996 |
| 223.60 | 100.00 | 5.23 | 4.84 | 0.39 | 5.23 | -4.22 | 26.30 | -95.36 | 7.40 | 0.996 |
| 223.70 | 100.00 | 5.26 | 4.83 | 0.43 | 5.26 | -4.26 | 26.34 | -95.05 | 7.37 | 0.996 |
| 223.80 | 100.00 | 5.28 | 4.83 | 0.45 | 5.28 | -4.29 | 26.39 | -94.75 | 7.34 | 0.996 |
| 223.90 | 100.00 | 5.30 | 4.82 | 0.48 | 5.30 | -4.33 | 26.43 | -94.44 | 7.31 | 0.996 |
| 224.00 | 100.00 | 5.33 | 4.81 | 0.52 | 5.33 | -4.36 | 26.47 | -94.14 | 7.27 | 0.996 |
| 224.10 | 100.00 | 5.35 | 4.80 | 0.55 | 5.35 | -4.39 | 26.52 | -93.83 | 7.24 | 0.996 |
| 224.20 | 100.00 | 5.37 | 4.80 | 0.57 | 5.37 | -4.43 | 26.56 | -93.53 | 7.21 | 0.996 |
| 224.30 | 100.00 | 5.39 | 4.79 | 0.60 | 5.39 | -4.46 | 26.61 | -93.22 | 7.17 | 0.996 |
| 224.40 | 100.00 | 5.42 | 4.78 | 0.64 | 5.42 | -4.49 | 26.66 | -92.92 | 7.14 | 0.996 |
| 224.50 | 100.00 | 5.44 | 4.77 | 0.67 | 5.44 | -4.52 | 26.71 | -92.61 | 7.11 | 0.996 |
| 224.60 | 100.00 | 5.46 | 4.76 | 0.70 | 5.46 | -4.56 | 26.75 | -92.31 | 7.08 | 0.996 |
| 224.70 | 100.00 | 5.48 | 4.76 | 0.72 | 5.48 | -4.59 | 26.80 | -92.00 | 7.04 | 0.996 |
| 224.80 | 100.00 | 5.51 | 4.75 | 0.76 | 5.51 | -4.62 | 26.85 | -91.70 | 7.01 | 0.996 |
| 224.90 | 100.00 | 5.53 | 4.74 | 0.79 | 5.53 | -4.65 | 26.90 | -91.39 | 6.98 | 0.996 |
| 225.00 | 100.00 | 5.55 | 4.73 | 0.82 | 5.55 | -4.68 | 26.95 | -91.08 | 6.94 | 0.996 |
| 225.10 | 100.00 | 5.57 | 4.72 | 0.85 | 5.57 | -4.71 | 27.01 | -90.78 | 6.91 | 0.996 |
| 225.20 | 100.00 | 5.60 | 4.71 | 0.89 | 5.60 | -4.74 | 27.06 | -90.47 | 6.88 | 0.996 |
| 225.30 | 100.00 | 5.62 | 4.70 | 0.92 | 5.62 | -4.77 | 27.11 | -90.16 | 6.85 | 0.996 |
| 225.40 | 100.00 | 5.64 | 4.69 | 0.95 | 5.64 | -4.80 | 27.16 | -89.86 | 6.81 | 0.996 |
| 225.50 | 100.00 | 5.66 | 4.68 | 0.98 | 5.66 | -4.83 | 27.22 | -89.55 | 6.78 | 0.996 |
| 225.60 | 100.00 | 5.68 | 4.67 | 1.01 | 5.68 | -4.86 | 27.27 | -89.24 | 6.75 | 0.996 |
| 225.70 | 100.00 | 5.70 | 4.66 | 1.04 | 5.70 | -4.89 | 27.33 | -88.93 | 6.72 | 0.996 |
| 225.80 | 100.00 | 5.73 | 4.65 | 1.08 | 5.73 | -4.92 | 27.38 | -88.63 | 6.68 | 0.996 |
| 225.90 | 100.00 | 5.75 | 4.64 | 1.11 | 5.75 | -4.95 | 27.44 | -88.32 | 6.65 | 0.996 |
| 226.00 | 100.00 | 5.77 | 4.62 | 1.15 | 5.77 | -4.97 | 27.50 | -88.01 | 6.62 | 0.996 |
| 226.10 | 100.00 | 5.79 | 4.61 | 1.18 | 5.79 | -5.00 | 27.56 | -87.70 | 6.58 | 0.996 |
| 226.20 | 100.00 | 5.81 | 4.60 | 1.21 | 5.81 | -5.03 | 27.62 | -87.39 | 6.55 | 0.996 |
| 226.30 | 100.00 | 5.83 | 4.59 | 1.24 | 5.83 | -5.05 | 27.68 | -87.09 | 6.52 | 0.996 |
| 226.40 | 100.00 | 5.86 | 4.58 | 1.28 | 5.86 | -5.08 | 27.74 | -86.78 | 6.49 | 0.996 |
| 226.50 | 100.00 | 5.88 | 4.56 | 1.32 | 5.88 | -5.11 | 27.80 | -86.47 | 6.45 | 0.996 |
| 226.60 | 100.00 | 5.90 | 4.55 | 1.35 | 5.90 | -5.15 | 27.86 | -86.16 | 6.42 | 0.995 |
| 226.70 | 100.00 | 5.92 | 4.54 | 1.38 | 5.92 | -5.19 | 27.92 | -85.85 | 6.39 | 0.995 |
| 226.80 | 100.00 | 5.94 | 4.53 | 1.41 | 5.94 | -5.22 | 27.99 | -85.54 | 6.36 | 0.995 |
| 226.90 | 100.00 | 5.96 | 4.51 | 1.45 | 5.96 | -5.26 | 28.05 | -85.23 | 6.33 | 0.995 |
| 227.00 | 100.00 | 5.98 | 4.50 | 1.48 | 5.98 | -5.30 | 28.11 | -84.92 | 6.29 | 0.995 |
| 227.10 | 100.00 | 6.00 | 4.49 | 1.51 | 6.00 | -5.33 | 28.18 | -84.62 | 6.26 | 0.995 |
| 227.20 | 100.00 | 6.02 | 4.47 | 1.55 | 6.02 | -5.37 | 28.25 | -84.31 | 6.23 | 0.995 |
| 227.30 | 100.00 | 6.04 | 4.46 | 1.58 | 6.04 | -5.41 | 28.31 | -84.00 | 6.20 | 0.995 |
| 227.40 | 100.00 | 6.06 | 4.44 | 1.62 | 6.06 | -5.44 | 28.38 | -83.69 | 6.16 | 0.995 |
| 227.50 | 100.00 | 6.08 | 4.43 | 1.65 | 6.08 | -5.48 | 28.45 | -83.38 | 6.13 | 0.995 |
| 227.60 | 100.00 | 6.10 | 4.41 | 1.69 | 6.10 | -5.51 | 28.52 | -83.07 | 6.10 | 0.995 |
| 227.70 | 100.00 | 6.12 | 4.40 | 1.72 | 6.12 | -5.54 | 28.59 | -82.76 | 6.07 | 0.995 |
| 227.80 | 100.00 | 6.14 | 4.38 | 1.76 | 6.14 | -5.58 | 28.66 | -82.45 | 6.04 | 0.995 |
| 227.90 | 100.00 | 6.16 | 4.37 | 1.79 | 6.16 | -5.61 | 28.73 | -82.14 | 6.00 | 0.995 |
| 228.00 | 100.00 | 6.18 | 4.35 | 1.83 | 6.18 | -5.64 | 28.81 | -81.84 | 5.97 | 0.995 |
| 228.10 | 100.00 | 6.20 | 4.34 | 1.86 | 6.20 | -5.67 | 28.88 | -81.53 | 5.94 | 0.995 |
| 228.20 | 100.00 | 6.22 | 4.32 | 1.90 | 6.22 | -5.70 | 28.95 | -81.22 | 5.91 | 0.995 |
| 228.30 | 100.00 | 6.24 | 4.30 | 1.94 | 6.24 | -5.73 | 29.03 | -80.91 | 5.88 | 0.995 |
| 228.40 | 100.00 | 6.26 | 4.29 | 1.97 | 6.26 | -5.76 | 29.11 | -80.60 | 5.85 | 0.995 |
| 228.50 | 100.00 | 6.28 | 4.27 | 2.01 | 6.28 | -5.79 | 29.18 | -80.29 | 5.82 | 0.995 |
| 228.60 | 100.00 | 6.30 | 4.25 | 2.05 | 6.30 | -5.82 | 29.26 | -79.98 | 5.78 | 0.995 |
| 228.70 | 100.00 | 6.32 | 4.24 | 2.08 | 6.32 | -5.85 | 29.34 | -79.67 | 5.75 | 0.995 |
| 228.80 | 100.00 | 6.33 | 4.22 | 2.11 | 6.33 | -5.87 | 29.42 | -79.37 | 5.72 | 0.995 |
| 228.90 | 100.00 | 6.35 | 4.20 | 2.15 | 6.35 | -5.90 | 29.50 | -79.06 | 5.69 | 0.995 |
| 229.00 | 100.00 | 6.37 | 4.18 | 2.19 | 6.37 | -5.92 | 29.58 | -78.75 | 5.66 | 0.995 |
| 229.10 | 100.00 | 6.39 | 4.17 | 2.22 | 6.39 | -5.95 | 29.66 | -78.44 | 5.63 | 0.995 |
| 229.20 | 100.00 | 6.41 | 4.15 | 2.26 | 6.41 | -5.97 | 29.74 | -78.13 | 5.60 | 0.995 |
| 229.30 | 100.00 | 6.43 | 4.13 | 2.30 | 6.43 | -5.99 | 29.83 | -77.83 | 5.57 | 0.995 |
| 229.40 | 100.00 | 6.44 | 4.11 | 2.33 | 6.44 | -6.01 | 29.91 | -77.52 | 5.54 | 0.995 |
| 229.50 | 100.00 | 6.46 | 4.09 | 2.37 | 6.46 | -6.03 | 30.00 | -77.21 | 5.51 | 0.995 |
| 229.60 | 100.00 | 6.48 | 4.07 | 2.41 | 6.48 | -6.05 | 30.08 | -76.90 | 5.47 | 0.995 |
| 229.70 | 100.00 | 6.50 | 4.05 | 2.45 | 6.50 | -6.07 | 30.17 | -76.60 | 5.44 | 0.995 |
| 229.80 | 100.00 | 6.51 | 4.03 | 2.48 | 6.51 | -6.08 | 30.26 | -76.29 | 5.41 | 0.995 |
| 229.90 | 100.00 | 6.53 | 4.01 | 2.52 | 6.53 | -6.10 | 30.35 | -75.98 | 5.38 | 0.995 |
| 230.00 | 100.00 | 6.55 | 3.99 | 2.56 | 6.55 | -6.11 | 30.43 | -75.68 | 5.35 | 0.995 |
| 230.10 | 100.00 | 6.56 | 3.97 | 2.59 | 6.56 | -6.13 | 30.52 | -75.37 | 5.32 | 0.995 |
| 230.20 | 100.00 | 6.58 | 3.95 | 2.63 | 6.58 | -6.14 | 30.62 | -75.06 | 5.29 | 0.995 |
| 230.30 | 100.00 | 6.60 | 3.93 | 2.67 | 6.60 | -6.15 | 30.71 | -74.76 | 5.26 | 0.995 |
| 230.40 | 100.00 | 6.61 | 3.91 | 2.70 | 6.61 | -6.16 | 30.80 | -74.45 | 5.23 | 0.995 |
| 230.50 | 100.00 | 6.63 | 3.89 | 2.74 | 6.63 | -6.17 | 30.89 | -74.15 | 5.20 | 0.995 |
| 230.60 | 100.00 | 6.65 | 3.86 | 2.79 | 6.65 | -6.18 | 30.99 | -73.84 | 5.17 | 0.995 |
| 230.70 | 100.00 | 6.66 | 3.84 | 2.82 | 6.66 | -6.19 | 31.08 | -73.54 | 5.15 | 0.995 |
| 230.80 | 100.00 | 6.68 | 3.82 | 2.86 | 6.68 | -6.20 | 31.18 | -73.23 | 5.12 | 0.995 |
| 230.90 | 100.00 | 6.69 | 3.80 | 2.89 | 6.69 | -6.20 | 31.28 | -72.93 | 5.09 | 0.995 |
| 231.00 | 100.00 | 6.71 | 3.78 | 2.93 | 6.71 | -6.21 | 31.37 | -72.62 | 5.06 | 0.995 |
| 231.10 | 100.00 | 6.73 | 3.75 | 2.98 | 6.73 | -6.21 | 31.47 | -72.32 | 5.03 | 0.995 |
| 231.20 | 100.00 | 6.74 | 3.73 | 3.01 | 6.74 | -6.21 | 31.57 | -72.02 | 5.00 | 0.995 |
| 231.30 | 100.00 | 6.76 | 3.71 | 3.05 | 6.76 | -6.21 | 31.67 | -71.71 | 4.97 | 0.995 |
| 231.40 | 100.00 | 6.77 | 3.68 | 3.09 | 6.77 | -6.21 | 31.77 | -71.41 | 4.94 | 0.995 |
| 231.50 | 100.00 | 6.79 | 3.66 | 3.13 | 6.79 | -6.21 | 31.88 | -71.11 | 4.91 | 0.995 |
| 231.60 | 100.00 | 6.80 | 3.63 | 3.17 | 6.80 | -6.21 | 31.98 | -70.81 | 4.89 | 0.995 |
| 231.70 | 100.00 | 6.81 | 3.61 | 3.20 | 6.81 | -6.20 | 32.08 | -70.50 | 4.86 | 0.995 |
| 231.80 | 100.00 | 6.83 | 3.59 | 3.24 | 6.83 | -6.20 | 32.19 | -70.20 | 4.83 | 0.995 |
| 231.90 | 100.00 | 6.84 | 3.56 | 3.28 | 6.84 | -6.19 | 32.29 | -69.90 | 4.80 | 0.995 |
| 232.00 | 100.00 | 6.86 | 3.54 | 3.32 | 6.86 | -6.18 | 32.40 | -69.60 | 4.77 | 0.995 |
| 232.10 | 100.00 | 6.87 | 3.51 | 3.36 | 6.87 | -6.18 | 32.51 | -69.30 | 4.75 | 0.995 |
| 232.20 | 100.00 | 6.89 | 3.49 | 3.40 | 6.89 | -6.17 | 32.61 | -69.00 | 4.72 | 0.995 |
| 232.30 | 100.00 | 6.90 | 3.46 | 3.44 | 6.90 | -6.16 | 32.72 | -68.70 | 4.69 | 0.995 |
| 232.40 | 100.00 | 6.91 | 3.44 | 3.47 | 6.91 | -6.14 | 32.83 | -68.40 | 4.66 | 0.995 |
| 232.50 | 100.00 | 6.93 | 3.41 | 3.52 | 6.93 | -6.13 | 32.94 | -68.11 | 4.64 | 0.995 |
| 232.60 | 100.00 | 6.94 | 3.38 | 3.56 | 6.94 | -6.12 | 33.05 | -67.81 | 4.61 | 0.995 |
| 232.70 | 100.00 | 6.95 | 3.36 | 3.59 | 6.95 | -6.10 | 33.17 | -67.51 | 4.58 | 0.995 |
| 232.80 | 100.00 | 6.96 | 3.33 | 3.63 | 6.96 | -6.09 | 33.28 | -67.21 | 4.55 | 0.995 |
| 232.90 | 100.00 | 6.98 | 3.30 | 3.68 | 6.98 | -6.07 | 33.39 | -66.92 | 4.53 | 0.995 |
| 233.00 | 100.00 | 6.99 | 3.28 | 3.71 | 6.99 | -6.05 | 33.51 | -66.62 | 4.50 | 0.995 |
| 233.10 | 100.00 | 7.00 | 3.25 | 3.75 | 7.00 | -6.04 | 33.62 | -66.33 | 4.48 | 0.995 |
| 233.20 | 100.00 | 7.01 | 3.22 | 3.79 | 7.01 | -6.02 | 33.74 | -66.03 | 4.45 | 0.995 |
| 233.30 | 100.00 | 7.03 | 3.19 | 3.84 | 7.03 | -6.00 | 33.86 | -65.74 | 4.42 | 0.995 |
| 233.40 | 100.00 | 7.04 | 3.17 | 3.87 | 7.04 | -5.97 | 33.97 | -65.44 | 4.40 | 0.995 |
| 233.50 | 100.00 | 7.05 | 3.14 | 3.91 | 7.05 | -5.95 | 34.09 | -65.15 | 4.37 | 0.995 |
| 233.60 | 100.00 | 7.06 | 3.11 | 3.95 | 7.06 | -5.93 | 34.21 | -64.86 | 4.35 | 0.995 |
| 233.70 | 100.00 | 7.07 | 3.08 | 3.99 | 7.07 | -5.91 | 34.33 | -64.57 | 4.32 | 0.995 |
| 233.80 | 100.00 | 7.08 | 3.05 | 4.03 | 7.08 | -5.88 | 34.45 | -64.28 | 4.29 | 0.995 |
| 233.90 | 100.00 | 7.10 | 3.02 | 4.08 | 7.10 | -5.86 | 34.57 | -63.98 | 4.27 | 0.995 |
| 234.00 | 100.00 | 7.11 | 2.99 | 4.12 | 7.11 | -5.83 | 34.70 | -63.69 | 4.24 | 0.995 |
| 234.10 | 100.00 | 7.12 | 2.97 | 4.15 | 7.12 | -5.80 | 34.82 | -63.41 | 4.22 | 0.995 |
| 234.20 | 100.00 | 7.13 | 2.94 | 4.19 | 7.13 | -5.78 | 34.94 | -63.12 | 4.19 | 0.995 |
| 234.30 | 100.00 | 7.14 | 2.91 | 4.23 | 7.14 | -5.75 | 35.07 | -62.83 | 4.17 | 0.995 |
| 234.40 | 100.00 | 7.15 | 2.88 | 4.27 | 7.15 | -5.72 | 35.19 | -62.54 | 4.15 | 0.995 |
| 234.50 | 100.00 | 7.16 | 2.85 | 4.31 | 7.16 | -5.69 | 35.32 | -62.25 | 4.12 | 0.995 |
| 234.60 | 100.00 | 7.17 | 2.82 | 4.35 | 7.17 | -5.66 | 35.45 | -61.97 | 4.10 | 0.995 |
| 234.70 | 100.00 | 7.18 | 2.79 | 4.39 | 7.18 | -5.63 | 35.57 | -61.68 | 4.07 | 0.995 |
| 234.80 | 100.00 | 7.19 | 2.76 | 4.43 | 7.19 | -5.60 | 35.70 | -61.40 | 4.05 | 0.995 |
| 234.90 | 100.00 | 7.20 | 2.72 | 4.48 | 7.20 | -5.57 | 35.83 | -61.11 | 4.03 | 0.995 |
| 235.00 | 100.00 | 7.21 | 2.69 | 4.52 | 7.21 | -5.53 | 35.96 | -60.83 | 4.00 | 0.995 |
| 235.10 | 100.00 | 7.22 | 2.66 | 4.56 | 7.22 | -5.50 | 36.09 | -60.55 | 3.98 | 0.995 |
| 235.20 | 100.00 | 7.23 | 2.63 | 4.60 | 7.23 | -5.47 | 36.22 | -60.26 | 3.96 | 0.995 |
| 235.30 | 100.00 | 7.24 | 2.60 | 4.64 | 7.24 | -5.43 | 36.35 | -59.98 | 3.93 | 0.995 |
| 235.40 | 100.00 | 7.24 | 2.57 | 4.67 | 7.24 | -5.40 | 36.48 | -59.70 | 3.91 | 0.995 |
| 235.50 | 100.00 | 7.25 | 2.54 | 4.71 | 7.25 | -5.36 | 36.62 | -59.42 | 3.89 | 0.995 |
| 235.60 | 100.00 | 7.26 | 2.50 | 4.76 | 7.26 | -5.33 | 36.75 | -59.14 | 3.86 | 0.995 |
| 235.70 | 100.00 | 7.27 | 2.47 | 4.80 | 7.27 | -5.29 | 36.88 | -58.86 | 3.84 | 0.995 |
| 235.80 | 100.00 | 7.28 | 2.44 | 4.84 | 7.28 | -5.26 | 37.02 | -58.59 | 3.82 | 0.995 |
| 235.90 | 100.00 | 7.29 | 2.41 | 4.88 | 7.29 | -5.22 | 37.15 | -58.31 | 3.80 | 0.995 |
| 236.00 | 100.00 | 7.30 | 2.37 | 4.93 | 7.30 | -5.18 | 37.29 | -58.03 | 3.78 | 0.995 |
| 236.10 | 100.00 | 7.30 | 2.34 | 4.96 | 7.30 | -5.15 | 37.43 | -57.76 | 3.75 | 0.995 |
| 236.20 | 100.00 | 7.31 | 2.31 | 5.00 | 7.31 | -5.11 | 37.56 | -57.48 | 3.73 | 0.995 |
| 236.30 | 100.00 | 7.32 | 2.27 | 5.05 | 7.32 | -5.07 | 37.70 | -57.21 | 3.71 | 0.995 |
| 236.40 | 100.00 | 7.33 | 2.24 | 5.09 | 7.33 | -5.03 | 37.84 | -56.94 | 3.69 | 0.995 |
| 236.50 | 100.00 | 7.33 | 2.21 | 5.12 | 7.33 | -5.00 | 37.98 | -56.67 | 3.67 | 0.995 |
| 236.60 | 100.00 | 7.34 | 2.17 | 5.17 | 7.34 | -4.96 | 38.12 | -56.39 | 3.65 | 0.995 |
| 236.70 | 100.00 | 7.35 | 2.14 | 5.21 | 7.35 | -4.92 | 38.26 | -56.12 | 3.63 | 0.995 |
| 236.80 | 100.00 | 7.35 | 2.11 | 5.24 | 7.35 | -4.88 | 38.40 | -55.85 | 3.61 | 0.995 |
| 236.90 | 100.00 | 7.36 | 2.07 | 5.29 | 7.36 | -4.84 | 38.54 | -55.59 | 3.59 | 0.995 |
| 237.00 | 100.00 | 7.37 | 2.04 | 5.33 | 7.37 | -4.80 | 38.68 | -55.32 | 3.57 | 0.995 |
| 237.10 | 100.00 | 7.38 | 2.00 | 5.38 | 7.38 | -4.77 | 38.82 | -55.05 | 3.55 | 0.995 |
| 237.20 | 100.00 | 7.38 | 1.97 | 5.41 | 7.38 | -4.73 | 38.97 | -54.78 | 3.53 | 0.995 |
| 237.30 | 100.00 | 7.39 | 1.93 | 5.46 | 7.39 | -4.69 | 39.11 | -54.52 | 3.51 | 0.995 |
| 237.40 | 100.00 | 7.39 | 1.90 | 5.49 | 7.39 | -4.65 | 39.25 | -54.25 | 3.49 | 0.995 |
| 237.50 | 100.00 | 7.40 | 1.86 | 5.54 | 7.40 | -4.61 | 39.40 | -53.99 | 3.47 | 0.995 |
| 237.60 | 100.00 | 7.41 | 1.83 | 5.58 | 7.41 | -4.57 | 39.54 | -53.73 | 3.45 | 0.995 |
| 237.70 | 100.00 | 7.41 | 1.79 | 5.62 | 7.41 | -4.53 | 39.69 | -53.46 | 3.43 | 0.995 |
| 237.80 | 100.00 | 7.42 | 1.76 | 5.66 | 7.42 | -4.49 | 39.83 | -53.20 | 3.41 | 0.995 |
| 237.90 | 100.00 | 7.42 | 1.72 | 5.70 | 7.42 | -4.45 | 39.98 | -52.94 | 3.39 | 0.995 |
| 238.00 | 100.00 | 7.43 | 1.69 | 5.74 | 7.43 | -4.41 | 40.12 | -52.68 | 3.37 | 0.995 |
| 238.10 | 100.00 | 7.43 | 1.65 | 5.78 | 7.43 | -4.37 | 40.27 | -52.43 | 3.35 | 0.995 |
| 238.20 | 100.00 | 7.44 | 1.62 | 5.82 | 7.44 | -4.33 | 40.42 | -52.17 | 3.33 | 0.995 |
| 238.30 | 100.00 | 7.44 | 1.58 | 5.86 | 7.44 | -4.29 | 40.56 | -51.91 | 3.32 | 0.995 |
| 238.40 | 100.00 | 7.45 | 1.54 | 5.91 | 7.45 | -4.25 | 40.71 | -51.65 | 3.30 | 0.995 |
| 238.50 | 100.00 | 7.45 | 1.51 | 5.94 | 7.45 | -4.22 | 40.86 | -51.40 | 3.28 | 0.995 |
| 238.60 | 100.00 | 7.46 | 1.47 | 5.99 | 7.46 | -4.18 | 41.01 | -51.15 | 3.26 | 0.995 |
| 238.70 | 100.00 | 7.46 | 1.44 | 6.02 | 7.46 | -4.14 | 41.16 | -50.89 | 3.24 | 0.995 |
| 238.80 | 100.00 | 7.47 | 1.40 | 6.07 | 7.47 | -4.10 | 41.31 | -50.64 | 3.23 | 0.995 |
| 238.90 | 100.00 | 7.47 | 1.36 | 6.11 | 7.47 | -4.06 | 41.45 | -50.39 | 3.21 | 0.995 |
| 239.00 | 100.00 | 7.48 | 1.33 | 6.15 | 7.48 | -4.02 | 41.60 | -50.14 | 3.19 | 0.995 |
| 239.10 | 100.00 | 7.48 | 1.29 | 6.19 | 7.48 | -3.98 | 41.75 | -49.89 | 3.17 | 0.995 |
| 239.20 | 100.00 | 7.49 | 1.25 | 6.24 | 7.49 | -3.94 | 41.90 | -49.64 | 3.16 | 0.995 |
| 239.30 | 100.00 | 7.49 | 1.21 | 6.28 | 7.49 | -3.90 | 42.06 | -49.39 | 3.14 | 0.995 |
| 239.40 | 100.00 | 7.49 | 1.18 | 6.31 | 7.49 | -3.87 | 42.21 | -49.15 | 3.12 | 0.995 |
| 239.50 | 100.00 | 7.50 | 1.14 | 6.36 | 7.50 | -3.83 | 42.36 | -48.90 | 3.11 | 0.995 |
| 239.60 | 100.00 | 7.50 | 1.10 | 6.40 | 7.50 | -3.79 | 42.51 | -48.65 | 3.09 | 0.995 |
| 239.70 | 100.00 | 7.51 | 1.07 | 6.44 | 7.51 | -3.75 | 42.66 | -48.41 | 3.07 | 0.995 |
| 239.80 | 100.00 | 7.51 | 1.03 | 6.48 | 7.51 | -3.71 | 42.81 | -48.17 | 3.06 | 0.995 |
| 239.90 | 100.00 | 7.51 | 0.99 | 6.52 | 7.51 | -3.68 | 42.96 | -47.93 | 3.04 | 0.995 |
| 240.00 | 100.00 | 7.52 | 0.95 | 6.57 | 7.52 | -3.64 | 43.12 | -47.68 | 3.03 | 0.995 |
| 240.10 | 100.00 | 7.52 | 0.92 | 6.60 | 7.52 | -3.60 | 43.27 | -47.44 | 3.01 | 0.995 |
| 240.20 | 100.00 | 7.52 | 0.88 | 6.64 | 7.52 | -3.56 | 43.42 | -47.20 | 3.00 | 0.995 |
| 240.30 | 100.00 | 7.53 | 0.84 | 6.69 | 7.53 | -3.53 | 43.57 | -46.97 | 2.98 | 0.995 |
| 240.40 | 100.00 | 7.53 | 0.80 | 6.73 | 7.53 | -3.49 | 43.73 | -46.73 | 2.97 | 0.995 |
| 240.50 | 100.00 | 7.53 | 0.76 | 6.77 | 7.53 | -3.45 | 43.88 | -46.49 | 2.95 | 0.995 |
| 240.60 | 100.00 | 7.53 | 0.73 | 6.80 | 7.53 | -3.42 | 44.03 | -46.26 | 2.94 | 0.995 |
| 240.70 | 100.00 | 7.54 | 0.69 | 6.85 | 7.54 | -3.38 | 44.19 | -46.02 | 2.92 | 0.995 |
| 240.80 | 100.00 | 7.54 | 0.65 | 6.89 | 7.54 | -3.34 | 44.34 | -45.79 | 2.91 | 0.995 |
| 240.90 | 100.00 | 7.54 | 0.61 | 6.93 | 7.54 | -3.31 | 44.49 | -45.55 | 2.89 | 0.995 |
| 241.00 | 100.00 | 7.54 | 0.57 | 6.97 | 7.54 | -3.27 | 44.65 | -45.32 | 2.88 | 0.995 |
| 241.10 | 100.00 | 7.55 | 0.54 | 7.01 | 7.55 | -3.24 | 44.80 | -45.09 | 2.86 | 0.995 |
| 241.20 | 100.00 | 7.55 | 0.50 | 7.05 | 7.55 | -3.20 | 44.96 | -44.86 | 2.85 | 0.995 |
| 241.30 | 100.00 | 7.55 | 0.46 | 7.09 | 7.55 | -3.17 | 45.11 | -44.63 | 2.83 | 0.995 |
| 241.40 | 100.00 | 7.55 | 0.42 | 7.13 | 7.55 | -3.13 | 45.26 | -44.40 | 2.82 | 0.995 |
| 241.50 | 100.00 | 7.56 | 0.38 | 7.18 | 7.56 | -3.10 | 45.42 | -44.17 | 2.81 | 0.995 |
| 241.60 | 100.00 | 7.56 | 0.34 | 7.22 | 7.56 | -3.06 | 45.57 | -43.95 | 2.79 | 0.995 |
| 241.70 | 100.00 | 7.56 | 0.30 | 7.26 | 7.56 | -3.03 | 45.73 | -43.72 | 2.78 | 0.995 |
| 241.80 | 100.00 | 7.56 | 0.27 | 7.29 | 7.56 | -2.99 | 45.88 | -43.50 | 2.77 | 0.995 |
| 241.90 | 100.00 | 7.56 | 0.23 | 7.33 | 7.56 | -2.96 | 46.03 | -43.27 | 2.75 | 0.995 |
| 242.00 | 100.00 | 7.56 | 0.19 | 7.37 | 7.56 | -2.93 | 46.19 | -43.05 | 2.74 | 0.995 |
| 242.10 | 100.00 | 7.57 | 0.15 | 7.42 | 7.57 | -2.89 | 46.34 | -42.83 | 2.73 | 0.995 |
| 242.20 | 100.00 | 7.57 | 0.11 | 7.46 | 7.57 | -2.86 | 46.50 | -42.60 | 2.71 | 0.995 |
| 242.30 | 100.00 | 7.57 | 0.07 | 7.50 | 7.57 | -2.83 | 46.65 | -42.38 | 2.70 | 0.995 |
| 242.40 | 100.00 | 7.57 | 0.03 | 7.54 | 7.57 | -2.79 | 46.80 | -42.16 | 2.69 | 0.995 |
| 242.50 | 100.00 | 7.57 | 0.00 | 7.57 | 7.57 | -2.76 | 46.96 | -41.94 | 2.67 | 0.995 |
| 242.60 | 100.00 | 7.57 | -0.04 | 7.61 | 7.57 | -2.73 | 47.11 | -41.73 | 2.66 | 0.995 |
| 242.70 | 100.00 | 7.57 | -0.08 | 7.65 | 7.57 | -2.69 | 47.27 | -41.51 | 2.65 | 0.995 |
| 242.80 | 100.00 | 7.57 | -0.12 | 7.69 | 7.57 | -2.66 | 47.42 | -41.29 | 2.64 | 0.995 |
| 242.90 | 100.00 | 7.58 | -0.16 | 7.74 | 7.58 | -2.63 | 47.57 | -41.08 | 2.63 | 0.995 |
| 243.00 | 100.00 | 7.58 | -0.20 | 7.78 | 7.58 | -2.60 | 47.73 | -40.86 | 2.61 | 0.995 |
| 243.10 | 100.00 | 7.58 | -0.24 | 7.82 | 7.58 | -2.57 | 47.88 | -40.65 | 2.60 | 0.995 |
| 243.20 | 100.00 | 7.58 | -0.28 | 7.86 | 7.58 | -2.54 | 48.03 | -40.44 | 2.59 | 0.995 |
| 243.30 | 100.00 | 7.58 | -0.31 | 7.89 | 7.58 | -2.51 | 48.19 | -40.22 | 2.58 | 0.995 |
| 243.40 | 100.00 | 7.58 | -0.35 | 7.93 | 7.58 | -2.47 | 48.34 | -40.01 | 2.57 | 0.995 |
| 243.50 | 100.00 | 7.58 | -0.39 | 7.97 | 7.58 | -2.44 | 48.49 | -39.80 | 2.56 | 0.995 |
| 243.60 | 100.00 | 7.58 | -0.43 | 8.01 | 7.58 | -2.41 | 48.65 | -39.59 | 2.54 | 0.995 |
| 243.70 | 100.00 | 7.58 | -0.47 | 8.05 | 7.58 | -2.38 | 48.80 | -39.38 | 2.53 | 0.995 |
| 243.80 | 100.00 | 7.58 | -0.51 | 8.09 | 7.58 | -2.35 | 48.95 | -39.18 | 2.52 | 0.995 |
| 243.90 | 100.00 | 7.58 | -0.55 | 8.13 | 7.58 | -2.32 | 49.11 | -38.97 | 2.51 | 0.995 |
| 244.00 | 100.00 | 7.58 | -0.58 | 8.16 | 7.58 | -2.30 | 49.26 | -38.76 | 2.50 | 0.995 |
| 244.10 | 100.00 | 7.58 | -0.62 | 8.20 | 7.58 | -2.27 | 49.41 | -38.56 | 2.49 | 0.995 |
| 244.20 | 100.00 | 7.58 | -0.66 | 8.24 | 7.58 | -2.24 | 49.56 | -38.35 | 2.48 | 0.995 |
| 244.30 | 100.00 | 7.58 | -0.70 | 8.28 | 7.58 | -2.21 | 49.71 | -38.15 | 2.47 | 0.995 |
| 244.40 | 100.00 | 7.58 | -0.74 | 8.32 | 7.58 | -2.18 | 49.87 | -37.94 | 2.46 | 0.995 |
| 244.50 | 100.00 | 7.58 | -0.78 | 8.36 | 7.58 | -2.15 | 50.02 | -37.74 | 2.45 | 0.995 |
| 244.60 | 100.00 | 7.58 | -0.82 | 8.40 | 7.58 | -2.12 | 50.17 | -37.54 | 2.44 | 0.996 |
| 244.70 | 100.00 | 7.58 | -0.85 | 8.43 | 7.58 | -2.09 | 50.32 | -37.34 | 2.43 | 0.996 |
| 244.80 | 100.00 | 7.58 | -0.89 | 8.47 | 7.58 | -2.07 | 50.47 | -37.14 | 2.42 | 0.996 |
| 244.90 | 100.00 | 7.58 | -0.93 | 8.51 | 7.58 | -2.04 | 50.62 | -36.94 | 2.41 | 0.996 |
| 245.00 | 100.00 | 7.58 | -0.97 | 8.55 | 7.58 | -2.01 | 50.77 | -36.74 | 2.40 | 0.996 |
| 245.10 | 100.00 | 7.58 | -1.01 | 8.59 | 7.58 | -1.99 | 50.92 | -36.54 | 2.39 | 0.996 |
| 245.20 | 100.00 | 7.58 | -1.05 | 8.63 | 7.58 | -1.96 | 51.07 | -36.34 | 2.38 | 0.996 |
| 245.30 | 100.00 | 7.58 | -1.08 | 8.66 | 7.58 | -1.93 | 51.22 | -36.15 | 2.37 | 0.996 |
| 245.40 | 100.00 | 7.58 | -1.12 | 8.70 | 7.58 | -1.90 | 51.37 | -35.95 | 2.36 | 0.996 |
| 245.50 | 100.00 | 7.58 | -1.16 | 8.74 | 7.58 | -1.88 | 51.52 | -35.76 | 2.35 | 0.996 |
| 245.60 | 100.00 | 7.58 | -1.20 | 8.78 | 7.58 | -1.85 | 51.67 | -35.56 | 2.34 | 0.996 |
| 245.70 | 100.00 | 7.58 | -1.24 | 8.82 | 7.58 | -1.83 | 51.82 | -35.37 | 2.33 | 0.996 |
| 245.80 | 100.00 | 7.58 | -1.27 | 8.85 | 7.58 | -1.80 | 51.97 | -35.18 | 2.32 | 0.996 |
| 245.90 | 100.00 | 7.58 | -1.31 | 8.89 | 7.58 | -1.78 | 52.12 | -34.98 | 2.31 | 0.996 |
| 246.00 | 100.00 | 7.58 | -1.35 | 8.93 | 7.58 | -1.75 | 52.26 | -34.79 | 2.30 | 0.996 |
| 246.10 | 100.00 | 7.58 | -1.39 | 8.97 | 7.58 | -1.72 | 52.41 | -34.60 | 2.29 | 0.996 |
| 246.20 | 100.00 | 7.58 | -1.42 | 9.00 | 7.58 | -1.70 | 52.56 | -34.41 | 2.28 | 0.996 |
| 246.30 | 100.00 | 7.58 | -1.46 | 9.04 | 7.58 | -1.68 | 52.71 | -34.22 | 2.27 | 0.996 |
| 246.40 | 100.00 | 7.58 | -1.50 | 9.08 | 7.58 | -1.65 | 52.85 | -34.03 | 2.26 | 0.996 |
| 246.50 | 100.00 | 7.58 | -1.54 | 9.12 | 7.58 | -1.63 | 53.00 | -33.84 | 2.26 | 0.996 |
| 246.60 | 100.00 | 7.58 | -1.57 | 9.15 | 7.58 | -1.60 | 53.14 | -33.65 | 2.25 | 0.996 |
| 246.70 | 100.00 | 7.58 | -1.61 | 9.19 | 7.58 | -1.58 | 53.29 | -33.47 | 2.24 | 0.996 |
| 246.80 | 100.00 | 7.57 | -1.65 | 9.22 | 7.57 | -1.55 | 53.44 | -33.28 | 2.23 | 0.996 |
| 246.90 | 100.00 | 7.57 | -1.68 | 9.25 | 7.57 | -1.53 | 53.58 | -33.09 | 2.22 | 0.996 |
| 247.00 | 100.00 | 7.57 | -1.72 | 9.29 | 7.57 | -1.51 | 53.73 | -32.91 | 2.21 | 0.996 |
| 247.10 | 100.00 | 7.57 | -1.76 | 9.33 | 7.57 | -1.48 | 53.87 | -32.72 | 2.21 | 0.996 |
| 247.20 | 100.00 | 7.57 | -1.79 | 9.36 | 7.57 | -1.46 | 54.02 | -32.54 | 2.20 | 0.996 |
| 247.30 | 100.00 | 7.57 | -1.83 | 9.40 | 7.57 | -1.44 | 54.16 | -32.36 | 2.19 | 0.996 |
| 247.40 | 100.00 | 7.57 | -1.87 | 9.44 | 7.57 | -1.42 | 54.30 | -32.17 | 2.18 | 0.996 |
| 247.50 | 100.00 | 7.57 | -1.90 | 9.47 | 7.57 | -1.39 | 54.45 | -31.99 | 2.17 | 0.996 |
| 247.60 | 100.00 | 7.57 | -1.94 | 9.51 | 7.57 | -1.37 | 54.59 | -31.81 | 2.17 | 0.996 |
| 247.70 | 100.00 | 7.57 | -1.97 | 9.54 | 7.57 | -1.35 | 54.73 | -31.63 | 2.16 | 0.996 |
| 247.80 | 100.00 | 7.56 | -2.01 | 9.57 | 7.56 | -1.33 | 54.87 | -31.45 | 2.15 | 0.996 |
| 247.90 | 100.00 | 7.56 | -2.05 | 9.61 | 7.56 | -1.30 | 55.01 | -31.27 | 2.14 | 0.996 |
| 248.00 | 100.00 | 7.56 | -2.08 | 9.64 | 7.56 | -1.28 | 55.16 | -31.09 | 2.13 | 0.996 |
| 248.10 | 100.00 | 7.56 | -2.12 | 9.68 | 7.56 | -1.26 | 55.30 | -30.91 | 2.13 | 0.996 |
| 248.20 | 100.00 | 7.56 | -2.15 | 9.71 | 7.56 | -1.24 | 55.44 | -30.73 | 2.12 | 0.996 |
| 248.30 | 100.00 | 7.56 | -2.19 | 9.75 | 7.56 | -1.22 | 55.58 | -30.55 | 2.11 | 0.996 |
| 248.40 | 100.00 | 7.56 | -2.22 | 9.78 | 7.56 | -1.20 | 55.72 | -30.37 | 2.11 | 0.996 |
| 248.50 | 100.00 | 7.56 | -2.26 | 9.82 | 7.56 | -1.18 | 55.86 | -30.20 | 2.10 | 0.996 |
| 248.60 | 100.00 | 7.56 | -2.29 | 9.85 | 7.56 | -1.16 | 56.00 | -30.02 | 2.09 | 0.996 |
| 248.70 | 100.00 | 7.55 | -2.33 | 9.88 | 7.55 | -1.14 | 56.14 | -29.85 | 2.08 | 0.996 |
| 248.80 | 100.00 | 7.55 | -2.36 | 9.91 | 7.55 | -1.11 | 56.28 | -29.67 | 2.08 | 0.996 |
| 248.90 | 100.00 | 7.55 | -2.40 | 9.95 | 7.55 | -1.09 | 56.41 | -29.50 | 2.07 | 0.996 |
| 249.00 | 100.00 | 7.55 | -2.43 | 9.98 | 7.55 | -1.07 | 56.55 | -29.32 | 2.06 | 0.996 |
| 249.10 | 100.00 | 7.55 | -2.46 | 10.01 | 7.55 | -1.05 | 56.69 | -29.15 | 2.06 | 0.996 |
| 249.20 | 100.00 | 7.55 | -2.50 | 10.05 | 7.55 | -1.04 | 56.83 | -28.97 | 2.05 | 0.996 |
| 249.30 | 100.00 | 7.55 | -2.53 | 10.08 | 7.55 | -1.02 | 56.96 | -28.80 | 2.04 | 0.996 |
| 249.40 | 100.00 | 7.54 | -2.57 | 10.11 | 7.54 | -1.00 | 57.10 | -28.63 | 2.04 | 0.996 |
| 249.50 | 100.00 | 7.54 | -2.60 | 10.14 | 7.54 | -0.98 | 57.23 | -28.46 | 2.03 | 0.996 |
| 249.60 | 100.00 | 7.54 | -2.63 | 10.17 | 7.54 | -0.96 | 57.37 | -28.28 | 2.02 | 0.996 |
| 249.70 | 100.00 | 7.54 | -2.67 | 10.21 | 7.54 | -0.94 | 57.51 | -28.11 | 2.02 | 0.996 |
| 249.80 | 100.00 | 7.54 | -2.70 | 10.24 | 7.54 | -0.92 | 57.64 | -27.94 | 2.01 | 0.996 |
| 249.90 | 100.00 | 7.54 | -2.73 | 10.27 | 7.54 | -0.90 | 57.77 | -27.77 | 2.00 | 0.996 |
| 250.00 | 100.00 | 7.54 | -2.76 | 10.30 | 7.54 | -0.88 | 57.91 | -27.60 | 2.00 | 0.996 |
| 250.10 | 100.00 | 7.53 | -2.80 | 10.33 | 7.53 | -0.86 | 58.04 | -27.43 | 1.99 | 0.996 |
| 250.20 | 100.00 | 7.53 | -2.83 | 10.36 | 7.53 | -0.85 | 58.18 | -27.26 | 1.98 | 0.996 |
| 250.30 | 100.00 | 7.53 | -2.86 | 10.39 | 7.53 | -0.83 | 58.31 | -27.09 | 1.98 | 0.996 |
| 250.40 | 100.00 | 7.53 | -2.89 | 10.42 | 7.53 | -0.81 | 58.44 | -26.92 | 1.97 | 0.996 |
| 250.50 | 100.00 | 7.53 | -2.92 | 10.45 | 7.53 | -0.79 | 58.57 | -26.76 | 1.96 | 0.996 |
| 250.60 | 100.00 | 7.53 | -2.96 | 10.49 | 7.53 | -0.77 | 58.70 | -26.59 | 1.96 | 0.996 |
| 250.70 | 100.00 | 7.53 | -2.99 | 10.52 | 7.53 | -0.76 | 58.84 | -26.42 | 1.95 | 0.996 |
| 250.80 | 100.00 | 7.52 | -3.02 | 10.54 | 7.52 | -0.74 | 58.97 | -26.25 | 1.95 | 0.996 |
| 250.90 | 100.00 | 7.52 | -3.05 | 10.57 | 7.52 | -0.72 | 59.10 | -26.09 | 1.94 | 0.996 |
| 251.00 | 100.00 | 7.52 | -3.08 | 10.60 | 7.52 | -0.70 | 59.23 | -25.92 | 1.93 | 0.996 |
| 251.10 | 100.00 | 7.52 | -3.11 | 10.63 | 7.52 | -0.69 | 59.36 | -25.76 | 1.93 | 0.996 |
| 251.20 | 100.00 | 7.52 | -3.14 | 10.66 | 7.52 | -0.67 | 59.49 | -25.59 | 1.92 | 0.996 |
| 251.30 | 100.00 | 7.52 | -3.17 | 10.69 | 7.52 | -0.65 | 59.62 | -25.43 | 1.92 | 0.996 |
| 251.40 | 100.00 | 7.51 | -3.20 | 10.71 | 7.51 | -0.64 | 59.74 | -25.26 | 1.91 | 0.996 |
| 251.50 | 100.00 | 7.51 | -3.23 | 10.74 | 7.51 | -0.62 | 59.87 | -25.10 | 1.91 | 0.996 |
| 251.60 | 100.00 | 7.51 | -3.26 | 10.77 | 7.51 | -0.60 | 60.00 | -24.93 | 1.90 | 0.996 |
| 251.70 | 100.00 | 7.51 | -3.29 | 10.80 | 7.51 | -0.59 | 60.13 | -24.77 | 1.89 | 0.996 |
| 251.80 | 100.00 | 7.51 | -3.32 | 10.83 | 7.51 | -0.57 | 60.26 | -24.61 | 1.89 | 0.996 |
| 251.90 | 100.00 | 7.51 | -3.35 | 10.86 | 7.51 | -0.56 | 60.38 | -24.44 | 1.88 | 0.996 |
| 252.00 | 100.00 | 7.50 | -3.38 | 10.88 | 7.50 | -0.54 | 60.51 | -24.28 | 1.88 | 0.996 |
| 252.10 | 100.00 | 7.50 | -3.41 | 10.91 | 7.50 | -0.52 | 60.64 | -24.12 | 1.87 | 0.996 |
| 252.20 | 100.00 | 7.50 | -3.43 | 10.93 | 7.50 | -0.51 | 60.76 | -23.95 | 1.87 | 0.996 |
| 252.30 | 100.00 | 7.50 | -3.46 | 10.96 | 7.50 | -0.49 | 60.89 | -23.79 | 1.86 | 0.996 |
| 252.40 | 100.00 | 7.50 | -3.49 | 10.99 | 7.50 | -0.48 | 61.01 | -23.63 | 1.86 | 0.996 |
| 252.50 | 100.00 | 7.50 | -3.52 | 11.02 | 7.50 | -0.46 | 61.14 | -23.47 | 1.85 | 0.996 |
| 252.60 | 100.00 | 7.49 | -3.54 | 11.03 | 7.49 | -0.45 | 61.26 | -23.31 | 1.85 | 0.996 |
| 252.70 | 100.00 | 7.49 | -3.57 | 11.06 | 7.49 | -0.43 | 61.38 | -23.15 | 1.84 | 0.996 |
| 252.80 | 100.00 | 7.49 | -3.60 | 11.09 | 7.49 | -0.42 | 61.51 | -22.99 | 1.84 | 0.996 |
| 252.90 | 100.00 | 7.49 | -3.63 | 11.12 | 7.49 | -0.40 | 61.63 | -22.83 | 1.83 | 0.996 |
| 253.00 | 100.00 | 7.49 | -3.65 | 11.14 | 7.49 | -0.39 | 61.75 | -22.67 | 1.83 | 0.996 |
| 253.10 | 100.00 | 7.49 | -3.68 | 11.17 | 7.49 | -0.37 | 61.88 | -22.51 | 1.82 | 0.996 |
| 253.20 | 100.00 | 7.48 | -3.70 | 11.18 | 7.48 | -0.36 | 62.00 | -22.35 | 1.82 | 0.996 |
| 253.30 | 100.00 | 7.48 | -3.73 | 11.21 | 7.48 | -0.34 | 62.12 | -22.19 | 1.81 | 0.996 |
| 253.40 | 100.00 | 7.48 | -3.76 | 11.24 | 7.48 | -0.33 | 62.24 | -22.03 | 1.81 | 0.996 |
| 253.50 | 100.00 | 7.48 | -3.78 | 11.26 | 7.48 | -0.31 | 62.36 | -21.87 | 1.80 | 0.996 |
| 253.60 | 100.00 | 7.48 | -3.81 | 11.29 | 7.48 | -0.30 | 62.48 | -21.71 | 1.80 | 0.996 |
| 253.70 | 100.00 | 7.48 | -3.83 | 11.31 | 7.48 | -0.29 | 62.60 | -21.55 | 1.79 | 0.996 |
| 253.80 | 100.00 | 7.47 | -3.86 | 11.33 | 7.47 | -0.27 | 62.72 | -21.39 | 1.79 | 0.996 |
| 253.90 | 100.00 | 7.47 | -3.88 | 11.35 | 7.47 | -0.26 | 62.84 | -21.24 | 1.78 | 0.996 |
| 254.00 | 100.00 | 7.47 | -3.90 | 11.37 | 7.47 | -0.24 | 62.96 | -21.08 | 1.78 | 0.996 |
| 254.10 | 100.00 | 7.47 | -3.93 | 11.40 | 7.47 | -0.23 | 63.08 | -20.92 | 1.77 | 0.996 |
| 254.20 | 100.00 | 7.47 | -3.95 | 11.42 | 7.47 | -0.22 | 63.20 | -20.76 | 1.77 | 0.996 |
| 254.30 | 100.00 | 7.47 | -3.98 | 11.45 | 7.47 | -0.20 | 63.31 | -20.61 | 1.76 | 0.996 |
| 254.40 | 100.00 | 7.46 | -4.00 | 11.46 | 7.46 | -0.19 | 63.43 | -20.45 | 1.76 | 0.996 |
| 254.50 | 100.00 | 7.46 | -4.02 | 11.48 | 7.46 | -0.18 | 63.55 | -20.29 | 1.76 | 0.996 |
| 254.60 | 100.00 | 7.46 | -4.04 | 11.50 | 7.46 | -0.16 | 63.67 | -20.14 | 1.75 | 0.996 |
| 254.70 | 100.00 | 7.46 | -4.07 | 11.53 | 7.46 | -0.15 | 63.78 | -19.98 | 1.75 | 0.996 |
| 254.80 | 100.00 | 7.46 | -4.09 | 11.55 | 7.46 | -0.14 | 63.90 | -19.82 | 1.74 | 0.996 |
| 254.90 | 100.00 | 7.46 | -4.11 | 11.57 | 7.46 | -0.12 | 64.02 | -19.67 | 1.74 | 0.996 |
| 255.00 | 100.00 | 7.45 | -4.13 | 11.58 | 7.45 | -0.11 | 64.13 | -19.51 | 1.73 | 0.996 |
| 255.10 | 100.00 | 7.45 | -4.15 | 11.60 | 7.45 | -0.10 | 64.25 | -19.36 | 1.73 | 0.996 |
| 255.20 | 100.00 | 7.45 | -4.18 | 11.63 | 7.45 | -0.09 | 64.36 | -19.20 | 1.73 | 0.996 |
| 255.30 | 100.00 | 7.45 | -4.20 | 11.65 | 7.45 | -0.07 | 64.48 | -19.04 | 1.72 | 0.996 |
| 255.40 | 100.00 | 7.45 | -4.22 | 11.67 | 7.45 | -0.06 | 64.59 | -18.89 | 1.72 | 0.996 |
| 255.50 | 100.00 | 7.45 | -4.24 | 11.69 | 7.45 | -0.05 | 64.70 | -18.73 | 1.71 | 0.996 |
| 255.60 | 100.00 | 7.44 | -4.26 | 11.70 | 7.44 | -0.04 | 64.82 | -18.58 | 1.71 | 0.996 |
| 255.70 | 100.00 | 7.44 | -4.28 | 11.72 | 7.44 | -0.02 | 64.93 | -18.43 | 1.70 | 0.996 |
| 255.80 | 100.00 | 7.44 | -4.30 | 11.74 | 7.44 | -0.01 | 65.04 | -18.27 | 1.70 | 0.996 |
| 255.90 | 100.00 | 7.44 | -4.32 | 11.76 | 7.44 | 0.00 | 65.16 | -18.12 | 1.70 | 0.996 |
| 256.00 | 100.00 | 7.44 | -4.34 | 11.78 | 7.44 | 0.01 | 65.27 | -17.96 | 1.69 | 0.996 |
| 256.10 | 100.00 | 7.44 | -4.35 | 11.79 | 7.44 | 0.02 | 65.38 | -17.81 | 1.69 | 0.996 |
| 256.20 | 100.00 | 7.44 | -4.37 | 11.81 | 7.44 | 0.04 | 65.49 | -17.65 | 1.68 | 0.996 |
| 256.30 | 100.00 | 7.43 | -4.39 | 11.82 | 7.43 | 0.05 | 65.61 | -17.50 | 1.68 | 0.996 |
| 256.40 | 100.00 | 7.43 | -4.41 | 11.84 | 7.43 | 0.06 | 65.72 | -17.35 | 1.68 | 0.996 |
| 256.50 | 100.00 | 7.43 | -4.43 | 11.86 | 7.43 | 0.07 | 65.83 | -17.19 | 1.67 | 0.996 |
| 256.60 | 100.00 | 7.43 | -4.44 | 11.87 | 7.43 | 0.08 | 65.94 | -17.04 | 1.67 | 0.996 |
| 256.70 | 100.00 | 7.43 | -4.46 | 11.89 | 7.43 | 0.09 | 66.05 | -16.89 | 1.66 | 0.996 |
| 256.80 | 100.00 | 7.43 | -4.48 | 11.91 | 7.43 | 0.11 | 66.16 | -16.73 | 1.66 | 0.996 |
| 256.90 | 100.00 | 7.42 | -4.50 | 11.92 | 7.42 | 0.12 | 66.27 | -16.58 | 1.66 | 0.996 |
| 257.00 | 100.00 | 7.42 | -4.51 | 11.93 | 7.42 | 0.13 | 66.38 | -16.43 | 1.65 | 0.996 |
| 257.10 | 100.00 | 7.42 | -4.53 | 11.95 | 7.42 | 0.14 | 66.49 | -16.27 | 1.65 | 0.996 |
| 257.20 | 100.00 | 7.42 | -4.54 | 11.96 | 7.42 | 0.15 | 66.60 | -16.12 | 1.65 | 0.996 |
| 257.30 | 100.00 | 7.42 | -4.56 | 11.98 | 7.42 | 0.16 | 66.70 | -15.97 | 1.64 | 0.996 |
| 257.40 | 100.00 | 7.42 | -4.58 | 12.00 | 7.42 | 0.17 | 66.81 | -15.81 | 1.64 | 0.996 |
| 257.50 | 100.00 | 7.42 | -4.59 | 12.01 | 7.42 | 0.18 | 66.92 | -15.66 | 1.64 | 0.996 |
| 257.60 | 100.00 | 7.41 | -4.61 | 12.02 | 7.41 | 0.19 | 67.03 | -15.51 | 1.63 | 0.996 |
| 257.70 | 100.00 | 7.41 | -4.62 | 12.03 | 7.41 | 0.20 | 67.14 | -15.36 | 1.63 | 0.996 |
| 257.80 | 100.00 | 7.41 | -4.64 | 12.05 | 7.41 | 0.21 | 67.24 | -15.20 | 1.62 | 0.996 |
| 257.90 | 100.00 | 7.41 | -4.65 | 12.06 | 7.41 | 0.23 | 67.35 | -15.05 | 1.62 | 0.996 |
| 258.00 | 100.00 | 7.41 | -4.66 | 12.07 | 7.41 | 0.24 | 67.46 | -14.90 | 1.62 | 0.996 |
| 258.10 | 100.00 | 7.41 | -4.68 | 12.09 | 7.41 | 0.25 | 67.56 | -14.75 | 1.61 | 0.996 |
| 258.20 | 100.00 | 7.41 | -4.69 | 12.10 | 7.41 | 0.26 | 67.67 | -14.60 | 1.61 | 0.996 |
| 258.30 | 100.00 | 7.40 | -4.70 | 12.10 | 7.40 | 0.27 | 67.78 | -14.44 | 1.61 | 0.996 |
| 258.40 | 100.00 | 7.40 | -4.72 | 12.12 | 7.40 | 0.28 | 67.88 | -14.29 | 1.60 | 0.996 |
| 258.50 | 100.00 | 7.40 | -4.73 | 12.13 | 7.40 | 0.29 | 67.99 | -14.14 | 1.60 | 0.996 |
| 258.60 | 100.00 | 7.40 | -4.74 | 12.14 | 7.40 | 0.30 | 68.09 | -13.99 | 1.60 | 0.996 |
| 258.70 | 100.00 | 7.40 | -4.75 | 12.15 | 7.40 | 0.31 | 68.20 | -13.84 | 1.59 | 0.996 |
| 258.80 | 100.00 | 7.40 | -4.77 | 12.17 | 7.40 | 0.32 | 68.30 | -13.69 | 1.59 | 0.996 |
| 258.90 | 100.00 | 7.40 | -4.78 | 12.18 | 7.40 | 0.33 | 68.40 | -13.54 | 1.59 | 0.996 |
| 259.00 | 100.00 | 7.39 | -4.79 | 12.18 | 7.39 | 0.34 | 68.51 | -13.38 | 1.58 | 0.996 |
| 259.10 | 100.00 | 7.39 | -4.80 | 12.19 | 7.39 | 0.35 | 68.61 | -13.23 | 1.58 | 0.996 |
| 259.20 | 100.00 | 7.39 | -4.81 | 12.20 | 7.39 | 0.36 | 68.72 | -13.08 | 1.58 | 0.996 |
| 259.30 | 100.00 | 7.39 | -4.82 | 12.21 | 7.39 | 0.37 | 68.82 | -12.93 | 1.57 | 0.996 |
| 259.40 | 100.00 | 7.39 | -4.83 | 12.22 | 7.39 | 0.38 | 68.92 | -12.78 | 1.57 | 0.996 |
| 259.50 | 100.00 | 7.39 | -4.84 | 12.23 | 7.39 | 0.39 | 69.03 | -12.63 | 1.57 | 0.996 |
| 259.60 | 100.00 | 7.39 | -4.85 | 12.24 | 7.39 | 0.39 | 69.13 | -12.48 | 1.56 | 0.996 |
| 259.70 | 100.00 | 7.38 | -4.86 | 12.24 | 7.38 | 0.40 | 69.23 | -12.33 | 1.56 | 0.996 |
| 259.80 | 100.00 | 7.38 | -4.87 | 12.25 | 7.38 | 0.41 | 69.33 | -12.18 | 1.56 | 0.996 |
| 259.90 | 100.00 | 7.38 | -4.88 | 12.26 | 7.38 | 0.42 | 69.44 | -12.02 | 1.56 | 0.996 |
| 260.00 | 100.00 | 7.38 | -4.89 | 12.27 | 7.38 | 0.43 | 69.54 | -11.87 | 1.55 | 0.996 |
| 260.10 | 100.00 | 7.38 | -4.90 | 12.28 | 7.38 | 0.44 | 69.64 | -11.72 | 1.55 | 0.996 |
| 260.20 | 100.00 | 7.38 | -4.91 | 12.29 | 7.38 | 0.45 | 69.74 | -11.57 | 1.55 | 0.996 |
| 260.30 | 100.00 | 7.38 | -4.91 | 12.29 | 7.38 | 0.46 | 69.84 | -11.42 | 1.54 | 0.996 |
| 260.40 | 100.00 | 7.38 | -4.92 | 12.30 | 7.38 | 0.47 | 69.94 | -11.27 | 1.54 | 0.996 |
| 260.50 | 100.00 | 7.37 | -4.93 | 12.30 | 7.37 | 0.48 | 70.04 | -11.12 | 1.54 | 0.996 |
| 260.60 | 100.00 | 7.37 | -4.94 | 12.31 | 7.37 | 0.49 | 70.14 | -10.97 | 1.53 | 0.996 |
| 260.70 | 100.00 | 7.37 | -4.95 | 12.32 | 7.37 | 0.50 | 70.24 | -10.82 | 1.53 | 0.996 |
| 260.80 | 100.00 | 7.37 | -4.95 | 12.32 | 7.37 | 0.50 | 70.34 | -10.67 | 1.53 | 0.996 |
| 260.90 | 100.00 | 7.37 | -4.96 | 12.33 | 7.37 | 0.51 | 70.44 | -10.52 | 1.53 | 0.996 |
| 261.00 | 100.00 | 7.37 | -4.97 | 12.34 | 7.37 | 0.52 | 70.54 | -10.37 | 1.52 | 0.996 |
| 261.10 | 100.00 | 7.37 | -4.97 | 12.34 | 7.37 | 0.53 | 70.64 | -10.22 | 1.52 | 0.996 |
| 261.20 | 100.00 | 7.37 | -4.98 | 12.35 | 7.37 | 0.54 | 70.74 | -10.07 | 1.52 | 0.996 |
| 261.30 | 100.00 | 7.37 | -4.98 | 12.35 | 7.37 | 0.55 | 70.84 | -9.92 | 1.51 | 0.996 |
| 261.40 | 100.00 | 7.36 | -4.99 | 12.35 | 7.36 | 0.56 | 70.94 | -9.77 | 1.51 | 0.996 |
| 261.50 | 100.00 | 7.36 | -5.00 | 12.36 | 7.36 | 0.56 | 71.04 | -9.62 | 1.51 | 0.996 |
| 261.60 | 100.00 | 7.36 | -5.00 | 12.36 | 7.36 | 0.57 | 71.14 | -9.47 | 1.51 | 0.996 |
| 261.70 | 100.00 | 7.36 | -5.01 | 12.37 | 7.36 | 0.58 | 71.24 | -9.32 | 1.50 | 0.996 |
| 261.80 | 100.00 | 7.36 | -5.01 | 12.37 | 7.36 | 0.59 | 71.33 | -9.17 | 1.50 | 0.996 |
| 261.90 | 100.00 | 7.36 | -5.01 | 12.37 | 7.36 | 0.60 | 71.43 | -9.02 | 1.50 | 0.996 |
| 262.00 | 100.00 | 7.36 | -5.02 | 12.38 | 7.36 | 0.61 | 71.53 | -8.87 | 1.50 | 0.996 |
| 262.10 | 100.00 | 7.36 | -5.02 | 12.38 | 7.36 | 0.61 | 71.63 | -8.72 | 1.49 | 0.996 |
| 262.20 | 100.00 | 7.36 | -5.03 | 12.39 | 7.36 | 0.62 | 71.73 | -8.57 | 1.49 | 0.996 |
| 262.30 | 100.00 | 7.36 | -5.03 | 12.39 | 7.36 | 0.63 | 71.82 | -8.42 | 1.49 | 0.996 |
| 262.40 | 100.00 | 7.35 | -5.03 | 12.38 | 7.35 | 0.64 | 71.92 | -8.27 | 1.49 | 0.996 |
| 262.50 | 100.00 | 7.35 | -5.04 | 12.39 | 7.35 | 0.65 | 72.02 | -8.12 | 1.48 | 0.996 |
| 262.60 | 100.00 | 7.35 | -5.04 | 12.39 | 7.35 | 0.65 | 72.11 | -7.97 | 1.48 | 0.996 |
| 262.70 | 100.00 | 7.35 | -5.04 | 12.39 | 7.35 | 0.66 | 72.21 | -7.82 | 1.48 | 0.996 |
| 262.80 | 100.00 | 7.35 | -5.05 | 12.40 | 7.35 | 0.67 | 72.31 | -7.67 | 1.48 | 0.996 |
| 262.90 | 100.00 | 7.35 | -5.05 | 12.40 | 7.35 | 0.68 | 72.40 | -7.52 | 1.47 | 0.996 |
| 263.00 | 100.00 | 7.35 | -5.05 | 12.40 | 7.35 | 0.69 | 72.50 | -7.37 | 1.47 | 0.996 |
| 263.10 | 100.00 | 7.35 | -5.05 | 12.40 | 7.35 | 0.69 | 72.60 | -7.22 | 1.47 | 0.996 |
| 263.20 | 100.00 | 7.35 | -5.05 | 12.40 | 7.35 | 0.70 | 72.69 | -7.07 | 1.47 | 0.996 |
| 263.30 | 100.00 | 7.35 | -5.06 | 12.41 | 7.35 | 0.71 | 72.79 | -6.92 | 1.46 | 0.996 |
| 263.40 | 100.00 | 7.35 | -5.06 | 12.41 | 7.35 | 0.72 | 72.88 | -6.77 | 1.46 | 0.996 |
| 263.50 | 100.00 | 7.34 | -5.06 | 12.40 | 7.34 | 0.72 | 72.98 | -6.62 | 1.46 | 0.996 |
| 263.60 | 100.00 | 7.34 | -5.06 | 12.40 | 7.34 | 0.73 | 73.07 | -6.47 | 1.46 | 0.996 |
| 263.70 | 100.00 | 7.34 | -5.06 | 12.40 | 7.34 | 0.74 | 73.17 | -6.32 | 1.45 | 0.996 |
| 263.80 | 100.00 | 7.34 | -5.06 | 12.40 | 7.34 | 0.75 | 73.27 | -6.17 | 1.45 | 0.996 |
| 263.90 | 100.00 | 7.34 | -5.06 | 12.40 | 7.34 | 0.75 | 73.36 | -6.02 | 1.45 | 0.996 |
| 264.00 | 100.00 | 7.34 | -5.06 | 12.40 | 7.34 | 0.76 | 73.45 | -5.87 | 1.45 | 0.996 |
| 264.10 | 100.00 | 7.34 | -5.06 | 12.40 | 7.34 | 0.77 | 73.55 | -5.72 | 1.44 | 0.996 |
| 264.20 | 100.00 | 7.34 | -5.06 | 12.40 | 7.34 | 0.78 | 73.64 | -5.58 | 1.44 | 0.996 |
| 264.30 | 100.00 | 7.34 | -5.06 | 12.40 | 7.34 | 0.78 | 73.74 | -5.43 | 1.44 | 0.996 |
| 264.40 | 100.00 | 7.34 | -5.06 | 12.40 | 7.34 | 0.79 | 73.83 | -5.28 | 1.44 | 0.996 |
| 264.50 | 100.00 | 7.34 | -5.06 | 12.40 | 7.34 | 0.80 | 73.93 | -5.13 | 1.44 | 0.996 |
| 264.60 | 100.00 | 7.34 | -5.06 | 12.40 | 7.34 | 0.80 | 74.02 | -4.98 | 1.43 | 0.996 |
| 264.70 | 100.00 | 7.33 | -5.06 | 12.39 | 7.33 | 0.81 | 74.12 | -4.83 | 1.43 | 0.996 |
| 264.80 | 100.00 | 7.33 | -5.06 | 12.39 | 7.33 | 0.82 | 74.21 | -4.68 | 1.43 | 0.996 |
| 264.90 | 100.00 | 7.33 | -5.06 | 12.39 | 7.33 | 0.83 | 74.30 | -4.53 | 1.43 | 0.996 |
| 265.00 | 100.00 | 7.33 | -5.06 | 12.39 | 7.33 | 0.83 | 74.40 | -4.38 | 1.42 | 0.996 |
| 265.10 | 100.00 | 7.33 | -5.06 | 12.39 | 7.33 | 0.84 | 74.49 | -4.23 | 1.42 | 0.996 |
| 265.20 | 100.00 | 7.33 | -5.05 | 12.38 | 7.33 | 0.85 | 74.58 | -4.08 | 1.42 | 0.996 |
| 265.30 | 100.00 | 7.33 | -5.05 | 12.38 | 7.33 | 0.85 | 74.68 | -3.93 | 1.42 | 0.996 |
| 265.40 | 100.00 | 7.33 | -5.05 | 12.38 | 7.33 | 0.86 | 74.77 | -3.78 | 1.42 | 0.996 |
| 265.50 | 100.00 | 7.33 | -5.05 | 12.38 | 7.33 | 0.87 | 74.86 | -3.63 | 1.41 | 0.996 |
| 265.60 | 100.00 | 7.33 | -5.05 | 12.38 | 7.33 | 0.87 | 74.96 | -3.48 | 1.41 | 0.996 |
| 265.70 | 100.00 | 7.33 | -5.04 | 12.37 | 7.33 | 0.88 | 75.05 | -3.34 | 1.41 | 0.996 |
| 265.80 | 100.00 | 7.33 | -5.04 | 12.37 | 7.33 | 0.89 | 75.14 | -3.19 | 1.41 | 0.996 |
| 265.90 | 100.00 | 7.33 | -5.04 | 12.37 | 7.33 | 0.89 | 75.23 | -3.04 | 1.41 | 0.996 |
| 266.00 | 100.00 | 7.33 | -5.04 | 12.37 | 7.33 | 0.90 | 75.33 | -2.89 | 1.40 | 0.996 |
| 266.10 | 100.00 | 7.33 | -5.03 | 12.36 | 7.33 | 0.91 | 75.42 | -2.74 | 1.40 | 0.996 |
| 266.20 | 100.00 | 7.33 | -5.03 | 12.36 | 7.33 | 0.91 | 75.51 | -2.59 | 1.40 | 0.996 |
| 266.30 | 100.00 | 7.33 | -5.03 | 12.36 | 7.33 | 0.92 | 75.60 | -2.44 | 1.40 | 0.996 |
| 266.40 | 100.00 | 7.32 | -5.02 | 12.34 | 7.32 | 0.93 | 75.70 | -2.29 | 1.40 | 0.996 |
| 266.50 | 100.00 | 7.32 | -5.02 | 12.34 | 7.32 | 0.93 | 75.79 | -2.14 | 1.40 | 0.996 |
| 266.60 | 100.00 | 7.32 | -5.02 | 12.34 | 7.32 | 0.94 | 75.88 | -1.99 | 1.39 | 0.996 |
| 266.70 | 100.00 | 7.32 | -5.01 | 12.33 | 7.32 | 0.95 | 75.97 | -1.84 | 1.39 | 0.996 |
| 266.80 | 100.00 | 7.32 | -5.01 | 12.33 | 7.32 | 0.95 | 76.06 | -1.70 | 1.39 | 0.996 |
| 266.90 | 100.00 | 7.32 | -5.00 | 12.32 | 7.32 | 0.96 | 76.16 | -1.55 | 1.39 | 0.996 |
| 267.00 | 100.00 | 7.32 | -5.00 | 12.32 | 7.32 | 0.96 | 76.25 | -1.40 | 1.39 | 0.996 |
| 267.10 | 100.00 | 7.32 | -5.00 | 12.32 | 7.32 | 0.97 | 76.34 | -1.25 | 1.38 | 0.996 |
| 267.20 | 100.00 | 7.32 | -4.99 | 12.31 | 7.32 | 0.98 | 76.43 | -1.10 | 1.38 | 0.996 |
| 267.30 | 100.00 | 7.32 | -4.99 | 12.31 | 7.32 | 0.98 | 76.52 | -0.95 | 1.38 | 0.996 |
| 267.40 | 100.00 | 7.32 | -4.98 | 12.30 | 7.32 | 0.99 | 76.61 | -0.80 | 1.38 | 0.996 |
| 267.50 | 100.00 | 7.32 | -4.98 | 12.30 | 7.32 | 1.00 | 76.71 | -0.65 | 1.38 | 0.996 |
| 267.60 | 100.00 | 7.32 | -4.97 | 12.29 | 7.32 | 1.00 | 76.80 | -0.50 | 1.38 | 0.996 |
| 267.70 | 100.00 | 7.32 | -4.97 | 12.29 | 7.32 | 1.01 | 76.89 | -0.36 | 1.37 | 0.996 |
| 267.80 | 100.00 | 7.32 | -4.96 | 12.28 | 7.32 | 1.01 | 76.98 | -0.21 | 1.37 | 0.996 |
| 267.90 | 100.00 | 7.32 | -4.96 | 12.28 | 7.32 | 1.02 | 77.07 | -0.06 | 1.37 | 0.996 |
| 268.00 | 100.00 | 7.32 | -4.95 | 12.27 | 7.32 | 1.03 | 77.16 | 0.09 | 1.37 | 0.996 |
| 268.10 | 100.00 | 7.32 | -4.95 | 12.27 | 7.32 | 1.03 | 77.25 | 0.24 | 1.37 | 0.996 |
| 268.20 | 100.00 | 7.32 | -4.94 | 12.26 | 7.32 | 1.04 | 77.34 | 0.39 | 1.37 | 0.996 |
| 268.30 | 100.00 | 7.32 | -4.93 | 12.25 | 7.32 | 1.04 | 77.43 | 0.54 | 1.36 | 0.996 |
| 268.40 | 100.00 | 7.32 | -4.93 | 12.25 | 7.32 | 1.05 | 77.52 | 0.68 | 1.36 | 0.996 |
| 268.50 | 100.00 | 7.32 | -4.92 | 12.24 | 7.32 | 1.06 | 77.62 | 0.83 | 1.36 | 0.996 |
| 268.60 | 100.00 | 7.32 | -4.92 | 12.24 | 7.32 | 1.06 | 77.71 | 0.98 | 1.36 | 0.996 |
| 268.70 | 100.00 | 7.32 | -4.91 | 12.23 | 7.32 | 1.07 | 77.80 | 1.13 | 1.36 | 0.996 |
| 268.80 | 100.00 | 7.32 | -4.90 | 12.22 | 7.32 | 1.07 | 77.89 | 1.28 | 1.36 | 0.996 |
| 268.90 | 100.00 | 7.32 | -4.90 | 12.22 | 7.32 | 1.08 | 77.98 | 1.43 | 1.36 | 0.996 |
| 269.00 | 100.00 | 7.32 | -4.89 | 12.21 | 7.32 | 1.08 | 78.07 | 1.58 | 1.35 | 0.996 |
| 269.10 | 100.00 | 7.32 | -4.88 | 12.20 | 7.32 | 1.09 | 78.16 | 1.72 | 1.35 | 0.996 |
| 269.20 | 100.00 | 7.32 | -4.88 | 12.20 | 7.32 | 1.10 | 78.25 | 1.87 | 1.35 | 0.996 |
| 269.30 | 100.00 | 7.32 | -4.87 | 12.19 | 7.32 | 1.10 | 78.34 | 2.02 | 1.35 | 0.996 |
| 269.40 | 100.00 | 7.32 | -4.86 | 12.18 | 7.32 | 1.11 | 78.43 | 2.17 | 1.35 | 0.996 |
| 269.50 | 100.00 | 7.32 | -4.86 | 12.18 | 7.32 | 1.11 | 78.52 | 2.32 | 1.35 | 0.996 |
| 269.60 | 100.00 | 7.32 | -4.85 | 12.17 | 7.32 | 1.12 | 78.61 | 2.47 | 1.35 | 0.996 |
| 269.70 | 100.00 | 7.32 | -4.84 | 12.16 | 7.32 | 1.12 | 78.70 | 2.61 | 1.34 | 0.996 |
| 269.80 | 100.00 | 7.32 | -4.84 | 12.16 | 7.32 | 1.13 | 78.79 | 2.76 | 1.34 | 0.996 |
| 269.90 | 100.00 | 7.32 | -4.83 | 12.15 | 7.32 | 1.14 | 78.88 | 2.91 | 1.34 | 0.996 |
| 270.00 | 100.00 | 7.32 | -4.82 | 12.14 | 7.32 | 1.14 | 78.97 | 3.06 | 1.34 | 0.996 |
| 270.10 | 100.00 | 7.32 | -4.82 | 12.14 | 7.32 | 1.15 | 79.06 | 3.21 | 1.34 | 0.996 |
| 270.20 | 100.00 | 7.32 | -4.81 | 12.13 | 7.32 | 1.15 | 79.15 | 3.35 | 1.34 | 0.996 |
| 270.30 | 100.00 | 7.32 | -4.80 | 12.12 | 7.32 | 1.16 | 79.24 | 3.50 | 1.34 | 0.996 |
| 270.40 | 100.00 | 7.32 | -4.79 | 12.11 | 7.32 | 1.16 | 79.33 | 3.65 | 1.34 | 0.996 |
| 270.50 | 100.00 | 7.32 | -4.79 | 12.11 | 7.32 | 1.17 | 79.42 | 3.80 | 1.33 | 0.996 |
| 270.60 | 100.00 | 7.32 | -4.78 | 12.10 | 7.32 | 1.17 | 79.51 | 3.95 | 1.33 | 0.996 |
| 270.70 | 100.00 | 7.32 | -4.77 | 12.09 | 7.32 | 1.18 | 79.60 | 4.09 | 1.33 | 0.996 |
| 270.80 | 100.00 | 7.32 | -4.76 | 12.08 | 7.32 | 1.18 | 79.69 | 4.24 | 1.33 | 0.996 |
| 270.90 | 100.00 | 7.32 | -4.75 | 12.07 | 7.32 | 1.19 | 79.78 | 4.39 | 1.33 | 0.996 |
| 271.00 | 100.00 | 7.32 | -4.75 | 12.07 | 7.32 | 1.19 | 79.87 | 4.54 | 1.33 | 0.996 |
| 271.10 | 100.00 | 7.32 | -4.74 | 12.06 | 7.32 | 1.20 | 79.96 | 4.69 | 1.33 | 0.996 |
| 271.20 | 100.00 | 7.32 | -4.73 | 12.05 | 7.32 | 1.21 | 80.05 | 4.83 | 1.33 | 0.996 |
| 271.30 | 100.00 | 7.32 | -4.72 | 12.04 | 7.32 | 1.21 | 80.14 | 4.98 | 1.33 | 0.996 |
| 271.40 | 100.00 | 7.32 | -4.71 | 12.03 | 7.32 | 1.22 | 80.23 | 5.13 | 1.32 | 0.996 |
| 271.50 | 100.00 | 7.32 | -4.71 | 12.03 | 7.32 | 1.22 | 80.32 | 5.28 | 1.32 | 0.996 |
| 271.60 | 100.00 | 7.32 | -4.70 | 12.02 | 7.32 | 1.23 | 80.41 | 5.42 | 1.32 | 0.996 |
| 271.70 | 100.00 | 7.32 | -4.69 | 12.01 | 7.32 | 1.23 | 80.50 | 5.57 | 1.32 | 0.996 |
| 271.80 | 100.00 | 7.32 | -4.68 | 12.00 | 7.32 | 1.24 | 80.59 | 5.72 | 1.32 | 0.996 |
| 271.90 | 100.00 | 7.32 | -4.67 | 11.99 | 7.32 | 1.24 | 80.68 | 5.87 | 1.32 | 0.996 |
| 272.00 | 100.00 | 7.32 | -4.66 | 11.98 | 7.32 | 1.25 | 80.77 | 6.01 | 1.32 | 0.996 |
| 272.10 | 100.00 | 7.32 | -4.65 | 11.97 | 7.32 | 1.25 | 80.86 | 6.16 | 1.32 | 0.996 |
| 272.20 | 100.00 | 7.32 | -4.65 | 11.97 | 7.32 | 1.26 | 80.95 | 6.31 | 1.32 | 0.996 |
| 272.30 | 100.00 | 7.32 | -4.64 | 11.96 | 7.32 | 1.26 | 81.04 | 6.45 | 1.32 | 0.996 |
| 272.40 | 100.00 | 7.32 | -4.63 | 11.95 | 7.32 | 1.27 | 81.13 | 6.60 | 1.31 | 0.996 |
| 272.50 | 100.00 | 7.32 | -4.62 | 11.94 | 7.32 | 1.27 | 81.22 | 6.75 | 1.31 | 0.996 |
| 272.60 | 100.00 | 7.32 | -4.61 | 11.93 | 7.32 | 1.28 | 81.30 | 6.90 | 1.31 | 0.996 |
| 272.70 | 100.00 | 7.32 | -4.60 | 11.92 | 7.32 | 1.28 | 81.39 | 7.04 | 1.31 | 0.996 |
| 272.80 | 100.00 | 7.32 | -4.59 | 11.91 | 7.32 | 1.29 | 81.48 | 7.19 | 1.31 | 0.996 |
| 272.90 | 100.00 | 7.32 | -4.58 | 11.90 | 7.32 | 1.29 | 81.57 | 7.34 | 1.31 | 0.996 |
| 273.00 | 100.00 | 7.32 | -4.57 | 11.89 | 7.32 | 1.30 | 81.66 | 7.48 | 1.31 | 0.996 |
| 273.10 | 100.00 | 7.32 | -4.57 | 11.89 | 7.32 | 1.30 | 81.75 | 7.63 | 1.31 | 0.996 |
| 273.20 | 100.00 | 7.32 | -4.56 | 11.88 | 7.32 | 1.30 | 81.84 | 7.78 | 1.31 | 0.996 |
| 273.30 | 100.00 | 7.32 | -4.55 | 11.87 | 7.32 | 1.31 | 81.93 | 7.92 | 1.31 | 0.996 |
| 273.40 | 100.00 | 7.32 | -4.54 | 11.86 | 7.32 | 1.31 | 82.02 | 8.07 | 1.31 | 0.996 |
| 273.50 | 100.00 | 7.32 | -4.53 | 11.85 | 7.32 | 1.32 | 82.11 | 8.21 | 1.31 | 0.996 |
| 273.60 | 100.00 | 7.32 | -4.52 | 11.84 | 7.32 | 1.32 | 82.20 | 8.36 | 1.31 | 0.996 |
| 273.70 | 100.00 | 7.32 | -4.51 | 11.83 | 7.32 | 1.33 | 82.29 | 8.51 | 1.30 | 0.996 |
| 273.80 | 100.00 | 7.32 | -4.50 | 11.82 | 7.32 | 1.33 | 82.38 | 8.65 | 1.30 | 0.996 |
| 273.90 | 100.00 | 7.32 | -4.49 | 11.81 | 7.32 | 1.34 | 82.47 | 8.80 | 1.30 | 0.996 |
| 274.00 | 100.00 | 7.32 | -4.48 | 11.80 | 7.32 | 1.34 | 82.56 | 8.95 | 1.30 | 0.996 |
| 274.10 | 100.00 | 7.32 | -4.47 | 11.79 | 7.32 | 1.35 | 82.65 | 9.09 | 1.30 | 0.996 |
| 274.20 | 100.00 | 7.32 | -4.46 | 11.78 | 7.32 | 1.35 | 82.74 | 9.24 | 1.30 | 0.996 |
| 274.30 | 100.00 | 7.32 | -4.45 | 11.77 | 7.32 | 1.36 | 82.83 | 9.38 | 1.30 | 0.996 |
| 274.40 | 100.00 | 7.32 | -4.44 | 11.76 | 7.32 | 1.36 | 82.92 | 9.53 | 1.30 | 0.996 |
| 274.50 | 100.00 | 7.33 | -4.44 | 11.77 | 7.33 | 1.37 | 83.01 | 9.67 | 1.30 | 0.996 |
| 274.60 | 100.00 | 7.33 | -4.43 | 11.76 | 7.33 | 1.37 | 83.09 | 9.82 | 1.30 | 0.996 |
| 274.70 | 100.00 | 7.33 | -4.42 | 11.75 | 7.33 | 1.37 | 83.18 | 9.96 | 1.30 | 0.996 |
| 274.80 | 100.00 | 7.33 | -4.41 | 11.74 | 7.33 | 1.38 | 83.27 | 10.11 | 1.30 | 0.996 |
| 274.90 | 100.00 | 7.33 | -4.40 | 11.73 | 7.33 | 1.38 | 83.36 | 10.26 | 1.30 | 0.996 |
| 275.00 | 100.00 | 7.33 | -4.39 | 11.72 | 7.33 | 1.39 | 83.45 | 10.40 | 1.30 | 0.996 |
| 275.10 | 100.00 | 7.33 | -4.38 | 11.71 | 7.33 | 1.39 | 83.54 | 10.55 | 1.30 | 0.996 |
| 275.20 | 100.00 | 7.33 | -4.37 | 11.70 | 7.33 | 1.40 | 83.63 | 10.69 | 1.30 | 0.996 |
| 275.30 | 100.00 | 7.33 | -4.36 | 11.69 | 7.33 | 1.40 | 83.72 | 10.84 | 1.30 | 0.996 |
| 275.40 | 100.00 | 7.33 | -4.35 | 11.68 | 7.33 | 1.41 | 83.81 | 10.98 | 1.30 | 0.996 |
| 275.50 | 100.00 | 7.33 | -4.34 | 11.67 | 7.33 | 1.41 | 83.90 | 11.13 | 1.30 | 0.996 |
| 275.60 | 100.00 | 7.33 | -4.33 | 11.66 | 7.33 | 1.41 | 83.99 | 11.27 | 1.29 | 0.996 |
| 275.70 | 100.00 | 7.33 | -4.32 | 11.65 | 7.33 | 1.42 | 84.08 | 11.41 | 1.29 | 0.996 |
| 275.80 | 100.00 | 7.33 | -4.31 | 11.64 | 7.33 | 1.42 | 84.17 | 11.56 | 1.29 | 0.996 |
| 275.90 | 100.00 | 7.33 | -4.30 | 11.63 | 7.33 | 1.43 | 84.26 | 11.70 | 1.29 | 0.996 |
| 276.00 | 100.00 | 7.33 | -4.29 | 11.62 | 7.33 | 1.43 | 84.35 | 11.85 | 1.29 | 0.996 |
| 276.10 | 100.00 | 7.33 | -4.28 | 11.61 | 7.33 | 1.44 | 84.44 | 11.99 | 1.29 | 0.996 |
| 276.20 | 100.00 | 7.34 | -4.27 | 11.61 | 7.34 | 1.44 | 84.53 | 12.14 | 1.29 | 0.996 |
| 276.30 | 100.00 | 7.34 | -4.26 | 11.60 | 7.34 | 1.44 | 84.62 | 12.28 | 1.29 | 0.996 |
| 276.40 | 100.00 | 7.34 | -4.25 | 11.59 | 7.34 | 1.45 | 84.71 | 12.42 | 1.29 | 0.996 |
| 276.50 | 100.00 | 7.34 | -4.24 | 11.58 | 7.34 | 1.45 | 84.80 | 12.57 | 1.29 | 0.996 |
| 276.60 | 100.00 | 7.34 | -4.23 | 11.57 | 7.34 | 1.46 | 84.89 | 12.71 | 1.29 | 0.996 |
| 276.70 | 100.00 | 7.34 | -4.22 | 11.56 | 7.34 | 1.46 | 84.98 | 12.86 | 1.29 | 0.996 |
| 276.80 | 100.00 | 7.34 | -4.21 | 11.55 | 7.34 | 1.47 | 85.07 | 13.00 | 1.29 | 0.996 |
| 276.90 | 100.00 | 7.34 | -4.20 | 11.54 | 7.34 | 1.47 | 85.16 | 13.14 | 1.29 | 0.996 |
| 277.00 | 100.00 | 7.34 | -4.19 | 11.53 | 7.34 | 1.47 | 85.25 | 13.29 | 1.29 | 0.996 |
| 277.10 | 100.00 | 7.34 | -4.18 | 11.52 | 7.34 | 1.48 | 85.34 | 13.43 | 1.29 | 0.996 |
| 277.20 | 100.00 | 7.34 | -4.17 | 11.51 | 7.34 | 1.48 | 85.43 | 13.57 | 1.29 | 0.996 |
| 277.30 | 100.00 | 7.34 | -4.16 | 11.50 | 7.34 | 1.49 | 85.52 | 13.71 | 1.29 | 0.996 |
| 277.40 | 100.00 | 7.34 | -4.15 | 11.49 | 7.34 | 1.49 | 85.61 | 13.86 | 1.29 | 0.996 |
| 277.50 | 100.00 | 7.34 | -4.14 | 11.48 | 7.34 | 1.49 | 85.70 | 14.00 | 1.29 | 0.996 |
| 277.60 | 100.00 | 7.35 | -4.13 | 11.48 | 7.35 | 1.50 | 85.79 | 14.14 | 1.29 | 0.996 |
| 277.70 | 100.00 | 7.35 | -4.12 | 11.47 | 7.35 | 1.50 | 85.88 | 14.28 | 1.29 | 0.996 |
| 277.80 | 100.00 | 7.35 | -4.11 | 11.46 | 7.35 | 1.51 | 85.97 | 14.43 | 1.29 | 0.996 |
| 277.90 | 100.00 | 7.35 | -4.10 | 11.45 | 7.35 | 1.51 | 86.06 | 14.57 | 1.29 | 0.996 |
| 278.00 | 100.00 | 7.35 | -4.09 | 11.44 | 7.35 | 1.51 | 86.15 | 14.71 | 1.29 | 0.996 |
| 278.10 | 100.00 | 7.35 | -4.08 | 11.43 | 7.35 | 1.51 | 86.24 | 14.85 | 1.29 | 0.996 |
| 278.20 | 100.00 | 7.35 | -4.07 | 11.42 | 7.35 | 1.51 | 86.33 | 15.00 | 1.29 | 0.996 |
| 278.30 | 100.00 | 7.35 | -4.06 | 11.41 | 7.35 | 1.51 | 86.42 | 15.14 | 1.29 | 0.996 |
| 278.40 | 100.00 | 7.35 | -4.05 | 11.40 | 7.35 | 1.51 | 86.51 | 15.28 | 1.29 | 0.996 |
| 278.50 | 100.00 | 7.35 | -4.04 | 11.39 | 7.35 | 1.50 | 86.60 | 15.42 | 1.29 | 0.996 |
| 278.60 | 100.00 | 7.35 | -4.03 | 11.38 | 7.35 | 1.50 | 86.69 | 15.56 | 1.29 | 0.996 |
| 278.70 | 100.00 | 7.36 | -4.02 | 11.38 | 7.36 | 1.50 | 86.78 | 15.70 | 1.29 | 0.996 |
| 278.80 | 100.00 | 7.36 | -4.01 | 11.37 | 7.36 | 1.50 | 86.87 | 15.84 | 1.29 | 0.996 |
| 278.90 | 100.00 | 7.36 | -4.00 | 11.36 | 7.36 | 1.50 | 86.96 | 15.98 | 1.29 | 0.996 |
| 279.00 | 100.00 | 7.36 | -3.99 | 11.35 | 7.36 | 1.50 | 87.05 | 16.13 | 1.29 | 0.996 |
| 279.10 | 100.00 | 7.36 | -3.98 | 11.34 | 7.36 | 1.50 | 87.14 | 16.27 | 1.29 | 0.996 |
| 279.20 | 100.00 | 7.36 | -3.97 | 11.33 | 7.36 | 1.50 | 87.23 | 16.41 | 1.29 | 0.996 |
| 279.30 | 100.00 | 7.36 | -3.96 | 11.32 | 7.36 | 1.50 | 87.32 | 16.55 | 1.29 | 0.996 |
| 279.40 | 100.00 | 7.36 | -3.95 | 11.31 | 7.36 | 1.49 | 87.41 | 16.69 | 1.29 | 0.996 |
| 279.50 | 100.00 | 7.36 | -3.94 | 11.30 | 7.36 | 1.49 | 87.50 | 16.83 | 1.29 | 0.996 |
| 279.60 | 100.00 | 7.36 | -3.93 | 11.29 | 7.36 | 1.49 | 87.59 | 16.97 | 1.29 | 0.996 |
| 279.70 | 100.00 | 7.36 | -3.93 | 11.29 | 7.36 | 1.49 | 87.68 | 17.11 | 1.29 | 0.996 |
| 279.80 | 100.00 | 7.37 | -3.92 | 11.29 | 7.37 | 1.49 | 87.77 | 17.25 | 1.29 | 0.996 |
| 279.90 | 100.00 | 7.37 | -3.91 | 11.28 | 7.37 | 1.49 | 87.86 | 17.39 | 1.29 | 0.996 |
| 280.00 | 100.00 | 7.37 | -3.90 | 11.27 | 7.37 | 1.49 | 87.95 | 17.52 | 1.29 | 0.996 |
| 280.10 | 100.00 | 7.37 | -3.89 | 11.26 | 7.37 | 1.49 | 88.04 | 17.66 | 1.29 | 0.996 |
| 280.20 | 100.00 | 7.37 | -3.88 | 11.25 | 7.37 | 1.49 | 88.13 | 17.80 | 1.29 | 0.996 |
| 280.30 | 100.00 | 7.37 | -3.87 | 11.24 | 7.37 | 1.48 | 88.22 | 17.94 | 1.29 | 0.996 |
| 280.40 | 100.00 | 7.37 | -3.86 | 11.23 | 7.37 | 1.48 | 88.31 | 18.08 | 1.30 | 0.996 |
| 280.50 | 100.00 | 7.37 | -3.85 | 11.22 | 7.37 | 1.48 | 88.40 | 18.22 | 1.30 | 0.996 |
| 280.60 | 100.00 | 7.37 | -3.84 | 11.21 | 7.37 | 1.48 | 88.49 | 18.36 | 1.30 | 0.996 |
| 280.70 | 100.00 | 7.38 | -3.83 | 11.21 | 7.38 | 1.48 | 88.58 | 18.50 | 1.30 | 0.996 |
| 280.80 | 100.00 | 7.38 | -3.82 | 11.20 | 7.38 | 1.48 | 88.67 | 18.63 | 1.30 | 0.996 |
| 280.90 | 100.00 | 7.38 | -3.81 | 11.19 | 7.38 | 1.48 | 88.76 | 18.77 | 1.30 | 0.996 |
| 281.00 | 100.00 | 7.38 | -3.80 | 11.18 | 7.38 | 1.48 | 88.85 | 18.91 | 1.30 | 0.996 |
| 281.10 | 100.00 | 7.38 | -3.79 | 11.17 | 7.38 | 1.48 | 88.94 | 19.05 | 1.30 | 0.996 |
| 281.20 | 100.00 | 7.38 | -3.78 | 11.16 | 7.38 | 1.47 | 89.03 | 19.18 | 1.30 | 0.996 |
| 281.30 | 100.00 | 7.38 | -3.77 | 11.15 | 7.38 | 1.47 | 89.12 | 19.32 | 1.30 | 0.996 |
| 281.40 | 100.00 | 7.38 | -3.76 | 11.14 | 7.38 | 1.47 | 89.22 | 19.46 | 1.30 | 0.996 |
| 281.50 | 100.00 | 7.38 | -3.75 | 11.13 | 7.38 | 1.47 | 89.31 | 19.60 | 1.30 | 0.996 |
| 281.60 | 100.00 | 7.39 | -3.74 | 11.13 | 7.39 | 1.47 | 89.40 | 19.73 | 1.30 | 0.996 |
| 281.70 | 100.00 | 7.39 | -3.73 | 11.12 | 7.39 | 1.47 | 89.49 | 19.87 | 1.30 | 0.996 |
| 281.80 | 100.00 | 7.39 | -3.72 | 11.11 | 7.39 | 1.47 | 89.58 | 20.00 | 1.30 | 0.996 |
| 281.90 | 100.00 | 7.39 | -3.71 | 11.10 | 7.39 | 1.47 | 89.67 | 20.14 | 1.30 | 0.996 |
| 282.00 | 100.00 | 7.39 | -3.70 | 11.09 | 7.39 | 1.46 | 89.76 | 20.28 | 1.30 | 0.996 |
| 282.10 | 100.00 | 7.39 | -3.70 | 11.09 | 7.39 | 1.46 | 89.85 | 20.41 | 1.30 | 0.996 |
| 282.20 | 100.00 | 7.39 | -3.69 | 11.08 | 7.39 | 1.46 | 89.94 | 20.55 | 1.30 | 0.996 |
| 282.30 | 100.00 | 7.39 | -3.68 | 11.07 | 7.39 | 1.46 | 90.03 | 20.68 | 1.30 | 0.996 |
| 282.40 | 100.00 | 7.40 | -3.67 | 11.07 | 7.40 | 1.46 | 90.12 | 20.82 | 1.30 | 0.996 |
| 282.50 | 100.00 | 7.40 | -3.66 | 11.06 | 7.40 | 1.46 | 90.21 | 20.95 | 1.30 | 0.996 |
| 282.60 | 100.00 | 7.40 | -3.65 | 11.05 | 7.40 | 1.46 | 90.30 | 21.09 | 1.30 | 0.996 |
| 282.70 | 100.00 | 7.40 | -3.64 | 11.04 | 7.40 | 1.46 | 90.39 | 21.22 | 1.31 | 0.996 |
| 282.80 | 100.00 | 7.40 | -3.63 | 11.03 | 7.40 | 1.46 | 90.48 | 21.36 | 1.31 | 0.996 |
| 282.90 | 100.00 | 7.40 | -3.62 | 11.02 | 7.40 | 1.45 | 90.57 | 21.49 | 1.31 | 0.996 |
| 283.00 | 100.00 | 7.40 | -3.61 | 11.01 | 7.40 | 1.45 | 90.66 | 21.63 | 1.31 | 0.996 |
| 283.10 | 100.00 | 7.41 | -3.60 | 11.01 | 7.41 | 1.45 | 90.76 | 21.76 | 1.31 | 0.996 |
| 283.20 | 100.00 | 7.41 | -3.59 | 11.00 | 7.41 | 1.45 | 90.85 | 21.89 | 1.31 | 0.996 |
| 283.30 | 100.00 | 7.41 | -3.58 | 10.99 | 7.41 | 1.45 | 90.94 | 22.03 | 1.31 | 0.996 |
| 283.40 | 100.00 | 7.41 | -3.58 | 10.99 | 7.41 | 1.45 | 91.03 | 22.16 | 1.31 | 0.996 |
| 283.50 | 100.00 | 7.41 | -3.57 | 10.98 | 7.41 | 1.45 | 91.12 | 22.29 | 1.31 | 0.996 |
| 283.60 | 100.00 | 7.41 | -3.56 | 10.97 | 7.41 | 1.45 | 91.21 | 22.43 | 1.31 | 0.996 |
| 283.70 | 100.00 | 7.41 | -3.55 | 10.96 | 7.41 | 1.45 | 91.30 | 22.56 | 1.31 | 0.996 |
| 283.80 | 100.00 | 7.42 | -3.54 | 10.96 | 7.42 | 1.44 | 91.39 | 22.69 | 1.31 | 0.996 |
| 283.90 | 100.00 | 7.42 | -3.53 | 10.95 | 7.42 | 1.44 | 91.48 | 22.82 | 1.31 | 0.996 |
| 284.00 | 100.00 | 7.42 | -3.52 | 10.94 | 7.42 | 1.44 | 91.57 | 22.95 | 1.31 | 0.996 |
| 284.10 | 100.00 | 7.42 | -3.51 | 10.93 | 7.42 | 1.44 | 91.66 | 23.09 | 1.31 | 0.996 |
| 284.20 | 100.00 | 7.42 | -3.50 | 10.92 | 7.42 | 1.44 | 91.75 | 23.22 | 1.32 | 0.996 |
| 284.30 | 100.00 | 7.42 | -3.50 | 10.92 | 7.42 | 1.44 | 91.84 | 23.35 | 1.32 | 0.996 |
| 284.40 | 100.00 | 7.42 | -3.49 | 10.91 | 7.42 | 1.44 | 91.93 | 23.48 | 1.32 | 0.996 |
| 284.50 | 100.00 | 7.43 | -3.48 | 10.91 | 7.43 | 1.44 | 92.02 | 23.61 | 1.32 | 0.996 |
| 284.60 | 100.00 | 7.43 | -3.47 | 10.90 | 7.43 | 1.43 | 92.11 | 23.74 | 1.32 | 0.996 |
| 284.70 | 100.00 | 7.43 | -3.46 | 10.89 | 7.43 | 1.43 | 92.20 | 23.87 | 1.32 | 0.996 |
| 284.80 | 100.00 | 7.43 | -3.45 | 10.88 | 7.43 | 1.43 | 92.29 | 24.00 | 1.32 | 0.996 |
| 284.90 | 100.00 | 7.43 | -3.44 | 10.87 | 7.43 | 1.42 | 92.39 | 24.13 | 1.32 | 0.996 |
| 285.00 | 100.00 | 7.43 | -3.44 | 10.87 | 7.43 | 1.42 | 92.48 | 24.26 | 1.32 | 0.996 |
| 285.10 | 100.00 | 7.43 | -3.43 | 10.86 | 7.43 | 1.41 | 92.57 | 24.39 | 1.32 | 0.996 |
| 285.20 | 100.00 | 7.44 | -3.42 | 10.86 | 7.44 | 1.41 | 92.66 | 24.52 | 1.32 | 0.996 |
| 285.30 | 100.00 | 7.44 | -3.41 | 10.85 | 7.44 | 1.41 | 92.75 | 24.65 | 1.32 | 0.996 |
| 285.40 | 100.00 | 7.44 | -3.40 | 10.84 | 7.44 | 1.40 | 92.84 | 24.78 | 1.32 | 0.996 |
| 285.50 | 100.00 | 7.44 | -3.39 | 10.83 | 7.44 | 1.40 | 92.93 | 24.91 | 1.32 | 0.996 |
| 285.60 | 100.00 | 7.44 | -3.39 | 10.83 | 7.44 | 1.39 | 93.02 | 25.04 | 1.33 | 0.996 |
| 285.70 | 100.00 | 7.44 | -3.38 | 10.82 | 7.44 | 1.39 | 93.11 | 25.16 | 1.33 | 0.996 |
| 285.80 | 100.00 | 7.45 | -3.37 | 10.82 | 7.45 | 1.39 | 93.20 | 25.29 | 1.33 | 0.996 |
| 285.90 | 100.00 | 7.45 | -3.36 | 10.81 | 7.45 | 1.38 | 93.29 | 25.42 | 1.33 | 0.996 |
| 286.00 | 100.00 | 7.45 | -3.35 | 10.80 | 7.45 | 1.38 | 93.38 | 25.55 | 1.33 | 0.996 |
| 286.10 | 100.00 | 7.45 | -3.34 | 10.79 | 7.45 | 1.37 | 93.47 | 25.67 | 1.33 | 0.996 |
| 286.20 | 100.00 | 7.45 | -3.34 | 10.79 | 7.45 | 1.37 | 93.56 | 25.80 | 1.33 | 0.996 |
| 286.30 | 100.00 | 7.45 | -3.33 | 10.78 | 7.45 | 1.37 | 93.65 | 25.93 | 1.33 | 0.996 |
| 286.40 | 100.00 | 7.46 | -3.32 | 10.78 | 7.46 | 1.36 | 93.74 | 26.05 | 1.33 | 0.996 |
| 286.50 | 100.00 | 7.46 | -3.31 | 10.77 | 7.46 | 1.36 | 93.83 | 26.18 | 1.33 | 0.996 |
| 286.60 | 100.00 | 7.46 | -3.31 | 10.77 | 7.46 | 1.35 | 93.92 | 26.30 | 1.33 | 0.996 |
| 286.70 | 100.00 | 7.46 | -3.30 | 10.76 | 7.46 | 1.35 | 94.01 | 26.43 | 1.33 | 0.996 |
| 286.80 | 100.00 | 7.46 | -3.29 | 10.75 | 7.46 | 1.34 | 94.10 | 26.56 | 1.34 | 0.996 |
| 286.90 | 100.00 | 7.46 | -3.28 | 10.74 | 7.46 | 1.34 | 94.19 | 26.68 | 1.34 | 0.996 |
| 287.00 | 100.00 | 7.47 | -3.27 | 10.74 | 7.47 | 1.34 | 94.28 | 26.80 | 1.34 | 0.996 |
| 287.10 | 100.00 | 7.47 | -3.27 | 10.74 | 7.47 | 1.33 | 94.37 | 26.93 | 1.34 | 0.996 |
| 287.20 | 100.00 | 7.47 | -3.26 | 10.73 | 7.47 | 1.33 | 94.46 | 27.05 | 1.34 | 0.996 |
| 287.30 | 100.00 | 7.47 | -3.25 | 10.72 | 7.47 | 1.32 | 94.55 | 27.18 | 1.34 | 0.996 |
| 287.40 | 100.00 | 7.47 | -3.24 | 10.71 | 7.47 | 1.32 | 94.64 | 27.30 | 1.34 | 0.996 |
| 287.50 | 100.00 | 7.47 | -3.24 | 10.71 | 7.47 | 1.32 | 94.73 | 27.42 | 1.34 | 0.996 |
| 287.60 | 100.00 | 7.48 | -3.23 | 10.71 | 7.48 | 1.31 | 94.82 | 27.55 | 1.34 | 0.996 |
| 287.70 | 100.00 | 7.48 | -3.22 | 10.70 | 7.48 | 1.31 | 94.90 | 27.67 | 1.34 | 0.996 |
| 287.80 | 100.00 | 7.48 | -3.21 | 10.69 | 7.48 | 1.30 | 94.99 | 27.79 | 1.34 | 0.996 |
| 287.90 | 100.00 | 7.48 | -3.21 | 10.69 | 7.48 | 1.30 | 95.08 | 27.92 | 1.35 | 0.996 |
| 288.00 | 100.00 | 7.48 | -3.20 | 10.68 | 7.48 | 1.30 | 95.17 | 28.04 | 1.35 | 0.996 |
| 288.10 | 100.00 | 7.48 | -3.19 | 10.67 | 7.48 | 1.29 | 95.26 | 28.16 | 1.35 | 0.996 |
| 288.20 | 100.00 | 7.49 | -3.18 | 10.67 | 7.49 | 1.29 | 95.35 | 28.28 | 1.35 | 0.996 |
| 288.30 | 100.00 | 7.49 | -3.18 | 10.67 | 7.49 | 1.28 | 95.44 | 28.40 | 1.35 | 0.996 |
| 288.40 | 100.00 | 7.49 | -3.17 | 10.66 | 7.49 | 1.28 | 95.53 | 28.52 | 1.35 | 0.996 |
| 288.50 | 100.00 | 7.49 | -3.16 | 10.65 | 7.49 | 1.28 | 95.62 | 28.64 | 1.35 | 0.996 |
| 288.60 | 100.00 | 7.49 | -3.16 | 10.65 | 7.49 | 1.27 | 95.71 | 28.76 | 1.35 | 0.996 |
| 288.70 | 100.00 | 7.50 | -3.15 | 10.65 | 7.50 | 1.27 | 95.79 | 28.88 | 1.35 | 0.996 |
| 288.80 | 100.00 | 7.50 | -3.14 | 10.64 | 7.50 | 1.26 | 95.88 | 29.00 | 1.35 | 0.996 |
| 288.90 | 100.00 | 7.50 | -3.14 | 10.64 | 7.50 | 1.26 | 95.97 | 29.12 | 1.35 | 0.996 |
| 289.00 | 100.00 | 7.50 | -3.13 | 10.63 | 7.50 | 1.25 | 96.06 | 29.24 | 1.36 | 0.996 |
| 289.10 | 100.00 | 7.50 | -3.12 | 10.62 | 7.50 | 1.25 | 96.15 | 29.36 | 1.36 | 0.996 |
| 289.20 | 100.00 | 7.51 | -3.12 | 10.63 | 7.51 | 1.25 | 96.24 | 29.48 | 1.36 | 0.996 |
| 289.30 | 100.00 | 7.51 | -3.11 | 10.62 | 7.51 | 1.24 | 96.32 | 29.60 | 1.36 | 0.996 |
| 289.40 | 100.00 | 7.51 | -3.10 | 10.61 | 7.51 | 1.24 | 96.41 | 29.72 | 1.36 | 0.996 |
| 289.50 | 100.00 | 7.51 | -3.10 | 10.61 | 7.51 | 1.23 | 96.50 | 29.83 | 1.36 | 0.996 |
| 289.60 | 100.00 | 7.51 | -3.09 | 10.60 | 7.51 | 1.23 | 96.59 | 29.95 | 1.36 | 0.996 |
| 289.70 | 100.00 | 7.51 | -3.08 | 10.59 | 7.51 | 1.23 | 96.68 | 30.07 | 1.36 | 0.996 |
| 289.80 | 100.00 | 7.52 | -3.08 | 10.60 | 7.52 | 1.22 | 96.76 | 30.18 | 1.36 | 0.996 |
| 289.90 | 100.00 | 7.52 | -3.07 | 10.59 | 7.52 | 1.22 | 96.85 | 30.30 | 1.36 | 0.996 |
| 290.00 | 100.00 | 7.52 | -3.06 | 10.58 | 7.52 | 1.21 | 96.94 | 30.42 | 1.37 | 0.996 |
| 290.10 | 100.00 | 7.52 | -3.06 | 10.58 | 7.52 | 1.21 | 97.02 | 30.53 | 1.37 | 0.996 |
| 290.20 | 100.00 | 7.52 | -3.05 | 10.57 | 7.52 | 1.20 | 97.11 | 30.65 | 1.37 | 0.996 |
| 290.30 | 100.00 | 7.53 | -3.04 | 10.57 | 7.53 | 1.20 | 97.20 | 30.76 | 1.37 | 0.996 |
| 290.40 | 100.00 | 7.53 | -3.04 | 10.57 | 7.53 | 1.20 | 97.29 | 30.88 | 1.37 | 0.996 |
| 290.50 | 100.00 | 7.53 | -3.03 | 10.56 | 7.53 | 1.19 | 97.37 | 30.99 | 1.37 | 0.996 |
| 290.60 | 100.00 | 7.53 | -3.03 | 10.56 | 7.53 | 1.19 | 97.46 | 31.11 | 1.37 | 0.996 |
| 290.70 | 100.00 | 7.54 | -3.02 | 10.56 | 7.54 | 1.18 | 97.55 | 31.22 | 1.37 | 0.996 |
| 290.80 | 100.00 | 7.54 | -3.01 | 10.55 | 7.54 | 1.18 | 97.63 | 31.33 | 1.37 | 0.996 |
| 290.90 | 100.00 | 7.54 | -3.01 | 10.55 | 7.54 | 1.17 | 97.72 | 31.45 | 1.37 | 0.996 |
| 291.00 | 100.00 | 7.54 | -3.00 | 10.54 | 7.54 | 1.17 | 97.80 | 31.56 | 1.38 | 0.996 |
| 291.10 | 100.00 | 7.54 | -3.00 | 10.54 | 7.54 | 1.17 | 97.89 | 31.67 | 1.38 | 0.996 |
| 291.20 | 100.00 | 7.55 | -2.99 | 10.54 | 7.55 | 1.16 | 97.98 | 31.79 | 1.38 | 0.996 |
| 291.30 | 100.00 | 7.55 | -2.99 | 10.54 | 7.55 | 1.16 | 98.06 | 31.90 | 1.38 | 0.996 |
| 291.40 | 100.00 | 7.55 | -2.98 | 10.53 | 7.55 | 1.15 | 98.15 | 32.01 | 1.38 | 0.996 |
| 291.50 | 100.00 | 7.55 | -2.98 | 10.53 | 7.55 | 1.15 | 98.23 | 32.12 | 1.38 | 0.996 |
| 291.60 | 100.00 | 7.55 | -2.97 | 10.52 | 7.55 | 1.15 | 98.32 | 32.23 | 1.38 | 0.996 |
| 291.70 | 100.00 | 7.56 | -2.96 | 10.52 | 7.56 | 1.14 | 98.40 | 32.34 | 1.38 | 0.996 |
| 291.80 | 100.00 | 7.56 | -2.96 | 10.52 | 7.56 | 1.14 | 98.49 | 32.45 | 1.38 | 0.996 |
| 291.90 | 100.00 | 7.56 | -2.95 | 10.51 | 7.56 | 1.13 | 98.57 | 32.56 | 1.38 | 0.996 |
| 292.00 | 100.00 | 7.56 | -2.95 | 10.51 | 7.56 | 1.13 | 98.66 | 32.67 | 1.39 | 0.996 |
| 292.10 | 100.00 | 7.56 | -2.94 | 10.50 | 7.56 | 1.12 | 98.74 | 32.78 | 1.39 | 0.996 |
| 292.20 | 100.00 | 7.57 | -2.94 | 10.51 | 7.57 | 1.12 | 98.83 | 32.89 | 1.39 | 0.996 |
| 292.30 | 100.00 | 7.57 | -2.93 | 10.50 | 7.57 | 1.12 | 98.91 | 33.00 | 1.39 | 0.996 |
| 292.40 | 100.00 | 7.57 | -2.93 | 10.50 | 7.57 | 1.11 | 98.99 | 33.11 | 1.39 | 0.996 |
| 292.50 | 100.00 | 7.57 | -2.92 | 10.49 | 7.57 | 1.11 | 99.08 | 33.22 | 1.39 | 0.996 |
| 292.60 | 100.00 | 7.58 | -2.92 | 10.50 | 7.58 | 1.10 | 99.16 | 33.32 | 1.39 | 0.996 |
| 292.70 | 100.00 | 7.58 | -2.91 | 10.49 | 7.58 | 1.10 | 99.25 | 33.43 | 1.39 | 0.996 |
| 292.80 | 100.00 | 7.58 | -2.91 | 10.49 | 7.58 | 1.09 | 99.33 | 33.54 | 1.39 | 0.996 |
| 292.90 | 100.00 | 7.58 | -2.90 | 10.48 | 7.58 | 1.09 | 99.41 | 33.64 | 1.39 | 0.996 |
| 293.00 | 100.00 | 7.59 | -2.90 | 10.49 | 7.59 | 1.09 | 99.49 | 33.75 | 1.40 | 0.996 |
| 293.10 | 100.00 | 7.59 | -2.90 | 10.49 | 7.59 | 1.08 | 99.58 | 33.86 | 1.40 | 0.996 |
| 293.20 | 100.00 | 7.59 | -2.89 | 10.48 | 7.59 | 1.08 | 99.66 | 33.96 | 1.40 | 0.996 |
| 293.30 | 100.00 | 7.59 | -2.89 | 10.48 | 7.59 | 1.07 | 99.74 | 34.07 | 1.40 | 0.996 |
| 293.40 | 100.00 | 7.59 | -2.88 | 10.47 | 7.59 | 1.07 | 99.82 | 34.17 | 1.40 | 0.996 |
| 293.50 | 100.00 | 7.60 | -2.88 | 10.48 | 7.60 | 1.06 | 99.91 | 34.28 | 1.40 | 0.996 |
| 293.60 | 100.00 | 7.60 | -2.87 | 10.47 | 7.60 | 1.06 | 99.99 | 34.38 | 1.40 | 0.996 |
| 293.70 | 100.00 | 7.60 | -2.87 | 10.47 | 7.60 | 1.06 | 100.07 | 34.48 | 1.40 | 0.996 |
| 293.80 | 100.00 | 7.60 | -2.87 | 10.47 | 7.60 | 1.05 | 100.15 | 34.59 | 1.40 | 0.996 |
| 293.90 | 100.00 | 7.61 | -2.86 | 10.47 | 7.61 | 1.05 | 100.23 | 34.69 | 1.40 | 0.996 |
| 294.00 | 100.00 | 7.61 | -2.86 | 10.47 | 7.61 | 1.04 | 100.31 | 34.79 | 1.41 | 0.996 |
| 294.10 | 100.00 | 7.61 | -2.85 | 10.46 | 7.61 | 1.04 | 100.39 | 34.90 | 1.41 | 0.996 |
| 294.20 | 100.00 | 7.61 | -2.85 | 10.46 | 7.61 | 1.03 | 100.47 | 35.00 | 1.41 | 0.996 |
| 294.30 | 100.00 | 7.62 | -2.85 | 10.47 | 7.62 | 1.03 | 100.55 | 35.10 | 1.41 | 0.996 |
| 294.40 | 100.00 | 7.62 | -2.84 | 10.46 | 7.62 | 1.03 | 100.63 | 35.20 | 1.41 | 0.996 |
| 294.50 | 100.00 | 7.62 | -2.84 | 10.46 | 7.62 | 1.02 | 100.71 | 35.30 | 1.41 | 0.996 |
| 294.60 | 100.00 | 7.62 | -2.84 | 10.46 | 7.62 | 1.02 | 100.79 | 35.40 | 1.41 | 0.996 |
| 294.70 | 100.00 | 7.63 | -2.83 | 10.46 | 7.63 | 1.01 | 100.87 | 35.50 | 1.41 | 0.996 |
| 294.80 | 100.00 | 7.63 | -2.83 | 10.46 | 7.63 | 1.01 | 100.95 | 35.60 | 1.41 | 0.996 |
| 294.90 | 100.00 | 7.63 | -2.83 | 10.46 | 7.63 | 1.00 | 101.02 | 35.70 | 1.41 | 0.996 |
| 295.00 | 100.00 | 7.63 | -2.82 | 10.45 | 7.63 | 1.00 | 101.10 | 35.80 | 1.42 | 0.996 |
| 295.10 | 100.00 | 7.64 | -2.82 | 10.46 | 7.64 | 1.00 | 101.18 | 35.90 | 1.42 | 0.996 |
| 295.20 | 100.00 | 7.64 | -2.82 | 10.46 | 7.64 | 0.99 | 101.26 | 36.00 | 1.42 | 0.996 |
| 295.30 | 100.00 | 7.64 | -2.81 | 10.45 | 7.64 | 0.99 | 101.34 | 36.10 | 1.42 | 0.996 |
| 295.40 | 100.00 | 7.64 | -2.81 | 10.45 | 7.64 | 0.98 | 101.41 | 36.20 | 1.42 | 0.996 |
| 295.50 | 100.00 | 7.65 | -2.81 | 10.46 | 7.65 | 0.98 | 101.49 | 36.29 | 1.42 | 0.996 |
| 295.60 | 100.00 | 7.65 | -2.80 | 10.45 | 7.65 | 0.97 | 101.57 | 36.39 | 1.42 | 0.996 |
| 295.70 | 100.00 | 7.65 | -2.80 | 10.45 | 7.65 | 0.97 | 101.64 | 36.49 | 1.42 | 0.996 |
| 295.80 | 100.00 | 7.65 | -2.80 | 10.45 | 7.65 | 0.97 | 101.72 | 36.58 | 1.42 | 0.996 |
| 295.90 | 100.00 | 7.66 | -2.80 | 10.46 | 7.66 | 0.96 | 101.79 | 36.68 | 1.42 | 0.996 |
| 296.00 | 100.00 | 7.66 | -2.79 | 10.45 | 7.66 | 0.96 | 101.87 | 36.78 | 1.43 | 0.996 |
| 296.10 | 100.00 | 7.66 | -2.79 | 10.45 | 7.66 | 0.95 | 101.94 | 36.87 | 1.43 | 0.996 |
| 296.20 | 100.00 | 7.66 | -2.79 | 10.45 | 7.66 | 0.95 | 102.02 | 36.97 | 1.43 | 0.996 |
| 296.30 | 100.00 | 7.67 | -2.79 | 10.46 | 7.67 | 0.94 | 102.09 | 37.06 | 1.43 | 0.996 |
| 296.40 | 100.00 | 7.67 | -2.78 | 10.45 | 7.67 | 0.94 | 102.17 | 37.15 | 1.43 | 0.996 |
| 296.50 | 100.00 | 7.67 | -2.78 | 10.45 | 7.67 | 0.94 | 102.24 | 37.25 | 1.43 | 0.996 |
| 296.60 | 100.00 | 7.67 | -2.78 | 10.45 | 7.67 | 0.93 | 102.31 | 37.34 | 1.43 | 0.996 |
| 296.70 | 100.00 | 7.68 | -2.78 | 10.46 | 7.68 | 0.93 | 102.39 | 37.43 | 1.43 | 0.996 |
| 296.80 | 100.00 | 7.68 | -2.78 | 10.46 | 7.68 | 0.92 | 102.46 | 37.53 | 1.43 | 0.996 |
| 296.90 | 100.00 | 7.68 | -2.77 | 10.45 | 7.68 | 0.92 | 102.53 | 37.62 | 1.43 | 0.996 |
| 297.00 | 100.00 | 7.69 | -2.77 | 10.46 | 7.69 | 0.91 | 102.60 | 37.71 | 1.44 | 0.996 |
| 297.10 | 100.00 | 7.69 | -2.77 | 10.46 | 7.69 | 0.91 | 102.67 | 37.80 | 1.44 | 0.996 |
| 297.20 | 100.00 | 7.69 | -2.77 | 10.46 | 7.69 | 0.91 | 102.74 | 37.89 | 1.44 | 0.996 |
| 297.30 | 100.00 | 7.69 | -2.77 | 10.46 | 7.69 | 0.90 | 102.82 | 37.99 | 1.44 | 0.996 |
| 297.40 | 100.00 | 7.70 | -2.77 | 10.47 | 7.70 | 0.90 | 102.89 | 38.08 | 1.44 | 0.996 |
| 297.50 | 100.00 | 7.70 | -2.76 | 10.46 | 7.70 | 0.89 | 102.96 | 38.17 | 1.44 | 0.996 |
| 297.60 | 100.00 | 7.70 | -2.76 | 10.46 | 7.70 | 0.89 | 103.03 | 38.26 | 1.44 | 0.996 |
| 297.70 | 100.00 | 7.71 | -2.76 | 10.47 | 7.71 | 0.89 | 103.10 | 38.35 | 1.44 | 0.996 |
| 297.80 | 100.00 | 7.71 | -2.76 | 10.47 | 7.71 | 0.88 | 103.16 | 38.43 | 1.44 | 0.996 |
| 297.90 | 100.00 | 7.71 | -2.76 | 10.47 | 7.71 | 0.88 | 103.23 | 38.52 | 1.44 | 0.996 |
| 298.00 | 100.00 | 7.71 | -2.76 | 10.47 | 7.71 | 0.87 | 103.30 | 38.61 | 1.45 | 0.996 |
| 298.10 | 100.00 | 7.72 | -2.76 | 10.48 | 7.72 | 0.87 | 103.37 | 38.70 | 1.45 | 0.996 |
| 298.20 | 100.00 | 7.72 | -2.76 | 10.48 | 7.72 | 0.86 | 103.44 | 38.79 | 1.45 | 0.996 |
| 298.30 | 100.00 | 7.72 | -2.76 | 10.48 | 7.72 | 0.86 | 103.50 | 38.87 | 1.45 | 0.996 |
| 298.40 | 100.00 | 7.73 | -2.76 | 10.49 | 7.73 | 0.86 | 103.57 | 38.96 | 1.45 | 0.996 |
| 298.50 | 100.00 | 7.73 | -2.75 | 10.48 | 7.73 | 0.85 | 103.64 | 39.05 | 1.45 | 0.996 |
| 298.60 | 100.00 | 7.73 | -2.75 | 10.48 | 7.73 | 0.85 | 103.70 | 39.13 | 1.45 | 0.996 |
| 298.70 | 100.00 | 7.73 | -2.75 | 10.48 | 7.73 | 0.84 | 103.76 | 39.22 | 1.45 | 0.996 |
| 298.80 | 100.00 | 7.74 | -2.75 | 10.49 | 7.74 | 0.84 | 103.83 | 39.31 | 1.45 | 0.996 |
| 298.90 | 100.00 | 7.74 | -2.75 | 10.49 | 7.74 | 0.83 | 103.89 | 39.39 | 1.45 | 0.996 |
| 299.00 | 100.00 | 7.74 | -2.75 | 10.49 | 7.74 | 0.83 | 103.96 | 39.48 | 1.45 | 0.996 |
| 299.10 | 100.00 | 7.75 | -2.75 | 10.50 | 7.75 | 0.82 | 104.02 | 39.56 | 1.46 | 0.996 |
| 299.20 | 100.00 | 7.75 | -2.75 | 10.50 | 7.75 | 0.81 | 104.08 | 39.64 | 1.46 | 0.996 |
| 299.30 | 100.00 | 7.75 | -2.75 | 10.50 | 7.75 | 0.81 | 104.15 | 39.73 | 1.46 | 0.996 |
| 299.40 | 100.00 | 7.76 | -2.75 | 10.51 | 7.76 | 0.80 | 104.21 | 39.81 | 1.46 | 0.996 |
| 299.50 | 100.00 | 7.76 | -2.75 | 10.51 | 7.76 | 0.79 | 104.27 | 39.89 | 1.46 | 0.996 |
| 299.60 | 100.00 | 7.76 | -2.75 | 10.51 | 7.76 | 0.78 | 104.33 | 39.98 | 1.46 | 0.996 |
| 299.70 | 100.00 | 7.76 | -2.76 | 10.52 | 7.76 | 0.77 | 104.39 | 40.06 | 1.46 | 0.996 |
| 299.80 | 100.00 | 7.77 | -2.76 | 10.53 | 7.77 | 0.76 | 104.45 | 40.14 | 1.46 | 0.996 |
| 299.90 | 100.00 | 7.77 | -2.76 | 10.53 | 7.77 | 0.75 | 104.51 | 40.22 | 1.46 | 0.996 |
| 300.00 | 100.00 | 7.77 | -2.76 | 10.53 | 7.77 | 0.75 | 104.57 | 40.30 | 1.46 | 0.996 |
| 300.10 | 100.00 | 7.78 | -2.76 | 10.54 | 7.78 | 0.74 | 104.63 | 40.39 | 1.47 | 0.996 |
| 300.20 | 100.00 | 7.78 | -2.76 | 10.54 | 7.78 | 0.73 | 104.69 | 40.47 | 1.47 | 0.996 |
| 300.30 | 100.00 | 7.78 | -2.76 | 10.54 | 7.78 | 0.72 | 104.74 | 40.55 | 1.47 | 0.996 |
| 300.40 | 100.00 | 7.79 | -2.76 | 10.55 | 7.79 | 0.71 | 104.80 | 40.63 | 1.47 | 0.996 |
| 300.50 | 100.00 | 7.79 | -2.76 | 10.55 | 7.79 | 0.70 | 104.86 | 40.71 | 1.47 | 0.996 |
| 300.60 | 100.00 | 7.79 | -2.76 | 10.55 | 7.79 | 0.69 | 104.91 | 40.79 | 1.47 | 0.996 |
| 300.70 | 100.00 | 7.80 | -2.77 | 10.57 | 7.80 | 0.69 | 104.97 | 40.86 | 1.47 | 0.996 |
| 300.80 | 100.00 | 7.80 | -2.77 | 10.57 | 7.80 | 0.68 | 105.02 | 40.94 | 1.47 | 0.996 |
| 300.90 | 100.00 | 7.80 | -2.77 | 10.57 | 7.80 | 0.67 | 105.08 | 41.02 | 1.47 | 0.996 |
| 301.00 | 100.00 | 7.81 | -2.77 | 10.58 | 7.81 | 0.66 | 105.13 | 41.10 | 1.47 | 0.996 |
| 301.10 | 100.00 | 7.81 | -2.77 | 10.58 | 7.81 | 0.65 | 105.18 | 41.18 | 1.47 | 0.996 |
| 301.20 | 100.00 | 7.81 | -2.77 | 10.58 | 7.81 | 0.64 | 105.24 | 41.26 | 1.48 | 0.996 |
| 301.30 | 100.00 | 7.82 | -2.78 | 10.60 | 7.82 | 0.63 | 105.29 | 41.33 | 1.48 | 0.996 |
| 301.40 | 100.00 | 7.82 | -2.78 | 10.60 | 7.82 | 0.62 | 105.34 | 41.41 | 1.48 | 0.996 |
| 301.50 | 100.00 | 7.82 | -2.78 | 10.60 | 7.82 | 0.62 | 105.39 | 41.49 | 1.48 | 0.996 |
| 301.60 | 100.00 | 7.83 | -2.78 | 10.61 | 7.83 | 0.61 | 105.44 | 41.56 | 1.48 | 0.996 |
| 301.70 | 100.00 | 7.83 | -2.79 | 10.62 | 7.83 | 0.60 | 105.49 | 41.64 | 1.48 | 0.996 |
| 301.80 | 100.00 | 7.83 | -2.79 | 10.62 | 7.83 | 0.59 | 105.54 | 41.71 | 1.48 | 0.996 |
| 301.90 | 100.00 | 7.84 | -2.79 | 10.63 | 7.84 | 0.58 | 105.59 | 41.79 | 1.48 | 0.996 |
| 302.00 | 100.00 | 7.84 | -2.79 | 10.63 | 7.84 | 0.57 | 105.64 | 41.86 | 1.48 | 0.996 |
| 302.10 | 100.00 | 7.84 | -2.80 | 10.64 | 7.84 | 0.56 | 105.68 | 41.94 | 1.48 | 0.996 |
| 302.20 | 100.00 | 7.85 | -2.80 | 10.65 | 7.85 | 0.55 | 105.73 | 42.01 | 1.48 | 0.996 |
| 302.30 | 100.00 | 7.85 | -2.80 | 10.65 | 7.85 | 0.54 | 105.78 | 42.09 | 1.49 | 0.996 |
| 302.40 | 100.00 | 7.85 | -2.81 | 10.66 | 7.85 | 0.53 | 105.82 | 42.16 | 1.49 | 0.996 |
| 302.50 | 100.00 | 7.86 | -2.81 | 10.67 | 7.86 | 0.53 | 105.87 | 42.23 | 1.49 | 0.996 |
| 302.60 | 100.00 | 7.86 | -2.81 | 10.67 | 7.86 | 0.52 | 105.91 | 42.31 | 1.49 | 0.996 |
| 302.70 | 100.00 | 7.87 | -2.82 | 10.69 | 7.87 | 0.51 | 105.96 | 42.38 | 1.49 | 0.996 |
| 302.80 | 100.00 | 7.87 | -2.82 | 10.69 | 7.87 | 0.50 | 106.00 | 42.45 | 1.49 | 0.996 |
| 302.90 | 100.00 | 7.87 | -2.82 | 10.69 | 7.87 | 0.49 | 106.04 | 42.52 | 1.49 | 0.996 |
| 303.00 | 100.00 | 7.88 | -2.83 | 10.71 | 7.88 | 0.48 | 106.08 | 42.60 | 1.49 | 0.996 |
| 303.10 | 100.00 | 7.88 | -2.83 | 10.71 | 7.88 | 0.47 | 106.12 | 42.67 | 1.49 | 0.996 |
| 303.20 | 100.00 | 7.88 | -2.84 | 10.72 | 7.88 | 0.46 | 106.16 | 42.74 | 1.49 | 0.996 |
| 303.30 | 100.00 | 7.89 | -2.84 | 10.73 | 7.89 | 0.45 | 106.20 | 42.81 | 1.49 | 0.996 |
| 303.40 | 100.00 | 7.89 | -2.85 | 10.74 | 7.89 | 0.44 | 106.24 | 42.88 | 1.49 | 0.996 |
| 303.50 | 100.00 | 7.89 | -2.85 | 10.74 | 7.89 | 0.43 | 106.28 | 42.95 | 1.50 | 0.996 |
| 303.60 | 100.00 | 7.90 | -2.85 | 10.75 | 7.90 | 0.42 | 106.32 | 43.02 | 1.50 | 0.996 |
| 303.70 | 100.00 | 7.90 | -2.86 | 10.76 | 7.90 | 0.42 | 106.35 | 43.09 | 1.50 | 0.996 |
| 303.80 | 100.00 | 7.91 | -2.86 | 10.77 | 7.91 | 0.41 | 106.39 | 43.16 | 1.50 | 0.996 |
| 303.90 | 100.00 | 7.91 | -2.87 | 10.78 | 7.91 | 0.40 | 106.42 | 43.23 | 1.50 | 0.996 |
| 304.00 | 100.00 | 7.91 | -2.87 | 10.78 | 7.91 | 0.39 | 106.46 | 43.30 | 1.50 | 0.996 |
| 304.10 | 100.00 | 7.92 | -2.88 | 10.80 | 7.92 | 0.38 | 106.49 | 43.37 | 1.50 | 0.996 |
| 304.20 | 100.00 | 7.92 | -2.88 | 10.80 | 7.92 | 0.37 | 106.52 | 43.44 | 1.50 | 0.996 |
| 304.30 | 100.00 | 7.93 | -2.89 | 10.82 | 7.93 | 0.36 | 106.56 | 43.51 | 1.50 | 0.996 |
| 304.40 | 100.00 | 7.93 | -2.90 | 10.83 | 7.93 | 0.35 | 106.59 | 43.58 | 1.50 | 0.996 |
| 304.50 | 100.00 | 7.93 | -2.90 | 10.83 | 7.93 | 0.34 | 106.62 | 43.65 | 1.50 | 0.996 |
| 304.60 | 100.00 | 7.94 | -2.91 | 10.85 | 7.94 | 0.33 | 106.65 | 43.72 | 1.51 | 0.996 |
| 304.70 | 100.00 | 7.94 | -2.91 | 10.85 | 7.94 | 0.32 | 106.68 | 43.78 | 1.51 | 0.996 |
| 304.80 | 100.00 | 7.95 | -2.92 | 10.87 | 7.95 | 0.31 | 106.71 | 43.85 | 1.51 | 0.996 |
| 304.90 | 100.00 | 7.95 | -2.93 | 10.88 | 7.95 | 0.30 | 106.74 | 43.92 | 1.51 | 0.996 |
| 305.00 | 100.00 | 7.95 | -2.93 | 10.88 | 7.95 | 0.29 | 106.76 | 43.99 | 1.51 | 0.996 |
| 305.10 | 100.00 | 7.96 | -2.94 | 10.90 | 7.96 | 0.28 | 106.79 | 44.05 | 1.51 | 0.996 |
| 305.20 | 100.00 | 7.96 | -2.95 | 10.91 | 7.96 | 0.27 | 106.81 | 44.12 | 1.51 | 0.996 |
| 305.30 | 100.00 | 7.97 | -2.95 | 10.92 | 7.97 | 0.26 | 106.84 | 44.19 | 1.51 | 0.996 |
| 305.40 | 100.00 | 7.97 | -2.96 | 10.93 | 7.97 | 0.25 | 106.86 | 44.26 | 1.51 | 0.996 |
| 305.50 | 100.00 | 7.97 | -2.97 | 10.94 | 7.97 | 0.24 | 106.88 | 44.32 | 1.51 | 0.996 |
| 305.60 | 100.00 | 7.98 | -2.97 | 10.95 | 7.98 | 0.23 | 106.91 | 44.39 | 1.51 | 0.996 |
| 305.70 | 100.00 | 7.98 | -2.98 | 10.96 | 7.98 | 0.22 | 106.93 | 44.45 | 1.52 | 0.996 |
| 305.80 | 100.00 | 7.99 | -2.99 | 10.98 | 7.99 | 0.21 | 106.95 | 44.52 | 1.52 | 0.996 |
| 305.90 | 100.00 | 7.99 | -3.00 | 10.99 | 7.99 | 0.21 | 106.97 | 44.59 | 1.52 | 0.996 |
| 306.00 | 100.00 | 8.00 | -3.00 | 11.00 | 8.00 | 0.20 | 106.99 | 44.65 | 1.52 | 0.996 |
| 306.10 | 100.00 | 8.00 | -3.01 | 11.01 | 8.00 | 0.19 | 107.00 | 44.72 | 1.52 | 0.996 |
| 306.20 | 100.00 | 8.00 | -3.02 | 11.02 | 8.00 | 0.18 | 107.02 | 44.78 | 1.52 | 0.996 |
| 306.30 | 100.00 | 8.01 | -3.03 | 11.04 | 8.01 | 0.17 | 107.04 | 44.85 | 1.52 | 0.996 |
| 306.40 | 100.00 | 8.01 | -3.04 | 11.05 | 8.01 | 0.16 | 107.05 | 44.91 | 1.52 | 0.996 |
| 306.50 | 100.00 | 8.02 | -3.05 | 11.07 | 8.02 | 0.15 | 107.07 | 44.98 | 1.52 | 0.996 |
| 306.60 | 100.00 | 8.02 | -3.05 | 11.07 | 8.02 | 0.14 | 107.08 | 45.04 | 1.52 | 0.996 |
| 306.70 | 100.00 | 8.03 | -3.06 | 11.09 | 8.03 | 0.13 | 107.10 | 45.11 | 1.52 | 0.996 |
| 306.80 | 100.00 | 8.03 | -3.07 | 11.10 | 8.03 | 0.12 | 107.11 | 45.17 | 1.52 | 0.996 |
| 306.90 | 100.00 | 8.03 | -3.08 | 11.11 | 8.03 | 0.11 | 107.12 | 45.24 | 1.53 | 0.996 |
| 307.00 | 100.00 | 8.04 | -3.09 | 11.13 | 8.04 | 0.10 | 107.13 | 45.30 | 1.53 | 0.996 |
| 307.10 | 100.00 | 8.04 | -3.10 | 11.14 | 8.04 | 0.09 | 107.14 | 45.37 | 1.53 | 0.996 |
| 307.20 | 100.00 | 8.05 | -3.11 | 11.16 | 8.05 | 0.08 | 107.14 | 45.43 | 1.53 | 0.996 |
| 307.30 | 100.00 | 8.05 | -3.12 | 11.17 | 8.05 | 0.07 | 107.15 | 45.50 | 1.53 | 0.996 |
| 307.40 | 100.00 | 8.06 | -3.13 | 11.19 | 8.06 | 0.06 | 107.16 | 45.56 | 1.53 | 0.996 |
| 307.50 | 100.00 | 8.06 | -3.14 | 11.20 | 8.06 | 0.05 | 107.16 | 45.62 | 1.53 | 0.996 |
| 307.60 | 100.00 | 8.07 | -3.15 | 11.22 | 8.07 | 0.04 | 107.17 | 45.69 | 1.53 | 0.996 |
| 307.70 | 100.00 | 8.07 | -3.16 | 11.23 | 8.07 | 0.03 | 107.17 | 45.75 | 1.53 | 0.996 |
| 307.80 | 100.00 | 8.08 | -3.17 | 11.25 | 8.08 | 0.02 | 107.18 | 45.82 | 1.53 | 0.996 |
| 307.90 | 100.00 | 8.08 | -3.18 | 11.26 | 8.08 | 0.01 | 107.18 | 45.88 | 1.53 | 0.996 |
| 308.00 | 100.00 | 8.09 | -3.19 | 11.28 | 8.09 | 0.00 | 107.18 | 45.95 | 1.54 | 0.996 |
| 308.10 | 100.00 | 8.09 | -3.20 | 11.29 | 8.09 | -0.01 | 107.18 | 46.01 | 1.54 | 0.996 |
| 308.20 | 100.00 | 8.10 | -3.21 | 11.31 | 8.10 | -0.02 | 107.18 | 46.07 | 1.54 | 0.996 |
| 308.30 | 100.00 | 8.10 | -3.23 | 11.33 | 8.10 | -0.04 | 107.17 | 46.14 | 1.54 | 0.996 |
| 308.40 | 100.00 | 8.11 | -3.24 | 11.35 | 8.11 | -0.05 | 107.17 | 46.20 | 1.54 | 0.996 |
| 308.50 | 100.00 | 8.11 | -3.25 | 11.36 | 8.11 | -0.06 | 107.17 | 46.27 | 1.54 | 0.996 |
| 308.60 | 100.00 | 8.12 | -3.26 | 11.38 | 8.12 | -0.07 | 107.16 | 46.33 | 1.54 | 0.996 |
| 308.70 | 100.00 | 8.12 | -3.27 | 11.39 | 8.12 | -0.08 | 107.16 | 46.40 | 1.54 | 0.996 |
| 308.80 | 100.00 | 8.13 | -3.29 | 11.42 | 8.13 | -0.09 | 107.15 | 46.46 | 1.54 | 0.996 |
| 308.90 | 100.00 | 8.13 | -3.30 | 11.43 | 8.13 | -0.10 | 107.14 | 46.52 | 1.54 | 0.996 |
| 309.00 | 100.00 | 8.14 | -3.31 | 11.45 | 8.14 | -0.11 | 107.13 | 46.59 | 1.54 | 0.996 |
| 309.10 | 100.00 | 8.14 | -3.32 | 11.46 | 8.14 | -0.12 | 107.12 | 46.65 | 1.55 | 0.996 |
| 309.20 | 100.00 | 8.15 | -3.34 | 11.49 | 8.15 | -0.13 | 107.11 | 46.72 | 1.55 | 0.996 |
| 309.30 | 100.00 | 8.15 | -3.35 | 11.50 | 8.15 | -0.14 | 107.10 | 46.78 | 1.55 | 0.996 |
| 309.40 | 100.00 | 8.16 | -3.36 | 11.52 | 8.16 | -0.15 | 107.08 | 46.85 | 1.55 | 0.996 |
| 309.50 | 100.00 | 8.16 | -3.38 | 11.54 | 8.16 | -0.16 | 107.07 | 46.91 | 1.55 | 0.996 |
| 309.60 | 100.00 | 8.17 | -3.39 | 11.56 | 8.17 | -0.17 | 107.06 | 46.98 | 1.55 | 0.996 |
| 309.70 | 100.00 | 8.17 | -3.41 | 11.58 | 8.17 | -0.18 | 107.04 | 47.04 | 1.55 | 0.996 |
| 309.80 | 100.00 | 8.18 | -3.42 | 11.60 | 8.18 | -0.19 | 107.02 | 47.11 | 1.55 | 0.996 |
| 309.90 | 100.00 | 8.18 | -3.44 | 11.62 | 8.18 | -0.20 | 107.00 | 47.17 | 1.55 | 0.996 |
| 310.00 | 100.00 | 8.19 | -3.45 | 11.64 | 8.19 | -0.21 | 106.98 | 47.24 | 1.55 | 0.996 |
| 310.10 | 100.00 | 8.19 | -3.46 | 11.65 | 8.19 | -0.22 | 106.96 | 47.30 | 1.56 | 0.996 |
| 310.20 | 100.00 | 8.20 | -3.48 | 11.68 | 8.20 | -0.23 | 106.94 | 47.37 | 1.56 | 0.996 |
| 310.30 | 100.00 | 8.20 | -3.50 | 11.70 | 8.20 | -0.24 | 106.92 | 47.43 | 1.56 | 0.996 |
| 310.40 | 100.00 | 8.21 | -3.51 | 11.72 | 8.21 | -0.25 | 106.90 | 47.50 | 1.56 | 0.996 |
| 310.50 | 100.00 | 8.21 | -3.53 | 11.74 | 8.21 | -0.26 | 106.87 | 47.57 | 1.56 | 0.996 |
| 310.60 | 100.00 | 8.22 | -3.54 | 11.76 | 8.22 | -0.27 | 106.85 | 47.63 | 1.56 | 0.996 |
| 310.70 | 100.00 | 8.23 | -3.56 | 11.79 | 8.23 | -0.29 | 106.82 | 47.70 | 1.56 | 0.996 |
| 310.80 | 100.00 | 8.23 | -3.58 | 11.81 | 8.23 | -0.30 | 106.79 | 47.77 | 1.56 | 0.996 |
| 310.90 | 100.00 | 8.24 | -3.59 | 11.83 | 8.24 | -0.31 | 106.76 | 47.83 | 1.56 | 0.996 |
| 311.00 | 100.00 | 8.24 | -3.61 | 11.85 | 8.24 | -0.32 | 106.73 | 47.90 | 1.56 | 0.996 |
| 311.10 | 100.00 | 8.25 | -3.63 | 11.88 | 8.25 | -0.33 | 106.70 | 47.97 | 1.57 | 0.996 |
| 311.20 | 100.00 | 8.25 | -3.64 | 11.89 | 8.25 | -0.34 | 106.67 | 48.04 | 1.57 | 0.996 |
| 311.30 | 100.00 | 8.26 | -3.66 | 11.92 | 8.26 | -0.35 | 106.63 | 48.10 | 1.57 | 0.996 |
| 311.40 | 100.00 | 8.27 | -3.68 | 11.95 | 8.27 | -0.36 | 106.60 | 48.17 | 1.57 | 0.996 |
| 311.50 | 100.00 | 8.27 | -3.70 | 11.97 | 8.27 | -0.37 | 106.56 | 48.24 | 1.57 | 0.996 |
| 311.60 | 100.00 | 8.28 | -3.72 | 12.00 | 8.28 | -0.38 | 106.53 | 48.31 | 1.57 | 0.996 |
| 311.70 | 100.00 | 8.28 | -3.73 | 12.01 | 8.28 | -0.39 | 106.49 | 48.38 | 1.57 | 0.996 |
| 311.80 | 100.00 | 8.29 | -3.75 | 12.04 | 8.29 | -0.40 | 106.45 | 48.45 | 1.57 | 0.996 |
| 311.90 | 100.00 | 8.30 | -3.77 | 12.07 | 8.30 | -0.41 | 106.41 | 48.52 | 1.57 | 0.996 |
| 312.00 | 100.00 | 8.30 | -3.79 | 12.09 | 8.30 | -0.42 | 106.37 | 48.59 | 1.58 | 0.996 |
| 312.10 | 100.00 | 8.31 | -3.81 | 12.12 | 8.31 | -0.44 | 106.33 | 48.66 | 1.58 | 0.996 |
| 312.20 | 100.00 | 8.31 | -3.83 | 12.14 | 8.31 | -0.45 | 106.28 | 48.73 | 1.58 | 0.996 |
| 312.30 | 100.00 | 8.32 | -3.85 | 12.17 | 8.32 | -0.46 | 106.24 | 48.80 | 1.58 | 0.996 |
| 312.40 | 100.00 | 8.33 | -3.87 | 12.20 | 8.33 | -0.47 | 106.19 | 48.87 | 1.58 | 0.996 |
| 312.50 | 100.00 | 8.33 | -3.89 | 12.22 | 8.33 | -0.48 | 106.14 | 48.94 | 1.58 | 0.996 |
| 312.60 | 100.00 | 8.34 | -3.91 | 12.25 | 8.34 | -0.49 | 106.10 | 49.02 | 1.58 | 0.996 |
| 312.70 | 100.00 | 8.34 | -3.94 | 12.28 | 8.34 | -0.50 | 106.05 | 49.09 | 1.58 | 0.996 |
| 312.80 | 100.00 | 8.35 | -3.96 | 12.31 | 8.35 | -0.51 | 106.00 | 49.16 | 1.58 | 0.996 |
| 312.90 | 100.00 | 8.36 | -3.98 | 12.34 | 8.36 | -0.52 | 105.94 | 49.24 | 1.59 | 0.996 |
| 313.00 | 100.00 | 8.36 | -4.00 | 12.36 | 8.36 | -0.53 | 105.89 | 49.31 | 1.59 | 0.996 |
| 313.10 | 100.00 | 8.37 | -4.02 | 12.39 | 8.37 | -0.54 | 105.84 | 49.39 | 1.59 | 0.996 |
| 313.20 | 100.00 | 8.38 | -4.05 | 12.43 | 8.38 | -0.55 | 105.78 | 49.46 | 1.59 | 0.996 |
| 313.30 | 100.00 | 8.38 | -4.07 | 12.45 | 8.38 | -0.57 | 105.73 | 49.54 | 1.59 | 0.996 |
| 313.40 | 100.00 | 8.39 | -4.09 | 12.48 | 8.39 | -0.58 | 105.67 | 49.61 | 1.59 | 0.996 |
| 313.50 | 100.00 | 8.40 | -4.12 | 12.52 | 8.40 | -0.59 | 105.61 | 49.69 | 1.59 | 0.996 |
| 313.60 | 100.00 | 8.40 | -4.14 | 12.54 | 8.40 | -0.60 | 105.55 | 49.77 | 1.60 | 0.996 |
| 313.70 | 100.00 | 8.41 | -4.16 | 12.57 | 8.41 | -0.61 | 105.49 | 49.84 | 1.60 | 0.996 |
| 313.80 | 100.00 | 8.42 | -4.19 | 12.61 | 8.42 | -0.62 | 105.43 | 49.92 | 1.60 | 0.996 |
| 313.90 | 100.00 | 8.42 | -4.21 | 12.63 | 8.42 | -0.63 | 105.36 | 50.00 | 1.60 | 0.996 |
| 314.00 | 100.00 | 8.43 | -4.24 | 12.67 | 8.43 | -0.64 | 105.30 | 50.08 | 1.60 | 0.996 |
| 314.10 | 100.00 | 8.44 | -4.26 | 12.70 | 8.44 | -0.65 | 105.23 | 50.16 | 1.60 | 0.996 |
| 314.20 | 100.00 | 8.44 | -4.29 | 12.73 | 8.44 | -0.66 | 105.16 | 50.24 | 1.60 | 0.996 |
| 314.30 | 100.00 | 8.45 | -4.32 | 12.77 | 8.45 | -0.67 | 105.10 | 50.32 | 1.61 | 0.996 |
| 314.40 | 100.00 | 8.46 | -4.34 | 12.80 | 8.46 | -0.69 | 105.03 | 50.41 | 1.61 | 0.996 |
| 314.50 | 100.00 | 8.46 | -4.37 | 12.83 | 8.46 | -0.70 | 104.96 | 50.49 | 1.61 | 0.996 |
| 314.60 | 100.00 | 8.47 | -4.40 | 12.87 | 8.47 | -0.71 | 104.88 | 50.57 | 1.61 | 0.996 |
| 314.70 | 100.00 | 8.48 | -4.43 | 12.91 | 8.48 | -0.72 | 104.81 | 50.65 | 1.61 | 0.996 |
| 314.80 | 100.00 | 8.48 | -4.45 | 12.93 | 8.48 | -0.73 | 104.74 | 50.74 | 1.61 | 0.996 |
| 314.90 | 100.00 | 8.49 | -4.48 | 12.97 | 8.49 | -0.74 | 104.66 | 50.82 | 1.61 | 0.996 |
| 315.00 | 100.00 | 8.50 | -4.51 | 13.01 | 8.50 | -0.75 | 104.59 | 50.91 | 1.62 | 0.996 |
| 315.10 | 100.00 | 8.51 | -4.54 | 13.05 | 8.51 | -0.76 | 104.51 | 51.00 | 1.62 | 0.996 |
| 315.20 | 100.00 | 8.51 | -4.57 | 13.08 | 8.51 | -0.77 | 104.43 | 51.08 | 1.62 | 0.996 |
| 315.30 | 100.00 | 8.52 | -4.60 | 13.12 | 8.52 | -0.78 | 104.35 | 51.17 | 1.62 | 0.996 |
| 315.40 | 100.00 | 8.53 | -4.63 | 13.16 | 8.53 | -0.79 | 104.27 | 51.26 | 1.62 | 0.996 |
| 315.50 | 100.00 | 8.53 | -4.66 | 13.19 | 8.53 | -0.81 | 104.19 | 51.35 | 1.62 | 0.996 |
| 315.60 | 100.00 | 8.54 | -4.69 | 13.23 | 8.54 | -0.82 | 104.10 | 51.44 | 1.63 | 0.996 |
| 315.70 | 100.00 | 8.55 | -4.72 | 13.27 | 8.55 | -0.83 | 104.02 | 51.53 | 1.63 | 0.996 |
| 315.80 | 100.00 | 8.56 | -4.75 | 13.31 | 8.56 | -0.84 | 103.93 | 51.62 | 1.63 | 0.996 |
| 315.90 | 100.00 | 8.56 | -4.79 | 13.35 | 8.56 | -0.85 | 103.84 | 51.72 | 1.63 | 0.996 |
| 316.00 | 100.00 | 8.57 | -4.82 | 13.39 | 8.57 | -0.86 | 103.76 | 51.81 | 1.63 | 0.996 |
| 316.10 | 100.00 | 8.58 | -4.85 | 13.43 | 8.58 | -0.87 | 103.67 | 51.90 | 1.63 | 0.996 |
| 316.20 | 100.00 | 8.59 | -4.89 | 13.48 | 8.59 | -0.88 | 103.58 | 52.00 | 1.64 | 0.996 |
| 316.30 | 100.00 | 8.60 | -4.92 | 13.52 | 8.60 | -0.89 | 103.48 | 52.10 | 1.64 | 0.996 |
| 316.40 | 100.00 | 8.60 | -4.96 | 13.56 | 8.60 | -0.90 | 103.39 | 52.19 | 1.64 | 0.996 |
| 316.50 | 100.00 | 8.61 | -4.99 | 13.60 | 8.61 | -0.91 | 103.30 | 52.29 | 1.64 | 0.996 |
| 316.60 | 100.00 | 8.62 | -5.03 | 13.65 | 8.62 | -0.92 | 103.20 | 52.39 | 1.64 | 0.996 |
| 316.70 | 100.00 | 8.63 | -5.06 | 13.69 | 8.63 | -0.94 | 103.10 | 52.49 | 1.65 | 0.996 |
| 316.80 | 100.00 | 8.63 | -5.10 | 13.73 | 8.63 | -0.96 | 103.00 | 52.59 | 1.65 | 0.996 |
| 316.90 | 100.00 | 8.64 | -5.14 | 13.78 | 8.64 | -0.99 | 102.91 | 52.69 | 1.65 | 0.996 |
| 317.00 | 100.00 | 8.65 | -5.17 | 13.82 | 8.65 | -1.01 | 102.81 | 52.80 | 1.65 | 0.996 |
| 317.10 | 100.00 | 8.66 | -5.21 | 13.87 | 8.66 | -1.03 | 102.70 | 52.90 | 1.65 | 0.996 |
| 317.20 | 100.00 | 8.67 | -5.25 | 13.92 | 8.67 | -1.05 | 102.60 | 53.00 | 1.66 | 0.996 |
| 317.30 | 100.00 | 8.67 | -5.29 | 13.96 | 8.67 | -1.07 | 102.50 | 53.11 | 1.66 | 0.996 |
| 317.40 | 100.00 | 8.68 | -5.33 | 14.01 | 8.68 | -1.09 | 102.39 | 53.22 | 1.66 | 0.996 |
| 317.50 | 100.00 | 8.69 | -5.37 | 14.06 | 8.69 | -1.11 | 102.29 | 53.33 | 1.66 | 0.996 |
| 317.60 | 100.00 | 8.70 | -5.41 | 14.11 | 8.70 | -1.13 | 102.18 | 53.44 | 1.67 | 0.996 |
| 317.70 | 100.00 | 8.71 | -5.45 | 14.16 | 8.71 | -1.16 | 102.07 | 53.55 | 1.67 | 0.996 |
| 317.80 | 100.00 | 8.72 | -5.49 | 14.21 | 8.72 | -1.18 | 101.96 | 53.66 | 1.67 | 0.996 |
| 317.90 | 100.00 | 8.72 | -5.53 | 14.25 | 8.72 | -1.20 | 101.85 | 53.77 | 1.67 | 0.996 |
| 318.00 | 100.00 | 8.73 | -5.58 | 14.31 | 8.73 | -1.22 | 101.74 | 53.88 | 1.67 | 0.996 |
| 318.10 | 100.00 | 8.74 | -5.62 | 14.36 | 8.74 | -1.24 | 101.63 | 54.00 | 1.68 | 0.996 |
| 318.20 | 100.00 | 8.75 | -5.67 | 14.42 | 8.75 | -1.27 | 101.51 | 54.11 | 1.68 | 0.996 |
| 318.30 | 100.00 | 8.76 | -5.71 | 14.47 | 8.76 | -1.29 | 101.40 | 54.23 | 1.68 | 0.996 |
| 318.40 | 100.00 | 8.77 | -5.76 | 14.53 | 8.77 | -1.31 | 101.28 | 54.35 | 1.68 | 0.996 |
| 318.50 | 100.00 | 8.78 | -5.80 | 14.58 | 8.78 | -1.33 | 101.16 | 54.47 | 1.69 | 0.996 |
| 318.60 | 100.00 | 8.79 | -5.85 | 14.64 | 8.79 | -1.36 | 101.04 | 54.59 | 1.69 | 0.996 |
| 318.70 | 100.00 | 8.79 | -5.90 | 14.69 | 8.79 | -1.38 | 100.93 | 54.71 | 1.69 | 0.996 |
| 318.80 | 100.00 | 8.80 | -5.94 | 14.74 | 8.80 | -1.40 | 100.80 | 54.83 | 1.69 | 0.996 |
| 318.90 | 100.00 | 8.81 | -5.99 | 14.80 | 8.81 | -1.43 | 100.68 | 54.96 | 1.70 | 0.996 |
| 319.00 | 100.00 | 8.82 | -6.04 | 14.86 | 8.82 | -1.45 | 100.56 | 55.08 | 1.70 | 0.996 |
| 319.10 | 100.00 | 8.83 | -6.09 | 14.92 | 8.83 | -1.47 | 100.44 | 55.21 | 1.70 | 0.996 |
| 319.20 | 100.00 | 8.84 | -6.14 | 14.98 | 8.84 | -1.50 | 100.31 | 55.34 | 1.71 | 0.996 |
| 319.30 | 100.00 | 8.85 | -6.20 | 15.05 | 8.85 | -1.52 | 100.18 | 55.47 | 1.71 | 0.995 |
| 319.40 | 100.00 | 8.86 | -6.25 | 15.11 | 8.86 | -1.54 | 100.06 | 55.60 | 1.71 | 0.995 |
| 319.50 | 100.00 | 8.87 | -6.30 | 15.17 | 8.87 | -1.57 | 99.93 | 55.73 | 1.71 | 0.995 |
| 319.60 | 100.00 | 8.88 | -6.36 | 15.24 | 8.88 | -1.59 | 99.80 | 55.87 | 1.72 | 0.995 |
| 319.70 | 100.00 | 8.89 | -6.41 | 15.30 | 8.89 | -1.61 | 99.67 | 56.00 | 1.72 | 0.995 |
| 319.80 | 100.00 | 8.90 | -6.47 | 15.37 | 8.90 | -1.64 | 99.54 | 56.14 | 1.72 | 0.995 |
| 319.90 | 100.00 | 8.90 | -6.52 | 15.42 | 8.90 | -1.66 | 99.41 | 56.28 | 1.73 | 0.995 |
| 320.00 | 100.00 | 8.91 | -6.58 | 15.49 | 8.91 | -1.69 | 99.27 | 56.42 | 1.73 | 0.995 |
| 320.10 | 100.00 | 8.92 | -6.64 | 15.56 | 8.92 | -1.71 | 99.14 | 56.56 | 1.73 | 0.995 |
| 320.20 | 100.00 | 8.93 | -6.70 | 15.63 | 8.93 | -1.74 | 99.00 | 56.70 | 1.74 | 0.995 |
| 320.30 | 100.00 | 8.94 | -6.76 | 15.70 | 8.94 | -1.76 | 98.87 | 56.85 | 1.74 | 0.995 |
| 320.40 | 100.00 | 8.95 | -6.82 | 15.77 | 8.95 | -1.79 | 98.73 | 56.99 | 1.74 | 0.995 |
| 320.50 | 100.00 | 8.96 | -6.88 | 15.84 | 8.96 | -1.81 | 98.59 | 57.14 | 1.75 | 0.995 |
| 320.60 | 100.00 | 8.97 | -6.94 | 15.91 | 8.97 | -1.84 | 98.45 | 57.29 | 1.75 | 0.995 |
| 320.70 | 100.00 | 8.98 | -7.01 | 15.99 | 8.98 | -1.86 | 98.31 | 57.44 | 1.75 | 0.995 |
| 320.80 | 100.00 | 8.99 | -7.07 | 16.06 | 8.99 | -1.89 | 98.17 | 57.59 | 1.76 | 0.995 |
| 320.90 | 100.00 | 9.00 | -7.14 | 16.14 | 9.00 | -1.92 | 98.03 | 57.74 | 1.76 | 0.995 |
| 321.00 | 100.00 | 9.01 | -7.21 | 16.22 | 9.01 | -1.94 | 97.88 | 57.90 | 1.76 | 0.995 |
| 321.10 | 100.00 | 9.02 | -7.27 | 16.29 | 9.02 | -1.97 | 97.74 | 58.06 | 1.77 | 0.995 |
| 321.20 | 100.00 | 9.03 | -7.34 | 16.37 | 9.03 | -1.99 | 97.60 | 58.21 | 1.77 | 0.995 |
| 321.30 | 100.00 | 9.04 | -7.41 | 16.45 | 9.04 | -2.02 | 97.45 | 58.37 | 1.77 | 0.995 |
| 321.40 | 100.00 | 9.05 | -7.48 | 16.53 | 9.05 | -2.05 | 97.30 | 58.54 | 1.78 | 0.995 |
| 321.50 | 100.00 | 9.06 | -7.56 | 16.62 | 9.06 | -2.07 | 97.16 | 58.70 | 1.78 | 0.995 |
| 321.60 | 100.00 | 9.07 | -7.63 | 16.70 | 9.07 | -2.10 | 97.01 | 58.87 | 1.79 | 0.995 |
| 321.70 | 100.00 | 9.08 | -7.71 | 16.79 | 9.08 | -2.13 | 96.86 | 59.03 | 1.79 | 0.995 |
| 321.80 | 100.00 | 9.09 | -7.78 | 16.87 | 9.09 | -2.15 | 96.71 | 59.20 | 1.79 | 0.995 |
| 321.90 | 100.00 | 9.11 | -7.86 | 16.97 | 9.11 | -2.18 | 96.56 | 59.37 | 1.80 | 0.995 |
| 322.00 | 100.00 | 9.12 | -7.94 | 17.06 | 9.12 | -2.21 | 96.40 | 59.54 | 1.80 | 0.995 |
| 322.10 | 100.00 | 9.13 | -8.02 | 17.15 | 9.13 | -2.24 | 96.25 | 59.72 | 1.81 | 0.995 |
| 322.20 | 100.00 | 9.14 | -8.10 | 17.24 | 9.14 | -2.26 | 96.10 | 59.89 | 1.81 | 0.995 |
| 322.30 | 100.00 | 9.15 | -8.19 | 17.34 | 9.15 | -2.29 | 95.94 | 60.07 | 1.81 | 0.995 |
| 322.40 | 100.00 | 9.16 | -8.27 | 17.43 | 9.16 | -2.32 | 95.79 | 60.25 | 1.82 | 0.995 |
| 322.50 | 100.00 | 9.17 | -8.36 | 17.53 | 9.17 | -2.35 | 95.63 | 60.43 | 1.82 | 0.995 |
| 322.60 | 100.00 | 9.18 | -8.44 | 17.62 | 9.18 | -2.38 | 95.48 | 60.62 | 1.83 | 0.995 |
| 322.70 | 100.00 | 9.19 | -8.53 | 17.72 | 9.19 | -2.40 | 95.32 | 60.80 | 1.83 | 0.995 |
| 322.80 | 100.00 | 9.20 | -8.63 | 17.83 | 9.20 | -2.43 | 95.16 | 60.99 | 1.84 | 0.995 |
| 322.90 | 100.00 | 9.21 | -8.72 | 17.93 | 9.21 | -2.46 | 95.00 | 61.18 | 1.84 | 0.995 |
| 323.00 | 100.00 | 9.22 | -8.81 | 18.03 | 9.22 | -2.49 | 94.85 | 61.37 | 1.85 | 0.995 |
| 323.10 | 100.00 | 9.24 | -8.91 | 18.15 | 9.24 | -2.52 | 94.69 | 61.56 | 1.85 | 0.995 |
| 323.20 | 100.00 | 9.25 | -9.01 | 18.26 | 9.25 | -2.55 | 94.52 | 61.76 | 1.86 | 0.995 |
| 323.30 | 100.00 | 9.26 | -9.11 | 18.37 | 9.26 | -2.58 | 94.36 | 61.96 | 1.86 | 0.995 |
| 323.40 | 100.00 | 9.27 | -9.21 | 18.48 | 9.27 | -2.61 | 94.20 | 62.15 | 1.87 | 0.995 |
| 323.50 | 100.00 | 9.28 | -9.31 | 18.59 | 9.28 | -2.64 | 94.04 | 62.36 | 1.87 | 0.995 |
| 323.60 | 100.00 | 9.29 | -9.42 | 18.71 | 9.29 | -2.67 | 93.88 | 62.56 | 1.88 | 0.995 |
| 323.70 | 100.00 | 9.30 | -9.53 | 18.83 | 9.30 | -2.69 | 93.71 | 62.76 | 1.88 | 0.995 |
| 323.80 | 100.00 | 9.32 | -9.64 | 18.96 | 9.32 | -2.72 | 93.55 | 62.97 | 1.89 | 0.995 |
| 323.90 | 100.00 | 9.33 | -9.75 | 19.08 | 9.33 | -2.75 | 93.38 | 63.18 | 1.89 | 0.995 |
| 324.00 | 100.00 | 9.34 | -9.87 | 19.21 | 9.34 | -2.78 | 93.22 | 63.39 | 1.90 | 0.995 |
| 324.10 | 100.00 | 9.35 | -9.98 | 19.33 | 9.35 | -2.81 | 93.05 | 63.61 | 1.90 | 0.995 |
| 324.20 | 100.00 | 9.36 | -10.10 | 19.46 | 9.36 | -2.85 | 92.89 | 63.82 | 1.91 | 0.995 |
| 324.30 | 100.00 | 9.37 | -10.23 | 19.60 | 9.37 | -2.88 | 92.72 | 64.04 | 1.91 | 0.995 |
| 324.40 | 100.00 | 9.39 | -10.35 | 19.74 | 9.39 | -2.91 | 92.56 | 64.26 | 1.92 | 0.995 |
| 324.50 | 100.00 | 9.40 | -10.48 | 19.88 | 9.40 | -2.94 | 92.39 | 64.48 | 1.92 | 0.995 |
| 324.60 | 100.00 | 9.41 | -10.61 | 20.02 | 9.41 | -2.97 | 92.22 | 64.71 | 1.93 | 0.995 |
| 324.70 | 100.00 | 9.42 | -10.74 | 20.16 | 9.42 | -3.00 | 92.05 | 64.93 | 1.94 | 0.995 |
| 324.80 | 100.00 | 9.43 | -10.88 | 20.31 | 9.43 | -3.03 | 91.88 | 65.16 | 1.94 | 0.995 |
| 324.90 | 100.00 | 9.45 | -11.02 | 20.47 | 9.45 | -3.06 | 91.71 | 65.39 | 1.95 | 0.995 |
| 325.00 | 100.00 | 9.46 | -11.16 | 20.62 | 9.46 | -3.09 | 91.55 | 65.63 | 1.95 | 0.995 |
| 325.10 | 100.00 | 9.47 | -11.31 | 20.78 | 9.47 | -3.12 | 91.38 | 65.86 | 1.96 | 0.995 |
| 325.20 | 100.00 | 9.48 | -11.46 | 20.94 | 9.48 | -3.15 | 91.21 | 66.10 | 1.97 | 0.995 |
| 325.30 | 100.00 | 9.49 | -11.61 | 21.10 | 9.49 | -3.19 | 91.04 | 66.34 | 1.97 | 0.995 |
| 325.40 | 100.00 | 9.51 | -11.77 | 21.28 | 9.51 | -3.22 | 90.87 | 66.58 | 1.98 | 0.995 |
| 325.50 | 100.00 | 9.52 | -11.93 | 21.45 | 9.52 | -3.25 | 90.70 | 66.83 | 1.99 | 0.995 |
| 325.60 | 100.00 | 9.53 | -12.09 | 21.62 | 9.53 | -3.28 | 90.53 | 67.08 | 1.99 | 0.995 |
| 325.70 | 100.00 | 9.54 | -12.26 | 21.80 | 9.54 | -3.31 | 90.36 | 67.33 | 2.00 | 0.995 |
| 325.80 | 100.00 | 9.56 | -12.44 | 22.00 | 9.56 | -3.34 | 90.18 | 67.58 | 2.01 | 0.995 |
| 325.90 | 100.00 | 9.57 | -12.61 | 22.18 | 9.57 | -3.38 | 90.01 | 67.83 | 2.01 | 0.995 |
| 326.00 | 100.00 | 9.58 | -12.80 | 22.38 | 9.58 | -3.41 | 89.84 | 68.09 | 2.02 | 0.995 |
| 326.10 | 100.00 | 9.60 | -12.98 | 22.58 | 9.60 | -3.44 | 89.67 | 68.35 | 2.03 | 0.995 |
| 326.20 | 100.00 | 9.61 | -13.18 | 22.79 | 9.61 | -3.47 | 89.50 | 68.61 | 2.03 | 0.995 |
| 326.30 | 100.00 | 9.62 | -13.37 | 22.99 | 9.62 | -3.50 | 89.33 | 68.88 | 2.04 | 0.995 |
| 326.40 | 100.00 | 9.63 | -13.58 | 23.21 | 9.63 | -3.54 | 89.16 | 69.14 | 2.05 | 0.995 |
| 326.50 | 100.00 | 9.65 | -13.79 | 23.44 | 9.65 | -3.57 | 88.99 | 69.41 | 2.06 | 0.995 |
| 326.60 | 100.00 | 9.66 | -14.00 | 23.66 | 9.66 | -3.60 | 88.82 | 69.68 | 2.06 | 0.995 |
| 326.70 | 100.00 | 9.67 | -14.22 | 23.89 | 9.67 | -3.63 | 88.65 | 69.96 | 2.07 | 0.995 |
| 326.80 | 100.00 | 9.69 | -14.45 | 24.14 | 9.69 | -3.67 | 88.48 | 70.23 | 2.08 | 0.995 |
| 326.90 | 100.00 | 9.70 | -14.69 | 24.39 | 9.70 | -3.70 | 88.31 | 70.51 | 2.09 | 0.995 |
| 327.00 | 100.00 | 9.71 | -14.93 | 24.64 | 9.71 | -3.73 | 88.13 | 70.79 | 2.09 | 0.995 |
| 327.10 | 100.00 | 9.72 | -15.18 | 24.90 | 9.72 | -3.76 | 87.96 | 71.08 | 2.10 | 0.995 |
| 327.20 | 100.00 | 9.74 | -15.43 | 25.17 | 9.74 | -3.80 | 87.79 | 71.37 | 2.11 | 0.995 |
| 327.30 | 100.00 | 9.75 | -15.70 | 25.45 | 9.75 | -3.83 | 87.63 | 71.65 | 2.12 | 0.995 |
| 327.40 | 100.00 | 9.76 | -15.97 | 25.73 | 9.76 | -3.86 | 87.46 | 71.95 | 2.13 | 0.995 |
| 327.50 | 100.00 | 9.78 | -16.25 | 26.03 | 9.78 | -3.89 | 87.29 | 72.24 | 2.13 | 0.995 |
| 327.60 | 100.00 | 9.79 | -16.54 | 26.33 | 9.79 | -3.92 | 87.12 | 72.54 | 2.14 | 0.995 |
| 327.70 | 100.00 | 9.80 | -16.83 | 26.63 | 9.80 | -3.96 | 86.95 | 72.84 | 2.15 | 0.995 |
| 327.80 | 100.00 | 9.82 | -17.14 | 26.96 | 9.82 | -3.99 | 86.78 | 73.14 | 2.16 | 0.995 |
| 327.90 | 100.00 | 9.83 | -17.45 | 27.28 | 9.83 | -4.02 | 86.61 | 73.45 | 2.17 | 0.995 |
| 328.00 | 100.00 | 9.84 | -17.78 | 27.62 | 9.84 | -4.05 | 86.45 | 73.75 | 2.18 | 0.995 |
| 328.10 | 100.00 | 9.86 | -18.11 | 27.97 | 9.86 | -4.08 | 86.28 | 74.06 | 2.19 | 0.995 |
| 328.20 | 100.00 | 9.87 | -18.45 | 28.32 | 9.87 | -4.11 | 86.12 | 74.38 | 2.19 | 0.995 |
| 328.30 | 100.00 | 9.89 | -18.79 | 28.68 | 9.89 | -4.14 | 85.95 | 74.69 | 2.20 | 0.995 |
| 328.40 | 100.00 | 9.90 | -19.14 | 29.04 | 9.90 | -4.18 | 85.79 | 75.01 | 2.21 | 0.995 |
| 328.50 | 100.00 | 9.91 | -19.49 | 29.40 | 9.91 | -4.21 | 85.62 | 75.33 | 2.22 | 0.995 |
| 328.60 | 100.00 | 9.93 | -19.85 | 29.78 | 9.93 | -4.24 | 85.46 | 75.66 | 2.23 | 0.995 |
| 328.70 | 100.00 | 9.94 | -20.20 | 30.14 | 9.94 | -4.27 | 85.30 | 75.98 | 2.24 | 0.995 |
| 328.80 | 100.00 | 9.95 | -20.55 | 30.50 | 9.95 | -4.30 | 85.13 | 76.31 | 2.25 | 0.995 |
| 328.90 | 100.00 | 9.97 | -20.89 | 30.86 | 9.97 | -4.33 | 84.97 | 76.64 | 2.26 | 0.995 |
| 329.00 | 100.00 | 9.98 | -21.21 | 31.19 | 9.98 | -4.36 | 84.81 | 76.98 | 2.27 | 0.995 |
| 329.10 | 100.00 | 10.00 | -21.51 | 31.51 | 10.00 | -4.39 | 84.65 | 77.31 | 2.28 | 0.995 |
| 329.20 | 100.00 | 10.01 | -21.78 | 31.79 | 10.01 | -4.42 | 84.49 | 77.65 | 2.29 | 0.995 |
| 329.30 | 100.00 | 10.02 | -22.00 | 32.02 | 10.02 | -4.44 | 84.34 | 78.00 | 2.30 | 0.995 |
| 329.40 | 100.00 | 10.04 | -22.18 | 32.22 | 10.04 | -4.47 | 84.18 | 78.34 | 2.31 | 0.995 |
| 329.50 | 100.00 | 10.05 | -22.30 | 32.35 | 10.05 | -4.50 | 84.02 | 78.69 | 2.32 | 0.995 |
| 329.60 | 100.00 | 10.06 | -22.35 | 32.41 | 10.06 | -4.53 | 83.87 | 79.04 | 2.33 | 0.995 |
| 329.70 | 100.00 | 10.08 | -22.34 | 32.42 | 10.08 | -4.55 | 83.72 | 79.40 | 2.34 | 0.995 |
| 329.80 | 100.00 | 10.09 | -22.26 | 32.35 | 10.09 | -4.58 | 83.56 | 79.75 | 2.35 | 0.995 |
| 329.90 | 100.00 | 10.11 | -22.12 | 32.23 | 10.11 | -4.61 | 83.41 | 80.11 | 2.36 | 0.995 |
| 330.00 | 100.00 | 10.12 | -21.91 | 32.03 | 10.12 | -4.63 | 83.26 | 80.47 | 2.37 | 0.995 |
| 330.10 | 100.00 | 10.13 | -21.65 | 31.78 | 10.13 | -4.66 | 83.11 | 80.84 | 2.38 | 0.995 |
| 330.20 | 100.00 | 10.15 | -21.35 | 31.50 | 10.15 | -4.68 | 82.97 | 81.21 | 2.39 | 0.995 |
| 330.30 | 100.00 | 10.16 | -21.01 | 31.17 | 10.16 | -4.71 | 82.82 | 81.58 | 2.40 | 0.995 |
| 330.40 | 100.00 | 10.18 | -20.64 | 30.82 | 10.18 | -4.73 | 82.68 | 81.95 | 2.42 | 0.995 |
| 330.50 | 100.00 | 10.19 | -20.25 | 30.44 | 10.19 | -4.75 | 82.53 | 82.33 | 2.43 | 0.995 |
| 330.60 | 100.00 | 10.20 | -19.85 | 30.05 | 10.20 | -4.78 | 82.39 | 82.70 | 2.44 | 0.995 |
| 330.70 | 100.00 | 10.22 | -19.44 | 29.66 | 10.22 | -4.80 | 82.25 | 83.09 | 2.45 | 0.995 |
| 330.80 | 100.00 | 10.23 | -19.03 | 29.26 | 10.23 | -4.82 | 82.11 | 83.47 | 2.46 | 0.995 |
| 330.90 | 100.00 | 10.25 | -18.61 | 28.86 | 10.25 | -4.84 | 81.97 | 83.86 | 2.47 | 0.995 |
| 331.00 | 100.00 | 10.26 | -18.20 | 28.46 | 10.26 | -4.86 | 81.84 | 84.25 | 2.48 | 0.995 |
| 331.10 | 100.00 | 10.27 | -17.79 | 28.06 | 10.27 | -4.88 | 81.70 | 84.64 | 2.50 | 0.995 |
| 331.20 | 100.00 | 10.29 | -17.39 | 27.68 | 10.29 | -4.90 | 81.57 | 85.04 | 2.51 | 0.995 |
| 331.30 | 100.00 | 10.30 | -17.00 | 27.30 | 10.30 | -4.91 | 81.44 | 85.43 | 2.52 | 0.995 |
| 331.40 | 100.00 | 10.32 | -16.62 | 26.94 | 10.32 | -4.93 | 81.31 | 85.83 | 2.53 | 0.995 |
| 331.50 | 100.00 | 10.33 | -16.24 | 26.57 | 10.33 | -4.95 | 81.19 | 86.24 | 2.54 | 0.995 |
| 331.60 | 100.00 | 10.34 | -15.87 | 26.21 | 10.34 | -4.96 | 81.06 | 86.65 | 2.56 | 0.995 |
| 331.70 | 100.00 | 10.36 | -15.51 | 25.87 | 10.36 | -5.04 | 80.94 | 87.05 | 2.57 | 0.995 |
| 331.80 | 100.00 | 10.37 | -15.16 | 25.53 | 10.37 | -5.12 | 80.82 | 87.47 | 2.58 | 0.995 |
| 331.90 | 100.00 | 10.39 | -14.82 | 25.21 | 10.39 | -5.20 | 80.70 | 87.88 | 2.59 | 0.995 |
| 332.00 | 100.00 | 10.40 | -14.49 | 24.89 | 10.40 | -5.28 | 80.58 | 88.30 | 2.61 | 0.995 |
| 332.10 | 100.00 | 10.41 | -14.16 | 24.57 | 10.41 | -5.36 | 80.47 | 88.72 | 2.62 | 0.995 |
| 332.20 | 100.00 | 10.43 | -13.84 | 24.27 | 10.43 | -5.44 | 80.36 | 89.14 | 2.63 | 0.995 |
| 332.30 | 100.00 | 10.44 | -13.53 | 23.97 | 10.44 | -5.52 | 80.25 | 89.57 | 2.64 | 0.995 |
| 332.40 | 100.00 | 10.45 | -13.23 | 23.68 | 10.45 | -5.61 | 80.14 | 90.00 | 2.66 | 0.995 |
| 332.50 | 100.00 | 10.47 | -12.94 | 23.41 | 10.47 | -5.69 | 80.04 | 90.43 | 2.67 | 0.995 |
| 332.60 | 100.00 | 10.48 | -12.65 | 23.13 | 10.48 | -5.78 | 79.93 | 90.86 | 2.68 | 0.995 |
| 332.70 | 100.00 | 10.50 | -12.37 | 22.87 | 10.50 | -5.87 | 79.83 | 91.30 | 2.70 | 0.995 |
| 332.80 | 100.00 | 10.51 | -12.09 | 22.60 | 10.51 | -5.96 | 79.74 | 91.74 | 2.71 | 0.995 |
| 332.90 | 100.00 | 10.52 | -11.82 | 22.34 | 10.52 | -6.05 | 79.64 | 92.18 | 2.72 | 0.995 |
| 333.00 | 100.00 | 10.54 | -11.56 | 22.10 | 10.54 | -6.14 | 79.55 | 92.62 | 2.74 | 0.995 |
| 333.10 | 100.00 | 10.55 | -11.30 | 21.85 | 10.55 | -6.24 | 79.46 | 93.07 | 2.75 | 0.995 |
| 333.20 | 100.00 | 10.56 | -11.05 | 21.61 | 10.56 | -6.33 | 79.37 | 93.52 | 2.76 | 0.995 |
| 333.30 | 100.00 | 10.58 | -10.81 | 21.39 | 10.58 | -6.43 | 79.29 | 93.97 | 2.78 | 0.995 |
| 333.40 | 100.00 | 10.59 | -10.56 | 21.15 | 10.59 | -6.52 | 79.21 | 94.43 | 2.79 | 0.995 |
| 333.50 | 100.00 | 10.60 | -10.33 | 20.93 | 10.60 | -6.62 | 79.13 | 94.88 | 2.80 | 0.995 |
| 333.60 | 100.00 | 10.62 | -10.10 | 20.72 | 10.62 | -6.72 | 79.06 | 95.34 | 2.82 | 0.995 |
| 333.70 | 100.00 | 10.63 | -9.87 | 20.50 | 10.63 | -6.82 | 78.99 | 95.81 | 2.83 | 0.995 |
| 333.80 | 100.00 | 10.64 | -9.65 | 20.29 | 10.64 | -6.92 | 78.92 | 96.27 | 2.85 | 0.995 |
| 333.90 | 100.00 | 10.65 | -9.43 | 20.08 | 10.65 | -7.03 | 78.85 | 96.74 | 2.86 | 0.995 |
| 334.00 | 100.00 | 10.67 | -9.22 | 19.89 | 10.67 | -7.13 | 78.79 | 97.21 | 2.87 | 0.995 |
| 334.10 | 100.00 | 10.68 | -9.01 | 19.69 | 10.68 | -7.24 | 78.73 | 97.68 | 2.89 | 0.995 |
| 334.20 | 100.00 | 10.69 | -8.80 | 19.49 | 10.69 | -7.35 | 78.68 | 98.15 | 2.90 | 0.995 |
| 334.30 | 100.00 | 10.71 | -8.60 | 19.31 | 10.71 | -7.46 | 78.63 | 98.63 | 2.92 | 0.995 |
| 334.40 | 100.00 | 10.72 | -8.40 | 19.12 | 10.72 | -7.57 | 78.58 | 99.11 | 2.93 | 0.995 |
| 334.50 | 100.00 | 10.73 | -8.21 | 18.94 | 10.73 | -7.68 | 78.54 | 99.59 | 2.94 | 0.995 |
| 334.60 | 100.00 | 10.74 | -8.02 | 18.76 | 10.74 | -7.79 | 78.50 | 100.07 | 2.96 | 0.995 |
| 334.70 | 100.00 | 10.76 | -7.83 | 18.59 | 10.76 | -7.90 | 78.46 | 100.56 | 2.97 | 0.995 |
| 334.80 | 100.00 | 10.77 | -7.64 | 18.41 | 10.77 | -8.02 | 78.43 | 101.05 | 2.99 | 0.995 |
| 334.90 | 100.00 | 10.78 | -7.46 | 18.24 | 10.78 | -8.14 | 78.40 | 101.54 | 3.00 | 0.995 |
| 335.00 | 100.00 | 10.79 | -7.28 | 18.07 | 10.79 | -8.25 | 78.37 | 102.03 | 3.02 | 0.995 |
| 335.10 | 100.00 | 10.80 | -7.11 | 17.91 | 10.80 | -8.37 | 78.35 | 102.52 | 3.03 | 0.995 |
| 335.20 | 100.00 | 10.82 | -6.93 | 17.75 | 10.82 | -8.49 | 78.34 | 103.02 | 3.04 | 0.995 |
| 335.30 | 100.00 | 10.83 | -6.76 | 17.59 | 10.83 | -8.61 | 78.33 | 103.52 | 3.06 | 0.995 |
| 335.40 | 100.00 | 10.84 | -6.60 | 17.44 | 10.84 | -8.74 | 78.32 | 104.02 | 3.07 | 0.995 |
| 335.50 | 100.00 | 10.85 | -6.43 | 17.28 | 10.85 | -8.86 | 78.32 | 104.52 | 3.09 | 0.995 |
| 335.60 | 100.00 | 10.86 | -6.27 | 17.13 | 10.86 | -8.98 | 78.32 | 105.02 | 3.10 | 0.995 |
| 335.70 | 100.00 | 10.87 | -6.11 | 16.98 | 10.87 | -9.11 | 78.32 | 105.53 | 3.12 | 0.995 |
| 335.80 | 100.00 | 10.89 | -5.95 | 16.84 | 10.89 | -9.23 | 78.33 | 106.03 | 3.13 | 0.995 |
| 335.90 | 100.00 | 10.90 | -5.80 | 16.70 | 10.90 | -9.36 | 78.35 | 106.54 | 3.14 | 0.995 |
| 336.00 | 100.00 | 10.91 | -5.65 | 16.56 | 10.91 | -9.49 | 78.37 | 107.05 | 3.16 | 0.995 |
| 336.10 | 100.00 | 10.92 | -5.50 | 16.42 | 10.92 | -9.61 | 78.39 | 107.56 | 3.17 | 0.995 |
| 336.20 | 100.00 | 10.93 | -5.35 | 16.28 | 10.93 | -9.74 | 78.42 | 108.08 | 3.19 | 0.995 |
| 336.30 | 100.00 | 10.94 | -5.20 | 16.14 | 10.94 | -9.87 | 78.46 | 108.59 | 3.20 | 0.995 |
| 336.40 | 100.00 | 10.95 | -5.06 | 16.01 | 10.95 | -9.99 | 78.50 | 109.11 | 3.21 | 0.995 |
| 336.50 | 100.00 | 10.96 | -4.92 | 15.88 | 10.96 | -10.12 | 78.54 | 109.62 | 3.23 | 0.995 |
| 336.60 | 100.00 | 10.97 | -4.78 | 15.75 | 10.97 | -10.24 | 78.59 | 110.14 | 3.24 | 0.995 |
| 336.70 | 100.00 | 10.98 | -4.64 | 15.62 | 10.98 | -10.37 | 78.65 | 110.66 | 3.25 | 0.995 |
| 336.80 | 100.00 | 10.99 | -4.50 | 15.49 | 10.99 | -10.49 | 78.71 | 111.18 | 3.27 | 0.995 |
| 336.90 | 100.00 | 11.00 | -4.37 | 15.37 | 11.00 | -10.61 | 78.78 | 111.70 | 3.28 | 0.995 |
| 337.00 | 100.00 | 11.01 | -4.24 | 15.25 | 11.01 | -10.73 | 78.85 | 112.22 | 3.29 | 0.995 |
| 337.10 | 100.00 | 11.02 | -4.11 | 15.13 | 11.02 | -10.85 | 78.93 | 112.75 | 3.31 | 0.995 |
| 337.20 | 100.00 | 11.03 | -3.98 | 15.01 | 11.03 | -10.96 | 79.01 | 113.27 | 3.32 | 0.995 |
| 337.30 | 100.00 | 11.04 | -3.86 | 14.90 | 11.04 | -11.08 | 79.11 | 113.80 | 3.33 | 0.995 |
| 337.40 | 100.00 | 11.05 | -3.73 | 14.78 | 11.05 | -11.19 | 79.20 | 114.32 | 3.35 | 0.995 |
| 337.50 | 100.00 | 11.06 | -3.61 | 14.67 | 11.06 | -11.29 | 79.30 | 114.84 | 3.36 | 0.995 |
| 337.60 | 100.00 | 11.07 | -3.49 | 14.56 | 11.07 | -11.39 | 79.41 | 115.37 | 3.37 | 0.995 |
| 337.70 | 100.00 | 11.07 | -3.37 | 14.44 | 11.07 | -11.49 | 79.53 | 115.90 | 3.38 | 0.995 |
| 337.80 | 100.00 | 11.08 | -3.25 | 14.33 | 11.08 | -11.58 | 79.65 | 116.42 | 3.40 | 0.995 |
| 337.90 | 100.00 | 11.09 | -3.14 | 14.23 | 11.09 | -11.66 | 79.78 | 116.95 | 3.41 | 0.995 |
| 338.00 | 100.00 | 11.10 | -3.02 | 14.12 | 11.10 | -11.74 | 79.91 | 117.47 | 3.42 | 0.995 |
| 338.10 | 100.00 | 11.11 | -2.91 | 14.02 | 11.11 | -11.81 | 80.06 | 118.00 | 3.43 | 0.995 |
| 338.20 | 100.00 | 11.11 | -2.80 | 13.91 | 11.11 | -11.88 | 80.20 | 118.52 | 3.44 | 0.995 |
| 338.30 | 100.00 | 11.12 | -2.69 | 13.81 | 11.12 | -11.94 | 80.36 | 119.04 | 3.46 | 0.995 |
| 338.40 | 100.00 | 11.13 | -2.58 | 13.71 | 11.13 | -11.99 | 80.52 | 119.57 | 3.47 | 0.995 |
| 338.50 | 100.00 | 11.14 | -2.47 | 13.61 | 11.14 | -12.03 | 80.69 | 120.09 | 3.48 | 0.995 |
| 338.60 | 100.00 | 11.14 | -2.37 | 13.51 | 11.14 | -12.06 | 80.87 | 120.61 | 3.49 | 0.995 |
| 338.70 | 100.00 | 11.15 | -2.26 | 13.41 | 11.15 | -12.09 | 81.05 | 121.13 | 3.50 | 0.995 |
| 338.80 | 100.00 | 11.16 | -2.16 | 13.32 | 11.16 | -12.10 | 81.24 | 121.65 | 3.51 | 0.995 |
| 338.90 | 100.00 | 11.16 | -2.06 | 13.22 | 11.16 | -12.11 | 81.44 | 122.16 | 3.52 | 0.995 |
| 339.00 | 100.00 | 11.17 | -1.96 | 13.13 | 11.17 | -12.10 | 81.65 | 122.68 | 3.53 | 0.995 |
| 339.10 | 100.00 | 11.17 | -1.87 | 13.04 | 11.17 | -12.09 | 81.86 | 123.19 | 3.54 | 0.995 |
| 339.20 | 100.00 | 11.18 | -1.77 | 12.95 | 11.18 | -12.07 | 82.09 | 123.70 | 3.55 | 0.995 |
| 339.30 | 100.00 | 11.19 | -1.67 | 12.86 | 11.19 | -12.04 | 82.32 | 124.21 | 3.56 | 0.995 |
| 339.40 | 100.00 | 11.19 | -1.58 | 12.77 | 11.19 | -11.99 | 82.55 | 124.72 | 3.56 | 0.995 |
| 339.50 | 100.00 | 11.20 | -1.49 | 12.69 | 11.20 | -11.94 | 82.80 | 125.22 | 3.57 | 0.995 |
| 339.60 | 100.00 | 11.20 | -1.40 | 12.60 | 11.20 | -11.88 | 83.06 | 125.73 | 3.58 | 0.995 |
| 339.70 | 100.00 | 11.21 | -1.31 | 12.52 | 11.21 | -11.82 | 83.32 | 126.22 | 3.59 | 0.995 |
| 339.80 | 100.00 | 11.21 | -1.22 | 12.43 | 11.21 | -11.74 | 83.59 | 126.72 | 3.60 | 0.995 |
| 339.90 | 100.00 | 11.22 | -1.13 | 12.35 | 11.22 | -11.66 | 83.87 | 127.21 | 3.60 | 0.995 |
| 340.00 | 100.00 | 11.22 | -1.05 | 12.27 | 11.22 | -11.57 | 84.16 | 127.70 | 3.61 | 0.995 |
| 340.10 | 100.00 | 11.22 | -0.96 | 12.18 | 11.22 | -11.47 | 84.45 | 128.18 | 3.62 | 0.995 |
| 340.20 | 100.00 | 11.23 | -0.88 | 12.11 | 11.23 | -11.36 | 84.76 | 128.66 | 3.62 | 0.995 |
| 340.30 | 100.00 | 11.23 | -0.80 | 12.03 | 11.23 | -11.25 | 85.08 | 129.13 | 3.63 | 0.995 |
| 340.40 | 100.00 | 11.24 | -0.72 | 11.96 | 11.24 | -11.14 | 85.40 | 129.60 | 3.63 | 0.995 |
| 340.50 | 100.00 | 11.24 | -0.64 | 11.88 | 11.24 | -11.01 | 85.73 | 130.06 | 3.64 | 0.995 |
| 340.60 | 100.00 | 11.24 | -0.56 | 11.80 | 11.24 | -10.89 | 86.07 | 130.52 | 3.64 | 0.995 |
| 340.70 | 100.00 | 11.24 | -0.48 | 11.72 | 11.24 | -10.76 | 86.42 | 130.97 | 3.65 | 0.995 |
| 340.80 | 100.00 | 11.25 | -0.41 | 11.66 | 11.25 | -10.62 | 86.78 | 131.42 | 3.65 | 0.995 |
| 340.90 | 100.00 | 11.25 | -0.33 | 11.58 | 11.25 | -10.49 | 87.15 | 131.86 | 3.65 | 0.995 |
| 341.00 | 100.00 | 11.25 | -0.26 | 11.51 | 11.25 | -10.35 | 87.53 | 132.29 | 3.65 | 0.995 |
| 341.10 | 100.00 | 11.25 | -0.19 | 11.44 | 11.25 | -10.20 | 87.92 | 132.72 | 3.66 | 0.995 |
| 341.20 | 100.00 | 11.26 | -0.12 | 11.38 | 11.26 | -10.06 | 88.31 | 133.14 | 3.66 | 0.995 |
| 341.30 | 100.00 | 11.26 | -0.05 | 11.31 | 11.26 | -9.91 | 88.72 | 133.55 | 3.66 | 0.995 |
| 341.40 | 100.00 | 11.26 | 0.02 | 11.24 | 11.26 | -9.92 | 89.14 | 133.95 | 3.66 | 0.995 |
| 341.50 | 100.00 | 11.26 | 0.09 | 11.17 | 11.26 | -10.14 | 89.56 | 134.34 | 3.66 | 0.995 |
| 341.60 | 100.00 | 11.26 | 0.15 | 11.11 | 11.26 | -10.36 | 90.00 | 134.72 | 3.66 | 0.995 |
| 341.70 | 100.00 | 11.26 | 0.22 | 11.04 | 11.26 | -10.60 | 90.44 | 135.10 | 3.66 | 0.995 |
| 341.80 | 100.00 | 11.26 | 0.28 | 10.98 | 11.26 | -10.83 | 90.89 | 135.46 | 3.66 | 0.995 |
| 341.90 | 100.00 | 11.26 | 0.34 | 10.92 | 11.26 | -11.08 | 91.36 | 135.81 | 3.66 | 0.995 |
| 342.00 | 100.00 | 11.27 | 0.40 | 10.87 | 11.27 | -11.33 | 91.83 | 136.15 | 3.66 | 0.995 |
| 342.10 | 100.00 | 11.27 | 0.46 | 10.81 | 11.27 | -11.59 | 92.31 | 136.48 | 3.65 | 0.995 |
| 342.20 | 100.00 | 11.27 | 0.52 | 10.75 | 11.27 | -11.86 | 92.80 | 136.80 | 3.65 | 0.995 |
| 342.30 | 100.00 | 11.27 | 0.58 | 10.69 | 11.27 | -12.13 | 93.30 | 137.11 | 3.65 | 0.995 |
| 342.40 | 100.00 | 11.27 | 0.64 | 10.63 | 11.27 | -12.41 | 93.81 | 137.40 | 3.64 | 0.995 |
| 342.50 | 100.00 | 11.27 | 0.69 | 10.58 | 11.27 | -12.70 | 94.33 | 137.68 | 3.64 | 0.995 |
| 342.60 | 100.00 | 11.27 | 0.75 | 10.52 | 11.27 | -13.00 | 94.86 | 137.94 | 3.64 | 0.995 |
| 342.70 | 100.00 | 11.26 | 0.80 | 10.46 | 11.26 | -13.30 | 95.39 | 138.19 | 3.63 | 0.995 |
| 342.80 | 100.00 | 11.26 | 0.85 | 10.41 | 11.26 | -13.61 | 95.93 | 138.42 | 3.62 | 0.995 |
| 342.90 | 100.00 | 11.26 | 0.90 | 10.36 | 11.26 | -13.94 | 96.49 | 138.64 | 3.62 | 0.995 |
| 343.00 | 100.00 | 11.26 | 0.95 | 10.31 | 11.26 | -14.27 | 97.05 | 138.84 | 3.61 | 0.995 |
| 343.10 | 100.00 | 11.26 | 1.00 | 10.26 | 11.26 | -14.60 | 97.61 | 139.02 | 3.60 | 0.995 |
| 343.20 | 100.00 | 11.26 | 1.05 | 10.21 | 11.26 | -14.95 | 98.19 | 139.19 | 3.59 | 0.995 |
| 343.30 | 100.00 | 11.26 | 1.09 | 10.17 | 11.26 | -15.30 | 98.77 | 139.33 | 3.59 | 0.995 |
| 343.40 | 100.00 | 11.26 | 1.14 | 10.12 | 11.26 | -15.66 | 99.36 | 139.46 | 3.58 | 0.995 |
| 343.50 | 100.00 | 11.26 | 1.18 | 10.08 | 11.26 | -16.02 | 99.96 | 139.56 | 3.57 | 0.995 |
| 343.60 | 100.00 | 11.25 | 1.23 | 10.02 | 11.25 | -16.38 | 100.56 | 139.64 | 3.56 | 0.995 |
| 343.70 | 100.00 | 11.25 | 1.27 | 9.98 | 11.25 | -16.75 | 101.17 | 139.71 | 3.55 | 0.995 |
| 343.80 | 100.00 | 11.25 | 1.31 | 9.94 | 11.25 | -17.11 | 101.78 | 139.75 | 3.53 | 0.995 |
| 343.90 | 100.00 | 11.25 | 1.35 | 9.90 | 11.25 | -17.46 | 102.40 | 139.76 | 3.52 | 0.995 |
| 344.00 | 100.00 | 11.25 | 1.39 | 9.86 | 11.25 | -17.79 | 103.02 | 139.75 | 3.51 | 0.995 |
| 344.10 | 100.00 | 11.24 | 1.42 | 9.82 | 11.24 | -18.11 | 103.64 | 139.72 | 3.50 | 0.995 |
| 344.20 | 100.00 | 11.24 | 1.46 | 9.78 | 11.24 | -18.39 | 104.27 | 139.66 | 3.48 | 0.995 |
| 344.30 | 100.00 | 11.24 | 1.49 | 9.75 | 11.24 | -18.64 | 104.89 | 139.57 | 3.47 | 0.995 |
| 344.40 | 100.00 | 11.24 | 1.53 | 9.71 | 11.24 | -18.85 | 105.52 | 139.46 | 3.46 | 0.995 |
| 344.50 | 100.00 | 11.23 | 1.56 | 9.67 | 11.23 | -19.00 | 106.15 | 139.32 | 3.44 | 0.995 |
| 344.60 | 100.00 | 11.23 | 1.59 | 9.64 | 11.23 | -19.09 | 106.78 | 139.15 | 3.42 | 0.995 |
| 344.70 | 100.00 | 11.23 | 1.62 | 9.61 | 11.23 | -19.12 | 107.41 | 138.94 | 3.41 | 0.995 |
| 344.80 | 100.00 | 11.23 | 1.65 | 9.58 | 11.23 | -19.08 | 108.03 | 138.71 | 3.39 | 0.995 |
| 344.90 | 100.00 | 11.23 | 1.68 | 9.55 | 11.23 | -18.98 | 108.65 | 138.45 | 3.37 | 0.995 |
| 345.00 | 100.00 | 11.22 | 1.71 | 9.51 | 11.22 | -18.81 | 109.26 | 138.15 | 3.36 | 0.995 |
| 345.10 | 100.00 | 11.22 | 1.73 | 9.49 | 11.22 | -18.58 | 109.87 | 137.82 | 3.34 | 0.995 |
| 345.20 | 100.00 | 11.22 | 1.76 | 9.46 | 11.22 | -18.31 | 110.46 | 137.46 | 3.32 | 0.995 |
| 345.30 | 100.00 | 11.22 | 1.78 | 9.44 | 11.22 | -17.99 | 111.05 | 137.06 | 3.30 | 0.995 |
| 345.40 | 100.00 | 11.21 | 1.80 | 9.41 | 11.21 | -17.65 | 111.63 | 136.63 | 3.28 | 0.995 |
| 345.50 | 100.00 | 11.21 | 1.82 | 9.39 | 11.21 | -17.28 | 112.20 | 136.16 | 3.26 | 0.995 |
| 345.60 | 100.00 | 11.21 | 1.84 | 9.37 | 11.21 | -16.89 | 112.75 | 135.66 | 3.24 | 0.995 |
| 345.70 | 100.00 | 11.21 | 1.86 | 9.35 | 11.21 | -16.49 | 113.28 | 135.12 | 3.22 | 0.995 |
| 345.80 | 100.00 | 11.20 | 1.88 | 9.32 | 11.20 | -16.09 | 113.79 | 134.54 | 3.20 | 0.995 |
| 345.90 | 100.00 | 11.20 | 1.89 | 9.31 | 11.20 | -15.68 | 114.29 | 133.93 | 3.18 | 0.995 |
| 346.00 | 100.00 | 11.20 | 1.91 | 9.29 | 11.20 | -15.27 | 114.76 | 133.28 | 3.16 | 0.995 |
| 346.10 | 100.00 | 11.20 | 1.92 | 9.28 | 11.20 | -14.86 | 115.21 | 132.59 | 3.13 | 0.995 |
| 346.20 | 100.00 | 11.20 | 1.93 | 9.27 | 11.20 | -14.46 | 115.63 | 131.86 | 3.11 | 0.995 |
| 346.30 | 100.00 | 11.20 | 1.95 | 9.25 | 11.20 | -14.07 | 116.02 | 131.09 | 3.09 | 0.995 |
| 346.40 | 100.00 | 11.19 | 1.96 | 9.23 | 11.19 | -13.68 | 116.38 | 130.29 | 3.06 | 0.995 |
| 346.50 | 100.00 | 11.19 | 1.96 | 9.23 | 11.19 | -13.30 | 116.70 | 129.45 | 3.04 | 0.995 |
| 346.60 | 100.00 | 11.19 | 1.97 | 9.22 | 11.19 | -12.93 | 116.99 | 128.57 | 3.02 | 0.995 |
| 346.70 | 100.00 | 11.19 | 1.98 | 9.21 | 11.19 | -12.56 | 117.24 | 127.66 | 2.99 | 0.995 |
| 346.80 | 100.00 | 11.19 | 1.98 | 9.21 | 11.19 | -12.20 | 117.44 | 126.71 | 2.97 | 0.995 |
| 346.90 | 100.00 | 11.19 | 1.98 | 9.21 | 11.19 | -11.85 | 117.60 | 125.73 | 2.94 | 0.995 |
| 347.00 | 100.00 | 11.19 | 1.98 | 9.21 | 11.19 | -11.51 | 117.71 | 124.72 | 2.92 | 0.995 |
| 347.10 | 100.00 | 11.19 | 1.98 | 9.21 | 11.19 | -11.17 | 117.78 | 123.67 | 2.90 | 0.995 |
| 347.20 | 100.00 | 11.19 | 1.98 | 9.21 | 11.19 | -10.84 | 117.78 | 122.60 | 2.87 | 0.995 |
| 347.30 | 100.00 | 11.19 | 1.98 | 9.21 | 11.19 | -10.52 | 117.74 | 121.50 | 2.85 | 0.995 |
| 347.40 | 100.00 | 11.19 | 1.98 | 9.21 | 11.19 | -10.20 | 117.63 | 120.37 | 2.82 | 0.994 |
| 347.50 | 100.00 | 11.19 | 1.97 | 9.22 | 11.19 | -9.89 | 117.46 | 119.22 | 2.80 | 0.994 |
| 347.60 | 100.00 | 11.19 | 1.96 | 9.23 | 11.19 | -9.59 | 117.22 | 118.05 | 2.78 | 0.994 |
| 347.70 | 100.00 | 11.19 | 1.95 | 9.24 | 11.19 | -9.29 | 116.92 | 116.86 | 2.75 | 0.994 |
| 347.80 | 100.00 | 11.19 | 1.94 | 9.25 | 11.19 | -8.99 | 116.56 | 115.66 | 2.73 | 0.994 |
| 347.90 | 100.00 | 11.19 | 1.93 | 9.26 | 11.19 | -8.70 | 116.11 | 114.44 | 2.71 | 0.994 |
| 348.00 | 100.00 | 11.20 | 1.92 | 9.28 | 11.20 | -8.42 | 115.60 | 113.22 | 2.69 | 0.994 |
| 348.10 | 100.00 | 11.20 | 1.90 | 9.30 | 11.20 | -8.14 | 115.01 | 112.00 | 2.66 | 0.994 |
| 348.20 | 100.00 | 11.20 | 1.88 | 9.32 | 11.20 | -7.87 | 114.34 | 110.77 | 2.64 | 0.994 |
| 348.30 | 100.00 | 11.20 | 1.86 | 9.34 | 11.20 | -7.81 | 113.60 | 109.55 | 2.62 | 0.994 |
| 348.40 | 100.00 | 11.21 | 1.84 | 9.37 | 11.21 | -7.81 | 112.77 | 108.34 | 2.61 | 0.994 |
| 348.50 | 100.00 | 11.21 | 1.82 | 9.39 | 11.21 | -7.80 | 111.87 | 107.14 | 2.59 | 0.994 |
| 348.60 | 100.00 | 11.22 | 1.79 | 9.43 | 11.22 | -7.79 | 110.88 | 105.96 | 2.57 | 0.994 |
| 348.70 | 100.00 | 11.22 | 1.76 | 9.46 | 11.22 | -7.78 | 109.81 | 104.81 | 2.56 | 0.994 |
| 348.80 | 100.00 | 11.22 | 1.73 | 9.49 | 11.22 | -7.76 | 108.66 | 103.68 | 2.54 | 0.994 |
| 348.90 | 100.00 | 11.23 | 1.70 | 9.53 | 11.23 | -7.75 | 107.44 | 102.59 | 2.53 | 0.994 |
| 349.00 | 100.00 | 11.24 | 1.67 | 9.57 | 11.24 | -7.73 | 106.13 | 101.53 | 2.52 | 0.994 |
| 349.10 | 100.00 | 11.24 | 1.63 | 9.61 | 11.24 | -7.71 | 104.74 | 100.52 | 2.52 | 0.994 |
| 349.20 | 100.00 | 11.25 | 1.59 | 9.66 | 11.25 | -7.68 | 103.28 | 99.56 | 2.51 | 0.994 |
| 349.30 | 100.00 | 11.25 | 1.55 | 9.70 | 11.25 | -7.66 | 101.75 | 98.65 | 2.51 | 0.994 |
| 349.40 | 100.00 | 11.26 | 1.50 | 9.76 | 11.26 | -7.70 | 100.14 | 97.80 | 2.51 | 0.994 |
| 349.50 | 100.00 | 11.27 | 1.46 | 9.81 | 11.27 | -7.92 | 98.46 | 97.01 | 2.51 | 0.994 |
| 349.60 | 100.00 | 11.28 | 1.41 | 9.87 | 11.28 | -8.14 | 96.72 | 96.29 | 2.52 | 0.994 |
| 349.70 | 100.00 | 11.29 | 1.35 | 9.94 | 11.29 | -8.38 | 94.92 | 95.64 | 2.53 | 0.994 |
| 349.80 | 100.00 | 11.30 | 1.30 | 10.00 | 11.30 | -8.63 | 93.06 | 95.07 | 2.54 | 0.994 |
| 349.90 | 100.00 | 11.31 | 1.24 | 10.07 | 11.31 | -8.89 | 91.15 | 94.57 | 2.56 | 0.994 |
| 350.00 | 100.00 | 11.32 | 1.18 | 10.14 | 11.32 | -9.17 | 89.19 | 94.16 | 2.58 | 0.994 |
| 350.10 | 100.00 | 11.33 | 1.11 | 10.22 | 11.33 | -9.46 | 87.18 | 93.83 | 2.61 | 0.994 |
| 350.20 | 100.00 | 11.34 | 1.04 | 10.30 | 11.34 | -9.77 | 85.14 | 93.59 | 2.64 | 0.994 |
| 350.30 | 100.00 | 11.35 | 0.97 | 10.38 | 11.35 | -10.10 | 83.06 | 93.44 | 2.68 | 0.994 |
| 350.40 | 100.00 | 11.36 | 0.89 | 10.47 | 11.36 | -10.45 | 80.95 | 93.38 | 2.72 | 0.994 |
| 350.50 | 100.00 | 11.37 | 0.81 | 10.56 | 11.37 | -10.83 | 78.83 | 93.42 | 2.77 | 0.994 |
| 350.60 | 100.00 | 11.38 | 0.73 | 10.65 | 11.38 | -11.22 | 76.68 | 93.54 | 2.83 | 0.994 |
| 350.70 | 100.00 | 11.40 | 0.64 | 10.76 | 11.40 | -11.65 | 74.52 | 93.76 | 2.89 | 0.994 |
| 350.80 | 100.00 | 11.41 | 0.55 | 10.86 | 11.41 | -12.11 | 72.36 | 94.08 | 2.97 | 0.994 |
| 350.90 | 100.00 | 11.42 | 0.45 | 10.97 | 11.42 | -12.59 | 70.20 | 94.49 | 3.05 | 0.994 |
| 351.00 | 100.00 | 11.43 | 0.35 | 11.08 | 11.43 | -13.12 | 68.04 | 94.99 | 3.13 | 0.994 |
| 351.10 | 100.00 | 11.45 | 0.24 | 11.21 | 11.45 | -13.68 | 65.89 | 95.59 | 3.23 | 0.994 |
| 351.20 | 100.00 | 11.46 | 0.13 | 11.33 | 11.46 | -14.27 | 63.75 | 96.27 | 3.34 | 0.994 |
| 351.30 | 100.00 | 11.48 | 0.01 | 11.47 | 11.48 | -14.90 | 61.64 | 97.04 | 3.45 | 0.994 |
| 351.40 | 100.00 | 11.49 | -0.11 | 11.60 | 11.49 | -15.57 | 59.54 | 97.90 | 3.58 | 0.994 |
| 351.50 | 100.00 | 11.50 | -0.23 | 11.73 | 11.50 | -16.25 | 57.48 | 98.85 | 3.72 | 0.994 |
| 351.60 | 100.00 | 11.52 | -0.37 | 11.89 | 11.52 | -16.92 | 55.45 | 99.87 | 3.87 | 0.994 |
| 351.70 | 100.00 | 11.53 | -0.50 | 12.03 | 11.53 | -17.54 | 53.45 | 100.97 | 4.04 | 0.994 |
| 351.80 | 100.00 | 11.54 | -0.64 | 12.18 | 11.54 | -18.04 | 51.49 | 102.15 | 4.22 | 0.994 |
| 351.90 | 100.00 | 11.56 | -0.78 | 12.34 | 11.56 | -18.35 | 49.58 | 103.40 | 4.42 | 0.994 |
| 352.00 | 100.00 | 11.57 | -0.93 | 12.50 | 11.57 | -18.39 | 47.71 | 104.72 | 4.63 | 0.994 |
| 352.10 | 100.00 | 11.58 | -1.08 | 12.66 | 11.58 | -18.13 | 45.88 | 106.10 | 4.85 | 0.994 |
| 352.20 | 100.00 | 11.59 | -1.23 | 12.82 | 11.59 | -17.58 | 44.11 | 107.54 | 5.10 | 0.994 |
| 352.30 | 100.00 | 11.60 | -1.39 | 12.99 | 11.60 | -16.83 | 42.39 | 109.05 | 5.36 | 0.994 |
| 352.40 | 100.00 | 11.61 | -1.54 | 13.15 | 11.61 | -15.94 | 40.73 | 110.60 | 5.65 | 0.994 |
| 352.50 | 100.00 | 11.61 | -1.70 | 13.31 | 11.61 | -14.99 | 39.12 | 112.21 | 5.95 | 0.994 |
| 352.60 | 100.00 | 11.62 | -1.84 | 13.46 | 11.62 | -14.02 | 37.56 | 113.87 | 6.28 | 0.994 |
| 352.70 | 100.00 | 11.62 | -1.99 | 13.61 | 11.62 | -13.08 | 36.06 | 115.57 | 6.63 | 0.993 |
| 352.80 | 100.00 | 11.63 | -2.12 | 13.75 | 11.63 | -12.16 | 34.62 | 117.31 | 7.00 | 0.993 |
| 352.90 | 100.00 | 11.63 | -2.24 | 13.87 | 11.63 | -11.69 | 33.24 | 119.08 | 7.40 | 0.993 |
| 353.00 | 100.00 | 11.62 | -2.35 | 13.97 | 11.62 | -12.69 | 31.92 | 120.89 | 7.82 | 0.993 |
| 353.10 | 100.00 | 11.62 | -2.43 | 14.05 | 11.62 | -13.83 | 30.65 | 122.73 | 8.27 | 0.993 |
| 353.20 | 100.00 | 11.61 | -2.49 | 14.10 | 11.61 | -15.14 | 29.45 | 124.60 | 8.74 | 0.994 |
| 353.30 | 100.00 | 11.59 | -2.53 | 14.12 | 11.59 | -16.66 | 28.30 | 126.49 | 9.24 | 0.994 |
| 353.40 | 100.00 | 11.58 | -2.53 | 14.11 | 11.58 | -18.38 | 27.21 | 128.41 | 9.77 | 0.994 |
| 353.50 | 100.00 | 11.56 | -2.50 | 14.06 | 11.56 | -20.13 | 26.18 | 130.34 | 10.33 | 0.994 |
| 353.60 | 100.00 | 11.53 | -2.43 | 13.96 | 11.53 | -21.36 | 25.20 | 132.29 | 10.91 | 0.994 |
| 353.70 | 100.00 | 11.50 | -2.33 | 13.83 | 11.50 | -21.23 | 24.28 | 134.25 | 11.52 | 0.994 |
| 353.80 | 100.00 | 11.46 | -2.18 | 13.64 | 11.46 | -19.76 | 23.41 | 136.23 | 12.15 | 0.994 |
| 353.90 | 100.00 | 11.42 | -2.00 | 13.42 | 11.42 | -17.78 | 22.60 | 138.21 | 12.81 | 0.994 |
| 354.00 | 100.00 | 11.37 | -1.79 | 13.16 | 11.37 | -15.86 | 21.84 | 140.20 | 13.49 | 0.994 |
| 354.10 | 100.00 | 11.31 | -1.54 | 12.85 | 11.31 | -14.13 | 21.14 | 142.20 | 14.18 | 0.994 |
| 354.20 | 100.00 | 11.25 | -1.27 | 12.52 | 11.25 | -12.61 | 20.48 | 144.20 | 14.90 | 0.994 |
| 354.30 | 100.00 | 11.18 | -0.97 | 12.15 | 11.18 | -11.27 | 19.87 | 146.20 | 15.62 | 0.994 |
| 354.40 | 100.00 | 11.10 | -0.66 | 11.76 | 11.10 | -10.56 | 19.31 | 148.21 | 16.36 | 0.994 |
| 354.50 | 100.00 | 11.01 | -0.34 | 11.35 | 11.01 | -11.91 | 18.80 | 150.21 | 17.10 | 0.994 |
| 354.60 | 100.00 | 10.92 | -0.01 | 10.93 | 10.92 | -13.53 | 18.34 | 152.20 | 17.84 | 0.994 |
| 354.70 | 100.00 | 10.81 | 0.33 | 10.48 | 10.81 | -15.50 | 17.91 | 154.20 | 18.58 | 0.994 |
| 354.80 | 100.00 | 10.70 | 0.67 | 10.03 | 10.70 | -17.95 | 17.54 | 156.19 | 19.31 | 0.995 |
| 354.90 | 100.00 | 10.57 | 1.00 | 9.57 | 10.57 | -20.95 | 17.20 | 158.17 | 20.02 | 0.995 |
| 355.00 | 100.00 | 10.44 | 1.33 | 9.11 | 10.44 | -23.60 | 16.90 | 160.14 | 20.72 | 0.995 |
| 355.10 | 100.00 | 10.29 | 1.66 | 8.63 | 10.29 | -22.95 | 16.65 | 162.11 | 21.40 | 0.995 |
| 355.20 | 100.00 | 10.14 | 1.97 | 8.17 | 10.14 | -19.92 | 16.43 | 164.07 | 22.05 | 0.995 |
| 355.30 | 100.00 | 9.98 | 2.27 | 7.71 | 9.98 | -17.07 | 16.25 | 166.02 | 22.66 | 0.995 |
| 355.40 | 100.00 | 9.80 | 2.56 | 7.24 | 9.80 | -14.76 | 16.11 | 167.96 | 23.24 | 0.995 |
| 355.50 | 100.00 | 9.62 | 2.84 | 6.78 | 9.62 | -12.90 | 16.00 | 169.88 | 23.79 | 0.996 |
| 355.60 | 100.00 | 9.42 | 3.10 | 6.32 | 9.42 | -11.36 | 15.92 | 171.80 | 24.29 | 0.996 |
| 355.70 | 100.00 | 9.22 | 3.36 | 5.86 | 9.22 | -10.06 | 15.88 | 173.70 | 24.75 | 0.996 |
| 355.80 | 100.00 | 9.00 | 3.59 | 5.41 | 9.00 | -8.94 | 15.87 | 175.59 | 25.16 | 0.996 |
| 355.90 | 100.00 | 8.78 | 3.82 | 4.96 | 8.78 | -7.97 | 15.89 | 177.47 | 25.52 | 0.996 |
| 356.00 | 100.00 | 8.55 | 4.03 | 4.52 | 8.55 | -7.11 | 15.94 | 179.34 | 25.84 | 0.996 |
| 356.10 | 100.00 | 8.31 | 4.23 | 4.08 | 8.31 | -6.34 | 16.02 | 181.19 | 26.11 | 0.997 |
| 356.20 | 100.00 | 8.07 | 4.41 | 3.66 | 8.07 | -5.66 | 16.12 | 183.02 | 26.34 | 0.997 |
| 356.30 | 100.00 | 7.81 | 4.59 | 3.22 | 7.81 | -5.26 | 16.25 | 184.85 | 26.52 | 0.997 |
| 356.40 | 100.00 | 7.55 | 4.75 | 2.80 | 7.55 | -5.10 | 16.41 | 186.66 | 26.66 | 0.997 |
| 356.50 | 100.00 | 7.29 | 4.90 | 2.39 | 7.29 | -4.90 | 16.59 | 188.45 | 26.75 | 0.997 |
| 356.60 | 100.00 | 7.02 | 5.04 | 1.98 | 7.02 | -4.67 | 16.80 | 190.23 | 26.81 | 0.997 |
| 356.70 | 100.00 | 6.74 | 5.17 | 1.57 | 6.74 | -4.42 | 17.02 | 192.00 | 26.83 | 0.997 |
| 356.80 | 100.00 | 6.46 | 5.29 | 1.17 | 6.68 | -4.16 | 17.27 | 193.75 | 26.81 | 0.997 |
| 356.90 | 100.00 | 6.17 | 5.40 | 0.77 | 6.61 | -3.90 | 17.55 | 195.48 | 26.77 | 0.997 |
| 357.00 | 100.00 | 5.88 | 5.51 | 0.37 | 6.55 | -3.63 | 17.84 | 197.20 | 26.69 | 0.998 |
| 357.10 | 100.00 | 5.59 | 5.60 | -0.01 | 6.47 | -3.37 | 18.15 | 198.91 | 26.59 | 0.998 |
| 357.20 | 100.00 | 5.30 | 5.69 | -0.39 | 6.40 | -3.11 | 18.48 | 200.60 | 26.46 | 0.998 |
| 357.30 | 100.00 | 5.00 | 5.77 | -0.77 | 6.32 | -3.42 | 18.84 | 202.28 | 26.31 | 0.998 |
| 357.40 | 100.00 | 4.70 | 5.84 | -1.14 | 6.25 | -3.84 | 19.20 | 203.94 | 26.15 | 0.998 |
| 357.50 | 100.00 | 4.40 | 5.91 | -1.51 | 6.17 | -4.26 | 19.59 | 205.59 | 25.96 | 0.998 |
| 357.60 | 100.00 | 4.10 | 5.97 | -1.87 | 6.09 | -4.68 | 19.99 | 207.23 | 25.77 | 0.998 |
| 357.70 | 100.00 | 3.81 | 6.03 | -2.22 | 6.01 | -5.09 | 20.41 | 208.85 | 25.56 | 0.998 |
| 357.80 | 100.00 | 3.51 | 6.08 | -2.57 | 5.93 | -5.49 | 20.85 | 210.45 | 25.34 | 0.998 |
| 357.90 | 100.00 | 3.21 | 6.13 | -2.92 | 5.85 | -5.88 | 21.30 | 212.04 | 25.11 | 0.998 |
| 358.00 | 100.00 | 2.91 | 6.17 | -3.26 | 5.77 | -6.26 | 21.76 | 213.62 | 24.87 | 0.998 |
| 358.10 | 100.00 | 2.62 | 6.21 | -3.59 | 5.69 | -6.63 | 22.24 | 215.18 | 24.63 | 0.998 |
| 358.20 | 100.00 | 2.33 | 6.24 | -3.91 | 5.61 | -7.12 | 22.73 | 216.73 | 24.38 | 0.998 |
| 358.30 | 100.00 | 2.04 | 6.27 | -4.23 | 5.53 | -7.86 | 23.24 | 218.27 | 24.13 | 0.998 |
| 358.40 | 100.00 | 1.76 | 6.30 | -4.54 | 5.45 | -8.64 | 23.76 | 219.79 | 23.88 | 0.998 |
| 358.50 | 100.00 | 1.48 | 6.33 | -4.85 | 5.37 | -9.47 | 24.29 | 221.30 | 23.62 | 0.998 |
| 358.60 | 100.00 | 1.21 | 6.35 | -5.14 | 5.29 | -10.35 | 24.83 | 222.79 | 23.37 | 0.998 |
| 358.70 | 100.00 | 0.94 | 6.37 | -5.43 | 5.21 | -11.29 | 25.39 | 224.27 | 23.11 | 0.998 |
| 358.80 | 100.00 | 0.67 | 6.39 | -5.72 | 5.13 | -12.30 | 25.95 | 225.74 | 22.86 | 0.998 |
| 358.90 | 100.00 | 0.41 | 6.40 | -5.99 | 5.06 | -13.40 | 26.53 | 227.19 | 22.60 | 0.998 |
| 359.00 | 100.00 | 0.16 | 6.42 | -6.26 | 4.98 | -14.61 | 27.12 | 228.63 | 22.35 | 0.998 |
| 359.10 | 100.00 | -0.09 | 6.43 | -6.52 | 4.91 | -15.95 | 27.72 | 230.06 | 22.11 | 0.998 |
| 359.20 | 100.00 | -0.32 | 6.44 | -6.76 | 4.84 | -17.46 | 28.32 | 231.48 | 21.86 | 0.998 |
| 359.30 | 100.00 | -0.55 | 6.45 | -7.00 | 4.77 | -19.20 | 28.94 | 232.88 | 21.62 | 0.998 |
| 359.40 | 100.00 | -0.78 | 6.46 | -7.24 | 4.70 | -21.26 | 29.57 | 234.28 | 21.38 | 0.998 |
| 359.50 | 100.00 | -0.99 | 6.46 | -7.45 | 4.63 | -23.76 | 30.20 | 235.66 | 21.14 | 0.998 |
| 359.60 | 100.00 | -1.20 | 6.47 | -7.67 | 4.56 | -26.82 | 30.85 | 237.02 | 20.91 | 0.998 |
| 359.70 | 100.00 | -1.40 | 6.47 | -7.87 | 4.49 | -30.12 | 31.50 | 238.38 | 20.68 | 0.998 |
| 359.80 | 100.00 | -1.58 | 6.48 | -8.06 | 4.43 | -31.07 | 32.16 | 239.72 | 20.45 | 0.998 |
| 359.90 | 100.00 | -1.76 | 6.48 | -8.24 | 4.36 | -28.51 | 32.83 | 241.05 | 20.23 | 0.998 |
| 360.00 | 100.00 | -1.93 | 6.48 | -8.41 | 4.30 | -25.57 | 33.51 | 242.37 | 20.02 | 0.998 |
| 360.10 | 100.00 | -2.09 | 6.48 | -8.57 | 4.24 | -23.20 | 34.19 | 243.68 | 19.80 | 0.998 |
| 360.20 | 100.00 | -2.25 | 6.48 | -8.73 | 4.17 | -21.32 | 34.88 | 244.98 | 19.59 | 0.998 |
| 360.30 | 100.00 | -2.39 | 6.48 | -8.87 | 4.11 | -19.80 | 35.58 | 246.27 | 19.39 | 0.998 |
| 360.40 | 100.00 | -2.52 | 6.48 | -9.00 | 4.06 | -18.53 | 36.29 | 247.54 | 19.19 | 0.998 |
| 360.50 | 100.00 | -2.65 | 6.47 | -9.12 | 4.00 | -17.46 | 37.00 | 248.81 | 18.99 | 0.998 |
| 360.60 | 100.00 | -2.76 | 6.47 | -9.23 | 3.94 | -16.53 | 37.72 | 250.06 | 18.80 | 0.998 |
| 360.70 | 100.00 | -2.87 | 6.47 | -9.34 | 3.89 | -15.72 | 38.44 | 251.30 | 18.61 | 0.998 |
| 360.80 | 100.00 | -2.97 | 6.46 | -9.43 | 3.83 | -15.00 | 39.18 | 252.54 | 18.42 | 0.998 |
| 360.90 | 100.00 | -3.06 | 6.46 | -9.52 | 3.78 | -14.35 | 39.91 | 253.76 | 18.24 | 0.998 |
| 361.00 | 100.00 | -3.14 | 6.45 | -9.59 | 3.73 | -13.77 | 40.66 | 254.97 | 18.06 | 0.998 |
| 361.10 | 100.00 | -3.22 | 6.45 | -9.67 | 3.68 | -13.25 | 41.41 | 256.17 | 17.89 | 0.998 |
| 361.20 | 100.00 | -3.28 | 6.44 | -9.72 | 3.63 | -12.77 | 42.16 | 257.36 | 17.72 | 0.998 |
| 361.30 | 100.00 | -3.34 | 6.44 | -9.78 | 3.59 | -12.33 | 42.92 | 258.55 | 17.55 | 0.998 |
| 361.40 | 100.00 | -3.40 | 6.43 | -9.83 | 3.58 | -11.93 | 43.69 | 259.72 | 17.39 | 0.998 |
| 361.50 | 100.00 | -3.45 | 6.43 | -9.88 | 3.58 | -11.56 | 44.46 | 260.88 | 17.23 | 0.998 |
| 361.60 | 100.00 | -3.49 | 6.42 | -9.91 | 3.57 | -11.21 | 45.24 | 262.03 | 17.07 | 0.998 |
| 361.70 | 100.00 | -3.53 | 6.41 | -9.94 | 3.56 | -10.89 | 46.02 | 263.17 | 16.92 | 0.998 |
| 361.80 | 100.00 | -3.56 | 6.41 | -9.97 | 3.55 | -10.59 | 46.80 | 264.31 | 16.77 | 0.998 |
| 361.90 | 100.00 | -3.59 | 6.40 | -9.99 | 3.54 | -10.31 | 47.59 | 265.43 | 16.62 | 0.998 |
| 362.00 | 100.00 | -3.61 | 6.39 | -10.00 | 3.53 | -10.05 | 48.39 | 266.55 | 16.48 | 0.998 |
| 362.10 | 100.00 | -3.63 | 6.38 | -10.01 | 3.52 | -9.81 | 49.19 | 267.65 | 16.34 | 0.998 |
| 362.20 | 100.00 | -3.65 | 6.38 | -10.03 | 3.51 | -9.58 | 49.99 | 268.75 | 16.20 | 0.998 |
| 362.30 | 100.00 | -3.66 | 6.37 | -10.03 | 3.50 | -9.36 | 50.80 | 269.84 | 16.07 | 0.998 |
| 362.40 | 100.00 | -3.67 | 6.36 | -10.03 | 3.50 | -9.15 | 51.61 | 270.91 | 15.93 | 0.998 |
| 362.50 | 100.00 | -3.67 | 6.35 | -10.02 | 3.49 | -8.96 | 52.43 | 271.98 | 15.80 | 0.998 |
| 362.60 | 100.00 | -3.68 | 6.34 | -10.02 | 3.48 | -8.78 | 53.25 | 273.04 | 15.68 | 0.998 |
| 362.70 | 100.00 | -3.68 | 6.33 | -10.01 | 3.47 | -8.60 | 54.08 | 274.10 | 15.55 | 0.998 |
| 362.80 | 100.00 | -3.68 | 6.33 | -10.01 | 3.46 | -8.44 | 54.90 | 275.14 | 15.43 | 0.998 |
| 362.90 | 100.00 | -3.67 | 6.32 | -9.99 | 3.46 | -8.28 | 55.74 | 276.17 | 15.31 | 0.998 |
| 363.00 | 100.00 | -3.67 | 6.31 | -9.98 | 3.45 | -8.13 | 56.57 | 277.20 | 15.19 | 0.998 |
| 363.10 | 100.00 | -3.67 | 6.30 | -9.97 | 3.44 | -7.99 | 57.41 | 278.22 | 15.08 | 0.998 |
| 363.20 | 100.00 | -3.66 | 6.29 | -9.95 | 3.43 | -7.86 | 58.25 | 279.23 | 14.97 | 0.998 |
| 363.30 | 100.00 | -3.65 | 6.28 | -9.93 | 3.43 | -7.73 | 59.10 | 280.23 | 14.86 | 0.998 |
| 363.40 | 100.00 | -3.64 | 6.27 | -9.91 | 3.42 | -7.61 | 59.95 | 281.22 | 14.75 | 0.998 |
| 363.50 | 100.00 | -3.63 | 6.27 | -9.90 | 3.41 | -7.49 | 60.80 | 282.21 | 14.64 | 0.998 |
| 363.60 | 100.00 | -3.62 | 6.26 | -9.88 | 3.41 | -7.38 | 61.66 | 283.19 | 14.54 | 0.998 |
| 363.70 | 100.00 | -3.61 | 6.25 | -9.86 | 3.40 | -7.27 | 62.52 | 284.16 | 14.44 | 0.998 |
| 363.80 | 100.00 | -3.60 | 6.24 | -9.84 | 3.39 | -7.17 | 63.38 | 285.12 | 14.34 | 0.998 |
| 363.90 | 100.00 | -3.58 | 6.23 | -9.81 | 3.39 | -7.07 | 64.24 | 286.07 | 14.24 | 0.998 |
| 364.00 | 100.00 | -3.57 | 6.22 | -9.79 | 3.38 | -6.98 | 65.11 | 287.02 | 14.14 | 0.998 |
| 364.10 | 100.00 | -3.56 | 6.21 | -9.77 | 3.38 | -6.89 | 65.98 | 287.96 | 14.05 | 0.998 |
| 364.20 | 100.00 | -3.54 | 6.20 | -9.74 | 3.37 | -6.81 | 66.86 | 288.89 | 13.96 | 0.998 |
| 364.30 | 100.00 | -3.53 | 6.19 | -9.72 | 3.37 | -6.72 | 67.73 | 289.81 | 13.87 | 0.998 |
| 364.40 | 100.00 | -3.52 | 6.19 | -9.71 | 3.36 | -6.65 | 68.61 | 290.72 | 13.78 | 0.998 |
| 364.50 | 100.00 | -3.50 | 6.18 | -9.68 | 3.36 | -6.57 | 69.49 | 291.63 | 13.69 | 0.998 |
| 364.60 | 100.00 | -3.49 | 6.17 | -9.66 | 3.35 | -6.50 | 70.38 | 292.53 | 13.60 | 0.998 |
| 364.70 | 100.00 | -3.48 | 6.16 | -9.64 | 3.35 | -6.43 | 71.26 | 293.43 | 13.52 | 0.998 |
| 364.80 | 100.00 | -3.46 | 6.15 | -9.61 | 3.35 | -6.36 | 72.15 | 294.31 | 13.44 | 0.998 |
| 364.90 | 100.00 | -3.45 | 6.14 | -9.59 | 3.34 | -6.30 | 73.04 | 295.19 | 13.36 | 0.998 |
| 365.00 | 100.00 | -3.44 | 6.13 | -9.57 | 3.34 | -6.24 | 73.94 | 296.06 | 13.28 | 0.998 |
| 365.10 | 100.00 | -3.42 | 6.12 | -9.54 | 3.34 | -6.18 | 74.83 | 296.93 | 13.20 | 0.998 |
| 365.20 | 100.00 | -3.41 | 6.11 | -9.52 | 3.33 | -6.12 | 75.73 | 297.79 | 13.12 | 0.998 |
| 365.30 | 100.00 | -3.40 | 6.10 | -9.50 | 3.33 | -6.06 | 76.63 | 298.64 | 13.04 | 0.998 |
| 365.40 | 100.00 | -3.38 | 6.10 | -9.48 | 3.33 | -6.01 | 77.53 | 299.48 | 12.97 | 0.998 |
| 365.50 | 100.00 | -3.37 | 6.09 | -9.46 | 3.33 | -5.96 | 78.44 | 300.32 | 12.90 | 0.998 |
| 365.60 | 100.00 | -3.36 | 6.08 | -9.44 | 3.32 | -5.91 | 79.35 | 301.15 | 12.82 | 0.998 |
| 365.70 | 100.00 | -3.35 | 6.07 | -9.42 | 3.32 | -5.86 | 80.25 | 301.97 | 12.75 | 0.998 |
| 365.80 | 100.00 | -3.34 | 6.06 | -9.40 | 3.32 | -5.82 | 81.16 | 302.79 | 12.68 | 0.998 |
| 365.90 | 100.00 | -3.33 | 6.05 | -9.38 | 3.32 | -5.77 | 82.08 | 303.60 | 12.62 | 0.998 |
| 366.00 | 100.00 | -3.32 | 6.04 | -9.36 | 3.32 | -5.73 | 82.99 | 304.40 | 12.55 | 0.998 |
| 366.10 | 100.00 | -3.30 | 6.03 | -9.33 | 3.31 | -5.69 | 83.91 | 305.20 | 12.48 | 0.998 |
| 366.20 | 100.00 | -3.29 | 6.02 | -9.31 | 3.31 | -5.65 | 84.82 | 305.98 | 12.42 | 0.998 |
| 366.30 | 100.00 | -3.28 | 6.02 | -9.30 | 3.31 | -5.61 | 85.74 | 306.77 | 12.35 | 0.998 |
| 366.40 | 100.00 | -3.28 | 6.01 | -9.29 | 3.31 | -5.58 | 86.67 | 307.54 | 12.29 | 0.998 |
| 366.50 | 100.00 | -3.27 | 6.00 | -9.27 | 3.31 | -5.54 | 87.59 | 308.31 | 12.23 | 0.998 |
| 366.60 | 100.00 | -3.26 | 5.99 | -9.25 | 3.31 | -5.51 | 88.51 | 309.08 | 12.17 | 0.998 |
| 366.70 | 100.00 | -3.25 | 5.98 | -9.23 | 3.31 | -5.48 | 89.44 | 309.83 | 12.11 | 0.998 |
| 366.80 | 100.00 | -3.24 | 5.97 | -9.21 | 3.30 | -5.45 | 90.37 | 310.58 | 12.05 | 0.998 |
| 366.90 | 100.00 | -3.23 | 5.96 | -9.19 | 3.30 | -5.41 | 91.30 | 311.33 | 11.99 | 0.998 |
| 367.00 | 100.00 | -3.23 | 5.96 | -9.19 | 3.30 | -5.39 | 92.23 | 312.06 | 11.93 | 0.998 |
| 367.10 | 100.00 | -3.22 | 5.95 | -9.17 | 3.30 | -5.36 | 93.16 | 312.80 | 11.87 | 0.998 |
| 367.20 | 100.00 | -3.21 | 5.94 | -9.15 | 3.30 | -5.33 | 94.09 | 313.52 | 11.82 | 0.998 |
| 367.30 | 100.00 | -3.21 | 5.93 | -9.14 | 3.30 | -5.30 | 95.02 | 314.24 | 11.76 | 0.998 |
| 367.40 | 100.00 | -3.20 | 5.92 | -9.12 | 3.30 | -5.28 | 95.96 | 314.95 | 11.71 | 0.998 |
| 367.50 | 100.00 | -3.19 | 5.91 | -9.10 | 3.30 | -5.25 | 96.90 | 315.66 | 11.66 | 0.998 |
| 367.60 | 100.00 | -3.19 | 5.90 | -9.09 | 3.30 | -5.23 | 97.84 | 316.36 | 11.60 | 0.998 |
| 367.70 | 100.00 | -3.18 | 5.90 | -9.08 | 3.30 | -5.21 | 98.78 | 317.06 | 11.55 | 0.998 |
| 367.80 | 100.00 | -3.18 | 5.89 | -9.07 | 3.30 | -5.19 | 99.72 | 317.74 | 11.50 | 0.998 |
| 367.90 | 100.00 | -3.17 | 5.88 | -9.05 | 3.30 | -5.16 | 100.66 | 318.43 | 11.45 | 0.998 |
| 368.00 | 100.00 | -3.17 | 5.87 | -9.04 | 3.30 | -5.14 | 101.60 | 319.10 | 11.40 | 0.998 |
| 368.10 | 100.00 | -3.17 | 5.86 | -9.03 | 3.30 | -5.12 | 102.54 | 319.77 | 11.35 | 0.998 |
| 368.20 | 100.00 | -3.16 | 5.85 | -9.01 | 3.30 | -5.10 | 103.49 | 320.44 | 11.30 | 0.998 |
| 368.30 | 100.00 | -3.16 | 5.85 | -9.01 | 3.30 | -5.09 | 104.44 | 321.10 | 11.26 | 0.998 |
| 368.40 | 100.00 | -3.16 | 5.84 | -9.00 | 3.30 | -5.07 | 105.38 | 321.75 | 11.21 | 0.998 |
| 368.50 | 100.00 | -3.15 | 5.83 | -8.98 | 3.30 | -5.05 | 106.33 | 322.40 | 11.16 | 0.998 |
| 368.60 | 100.00 | -3.15 | 5.82 | -8.97 | 3.30 | -5.04 | 107.28 | 323.04 | 11.12 | 0.998 |
| 368.70 | 100.00 | -3.15 | 5.81 | -8.96 | 3.30 | -5.02 | 108.23 | 323.68 | 11.07 | 0.998 |
| 368.80 | 100.00 | -3.15 | 5.80 | -8.95 | 3.30 | -5.00 | 109.18 | 324.30 | 11.03 | 0.998 |
| 368.90 | 100.00 | -3.14 | 5.80 | -8.94 | 3.30 | -4.99 | 110.13 | 324.93 | 10.99 | 0.998 |
| 369.00 | 100.00 | -3.14 | 5.79 | -8.93 | 3.30 | -4.98 | 111.08 | 325.55 | 10.94 | 0.998 |
| 369.10 | 100.00 | -3.14 | 5.78 | -8.92 | 3.30 | -4.96 | 112.03 | 326.16 | 10.90 | 0.998 |
| 369.20 | 100.00 | -3.14 | 5.77 | -8.91 | 3.30 | -4.95 | 112.99 | 326.77 | 10.86 | 0.998 |
| 369.30 | 100.00 | -3.14 | 5.76 | -8.90 | 3.30 | -4.94 | 113.94 | 327.37 | 10.82 | 0.998 |
| 369.40 | 100.00 | -3.14 | 5.76 | -8.90 | 3.30 | -4.92 | 114.89 | 327.97 | 10.78 | 0.998 |
| 369.50 | 100.00 | -3.14 | 5.75 | -8.89 | 3.30 | -4.91 | 115.85 | 328.56 | 10.74 | 0.998 |
| 369.60 | 100.00 | -3.14 | 5.74 | -8.88 | 3.31 | -4.90 | 116.80 | 329.14 | 10.70 | 0.998 |
| 369.70 | 100.00 | -3.14 | 5.73 | -8.87 | 3.31 | -4.89 | 117.76 | 329.72 | 10.66 | 0.998 |
| 369.80 | 100.00 | -3.14 | 5.72 | -8.86 | 3.31 | -4.88 | 118.72 | 330.30 | 10.62 | 0.998 |
| 369.90 | 100.00 | -3.14 | 5.72 | -8.86 | 3.31 | -4.87 | 119.67 | 330.87 | 10.58 | 0.998 |
| 370.00 | 100.00 | -3.14 | 5.71 | -8.85 | 3.31 | -4.86 | 120.63 | 331.43 | 10.54 | 0.998 |
| 370.10 | 100.00 | -3.14 | 5.70 | -8.84 | 3.31 | -4.85 | 121.59 | 331.99 | 10.51 | 0.998 |
| 370.20 | 100.00 | -3.14 | 5.69 | -8.83 | 3.31 | -4.84 | 122.55 | 332.54 | 10.47 | 0.998 |
| 370.30 | 100.00 | -3.14 | 5.68 | -8.82 | 3.31 | -4.83 | 123.51 | 333.09 | 10.43 | 0.998 |
| 370.40 | 100.00 | -3.14 | 5.68 | -8.82 | 3.31 | -4.83 | 124.47 | 333.63 | 10.40 | 0.998 |
| 370.50 | 100.00 | -3.15 | 5.67 | -8.82 | 3.31 | -4.82 | 125.43 | 334.17 | 10.36 | 0.998 |
| 370.60 | 100.00 | -3.15 | 5.66 | -8.81 | 3.32 | -4.81 | 126.38 | 334.70 | 10.33 | 0.998 |
| 370.70 | 100.00 | -3.15 | 5.65 | -8.80 | 3.32 | -4.80 | 127.34 | 335.23 | 10.29 | 0.998 |
| 370.80 | 100.00 | -3.15 | 5.65 | -8.80 | 3.32 | -4.80 | 128.31 | 335.75 | 10.26 | 0.998 |
| 370.90 | 100.00 | -3.16 | 5.64 | -8.80 | 3.32 | -4.79 | 129.27 | 336.26 | 10.22 | 0.998 |
| 371.00 | 100.00 | -3.16 | 5.63 | -8.79 | 3.32 | -4.78 | 130.23 | 336.78 | 10.19 | 0.998 |
| 371.10 | 100.00 | -3.16 | 5.62 | -8.78 | 3.32 | -4.78 | 131.19 | 337.28 | 10.16 | 0.998 |
| 371.20 | 100.00 | -3.16 | 5.62 | -8.78 | 3.32 | -4.77 | 132.15 | 337.78 | 10.12 | 0.998 |
| 371.30 | 100.00 | -3.17 | 5.61 | -8.78 | 3.32 | -4.77 | 133.11 | 338.28 | 10.09 | 0.998 |
| 371.40 | 100.00 | -3.17 | 5.60 | -8.77 | 3.33 | -4.76 | 134.07 | 338.77 | 10.06 | 0.998 |
| 371.50 | 100.00 | -3.17 | 5.59 | -8.76 | 3.33 | -4.76 | 135.03 | 339.26 | 10.03 | 0.998 |
| 371.60 | 100.00 | -3.18 | 5.59 | -8.77 | 3.33 | -4.75 | 135.99 | 339.74 | 10.00 | 0.998 |
| 371.70 | 100.00 | -3.18 | 5.58 | -8.76 | 3.33 | -4.75 | 136.95 | 340.22 | 9.97 | 0.998 |
| 371.80 | 100.00 | -3.18 | 5.57 | -8.75 | 3.33 | -4.74 | 137.92 | 340.69 | 9.94 | 0.998 |
| 371.90 | 100.00 | -3.19 | 5.56 | -8.75 | 3.33 | -4.74 | 138.88 | 341.15 | 9.91 | 0.998 |
| 372.00 | 100.00 | -3.19 | 5.56 | -8.75 | 3.34 | -4.74 | 139.84 | 341.62 | 9.88 | 0.998 |
| 372.10 | 100.00 | -3.20 | 5.55 | -8.75 | 3.34 | -4.73 | 140.80 | 342.07 | 9.85 | 0.998 |
| 372.20 | 100.00 | -3.20 | 5.54 | -8.74 | 3.34 | -4.73 | 141.76 | 342.52 | 9.82 | 0.998 |
| 372.30 | 100.00 | -3.21 | 5.54 | -8.75 | 3.34 | -4.73 | 142.72 | 342.97 | 9.79 | 0.998 |
| 372.40 | 100.00 | -3.21 | 5.53 | -8.74 | 3.34 | -4.73 | 143.68 | 343.42 | 9.76 | 0.998 |
| 372.50 | 100.00 | -3.22 | 5.52 | -8.74 | 3.34 | -4.72 | 144.64 | 343.85 | 9.73 | 0.998 |
| 372.60 | 100.00 | -3.22 | 5.51 | -8.73 | 3.35 | -4.72 | 145.60 | 344.29 | 9.71 | 0.998 |
| 372.70 | 100.00 | -3.22 | 5.51 | -8.73 | 3.35 | -4.72 | 146.56 | 344.72 | 9.68 | 0.998 |
| 372.80 | 100.00 | -3.23 | 5.50 | -8.73 | 3.35 | -4.72 | 147.52 | 345.14 | 9.65 | 0.998 |
| 372.90 | 100.00 | -3.24 | 5.49 | -8.73 | 3.35 | -4.71 | 148.48 | 345.56 | 9.63 | 0.998 |
| 373.00 | 100.00 | -3.24 | 5.49 | -8.73 | 3.35 | -4.71 | 149.44 | 345.97 | 9.60 | 0.998 |
| 373.10 | 100.00 | -3.25 | 5.48 | -8.73 | 3.36 | -4.71 | 150.40 | 346.38 | 9.57 | 0.998 |
| 373.20 | 100.00 | -3.25 | 5.47 | -8.72 | 3.36 | -4.71 | 151.36 | 346.79 | 9.55 | 0.998 |
| 373.30 | 100.00 | -3.26 | 5.46 | -8.72 | 3.36 | -4.71 | 152.32 | 347.19 | 9.52 | 0.998 |
| 373.40 | 100.00 | -3.26 | 5.46 | -8.72 | 3.36 | -4.71 | 153.27 | 347.59 | 9.50 | 0.998 |
| 373.50 | 100.00 | -3.27 | 5.45 | -8.72 | 3.36 | -4.71 | 154.23 | 347.98 | 9.47 | 0.998 |
| 373.60 | 100.00 | -3.27 | 5.44 | -8.71 | 3.36 | -4.71 | 155.19 | 348.37 | 9.45 | 0.998 |
| 373.70 | 100.00 | -3.28 | 5.44 | -8.72 | 3.37 | -4.71 | 156.14 | 348.75 | 9.42 | 0.998 |
| 373.80 | 100.00 | -3.29 | 5.43 | -8.72 | 3.37 | -4.71 | 157.10 | 349.13 | 9.40 | 0.998 |
| 373.90 | 100.00 | -3.29 | 5.42 | -8.71 | 3.37 | -4.70 | 158.06 | 349.50 | 9.37 | 0.998 |
| 374.00 | 100.00 | -3.30 | 5.42 | -8.72 | 3.37 | -4.70 | 159.01 | 349.87 | 9.35 | 0.998 |
| 374.10 | 100.00 | -3.31 | 5.41 | -8.72 | 3.37 | -4.70 | 159.97 | 350.24 | 9.33 | 0.998 |
| 374.20 | 100.00 | -3.31 | 5.40 | -8.71 | 3.38 | -4.70 | 160.92 | 350.60 | 9.30 | 0.998 |
| 374.30 | 100.00 | -3.32 | 5.40 | -8.72 | 3.38 | -4.70 | 161.88 | 350.96 | 9.28 | 0.998 |
| 374.40 | 100.00 | -3.32 | 5.39 | -8.71 | 3.38 | -4.71 | 162.83 | 351.31 | 9.26 | 0.998 |
| 374.50 | 100.00 | -3.33 | 5.38 | -8.71 | 3.38 | -4.71 | 163.78 | 351.66 | 9.23 | 0.998 |
| 374.60 | 100.00 | -3.34 | 5.37 | -8.71 | 3.39 | -4.71 | 164.73 | 352.00 | 9.21 | 0.998 |
| 374.70 | 100.00 | -3.35 | 5.37 | -8.72 | 3.39 | -4.71 | 165.68 | 352.34 | 9.19 | 0.998 |
| 374.80 | 100.00 | -3.35 | 5.36 | -8.71 | 3.39 | -4.71 | 166.64 | 352.68 | 9.17 | 0.998 |
| 374.90 | 100.00 | -3.36 | 5.35 | -8.71 | 3.39 | -4.71 | 167.58 | 353.01 | 9.15 | 0.998 |
| 375.00 | 100.00 | -3.37 | 5.35 | -8.72 | 3.39 | -4.71 | 168.53 | 353.34 | 9.12 | 0.998 |
| 375.10 | 100.00 | -3.37 | 5.34 | -8.71 | 3.40 | -4.71 | 169.48 | 353.66 | 9.10 | 0.998 |
| 375.20 | 100.00 | -3.38 | 5.33 | -8.71 | 3.40 | -4.71 | 170.43 | 353.98 | 9.08 | 0.998 |
| 375.30 | 100.00 | -3.39 | 5.33 | -8.72 | 3.40 | -4.71 | 171.38 | 354.30 | 9.06 | 0.998 |
| 375.40 | 100.00 | -3.39 | 5.32 | -8.71 | 3.40 | -4.71 | 172.32 | 354.61 | 9.04 | 0.998 |
| 375.50 | 100.00 | -3.40 | 5.32 | -8.72 | 3.40 | -4.72 | 173.27 | 354.91 | 9.02 | 0.998 |
| 375.60 | 100.00 | -3.41 | 5.31 | -8.72 | 3.41 | -4.72 | 174.22 | 355.22 | 9.00 | 0.998 |
| 375.70 | 100.00 | -3.42 | 5.30 | -8.72 | 3.41 | -4.72 | 175.16 | 355.52 | 8.98 | 0.998 |
| 375.80 | 100.00 | -3.42 | 5.30 | -8.72 | 3.41 | -4.72 | 176.10 | 355.81 | 8.96 | 0.998 |
| 375.90 | 100.00 | -3.43 | 5.29 | -8.72 | 3.41 | -4.72 | 177.04 | 356.10 | 8.94 | 0.998 |
| 376.00 | 100.00 | -3.44 | 5.28 | -8.72 | 3.42 | -4.73 | 177.99 | 356.39 | 8.92 | 0.998 |
| 376.10 | 100.00 | -3.45 | 5.28 | -8.73 | 3.42 | -4.73 | 178.93 | 356.67 | 8.90 | 0.998 |
| 376.20 | 100.00 | -3.45 | 5.27 | -8.72 | 3.42 | -4.73 | 179.87 | 356.95 | 8.88 | 0.998 |
| 376.30 | 100.00 | -3.46 | 5.26 | -8.72 | 3.42 | -4.73 | 180.81 | 357.23 | 8.86 | 0.998 |
| 376.40 | 100.00 | -3.47 | 5.26 | -8.73 | 3.43 | -4.73 | 181.75 | 357.50 | 8.85 | 0.998 |
| 376.50 | 100.00 | -3.48 | 5.25 | -8.73 | 3.43 | -4.74 | 182.68 | 357.77 | 8.83 | 0.998 |
| 376.60 | 100.00 | -3.49 | 5.24 | -8.73 | 3.43 | -4.74 | 183.62 | 358.04 | 8.81 | 0.998 |
| 376.70 | 100.00 | -3.49 | 5.24 | -8.73 | 3.43 | -4.74 | 184.56 | 358.30 | 8.79 | 0.998 |
| 376.80 | 100.00 | -3.50 | 5.23 | -8.73 | 3.43 | -4.74 | 185.49 | 358.55 | 8.77 | 0.998 |
| 376.90 | 100.00 | -3.51 | 5.23 | -8.74 | 3.44 | -4.75 | 186.42 | 358.81 | 8.75 | 0.998 |
| 377.00 | 100.00 | -3.52 | 5.22 | -8.74 | 3.44 | -4.75 | 187.36 | 359.06 | 8.74 | 0.998 |
| 377.10 | 100.00 | -3.53 | 5.21 | -8.74 | 3.44 | -4.75 | 188.29 | 359.30 | 8.72 | 0.998 |
| 377.20 | 100.00 | -3.53 | 5.21 | -8.74 | 3.44 | -4.75 | 189.22 | 359.54 | 8.70 | 0.998 |
| 377.30 | 100.00 | -3.54 | 5.20 | -8.74 | 3.45 | -4.76 | 190.15 | 359.78 | 8.68 | 0.998 |
| 377.40 | 100.00 | -3.55 | 5.20 | -8.75 | 3.45 | -4.76 | 191.08 | 360.02 | 8.67 | 0.998 |
| 377.50 | 100.00 | -3.56 | 5.19 | -8.75 | 3.45 | -4.76 | 192.00 | 360.25 | 8.65 | 0.998 |
| 377.60 | 100.00 | -3.57 | 5.18 | -8.75 | 3.45 | -4.77 | 192.93 | 360.48 | 8.63 | 0.998 |
| 377.70 | 100.00 | -3.58 | 5.18 | -8.76 | 3.46 | -4.77 | 193.86 | 360.70 | 8.62 | 0.998 |
| 377.80 | 100.00 | -3.58 | 5.17 | -8.75 | 3.46 | -4.77 | 194.78 | 360.92 | 8.60 | 0.998 |
| 377.90 | 100.00 | -3.59 | 5.16 | -8.75 | 3.46 | -4.77 | 195.70 | 361.14 | 8.58 | 0.998 |
| 378.00 | 100.00 | -3.60 | 5.16 | -8.76 | 3.46 | -4.78 | 196.63 | 361.36 | 8.57 | 0.998 |
| 378.10 | 100.00 | -3.61 | 5.15 | -8.76 | 3.47 | -4.78 | 197.55 | 361.57 | 8.55 | 0.998 |
| 378.20 | 100.00 | -3.62 | 5.15 | -8.77 | 3.47 | -4.78 | 198.47 | 361.78 | 8.54 | 0.998 |
| 378.30 | 100.00 | -3.63 | 5.14 | -8.77 | 3.47 | -4.79 | 199.39 | 361.98 | 8.52 | 0.998 |
| 378.40 | 100.00 | -3.63 | 5.14 | -8.77 | 3.47 | -4.79 | 200.30 | 362.18 | 8.51 | 0.998 |
| 378.50 | 100.00 | -3.64 | 5.13 | -8.77 | 3.48 | -4.79 | 201.22 | 362.38 | 8.49 | 0.998 |
| 378.60 | 100.00 | -3.65 | 5.12 | -8.77 | 3.48 | -4.80 | 202.14 | 362.57 | 8.47 | 0.998 |
| 378.70 | 100.00 | -3.66 | 5.12 | -8.78 | 3.48 | -4.80 | 203.05 | 362.76 | 8.46 | 0.998 |
| 378.80 | 100.00 | -3.67 | 5.11 | -8.78 | 3.48 | -4.80 | 203.96 | 362.95 | 8.44 | 0.998 |
| 378.90 | 100.00 | -3.68 | 5.11 | -8.79 | 3.49 | -4.81 | 204.88 | 363.13 | 8.43 | 0.998 |
| 379.00 | 100.00 | -3.69 | 5.10 | -8.79 | 3.49 | -4.81 | 205.79 | 363.31 | 8.41 | 0.998 |
| 379.10 | 100.00 | -3.70 | 5.09 | -8.79 | 3.49 | -4.82 | 206.70 | 363.49 | 8.40 | 0.998 |
| 379.20 | 100.00 | -3.70 | 5.09 | -8.79 | 3.49 | -4.82 | 207.60 | 363.66 | 8.39 | 0.998 |
| 379.30 | 100.00 | -3.71 | 5.08 | -8.79 | 3.50 | -4.82 | 208.51 | 363.84 | 8.37 | 0.998 |
| 379.40 | 100.00 | -3.72 | 5.08 | -8.80 | 3.50 | -4.83 | 209.42 | 364.00 | 8.36 | 0.998 |
| 379.50 | 100.00 | -3.73 | 5.07 | -8.80 | 3.50 | -4.83 | 210.32 | 364.17 | 8.34 | 0.998 |
| 379.60 | 100.00 | -3.74 | 5.07 | -8.81 | 3.50 | -4.83 | 211.22 | 364.33 | 8.33 | 0.998 |
| 379.70 | 100.00 | -3.75 | 5.06 | -8.81 | 3.51 | -4.84 | 212.13 | 364.49 | 8.31 | 0.998 |
| 379.80 | 100.00 | -3.76 | 5.05 | -8.81 | 3.51 | -4.84 | 213.03 | 364.64 | 8.30 | 0.998 |
| 379.90 | 100.00 | -3.77 | 5.05 | -8.82 | 3.51 | -4.85 | 213.93 | 364.80 | 8.29 | 0.998 |
| 380.00 | 100.00 | -3.77 | 5.04 | -8.81 | 3.51 | -4.85 | 214.82 | 364.95 | 8.27 | 0.998 |
| 380.10 | 100.00 | -3.78 | 5.04 | -8.82 | 3.51 | -4.85 | 215.72 | 365.09 | 8.26 | 0.998 |
| 380.20 | 100.00 | -3.79 | 5.03 | -8.82 | 3.52 | -4.86 | 216.62 | 365.24 | 8.25 | 0.998 |
| 380.30 | 100.00 | -3.80 | 5.03 | -8.83 | 3.52 | -4.86 | 217.51 | 365.38 | 8.23 | 0.998 |
| 380.40 | 100.00 | -3.81 | 5.02 | -8.83 | 3.52 | -4.87 | 218.40 | 365.52 | 8.22 | 0.998 |
| 380.50 | 100.00 | -3.82 | 5.01 | -8.83 | 3.53 | -4.87 | 219.30 | 365.65 | 8.21 | 0.998 |
| 380.60 | 100.00 | -3.83 | 5.01 | -8.84 | 3.53 | -4.87 | 220.19 | 365.78 | 8.19 | 0.998 |
| 380.70 | 100.00 | -3.84 | 5.00 | -8.84 | 3.53 | -4.88 | 221.08 | 365.91 | 8.18 | 0.998 |
| 380.80 | 100.00 | -3.85 | 5.00 | -8.85 | 3.53 | -4.88 | 221.96 | 366.04 | 8.17 | 0.998 |
| 380.90 | 100.00 | -3.86 | 4.99 | -8.85 | 3.54 | -4.89 | 222.85 | 366.16 | 8.16 | 0.998 |
| 381.00 | 100.00 | -3.86 | 4.99 | -8.85 | 3.54 | -4.89 | 223.73 | 366.28 | 8.14 | 0.998 |
| 381.10 | 100.00 | -3.87 | 4.98 | -8.85 | 3.54 | -4.90 | 224.62 | 366.40 | 8.13 | 0.998 |
| 381.20 | 100.00 | -3.88 | 4.98 | -8.86 | 3.54 | -4.90 | 225.50 | 366.51 | 8.12 | 0.998 |
| 381.30 | 100.00 | -3.89 | 4.97 | -8.86 | 3.55 | -4.90 | 226.38 | 366.62 | 8.11 | 0.998 |
| 381.40 | 100.00 | -3.90 | 4.97 | -8.87 | 3.55 | -4.91 | 227.26 | 366.73 | 8.09 | 0.998 |
| 381.50 | 100.00 | -3.91 | 4.96 | -8.87 | 3.55 | -4.91 | 228.14 | 366.84 | 8.08 | 0.998 |
| 381.60 | 100.00 | -3.92 | 4.95 | -8.87 | 3.55 | -4.92 | 229.01 | 366.94 | 8.07 | 0.998 |
| 381.70 | 100.00 | -3.93 | 4.95 | -8.88 | 3.56 | -4.92 | 229.89 | 367.04 | 8.06 | 0.998 |
| 381.80 | 100.00 | -3.94 | 4.94 | -8.88 | 3.56 | -4.93 | 230.76 | 367.14 | 8.05 | 0.998 |
| 381.90 | 100.00 | -3.95 | 4.94 | -8.89 | 3.56 | -4.93 | 231.64 | 367.24 | 8.04 | 0.998 |
| 382.00 | 100.00 | -3.96 | 4.93 | -8.89 | 3.56 | -4.94 | 232.51 | 367.33 | 8.02 | 0.998 |
| 382.10 | 100.00 | -3.97 | 4.93 | -8.90 | 3.57 | -4.94 | 233.38 | 367.42 | 8.01 | 0.998 |
| 382.20 | 100.00 | -3.97 | 4.92 | -8.89 | 3.57 | -4.94 | 234.24 | 367.51 | 8.00 | 0.998 |
| 382.30 | 100.00 | -3.98 | 4.92 | -8.90 | 3.57 | -4.95 | 235.11 | 367.59 | 7.99 | 0.998 |
| 382.40 | 100.00 | -3.99 | 4.91 | -8.90 | 3.57 | -4.95 | 235.98 | 367.68 | 7.98 | 0.998 |
| 382.50 | 100.00 | -4.00 | 4.91 | -8.91 | 3.58 | -4.96 | 236.84 | 367.76 | 7.97 | 0.998 |
| 382.60 | 100.00 | -4.01 | 4.90 | -8.91 | 3.58 | -4.96 | 237.70 | 367.83 | 7.96 | 0.998 |
| 382.70 | 100.00 | -4.02 | 4.90 | -8.92 | 3.58 | -4.97 | 238.56 | 367.91 | 7.95 | 0.998 |
| 382.80 | 100.00 | -4.03 | 4.89 | -8.92 | 3.58 | -4.97 | 239.42 | 367.98 | 7.94 | 0.998 |
| 382.90 | 100.00 | -4.04 | 4.89 | -8.93 | 3.59 | -4.98 | 240.28 | 368.05 | 7.92 | 0.998 |
| 383.00 | 100.00 | -4.05 | 4.88 | -8.93 | 3.59 | -4.98 | 241.14 | 368.12 | 7.91 | 0.998 |
| 383.10 | 100.00 | -4.06 | 4.88 | -8.94 | 3.59 | -4.99 | 241.99 | 368.18 | 7.90 | 0.998 |
| 383.20 | 100.00 | -4.07 | 4.87 | -8.94 | 3.59 | -4.99 | 242.84 | 368.25 | 7.89 | 0.998 |
| 383.30 | 100.00 | -4.08 | 4.87 | -8.95 | 3.60 | -4.99 | 243.70 | 368.31 | 7.88 | 0.998 |
| 383.40 | 100.00 | -4.08 | 4.86 | -8.94 | 3.60 | -5.00 | 244.55 | 368.36 | 7.87 | 0.998 |
| 383.50 | 100.00 | -4.09 | 4.86 | -8.95 | 3.60 | -5.00 | 245.40 | 368.42 | 7.86 | 0.998 |
| 383.60 | 100.00 | -4.10 | 4.85 | -8.95 | 3.60 | -5.01 | 246.24 | 368.47 | 7.85 | 0.998 |
| 383.70 | 100.00 | -4.11 | 4.85 | -8.96 | 3.61 | -5.01 | 247.09 | 368.52 | 7.84 | 0.998 |
| 383.80 | 100.00 | -4.12 | 4.84 | -8.96 | 3.61 | -5.02 | 247.94 | 368.57 | 7.83 | 0.998 |
| 383.90 | 100.00 | -4.13 | 4.84 | -8.97 | 3.61 | -5.02 | 248.78 | 368.62 | 7.82 | 0.998 |
| 384.00 | 100.00 | -4.14 | 4.83 | -8.97 | 3.61 | -5.03 | 249.62 | 368.66 | 7.81 | 0.998 |
| 384.10 | 100.00 | -4.15 | 4.83 | -8.98 | 3.62 | -5.03 | 250.46 | 368.70 | 7.80 | 0.998 |
| 384.20 | 100.00 | -4.16 | 4.82 | -8.98 | 3.62 | -5.04 | 251.30 | 368.74 | 7.79 | 0.998 |
| 384.30 | 100.00 | -4.17 | 4.82 | -8.99 | 3.62 | -5.04 | 252.14 | 368.78 | 7.78 | 0.998 |
| 384.40 | 100.00 | -4.18 | 4.81 | -8.99 | 3.62 | -5.05 | 252.97 | 368.82 | 7.77 | 0.998 |
| 384.50 | 100.00 | -4.19 | 4.81 | -9.00 | 3.63 | -5.05 | 253.80 | 368.85 | 7.76 | 0.998 |
| 384.60 | 100.00 | -4.20 | 4.80 | -9.00 | 3.63 | -5.06 | 254.64 | 368.88 | 7.75 | 0.998 |
| 384.70 | 100.00 | -4.21 | 4.80 | -9.01 | 3.63 | -5.06 | 255.47 | 368.91 | 7.75 | 0.998 |
| 384.80 | 100.00 | -4.21 | 4.79 | -9.00 | 3.63 | -5.07 | 256.30 | 368.94 | 7.74 | 0.998 |
| 384.90 | 100.00 | -4.22 | 4.79 | -9.01 | 3.64 | -5.07 | 257.13 | 368.96 | 7.73 | 0.998 |
| 385.00 | 100.00 | -4.23 | 4.78 | -9.01 | 3.64 | -5.08 | 257.95 | 368.98 | 7.72 | 0.998 |
| 385.10 | 100.00 | -4.24 | 4.78 | -9.02 | 3.64 | -5.08 | 258.78 | 369.00 | 7.71 | 0.998 |
| 385.20 | 100.00 | -4.25 | 4.77 | -9.02 | 3.64 | -5.09 | 259.60 | 369.02 | 7.70 | 0.998 |
| 385.30 | 100.00 | -4.26 | 4.77 | -9.03 | 3.65 | -5.09 | 260.42 | 369.04 | 7.69 | 0.998 |
| 385.40 | 100.00 | -4.27 | 4.76 | -9.03 | 3.65 | -5.10 | 261.24 | 369.05 | 7.68 | 0.998 |
| 385.50 | 100.00 | -4.28 | 4.76 | -9.04 | 3.65 | -5.10 | 262.06 | 369.06 | 7.67 | 0.998 |
| 385.60 | 100.00 | -4.29 | 4.75 | -9.04 | 3.66 | -5.11 | 262.88 | 369.07 | 7.66 | 0.998 |
| 385.70 | 100.00 | -4.30 | 4.75 | -9.05 | 3.66 | -5.11 | 263.70 | 369.08 | 7.66 | 0.998 |
| 385.80 | 100.00 | -4.31 | 4.74 | -9.05 | 3.66 | -5.12 | 264.51 | 369.08 | 7.65 | 0.998 |
| 385.90 | 100.00 | -4.32 | 4.74 | -9.06 | 3.66 | -5.12 | 265.32 | 369.09 | 7.64 | 0.998 |
| 386.00 | 100.00 | -4.33 | 4.73 | -9.06 | 3.67 | -5.13 | 266.14 | 369.09 | 7.63 | 0.998 |
| 386.10 | 100.00 | -4.34 | 4.73 | -9.07 | 3.67 | -5.13 | 266.95 | 369.09 | 7.62 | 0.998 |
| 386.20 | 100.00 | -4.34 | 4.72 | -9.06 | 3.67 | -5.14 | 267.75 | 369.09 | 7.61 | 0.998 |
| 386.30 | 100.00 | -4.35 | 4.72 | -9.07 | 3.67 | -5.14 | 268.56 | 369.09 | 7.61 | 0.998 |
| 386.40 | 100.00 | -4.36 | 4.72 | -9.08 | 3.68 | -5.15 | 269.37 | 369.08 | 7.60 | 0.998 |
| 386.50 | 100.00 | -4.37 | 4.71 | -9.08 | 3.68 | -5.15 | 270.17 | 369.07 | 7.59 | 0.998 |
| 386.60 | 100.00 | -4.38 | 4.71 | -9.09 | 3.68 | -5.16 | 270.97 | 369.06 | 7.58 | 0.998 |
| 386.70 | 100.00 | -4.39 | 4.70 | -9.09 | 3.68 | -5.16 | 271.77 | 369.05 | 7.57 | 0.998 |
| 386.80 | 100.00 | -4.40 | 4.70 | -9.10 | 3.69 | -5.17 | 272.57 | 369.04 | 7.57 | 0.998 |
| 386.90 | 100.00 | -4.41 | 4.69 | -9.10 | 3.69 | -5.17 | 273.37 | 369.02 | 7.56 | 0.998 |
| 387.00 | 100.00 | -4.42 | 4.69 | -9.11 | 3.69 | -5.18 | 274.17 | 369.01 | 7.55 | 0.998 |
| 387.10 | 100.00 | -4.43 | 4.68 | -9.11 | 3.69 | -5.18 | 274.96 | 368.99 | 7.54 | 0.998 |
| 387.20 | 100.00 | -4.44 | 4.68 | -9.12 | 3.70 | -5.19 | 275.76 | 368.97 | 7.53 | 0.998 |
| 387.30 | 100.00 | -4.45 | 4.67 | -9.12 | 3.70 | -5.19 | 276.55 | 368.95 | 7.53 | 0.998 |
| 387.40 | 100.00 | -4.46 | 4.67 | -9.13 | 3.70 | -5.20 | 277.34 | 368.92 | 7.52 | 0.998 |
| 387.50 | 100.00 | -4.46 | 4.67 | -9.13 | 3.70 | -5.20 | 278.12 | 368.90 | 7.51 | 0.998 |
| 387.60 | 100.00 | -4.47 | 4.66 | -9.13 | 3.71 | -5.21 | 278.91 | 368.87 | 7.50 | 0.998 |
| 387.70 | 100.00 | -4.48 | 4.66 | -9.14 | 3.71 | -5.21 | 279.70 | 368.84 | 7.50 | 0.998 |
| 387.80 | 100.00 | -4.49 | 4.65 | -9.14 | 3.71 | -5.22 | 280.48 | 368.81 | 7.49 | 0.998 |
| 387.90 | 100.00 | -4.50 | 4.65 | -9.15 | 3.71 | -5.22 | 281.26 | 368.78 | 7.48 | 0.998 |
| 388.00 | 100.00 | -4.51 | 4.64 | -9.15 | 3.72 | -5.23 | 282.05 | 368.75 | 7.47 | 0.998 |
| 388.10 | 100.00 | -4.52 | 4.64 | -9.16 | 3.72 | -5.23 | 282.82 | 368.71 | 7.47 | 0.998 |
| 388.20 | 100.00 | -4.53 | 4.63 | -9.16 | 3.72 | -5.24 | 283.60 | 368.68 | 7.46 | 0.998 |
| 388.30 | 100.00 | -4.54 | 4.63 | -9.17 | 3.72 | -5.24 | 284.38 | 368.64 | 7.45 | 0.998 |
| 388.40 | 100.00 | -4.55 | 4.63 | -9.18 | 3.73 | -5.25 | 285.16 | 368.60 | 7.45 | 0.998 |
| 388.50 | 100.00 | -4.56 | 4.62 | -9.18 | 3.73 | -5.25 | 285.93 | 368.56 | 7.44 | 0.998 |
| 388.60 | 100.00 | -4.57 | 4.62 | -9.19 | 3.73 | -5.26 | 286.70 | 368.51 | 7.43 | 0.998 |
| 388.70 | 100.00 | -4.58 | 4.61 | -9.19 | 3.73 | -5.26 | 287.47 | 368.47 | 7.42 | 0.998 |
| 388.80 | 100.00 | -4.58 | 4.61 | -9.19 | 3.74 | -5.27 | 288.24 | 368.42 | 7.42 | 0.998 |
| 388.90 | 100.00 | -4.59 | 4.60 | -9.19 | 3.74 | -5.27 | 289.01 | 368.37 | 7.41 | 0.998 |
| 389.00 | 100.00 | -4.60 | 4.60 | -9.20 | 3.74 | -5.28 | 289.77 | 368.32 | 7.40 | 0.998 |
| 389.10 | 100.00 | -4.61 | 4.60 | -9.21 | 3.74 | -5.28 | 290.54 | 368.27 | 7.40 | 0.998 |
| 389.20 | 100.00 | -4.62 | 4.59 | -9.21 | 3.75 | -5.29 | 291.30 | 368.22 | 7.39 | 0.998 |
| 389.30 | 100.00 | -4.63 | 4.59 | -9.22 | 3.75 | -5.29 | 292.06 | 368.17 | 7.38 | 0.998 |
| 389.40 | 100.00 | -4.64 | 4.58 | -9.22 | 3.75 | -5.30 | 292.82 | 368.11 | 7.38 | 0.998 |
| 389.50 | 100.00 | -4.65 | 4.58 | -9.23 | 3.76 | -5.30 | 293.58 | 368.06 | 7.37 | 0.998 |
| 389.60 | 100.00 | -4.66 | 4.58 | -9.24 | 3.76 | -5.31 | 294.34 | 368.00 | 7.36 | 0.998 |
| 389.70 | 100.00 | -4.67 | 4.57 | -9.24 | 3.76 | -5.31 | 295.10 | 367.94 | 7.36 | 0.998 |
| 389.80 | 100.00 | -4.68 | 4.57 | -9.25 | 3.76 | -5.32 | 295.85 | 367.88 | 7.35 | 0.998 |
| 389.90 | 100.00 | -4.69 | 4.56 | -9.25 | 3.77 | -5.32 | 296.60 | 367.82 | 7.35 | 0.998 |
| 390.00 | 100.00 | -4.69 | 4.56 | -9.25 | 3.77 | -5.33 | 297.35 | 367.75 | 7.34 | 0.998 |
| 390.10 | 100.00 | -4.70 | 4.55 | -9.25 | 3.77 | -5.33 | 298.10 | 367.69 | 7.33 | 0.998 |
| 390.20 | 100.00 | -4.71 | 4.55 | -9.26 | 3.77 | -5.34 | 298.85 | 367.62 | 7.33 | 0.998 |
| 390.30 | 100.00 | -4.72 | 4.55 | -9.27 | 3.78 | -5.34 | 299.60 | 367.55 | 7.32 | 0.998 |
| 390.40 | 100.00 | -4.73 | 4.54 | -9.27 | 3.78 | -5.35 | 300.34 | 367.48 | 7.31 | 0.998 |
| 390.50 | 100.00 | -4.74 | 4.54 | -9.28 | 3.78 | -5.35 | 301.09 | 367.41 | 7.31 | 0.998 |
| 390.60 | 100.00 | -4.75 | 4.53 | -9.28 | 3.78 | -5.36 | 301.83 | 367.34 | 7.30 | 0.998 |
| 390.70 | 100.00 | -4.76 | 4.53 | -9.29 | 3.79 | -5.36 | 302.57 | 367.27 | 7.30 | 0.998 |
| 390.80 | 100.00 | -4.77 | 4.53 | -9.30 | 3.79 | -5.37 | 303.31 | 367.20 | 7.29 | 0.998 |
| 390.90 | 100.00 | -4.78 | 4.52 | -9.30 | 3.79 | -5.37 | 304.05 | 367.12 | 7.28 | 0.998 |
| 391.00 | 100.00 | -4.79 | 4.52 | -9.31 | 3.79 | -5.38 | 304.79 | 367.04 | 7.28 | 0.998 |
| 391.10 | 100.00 | -4.79 | 4.51 | -9.30 | 3.80 | -5.38 | 305.52 | 366.96 | 7.27 | 0.998 |
| 391.20 | 100.00 | -4.80 | 4.51 | -9.31 | 3.80 | -5.39 | 306.26 | 366.88 | 7.27 | 0.998 |
| 391.30 | 100.00 | -4.81 | 4.51 | -9.32 | 3.80 | -5.39 | 306.99 | 366.80 | 7.26 | 0.998 |
| 391.40 | 100.00 | -4.82 | 4.50 | -9.32 | 3.80 | -5.40 | 307.72 | 366.72 | 7.26 | 0.998 |
| 391.50 | 100.00 | -4.83 | 4.50 | -9.33 | 3.81 | -5.40 | 308.45 | 366.64 | 7.25 | 0.998 |
| 391.60 | 100.00 | -4.84 | 4.49 | -9.33 | 3.81 | -5.41 | 309.18 | 366.56 | 7.24 | 0.998 |
| 391.70 | 100.00 | -4.85 | 4.49 | -9.34 | 3.81 | -5.41 | 309.90 | 366.47 | 7.24 | 0.998 |
| 391.80 | 100.00 | -4.86 | 4.49 | -9.35 | 3.81 | -5.42 | 310.63 | 366.38 | 7.23 | 0.998 |
| 391.90 | 100.00 | -4.87 | 4.48 | -9.35 | 3.82 | -5.42 | 311.35 | 366.30 | 7.23 | 0.998 |
| 392.00 | 100.00 | -4.88 | 4.48 | -9.36 | 3.82 | -5.43 | 312.08 | 366.21 | 7.22 | 0.998 |
| 392.10 | 100.00 | -4.88 | 4.47 | -9.35 | 3.82 | -5.43 | 312.80 | 366.12 | 7.22 | 0.998 |
| 392.20 | 100.00 | -4.89 | 4.47 | -9.36 | 3.82 | -5.44 | 313.52 | 366.03 | 7.21 | 0.998 |
| 392.30 | 100.00 | -4.90 | 4.47 | -9.37 | 3.83 | -5.44 | 314.24 | 365.94 | 7.21 | 0.998 |
| 392.40 | 100.00 | -4.91 | 4.46 | -9.37 | 3.83 | -5.45 | 314.95 | 365.84 | 7.20 | 0.998 |
| 392.50 | 100.00 | -4.92 | 4.46 | -9.38 | 3.83 | -5.45 | 315.67 | 365.75 | 7.20 | 0.998 |
| 392.60 | 100.00 | -4.93 | 4.46 | -9.39 | 3.83 | -5.46 | 316.38 | 365.65 | 7.19 | 0.998 |
| 392.70 | 100.00 | -4.94 | 4.45 | -9.39 | 3.84 | -5.46 | 317.10 | 365.56 | 7.18 | 0.998 |
| 392.80 | 100.00 | -4.95 | 4.45 | -9.40 | 3.84 | -5.47 | 317.81 | 365.46 | 7.18 | 0.998 |
| 392.90 | 100.00 | -4.96 | 4.44 | -9.40 | 3.84 | -5.47 | 318.52 | 365.36 | 7.17 | 0.998 |
| 393.00 | 100.00 | -4.97 | 4.44 | -9.41 | 3.84 | -5.48 | 319.22 | 365.26 | 7.17 | 0.998 |
| 393.10 | 100.00 | -4.97 | 4.44 | -9.41 | 3.85 | -5.49 | 319.93 | 365.16 | 7.16 | 0.998 |
| 393.20 | 100.00 | -4.98 | 4.43 | -9.41 | 3.85 | -5.49 | 320.64 | 365.06 | 7.16 | 0.998 |
| 393.30 | 100.00 | -4.99 | 4.43 | -9.42 | 3.85 | -5.50 | 321.34 | 364.95 | 7.15 | 0.998 |
| 393.40 | 100.00 | -5.00 | 4.43 | -9.43 | 3.85 | -5.50 | 322.05 | 364.85 | 7.15 | 0.998 |
| 393.50 | 100.00 | -5.01 | 4.42 | -9.43 | 3.86 | -5.51 | 322.75 | 364.74 | 7.14 | 0.998 |
| 393.60 | 100.00 | -5.02 | 4.42 | -9.44 | 3.86 | -5.51 | 323.45 | 364.64 | 7.14 | 0.998 |
| 393.70 | 100.00 | -5.03 | 4.41 | -9.44 | 3.86 | -5.52 | 324.15 | 364.53 | 7.13 | 0.998 |
| 393.80 | 100.00 | -5.04 | 4.41 | -9.45 | 3.86 | -5.52 | 324.85 | 364.42 | 7.13 | 0.998 |
| 393.90 | 100.00 | -5.05 | 4.41 | -9.46 | 3.87 | -5.53 | 325.54 | 364.32 | 7.13 | 0.998 |
| 394.00 | 100.00 | -5.05 | 4.40 | -9.45 | 3.87 | -5.53 | 326.24 | 364.21 | 7.12 | 0.997 |
| 394.10 | 100.00 | -5.06 | 4.40 | -9.46 | 3.87 | -5.54 | 326.93 | 364.10 | 7.12 | 0.997 |
| 394.20 | 100.00 | -5.07 | 4.40 | -9.47 | 3.87 | -5.54 | 327.63 | 363.98 | 7.11 | 0.997 |
| 394.30 | 100.00 | -5.08 | 4.39 | -9.47 | 3.88 | -5.55 | 328.32 | 363.87 | 7.11 | 0.997 |
| 394.40 | 100.00 | -5.09 | 4.39 | -9.48 | 3.88 | -5.55 | 329.01 | 363.76 | 7.10 | 0.997 |
| 394.50 | 100.00 | -5.10 | 4.39 | -9.49 | 3.88 | -5.56 | 329.70 | 363.64 | 7.10 | 0.997 |
| 394.60 | 100.00 | -5.11 | 4.38 | -9.49 | 3.88 | -5.56 | 330.38 | 363.53 | 7.09 | 0.997 |
| 394.70 | 100.00 | -5.12 | 4.38 | -9.50 | 3.89 | -5.57 | 331.07 | 363.41 | 7.09 | 0.997 |
| 394.80 | 100.00 | -5.12 | 4.38 | -9.50 | 3.89 | -5.57 | 331.75 | 363.30 | 7.08 | 0.997 |
| 394.90 | 100.00 | -5.13 | 4.37 | -9.50 | 3.89 | -5.58 | 332.44 | 363.18 | 7.08 | 0.997 |
| 395.00 | 100.00 | -5.14 | 4.37 | -9.51 | 3.89 | -5.58 | 333.12 | 363.06 | 7.07 | 0.997 |
| 395.10 | 100.00 | -5.15 | 4.36 | -9.51 | 3.90 | -5.59 | 333.80 | 362.94 | 7.07 | 0.997 |
| 395.20 | 100.00 | -5.16 | 4.36 | -9.52 | 3.90 | -5.59 | 334.48 | 362.82 | 7.07 | 0.997 |
| 395.30 | 100.00 | -5.17 | 4.36 | -9.53 | 3.90 | -5.60 | 335.16 | 362.70 | 7.06 | 0.997 |
| 395.40 | 100.00 | -5.18 | 4.35 | -9.53 | 3.90 | -5.60 | 335.84 | 362.58 | 7.06 | 0.997 |
| 395.50 | 100.00 | -5.19 | 4.35 | -9.54 | 3.91 | -5.61 | 336.51 | 362.45 | 7.05 | 0.997 |
| 395.60 | 100.00 | -5.19 | 4.35 | -9.54 | 3.91 | -5.61 | 337.19 | 362.33 | 7.05 | 0.997 |
| 395.70 | 100.00 | -5.20 | 4.34 | -9.54 | 3.91 | -5.62 | 337.86 | 362.20 | 7.04 | 0.997 |
| 395.80 | 100.00 | -5.21 | 4.34 | -9.55 | 3.91 | -5.62 | 338.53 | 362.08 | 7.04 | 0.997 |
| 395.90 | 100.00 | -5.22 | 4.34 | -9.56 | 3.92 | -5.63 | 339.20 | 361.95 | 7.04 | 0.997 |
| 396.00 | 100.00 | -5.23 | 4.33 | -9.56 | 3.92 | -5.63 | 339.87 | 361.83 | 7.03 | 0.997 |
| 396.10 | 100.00 | -5.24 | 4.33 | -9.57 | 3.92 | -5.64 | 340.54 | 361.70 | 7.03 | 0.997 |
| 396.20 | 100.00 | -5.25 | 4.33 | -9.58 | 3.92 | -5.64 | 341.21 | 361.57 | 7.02 | 0.997 |
| 396.30 | 100.00 | -5.26 | 4.32 | -9.58 | 3.93 | -5.65 | 341.88 | 361.44 | 7.02 | 0.997 |
| 396.40 | 100.00 | -5.26 | 4.32 | -9.58 | 3.93 | -5.65 | 342.54 | 361.31 | 7.02 | 0.997 |
| 396.50 | 100.00 | -5.27 | 4.32 | -9.59 | 3.93 | -5.66 | 343.20 | 361.18 | 7.01 | 0.997 |
| 396.60 | 100.00 | -5.28 | 4.31 | -9.59 | 3.93 | -5.66 | 343.87 | 361.05 | 7.01 | 0.997 |
| 396.70 | 100.00 | -5.29 | 4.31 | -9.60 | 3.94 | -5.67 | 344.53 | 360.92 | 7.00 | 0.997 |
| 396.80 | 100.00 | -5.30 | 4.31 | -9.61 | 3.94 | -5.67 | 345.19 | 360.78 | 7.00 | 0.997 |
| 396.90 | 100.00 | -5.31 | 4.30 | -9.61 | 3.94 | -5.68 | 345.85 | 360.65 | 7.00 | 0.997 |
| 397.00 | 100.00 | -5.32 | 4.30 | -9.62 | 3.94 | -5.68 | 346.51 | 360.52 | 6.99 | 0.997 |
| 397.10 | 100.00 | -5.32 | 4.30 | -9.62 | 3.94 | -5.69 | 347.16 | 360.38 | 6.99 | 0.997 |
| 397.20 | 100.00 | -5.33 | 4.29 | -9.62 | 3.95 | -5.69 | 347.82 | 360.25 | 6.99 | 0.997 |
| 397.30 | 100.00 | -5.34 | 4.29 | -9.63 | 3.95 | -5.70 | 348.47 | 360.11 | 6.98 | 0.997 |
| 397.40 | 100.00 | -5.35 | 4.29 | -9.64 | 3.95 | -5.70 | 349.12 | 359.97 | 6.98 | 0.997 |
| 397.50 | 100.00 | -5.36 | 4.28 | -9.64 | 3.95 | -5.71 | 349.78 | 359.83 | 6.97 | 0.997 |
| 397.60 | 100.00 | -5.37 | 4.28 | -9.65 | 3.96 | -5.71 | 350.43 | 359.70 | 6.97 | 0.997 |
| 397.70 | 100.00 | -5.38 | 4.28 | -9.66 | 3.96 | -5.72 | 351.08 | 359.56 | 6.97 | 0.997 |
| 397.80 | 100.00 | -5.38 | 4.27 | -9.65 | 3.96 | -5.72 | 351.72 | 359.42 | 6.96 | 0.997 |
| 397.90 | 100.00 | -5.39 | 4.27 | -9.66 | 3.96 | -5.73 | 352.37 | 359.28 | 6.96 | 0.997 |
| 398.00 | 100.00 | -5.40 | 4.27 | -9.67 | 3.97 | -5.73 | 353.02 | 359.14 | 6.96 | 0.997 |
| 398.10 | 100.00 | -5.41 | 4.27 | -9.68 | 3.97 | -5.74 | 353.66 | 358.99 | 6.95 | 0.997 |
| 398.20 | 100.00 | -5.42 | 4.26 | -9.68 | 3.97 | -5.75 | 354.31 | 358.85 | 6.95 | 0.997 |
| 398.30 | 100.00 | -5.43 | 4.26 | -9.69 | 3.97 | -5.75 | 354.95 | 358.71 | 6.95 | 0.997 |
| 398.40 | 100.00 | -5.44 | 4.26 | -9.70 | 3.98 | -5.76 | 355.59 | 358.56 | 6.94 | 0.997 |
| 398.50 | 100.00 | -5.44 | 4.25 | -9.69 | 3.98 | -5.76 | 356.23 | 358.42 | 6.94 | 0.997 |
| 398.60 | 100.00 | -5.45 | 4.25 | -9.70 | 3.98 | -5.77 | 356.87 | 358.28 | 6.93 | 0.997 |
| 398.70 | 100.00 | -5.46 | 4.25 | -9.71 | 3.98 | -5.78 | 357.51 | 358.13 | 6.93 | 0.997 |
| 398.80 | 100.00 | -5.47 | 4.24 | -9.71 | 3.99 | -5.78 | 358.15 | 357.98 | 6.93 | 0.997 |
| 398.90 | 100.00 | -5.48 | 4.24 | -9.72 | 3.99 | -5.79 | 358.78 | 357.84 | 6.92 | 0.997 |
| 399.00 | 100.00 | -5.49 | 4.24 | -9.73 | 3.99 | -5.80 | 359.42 | 357.69 | 6.92 | 0.997 |
| 399.10 | 100.00 | -5.49 | 4.23 | -9.72 | 3.99 | -5.80 | 360.05 | 357.54 | 6.92 | 0.997 |
| 399.20 | 100.00 | -5.50 | 4.23 | -9.73 | 4.00 | -5.81 | 360.68 | 357.39 | 6.91 | 0.997 |
| 399.30 | 100.00 | -5.51 | 4.23 | -9.74 | 4.00 | -5.82 | 361.32 | 357.24 | 6.91 | 0.997 |
| 399.40 | 100.00 | -5.52 | 4.22 | -9.74 | 4.00 | -5.82 | 361.95 | 357.10 | 6.91 | 0.997 |
| 399.50 | 100.00 | -5.53 | 4.22 | -9.75 | 4.00 | -5.83 | 362.58 | 356.94 | 6.91 | 0.997 |
| 399.60 | 100.00 | -5.54 | 4.22 | -9.76 | 4.01 | -5.83 | 363.21 | 356.80 | 6.90 | 0.997 |
| 399.70 | 100.00 | -5.55 | 4.22 | -9.77 | 4.01 | -5.84 | 363.83 | 356.64 | 6.90 | 0.997 |
| 399.80 | 100.00 | -5.55 | 4.21 | -9.76 | 4.01 | -5.85 | 364.46 | 356.49 | 6.90 | 0.997 |
| 399.90 | 100.00 | -5.56 | 4.21 | -9.77 | 4.01 | -5.85 | 365.09 | 356.34 | 6.89 | 0.997 |
| 400.00 | 100.00 | -5.57 | 4.21 | -9.78 | 4.01 | -5.86 | 365.71 | 356.19 | 6.89 | 0.997 |

**IMPORTANT NOTE: AVG PWR GAIN _MUST_ BE IN THE RANGE 0.8-1.2 FOR A VALID NEC MODEL (IDEALLY = 1)!
VALUES OUTSIDE THIS RANGE INDICATE A FLAWED MODEL, USUALLY A RESULT OF INCORRECT SEGMENTATION.
SEE DISCUSSION, p.101, NEC 4.1 USER'S MANUAL, LLNL UCRL-MA-109338 Pt. I (Gerald J. Burke), 1992.

# Appendix III. Fixed $Z_0$ Yagi Performance

```
CM File: YAGI.NEC
CM YAGI ARRAY IN FREE SPACE
CM Band center frequency, Fc = 299.8 MHz
CM Freq step = 50 MHz +/- Fc
CM Run ID: 07282011_124011
CM Fitness function:
CM .2*Gfwd(L)-4*VSWR(L)+1*Gfwd(M)-8*VSWR(M)+
CM 1*Gfwd(U)-.8*VSWR(U)
CM where L,M,U are lower/mid/upper frequencies
CM Zo=50 ohms
CM Note: All dimensions are in METERS.
CM File ID 07282011180440
CM Nd= 12, p= 44, j= 49
```

FREE SPACE YAGI: SUMMARY NEC DATA
================================

| F(MHz) | Rad Eff (%) | Fwd Gain (dBi) | Rear Gain (dBi) | FB Ratio (dB) | Max Gain (dBi) | Min Gain (dBi) | Rin (ohms) | Xin (ohms) | VSWR//50 | Avg Pwr Gain** |
|---|---|---|---|---|---|---|---|---|---|---|
| 200.00 | 100.00 | 2.71 | 1.99 | 0.72 | 2.71 | 1.45 | 24.58 | -177.96 | 28.26 | 0.996 |
| 200.10 | 100.00 | 2.71 | 1.99 | 0.72 | 2.71 | 1.45 | 24.61 | -177.71 | 28.15 | 0.996 |
| 200.20 | 100.00 | 2.72 | 1.99 | 0.73 | 2.72 | 1.45 | 24.64 | -177.45 | 28.04 | 0.996 |
| 200.30 | 100.00 | 2.72 | 1.99 | 0.73 | 2.72 | 1.45 | 24.67 | -177.20 | 27.94 | 0.996 |
| 200.40 | 100.00 | 2.72 | 1.99 | 0.73 | 2.72 | 1.45 | 24.70 | -176.94 | 27.83 | 0.996 |
| 200.50 | 100.00 | 2.72 | 1.99 | 0.73 | 2.72 | 1.45 | 24.73 | -176.69 | 27.73 | 0.996 |
| 200.60 | 100.00 | 2.72 | 1.99 | 0.73 | 2.72 | 1.45 | 24.76 | -176.44 | 27.62 | 0.996 |
| 200.70 | 100.00 | 2.73 | 1.99 | 0.74 | 2.73 | 1.45 | 24.79 | -176.18 | 27.52 | 0.996 |
| 200.80 | 100.00 | 2.73 | 1.99 | 0.74 | 2.73 | 1.45 | 24.82 | -175.93 | 27.41 | 0.996 |
| 200.90 | 100.00 | 2.73 | 1.99 | 0.74 | 2.73 | 1.45 | 24.85 | -175.68 | 27.31 | 0.996 |
| 201.00 | 100.00 | 2.73 | 1.99 | 0.74 | 2.73 | 1.45 | 24.88 | -175.43 | 27.20 | 0.996 |
| 201.10 | 100.00 | 2.73 | 1.99 | 0.74 | 2.73 | 1.45 | 24.91 | -175.17 | 27.10 | 0.996 |
| 201.20 | 100.00 | 2.73 | 1.99 | 0.74 | 2.73 | 1.44 | 24.94 | -174.92 | 27.00 | 0.996 |
| 201.30 | 100.00 | 2.74 | 1.99 | 0.75 | 2.74 | 1.44 | 24.98 | -174.67 | 26.90 | 0.996 |
| 201.40 | 100.00 | 2.74 | 2.00 | 0.74 | 2.74 | 1.44 | 25.01 | -174.42 | 26.79 | 0.996 |
| 201.50 | 100.00 | 2.74 | 2.00 | 0.74 | 2.74 | 1.44 | 25.04 | -174.16 | 26.69 | 0.996 |
| 201.60 | 100.00 | 2.74 | 2.00 | 0.74 | 2.74 | 1.44 | 25.07 | -173.91 | 26.59 | 0.996 |
| 201.70 | 100.00 | 2.74 | 2.00 | 0.74 | 2.74 | 1.44 | 25.10 | -173.66 | 26.49 | 0.996 |
| 201.80 | 100.00 | 2.75 | 2.00 | 0.75 | 2.75 | 1.44 | 25.13 | -173.41 | 26.39 | 0.996 |
| 201.90 | 100.00 | 2.75 | 2.00 | 0.75 | 2.75 | 1.44 | 25.16 | -173.16 | 26.29 | 0.996 |
| 202.00 | 100.00 | 2.75 | 2.00 | 0.75 | 2.75 | 1.44 | 25.19 | -172.91 | 26.19 | 0.996 |
| 202.10 | 100.00 | 2.75 | 2.00 | 0.75 | 2.75 | 1.44 | 25.22 | -172.65 | 26.09 | 0.996 |
| 202.20 | 100.00 | 2.75 | 2.00 | 0.75 | 2.75 | 1.44 | 25.25 | -172.40 | 25.99 | 0.996 |
| 202.30 | 100.00 | 2.75 | 2.00 | 0.75 | 2.75 | 1.44 | 25.28 | -172.15 | 25.89 | 0.996 |
| 202.40 | 100.00 | 2.76 | 2.00 | 0.76 | 2.76 | 1.44 | 25.31 | -171.90 | 25.79 | 0.996 |
| 202.50 | 100.00 | 2.76 | 2.00 | 0.76 | 2.76 | 1.44 | 25.34 | -171.65 | 25.70 | 0.996 |
| 202.60 | 100.00 | 2.76 | 2.00 | 0.76 | 2.76 | 1.44 | 25.37 | -171.40 | 25.60 | 0.996 |
| 202.70 | 100.00 | 2.76 | 2.00 | 0.76 | 2.76 | 1.44 | 25.40 | -171.15 | 25.50 | 0.996 |
| 202.80 | 100.00 | 2.76 | 2.00 | 0.76 | 2.76 | 1.44 | 25.43 | -170.90 | 25.40 | 0.996 |
| 202.90 | 100.00 | 2.77 | 2.00 | 0.77 | 2.77 | 1.44 | 25.46 | -170.65 | 25.31 | 0.996 |
| 203.00 | 100.00 | 2.77 | 2.00 | 0.77 | 2.77 | 1.43 | 25.49 | -170.40 | 25.21 | 0.996 |
| 203.10 | 100.00 | 2.77 | 2.00 | 0.77 | 2.77 | 1.43 | 25.52 | -170.15 | 25.11 | 0.996 |
| 203.20 | 100.00 | 2.77 | 2.00 | 0.77 | 2.77 | 1.43 | 25.56 | -169.90 | 25.02 | 0.996 |
| 203.30 | 100.00 | 2.77 | 2.01 | 0.76 | 2.77 | 1.43 | 25.59 | -169.65 | 24.92 | 0.996 |
| 203.40 | 100.00 | 2.78 | 2.01 | 0.77 | 2.78 | 1.43 | 25.62 | -169.40 | 24.83 | 0.996 |
| 203.50 | 100.00 | 2.78 | 2.01 | 0.77 | 2.78 | 1.43 | 25.65 | -169.15 | 24.73 | 0.996 |
| 203.60 | 100.00 | 2.78 | 2.01 | 0.77 | 2.78 | 1.43 | 25.68 | -168.90 | 24.64 | 0.996 |
| 203.70 | 100.00 | 2.78 | 2.01 | 0.77 | 2.78 | 1.43 | 25.71 | -168.65 | 24.55 | 0.996 |
| 203.80 | 100.00 | 2.78 | 2.01 | 0.77 | 2.78 | 1.43 | 25.74 | -168.40 | 24.45 | 0.996 |
| 203.90 | 100.00 | 2.78 | 2.01 | 0.77 | 2.78 | 1.43 | 25.77 | -168.16 | 24.36 | 0.996 |
| 204.00 | 100.00 | 2.79 | 2.01 | 0.78 | 2.79 | 1.43 | 25.80 | -167.91 | 24.26 | 0.996 |
| 204.10 | 100.00 | 2.79 | 2.01 | 0.78 | 2.79 | 1.43 | 25.83 | -167.66 | 24.17 | 0.996 |
| 204.20 | 100.00 | 2.79 | 2.01 | 0.78 | 2.79 | 1.43 | 25.86 | -167.41 | 24.08 | 0.996 |
| 204.30 | 100.00 | 2.79 | 2.01 | 0.78 | 2.79 | 1.43 | 25.90 | -167.16 | 23.99 | 0.996 |
| 204.40 | 100.00 | 2.79 | 2.01 | 0.78 | 2.79 | 1.43 | 25.93 | -166.91 | 23.90 | 0.996 |
| 204.50 | 100.00 | 2.80 | 2.01 | 0.79 | 2.80 | 1.43 | 25.96 | -166.66 | 23.80 | 0.996 |
| 204.60 | 100.00 | 2.80 | 2.01 | 0.79 | 2.80 | 1.42 | 25.99 | -166.42 | 23.71 | 0.996 |
| 204.70 | 100.00 | 2.80 | 2.01 | 0.79 | 2.80 | 1.42 | 26.02 | -166.17 | 23.62 | 0.996 |
| 204.80 | 100.00 | 2.80 | 2.01 | 0.79 | 2.80 | 1.42 | 26.05 | -165.92 | 23.53 | 0.996 |
| 204.90 | 100.00 | 2.80 | 2.01 | 0.79 | 2.80 | 1.42 | 26.08 | -165.67 | 23.44 | 0.996 |
| 205.00 | 100.00 | 2.81 | 2.01 | 0.80 | 2.81 | 1.42 | 26.11 | -165.43 | 23.35 | 0.996 |
| 205.10 | 100.00 | 2.81 | 2.02 | 0.79 | 2.81 | 1.42 | 26.15 | -165.18 | 23.26 | 0.996 |
| 205.20 | 100.00 | 2.81 | 2.02 | 0.79 | 2.81 | 1.42 | 26.18 | -164.93 | 23.17 | 0.996 |
| 205.30 | 100.00 | 2.81 | 2.02 | 0.79 | 2.81 | 1.42 | 26.21 | -164.68 | 23.09 | 0.996 |
| 205.40 | 100.00 | 2.81 | 2.02 | 0.79 | 2.81 | 1.42 | 26.24 | -164.44 | 23.00 | 0.996 |
| 205.50 | 100.00 | 2.82 | 2.02 | 0.80 | 2.82 | 1.42 | 26.27 | -164.19 | 22.91 | 0.996 |
| 205.60 | 100.00 | 2.82 | 2.02 | 0.80 | 2.82 | 1.42 | 26.30 | -163.94 | 22.82 | 0.996 |
| 205.70 | 100.00 | 2.82 | 2.02 | 0.80 | 2.82 | 1.42 | 26.33 | -163.70 | 22.73 | 0.996 |
| 205.80 | 100.00 | 2.82 | 2.02 | 0.80 | 2.82 | 1.42 | 26.37 | -163.45 | 22.65 | 0.996 |
| 205.90 | 100.00 | 2.82 | 2.02 | 0.80 | 2.82 | 1.42 | 26.40 | -163.20 | 22.56 | 0.996 |
| 206.00 | 100.00 | 2.83 | 2.02 | 0.81 | 2.83 | 1.42 | 26.43 | -162.96 | 22.47 | 0.996 |
| 206.10 | 100.00 | 2.83 | 2.02 | 0.81 | 2.83 | 1.41 | 26.46 | -162.71 | 22.39 | 0.996 |
| 206.20 | 100.00 | 2.83 | 2.02 | 0.81 | 2.83 | 1.41 | 26.49 | -162.47 | 22.30 | 0.996 |
| 206.30 | 100.00 | 2.83 | 2.02 | 0.81 | 2.83 | 1.41 | 26.52 | -162.22 | 22.21 | 0.996 |
| 206.40 | 100.00 | 2.83 | 2.02 | 0.81 | 2.83 | 1.41 | 26.56 | -161.98 | 22.13 | 0.996 |
| 206.50 | 100.00 | 2.84 | 2.02 | 0.82 | 2.84 | 1.41 | 26.59 | -161.73 | 22.04 | 0.996 |
| 206.60 | 100.00 | 2.84 | 2.03 | 0.81 | 2.84 | 1.41 | 26.62 | -161.48 | 21.96 | 0.996 |
| 206.70 | 100.00 | 2.84 | 2.03 | 0.81 | 2.84 | 1.41 | 26.65 | -161.24 | 21.87 | 0.996 |
| 206.80 | 100.00 | 2.84 | 2.03 | 0.81 | 2.84 | 1.41 | 26.68 | -160.99 | 21.79 | 0.996 |
| 206.90 | 100.00 | 2.84 | 2.03 | 0.81 | 2.84 | 1.41 | 26.71 | -160.75 | 21.71 | 0.996 |
| 207.00 | 100.00 | 2.85 | 2.03 | 0.82 | 2.85 | 1.41 | 26.75 | -160.50 | 21.62 | 0.996 |
| 207.10 | 100.00 | 2.85 | 2.03 | 0.82 | 2.85 | 1.41 | 26.78 | -160.26 | 21.54 | 0.996 |
| 207.20 | 100.00 | 2.85 | 2.03 | 0.82 | 2.85 | 1.41 | 26.81 | -160.02 | 21.46 | 0.996 |
| 207.30 | 100.00 | 2.85 | 2.03 | 0.82 | 2.85 | 1.41 | 26.84 | -159.77 | 21.37 | 0.996 |
| 207.40 | 100.00 | 2.85 | 2.03 | 0.82 | 2.85 | 1.41 | 26.87 | -159.53 | 21.29 | 0.996 |
| 207.50 | 100.00 | 2.86 | 2.03 | 0.83 | 2.86 | 1.41 | 26.91 | -159.28 | 21.21 | 0.996 |
| 207.60 | 100.00 | 2.86 | 2.03 | 0.83 | 2.86 | 1.40 | 26.94 | -159.04 | 21.13 | 0.996 |
| 207.70 | 100.00 | 2.86 | 2.03 | 0.83 | 2.86 | 1.40 | 26.97 | -158.79 | 21.04 | 0.996 |
| 207.80 | 100.00 | 2.86 | 2.03 | 0.83 | 2.86 | 1.40 | 27.00 | -158.55 | 20.96 | 0.996 |
| 207.90 | 100.00 | 2.86 | 2.03 | 0.83 | 2.86 | 1.40 | 27.03 | -158.31 | 20.88 | 0.996 |
| 208.00 | 100.00 | 2.87 | 2.03 | 0.84 | 2.87 | 1.40 | 27.07 | -158.06 | 20.80 | 0.996 |
| 208.10 | 100.00 | 2.87 | 2.04 | 0.83 | 2.87 | 1.40 | 27.10 | -157.82 | 20.72 | 0.996 |
| 208.20 | 100.00 | 2.87 | 2.04 | 0.83 | 2.87 | 1.40 | 27.13 | -157.58 | 20.64 | 0.996 |
| 208.30 | 100.00 | 2.87 | 2.04 | 0.83 | 2.87 | 1.40 | 27.16 | -157.33 | 20.56 | 0.996 |
| 208.40 | 100.00 | 2.87 | 2.04 | 0.83 | 2.87 | 1.40 | 27.19 | -157.09 | 20.48 | 0.996 |
| 208.50 | 100.00 | 2.88 | 2.04 | 0.84 | 2.88 | 1.40 | 27.23 | -156.84 | 20.40 | 0.996 |
| 208.60 | 100.00 | 2.88 | 2.04 | 0.84 | 2.88 | 1.40 | 27.26 | -156.60 | 20.32 | 0.996 |
| 208.70 | 100.00 | 2.88 | 2.04 | 0.84 | 2.88 | 1.40 | 27.29 | -156.36 | 20.24 | 0.996 |
| 208.80 | 100.00 | 2.88 | 2.04 | 0.84 | 2.88 | 1.40 | 27.32 | -156.12 | 20.17 | 0.996 |
| 208.90 | 100.00 | 2.88 | 2.04 | 0.84 | 2.88 | 1.40 | 27.36 | -155.87 | 20.09 | 0.996 |
| 209.00 | 100.00 | 2.89 | 2.04 | 0.85 | 2.89 | 1.40 | 27.39 | -155.63 | 20.01 | 0.996 |
| 209.10 | 100.00 | 2.89 | 2.04 | 0.85 | 2.89 | 1.39 | 27.42 | -155.39 | 19.93 | 0.996 |
| 209.20 | 100.00 | 2.89 | 2.04 | 0.85 | 2.89 | 1.39 | 27.45 | -155.15 | 19.86 | 0.996 |
| 209.30 | 100.00 | 2.89 | 2.04 | 0.85 | 2.89 | 1.39 | 27.49 | -154.90 | 19.78 | 0.996 |
| 209.40 | 100.00 | 2.89 | 2.05 | 0.84 | 2.89 | 1.39 | 27.52 | -154.66 | 19.70 | 0.996 |
| 209.50 | 100.00 | 2.90 | 2.05 | 0.85 | 2.90 | 1.39 | 27.55 | -154.42 | 19.62 | 0.996 |
| 209.60 | 100.00 | 2.90 | 2.05 | 0.85 | 2.90 | 1.39 | 27.58 | -154.18 | 19.55 | 0.996 |
| 209.70 | 100.00 | 2.90 | 2.05 | 0.85 | 2.90 | 1.39 | 27.62 | -153.94 | 19.47 | 0.996 |
| 209.80 | 100.00 | 2.90 | 2.05 | 0.85 | 2.90 | 1.39 | 27.65 | -153.70 | 19.40 | 0.996 |
| 209.90 | 100.00 | 2.91 | 2.05 | 0.86 | 2.91 | 1.39 | 27.68 | -153.46 | 19.32 | 0.996 |
| 210.00 | 100.00 | 2.91 | 2.05 | 0.86 | 2.91 | 1.39 | 27.72 | -153.21 | 19.25 | 0.996 |
| 210.10 | 100.00 | 2.91 | 2.05 | 0.86 | 2.91 | 1.39 | 27.75 | -152.97 | 19.17 | 0.996 |
| 210.20 | 100.00 | 2.91 | 2.05 | 0.86 | 2.91 | 1.39 | 27.78 | -152.73 | 19.10 | 0.996 |
| 210.30 | 100.00 | 2.91 | 2.05 | 0.86 | 2.91 | 1.39 | 27.81 | -152.49 | 19.02 | 0.996 |
| 210.40 | 100.00 | 2.92 | 2.05 | 0.87 | 2.92 | 1.39 | 27.85 | -152.25 | 18.95 | 0.996 |
| 210.50 | 100.00 | 2.92 | 2.05 | 0.87 | 2.92 | 1.38 | 27.88 | -152.01 | 18.87 | 0.996 |
| 210.60 | 100.00 | 2.92 | 2.05 | 0.87 | 2.92 | 1.38 | 27.91 | -151.77 | 18.80 | 0.996 |
| 210.70 | 100.00 | 2.92 | 2.06 | 0.86 | 2.92 | 1.38 | 27.95 | -151.53 | 18.73 | 0.996 |
| 210.80 | 100.00 | 2.92 | 2.06 | 0.86 | 2.92 | 1.38 | 27.98 | -151.29 | 18.65 | 0.996 |
| 210.90 | 100.00 | 2.93 | 2.06 | 0.87 | 2.93 | 1.38 | 28.01 | -151.05 | 18.58 | 0.996 |
| 211.00 | 100.00 | 2.93 | 2.06 | 0.87 | 2.93 | 1.38 | 28.04 | -150.81 | 18.51 | 0.996 |
| 211.10 | 100.00 | 2.93 | 2.06 | 0.87 | 2.93 | 1.38 | 28.08 | -150.57 | 18.44 | 0.996 |
| 211.20 | 100.00 | 2.93 | 2.06 | 0.87 | 2.93 | 1.38 | 28.11 | -150.33 | 18.36 | 0.996 |
| 211.30 | 100.00 | 2.94 | 2.06 | 0.88 | 2.94 | 1.38 | 28.14 | -150.09 | 18.29 | 0.996 |
| 211.40 | 100.00 | 2.94 | 2.06 | 0.88 | 2.94 | 1.38 | 28.18 | -149.85 | 18.22 | 0.996 |
| 211.50 | 100.00 | 2.94 | 2.06 | 0.88 | 2.94 | 1.38 | 28.21 | -149.61 | 18.15 | 0.996 |
| 211.60 | 100.00 | 2.94 | 2.06 | 0.88 | 2.94 | 1.38 | 28.24 | -149.37 | 18.08 | 0.996 |
| 211.70 | 100.00 | 2.94 | 2.06 | 0.88 | 2.94 | 1.38 | 28.28 | -149.13 | 18.01 | 0.996 |
| 211.80 | 100.00 | 2.95 | 2.06 | 0.89 | 2.95 | 1.38 | 28.31 | -148.89 | 17.94 | 0.996 |
| 211.90 | 100.00 | 2.95 | 2.07 | 0.88 | 2.95 | 1.37 | 28.34 | -148.65 | 17.87 | 0.996 |
| 212.00 | 100.00 | 2.95 | 2.07 | 0.88 | 2.95 | 1.37 | 28.38 | -148.41 | 17.80 | 0.996 |
| 212.10 | 100.00 | 2.95 | 2.07 | 0.88 | 2.95 | 1.37 | 28.41 | -148.17 | 17.73 | 0.996 |
| 212.20 | 100.00 | 2.95 | 2.07 | 0.88 | 2.95 | 1.37 | 28.44 | -147.93 | 17.66 | 0.996 |
| 212.30 | 100.00 | 2.96 | 2.07 | 0.89 | 2.96 | 1.37 | 28.48 | -147.70 | 17.59 | 0.996 |
| 212.40 | 100.00 | 2.96 | 2.07 | 0.89 | 2.96 | 1.37 | 28.51 | -147.46 | 17.52 | 0.996 |
| 212.50 | 100.00 | 2.96 | 2.07 | 0.89 | 2.96 | 1.37 | 28.54 | -147.22 | 17.45 | 0.996 |
| 212.60 | 100.00 | 2.96 | 2.07 | 0.89 | 2.96 | 1.37 | 28.58 | -146.98 | 17.38 | 0.996 |
| 212.70 | 100.00 | 2.97 | 2.07 | 0.90 | 2.97 | 1.37 | 28.61 | -146.74 | 17.31 | 0.996 |
| 212.80 | 100.00 | 2.97 | 2.07 | 0.90 | 2.97 | 1.37 | 28.64 | -146.50 | 17.25 | 0.996 |
| 212.90 | 100.00 | 2.97 | 2.07 | 0.90 | 2.97 | 1.37 | 28.68 | -146.26 | 17.18 | 0.996 |
| 213.00 | 100.00 | 2.97 | 2.08 | 0.89 | 2.97 | 1.37 | 28.71 | -146.03 | 17.11 | 0.996 |
| 213.10 | 100.00 | 2.97 | 2.08 | 0.89 | 2.97 | 1.37 | 28.75 | -145.79 | 17.04 | 0.996 |
| 213.20 | 100.00 | 2.98 | 2.08 | 0.90 | 2.98 | 1.36 | 28.78 | -145.55 | 16.98 | 0.996 |
| 213.30 | 100.00 | 2.98 | 2.08 | 0.90 | 2.98 | 1.36 | 28.81 | -145.31 | 16.91 | 0.996 |
| 213.40 | 100.00 | 2.98 | 2.08 | 0.90 | 2.98 | 1.36 | 28.85 | -145.08 | 16.84 | 0.996 |
| 213.50 | 100.00 | 2.98 | 2.08 | 0.90 | 2.98 | 1.36 | 28.88 | -144.84 | 16.78 | 0.996 |
| 213.60 | 100.00 | 2.99 | 2.08 | 0.91 | 2.99 | 1.36 | 28.91 | -144.60 | 16.71 | 0.996 |
| 213.70 | 100.00 | 2.99 | 2.08 | 0.91 | 2.99 | 1.36 | 28.95 | -144.36 | 16.65 | 0.996 |
| 213.80 | 100.00 | 2.99 | 2.08 | 0.91 | 2.99 | 1.36 | 28.98 | -144.13 | 16.58 | 0.996 |
| 213.90 | 100.00 | 2.99 | 2.08 | 0.91 | 2.99 | 1.36 | 29.02 | -143.89 | 16.51 | 0.996 |
| 214.00 | 100.00 | 2.99 | 2.08 | 0.91 | 2.99 | 1.36 | 29.05 | -143.65 | 16.45 | 0.996 |
| 214.10 | 100.00 | 3.00 | 2.09 | 0.91 | 3.00 | 1.36 | 29.08 | -143.42 | 16.38 | 0.996 |
| 214.20 | 100.00 | 3.00 | 2.09 | 0.91 | 3.00 | 1.36 | 29.12 | -143.18 | 16.32 | 0.996 |
| 214.30 | 100.00 | 3.00 | 2.09 | 0.91 | 3.00 | 1.36 | 29.15 | -142.94 | 16.26 | 0.996 |
| 214.40 | 100.00 | 3.00 | 2.09 | 0.91 | 3.00 | 1.36 | 29.19 | -142.71 | 16.19 | 0.996 |
| 214.50 | 100.00 | 3.01 | 2.09 | 0.92 | 3.01 | 1.35 | 29.22 | -142.47 | 16.13 | 0.996 |
| 214.60 | 100.00 | 3.01 | 2.09 | 0.92 | 3.01 | 1.35 | 29.25 | -142.24 | 16.06 | 0.996 |
| 214.70 | 100.00 | 3.01 | 2.09 | 0.92 | 3.01 | 1.35 | 29.29 | -142.00 | 16.00 | 0.996 |
| 214.80 | 100.00 | 3.01 | 2.09 | 0.92 | 3.01 | 1.35 | 29.32 | -141.76 | 15.94 | 0.996 |
| 214.90 | 100.00 | 3.02 | 2.09 | 0.93 | 3.02 | 1.35 | 29.36 | -141.53 | 15.87 | 0.996 |
| 215.00 | 100.00 | 3.02 | 2.09 | 0.93 | 3.02 | 1.35 | 29.39 | -141.29 | 15.81 | 0.996 |
| 215.10 | 100.00 | 3.02 | 2.10 | 0.92 | 3.02 | 1.35 | 29.43 | -141.06 | 15.75 | 0.996 |
| 215.20 | 100.00 | 3.02 | 2.10 | 0.92 | 3.02 | 1.35 | 29.46 | -140.82 | 15.69 | 0.996 |
| 215.30 | 100.00 | 3.02 | 2.10 | 0.92 | 3.02 | 1.35 | 29.49 | -140.58 | 15.62 | 0.996 |
| 215.40 | 100.00 | 3.03 | 2.10 | 0.93 | 3.03 | 1.35 | 29.53 | -140.35 | 15.56 | 0.996 |
| 215.50 | 100.00 | 3.03 | 2.10 | 0.93 | 3.03 | 1.35 | 29.56 | -140.12 | 15.50 | 0.996 |
| 215.60 | 100.00 | 3.03 | 2.10 | 0.93 | 3.03 | 1.35 | 29.60 | -139.88 | 15.44 | 0.996 |
| 215.70 | 100.00 | 3.03 | 2.10 | 0.93 | 3.03 | 1.35 | 29.63 | -139.64 | 15.38 | 0.996 |
| 215.80 | 100.00 | 3.04 | 2.10 | 0.94 | 3.04 | 1.34 | 29.67 | -139.41 | 15.32 | 0.996 |
| 215.90 | 100.00 | 3.04 | 2.10 | 0.94 | 3.04 | 1.34 | 29.70 | -139.18 | 15.25 | 0.996 |
| 216.00 | 100.00 | 3.04 | 2.10 | 0.94 | 3.04 | 1.34 | 29.74 | -138.94 | 15.19 | 0.996 |
| 216.10 | 100.00 | 3.04 | 2.11 | 0.93 | 3.04 | 1.34 | 29.77 | -138.71 | 15.13 | 0.996 |
| 216.20 | 100.00 | 3.05 | 2.11 | 0.94 | 3.05 | 1.34 | 29.81 | -138.47 | 15.07 | 0.996 |
| 216.30 | 100.00 | 3.05 | 2.11 | 0.94 | 3.05 | 1.34 | 29.84 | -138.24 | 15.01 | 0.996 |
| 216.40 | 100.00 | 3.05 | 2.11 | 0.94 | 3.05 | 1.34 | 29.88 | -138.00 | 14.95 | 0.996 |
| 216.50 | 100.00 | 3.05 | 2.11 | 0.94 | 3.05 | 1.34 | 29.91 | -137.77 | 14.89 | 0.996 |
| 216.60 | 100.00 | 3.05 | 2.11 | 0.94 | 3.05 | 1.34 | 29.95 | -137.53 | 14.83 | 0.996 |
| 216.70 | 100.00 | 3.06 | 2.11 | 0.95 | 3.06 | 1.34 | 29.98 | -137.30 | 14.78 | 0.996 |
| 216.80 | 100.00 | 3.06 | 2.11 | 0.95 | 3.06 | 1.34 | 30.02 | -137.07 | 14.72 | 0.996 |
| 216.90 | 100.00 | 3.06 | 2.11 | 0.95 | 3.06 | 1.34 | 30.05 | -136.83 | 14.66 | 0.996 |
| 217.00 | 100.00 | 3.06 | 2.12 | 0.94 | 3.06 | 1.34 | 30.09 | -136.60 | 14.60 | 0.996 |
| 217.10 | 100.00 | 3.07 | 2.12 | 0.95 | 3.07 | 1.33 | 30.12 | -136.36 | 14.54 | 0.996 |
| 217.20 | 100.00 | 3.07 | 2.12 | 0.95 | 3.07 | 1.33 | 30.16 | -136.13 | 14.48 | 0.996 |
| 217.30 | 100.00 | 3.07 | 2.12 | 0.95 | 3.07 | 1.33 | 30.19 | -135.90 | 14.43 | 0.996 |
| 217.40 | 100.00 | 3.07 | 2.12 | 0.95 | 3.07 | 1.33 | 30.23 | -135.66 | 14.37 | 0.996 |
| 217.50 | 100.00 | 3.08 | 2.12 | 0.96 | 3.08 | 1.33 | 30.26 | -135.43 | 14.31 | 0.996 |
| 217.60 | 100.00 | 3.08 | 2.12 | 0.96 | 3.08 | 1.33 | 30.30 | -135.20 | 14.25 | 0.996 |
| 217.70 | 100.00 | 3.08 | 2.12 | 0.96 | 3.08 | 1.33 | 30.33 | -134.97 | 14.20 | 0.996 |
| 217.80 | 100.00 | 3.08 | 2.12 | 0.96 | 3.08 | 1.33 | 30.37 | -134.73 | 14.14 | 0.996 |
| 217.90 | 100.00 | 3.09 | 2.13 | 0.96 | 3.09 | 1.33 | 30.40 | -134.50 | 14.08 | 0.996 |
| 218.00 | 100.00 | 3.09 | 2.13 | 0.96 | 3.09 | 1.33 | 30.44 | -134.27 | 14.03 | 0.996 |
| 218.10 | 100.00 | 3.09 | 2.13 | 0.96 | 3.09 | 1.33 | 30.47 | -134.04 | 13.97 | 0.996 |
| 218.20 | 100.00 | 3.09 | 2.13 | 0.96 | 3.09 | 1.33 | 30.51 | -133.80 | 13.91 | 0.996 |
| 218.30 | 100.00 | 3.10 | 2.13 | 0.97 | 3.10 | 1.32 | 30.54 | -133.57 | 13.86 | 0.996 |
| 218.40 | 100.00 | 3.10 | 2.13 | 0.97 | 3.10 | 1.32 | 30.58 | -133.34 | 13.80 | 0.996 |
| 218.50 | 100.00 | 3.10 | 2.13 | 0.97 | 3.10 | 1.32 | 30.61 | -133.10 | 13.75 | 0.996 |
| 218.60 | 100.00 | 3.10 | 2.13 | 0.97 | 3.10 | 1.32 | 30.65 | -132.87 | 13.69 | 0.996 |
| 218.70 | 100.00 | 3.10 | 2.13 | 0.97 | 3.10 | 1.32 | 30.69 | -132.64 | 13.64 | 0.996 |
| 218.80 | 100.00 | 3.11 | 2.14 | 0.97 | 3.11 | 1.32 | 30.72 | -132.41 | 13.58 | 0.996 |
| 218.90 | 100.00 | 3.11 | 2.14 | 0.97 | 3.11 | 1.32 | 30.76 | -132.18 | 13.53 | 0.996 |
| 219.00 | 100.00 | 3.11 | 2.14 | 0.97 | 3.11 | 1.32 | 30.79 | -131.95 | 13.47 | 0.996 |
| 219.10 | 100.00 | 3.11 | 2.14 | 0.97 | 3.11 | 1.32 | 30.83 | -131.71 | 13.42 | 0.996 |
| 219.20 | 100.00 | 3.12 | 2.14 | 0.98 | 3.12 | 1.32 | 30.86 | -131.48 | 13.37 | 0.996 |
| 219.30 | 100.00 | 3.12 | 2.14 | 0.98 | 3.12 | 1.32 | 30.90 | -131.25 | 13.31 | 0.996 |
| 219.40 | 100.00 | 3.12 | 2.14 | 0.98 | 3.12 | 1.32 | 30.94 | -131.02 | 13.26 | 0.996 |
| 219.50 | 100.00 | 3.12 | 2.14 | 0.98 | 3.12 | 1.31 | 30.97 | -130.79 | 13.20 | 0.996 |
| 219.60 | 100.00 | 3.13 | 2.14 | 0.99 | 3.13 | 1.31 | 31.01 | -130.56 | 13.15 | 0.996 |
| 219.70 | 100.00 | 3.13 | 2.15 | 0.98 | 3.13 | 1.31 | 31.04 | -130.33 | 13.10 | 0.996 |
| 219.80 | 100.00 | 3.13 | 2.15 | 0.98 | 3.13 | 1.31 | 31.08 | -130.10 | 13.05 | 0.996 |
| 219.90 | 100.00 | 3.13 | 2.15 | 0.98 | 3.13 | 1.31 | 31.12 | -129.86 | 12.99 | 0.996 |
| 220.00 | 100.00 | 3.14 | 2.15 | 0.99 | 3.14 | 1.31 | 31.15 | -129.63 | 12.94 | 0.996 |
| 220.10 | 100.00 | 3.14 | 2.15 | 0.99 | 3.14 | 1.31 | 31.19 | -129.40 | 12.89 | 0.996 |
| 220.20 | 100.00 | 3.14 | 2.15 | 0.99 | 3.14 | 1.31 | 31.22 | -129.17 | 12.84 | 0.996 |
| 220.30 | 100.00 | 3.14 | 2.15 | 0.99 | 3.14 | 1.31 | 31.26 | -128.94 | 12.78 | 0.996 |
| 220.40 | 100.00 | 3.15 | 2.15 | 1.00 | 3.15 | 1.31 | 31.30 | -128.71 | 12.73 | 0.996 |
| 220.50 | 100.00 | 3.15 | 2.15 | 1.00 | 3.15 | 1.31 | 31.33 | -128.48 | 12.68 | 0.996 |
| 220.60 | 100.00 | 3.15 | 2.16 | 0.99 | 3.15 | 1.30 | 31.37 | -128.25 | 12.63 | 0.996 |
| 220.70 | 100.00 | 3.15 | 2.16 | 0.99 | 3.15 | 1.30 | 31.40 | -128.02 | 12.58 | 0.996 |
| 220.80 | 100.00 | 3.16 | 2.16 | 1.00 | 3.16 | 1.30 | 31.44 | -127.79 | 12.53 | 0.996 |
| 220.90 | 100.00 | 3.16 | 2.16 | 1.00 | 3.16 | 1.30 | 31.48 | -127.56 | 12.48 | 0.996 |
| 221.00 | 100.00 | 3.16 | 2.16 | 1.00 | 3.16 | 1.30 | 31.51 | -127.33 | 12.43 | 0.996 |
| 221.10 | 100.00 | 3.16 | 2.16 | 1.00 | 3.16 | 1.30 | 31.55 | -127.10 | 12.38 | 0.996 |
| 221.20 | 100.00 | 3.17 | 2.16 | 1.01 | 3.17 | 1.30 | 31.59 | -126.87 | 12.33 | 0.996 |
| 221.30 | 100.00 | 3.17 | 2.16 | 1.01 | 3.17 | 1.30 | 31.62 | -126.64 | 12.28 | 0.996 |
| 221.40 | 100.00 | 3.17 | 2.17 | 1.00 | 3.17 | 1.30 | 31.66 | -126.41 | 12.23 | 0.996 |
| 221.50 | 100.00 | 3.17 | 2.17 | 1.00 | 3.17 | 1.30 | 31.70 | -126.18 | 12.18 | 0.996 |
| 221.60 | 100.00 | 3.18 | 2.17 | 1.01 | 3.18 | 1.30 | 31.73 | -125.96 | 12.13 | 0.996 |
| 221.70 | 100.00 | 3.18 | 2.17 | 1.01 | 3.18 | 1.30 | 31.77 | -125.73 | 12.08 | 0.996 |
| 221.80 | 100.00 | 3.18 | 2.17 | 1.01 | 3.18 | 1.29 | 31.81 | -125.50 | 12.03 | 0.996 |
| 221.90 | 100.00 | 3.18 | 2.17 | 1.01 | 3.18 | 1.29 | 31.84 | -125.27 | 11.98 | 0.996 |
| 222.00 | 100.00 | 3.19 | 2.17 | 1.02 | 3.19 | 1.29 | 31.88 | -125.04 | 11.93 | 0.996 |
| 222.10 | 100.00 | 3.19 | 2.17 | 1.02 | 3.19 | 1.29 | 31.92 | -124.81 | 11.88 | 0.996 |
| 222.20 | 100.00 | 3.19 | 2.18 | 1.01 | 3.19 | 1.29 | 31.95 | -124.58 | 11.83 | 0.996 |
| 222.30 | 100.00 | 3.19 | 2.18 | 1.01 | 3.19 | 1.29 | 31.99 | -124.35 | 11.79 | 0.996 |
| 222.40 | 100.00 | 3.20 | 2.18 | 1.02 | 3.20 | 1.29 | 32.03 | -124.12 | 11.74 | 0.996 |
| 222.50 | 100.00 | 3.20 | 2.18 | 1.02 | 3.20 | 1.29 | 32.06 | -123.90 | 11.69 | 0.996 |
| 222.60 | 100.00 | 3.20 | 2.18 | 1.02 | 3.20 | 1.29 | 32.10 | -123.67 | 11.64 | 0.996 |
| 222.70 | 100.00 | 3.21 | 2.18 | 1.03 | 3.21 | 1.29 | 32.14 | -123.44 | 11.59 | 0.996 |
| 222.80 | 100.00 | 3.21 | 2.18 | 1.03 | 3.21 | 1.29 | 32.18 | -123.21 | 11.55 | 0.996 |
| 222.90 | 100.00 | 3.21 | 2.18 | 1.03 | 3.21 | 1.28 | 32.21 | -122.98 | 11.50 | 0.996 |
| 223.00 | 100.00 | 3.21 | 2.19 | 1.02 | 3.21 | 1.28 | 32.25 | -122.76 | 11.45 | 0.996 |
| 223.10 | 100.00 | 3.22 | 2.19 | 1.03 | 3.22 | 1.28 | 32.29 | -122.53 | 11.41 | 0.996 |
| 223.20 | 100.00 | 3.22 | 2.19 | 1.03 | 3.22 | 1.28 | 32.32 | -122.30 | 11.36 | 0.996 |
| 223.30 | 100.00 | 3.22 | 2.19 | 1.03 | 3.22 | 1.28 | 32.36 | -122.07 | 11.31 | 0.996 |
| 223.40 | 100.00 | 3.22 | 2.19 | 1.03 | 3.22 | 1.28 | 32.40 | -121.84 | 11.27 | 0.996 |
| 223.50 | 100.00 | 3.23 | 2.19 | 1.04 | 3.23 | 1.28 | 32.44 | -121.62 | 11.22 | 0.996 |
| 223.60 | 100.00 | 3.23 | 2.19 | 1.04 | 3.23 | 1.28 | 32.47 | -121.39 | 11.17 | 0.996 |
| 223.70 | 100.00 | 3.23 | 2.19 | 1.04 | 3.23 | 1.28 | 32.51 | -121.16 | 11.13 | 0.996 |
| 223.80 | 100.00 | 3.23 | 2.20 | 1.03 | 3.23 | 1.28 | 32.55 | -120.94 | 11.08 | 0.996 |
| 223.90 | 100.00 | 3.24 | 2.20 | 1.04 | 3.24 | 1.28 | 32.59 | -120.71 | 11.04 | 0.996 |
| 224.00 | 100.00 | 3.24 | 2.20 | 1.04 | 3.24 | 1.27 | 32.62 | -120.48 | 10.99 | 0.996 |
| 224.10 | 100.00 | 3.24 | 2.20 | 1.04 | 3.24 | 1.27 | 32.66 | -120.25 | 10.95 | 0.996 |
| 224.20 | 100.00 | 3.24 | 2.20 | 1.04 | 3.24 | 1.27 | 32.70 | -120.03 | 10.90 | 0.996 |
| 224.30 | 100.00 | 3.25 | 2.20 | 1.05 | 3.25 | 1.27 | 32.74 | -119.80 | 10.86 | 0.996 |
| 224.40 | 100.00 | 3.25 | 2.20 | 1.05 | 3.25 | 1.27 | 32.77 | -119.57 | 10.81 | 0.996 |
| 224.50 | 100.00 | 3.25 | 2.21 | 1.04 | 3.25 | 1.27 | 32.81 | -119.35 | 10.77 | 0.996 |
| 224.60 | 100.00 | 3.25 | 2.21 | 1.04 | 3.25 | 1.27 | 32.85 | -119.12 | 10.72 | 0.996 |
| 224.70 | 100.00 | 3.26 | 2.21 | 1.05 | 3.26 | 1.27 | 32.89 | -118.89 | 10.68 | 0.996 |
| 224.80 | 100.00 | 3.26 | 2.21 | 1.05 | 3.26 | 1.27 | 32.93 | -118.67 | 10.64 | 0.996 |
| 224.90 | 100.00 | 3.26 | 2.21 | 1.05 | 3.26 | 1.27 | 32.96 | -118.44 | 10.59 | 0.996 |
| 225.00 | 100.00 | 3.27 | 2.21 | 1.06 | 3.27 | 1.26 | 33.00 | -118.22 | 10.55 | 0.996 |
| 225.10 | 100.00 | 3.27 | 2.21 | 1.06 | 3.27 | 1.26 | 33.04 | -117.99 | 10.51 | 0.996 |
| 225.20 | 100.00 | 3.27 | 2.21 | 1.06 | 3.27 | 1.26 | 33.08 | -117.76 | 10.46 | 0.996 |
| 225.30 | 100.00 | 3.27 | 2.22 | 1.05 | 3.27 | 1.26 | 33.12 | -117.54 | 10.42 | 0.996 |
| 225.40 | 100.00 | 3.28 | 2.22 | 1.06 | 3.28 | 1.26 | 33.15 | -117.31 | 10.38 | 0.996 |
| 225.50 | 100.00 | 3.28 | 2.22 | 1.06 | 3.28 | 1.26 | 33.19 | -117.08 | 10.33 | 0.996 |
| 225.60 | 100.00 | 3.28 | 2.22 | 1.06 | 3.28 | 1.26 | 33.23 | -116.86 | 10.29 | 0.996 |
| 225.70 | 100.00 | 3.28 | 2.22 | 1.06 | 3.28 | 1.26 | 33.27 | -116.63 | 10.25 | 0.996 |
| 225.80 | 100.00 | 3.29 | 2.22 | 1.07 | 3.29 | 1.26 | 33.31 | -116.41 | 10.21 | 0.996 |
| 225.90 | 100.00 | 3.29 | 2.22 | 1.07 | 3.29 | 1.26 | 33.35 | -116.18 | 10.16 | 0.996 |
| 226.00 | 100.00 | 3.29 | 2.23 | 1.06 | 3.29 | 1.26 | 33.38 | -115.96 | 10.12 | 0.996 |
| 226.10 | 100.00 | 3.30 | 2.23 | 1.07 | 3.30 | 1.25 | 33.42 | -115.73 | 10.08 | 0.996 |
| 226.20 | 100.00 | 3.30 | 2.23 | 1.07 | 3.30 | 1.25 | 33.46 | -115.51 | 10.04 | 0.996 |
| 226.30 | 100.00 | 3.30 | 2.23 | 1.07 | 3.30 | 1.25 | 33.50 | -115.28 | 10.00 | 0.996 |
| 226.40 | 100.00 | 3.30 | 2.23 | 1.07 | 3.30 | 1.25 | 33.54 | -115.06 | 9.96 | 0.996 |
| 226.50 | 100.00 | 3.31 | 2.23 | 1.08 | 3.31 | 1.25 | 33.58 | -114.83 | 9.91 | 0.996 |
| 226.60 | 100.00 | 3.31 | 2.23 | 1.08 | 3.31 | 1.25 | 33.61 | -114.61 | 9.87 | 0.996 |
| 226.70 | 100.00 | 3.31 | 2.24 | 1.07 | 3.31 | 1.25 | 33.65 | -114.38 | 9.83 | 0.996 |
| 226.80 | 100.00 | 3.31 | 2.24 | 1.07 | 3.31 | 1.25 | 33.69 | -114.16 | 9.79 | 0.996 |
| 226.90 | 100.00 | 3.32 | 2.24 | 1.08 | 3.32 | 1.25 | 33.73 | -113.93 | 9.75 | 0.996 |
| 227.00 | 100.00 | 3.32 | 2.24 | 1.08 | 3.32 | 1.25 | 33.77 | -113.71 | 9.71 | 0.996 |
| 227.10 | 100.00 | 3.32 | 2.24 | 1.08 | 3.32 | 1.24 | 33.81 | -113.48 | 9.67 | 0.996 |
| 227.20 | 100.00 | 3.33 | 2.24 | 1.09 | 3.33 | 1.24 | 33.85 | -113.26 | 9.63 | 0.996 |
| 227.30 | 100.00 | 3.33 | 2.24 | 1.09 | 3.33 | 1.24 | 33.89 | -113.04 | 9.59 | 0.996 |
| 227.40 | 100.00 | 3.33 | 2.24 | 1.09 | 3.33 | 1.24 | 33.92 | -112.81 | 9.55 | 0.996 |
| 227.50 | 100.00 | 3.33 | 2.25 | 1.08 | 3.33 | 1.24 | 33.96 | -112.59 | 9.51 | 0.996 |
| 227.60 | 100.00 | 3.34 | 2.25 | 1.09 | 3.34 | 1.24 | 34.00 | -112.36 | 9.47 | 0.996 |
| 227.70 | 100.00 | 3.34 | 2.25 | 1.09 | 3.34 | 1.24 | 34.04 | -112.14 | 9.43 | 0.996 |
| 227.80 | 100.00 | 3.34 | 2.25 | 1.09 | 3.34 | 1.24 | 34.08 | -111.92 | 9.39 | 0.996 |
| 227.90 | 100.00 | 3.35 | 2.25 | 1.10 | 3.35 | 1.24 | 34.12 | -111.69 | 9.35 | 0.996 |
| 228.00 | 100.00 | 3.35 | 2.25 | 1.10 | 3.35 | 1.24 | 34.16 | -111.47 | 9.31 | 0.996 |
| 228.10 | 100.00 | 3.35 | 2.25 | 1.10 | 3.35 | 1.23 | 34.20 | -111.25 | 9.28 | 0.996 |
| 228.20 | 100.00 | 3.35 | 2.26 | 1.09 | 3.35 | 1.23 | 34.24 | -111.02 | 9.24 | 0.996 |
| 228.30 | 100.00 | 3.36 | 2.26 | 1.10 | 3.36 | 1.23 | 34.28 | -110.80 | 9.20 | 0.996 |
| 228.40 | 100.00 | 3.36 | 2.26 | 1.10 | 3.36 | 1.23 | 34.32 | -110.58 | 9.16 | 0.996 |
| 228.50 | 100.00 | 3.36 | 2.26 | 1.10 | 3.36 | 1.23 | 34.36 | -110.35 | 9.12 | 0.996 |
| 228.60 | 100.00 | 3.36 | 2.26 | 1.10 | 3.36 | 1.23 | 34.40 | -110.13 | 9.08 | 0.996 |
| 228.70 | 100.00 | 3.37 | 2.26 | 1.11 | 3.37 | 1.23 | 34.44 | -109.91 | 9.05 | 0.996 |
| 228.80 | 100.00 | 3.37 | 2.26 | 1.11 | 3.37 | 1.23 | 34.47 | -109.68 | 9.01 | 0.996 |
| 228.90 | 100.00 | 3.37 | 2.27 | 1.10 | 3.37 | 1.23 | 34.51 | -109.46 | 8.97 | 0.996 |
| 229.00 | 100.00 | 3.38 | 2.27 | 1.11 | 3.38 | 1.23 | 34.55 | -109.24 | 8.93 | 0.996 |
| 229.10 | 100.00 | 3.38 | 2.27 | 1.11 | 3.38 | 1.22 | 34.59 | -109.02 | 8.90 | 0.996 |
| 229.20 | 100.00 | 3.38 | 2.27 | 1.11 | 3.38 | 1.22 | 34.63 | -108.79 | 8.86 | 0.996 |
| 229.30 | 100.00 | 3.38 | 2.27 | 1.11 | 3.38 | 1.22 | 34.67 | -108.57 | 8.82 | 0.996 |
| 229.40 | 100.00 | 3.39 | 2.27 | 1.12 | 3.39 | 1.22 | 34.71 | -108.35 | 8.78 | 0.996 |
| 229.50 | 100.00 | 3.39 | 2.27 | 1.12 | 3.39 | 1.22 | 34.75 | -108.12 | 8.75 | 0.996 |
| 229.60 | 100.00 | 3.39 | 2.28 | 1.11 | 3.39 | 1.22 | 34.79 | -107.90 | 8.71 | 0.996 |
| 229.70 | 100.00 | 3.40 | 2.28 | 1.12 | 3.40 | 1.22 | 34.83 | -107.68 | 8.67 | 0.996 |
| 229.80 | 100.00 | 3.40 | 2.28 | 1.12 | 3.40 | 1.22 | 34.87 | -107.46 | 8.64 | 0.996 |
| 229.90 | 100.00 | 3.40 | 2.28 | 1.12 | 3.40 | 1.22 | 34.91 | -107.24 | 8.60 | 0.996 |
| 230.00 | 100.00 | 3.40 | 2.28 | 1.12 | 3.40 | 1.22 | 34.95 | -107.01 | 8.57 | 0.996 |
| 230.10 | 100.00 | 3.41 | 2.28 | 1.13 | 3.41 | 1.21 | 34.99 | -106.79 | 8.53 | 0.996 |
| 230.20 | 100.00 | 3.41 | 2.28 | 1.13 | 3.41 | 1.21 | 35.03 | -106.57 | 8.49 | 0.996 |
| 230.30 | 100.00 | 3.41 | 2.29 | 1.12 | 3.41 | 1.21 | 35.07 | -106.35 | 8.46 | 0.996 |
| 230.40 | 100.00 | 3.42 | 2.29 | 1.13 | 3.42 | 1.21 | 35.11 | -106.13 | 8.42 | 0.996 |
| 230.50 | 100.00 | 3.42 | 2.29 | 1.13 | 3.42 | 1.21 | 35.15 | -105.90 | 8.39 | 0.996 |
| 230.60 | 100.00 | 3.42 | 2.29 | 1.13 | 3.42 | 1.21 | 35.19 | -105.68 | 8.35 | 0.996 |
| 230.70 | 100.00 | 3.43 | 2.29 | 1.14 | 3.43 | 1.21 | 35.23 | -105.46 | 8.32 | 0.996 |
| 230.80 | 100.00 | 3.43 | 2.29 | 1.14 | 3.43 | 1.21 | 35.27 | -105.24 | 8.28 | 0.996 |
| 230.90 | 100.00 | 3.43 | 2.29 | 1.14 | 3.43 | 1.21 | 35.32 | -105.02 | 8.25 | 0.996 |
| 231.00 | 100.00 | 3.43 | 2.30 | 1.13 | 3.43 | 1.20 | 35.36 | -104.80 | 8.21 | 0.996 |
| 231.10 | 100.00 | 3.44 | 2.30 | 1.14 | 3.44 | 1.20 | 35.40 | -104.58 | 8.18 | 0.996 |
| 231.20 | 100.00 | 3.44 | 2.30 | 1.14 | 3.44 | 1.20 | 35.44 | -104.36 | 8.14 | 0.996 |
| 231.30 | 100.00 | 3.44 | 2.30 | 1.14 | 3.44 | 1.20 | 35.48 | -104.13 | 8.11 | 0.996 |
| 231.40 | 100.00 | 3.45 | 2.30 | 1.15 | 3.45 | 1.20 | 35.52 | -103.91 | 8.07 | 0.996 |
| 231.50 | 100.00 | 3.45 | 2.30 | 1.15 | 3.45 | 1.20 | 35.56 | -103.69 | 8.04 | 0.996 |
| 231.60 | 100.00 | 3.45 | 2.30 | 1.15 | 3.45 | 1.20 | 35.60 | -103.47 | 8.01 | 0.996 |
| 231.70 | 100.00 | 3.45 | 2.31 | 1.14 | 3.45 | 1.20 | 35.64 | -103.25 | 7.97 | 0.996 |
| 231.80 | 100.00 | 3.46 | 2.31 | 1.15 | 3.46 | 1.20 | 35.68 | -103.03 | 7.94 | 0.996 |
| 231.90 | 100.00 | 3.46 | 2.31 | 1.15 | 3.46 | 1.19 | 35.72 | -102.81 | 7.91 | 0.996 |
| 232.00 | 100.00 | 3.46 | 2.31 | 1.15 | 3.46 | 1.19 | 35.76 | -102.59 | 7.87 | 0.996 |
| 232.10 | 100.00 | 3.47 | 2.31 | 1.16 | 3.47 | 1.19 | 35.80 | -102.37 | 7.84 | 0.996 |
| 232.20 | 100.00 | 3.47 | 2.31 | 1.16 | 3.47 | 1.19 | 35.85 | -102.15 | 7.81 | 0.996 |
| 232.30 | 100.00 | 3.47 | 2.32 | 1.15 | 3.47 | 1.19 | 35.89 | -101.93 | 7.77 | 0.996 |
| 232.40 | 100.00 | 3.48 | 2.32 | 1.16 | 3.48 | 1.19 | 35.93 | -101.71 | 7.74 | 0.996 |
| 232.50 | 100.00 | 3.48 | 2.32 | 1.16 | 3.48 | 1.19 | 35.97 | -101.49 | 7.71 | 0.996 |
| 232.60 | 100.00 | 3.48 | 2.32 | 1.16 | 3.48 | 1.19 | 36.01 | -101.27 | 7.67 | 0.996 |
| 232.70 | 100.00 | 3.48 | 2.32 | 1.16 | 3.48 | 1.19 | 36.05 | -101.05 | 7.64 | 0.996 |
| 232.80 | 100.00 | 3.49 | 2.32 | 1.17 | 3.49 | 1.19 | 36.09 | -100.83 | 7.61 | 0.996 |
| 232.90 | 100.00 | 3.49 | 2.32 | 1.17 | 3.49 | 1.18 | 36.13 | -100.61 | 7.58 | 0.996 |
| 233.00 | 100.00 | 3.49 | 2.33 | 1.16 | 3.49 | 1.18 | 36.18 | -100.39 | 7.54 | 0.996 |
| 233.10 | 100.00 | 3.50 | 2.33 | 1.17 | 3.50 | 1.18 | 36.22 | -100.17 | 7.51 | 0.996 |
| 233.20 | 100.00 | 3.50 | 2.33 | 1.17 | 3.50 | 1.18 | 36.26 | -99.95 | 7.48 | 0.996 |
| 233.30 | 100.00 | 3.50 | 2.33 | 1.17 | 3.50 | 1.18 | 36.30 | -99.73 | 7.45 | 0.996 |
| 233.40 | 100.00 | 3.51 | 2.33 | 1.18 | 3.51 | 1.18 | 36.34 | -99.51 | 7.42 | 0.996 |
| 233.50 | 100.00 | 3.51 | 2.33 | 1.18 | 3.51 | 1.18 | 36.38 | -99.29 | 7.39 | 0.996 |
| 233.60 | 100.00 | 3.51 | 2.33 | 1.18 | 3.51 | 1.18 | 36.43 | -99.07 | 7.35 | 0.996 |
| 233.70 | 100.00 | 3.52 | 2.34 | 1.18 | 3.52 | 1.18 | 36.47 | -98.85 | 7.32 | 0.996 |
| 233.80 | 100.00 | 3.52 | 2.34 | 1.18 | 3.52 | 1.17 | 36.51 | -98.63 | 7.29 | 0.996 |
| 233.90 | 100.00 | 3.52 | 2.34 | 1.18 | 3.52 | 1.17 | 36.55 | -98.42 | 7.26 | 0.996 |
| 234.00 | 100.00 | 3.52 | 2.34 | 1.18 | 3.52 | 1.17 | 36.59 | -98.20 | 7.23 | 0.996 |
| 234.10 | 100.00 | 3.53 | 2.34 | 1.19 | 3.53 | 1.17 | 36.63 | -97.98 | 7.20 | 0.996 |
| 234.20 | 100.00 | 3.53 | 2.34 | 1.19 | 3.53 | 1.17 | 36.68 | -97.76 | 7.17 | 0.996 |
| 234.30 | 100.00 | 3.53 | 2.34 | 1.19 | 3.53 | 1.17 | 36.72 | -97.54 | 7.14 | 0.996 |
| 234.40 | 100.00 | 3.54 | 2.35 | 1.19 | 3.54 | 1.17 | 36.76 | -97.32 | 7.11 | 0.996 |
| 234.50 | 100.00 | 3.54 | 2.35 | 1.19 | 3.54 | 1.17 | 36.80 | -97.10 | 7.08 | 0.996 |
| 234.60 | 100.00 | 3.54 | 2.35 | 1.19 | 3.54 | 1.17 | 36.84 | -96.88 | 7.05 | 0.996 |
| 234.70 | 100.00 | 3.55 | 2.35 | 1.20 | 3.55 | 1.16 | 36.89 | -96.67 | 7.02 | 0.996 |
| 234.80 | 100.00 | 3.55 | 2.35 | 1.20 | 3.55 | 1.16 | 36.93 | -96.45 | 6.99 | 0.996 |
| 234.90 | 100.00 | 3.55 | 2.35 | 1.20 | 3.55 | 1.16 | 36.97 | -96.23 | 6.96 | 0.996 |
| 235.00 | 100.00 | 3.56 | 2.35 | 1.21 | 3.56 | 1.16 | 37.01 | -96.01 | 6.93 | 0.996 |
| 235.10 | 100.00 | 3.56 | 2.36 | 1.20 | 3.56 | 1.16 | 37.06 | -95.79 | 6.90 | 0.996 |
| 235.20 | 100.00 | 3.56 | 2.36 | 1.20 | 3.56 | 1.16 | 37.10 | -95.58 | 6.87 | 0.996 |
| 235.30 | 100.00 | 3.56 | 2.36 | 1.20 | 3.56 | 1.16 | 37.14 | -95.36 | 6.84 | 0.996 |
| 235.40 | 100.00 | 3.57 | 2.36 | 1.21 | 3.57 | 1.16 | 37.18 | -95.14 | 6.81 | 0.996 |
| 235.50 | 100.00 | 3.57 | 2.36 | 1.21 | 3.57 | 1.15 | 37.23 | -94.92 | 6.78 | 0.996 |
| 235.60 | 100.00 | 3.57 | 2.36 | 1.21 | 3.57 | 1.15 | 37.27 | -94.70 | 6.75 | 0.996 |
| 235.70 | 100.00 | 3.58 | 2.37 | 1.21 | 3.58 | 1.15 | 37.31 | -94.49 | 6.72 | 0.996 |
| 235.80 | 100.00 | 3.58 | 2.37 | 1.21 | 3.58 | 1.15 | 37.35 | -94.27 | 6.69 | 0.996 |
| 235.90 | 100.00 | 3.58 | 2.37 | 1.21 | 3.58 | 1.15 | 37.40 | -94.05 | 6.67 | 0.996 |
| 236.00 | 100.00 | 3.59 | 2.37 | 1.22 | 3.59 | 1.15 | 37.44 | -93.83 | 6.64 | 0.996 |
| 236.10 | 100.00 | 3.59 | 2.37 | 1.22 | 3.59 | 1.15 | 37.48 | -93.62 | 6.61 | 0.996 |
| 236.20 | 100.00 | 3.59 | 2.37 | 1.22 | 3.59 | 1.15 | 37.53 | -93.40 | 6.58 | 0.996 |
| 236.30 | 100.00 | 3.60 | 2.37 | 1.23 | 3.60 | 1.15 | 37.57 | -93.18 | 6.55 | 0.996 |
| 236.40 | 100.00 | 3.60 | 2.38 | 1.22 | 3.60 | 1.14 | 37.61 | -92.97 | 6.52 | 0.996 |
| 236.50 | 100.00 | 3.60 | 2.38 | 1.22 | 3.60 | 1.14 | 37.66 | -92.75 | 6.50 | 0.996 |
| 236.60 | 100.00 | 3.61 | 2.38 | 1.23 | 3.61 | 1.14 | 37.70 | -92.53 | 6.47 | 0.996 |
| 236.70 | 100.00 | 3.61 | 2.38 | 1.23 | 3.61 | 1.14 | 37.74 | -92.31 | 6.44 | 0.996 |
| 236.80 | 100.00 | 3.61 | 2.38 | 1.23 | 3.61 | 1.14 | 37.79 | -92.10 | 6.41 | 0.996 |
| 236.90 | 100.00 | 3.62 | 2.38 | 1.24 | 3.62 | 1.14 | 37.83 | -91.88 | 6.39 | 0.996 |
| 237.00 | 100.00 | 3.62 | 2.38 | 1.24 | 3.62 | 1.14 | 37.87 | -91.66 | 6.36 | 0.996 |
| 237.10 | 100.00 | 3.62 | 2.39 | 1.23 | 3.62 | 1.14 | 37.92 | -91.45 | 6.33 | 0.996 |
| 237.20 | 100.00 | 3.63 | 2.39 | 1.24 | 3.63 | 1.13 | 37.96 | -91.23 | 6.30 | 0.996 |
| 237.30 | 100.00 | 3.63 | 2.39 | 1.24 | 3.63 | 1.13 | 38.00 | -91.01 | 6.28 | 0.996 |
| 237.40 | 100.00 | 3.63 | 2.39 | 1.24 | 3.63 | 1.13 | 38.05 | -90.80 | 6.25 | 0.996 |
| 237.50 | 100.00 | 3.64 | 2.39 | 1.25 | 3.64 | 1.13 | 38.09 | -90.58 | 6.22 | 0.996 |
| 237.60 | 100.00 | 3.64 | 2.39 | 1.25 | 3.64 | 1.13 | 38.13 | -90.36 | 6.20 | 0.996 |
| 237.70 | 100.00 | 3.64 | 2.39 | 1.25 | 3.64 | 1.13 | 38.18 | -90.15 | 6.17 | 0.996 |
| 237.80 | 100.00 | 3.65 | 2.40 | 1.25 | 3.65 | 1.13 | 38.22 | -89.93 | 6.14 | 0.996 |
| 237.90 | 100.00 | 3.65 | 2.40 | 1.25 | 3.65 | 1.13 | 38.26 | -89.72 | 6.12 | 0.996 |
| 238.00 | 100.00 | 3.65 | 2.40 | 1.25 | 3.65 | 1.13 | 38.31 | -89.50 | 6.09 | 0.996 |
| 238.10 | 100.00 | 3.66 | 2.40 | 1.26 | 3.66 | 1.12 | 38.35 | -89.28 | 6.06 | 0.996 |
| 238.20 | 100.00 | 3.66 | 2.40 | 1.26 | 3.66 | 1.12 | 38.40 | -89.07 | 6.04 | 0.996 |
| 238.30 | 100.00 | 3.66 | 2.40 | 1.26 | 3.66 | 1.12 | 38.44 | -88.85 | 6.01 | 0.996 |
| 238.40 | 100.00 | 3.67 | 2.41 | 1.26 | 3.67 | 1.12 | 38.48 | -88.64 | 5.98 | 0.996 |
| 238.50 | 100.00 | 3.67 | 2.41 | 1.26 | 3.67 | 1.12 | 38.53 | -88.42 | 5.96 | 0.996 |
| 238.60 | 100.00 | 3.67 | 2.41 | 1.26 | 3.67 | 1.12 | 38.57 | -88.21 | 5.93 | 0.996 |
| 238.70 | 100.00 | 3.67 | 2.41 | 1.26 | 3.67 | 1.12 | 38.62 | -87.99 | 5.91 | 0.996 |
| 238.80 | 100.00 | 3.68 | 2.41 | 1.27 | 3.68 | 1.12 | 38.66 | -87.78 | 5.88 | 0.996 |
| 238.90 | 100.00 | 3.68 | 2.41 | 1.27 | 3.68 | 1.11 | 38.71 | -87.56 | 5.86 | 0.996 |
| 239.00 | 100.00 | 3.68 | 2.41 | 1.27 | 3.68 | 1.11 | 38.75 | -87.34 | 5.83 | 0.996 |
| 239.10 | 100.00 | 3.69 | 2.42 | 1.27 | 3.69 | 1.11 | 38.79 | -87.13 | 5.81 | 0.996 |
| 239.20 | 100.00 | 3.69 | 2.42 | 1.27 | 3.69 | 1.11 | 38.84 | -86.91 | 5.78 | 0.996 |
| 239.30 | 100.00 | 3.70 | 2.42 | 1.28 | 3.70 | 1.11 | 38.88 | -86.70 | 5.76 | 0.996 |
| 239.40 | 100.00 | 3.70 | 2.42 | 1.28 | 3.70 | 1.11 | 38.93 | -86.48 | 5.73 | 0.996 |
| 239.50 | 100.00 | 3.70 | 2.42 | 1.28 | 3.70 | 1.11 | 38.97 | -86.27 | 5.71 | 0.996 |
| 239.60 | 100.00 | 3.71 | 2.42 | 1.29 | 3.71 | 1.10 | 39.02 | -86.05 | 5.68 | 0.996 |
| 239.70 | 100.00 | 3.71 | 2.42 | 1.29 | 3.71 | 1.10 | 39.06 | -85.84 | 5.66 | 0.996 |
| 239.80 | 100.00 | 3.71 | 2.43 | 1.28 | 3.71 | 1.10 | 39.11 | -85.62 | 5.63 | 0.996 |
| 239.90 | 100.00 | 3.72 | 2.43 | 1.29 | 3.72 | 1.10 | 39.15 | -85.41 | 5.61 | 0.996 |
| 240.00 | 100.00 | 3.72 | 2.43 | 1.29 | 3.72 | 1.10 | 39.20 | -85.20 | 5.58 | 0.996 |
| 240.10 | 100.00 | 3.72 | 2.43 | 1.29 | 3.72 | 1.10 | 39.24 | -84.98 | 5.56 | 0.996 |
| 240.20 | 100.00 | 3.73 | 2.43 | 1.30 | 3.73 | 1.10 | 39.29 | -84.77 | 5.54 | 0.996 |
| 240.30 | 100.00 | 3.73 | 2.43 | 1.30 | 3.73 | 1.09 | 39.33 | -84.55 | 5.51 | 0.996 |
| 240.40 | 100.00 | 3.73 | 2.43 | 1.30 | 3.73 | 1.09 | 39.38 | -84.34 | 5.49 | 0.996 |
| 240.50 | 100.00 | 3.74 | 2.44 | 1.30 | 3.74 | 1.09 | 39.42 | -84.12 | 5.46 | 0.996 |
| 240.60 | 100.00 | 3.74 | 2.44 | 1.30 | 3.74 | 1.09 | 39.47 | -83.91 | 5.44 | 0.996 |
| 240.70 | 100.00 | 3.74 | 2.44 | 1.30 | 3.74 | 1.09 | 39.51 | -83.70 | 5.42 | 0.996 |
| 240.80 | 100.00 | 3.75 | 2.44 | 1.31 | 3.75 | 1.09 | 39.56 | -83.48 | 5.39 | 0.996 |
| 240.90 | 100.00 | 3.75 | 2.44 | 1.31 | 3.75 | 1.08 | 39.60 | -83.27 | 5.37 | 0.996 |
| 241.00 | 100.00 | 3.75 | 2.44 | 1.31 | 3.75 | 1.08 | 39.65 | -83.05 | 5.35 | 0.996 |
| 241.10 | 100.00 | 3.76 | 2.44 | 1.32 | 3.76 | 1.08 | 39.69 | -82.84 | 5.32 | 0.996 |
| 241.20 | 100.00 | 3.76 | 2.45 | 1.31 | 3.76 | 1.08 | 39.74 | -82.63 | 5.30 | 0.996 |
| 241.30 | 100.00 | 3.76 | 2.45 | 1.31 | 3.76 | 1.08 | 39.79 | -82.41 | 5.28 | 0.996 |
| 241.40 | 100.00 | 3.77 | 2.45 | 1.32 | 3.77 | 1.08 | 39.83 | -82.20 | 5.25 | 0.996 |
| 241.50 | 100.00 | 3.77 | 2.45 | 1.32 | 3.77 | 1.08 | 39.88 | -81.99 | 5.23 | 0.996 |
| 241.60 | 100.00 | 3.77 | 2.45 | 1.32 | 3.77 | 1.07 | 39.92 | -81.77 | 5.21 | 0.996 |
| 241.70 | 100.00 | 3.78 | 2.45 | 1.33 | 3.78 | 1.07 | 39.97 | -81.56 | 5.19 | 0.996 |
| 241.80 | 100.00 | 3.78 | 2.45 | 1.33 | 3.78 | 1.07 | 40.01 | -81.34 | 5.16 | 0.996 |
| 241.90 | 100.00 | 3.78 | 2.46 | 1.32 | 3.78 | 1.07 | 40.06 | -81.13 | 5.14 | 0.996 |
| 242.00 | 100.00 | 3.79 | 2.46 | 1.33 | 3.79 | 1.07 | 40.11 | -80.92 | 5.12 | 0.996 |
| 242.10 | 100.00 | 3.79 | 2.46 | 1.33 | 3.79 | 1.07 | 40.15 | -80.71 | 5.10 | 0.996 |
| 242.20 | 100.00 | 3.79 | 2.46 | 1.33 | 3.79 | 1.06 | 40.20 | -80.49 | 5.07 | 0.996 |
| 242.30 | 100.00 | 3.80 | 2.46 | 1.34 | 3.80 | 1.06 | 40.24 | -80.28 | 5.05 | 0.996 |
| 242.40 | 100.00 | 3.80 | 2.46 | 1.34 | 3.80 | 1.06 | 40.29 | -80.07 | 5.03 | 0.996 |
| 242.50 | 100.00 | 3.81 | 2.46 | 1.35 | 3.81 | 1.06 | 40.34 | -79.85 | 5.01 | 0.996 |
| 242.60 | 100.00 | 3.81 | 2.47 | 1.34 | 3.81 | 1.06 | 40.38 | -79.64 | 4.99 | 0.996 |
| 242.70 | 100.00 | 3.81 | 2.47 | 1.34 | 3.81 | 1.06 | 40.43 | -79.43 | 4.96 | 0.996 |
| 242.80 | 100.00 | 3.82 | 2.47 | 1.35 | 3.82 | 1.06 | 40.48 | -79.22 | 4.94 | 0.996 |
| 242.90 | 100.00 | 3.82 | 2.47 | 1.35 | 3.82 | 1.05 | 40.52 | -79.00 | 4.92 | 0.996 |
| 243.00 | 100.00 | 3.82 | 2.47 | 1.35 | 3.82 | 1.05 | 40.57 | -78.79 | 4.90 | 0.996 |
| 243.10 | 100.00 | 3.83 | 2.47 | 1.36 | 3.83 | 1.05 | 40.62 | -78.58 | 4.88 | 0.996 |
| 243.20 | 100.00 | 3.83 | 2.47 | 1.36 | 3.83 | 1.05 | 40.66 | -78.37 | 4.86 | 0.996 |
| 243.30 | 100.00 | 3.83 | 2.47 | 1.36 | 3.83 | 1.05 | 40.71 | -78.15 | 4.84 | 0.996 |
| 243.40 | 100.00 | 3.84 | 2.48 | 1.36 | 3.84 | 1.05 | 40.76 | -77.94 | 4.82 | 0.996 |
| 243.50 | 100.00 | 3.84 | 2.48 | 1.36 | 3.84 | 1.04 | 40.80 | -77.73 | 4.79 | 0.996 |
| 243.60 | 100.00 | 3.84 | 2.48 | 1.36 | 3.84 | 1.04 | 40.85 | -77.52 | 4.77 | 0.996 |
| 243.70 | 100.00 | 3.85 | 2.48 | 1.37 | 3.85 | 1.04 | 40.90 | -77.30 | 4.75 | 0.996 |
| 243.80 | 100.00 | 3.85 | 2.48 | 1.37 | 3.85 | 1.04 | 40.94 | -77.09 | 4.73 | 0.996 |
| 243.90 | 100.00 | 3.86 | 2.48 | 1.38 | 3.86 | 1.04 | 40.99 | -76.88 | 4.71 | 0.996 |
| 244.00 | 100.00 | 3.86 | 2.48 | 1.38 | 3.86 | 1.04 | 41.04 | -76.67 | 4.69 | 0.996 |
| 244.10 | 100.00 | 3.86 | 2.49 | 1.37 | 3.86 | 1.03 | 41.08 | -76.46 | 4.67 | 0.996 |
| 244.20 | 100.00 | 3.87 | 2.49 | 1.38 | 3.87 | 1.03 | 41.13 | -76.25 | 4.65 | 0.996 |
| 244.30 | 100.00 | 3.87 | 2.49 | 1.38 | 3.87 | 1.03 | 41.18 | -76.03 | 4.63 | 0.996 |
| 244.40 | 100.00 | 3.87 | 2.49 | 1.38 | 3.87 | 1.03 | 41.23 | -75.82 | 4.61 | 0.996 |
| 244.50 | 100.00 | 3.88 | 2.49 | 1.39 | 3.88 | 1.03 | 41.27 | -75.61 | 4.59 | 0.996 |
| 244.60 | 100.00 | 3.88 | 2.49 | 1.39 | 3.88 | 1.03 | 41.32 | -75.40 | 4.57 | 0.996 |
| 244.70 | 100.00 | 3.88 | 2.49 | 1.39 | 3.88 | 1.03 | 41.37 | -75.19 | 4.55 | 0.996 |
| 244.80 | 100.00 | 3.89 | 2.49 | 1.40 | 3.89 | 1.02 | 41.42 | -74.98 | 4.53 | 0.996 |
| 244.90 | 100.00 | 3.89 | 2.50 | 1.39 | 3.89 | 1.02 | 41.46 | -74.77 | 4.51 | 0.996 |
| 245.00 | 100.00 | 3.90 | 2.50 | 1.40 | 3.90 | 1.02 | 41.51 | -74.55 | 4.49 | 0.996 |
| 245.10 | 100.00 | 3.90 | 2.50 | 1.40 | 3.90 | 1.02 | 41.56 | -74.34 | 4.47 | 0.996 |
| 245.20 | 100.00 | 3.90 | 2.50 | 1.40 | 3.90 | 1.02 | 41.61 | -74.13 | 4.45 | 0.996 |
| 245.30 | 100.00 | 3.91 | 2.50 | 1.41 | 3.91 | 1.02 | 41.65 | -73.92 | 4.43 | 0.996 |
| 245.40 | 100.00 | 3.91 | 2.50 | 1.41 | 3.91 | 1.01 | 41.70 | -73.71 | 4.41 | 0.996 |
| 245.50 | 100.00 | 3.91 | 2.50 | 1.41 | 3.91 | 1.01 | 41.75 | -73.50 | 4.39 | 0.996 |
| 245.60 | 100.00 | 3.92 | 2.50 | 1.42 | 3.92 | 1.01 | 41.80 | -73.29 | 4.37 | 0.996 |
| 245.70 | 100.00 | 3.92 | 2.51 | 1.41 | 3.92 | 1.01 | 41.85 | -73.08 | 4.35 | 0.996 |
| 245.80 | 100.00 | 3.93 | 2.51 | 1.42 | 3.93 | 1.01 | 41.89 | -72.87 | 4.34 | 0.996 |
| 245.90 | 100.00 | 3.93 | 2.51 | 1.42 | 3.93 | 1.01 | 41.94 | -72.66 | 4.32 | 0.996 |
| 246.00 | 100.00 | 3.93 | 2.51 | 1.42 | 3.93 | 1.00 | 41.99 | -72.45 | 4.30 | 0.996 |
| 246.10 | 100.00 | 3.94 | 2.51 | 1.43 | 3.94 | 1.00 | 42.04 | -72.24 | 4.28 | 0.996 |
| 246.20 | 100.00 | 3.94 | 2.51 | 1.43 | 3.94 | 1.00 | 42.09 | -72.03 | 4.26 | 0.996 |
| 246.30 | 100.00 | 3.94 | 2.51 | 1.43 | 3.94 | 1.00 | 42.14 | -71.82 | 4.24 | 0.996 |
| 246.40 | 100.00 | 3.95 | 2.51 | 1.44 | 3.95 | 1.00 | 42.18 | -71.61 | 4.22 | 0.996 |
| 246.50 | 100.00 | 3.95 | 2.52 | 1.43 | 3.95 | 1.00 | 42.23 | -71.40 | 4.20 | 0.996 |
| 246.60 | 100.00 | 3.96 | 2.52 | 1.44 | 3.96 | 0.99 | 42.28 | -71.19 | 4.19 | 0.996 |
| 246.70 | 100.00 | 3.96 | 2.52 | 1.44 | 3.96 | 0.99 | 42.33 | -70.98 | 4.17 | 0.996 |
| 246.80 | 100.00 | 3.96 | 2.52 | 1.44 | 3.96 | 0.99 | 42.38 | -70.77 | 4.15 | 0.996 |
| 246.90 | 100.00 | 3.97 | 2.52 | 1.45 | 3.97 | 0.99 | 42.43 | -70.56 | 4.13 | 0.996 |
| 247.00 | 100.00 | 3.97 | 2.52 | 1.45 | 3.97 | 0.99 | 42.48 | -70.35 | 4.11 | 0.996 |
| 247.10 | 100.00 | 3.97 | 2.52 | 1.45 | 3.97 | 0.99 | 42.52 | -70.14 | 4.10 | 0.996 |
| 247.20 | 100.00 | 3.98 | 2.52 | 1.46 | 3.98 | 0.98 | 42.57 | -69.93 | 4.08 | 0.996 |
| 247.30 | 100.00 | 3.98 | 2.53 | 1.45 | 3.98 | 0.98 | 42.62 | -69.72 | 4.06 | 0.996 |
| 247.40 | 100.00 | 3.99 | 2.53 | 1.46 | 3.99 | 0.98 | 42.67 | -69.51 | 4.04 | 0.996 |
| 247.50 | 100.00 | 3.99 | 2.53 | 1.46 | 3.99 | 0.98 | 42.72 | -69.30 | 4.02 | 0.996 |
| 247.60 | 100.00 | 3.99 | 2.53 | 1.46 | 3.99 | 0.98 | 42.77 | -69.09 | 4.01 | 0.996 |
| 247.70 | 100.00 | 4.00 | 2.53 | 1.47 | 4.00 | 0.98 | 42.82 | -68.88 | 3.99 | 0.996 |
| 247.80 | 100.00 | 4.00 | 2.53 | 1.47 | 4.00 | 0.97 | 42.87 | -68.67 | 3.97 | 0.996 |
| 247.90 | 100.00 | 4.00 | 2.53 | 1.47 | 4.00 | 0.97 | 42.92 | -68.46 | 3.95 | 0.996 |
| 248.00 | 100.00 | 4.01 | 2.53 | 1.48 | 4.01 | 0.97 | 42.97 | -68.25 | 3.94 | 0.996 |
| 248.10 | 100.00 | 4.01 | 2.54 | 1.47 | 4.01 | 0.97 | 43.02 | -68.04 | 3.92 | 0.996 |
| 248.20 | 100.00 | 4.02 | 2.54 | 1.48 | 4.02 | 0.97 | 43.07 | -67.83 | 3.90 | 0.996 |
| 248.30 | 100.00 | 4.02 | 2.54 | 1.48 | 4.02 | 0.97 | 43.12 | -67.62 | 3.89 | 0.996 |
| 248.40 | 100.00 | 4.02 | 2.54 | 1.48 | 4.02 | 0.96 | 43.17 | -67.42 | 3.87 | 0.996 |
| 248.50 | 100.00 | 4.03 | 2.54 | 1.49 | 4.03 | 0.96 | 43.21 | -67.21 | 3.85 | 0.996 |
| 248.60 | 100.00 | 4.03 | 2.54 | 1.49 | 4.03 | 0.96 | 43.26 | -67.00 | 3.84 | 0.996 |
| 248.70 | 100.00 | 4.04 | 2.54 | 1.50 | 4.04 | 0.96 | 43.31 | -66.79 | 3.82 | 0.996 |
| 248.80 | 100.00 | 4.04 | 2.54 | 1.50 | 4.04 | 0.96 | 43.36 | -66.58 | 3.80 | 0.996 |
| 248.90 | 100.00 | 4.04 | 2.54 | 1.50 | 4.04 | 0.96 | 43.41 | -66.37 | 3.79 | 0.996 |
| 249.00 | 100.00 | 4.05 | 2.55 | 1.50 | 4.05 | 0.95 | 43.46 | -66.16 | 3.77 | 0.996 |
| 249.10 | 100.00 | 4.05 | 2.55 | 1.50 | 4.05 | 0.95 | 43.51 | -65.96 | 3.75 | 0.996 |
| 249.20 | 100.00 | 4.06 | 2.55 | 1.51 | 4.06 | 0.95 | 43.56 | -65.75 | 3.74 | 0.996 |
| 249.30 | 100.00 | 4.06 | 2.55 | 1.51 | 4.06 | 0.95 | 43.61 | -65.54 | 3.72 | 0.996 |
| 249.40 | 100.00 | 4.06 | 2.55 | 1.51 | 4.06 | 0.95 | 43.66 | -65.33 | 3.70 | 0.996 |
| 249.50 | 100.00 | 4.07 | 2.55 | 1.52 | 4.07 | 0.94 | 43.71 | -65.12 | 3.69 | 0.996 |
| 249.60 | 100.00 | 4.07 | 2.55 | 1.52 | 4.07 | 0.94 | 43.77 | -64.91 | 3.67 | 0.996 |
| 249.70 | 100.00 | 4.08 | 2.55 | 1.53 | 4.08 | 0.94 | 43.82 | -64.71 | 3.66 | 0.996 |
| 249.80 | 100.00 | 4.08 | 2.55 | 1.53 | 4.08 | 0.94 | 43.87 | -64.50 | 3.64 | 0.996 |
| 249.90 | 100.00 | 4.08 | 2.55 | 1.53 | 4.08 | 0.94 | 43.92 | -64.29 | 3.62 | 0.996 |
| 250.00 | 100.00 | 4.09 | 2.56 | 1.53 | 4.09 | 0.94 | 43.97 | -64.08 | 3.61 | 0.996 |
| 250.10 | 100.00 | 4.09 | 2.56 | 1.53 | 4.09 | 0.93 | 44.02 | -63.87 | 3.59 | 0.996 |
| 250.20 | 100.00 | 4.10 | 2.56 | 1.54 | 4.10 | 0.93 | 44.07 | -63.67 | 3.58 | 0.996 |
| 250.30 | 100.00 | 4.10 | 2.56 | 1.54 | 4.10 | 0.93 | 44.12 | -63.46 | 3.56 | 0.996 |
| 250.40 | 100.00 | 4.10 | 2.56 | 1.54 | 4.10 | 0.93 | 44.17 | -63.25 | 3.54 | 0.996 |
| 250.50 | 100.00 | 4.11 | 2.56 | 1.55 | 4.11 | 0.93 | 44.22 | -63.04 | 3.53 | 0.996 |
| 250.60 | 100.00 | 4.11 | 2.56 | 1.55 | 4.11 | 0.93 | 44.27 | -62.84 | 3.51 | 0.996 |
| 250.70 | 100.00 | 4.12 | 2.56 | 1.56 | 4.12 | 0.92 | 44.32 | -62.63 | 3.50 | 0.996 |
| 250.80 | 100.00 | 4.12 | 2.56 | 1.56 | 4.12 | 0.92 | 44.37 | -62.42 | 3.48 | 0.996 |
| 250.90 | 100.00 | 4.12 | 2.56 | 1.56 | 4.12 | 0.92 | 44.42 | -62.22 | 3.47 | 0.996 |
| 251.00 | 100.00 | 4.13 | 2.57 | 1.56 | 4.13 | 0.92 | 44.48 | -62.01 | 3.45 | 0.996 |
| 251.10 | 100.00 | 4.13 | 2.57 | 1.56 | 4.13 | 0.92 | 44.53 | -61.80 | 3.44 | 0.996 |
| 251.20 | 100.00 | 4.14 | 2.57 | 1.57 | 4.14 | 0.92 | 44.58 | -61.59 | 3.42 | 0.996 |
| 251.30 | 100.00 | 4.14 | 2.57 | 1.57 | 4.14 | 0.91 | 44.63 | -61.39 | 3.41 | 0.996 |
| 251.40 | 100.00 | 4.14 | 2.57 | 1.57 | 4.14 | 0.91 | 44.68 | -61.18 | 3.39 | 0.996 |
| 251.50 | 100.00 | 4.15 | 2.57 | 1.58 | 4.15 | 0.91 | 44.73 | -60.97 | 3.38 | 0.996 |
| 251.60 | 100.00 | 4.15 | 2.57 | 1.58 | 4.15 | 0.91 | 44.78 | -60.77 | 3.36 | 0.996 |
| 251.70 | 100.00 | 4.16 | 2.57 | 1.59 | 4.16 | 0.91 | 44.84 | -60.56 | 3.35 | 0.996 |
| 251.80 | 100.00 | 4.16 | 2.57 | 1.59 | 4.16 | 0.90 | 44.89 | -60.35 | 3.33 | 0.996 |
| 251.90 | 100.00 | 4.17 | 2.57 | 1.60 | 4.17 | 0.90 | 44.94 | -60.15 | 3.32 | 0.996 |
| 252.00 | 100.00 | 4.17 | 2.58 | 1.59 | 4.17 | 0.90 | 44.99 | -59.94 | 3.31 | 0.996 |
| 252.10 | 100.00 | 4.17 | 2.58 | 1.59 | 4.17 | 0.90 | 45.04 | -59.73 | 3.29 | 0.996 |
| 252.20 | 100.00 | 4.18 | 2.58 | 1.60 | 4.18 | 0.90 | 45.09 | -59.53 | 3.28 | 0.996 |
| 252.30 | 100.00 | 4.18 | 2.58 | 1.60 | 4.18 | 0.90 | 45.15 | -59.32 | 3.26 | 0.996 |
| 252.40 | 100.00 | 4.19 | 2.58 | 1.61 | 4.19 | 0.89 | 45.20 | -59.12 | 3.25 | 0.996 |
| 252.50 | 100.00 | 4.19 | 2.58 | 1.61 | 4.19 | 0.89 | 45.25 | -58.91 | 3.23 | 0.996 |
| 252.60 | 100.00 | 4.19 | 2.58 | 1.61 | 4.19 | 0.89 | 45.30 | -58.70 | 3.22 | 0.996 |
| 252.70 | 100.00 | 4.20 | 2.58 | 1.62 | 4.20 | 0.89 | 45.36 | -58.50 | 3.21 | 0.996 |
| 252.80 | 100.00 | 4.20 | 2.58 | 1.62 | 4.20 | 0.89 | 45.41 | -58.29 | 3.19 | 0.996 |
| 252.90 | 100.00 | 4.21 | 2.58 | 1.63 | 4.21 | 0.89 | 45.46 | -58.09 | 3.18 | 0.996 |
| 253.00 | 100.00 | 4.21 | 2.58 | 1.63 | 4.21 | 0.88 | 45.51 | -57.88 | 3.17 | 0.996 |
| 253.10 | 100.00 | 4.22 | 2.58 | 1.64 | 4.22 | 0.88 | 45.56 | -57.67 | 3.15 | 0.996 |
| 253.20 | 100.00 | 4.22 | 2.59 | 1.63 | 4.22 | 0.88 | 45.62 | -57.47 | 3.14 | 0.996 |
| 253.30 | 100.00 | 4.22 | 2.59 | 1.63 | 4.22 | 0.88 | 45.67 | -57.26 | 3.12 | 0.996 |
| 253.40 | 100.00 | 4.23 | 2.59 | 1.64 | 4.23 | 0.88 | 45.72 | -57.06 | 3.11 | 0.996 |
| 253.50 | 100.00 | 4.23 | 2.59 | 1.64 | 4.23 | 0.87 | 45.78 | -56.85 | 3.10 | 0.996 |
| 253.60 | 100.00 | 4.24 | 2.59 | 1.65 | 4.24 | 0.87 | 45.83 | -56.65 | 3.08 | 0.996 |
| 253.70 | 100.00 | 4.24 | 2.59 | 1.65 | 4.24 | 0.87 | 45.88 | -56.44 | 3.07 | 0.996 |
| 253.80 | 100.00 | 4.24 | 2.59 | 1.65 | 4.24 | 0.87 | 45.93 | -56.24 | 3.06 | 0.996 |
| 253.90 | 100.00 | 4.25 | 2.59 | 1.66 | 4.25 | 0.87 | 45.99 | -56.03 | 3.04 | 0.996 |
| 254.00 | 100.00 | 4.25 | 2.59 | 1.66 | 4.25 | 0.87 | 46.04 | -55.83 | 3.03 | 0.996 |
| 254.10 | 100.00 | 4.26 | 2.59 | 1.67 | 4.26 | 0.86 | 46.09 | -55.62 | 3.02 | 0.996 |
| 254.20 | 100.00 | 4.26 | 2.59 | 1.67 | 4.26 | 0.86 | 46.15 | -55.42 | 3.00 | 0.996 |
| 254.30 | 100.00 | 4.27 | 2.59 | 1.68 | 4.27 | 0.86 | 46.20 | -55.21 | 2.99 | 0.996 |
| 254.40 | 100.00 | 4.27 | 2.59 | 1.68 | 4.27 | 0.86 | 46.25 | -55.01 | 2.98 | 0.996 |
| 254.50 | 100.00 | 4.28 | 2.59 | 1.69 | 4.28 | 0.86 | 46.31 | -54.80 | 2.97 | 0.996 |
| 254.60 | 100.00 | 4.28 | 2.60 | 1.68 | 4.28 | 0.85 | 46.36 | -54.60 | 2.95 | 0.996 |
| 254.70 | 100.00 | 4.28 | 2.60 | 1.68 | 4.28 | 0.85 | 46.41 | -54.39 | 2.94 | 0.996 |
| 254.80 | 100.00 | 4.29 | 2.60 | 1.69 | 4.29 | 0.85 | 46.47 | -54.19 | 2.93 | 0.996 |
| 254.90 | 100.00 | 4.29 | 2.60 | 1.69 | 4.29 | 0.85 | 46.52 | -53.98 | 2.92 | 0.996 |
| 255.00 | 100.00 | 4.30 | 2.60 | 1.70 | 4.30 | 0.85 | 46.57 | -53.78 | 2.90 | 0.996 |
| 255.10 | 100.00 | 4.30 | 2.60 | 1.70 | 4.30 | 0.85 | 46.63 | -53.57 | 2.89 | 0.996 |
| 255.20 | 100.00 | 4.31 | 2.60 | 1.71 | 4.31 | 0.84 | 46.68 | -53.37 | 2.88 | 0.996 |
| 255.30 | 100.00 | 4.31 | 2.60 | 1.71 | 4.31 | 0.84 | 46.73 | -53.17 | 2.87 | 0.996 |
| 255.40 | 100.00 | 4.31 | 2.60 | 1.71 | 4.31 | 0.84 | 46.79 | -52.96 | 2.85 | 0.996 |
| 255.50 | 100.00 | 4.32 | 2.60 | 1.72 | 4.32 | 0.84 | 46.84 | -52.76 | 2.84 | 0.996 |
| 255.60 | 100.00 | 4.32 | 2.60 | 1.72 | 4.32 | 0.84 | 46.90 | -52.55 | 2.83 | 0.996 |
| 255.70 | 100.00 | 4.33 | 2.60 | 1.73 | 4.33 | 0.83 | 46.95 | -52.35 | 2.82 | 0.996 |
| 255.80 | 100.00 | 4.33 | 2.60 | 1.73 | 4.33 | 0.83 | 47.00 | -52.15 | 2.80 | 0.996 |
| 255.90 | 100.00 | 4.34 | 2.60 | 1.74 | 4.34 | 0.83 | 47.06 | -51.94 | 2.79 | 0.996 |
| 256.00 | 100.00 | 4.34 | 2.60 | 1.74 | 4.34 | 0.83 | 47.11 | -51.74 | 2.78 | 0.996 |
| 256.10 | 100.00 | 4.35 | 2.60 | 1.75 | 4.35 | 0.83 | 47.17 | -51.53 | 2.77 | 0.996 |
| 256.20 | 100.00 | 4.35 | 2.61 | 1.74 | 4.35 | 0.82 | 47.22 | -51.33 | 2.76 | 0.996 |
| 256.30 | 100.00 | 4.35 | 2.61 | 1.74 | 4.35 | 0.82 | 47.28 | -51.13 | 2.74 | 0.996 |
| 256.40 | 100.00 | 4.36 | 2.61 | 1.75 | 4.36 | 0.82 | 47.33 | -50.92 | 2.73 | 0.996 |
| 256.50 | 100.00 | 4.36 | 2.61 | 1.75 | 4.36 | 0.82 | 47.38 | -50.72 | 2.72 | 0.996 |
| 256.60 | 100.00 | 4.37 | 2.61 | 1.76 | 4.37 | 0.82 | 47.44 | -50.52 | 2.71 | 0.996 |
| 256.70 | 100.00 | 4.37 | 2.61 | 1.76 | 4.37 | 0.82 | 47.49 | -50.31 | 2.70 | 0.996 |
| 256.80 | 100.00 | 4.38 | 2.61 | 1.77 | 4.38 | 0.81 | 47.55 | -50.11 | 2.69 | 0.996 |
| 256.90 | 100.00 | 4.38 | 2.61 | 1.77 | 4.38 | 0.81 | 47.60 | -49.91 | 2.68 | 0.996 |
| 257.00 | 100.00 | 4.39 | 2.61 | 1.78 | 4.39 | 0.81 | 47.66 | -49.70 | 2.66 | 0.996 |
| 257.10 | 100.00 | 4.39 | 2.61 | 1.78 | 4.39 | 0.81 | 47.71 | -49.50 | 2.65 | 0.996 |
| 257.20 | 100.00 | 4.39 | 2.61 | 1.78 | 4.39 | 0.81 | 47.77 | -49.30 | 2.64 | 0.996 |
| 257.30 | 100.00 | 4.40 | 2.61 | 1.79 | 4.40 | 0.80 | 47.82 | -49.10 | 2.63 | 0.996 |
| 257.40 | 100.00 | 4.40 | 2.61 | 1.79 | 4.40 | 0.80 | 47.88 | -48.89 | 2.62 | 0.996 |
| 257.50 | 100.00 | 4.41 | 2.61 | 1.80 | 4.41 | 0.80 | 47.93 | -48.69 | 2.61 | 0.996 |
| 257.60 | 100.00 | 4.41 | 2.61 | 1.80 | 4.41 | 0.80 | 47.99 | -48.49 | 2.60 | 0.996 |
| 257.70 | 100.00 | 4.42 | 2.61 | 1.81 | 4.42 | 0.80 | 48.04 | -48.29 | 2.59 | 0.996 |
| 257.80 | 100.00 | 4.42 | 2.61 | 1.81 | 4.42 | 0.80 | 48.10 | -48.08 | 2.57 | 0.996 |
| 257.90 | 100.00 | 4.43 | 2.61 | 1.82 | 4.43 | 0.79 | 48.15 | -47.88 | 2.56 | 0.996 |
| 258.00 | 100.00 | 4.43 | 2.61 | 1.82 | 4.43 | 0.79 | 48.21 | -47.68 | 2.55 | 0.996 |
| 258.10 | 100.00 | 4.44 | 2.61 | 1.83 | 4.44 | 0.79 | 48.27 | -47.48 | 2.54 | 0.996 |
| 258.20 | 100.00 | 4.44 | 2.61 | 1.83 | 4.44 | 0.79 | 48.32 | -47.27 | 2.53 | 0.996 |
| 258.30 | 100.00 | 4.44 | 2.61 | 1.83 | 4.44 | 0.79 | 48.38 | -47.07 | 2.52 | 0.996 |
| 258.40 | 100.00 | 4.45 | 2.61 | 1.84 | 4.45 | 0.78 | 48.43 | -46.87 | 2.51 | 0.996 |
| 258.50 | 100.00 | 4.45 | 2.61 | 1.84 | 4.45 | 0.78 | 48.49 | -46.67 | 2.50 | 0.996 |
| 258.60 | 100.00 | 4.46 | 2.61 | 1.85 | 4.46 | 0.78 | 48.54 | -46.47 | 2.49 | 0.996 |
| 258.70 | 100.00 | 4.46 | 2.61 | 1.85 | 4.46 | 0.78 | 48.60 | -46.26 | 2.48 | 0.996 |
| 258.80 | 100.00 | 4.47 | 2.62 | 1.85 | 4.47 | 0.78 | 48.66 | -46.06 | 2.47 | 0.996 |
| 258.90 | 100.00 | 4.47 | 2.62 | 1.85 | 4.47 | 0.77 | 48.71 | -45.86 | 2.46 | 0.996 |
| 259.00 | 100.00 | 4.48 | 2.62 | 1.86 | 4.48 | 0.77 | 48.77 | -45.66 | 2.45 | 0.996 |
| 259.10 | 100.00 | 4.48 | 2.62 | 1.86 | 4.48 | 0.77 | 48.82 | -45.46 | 2.44 | 0.996 |
| 259.20 | 100.00 | 4.49 | 2.62 | 1.87 | 4.49 | 0.77 | 48.88 | -45.26 | 2.43 | 0.996 |
| 259.30 | 100.00 | 4.49 | 2.62 | 1.87 | 4.49 | 0.77 | 48.94 | -45.05 | 2.42 | 0.996 |
| 259.40 | 100.00 | 4.50 | 2.62 | 1.88 | 4.50 | 0.77 | 48.99 | -44.85 | 2.41 | 0.996 |
| 259.50 | 100.00 | 4.50 | 2.62 | 1.88 | 4.50 | 0.76 | 49.05 | -44.65 | 2.40 | 0.996 |
| 259.60 | 100.00 | 4.51 | 2.62 | 1.89 | 4.51 | 0.76 | 49.11 | -44.45 | 2.39 | 0.996 |
| 259.70 | 100.00 | 4.51 | 2.62 | 1.89 | 4.51 | 0.76 | 49.16 | -44.25 | 2.38 | 0.996 |
| 259.80 | 100.00 | 4.52 | 2.62 | 1.90 | 4.52 | 0.76 | 49.22 | -44.05 | 2.37 | 0.996 |
| 259.90 | 100.00 | 4.52 | 2.62 | 1.90 | 4.52 | 0.76 | 49.28 | -43.85 | 2.36 | 0.996 |
| 260.00 | 100.00 | 4.52 | 2.62 | 1.90 | 4.52 | 0.75 | 49.33 | -43.65 | 2.35 | 0.996 |
| 260.10 | 100.00 | 4.53 | 2.62 | 1.91 | 4.53 | 0.75 | 49.39 | -43.45 | 2.34 | 0.996 |
| 260.20 | 100.00 | 4.53 | 2.62 | 1.91 | 4.53 | 0.75 | 49.45 | -43.24 | 2.33 | 0.996 |
| 260.30 | 100.00 | 4.54 | 2.62 | 1.92 | 4.54 | 0.75 | 49.50 | -43.04 | 2.32 | 0.996 |
| 260.40 | 100.00 | 4.54 | 2.62 | 1.92 | 4.54 | 0.75 | 49.56 | -42.84 | 2.31 | 0.996 |
| 260.50 | 100.00 | 4.55 | 2.62 | 1.93 | 4.55 | 0.74 | 49.62 | -42.64 | 2.30 | 0.996 |
| 260.60 | 100.00 | 4.55 | 2.62 | 1.93 | 4.55 | 0.74 | 49.67 | -42.44 | 2.29 | 0.996 |
| 260.70 | 100.00 | 4.56 | 2.62 | 1.94 | 4.56 | 0.74 | 49.73 | -42.24 | 2.28 | 0.996 |
| 260.80 | 100.00 | 4.56 | 2.62 | 1.94 | 4.56 | 0.74 | 49.79 | -42.04 | 2.27 | 0.996 |
| 260.90 | 100.00 | 4.57 | 2.62 | 1.95 | 4.57 | 0.74 | 49.85 | -41.84 | 2.26 | 0.996 |
| 261.00 | 100.00 | 4.57 | 2.62 | 1.95 | 4.57 | 0.74 | 49.90 | -41.64 | 2.25 | 0.996 |
| 261.10 | 100.00 | 4.58 | 2.62 | 1.96 | 4.58 | 0.73 | 49.96 | -41.44 | 2.24 | 0.996 |
| 261.20 | 100.00 | 4.58 | 2.62 | 1.96 | 4.58 | 0.73 | 50.02 | -41.24 | 2.23 | 0.996 |
| 261.30 | 100.00 | 4.59 | 2.62 | 1.97 | 4.59 | 0.73 | 50.07 | -41.04 | 2.22 | 0.996 |
| 261.40 | 100.00 | 4.59 | 2.62 | 1.97 | 4.59 | 0.73 | 50.13 | -40.84 | 2.21 | 0.996 |
| 261.50 | 100.00 | 4.60 | 2.62 | 1.98 | 4.60 | 0.73 | 50.19 | -40.64 | 2.20 | 0.996 |
| 261.60 | 100.00 | 4.60 | 2.62 | 1.98 | 4.60 | 0.72 | 50.25 | -40.44 | 2.20 | 0.996 |
| 261.70 | 100.00 | 4.61 | 2.62 | 1.99 | 4.61 | 0.72 | 50.31 | -40.24 | 2.19 | 0.996 |
| 261.80 | 100.00 | 4.61 | 2.62 | 1.99 | 4.61 | 0.72 | 50.36 | -40.04 | 2.18 | 0.996 |
| 261.90 | 100.00 | 4.62 | 2.62 | 2.00 | 4.62 | 0.72 | 50.42 | -39.84 | 2.17 | 0.996 |
| 262.00 | 100.00 | 4.62 | 2.62 | 2.00 | 4.62 | 0.72 | 50.48 | -39.64 | 2.16 | 0.996 |
| 262.10 | 100.00 | 4.63 | 2.62 | 2.01 | 4.63 | 0.71 | 50.54 | -39.44 | 2.15 | 0.996 |
| 262.20 | 100.00 | 4.63 | 2.62 | 2.01 | 4.63 | 0.71 | 50.60 | -39.24 | 2.14 | 0.996 |
| 262.30 | 100.00 | 4.64 | 2.62 | 2.02 | 4.64 | 0.71 | 50.65 | -39.04 | 2.13 | 0.996 |
| 262.40 | 100.00 | 4.64 | 2.62 | 2.02 | 4.64 | 0.71 | 50.71 | -38.85 | 2.12 | 0.996 |
| 262.50 | 100.00 | 4.65 | 2.62 | 2.03 | 4.65 | 0.71 | 50.77 | -38.65 | 2.12 | 0.996 |
| 262.60 | 100.00 | 4.65 | 2.62 | 2.03 | 4.65 | 0.70 | 50.83 | -38.45 | 2.11 | 0.996 |
| 262.70 | 100.00 | 4.66 | 2.62 | 2.04 | 4.66 | 0.70 | 50.89 | -38.25 | 2.10 | 0.996 |
| 262.80 | 100.00 | 4.66 | 2.61 | 2.05 | 4.66 | 0.70 | 50.94 | -38.05 | 2.09 | 0.996 |
| 262.90 | 100.00 | 4.67 | 2.61 | 2.06 | 4.67 | 0.70 | 51.00 | -37.85 | 2.08 | 0.996 |
| 263.00 | 100.00 | 4.67 | 2.61 | 2.06 | 4.67 | 0.70 | 51.06 | -37.65 | 2.07 | 0.996 |
| 263.10 | 100.00 | 4.68 | 2.61 | 2.07 | 4.68 | 0.70 | 51.12 | -37.45 | 2.07 | 0.996 |
| 263.20 | 100.00 | 4.68 | 2.61 | 2.07 | 4.68 | 0.69 | 51.18 | -37.25 | 2.06 | 0.996 |
| 263.30 | 100.00 | 4.69 | 2.61 | 2.08 | 4.69 | 0.69 | 51.24 | -37.06 | 2.05 | 0.996 |
| 263.40 | 100.00 | 4.69 | 2.61 | 2.08 | 4.69 | 0.69 | 51.30 | -36.86 | 2.04 | 0.996 |
| 263.50 | 100.00 | 4.70 | 2.61 | 2.09 | 4.70 | 0.69 | 51.35 | -36.66 | 2.03 | 0.996 |
| 263.60 | 100.00 | 4.70 | 2.61 | 2.09 | 4.70 | 0.69 | 51.41 | -36.46 | 2.02 | 0.996 |
| 263.70 | 100.00 | 4.71 | 2.61 | 2.10 | 4.71 | 0.68 | 51.47 | -36.26 | 2.02 | 0.996 |
| 263.80 | 100.00 | 4.71 | 2.61 | 2.10 | 4.71 | 0.68 | 51.53 | -36.06 | 2.01 | 0.996 |
| 263.90 | 100.00 | 4.72 | 2.61 | 2.11 | 4.72 | 0.68 | 51.59 | -35.87 | 2.00 | 0.996 |
| 264.00 | 100.00 | 4.72 | 2.61 | 2.11 | 4.72 | 0.68 | 51.65 | -35.67 | 1.99 | 0.996 |
| 264.10 | 100.00 | 4.73 | 2.61 | 2.12 | 4.73 | 0.68 | 51.71 | -35.47 | 1.98 | 0.996 |
| 264.20 | 100.00 | 4.73 | 2.61 | 2.12 | 4.73 | 0.67 | 51.77 | -35.27 | 1.98 | 0.996 |
| 264.30 | 100.00 | 4.74 | 2.61 | 2.13 | 4.74 | 0.67 | 51.83 | -35.07 | 1.97 | 0.996 |
| 264.40 | 100.00 | 4.74 | 2.61 | 2.13 | 4.74 | 0.67 | 51.89 | -34.88 | 1.96 | 0.996 |
| 264.50 | 100.00 | 4.75 | 2.61 | 2.14 | 4.75 | 0.67 | 51.95 | -34.68 | 1.95 | 0.996 |
| 264.60 | 100.00 | 4.75 | 2.61 | 2.14 | 4.75 | 0.67 | 52.01 | -34.48 | 1.94 | 0.996 |
| 264.70 | 100.00 | 4.76 | 2.61 | 2.15 | 4.76 | 0.67 | 52.07 | -34.28 | 1.94 | 0.996 |
| 264.80 | 100.00 | 4.76 | 2.61 | 2.15 | 4.76 | 0.66 | 52.12 | -34.09 | 1.93 | 0.996 |
| 264.90 | 100.00 | 4.77 | 2.61 | 2.16 | 4.77 | 0.66 | 52.18 | -33.89 | 1.92 | 0.996 |
| 265.00 | 100.00 | 4.77 | 2.61 | 2.16 | 4.77 | 0.66 | 52.24 | -33.69 | 1.91 | 0.996 |
| 265.10 | 100.00 | 4.78 | 2.60 | 2.18 | 4.78 | 0.66 | 52.30 | -33.49 | 1.91 | 0.996 |
| 265.20 | 100.00 | 4.78 | 2.60 | 2.18 | 4.78 | 0.66 | 52.36 | -33.30 | 1.90 | 0.996 |
| 265.30 | 100.00 | 4.79 | 2.60 | 2.19 | 4.79 | 0.65 | 52.42 | -33.10 | 1.89 | 0.996 |
| 265.40 | 100.00 | 4.79 | 2.60 | 2.19 | 4.79 | 0.65 | 52.48 | -32.90 | 1.88 | 0.996 |
| 265.50 | 100.00 | 4.80 | 2.60 | 2.20 | 4.80 | 0.65 | 52.54 | -32.71 | 1.88 | 0.996 |
| 265.60 | 100.00 | 4.80 | 2.60 | 2.20 | 4.80 | 0.65 | 52.60 | -32.51 | 1.87 | 0.996 |
| 265.70 | 100.00 | 4.81 | 2.60 | 2.21 | 4.81 | 0.65 | 52.66 | -32.31 | 1.86 | 0.996 |
| 265.80 | 100.00 | 4.81 | 2.60 | 2.21 | 4.81 | 0.64 | 52.72 | -32.12 | 1.86 | 0.996 |
| 265.90 | 100.00 | 4.82 | 2.60 | 2.22 | 4.82 | 0.64 | 52.78 | -31.92 | 1.85 | 0.996 |
| 266.00 | 100.00 | 4.83 | 2.60 | 2.23 | 4.83 | 0.64 | 52.84 | -31.72 | 1.84 | 0.996 |
| 266.10 | 100.00 | 4.83 | 2.60 | 2.23 | 4.83 | 0.64 | 52.90 | -31.53 | 1.83 | 0.996 |
| 266.20 | 100.00 | 4.84 | 2.60 | 2.24 | 4.84 | 0.64 | 52.96 | -31.33 | 1.83 | 0.996 |
| 266.30 | 100.00 | 4.84 | 2.60 | 2.24 | 4.84 | 0.64 | 53.02 | -31.13 | 1.82 | 0.996 |
| 266.40 | 100.00 | 4.85 | 2.60 | 2.25 | 4.85 | 0.63 | 53.09 | -30.94 | 1.81 | 0.996 |
| 266.50 | 100.00 | 4.85 | 2.59 | 2.26 | 4.85 | 0.63 | 53.15 | -30.74 | 1.81 | 0.996 |
| 266.60 | 100.00 | 4.86 | 2.59 | 2.27 | 4.86 | 0.63 | 53.21 | -30.55 | 1.80 | 0.996 |
| 266.70 | 100.00 | 4.86 | 2.59 | 2.27 | 4.86 | 0.63 | 53.27 | -30.35 | 1.79 | 0.996 |
| 266.80 | 100.00 | 4.87 | 2.59 | 2.28 | 4.87 | 0.63 | 53.33 | -30.15 | 1.78 | 0.996 |
| 266.90 | 100.00 | 4.87 | 2.59 | 2.28 | 4.87 | 0.62 | 53.39 | -29.96 | 1.78 | 0.996 |
| 267.00 | 100.00 | 4.88 | 2.59 | 2.29 | 4.88 | 0.62 | 53.45 | -29.76 | 1.77 | 0.996 |
| 267.10 | 100.00 | 4.88 | 2.59 | 2.29 | 4.88 | 0.62 | 53.51 | -29.57 | 1.76 | 0.996 |
| 267.20 | 100.00 | 4.89 | 2.59 | 2.30 | 4.89 | 0.62 | 53.57 | -29.37 | 1.76 | 0.996 |
| 267.30 | 100.00 | 4.89 | 2.59 | 2.30 | 4.89 | 0.62 | 53.63 | -29.18 | 1.75 | 0.996 |
| 267.40 | 100.00 | 4.90 | 2.59 | 2.31 | 4.90 | 0.61 | 53.69 | -28.98 | 1.74 | 0.996 |
| 267.50 | 100.00 | 4.91 | 2.58 | 2.33 | 4.91 | 0.61 | 53.75 | -28.78 | 1.74 | 0.996 |
| 267.60 | 100.00 | 4.91 | 2.58 | 2.33 | 4.91 | 0.61 | 53.82 | -28.59 | 1.73 | 0.996 |
| 267.70 | 100.00 | 4.92 | 2.58 | 2.34 | 4.92 | 0.61 | 53.88 | -28.39 | 1.73 | 0.996 |
| 267.80 | 100.00 | 4.92 | 2.58 | 2.34 | 4.92 | 0.61 | 53.94 | -28.20 | 1.72 | 0.996 |
| 267.90 | 100.00 | 4.93 | 2.58 | 2.35 | 4.93 | 0.61 | 54.00 | -28.00 | 1.71 | 0.996 |
| 268.00 | 100.00 | 4.93 | 2.58 | 2.35 | 4.93 | 0.60 | 54.06 | -27.81 | 1.71 | 0.996 |
| 268.10 | 100.00 | 4.94 | 2.58 | 2.36 | 4.94 | 0.60 | 54.12 | -27.61 | 1.70 | 0.996 |
| 268.20 | 100.00 | 4.94 | 2.58 | 2.36 | 4.94 | 0.60 | 54.18 | -27.42 | 1.69 | 0.996 |
| 268.30 | 100.00 | 4.95 | 2.58 | 2.37 | 4.95 | 0.60 | 54.24 | -27.22 | 1.69 | 0.996 |
| 268.40 | 100.00 | 4.95 | 2.58 | 2.37 | 4.95 | 0.60 | 54.31 | -27.03 | 1.68 | 0.996 |
| 268.50 | 100.00 | 4.96 | 2.57 | 2.39 | 4.96 | 0.59 | 54.37 | -26.84 | 1.67 | 0.996 |
| 268.60 | 100.00 | 4.97 | 2.57 | 2.40 | 4.97 | 0.59 | 54.43 | -26.64 | 1.67 | 0.996 |
| 268.70 | 100.00 | 4.97 | 2.57 | 2.40 | 4.97 | 0.59 | 54.49 | -26.45 | 1.66 | 0.996 |
| 268.80 | 100.00 | 4.98 | 2.57 | 2.41 | 4.98 | 0.59 | 54.55 | -26.25 | 1.66 | 0.996 |
| 268.90 | 100.00 | 4.98 | 2.57 | 2.41 | 4.98 | 0.59 | 54.61 | -26.06 | 1.65 | 0.996 |
| 269.00 | 100.00 | 4.99 | 2.57 | 2.42 | 4.99 | 0.59 | 54.68 | -25.86 | 1.64 | 0.996 |
| 269.10 | 100.00 | 4.99 | 2.57 | 2.42 | 4.99 | 0.58 | 54.74 | -25.67 | 1.64 | 0.996 |
| 269.20 | 100.00 | 5.00 | 2.57 | 2.43 | 5.00 | 0.58 | 54.80 | -25.48 | 1.63 | 0.996 |
| 269.30 | 100.00 | 5.00 | 2.56 | 2.44 | 5.00 | 0.58 | 54.86 | -25.28 | 1.63 | 0.996 |
| 269.40 | 100.00 | 5.01 | 2.56 | 2.45 | 5.01 | 0.58 | 54.92 | -25.09 | 1.62 | 0.996 |
| 269.50 | 100.00 | 5.02 | 2.56 | 2.46 | 5.02 | 0.58 | 54.99 | -24.89 | 1.62 | 0.996 |
| 269.60 | 100.00 | 5.02 | 2.56 | 2.46 | 5.02 | 0.57 | 55.05 | -24.70 | 1.61 | 0.996 |
| 269.70 | 100.00 | 5.03 | 2.56 | 2.47 | 5.03 | 0.57 | 55.11 | -24.51 | 1.60 | 0.996 |
| 269.80 | 100.00 | 5.03 | 2.56 | 2.47 | 5.03 | 0.57 | 55.17 | -24.31 | 1.60 | 0.996 |
| 269.90 | 100.00 | 5.04 | 2.56 | 2.48 | 5.04 | 0.57 | 55.24 | -24.12 | 1.59 | 0.996 |
| 270.00 | 100.00 | 5.04 | 2.55 | 2.49 | 5.04 | 0.57 | 55.30 | -23.93 | 1.59 | 0.996 |
| 270.10 | 100.00 | 5.05 | 2.55 | 2.50 | 5.05 | 0.57 | 55.36 | -23.73 | 1.58 | 0.996 |
| 270.20 | 100.00 | 5.06 | 2.55 | 2.51 | 5.06 | 0.56 | 55.42 | -23.54 | 1.58 | 0.996 |
| 270.30 | 100.00 | 5.06 | 2.55 | 2.51 | 5.06 | 0.56 | 55.48 | -23.35 | 1.57 | 0.996 |
| 270.40 | 100.00 | 5.07 | 2.55 | 2.52 | 5.07 | 0.56 | 55.55 | -23.15 | 1.57 | 0.996 |
| 270.50 | 100.00 | 5.07 | 2.55 | 2.52 | 5.07 | 0.56 | 55.61 | -22.96 | 1.56 | 0.996 |
| 270.60 | 100.00 | 5.08 | 2.54 | 2.54 | 5.08 | 0.56 | 55.67 | -22.77 | 1.55 | 0.996 |
| 270.70 | 100.00 | 5.08 | 2.54 | 2.54 | 5.08 | 0.55 | 55.74 | -22.57 | 1.55 | 0.996 |
| 270.80 | 100.00 | 5.09 | 2.54 | 2.55 | 5.09 | 0.55 | 55.80 | -22.38 | 1.54 | 0.996 |
| 270.90 | 100.00 | 5.10 | 2.54 | 2.56 | 5.10 | 0.55 | 55.86 | -22.19 | 1.54 | 0.996 |
| 271.00 | 100.00 | 5.10 | 2.54 | 2.56 | 5.10 | 0.55 | 55.92 | -22.00 | 1.53 | 0.996 |
| 271.10 | 100.00 | 5.11 | 2.54 | 2.57 | 5.11 | 0.55 | 55.99 | -21.80 | 1.53 | 0.996 |
| 271.20 | 100.00 | 5.11 | 2.53 | 2.58 | 5.11 | 0.55 | 56.05 | -21.61 | 1.52 | 0.996 |
| 271.30 | 100.00 | 5.12 | 2.53 | 2.59 | 5.12 | 0.54 | 56.11 | -21.42 | 1.52 | 0.996 |
| 271.40 | 100.00 | 5.12 | 2.53 | 2.59 | 5.12 | 0.54 | 56.18 | -21.23 | 1.51 | 0.996 |
| 271.50 | 100.00 | 5.13 | 2.53 | 2.60 | 5.13 | 0.54 | 56.24 | -21.03 | 1.51 | 0.996 |
| 271.60 | 100.00 | 5.14 | 2.53 | 2.61 | 5.14 | 0.54 | 56.30 | -20.84 | 1.50 | 0.996 |
| 271.70 | 100.00 | 5.14 | 2.53 | 2.61 | 5.14 | 0.54 | 56.36 | -20.65 | 1.50 | 0.996 |
| 271.80 | 100.00 | 5.15 | 2.52 | 2.63 | 5.15 | 0.54 | 56.43 | -20.46 | 1.49 | 0.996 |
| 271.90 | 100.00 | 5.15 | 2.52 | 2.63 | 5.15 | 0.53 | 56.49 | -20.27 | 1.49 | 0.996 |
| 272.00 | 100.00 | 5.16 | 2.52 | 2.64 | 5.16 | 0.53 | 56.55 | -20.07 | 1.48 | 0.996 |
| 272.10 | 100.00 | 5.17 | 2.52 | 2.65 | 5.17 | 0.53 | 56.62 | -19.88 | 1.48 | 0.996 |
| 272.20 | 100.00 | 5.17 | 2.52 | 2.65 | 5.17 | 0.53 | 56.68 | -19.69 | 1.47 | 0.996 |
| 272.30 | 100.00 | 5.18 | 2.51 | 2.67 | 5.18 | 0.53 | 56.74 | -19.50 | 1.47 | 0.996 |
| 272.40 | 100.00 | 5.18 | 2.51 | 2.67 | 5.18 | 0.53 | 56.81 | -19.31 | 1.46 | 0.996 |
| 272.50 | 100.00 | 5.19 | 2.51 | 2.68 | 5.19 | 0.52 | 56.87 | -19.12 | 1.46 | 0.996 |
| 272.60 | 100.00 | 5.19 | 2.51 | 2.68 | 5.19 | 0.52 | 56.93 | -18.92 | 1.46 | 0.996 |
| 272.70 | 100.00 | 5.20 | 2.51 | 2.69 | 5.20 | 0.52 | 57.00 | -18.73 | 1.45 | 0.996 |
| 272.80 | 100.00 | 5.21 | 2.50 | 2.71 | 5.21 | 0.52 | 57.06 | -18.54 | 1.45 | 0.996 |
| 272.90 | 100.00 | 5.21 | 2.50 | 2.71 | 5.21 | 0.52 | 57.12 | -18.35 | 1.44 | 0.996 |
| 273.00 | 100.00 | 5.22 | 2.50 | 2.72 | 5.22 | 0.52 | 57.19 | -18.16 | 1.44 | 0.996 |
| 273.10 | 100.00 | 5.22 | 2.50 | 2.72 | 5.22 | 0.51 | 57.25 | -17.97 | 1.43 | 0.996 |
| 273.20 | 100.00 | 5.23 | 2.50 | 2.73 | 5.23 | 0.51 | 57.32 | -17.78 | 1.43 | 0.996 |
| 273.30 | 100.00 | 5.24 | 2.49 | 2.75 | 5.24 | 0.51 | 57.38 | -17.59 | 1.43 | 0.996 |
| 273.40 | 100.00 | 5.24 | 2.49 | 2.75 | 5.24 | 0.51 | 57.44 | -17.40 | 1.42 | 0.996 |
| 273.50 | 100.00 | 5.25 | 2.49 | 2.76 | 5.25 | 0.51 | 57.51 | -17.21 | 1.42 | 0.996 |
| 273.60 | 100.00 | 5.25 | 2.49 | 2.76 | 5.25 | 0.50 | 57.57 | -17.02 | 1.41 | 0.996 |
| 273.70 | 100.00 | 5.26 | 2.49 | 2.77 | 5.26 | 0.50 | 57.64 | -16.82 | 1.41 | 0.996 |
| 273.80 | 100.00 | 5.27 | 2.48 | 2.79 | 5.27 | 0.50 | 57.70 | -16.63 | 1.40 | 0.996 |
| 273.90 | 100.00 | 5.27 | 2.48 | 2.79 | 5.27 | 0.50 | 57.76 | -16.44 | 1.40 | 0.996 |
| 274.00 | 100.00 | 5.28 | 2.48 | 2.80 | 5.28 | 0.50 | 57.83 | -16.25 | 1.40 | 0.996 |
| 274.10 | 100.00 | 5.28 | 2.48 | 2.80 | 5.28 | 0.50 | 57.89 | -16.06 | 1.39 | 0.996 |
| 274.20 | 100.00 | 5.29 | 2.47 | 2.82 | 5.29 | 0.49 | 57.96 | -15.87 | 1.39 | 0.996 |
| 274.30 | 100.00 | 5.30 | 2.47 | 2.83 | 5.30 | 0.49 | 58.02 | -15.68 | 1.38 | 0.996 |
| 274.40 | 100.00 | 5.30 | 2.47 | 2.83 | 5.30 | 0.49 | 58.08 | -15.49 | 1.38 | 0.996 |
| 274.50 | 100.00 | 5.31 | 2.47 | 2.84 | 5.31 | 0.49 | 58.15 | -15.30 | 1.38 | 0.996 |
| 274.60 | 100.00 | 5.31 | 2.46 | 2.85 | 5.31 | 0.49 | 58.21 | -15.11 | 1.37 | 0.996 |
| 274.70 | 100.00 | 5.32 | 2.46 | 2.86 | 5.32 | 0.49 | 58.28 | -14.92 | 1.37 | 0.996 |
| 274.80 | 100.00 | 5.33 | 2.46 | 2.87 | 5.33 | 0.49 | 58.34 | -14.73 | 1.37 | 0.996 |
| 274.90 | 100.00 | 5.33 | 2.46 | 2.87 | 5.33 | 0.48 | 58.41 | -14.54 | 1.36 | 0.996 |
| 275.00 | 100.00 | 5.34 | 2.45 | 2.89 | 5.34 | 0.48 | 58.47 | -14.36 | 1.36 | 0.996 |
| 275.10 | 100.00 | 5.35 | 2.45 | 2.90 | 5.35 | 0.48 | 58.53 | -14.17 | 1.36 | 0.996 |
| 275.20 | 100.00 | 5.35 | 2.45 | 2.90 | 5.35 | 0.48 | 58.60 | -13.98 | 1.35 | 0.996 |
| 275.30 | 100.00 | 5.36 | 2.45 | 2.91 | 5.36 | 0.48 | 58.66 | -13.79 | 1.35 | 0.996 |
| 275.40 | 100.00 | 5.36 | 2.44 | 2.92 | 5.36 | 0.48 | 58.73 | -13.60 | 1.35 | 0.996 |
| 275.50 | 100.00 | 5.37 | 2.44 | 2.93 | 5.37 | 0.47 | 58.79 | -13.41 | 1.34 | 0.996 |
| 275.60 | 100.00 | 5.38 | 2.44 | 2.94 | 5.38 | 0.47 | 58.86 | -13.22 | 1.34 | 0.996 |
| 275.70 | 100.00 | 5.38 | 2.44 | 2.94 | 5.38 | 0.47 | 58.92 | -13.03 | 1.34 | 0.996 |
| 275.80 | 100.00 | 5.39 | 2.43 | 2.96 | 5.39 | 0.47 | 58.99 | -12.84 | 1.33 | 0.996 |
| 275.90 | 100.00 | 5.40 | 2.43 | 2.97 | 5.40 | 0.47 | 59.05 | -12.65 | 1.33 | 0.996 |
| 276.00 | 100.00 | 5.40 | 2.43 | 2.97 | 5.40 | 0.47 | 59.12 | -12.47 | 1.33 | 0.996 |
| 276.10 | 100.00 | 5.41 | 2.43 | 2.98 | 5.41 | 0.46 | 59.18 | -12.28 | 1.32 | 0.996 |
| 276.20 | 100.00 | 5.41 | 2.42 | 2.99 | 5.41 | 0.46 | 59.24 | -12.09 | 1.32 | 0.996 |
| 276.30 | 100.00 | 5.42 | 2.42 | 3.00 | 5.42 | 0.46 | 59.31 | -11.90 | 1.32 | 0.996 |
| 276.40 | 100.00 | 5.43 | 2.42 | 3.01 | 5.43 | 0.46 | 59.37 | -11.71 | 1.32 | 0.996 |
| 276.50 | 100.00 | 5.43 | 2.41 | 3.02 | 5.43 | 0.46 | 59.44 | -11.52 | 1.31 | 0.996 |
| 276.60 | 100.00 | 5.44 | 2.41 | 3.03 | 5.44 | 0.46 | 59.50 | -11.33 | 1.31 | 0.996 |
| 276.70 | 100.00 | 5.45 | 2.41 | 3.04 | 5.45 | 0.46 | 59.57 | -11.15 | 1.31 | 0.996 |
| 276.80 | 100.00 | 5.45 | 2.41 | 3.04 | 5.45 | 0.45 | 59.63 | -10.96 | 1.31 | 0.996 |
| 276.90 | 100.00 | 5.46 | 2.40 | 3.06 | 5.46 | 0.45 | 59.70 | -10.77 | 1.30 | 0.996 |
| 277.00 | 100.00 | 5.46 | 2.40 | 3.06 | 5.46 | 0.45 | 59.76 | -10.58 | 1.30 | 0.996 |
| 277.10 | 100.00 | 5.47 | 2.40 | 3.07 | 5.47 | 0.45 | 59.83 | -10.39 | 1.30 | 0.996 |
| 277.20 | 100.00 | 5.48 | 2.39 | 3.09 | 5.48 | 0.45 | 59.89 | -10.21 | 1.30 | 0.996 |
| 277.30 | 100.00 | 5.48 | 2.39 | 3.09 | 5.48 | 0.45 | 59.96 | -10.02 | 1.29 | 0.996 |
| 277.40 | 100.00 | 5.49 | 2.39 | 3.10 | 5.49 | 0.44 | 60.02 | -9.83 | 1.29 | 0.996 |
| 277.50 | 100.00 | 5.50 | 2.39 | 3.11 | 5.50 | 0.44 | 60.09 | -9.64 | 1.29 | 0.996 |
| 277.60 | 100.00 | 5.50 | 2.38 | 3.12 | 5.50 | 0.44 | 60.15 | -9.46 | 1.29 | 0.996 |
| 277.70 | 100.00 | 5.51 | 2.38 | 3.13 | 5.51 | 0.44 | 60.22 | -9.27 | 1.29 | 0.996 |
| 277.80 | 100.00 | 5.52 | 2.38 | 3.14 | 5.52 | 0.44 | 60.28 | -9.08 | 1.28 | 0.996 |
| 277.90 | 100.00 | 5.52 | 2.37 | 3.15 | 5.52 | 0.44 | 60.35 | -8.89 | 1.28 | 0.996 |
| 278.00 | 100.00 | 5.53 | 2.37 | 3.16 | 5.53 | 0.44 | 60.42 | -8.71 | 1.28 | 0.996 |
| 278.10 | 100.00 | 5.54 | 2.37 | 3.17 | 5.54 | 0.43 | 60.48 | -8.52 | 1.28 | 0.996 |
| 278.20 | 100.00 | 5.54 | 2.36 | 3.18 | 5.54 | 0.43 | 60.55 | -8.33 | 1.28 | 0.996 |
| 278.30 | 100.00 | 5.55 | 2.36 | 3.19 | 5.55 | 0.43 | 60.61 | -8.15 | 1.27 | 0.996 |
| 278.40 | 100.00 | 5.55 | 2.36 | 3.19 | 5.55 | 0.43 | 60.68 | -7.96 | 1.27 | 0.996 |
| 278.50 | 100.00 | 5.56 | 2.35 | 3.21 | 5.56 | 0.42 | 60.74 | -7.77 | 1.27 | 0.996 |
| 278.60 | 100.00 | 5.57 | 2.35 | 3.22 | 5.57 | 0.42 | 60.81 | -7.59 | 1.27 | 0.996 |
| 278.70 | 100.00 | 5.57 | 2.35 | 3.22 | 5.57 | 0.42 | 60.87 | -7.40 | 1.27 | 0.996 |
| 278.80 | 100.00 | 5.58 | 2.34 | 3.24 | 5.58 | 0.42 | 60.94 | -7.21 | 1.27 | 0.996 |
| 278.90 | 100.00 | 5.59 | 2.34 | 3.25 | 5.59 | 0.42 | 61.00 | -7.03 | 1.27 | 0.996 |
| 279.00 | 100.00 | 5.59 | 2.34 | 3.25 | 5.59 | 0.41 | 61.07 | -6.84 | 1.26 | 0.996 |
| 279.10 | 100.00 | 5.60 | 2.33 | 3.27 | 5.60 | 0.41 | 61.13 | -6.65 | 1.26 | 0.996 |
| 279.20 | 100.00 | 5.61 | 2.33 | 3.28 | 5.61 | 0.41 | 61.20 | -6.47 | 1.26 | 0.996 |
| 279.30 | 100.00 | 5.61 | 2.33 | 3.28 | 5.61 | 0.41 | 61.26 | -6.28 | 1.26 | 0.996 |
| 279.40 | 100.00 | 5.62 | 2.32 | 3.30 | 5.62 | 0.41 | 61.33 | -6.10 | 1.26 | 0.996 |
| 279.50 | 100.00 | 5.63 | 2.32 | 3.31 | 5.63 | 0.40 | 61.40 | -5.91 | 1.26 | 0.996 |
| 279.60 | 100.00 | 5.63 | 2.32 | 3.31 | 5.63 | 0.40 | 61.46 | -5.72 | 1.26 | 0.996 |
| 279.70 | 100.00 | 5.64 | 2.31 | 3.33 | 5.64 | 0.40 | 61.53 | -5.54 | 1.26 | 0.996 |
| 279.80 | 100.00 | 5.65 | 2.31 | 3.34 | 5.65 | 0.40 | 61.59 | -5.35 | 1.26 | 0.996 |
| 279.90 | 100.00 | 5.65 | 2.30 | 3.35 | 5.65 | 0.40 | 61.66 | -5.17 | 1.26 | 0.996 |
| 280.00 | 100.00 | 5.66 | 2.30 | 3.36 | 5.66 | 0.39 | 61.72 | -4.98 | 1.26 | 0.996 |
| 280.10 | 100.00 | 5.67 | 2.30 | 3.37 | 5.67 | 0.39 | 61.79 | -4.79 | 1.26 | 0.996 |
| 280.20 | 100.00 | 5.67 | 2.29 | 3.38 | 5.67 | 0.39 | 61.85 | -4.61 | 1.26 | 0.996 |
| 280.30 | 100.00 | 5.68 | 2.29 | 3.39 | 5.68 | 0.39 | 61.92 | -4.42 | 1.26 | 0.996 |
| 280.40 | 100.00 | 5.69 | 2.29 | 3.40 | 5.69 | 0.38 | 61.99 | -4.24 | 1.26 | 0.996 |
| 280.50 | 100.00 | 5.69 | 2.28 | 3.41 | 5.69 | 0.38 | 62.05 | -4.05 | 1.26 | 0.996 |
| 280.60 | 100.00 | 5.70 | 2.28 | 3.42 | 5.70 | 0.38 | 62.12 | -3.87 | 1.26 | 0.996 |
| 280.70 | 100.00 | 5.71 | 2.27 | 3.44 | 5.71 | 0.38 | 62.18 | -3.68 | 1.26 | 0.996 |
| 280.80 | 100.00 | 5.71 | 2.27 | 3.44 | 5.71 | 0.38 | 62.25 | -3.50 | 1.26 | 0.996 |
| 280.90 | 100.00 | 5.72 | 2.27 | 3.45 | 5.72 | 0.37 | 62.31 | -3.31 | 1.26 | 0.996 |
| 281.00 | 100.00 | 5.73 | 2.26 | 3.47 | 5.73 | 0.37 | 62.38 | -3.13 | 1.26 | 0.996 |
| 281.10 | 100.00 | 5.73 | 2.26 | 3.47 | 5.73 | 0.37 | 62.45 | -2.94 | 1.26 | 0.996 |
| 281.20 | 100.00 | 5.74 | 2.26 | 3.48 | 5.74 | 0.37 | 62.51 | -2.76 | 1.26 | 0.996 |
| 281.30 | 100.00 | 5.75 | 2.25 | 3.50 | 5.75 | 0.37 | 62.58 | -2.57 | 1.26 | 0.996 |
| 281.40 | 100.00 | 5.75 | 2.25 | 3.50 | 5.75 | 0.36 | 62.64 | -2.39 | 1.26 | 0.996 |
| 281.50 | 100.00 | 5.76 | 2.24 | 3.52 | 5.76 | 0.36 | 62.71 | -2.20 | 1.26 | 0.996 |
| 281.60 | 100.00 | 5.77 | 2.24 | 3.53 | 5.77 | 0.36 | 62.77 | -2.02 | 1.26 | 0.996 |
| 281.70 | 100.00 | 5.78 | 2.23 | 3.55 | 5.78 | 0.36 | 62.84 | -1.83 | 1.26 | 0.996 |
| 281.80 | 100.00 | 5.78 | 2.23 | 3.55 | 5.78 | 0.36 | 62.90 | -1.65 | 1.26 | 0.996 |
| 281.90 | 100.00 | 5.79 | 2.23 | 3.56 | 5.79 | 0.35 | 62.97 | -1.47 | 1.26 | 0.996 |
| 282.00 | 100.00 | 5.80 | 2.22 | 3.58 | 5.80 | 0.35 | 63.04 | -1.28 | 1.26 | 0.996 |
| 282.10 | 100.00 | 5.80 | 2.22 | 3.58 | 5.80 | 0.35 | 63.10 | -1.10 | 1.26 | 0.996 |
| 282.20 | 100.00 | 5.81 | 2.21 | 3.60 | 5.81 | 0.35 | 63.17 | -0.91 | 1.26 | 0.996 |
| 282.30 | 100.00 | 5.82 | 2.21 | 3.61 | 5.82 | 0.35 | 63.23 | -0.73 | 1.27 | 0.996 |
| 282.40 | 100.00 | 5.82 | 2.21 | 3.61 | 5.82 | 0.34 | 63.30 | -0.55 | 1.27 | 0.996 |
| 282.50 | 100.00 | 5.83 | 2.20 | 3.63 | 5.83 | 0.34 | 63.36 | -0.36 | 1.27 | 0.996 |
| 282.60 | 100.00 | 5.84 | 2.20 | 3.64 | 5.84 | 0.34 | 63.43 | -0.18 | 1.27 | 0.996 |
| 282.70 | 100.00 | 5.84 | 2.19 | 3.65 | 5.84 | 0.34 | 63.50 | 0.01 | 1.27 | 0.996 |
| 282.80 | 100.00 | 5.85 | 2.19 | 3.66 | 5.85 | 0.34 | 63.56 | 0.19 | 1.27 | 0.996 |
| 282.90 | 100.00 | 5.86 | 2.18 | 3.68 | 5.86 | 0.33 | 63.63 | 0.37 | 1.27 | 0.996 |
| 283.00 | 100.00 | 5.87 | 2.18 | 3.69 | 5.87 | 0.33 | 63.69 | 0.56 | 1.27 | 0.996 |
| 283.10 | 100.00 | 5.87 | 2.18 | 3.69 | 5.87 | 0.33 | 63.76 | 0.74 | 1.28 | 0.996 |
| 283.20 | 100.00 | 5.88 | 2.17 | 3.71 | 5.88 | 0.33 | 63.82 | 0.92 | 1.28 | 0.996 |
| 283.30 | 100.00 | 5.89 | 2.17 | 3.72 | 5.89 | 0.33 | 63.89 | 1.11 | 1.28 | 0.996 |
| 283.40 | 100.00 | 5.89 | 2.16 | 3.73 | 5.89 | 0.32 | 63.96 | 1.29 | 1.28 | 0.996 |
| 283.50 | 100.00 | 5.90 | 2.16 | 3.74 | 5.90 | 0.32 | 64.02 | 1.48 | 1.28 | 0.996 |
| 283.60 | 100.00 | 5.91 | 2.15 | 3.76 | 5.91 | 0.32 | 64.09 | 1.66 | 1.28 | 0.996 |
| 283.70 | 100.00 | 5.91 | 2.15 | 3.76 | 5.91 | 0.32 | 64.15 | 1.84 | 1.29 | 0.996 |
| 283.80 | 100.00 | 5.92 | 2.14 | 3.78 | 5.92 | 0.32 | 64.22 | 2.02 | 1.29 | 0.996 |
| 283.90 | 100.00 | 5.93 | 2.14 | 3.79 | 5.93 | 0.31 | 64.28 | 2.21 | 1.29 | 0.996 |
| 284.00 | 100.00 | 5.94 | 2.13 | 3.81 | 5.94 | 0.31 | 64.35 | 2.39 | 1.29 | 0.996 |
| 284.10 | 100.00 | 5.94 | 2.13 | 3.81 | 5.94 | 0.31 | 64.42 | 2.57 | 1.29 | 0.996 |
| 284.20 | 100.00 | 5.95 | 2.13 | 3.82 | 5.95 | 0.31 | 64.48 | 2.76 | 1.30 | 0.996 |
| 284.30 | 100.00 | 5.96 | 2.12 | 3.84 | 5.96 | 0.31 | 64.55 | 2.94 | 1.30 | 0.996 |
| 284.40 | 100.00 | 5.96 | 2.12 | 3.84 | 5.96 | 0.30 | 64.61 | 3.12 | 1.30 | 0.996 |
| 284.50 | 100.00 | 5.97 | 2.11 | 3.86 | 5.97 | 0.30 | 64.68 | 3.31 | 1.30 | 0.996 |
| 284.60 | 100.00 | 5.98 | 2.11 | 3.87 | 5.98 | 0.30 | 64.74 | 3.49 | 1.30 | 0.996 |
| 284.70 | 100.00 | 5.99 | 2.10 | 3.89 | 5.99 | 0.30 | 64.81 | 3.67 | 1.31 | 0.996 |
| 284.80 | 100.00 | 5.99 | 2.10 | 3.89 | 5.99 | 0.30 | 64.87 | 3.85 | 1.31 | 0.996 |
| 284.90 | 100.00 | 6.00 | 2.09 | 3.91 | 6.00 | 0.30 | 64.94 | 4.04 | 1.31 | 0.996 |
| 285.00 | 100.00 | 6.01 | 2.09 | 3.92 | 6.01 | 0.29 | 65.01 | 4.22 | 1.31 | 0.996 |
| 285.10 | 100.00 | 6.02 | 2.08 | 3.94 | 6.02 | 0.29 | 65.07 | 4.40 | 1.32 | 0.996 |
| 285.20 | 100.00 | 6.02 | 2.08 | 3.94 | 6.02 | 0.29 | 65.14 | 4.58 | 1.32 | 0.996 |
| 285.30 | 100.00 | 6.03 | 2.07 | 3.96 | 6.03 | 0.29 | 65.20 | 4.77 | 1.32 | 0.996 |
| 285.40 | 100.00 | 6.04 | 2.07 | 3.97 | 6.04 | 0.29 | 65.27 | 4.95 | 1.32 | 0.996 |
| 285.50 | 100.00 | 6.04 | 2.06 | 3.98 | 6.04 | 0.28 | 65.33 | 5.13 | 1.33 | 0.996 |
| 285.60 | 100.00 | 6.05 | 2.06 | 3.99 | 6.05 | 0.28 | 65.40 | 5.31 | 1.33 | 0.996 |
| 285.70 | 100.00 | 6.06 | 2.05 | 4.01 | 6.06 | 0.28 | 65.46 | 5.50 | 1.33 | 0.996 |
| 285.80 | 100.00 | 6.07 | 2.05 | 4.02 | 6.07 | 0.28 | 65.53 | 5.68 | 1.33 | 0.996 |
| 285.90 | 100.00 | 6.07 | 2.04 | 4.03 | 6.07 | 0.28 | 65.59 | 5.86 | 1.34 | 0.996 |
| 286.00 | 100.00 | 6.08 | 2.04 | 4.04 | 6.08 | 0.28 | 65.66 | 6.04 | 1.34 | 0.996 |
| 286.10 | 100.00 | 6.09 | 2.03 | 4.06 | 6.09 | 0.27 | 65.72 | 6.22 | 1.34 | 0.996 |
| 286.20 | 100.00 | 6.10 | 2.03 | 4.07 | 6.10 | 0.27 | 65.79 | 6.41 | 1.34 | 0.996 |
| 286.30 | 100.00 | 6.10 | 2.02 | 4.08 | 6.10 | 0.27 | 65.86 | 6.59 | 1.35 | 0.996 |
| 286.40 | 100.00 | 6.11 | 2.02 | 4.09 | 6.11 | 0.27 | 65.92 | 6.77 | 1.35 | 0.996 |
| 286.50 | 100.00 | 6.12 | 2.01 | 4.11 | 6.12 | 0.27 | 65.99 | 6.95 | 1.35 | 0.996 |
| 286.60 | 100.00 | 6.13 | 2.01 | 4.12 | 6.13 | 0.26 | 66.05 | 7.13 | 1.36 | 0.996 |
| 286.70 | 100.00 | 6.13 | 2.00 | 4.13 | 6.13 | 0.26 | 66.12 | 7.32 | 1.36 | 0.996 |
| 286.80 | 100.00 | 6.14 | 2.00 | 4.14 | 6.14 | 0.26 | 66.18 | 7.50 | 1.36 | 0.996 |
| 286.90 | 100.00 | 6.15 | 1.99 | 4.16 | 6.15 | 0.26 | 66.25 | 7.68 | 1.36 | 0.996 |
| 287.00 | 100.00 | 6.15 | 1.99 | 4.16 | 6.15 | 0.26 | 66.31 | 7.86 | 1.37 | 0.996 |
| 287.10 | 100.00 | 6.16 | 1.98 | 4.18 | 6.16 | 0.26 | 66.38 | 8.04 | 1.37 | 0.996 |
| 287.20 | 100.00 | 6.17 | 1.97 | 4.20 | 6.17 | 0.25 | 66.44 | 8.22 | 1.37 | 0.996 |
| 287.30 | 100.00 | 6.18 | 1.97 | 4.21 | 6.18 | 0.25 | 66.51 | 8.41 | 1.38 | 0.996 |
| 287.40 | 100.00 | 6.18 | 1.96 | 4.22 | 6.18 | 0.25 | 66.57 | 8.59 | 1.38 | 0.996 |
| 287.50 | 100.00 | 6.19 | 1.96 | 4.23 | 6.19 | 0.25 | 66.64 | 8.77 | 1.38 | 0.996 |
| 287.60 | 100.00 | 6.20 | 1.95 | 4.25 | 6.20 | 0.25 | 66.70 | 8.95 | 1.39 | 0.996 |
| 287.70 | 100.00 | 6.21 | 1.95 | 4.26 | 6.21 | 0.24 | 66.77 | 9.13 | 1.39 | 0.996 |
| 287.80 | 100.00 | 6.21 | 1.94 | 4.27 | 6.21 | 0.24 | 66.83 | 9.31 | 1.39 | 0.996 |
| 287.90 | 100.00 | 6.22 | 1.94 | 4.28 | 6.22 | 0.24 | 66.90 | 9.50 | 1.40 | 0.996 |
| 288.00 | 100.00 | 6.23 | 1.93 | 4.30 | 6.23 | 0.24 | 66.96 | 9.68 | 1.40 | 0.996 |
| 288.10 | 100.00 | 6.24 | 1.93 | 4.31 | 6.24 | 0.24 | 67.03 | 9.86 | 1.40 | 0.996 |
| 288.20 | 100.00 | 6.25 | 1.92 | 4.33 | 6.25 | 0.24 | 67.09 | 10.04 | 1.41 | 0.996 |
| 288.30 | 100.00 | 6.25 | 1.91 | 4.34 | 6.25 | 0.23 | 67.16 | 10.22 | 1.41 | 0.996 |
| 288.40 | 100.00 | 6.26 | 1.91 | 4.35 | 6.26 | 0.23 | 67.22 | 10.40 | 1.41 | 0.996 |
| 288.50 | 100.00 | 6.27 | 1.90 | 4.37 | 6.27 | 0.23 | 67.29 | 10.58 | 1.42 | 0.996 |
| 288.60 | 100.00 | 6.28 | 1.90 | 4.38 | 6.28 | 0.23 | 67.35 | 10.76 | 1.42 | 0.996 |
| 288.70 | 100.00 | 6.28 | 1.89 | 4.39 | 6.28 | 0.23 | 67.42 | 10.95 | 1.42 | 0.996 |
| 288.80 | 100.00 | 6.29 | 1.89 | 4.40 | 6.29 | 0.23 | 67.48 | 11.13 | 1.43 | 0.996 |
| 288.90 | 100.00 | 6.30 | 1.88 | 4.42 | 6.30 | 0.22 | 67.54 | 11.31 | 1.43 | 0.996 |
| 289.00 | 100.00 | 6.31 | 1.87 | 4.44 | 6.31 | 0.22 | 67.61 | 11.49 | 1.43 | 0.996 |
| 289.10 | 100.00 | 6.31 | 1.87 | 4.44 | 6.31 | 0.22 | 67.67 | 11.67 | 1.44 | 0.996 |
| 289.20 | 100.00 | 6.32 | 1.86 | 4.46 | 6.32 | 0.22 | 67.74 | 11.85 | 1.44 | 0.996 |
| 289.30 | 100.00 | 6.33 | 1.86 | 4.47 | 6.33 | 0.22 | 67.80 | 12.03 | 1.44 | 0.996 |
| 289.40 | 100.00 | 6.34 | 1.85 | 4.49 | 6.34 | 0.22 | 67.87 | 12.21 | 1.45 | 0.996 |
| 289.50 | 100.00 | 6.34 | 1.85 | 4.49 | 6.34 | 0.21 | 67.93 | 12.40 | 1.45 | 0.996 |
| 289.60 | 100.00 | 6.35 | 1.84 | 4.51 | 6.35 | 0.21 | 68.00 | 12.58 | 1.45 | 0.996 |
| 289.70 | 100.00 | 6.36 | 1.83 | 4.53 | 6.36 | 0.21 | 68.06 | 12.76 | 1.46 | 0.996 |
| 289.80 | 100.00 | 6.37 | 1.83 | 4.54 | 6.37 | 0.21 | 68.12 | 12.94 | 1.46 | 0.996 |
| 289.90 | 100.00 | 6.38 | 1.82 | 4.56 | 6.38 | 0.21 | 68.19 | 13.12 | 1.46 | 0.996 |
| 290.00 | 100.00 | 6.38 | 1.82 | 4.56 | 6.38 | 0.21 | 68.25 | 13.30 | 1.47 | 0.996 |
| 290.10 | 100.00 | 6.39 | 1.81 | 4.58 | 6.39 | 0.21 | 68.32 | 13.48 | 1.47 | 0.996 |
| 290.20 | 100.00 | 6.40 | 1.80 | 4.60 | 6.40 | 0.20 | 68.38 | 13.66 | 1.48 | 0.996 |
| 290.30 | 100.00 | 6.41 | 1.80 | 4.61 | 6.41 | 0.20 | 68.45 | 13.84 | 1.48 | 0.996 |
| 290.40 | 100.00 | 6.41 | 1.79 | 4.62 | 6.41 | 0.20 | 68.51 | 14.03 | 1.48 | 0.996 |
| 290.50 | 100.00 | 6.42 | 1.79 | 4.63 | 6.42 | 0.20 | 68.57 | 14.21 | 1.49 | 0.996 |
| 290.60 | 100.00 | 6.43 | 1.78 | 4.65 | 6.43 | 0.20 | 68.64 | 14.39 | 1.49 | 0.996 |
| 290.70 | 100.00 | 6.44 | 1.78 | 4.66 | 6.44 | 0.20 | 68.70 | 14.57 | 1.49 | 0.996 |
| 290.80 | 100.00 | 6.45 | 1.77 | 4.68 | 6.45 | 0.19 | 68.77 | 14.75 | 1.50 | 0.996 |
| 290.90 | 100.00 | 6.45 | 1.76 | 4.69 | 6.45 | 0.18 | 68.83 | 14.93 | 1.50 | 0.996 |
| 291.00 | 100.00 | 6.46 | 1.76 | 4.70 | 6.46 | 0.18 | 68.89 | 15.11 | 1.51 | 0.996 |
| 291.10 | 100.00 | 6.47 | 1.75 | 4.72 | 6.47 | 0.17 | 68.96 | 15.29 | 1.51 | 0.996 |
| 291.20 | 100.00 | 6.48 | 1.75 | 4.73 | 6.48 | 0.17 | 69.02 | 15.47 | 1.51 | 0.996 |
| 291.30 | 100.00 | 6.49 | 1.74 | 4.75 | 6.49 | 0.16 | 69.09 | 15.66 | 1.52 | 0.996 |
| 291.40 | 100.00 | 6.49 | 1.73 | 4.76 | 6.49 | 0.15 | 69.15 | 15.84 | 1.52 | 0.996 |
| 291.50 | 100.00 | 6.50 | 1.73 | 4.77 | 6.50 | 0.15 | 69.21 | 16.02 | 1.53 | 0.996 |
| 291.60 | 100.00 | 6.51 | 1.72 | 4.79 | 6.51 | 0.14 | 69.28 | 16.20 | 1.53 | 0.996 |
| 291.70 | 100.00 | 6.52 | 1.71 | 4.81 | 6.52 | 0.14 | 69.34 | 16.38 | 1.53 | 0.996 |
| 291.80 | 100.00 | 6.53 | 1.71 | 4.82 | 6.53 | 0.13 | 69.40 | 16.56 | 1.54 | 0.996 |
| 291.90 | 100.00 | 6.53 | 1.70 | 4.83 | 6.53 | 0.12 | 69.47 | 16.74 | 1.54 | 0.996 |
| 292.00 | 100.00 | 6.54 | 1.70 | 4.84 | 6.54 | 0.12 | 69.53 | 16.92 | 1.54 | 0.996 |
| 292.10 | 100.00 | 6.55 | 1.69 | 4.86 | 6.55 | 0.11 | 69.59 | 17.11 | 1.55 | 0.996 |
| 292.20 | 100.00 | 6.56 | 1.68 | 4.88 | 6.56 | 0.10 | 69.66 | 17.29 | 1.55 | 0.996 |
| 292.30 | 100.00 | 6.57 | 1.68 | 4.89 | 6.57 | 0.10 | 69.72 | 17.47 | 1.56 | 0.996 |
| 292.40 | 100.00 | 6.57 | 1.67 | 4.90 | 6.57 | 0.09 | 69.78 | 17.65 | 1.56 | 0.996 |
| 292.50 | 100.00 | 6.58 | 1.67 | 4.91 | 6.58 | 0.09 | 69.85 | 17.83 | 1.56 | 0.996 |
| 292.60 | 100.00 | 6.59 | 1.66 | 4.93 | 6.59 | 0.08 | 69.91 | 18.01 | 1.57 | 0.996 |
| 292.70 | 100.00 | 6.60 | 1.65 | 4.95 | 6.60 | 0.07 | 69.97 | 18.19 | 1.57 | 0.996 |
| 292.80 | 100.00 | 6.61 | 1.65 | 4.96 | 6.61 | 0.07 | 70.04 | 18.37 | 1.58 | 0.996 |
| 292.90 | 100.00 | 6.61 | 1.64 | 4.97 | 6.61 | 0.06 | 70.10 | 18.56 | 1.58 | 0.996 |
| 293.00 | 100.00 | 6.62 | 1.63 | 4.99 | 6.62 | 0.05 | 70.16 | 18.74 | 1.59 | 0.996 |
| 293.10 | 100.00 | 6.63 | 1.63 | 5.00 | 6.63 | 0.05 | 70.23 | 18.92 | 1.59 | 0.996 |
| 293.20 | 100.00 | 6.64 | 1.62 | 5.02 | 6.64 | 0.04 | 70.29 | 19.10 | 1.59 | 0.996 |
| 293.30 | 100.00 | 6.65 | 1.62 | 5.03 | 6.65 | 0.03 | 70.35 | 19.28 | 1.60 | 0.996 |
| 293.40 | 100.00 | 6.65 | 1.61 | 5.04 | 6.65 | 0.03 | 70.41 | 19.46 | 1.60 | 0.996 |
| 293.50 | 100.00 | 6.66 | 1.60 | 5.06 | 6.66 | 0.02 | 70.48 | 19.65 | 1.61 | 0.996 |
| 293.60 | 100.00 | 6.67 | 1.60 | 5.07 | 6.67 | 0.01 | 70.54 | 19.83 | 1.61 | 0.996 |
| 293.70 | 100.00 | 6.68 | 1.59 | 5.09 | 6.68 | 0.01 | 70.60 | 20.01 | 1.61 | 0.996 |
| 293.80 | 100.00 | 6.69 | 1.59 | 5.10 | 6.69 | 0.00 | 70.67 | 20.19 | 1.62 | 0.996 |
| 293.90 | 100.00 | 6.70 | 1.58 | 5.12 | 6.70 | -0.01 | 70.73 | 20.37 | 1.62 | 0.996 |
| 294.00 | 100.00 | 6.70 | 1.57 | 5.13 | 6.70 | -0.01 | 70.79 | 20.56 | 1.63 | 0.996 |
| 294.10 | 100.00 | 6.71 | 1.57 | 5.14 | 6.71 | -0.02 | 70.85 | 20.74 | 1.63 | 0.996 |
| 294.20 | 100.00 | 6.72 | 1.56 | 5.16 | 6.72 | -0.03 | 70.92 | 20.92 | 1.64 | 0.996 |
| 294.30 | 100.00 | 6.73 | 1.55 | 5.18 | 6.73 | -0.03 | 70.98 | 21.10 | 1.64 | 0.996 |
| 294.40 | 100.00 | 6.74 | 1.55 | 5.19 | 6.74 | -0.04 | 71.04 | 21.28 | 1.64 | 0.996 |
| 294.50 | 100.00 | 6.75 | 1.54 | 5.21 | 6.75 | -0.05 | 71.10 | 21.47 | 1.65 | 0.996 |
| 294.60 | 100.00 | 6.75 | 1.54 | 5.21 | 6.75 | -0.05 | 71.17 | 21.65 | 1.65 | 0.996 |
| 294.70 | 100.00 | 6.76 | 1.53 | 5.23 | 6.76 | -0.06 | 71.23 | 21.83 | 1.66 | 0.996 |
| 294.80 | 100.00 | 6.77 | 1.52 | 5.25 | 6.77 | -0.07 | 71.29 | 22.01 | 1.66 | 0.996 |
| 294.90 | 100.00 | 6.78 | 1.52 | 5.26 | 6.78 | -0.08 | 71.35 | 22.20 | 1.67 | 0.996 |
| 295.00 | 100.00 | 6.79 | 1.51 | 5.28 | 6.79 | -0.08 | 71.42 | 22.38 | 1.67 | 0.996 |
| 295.10 | 100.00 | 6.80 | 1.50 | 5.30 | 6.80 | -0.09 | 71.48 | 22.56 | 1.67 | 0.996 |
| 295.20 | 100.00 | 6.80 | 1.50 | 5.30 | 6.80 | -0.10 | 71.54 | 22.74 | 1.68 | 0.996 |
| 295.30 | 100.00 | 6.81 | 1.49 | 5.32 | 6.81 | -0.10 | 71.60 | 22.93 | 1.68 | 0.996 |
| 295.40 | 100.00 | 6.82 | 1.49 | 5.33 | 6.82 | -0.11 | 71.66 | 23.11 | 1.69 | 0.996 |
| 295.50 | 100.00 | 6.83 | 1.48 | 5.35 | 6.83 | -0.12 | 71.73 | 23.29 | 1.69 | 0.996 |
| 295.60 | 100.00 | 6.84 | 1.47 | 5.37 | 6.84 | -0.13 | 71.79 | 23.48 | 1.70 | 0.996 |
| 295.70 | 100.00 | 6.85 | 1.47 | 5.38 | 6.85 | -0.13 | 71.85 | 23.66 | 1.70 | 0.996 |
| 295.80 | 100.00 | 6.85 | 1.46 | 5.39 | 6.85 | -0.14 | 71.91 | 23.84 | 1.71 | 0.996 |
| 295.90 | 100.00 | 6.86 | 1.46 | 5.40 | 6.86 | -0.15 | 71.97 | 24.03 | 1.71 | 0.996 |
| 296.00 | 100.00 | 6.87 | 1.45 | 5.42 | 6.87 | -0.16 | 72.04 | 24.21 | 1.71 | 0.996 |
| 296.10 | 100.00 | 6.88 | 1.44 | 5.44 | 6.88 | -0.16 | 72.10 | 24.39 | 1.72 | 0.996 |
| 296.20 | 100.00 | 6.89 | 1.44 | 5.45 | 6.89 | -0.17 | 72.16 | 24.58 | 1.72 | 0.996 |
| 296.30 | 100.00 | 6.90 | 1.43 | 5.47 | 6.90 | -0.18 | 72.22 | 24.76 | 1.73 | 0.996 |
| 296.40 | 100.00 | 6.91 | 1.42 | 5.49 | 6.91 | -0.19 | 72.28 | 24.94 | 1.73 | 0.996 |
| 296.50 | 100.00 | 6.91 | 1.42 | 5.49 | 6.91 | -0.19 | 72.34 | 25.13 | 1.74 | 0.996 |
| 296.60 | 100.00 | 6.92 | 1.41 | 5.51 | 6.92 | -0.20 | 72.41 | 25.31 | 1.74 | 0.996 |
| 296.70 | 100.00 | 6.93 | 1.41 | 5.52 | 6.93 | -0.21 | 72.47 | 25.50 | 1.75 | 0.996 |
| 296.80 | 100.00 | 6.94 | 1.40 | 5.54 | 6.94 | -0.22 | 72.53 | 25.68 | 1.75 | 0.996 |
| 296.90 | 100.00 | 6.95 | 1.40 | 5.55 | 6.95 | -0.23 | 72.59 | 25.86 | 1.76 | 0.996 |
| 297.00 | 100.00 | 6.96 | 1.39 | 5.57 | 6.96 | -0.23 | 72.65 | 26.05 | 1.76 | 0.996 |
| 297.10 | 100.00 | 6.97 | 1.38 | 5.59 | 6.97 | -0.24 | 72.71 | 26.23 | 1.76 | 0.996 |
| 297.20 | 100.00 | 6.97 | 1.38 | 5.59 | 6.97 | -0.25 | 72.77 | 26.42 | 1.77 | 0.996 |
| 297.30 | 100.00 | 6.98 | 1.37 | 5.61 | 6.98 | -0.26 | 72.84 | 26.60 | 1.77 | 0.996 |
| 297.40 | 100.00 | 6.99 | 1.37 | 5.62 | 6.99 | -0.27 | 72.90 | 26.79 | 1.78 | 0.996 |
| 297.50 | 100.00 | 7.00 | 1.36 | 5.64 | 7.00 | -0.27 | 72.96 | 26.97 | 1.78 | 0.996 |
| 297.60 | 100.00 | 7.01 | 1.35 | 5.66 | 7.01 | -0.28 | 73.02 | 27.16 | 1.79 | 0.996 |
| 297.70 | 100.00 | 7.02 | 1.35 | 5.67 | 7.02 | -0.29 | 73.08 | 27.34 | 1.79 | 0.996 |
| 297.80 | 100.00 | 7.03 | 1.34 | 5.69 | 7.03 | -0.30 | 73.14 | 27.53 | 1.80 | 0.996 |
| 297.90 | 100.00 | 7.03 | 1.34 | 5.69 | 7.03 | -0.31 | 73.20 | 27.71 | 1.80 | 0.996 |
| 298.00 | 100.00 | 7.04 | 1.33 | 5.71 | 7.04 | -0.32 | 73.27 | 27.90 | 1.81 | 0.996 |
| 298.10 | 100.00 | 7.05 | 1.33 | 5.72 | 7.05 | -0.32 | 73.33 | 28.09 | 1.81 | 0.996 |
| 298.20 | 100.00 | 7.06 | 1.32 | 5.74 | 7.06 | -0.33 | 73.39 | 28.27 | 1.82 | 0.996 |
| 298.30 | 100.00 | 7.07 | 1.31 | 5.76 | 7.07 | -0.34 | 73.45 | 28.46 | 1.82 | 0.996 |
| 298.40 | 100.00 | 7.08 | 1.31 | 5.77 | 7.08 | -0.35 | 73.51 | 28.64 | 1.83 | 0.996 |
| 298.50 | 100.00 | 7.09 | 1.30 | 5.79 | 7.09 | -0.36 | 73.57 | 28.83 | 1.83 | 0.996 |
| 298.60 | 100.00 | 7.10 | 1.30 | 5.80 | 7.10 | -0.37 | 73.63 | 29.02 | 1.84 | 0.996 |
| 298.70 | 100.00 | 7.10 | 1.29 | 5.81 | 7.10 | -0.38 | 73.69 | 29.20 | 1.84 | 0.996 |
| 298.80 | 100.00 | 7.11 | 1.29 | 5.82 | 7.11 | -0.38 | 73.76 | 29.39 | 1.85 | 0.996 |
| 298.90 | 100.00 | 7.12 | 1.28 | 5.84 | 7.12 | -0.39 | 73.82 | 29.58 | 1.85 | 0.996 |
| 299.00 | 100.00 | 7.13 | 1.28 | 5.85 | 7.13 | -0.40 | 73.88 | 29.76 | 1.86 | 0.996 |
| 299.10 | 100.00 | 7.14 | 1.27 | 5.87 | 7.14 | -0.41 | 73.94 | 29.95 | 1.86 | 0.996 |
| 299.20 | 100.00 | 7.15 | 1.27 | 5.88 | 7.15 | -0.42 | 74.00 | 30.14 | 1.86 | 0.996 |
| 299.30 | 100.00 | 7.16 | 1.26 | 5.90 | 7.16 | -0.43 | 74.06 | 30.33 | 1.87 | 0.996 |
| 299.40 | 100.00 | 7.17 | 1.26 | 5.91 | 7.17 | -0.44 | 74.12 | 30.51 | 1.87 | 0.996 |
| 299.50 | 100.00 | 7.17 | 1.25 | 5.92 | 7.17 | -0.45 | 74.18 | 30.70 | 1.88 | 0.996 |
| 299.60 | 100.00 | 7.18 | 1.25 | 5.93 | 7.18 | -0.46 | 74.24 | 30.89 | 1.88 | 0.996 |
| 299.70 | 100.00 | 7.19 | 1.24 | 5.95 | 7.19 | -0.47 | 74.30 | 31.08 | 1.89 | 0.996 |
| 299.80 | 100.00 | 7.20 | 1.24 | 5.96 | 7.20 | -0.47 | 74.37 | 31.27 | 1.89 | 0.996 |
| 299.90 | 100.00 | 7.21 | 1.23 | 5.98 | 7.21 | -0.48 | 74.43 | 31.45 | 1.90 | 0.996 |
| 300.00 | 100.00 | 7.22 | 1.23 | 5.99 | 7.22 | -0.49 | 74.49 | 31.64 | 1.90 | 0.996 |
| 300.10 | 100.00 | 7.23 | 1.22 | 6.01 | 7.23 | -0.50 | 74.55 | 31.83 | 1.91 | 0.996 |
| 300.20 | 100.00 | 7.24 | 1.22 | 6.02 | 7.24 | -0.51 | 74.61 | 32.02 | 1.92 | 0.996 |
| 300.30 | 100.00 | 7.25 | 1.21 | 6.04 | 7.25 | -0.52 | 74.67 | 32.21 | 1.92 | 0.996 |
| 300.40 | 100.00 | 7.26 | 1.21 | 6.05 | 7.26 | -0.53 | 74.73 | 32.40 | 1.93 | 0.996 |
| 300.50 | 100.00 | 7.26 | 1.20 | 6.06 | 7.26 | -0.54 | 74.79 | 32.59 | 1.93 | 0.996 |
| 300.60 | 100.00 | 7.27 | 1.20 | 6.07 | 7.27 | -0.55 | 74.85 | 32.78 | 1.94 | 0.996 |
| 300.70 | 100.00 | 7.28 | 1.19 | 6.09 | 7.28 | -0.56 | 74.92 | 32.97 | 1.94 | 0.996 |
| 300.80 | 100.00 | 7.29 | 1.19 | 6.10 | 7.29 | -0.57 | 74.98 | 33.16 | 1.95 | 0.996 |
| 300.90 | 100.00 | 7.30 | 1.18 | 6.12 | 7.30 | -0.58 | 75.04 | 33.35 | 1.95 | 0.996 |
| 301.00 | 100.00 | 7.31 | 1.18 | 6.13 | 7.31 | -0.59 | 75.10 | 33.54 | 1.96 | 0.996 |
| 301.10 | 100.00 | 7.32 | 1.18 | 6.14 | 7.32 | -0.60 | 75.16 | 33.73 | 1.96 | 0.996 |
| 301.20 | 100.00 | 7.33 | 1.17 | 6.16 | 7.33 | -0.61 | 75.22 | 33.92 | 1.97 | 0.996 |
| 301.30 | 100.00 | 7.34 | 1.17 | 6.17 | 7.34 | -0.62 | 75.28 | 34.11 | 1.97 | 0.996 |
| 301.40 | 100.00 | 7.35 | 1.16 | 6.19 | 7.35 | -0.63 | 75.34 | 34.31 | 1.98 | 0.996 |
| 301.50 | 100.00 | 7.35 | 1.16 | 6.19 | 7.35 | -0.64 | 75.41 | 34.50 | 1.98 | 0.996 |
| 301.60 | 100.00 | 7.36 | 1.16 | 6.20 | 7.36 | -0.65 | 75.47 | 34.69 | 1.99 | 0.996 |
| 301.70 | 100.00 | 7.37 | 1.15 | 6.22 | 7.37 | -0.66 | 75.53 | 34.88 | 1.99 | 0.996 |
| 301.80 | 100.00 | 7.38 | 1.15 | 6.23 | 7.38 | -0.67 | 75.59 | 35.07 | 2.00 | 0.996 |
| 301.90 | 100.00 | 7.39 | 1.14 | 6.25 | 7.39 | -0.68 | 75.65 | 35.27 | 2.00 | 0.996 |
| 302.00 | 100.00 | 7.40 | 1.14 | 6.26 | 7.40 | -0.69 | 75.71 | 35.46 | 2.01 | 0.996 |
| 302.10 | 100.00 | 7.41 | 1.14 | 6.27 | 7.41 | -0.71 | 75.77 | 35.65 | 2.01 | 0.996 |
| 302.20 | 100.00 | 7.42 | 1.13 | 6.29 | 7.42 | -0.72 | 75.84 | 35.85 | 2.02 | 0.996 |
| 302.30 | 100.00 | 7.43 | 1.13 | 6.30 | 7.43 | -0.73 | 75.90 | 36.04 | 2.03 | 0.996 |
| 302.40 | 100.00 | 7.44 | 1.13 | 6.31 | 7.44 | -0.74 | 75.96 | 36.23 | 2.03 | 0.996 |
| 302.50 | 100.00 | 7.45 | 1.12 | 6.33 | 7.45 | -0.75 | 76.02 | 36.43 | 2.04 | 0.996 |
| 302.60 | 100.00 | 7.45 | 1.12 | 6.33 | 7.45 | -0.76 | 76.08 | 36.62 | 2.04 | 0.996 |
| 302.70 | 100.00 | 7.46 | 1.12 | 6.34 | 7.46 | -0.77 | 76.14 | 36.82 | 2.05 | 0.996 |
| 302.80 | 100.00 | 7.47 | 1.12 | 6.35 | 7.47 | -0.78 | 76.21 | 37.01 | 2.05 | 0.996 |
| 302.90 | 100.00 | 7.48 | 1.11 | 6.37 | 7.48 | -0.79 | 76.27 | 37.21 | 2.06 | 0.996 |
| 303.00 | 100.00 | 7.49 | 1.11 | 6.38 | 7.49 | -0.81 | 76.33 | 37.40 | 2.06 | 0.996 |
| 303.10 | 100.00 | 7.50 | 1.11 | 6.39 | 7.50 | -0.82 | 76.39 | 37.60 | 2.07 | 0.996 |
| 303.20 | 100.00 | 7.51 | 1.10 | 6.41 | 7.51 | -0.83 | 76.45 | 37.79 | 2.07 | 0.996 |
| 303.30 | 100.00 | 7.52 | 1.10 | 6.42 | 7.52 | -0.84 | 76.52 | 37.99 | 2.08 | 0.996 |
| 303.40 | 100.00 | 7.53 | 1.10 | 6.43 | 7.53 | -0.85 | 76.58 | 38.19 | 2.09 | 0.996 |
| 303.50 | 100.00 | 7.54 | 1.10 | 6.44 | 7.54 | -0.86 | 76.64 | 38.38 | 2.09 | 0.996 |
| 303.60 | 100.00 | 7.55 | 1.10 | 6.45 | 7.55 | -0.87 | 76.70 | 38.58 | 2.10 | 0.996 |
| 303.70 | 100.00 | 7.56 | 1.09 | 6.47 | 7.56 | -0.89 | 76.77 | 38.78 | 2.10 | 0.996 |
| 303.80 | 100.00 | 7.57 | 1.09 | 6.48 | 7.57 | -0.90 | 76.83 | 38.97 | 2.11 | 0.996 |
| 303.90 | 100.00 | 7.58 | 1.09 | 6.49 | 7.58 | -0.91 | 76.89 | 39.17 | 2.11 | 0.996 |
| 304.00 | 100.00 | 7.58 | 1.09 | 6.49 | 7.58 | -0.92 | 76.96 | 39.37 | 2.12 | 0.996 |
| 304.10 | 100.00 | 7.59 | 1.09 | 6.50 | 7.59 | -0.94 | 77.02 | 39.57 | 2.13 | 0.996 |
| 304.20 | 100.00 | 7.60 | 1.08 | 6.52 | 7.60 | -0.95 | 77.08 | 39.77 | 2.13 | 0.996 |
| 304.30 | 100.00 | 7.61 | 1.08 | 6.53 | 7.61 | -0.96 | 77.14 | 39.97 | 2.14 | 0.996 |
| 304.40 | 100.00 | 7.62 | 1.08 | 6.54 | 7.62 | -0.97 | 77.21 | 40.16 | 2.14 | 0.996 |
| 304.50 | 100.00 | 7.63 | 1.08 | 6.55 | 7.63 | -0.98 | 77.27 | 40.36 | 2.15 | 0.996 |
| 304.60 | 100.00 | 7.64 | 1.08 | 6.56 | 7.64 | -1.00 | 77.34 | 40.56 | 2.15 | 0.996 |
| 304.70 | 100.00 | 7.65 | 1.08 | 6.57 | 7.65 | -1.01 | 77.40 | 40.76 | 2.16 | 0.996 |
| 304.80 | 100.00 | 7.66 | 1.08 | 6.58 | 7.66 | -1.02 | 77.46 | 40.96 | 2.17 | 0.996 |
| 304.90 | 100.00 | 7.67 | 1.08 | 6.59 | 7.67 | -1.04 | 77.53 | 41.17 | 2.17 | 0.996 |
| 305.00 | 100.00 | 7.68 | 1.08 | 6.60 | 7.68 | -1.05 | 77.59 | 41.37 | 2.18 | 0.996 |
| 305.10 | 100.00 | 7.69 | 1.08 | 6.61 | 7.69 | -1.06 | 77.66 | 41.57 | 2.18 | 0.996 |
| 305.20 | 100.00 | 7.70 | 1.08 | 6.62 | 7.70 | -1.08 | 77.72 | 41.77 | 2.19 | 0.995 |
| 305.30 | 100.00 | 7.71 | 1.08 | 6.63 | 7.71 | -1.09 | 77.78 | 41.97 | 2.20 | 0.995 |
| 305.40 | 100.00 | 7.72 | 1.08 | 6.64 | 7.72 | -1.10 | 77.85 | 42.17 | 2.20 | 0.995 |
| 305.50 | 100.00 | 7.73 | 1.08 | 6.65 | 7.73 | -1.12 | 77.91 | 42.38 | 2.21 | 0.995 |
| 305.60 | 100.00 | 7.74 | 1.08 | 6.66 | 7.74 | -1.13 | 77.98 | 42.58 | 2.21 | 0.995 |
| 305.70 | 100.00 | 7.74 | 1.08 | 6.66 | 7.74 | -1.14 | 78.04 | 42.78 | 2.22 | 0.995 |
| 305.80 | 100.00 | 7.75 | 1.08 | 6.67 | 7.75 | -1.16 | 78.11 | 42.99 | 2.23 | 0.995 |
| 305.90 | 100.00 | 7.76 | 1.08 | 6.68 | 7.76 | -1.17 | 78.17 | 43.19 | 2.23 | 0.995 |
| 306.00 | 100.00 | 7.77 | 1.08 | 6.69 | 7.77 | -1.18 | 78.24 | 43.39 | 2.24 | 0.995 |
| 306.10 | 100.00 | 7.78 | 1.08 | 6.70 | 7.78 | -1.20 | 78.31 | 43.60 | 2.24 | 0.995 |
| 306.20 | 100.00 | 7.79 | 1.08 | 6.71 | 7.79 | -1.21 | 78.37 | 43.80 | 2.25 | 0.995 |
| 306.30 | 100.00 | 7.80 | 1.08 | 6.72 | 7.80 | -1.23 | 78.44 | 44.01 | 2.26 | 0.995 |
| 306.40 | 100.00 | 7.81 | 1.08 | 6.73 | 7.81 | -1.24 | 78.51 | 44.21 | 2.26 | 0.995 |
| 306.50 | 100.00 | 7.82 | 1.08 | 6.74 | 7.82 | -1.26 | 78.57 | 44.42 | 2.27 | 0.995 |
| 306.60 | 100.00 | 7.83 | 1.09 | 6.74 | 7.83 | -1.27 | 78.64 | 44.63 | 2.28 | 0.995 |
| 306.70 | 100.00 | 7.84 | 1.09 | 6.75 | 7.84 | -1.28 | 78.71 | 44.83 | 2.28 | 0.995 |
| 306.80 | 100.00 | 7.85 | 1.09 | 6.76 | 7.85 | -1.30 | 78.77 | 45.04 | 2.29 | 0.995 |
| 306.90 | 100.00 | 7.86 | 1.09 | 6.77 | 7.86 | -1.31 | 78.84 | 45.25 | 2.29 | 0.995 |
| 307.00 | 100.00 | 7.87 | 1.09 | 6.78 | 7.87 | -1.33 | 78.91 | 45.46 | 2.30 | 0.995 |
| 307.10 | 100.00 | 7.88 | 1.10 | 6.78 | 7.88 | -1.34 | 78.98 | 45.66 | 2.31 | 0.995 |
| 307.20 | 100.00 | 7.89 | 1.10 | 6.79 | 7.89 | -1.36 | 79.04 | 45.87 | 2.31 | 0.995 |
| 307.30 | 100.00 | 7.90 | 1.10 | 6.80 | 7.90 | -1.37 | 79.11 | 46.08 | 2.32 | 0.995 |
| 307.40 | 100.00 | 7.91 | 1.10 | 6.81 | 7.91 | -1.39 | 79.18 | 46.29 | 2.33 | 0.995 |
| 307.50 | 100.00 | 7.92 | 1.11 | 6.81 | 7.92 | -1.41 | 79.25 | 46.50 | 2.33 | 0.995 |
| 307.60 | 100.00 | 7.93 | 1.11 | 6.82 | 7.93 | -1.42 | 79.32 | 46.71 | 2.34 | 0.995 |
| 307.70 | 100.00 | 7.94 | 1.11 | 6.83 | 7.94 | -1.44 | 79.39 | 46.92 | 2.35 | 0.995 |
| 307.80 | 100.00 | 7.95 | 1.12 | 6.83 | 7.95 | -1.45 | 79.46 | 47.13 | 2.35 | 0.995 |
| 307.90 | 100.00 | 7.96 | 1.12 | 6.84 | 7.96 | -1.47 | 79.53 | 47.34 | 2.36 | 0.995 |
| 308.00 | 100.00 | 7.97 | 1.12 | 6.85 | 7.97 | -1.49 | 79.60 | 47.55 | 2.37 | 0.995 |
| 308.10 | 100.00 | 7.98 | 1.13 | 6.85 | 7.98 | -1.50 | 79.67 | 47.77 | 2.37 | 0.995 |
| 308.20 | 100.00 | 7.99 | 1.13 | 6.86 | 7.99 | -1.52 | 79.74 | 47.98 | 2.38 | 0.995 |
| 308.30 | 100.00 | 8.00 | 1.14 | 6.86 | 8.00 | -1.53 | 79.81 | 48.19 | 2.39 | 0.995 |
| 308.40 | 100.00 | 8.00 | 1.14 | 6.86 | 8.00 | -1.55 | 79.89 | 48.40 | 2.39 | 0.995 |
| 308.50 | 100.00 | 8.01 | 1.15 | 6.86 | 8.01 | -1.57 | 79.96 | 48.62 | 2.40 | 0.995 |
| 308.60 | 100.00 | 8.02 | 1.15 | 6.87 | 8.02 | -1.59 | 80.03 | 48.83 | 2.41 | 0.995 |
| 308.70 | 100.00 | 8.03 | 1.15 | 6.88 | 8.03 | -1.60 | 80.10 | 49.04 | 2.41 | 0.995 |
| 308.80 | 100.00 | 8.04 | 1.16 | 6.88 | 8.04 | -1.62 | 80.17 | 49.26 | 2.42 | 0.995 |
| 308.90 | 100.00 | 8.05 | 1.17 | 6.88 | 8.05 | -1.64 | 80.25 | 49.47 | 2.43 | 0.995 |
| 309.00 | 100.00 | 8.06 | 1.17 | 6.89 | 8.06 | -1.65 | 80.32 | 49.69 | 2.43 | 0.995 |
| 309.10 | 100.00 | 8.07 | 1.18 | 6.89 | 8.07 | -1.67 | 80.40 | 49.91 | 2.44 | 0.995 |
| 309.20 | 100.00 | 8.08 | 1.18 | 6.90 | 8.08 | -1.69 | 80.47 | 50.12 | 2.45 | 0.995 |
| 309.30 | 100.00 | 8.09 | 1.19 | 6.90 | 8.09 | -1.71 | 80.54 | 50.34 | 2.45 | 0.995 |
| 309.40 | 100.00 | 8.10 | 1.19 | 6.91 | 8.10 | -1.73 | 80.62 | 50.56 | 2.46 | 0.995 |
| 309.50 | 100.00 | 8.11 | 1.20 | 6.91 | 8.11 | -1.74 | 80.70 | 50.77 | 2.47 | 0.995 |
| 309.60 | 100.00 | 8.12 | 1.21 | 6.91 | 8.12 | -1.76 | 80.77 | 50.99 | 2.47 | 0.995 |
| 309.70 | 100.00 | 8.13 | 1.21 | 6.92 | 8.13 | -1.78 | 80.85 | 51.21 | 2.48 | 0.995 |
| 309.80 | 100.00 | 8.14 | 1.22 | 6.92 | 8.14 | -1.80 | 80.92 | 51.43 | 2.49 | 0.995 |
| 309.90 | 100.00 | 8.15 | 1.23 | 6.92 | 8.15 | -1.82 | 81.00 | 51.65 | 2.50 | 0.995 |
| 310.00 | 100.00 | 8.16 | 1.24 | 6.92 | 8.16 | -1.84 | 81.08 | 51.87 | 2.50 | 0.995 |
| 310.10 | 100.00 | 8.17 | 1.24 | 6.93 | 8.17 | -1.86 | 81.16 | 52.09 | 2.51 | 0.995 |
| 310.20 | 100.00 | 8.18 | 1.25 | 6.93 | 8.18 | -1.87 | 81.24 | 52.31 | 2.52 | 0.995 |
| 310.30 | 100.00 | 8.19 | 1.26 | 6.93 | 8.19 | -1.89 | 81.31 | 52.53 | 2.52 | 0.995 |
| 310.40 | 100.00 | 8.20 | 1.27 | 6.93 | 8.20 | -1.91 | 81.39 | 52.75 | 2.53 | 0.995 |
| 310.50 | 100.00 | 8.21 | 1.28 | 6.93 | 8.21 | -1.93 | 81.47 | 52.97 | 2.54 | 0.995 |
| 310.60 | 100.00 | 8.22 | 1.28 | 6.94 | 8.22 | -1.95 | 81.55 | 53.19 | 2.55 | 0.995 |
| 310.70 | 100.00 | 8.23 | 1.29 | 6.94 | 8.23 | -1.97 | 81.64 | 53.42 | 2.55 | 0.995 |
| 310.80 | 100.00 | 8.24 | 1.30 | 6.94 | 8.24 | -1.99 | 81.72 | 53.64 | 2.56 | 0.995 |
| 310.90 | 100.00 | 8.25 | 1.31 | 6.94 | 8.25 | -2.01 | 81.80 | 53.86 | 2.57 | 0.995 |
| 311.00 | 100.00 | 8.26 | 1.32 | 6.94 | 8.26 | -2.03 | 81.88 | 54.09 | 2.57 | 0.995 |
| 311.10 | 100.00 | 8.27 | 1.33 | 6.94 | 8.27 | -2.05 | 81.96 | 54.31 | 2.58 | 0.995 |
| 311.20 | 100.00 | 8.28 | 1.34 | 6.94 | 8.28 | -2.08 | 82.05 | 54.53 | 2.59 | 0.995 |
| 311.30 | 100.00 | 8.29 | 1.35 | 6.94 | 8.29 | -2.10 | 82.13 | 54.76 | 2.60 | 0.995 |
| 311.40 | 100.00 | 8.30 | 1.36 | 6.94 | 8.30 | -2.12 | 82.22 | 54.99 | 2.60 | 0.995 |
| 311.50 | 100.00 | 8.31 | 1.37 | 6.94 | 8.31 | -2.14 | 82.30 | 55.21 | 2.61 | 0.995 |
| 311.60 | 100.00 | 8.32 | 1.38 | 6.94 | 8.32 | -2.16 | 82.39 | 55.44 | 2.62 | 0.995 |
| 311.70 | 100.00 | 8.33 | 1.39 | 6.94 | 8.33 | -2.18 | 82.47 | 55.66 | 2.63 | 0.995 |
| 311.80 | 100.00 | 8.34 | 1.40 | 6.94 | 8.34 | -2.20 | 82.56 | 55.89 | 2.63 | 0.995 |
| 311.90 | 100.00 | 8.35 | 1.41 | 6.94 | 8.35 | -2.23 | 82.65 | 56.12 | 2.64 | 0.995 |
| 312.00 | 100.00 | 8.36 | 1.42 | 6.94 | 8.36 | -2.25 | 82.74 | 56.35 | 2.65 | 0.995 |
| 312.10 | 100.00 | 8.37 | 1.43 | 6.94 | 8.37 | -2.27 | 82.82 | 56.58 | 2.66 | 0.995 |
| 312.20 | 100.00 | 8.38 | 1.45 | 6.93 | 8.38 | -2.29 | 82.91 | 56.81 | 2.66 | 0.995 |
| 312.30 | 100.00 | 8.39 | 1.46 | 6.93 | 8.39 | -2.32 | 83.00 | 57.04 | 2.67 | 0.995 |
| 312.40 | 100.00 | 8.40 | 1.47 | 6.93 | 8.40 | -2.34 | 83.09 | 57.27 | 2.68 | 0.995 |
| 312.50 | 100.00 | 8.41 | 1.48 | 6.93 | 8.41 | -2.36 | 83.18 | 57.50 | 2.69 | 0.995 |
| 312.60 | 100.00 | 8.42 | 1.49 | 6.93 | 8.42 | -2.39 | 83.28 | 57.73 | 2.70 | 0.995 |
| 312.70 | 100.00 | 8.43 | 1.51 | 6.92 | 8.43 | -2.41 | 83.37 | 57.96 | 2.70 | 0.995 |
| 312.80 | 100.00 | 8.44 | 1.52 | 6.92 | 8.44 | -2.43 | 83.46 | 58.19 | 2.71 | 0.995 |
| 312.90 | 100.00 | 8.45 | 1.53 | 6.92 | 8.45 | -2.46 | 83.56 | 58.42 | 2.72 | 0.995 |
| 313.00 | 100.00 | 8.46 | 1.55 | 6.91 | 8.46 | -2.48 | 83.65 | 58.65 | 2.73 | 0.995 |
| 313.10 | 100.00 | 8.47 | 1.56 | 6.91 | 8.47 | -2.51 | 83.75 | 58.89 | 2.73 | 0.995 |
| 313.20 | 100.00 | 8.48 | 1.57 | 6.91 | 8.48 | -2.53 | 83.84 | 59.12 | 2.74 | 0.995 |
| 313.30 | 100.00 | 8.49 | 1.59 | 6.90 | 8.49 | -2.56 | 83.94 | 59.35 | 2.75 | 0.995 |
| 313.40 | 100.00 | 8.50 | 1.60 | 6.90 | 8.50 | -2.58 | 84.04 | 59.59 | 2.76 | 0.995 |
| 313.50 | 100.00 | 8.51 | 1.62 | 6.89 | 8.51 | -2.61 | 84.14 | 59.82 | 2.77 | 0.995 |
| 313.60 | 100.00 | 8.52 | 1.63 | 6.89 | 8.52 | -2.63 | 84.24 | 60.06 | 2.77 | 0.995 |
| 313.70 | 100.00 | 8.53 | 1.64 | 6.89 | 8.53 | -2.66 | 84.34 | 60.29 | 2.78 | 0.995 |
| 313.80 | 100.00 | 8.54 | 1.66 | 6.88 | 8.54 | -2.69 | 84.44 | 60.53 | 2.79 | 0.995 |
| 313.90 | 100.00 | 8.55 | 1.67 | 6.88 | 8.55 | -2.71 | 84.54 | 60.77 | 2.80 | 0.995 |
| 314.00 | 100.00 | 8.56 | 1.69 | 6.87 | 8.56 | -2.74 | 84.64 | 61.00 | 2.81 | 0.995 |
| 314.10 | 100.00 | 8.57 | 1.70 | 6.87 | 8.57 | -2.77 | 84.75 | 61.24 | 2.81 | 0.995 |
| 314.20 | 100.00 | 8.58 | 1.72 | 6.86 | 8.58 | -2.79 | 84.85 | 61.48 | 2.82 | 0.995 |
| 314.30 | 100.00 | 8.59 | 1.74 | 6.85 | 8.59 | -2.82 | 84.96 | 61.72 | 2.83 | 0.995 |
| 314.40 | 100.00 | 8.60 | 1.75 | 6.85 | 8.60 | -2.85 | 85.06 | 61.95 | 2.84 | 0.995 |
| 314.50 | 100.00 | 8.61 | 1.77 | 6.84 | 8.61 | -2.87 | 85.17 | 62.19 | 2.85 | 0.995 |
| 314.60 | 100.00 | 8.62 | 1.78 | 6.84 | 8.62 | -2.90 | 85.28 | 62.43 | 2.86 | 0.995 |
| 314.70 | 100.00 | 8.63 | 1.80 | 6.83 | 8.63 | -2.93 | 85.39 | 62.67 | 2.86 | 0.995 |
| 314.80 | 100.00 | 8.64 | 1.82 | 6.82 | 8.64 | -2.96 | 85.50 | 62.91 | 2.87 | 0.995 |
| 314.90 | 100.00 | 8.65 | 1.83 | 6.82 | 8.65 | -2.99 | 85.61 | 63.15 | 2.88 | 0.995 |
| 315.00 | 100.00 | 8.66 | 1.85 | 6.81 | 8.66 | -3.02 | 85.72 | 63.39 | 2.89 | 0.995 |
| 315.10 | 100.00 | 8.67 | 1.87 | 6.80 | 8.67 | -3.05 | 85.83 | 63.64 | 2.90 | 0.995 |
| 315.20 | 100.00 | 8.68 | 1.89 | 6.79 | 8.68 | -3.08 | 85.95 | 63.88 | 2.91 | 0.995 |
| 315.30 | 100.00 | 8.69 | 1.90 | 6.79 | 8.69 | -3.11 | 86.06 | 64.12 | 2.91 | 0.995 |
| 315.40 | 100.00 | 8.70 | 1.92 | 6.78 | 8.70 | -3.14 | 86.18 | 64.36 | 2.92 | 0.995 |
| 315.50 | 100.00 | 8.71 | 1.94 | 6.77 | 8.71 | -3.17 | 86.29 | 64.60 | 2.93 | 0.995 |
| 315.60 | 100.00 | 8.72 | 1.96 | 6.76 | 8.72 | -3.20 | 86.41 | 64.85 | 2.94 | 0.995 |
| 315.70 | 100.00 | 8.73 | 1.98 | 6.75 | 8.73 | -3.23 | 86.53 | 65.09 | 2.95 | 0.995 |
| 315.80 | 100.00 | 8.74 | 1.99 | 6.75 | 8.74 | -3.26 | 86.65 | 65.33 | 2.96 | 0.995 |
| 315.90 | 100.00 | 8.75 | 2.01 | 6.74 | 8.75 | -3.29 | 86.77 | 65.58 | 2.97 | 0.995 |
| 316.00 | 100.00 | 8.76 | 2.03 | 6.73 | 8.76 | -3.33 | 86.90 | 65.82 | 2.97 | 0.995 |
| 316.10 | 100.00 | 8.77 | 2.05 | 6.72 | 8.77 | -3.36 | 87.02 | 66.07 | 2.98 | 0.995 |
| 316.20 | 100.00 | 8.78 | 2.07 | 6.71 | 8.78 | -3.39 | 87.15 | 66.31 | 2.99 | 0.995 |
| 316.30 | 100.00 | 8.79 | 2.09 | 6.70 | 8.79 | -3.42 | 87.27 | 66.56 | 3.00 | 0.995 |
| 316.40 | 100.00 | 8.80 | 2.11 | 6.69 | 8.80 | -3.46 | 87.40 | 66.81 | 3.01 | 0.995 |
| 316.50 | 100.00 | 8.81 | 2.13 | 6.68 | 8.81 | -3.49 | 87.53 | 67.05 | 3.02 | 0.995 |
| 316.60 | 100.00 | 8.82 | 2.15 | 6.67 | 8.82 | -3.52 | 87.66 | 67.30 | 3.03 | 0.995 |
| 316.70 | 100.00 | 8.83 | 2.17 | 6.66 | 8.83 | -3.56 | 87.79 | 67.54 | 3.04 | 0.995 |
| 316.80 | 100.00 | 8.84 | 2.19 | 6.65 | 8.84 | -3.59 | 87.92 | 67.79 | 3.04 | 0.995 |
| 316.90 | 100.00 | 8.85 | 2.21 | 6.64 | 8.85 | -3.63 | 88.06 | 68.04 | 3.05 | 0.995 |
| 317.00 | 100.00 | 8.86 | 2.23 | 6.63 | 8.86 | -3.66 | 88.19 | 68.29 | 3.06 | 0.995 |
| 317.10 | 100.00 | 8.87 | 2.25 | 6.62 | 8.87 | -3.70 | 88.33 | 68.54 | 3.07 | 0.995 |
| 317.20 | 100.00 | 8.88 | 2.27 | 6.61 | 8.88 | -3.74 | 88.47 | 68.78 | 3.08 | 0.995 |
| 317.30 | 100.00 | 8.89 | 2.29 | 6.60 | 8.89 | -3.77 | 88.60 | 69.03 | 3.09 | 0.995 |
| 317.40 | 100.00 | 8.90 | 2.31 | 6.59 | 8.90 | -3.81 | 88.75 | 69.28 | 3.10 | 0.995 |
| 317.50 | 100.00 | 8.91 | 2.34 | 6.57 | 8.91 | -3.85 | 88.89 | 69.53 | 3.11 | 0.995 |
| 317.60 | 100.00 | 8.92 | 2.36 | 6.56 | 8.92 | -3.88 | 89.03 | 69.78 | 3.11 | 0.995 |
| 317.70 | 100.00 | 8.93 | 2.38 | 6.55 | 8.93 | -3.92 | 89.18 | 70.03 | 3.12 | 0.995 |
| 317.80 | 100.00 | 8.94 | 2.40 | 6.54 | 8.94 | -3.96 | 89.32 | 70.28 | 3.13 | 0.995 |
| 317.90 | 100.00 | 8.95 | 2.42 | 6.53 | 8.95 | -4.00 | 89.47 | 70.53 | 3.14 | 0.995 |
| 318.00 | 100.00 | 8.96 | 2.44 | 6.52 | 8.96 | -4.04 | 89.62 | 70.78 | 3.15 | 0.995 |
| 318.10 | 100.00 | 8.97 | 2.47 | 6.50 | 8.97 | -4.08 | 89.77 | 71.03 | 3.16 | 0.995 |
| 318.20 | 100.00 | 8.98 | 2.49 | 6.49 | 8.98 | -4.12 | 89.92 | 71.28 | 3.17 | 0.995 |
| 318.30 | 100.00 | 8.99 | 2.51 | 6.48 | 8.99 | -4.16 | 90.08 | 71.53 | 3.18 | 0.995 |
| 318.40 | 100.00 | 9.00 | 2.53 | 6.47 | 9.00 | -4.20 | 90.23 | 71.78 | 3.19 | 0.995 |
| 318.50 | 100.00 | 9.01 | 2.56 | 6.45 | 9.01 | -4.24 | 90.39 | 72.03 | 3.20 | 0.995 |
| 318.60 | 100.00 | 9.02 | 2.58 | 6.44 | 9.02 | -4.28 | 90.55 | 72.28 | 3.21 | 0.995 |
| 318.70 | 100.00 | 9.03 | 2.60 | 6.43 | 9.03 | -4.32 | 90.71 | 72.53 | 3.21 | 0.995 |
| 318.80 | 100.00 | 9.04 | 2.63 | 6.41 | 9.04 | -4.37 | 90.87 | 72.78 | 3.22 | 0.995 |
| 318.90 | 100.00 | 9.05 | 2.65 | 6.40 | 9.05 | -4.41 | 91.04 | 73.03 | 3.23 | 0.995 |
| 319.00 | 100.00 | 9.06 | 2.67 | 6.39 | 9.06 | -4.45 | 91.20 | 73.29 | 3.24 | 0.995 |
| 319.10 | 100.00 | 9.07 | 2.69 | 6.38 | 9.07 | -4.50 | 91.37 | 73.54 | 3.25 | 0.995 |
| 319.20 | 100.00 | 9.08 | 2.72 | 6.36 | 9.08 | -4.54 | 91.54 | 73.79 | 3.26 | 0.995 |
| 319.30 | 100.00 | 9.09 | 2.74 | 6.35 | 9.09 | -4.58 | 91.71 | 74.04 | 3.27 | 0.995 |
| 319.40 | 100.00 | 9.10 | 2.77 | 6.33 | 9.10 | -4.63 | 91.88 | 74.29 | 3.28 | 0.995 |
| 319.50 | 100.00 | 9.11 | 2.79 | 6.32 | 9.11 | -4.68 | 92.06 | 74.54 | 3.29 | 0.995 |
| 319.60 | 100.00 | 9.12 | 2.81 | 6.31 | 9.12 | -4.72 | 92.23 | 74.79 | 3.30 | 0.995 |
| 319.70 | 100.00 | 9.13 | 2.84 | 6.29 | 9.13 | -4.77 | 92.41 | 75.04 | 3.31 | 0.995 |
| 319.80 | 100.00 | 9.14 | 2.86 | 6.28 | 9.14 | -4.82 | 92.59 | 75.30 | 3.31 | 0.995 |
| 319.90 | 100.00 | 9.15 | 2.89 | 6.26 | 9.15 | -4.86 | 92.77 | 75.55 | 3.32 | 0.995 |
| 320.00 | 100.00 | 9.16 | 2.91 | 6.25 | 9.16 | -4.91 | 92.96 | 75.80 | 3.33 | 0.995 |
| 320.10 | 100.00 | 9.17 | 2.94 | 6.23 | 9.17 | -4.96 | 93.14 | 76.05 | 3.34 | 0.995 |
| 320.20 | 100.00 | 9.18 | 2.96 | 6.22 | 9.18 | -5.01 | 93.33 | 76.30 | 3.35 | 0.995 |
| 320.30 | 100.00 | 9.19 | 2.98 | 6.21 | 9.19 | -5.06 | 93.52 | 76.55 | 3.36 | 0.995 |
| 320.40 | 100.00 | 9.20 | 3.01 | 6.19 | 9.20 | -5.11 | 93.72 | 76.80 | 3.37 | 0.995 |
| 320.50 | 100.00 | 9.21 | 3.03 | 6.18 | 9.21 | -5.16 | 93.91 | 77.05 | 3.38 | 0.995 |
| 320.60 | 100.00 | 9.22 | 3.06 | 6.16 | 9.22 | -5.21 | 94.11 | 77.30 | 3.39 | 0.995 |
| 320.70 | 100.00 | 9.23 | 3.08 | 6.15 | 9.23 | -5.26 | 94.30 | 77.55 | 3.40 | 0.995 |
| 320.80 | 100.00 | 9.24 | 3.11 | 6.13 | 9.24 | -5.32 | 94.51 | 77.80 | 3.41 | 0.995 |
| 320.90 | 100.00 | 9.25 | 3.13 | 6.12 | 9.25 | -5.37 | 94.71 | 78.05 | 3.42 | 0.995 |
| 321.00 | 100.00 | 9.26 | 3.16 | 6.10 | 9.26 | -5.42 | 94.91 | 78.30 | 3.42 | 0.995 |
| 321.10 | 100.00 | 9.27 | 3.18 | 6.09 | 9.27 | -5.48 | 95.12 | 78.54 | 3.43 | 0.995 |
| 321.20 | 100.00 | 9.28 | 3.21 | 6.07 | 9.28 | -5.53 | 95.33 | 78.79 | 3.44 | 0.995 |
| 321.30 | 100.00 | 9.29 | 3.24 | 6.05 | 9.29 | -5.59 | 95.54 | 79.04 | 3.45 | 0.995 |
| 321.40 | 100.00 | 9.30 | 3.26 | 6.04 | 9.30 | -5.64 | 95.76 | 79.29 | 3.46 | 0.995 |
| 321.50 | 100.00 | 9.31 | 3.29 | 6.02 | 9.31 | -5.70 | 95.97 | 79.53 | 3.47 | 0.995 |
| 321.60 | 100.00 | 9.32 | 3.31 | 6.01 | 9.32 | -5.76 | 96.19 | 79.78 | 3.48 | 0.995 |
| 321.70 | 100.00 | 9.33 | 3.34 | 5.99 | 9.33 | -5.82 | 96.41 | 80.02 | 3.49 | 0.995 |
| 321.80 | 100.00 | 9.34 | 3.36 | 5.98 | 9.34 | -5.88 | 96.64 | 80.27 | 3.50 | 0.995 |
| 321.90 | 100.00 | 9.35 | 3.39 | 5.96 | 9.35 | -5.94 | 96.86 | 80.51 | 3.51 | 0.995 |
| 322.00 | 100.00 | 9.36 | 3.42 | 5.94 | 9.36 | -6.00 | 97.09 | 80.76 | 3.52 | 0.995 |
| 322.10 | 100.00 | 9.36 | 3.44 | 5.92 | 9.36 | -6.06 | 97.32 | 81.00 | 3.52 | 0.995 |
| 322.20 | 100.00 | 9.37 | 3.47 | 5.90 | 9.37 | -6.12 | 97.56 | 81.24 | 3.53 | 0.995 |
| 322.30 | 100.00 | 9.38 | 3.49 | 5.89 | 9.38 | -6.18 | 97.79 | 81.48 | 3.54 | 0.995 |
| 322.40 | 100.00 | 9.39 | 3.52 | 5.87 | 9.39 | -6.24 | 98.03 | 81.72 | 3.55 | 0.995 |
| 322.50 | 100.00 | 9.40 | 3.55 | 5.85 | 9.40 | -6.31 | 98.27 | 81.96 | 3.56 | 0.995 |
| 322.60 | 100.00 | 9.41 | 3.57 | 5.84 | 9.41 | -6.37 | 98.52 | 82.20 | 3.57 | 0.995 |
| 322.70 | 100.00 | 9.42 | 3.60 | 5.82 | 9.42 | -6.44 | 98.76 | 82.44 | 3.58 | 0.995 |
| 322.80 | 100.00 | 9.43 | 3.62 | 5.81 | 9.43 | -6.50 | 99.01 | 82.68 | 3.59 | 0.995 |
| 322.90 | 100.00 | 9.44 | 3.65 | 5.79 | 9.44 | -6.57 | 99.27 | 82.91 | 3.60 | 0.995 |
| 323.00 | 100.00 | 9.45 | 3.68 | 5.77 | 9.45 | -6.64 | 99.52 | 83.15 | 3.60 | 0.995 |
| 323.10 | 100.00 | 9.46 | 3.70 | 5.76 | 9.46 | -6.70 | 99.78 | 83.38 | 3.61 | 0.995 |
| 323.20 | 100.00 | 9.47 | 3.73 | 5.74 | 9.47 | -6.77 | 100.04 | 83.62 | 3.62 | 0.995 |
| 323.30 | 100.00 | 9.48 | 3.75 | 5.73 | 9.48 | -6.84 | 100.30 | 83.85 | 3.63 | 0.995 |
| 323.40 | 100.00 | 9.49 | 3.78 | 5.71 | 9.49 | -6.91 | 100.57 | 84.08 | 3.64 | 0.995 |
| 323.50 | 100.00 | 9.50 | 3.81 | 5.69 | 9.50 | -6.98 | 100.84 | 84.31 | 3.65 | 0.995 |
| 323.60 | 100.00 | 9.50 | 3.83 | 5.67 | 9.50 | -7.06 | 101.11 | 84.54 | 3.66 | 0.995 |
| 323.70 | 100.00 | 9.51 | 3.86 | 5.65 | 9.51 | -7.13 | 101.38 | 84.76 | 3.67 | 0.995 |
| 323.80 | 100.00 | 9.52 | 3.89 | 5.63 | 9.52 | -7.20 | 101.66 | 84.99 | 3.67 | 0.995 |
| 323.90 | 100.00 | 9.53 | 3.91 | 5.62 | 9.53 | -7.28 | 101.94 | 85.21 | 3.68 | 0.995 |
| 324.00 | 100.00 | 9.54 | 3.94 | 5.60 | 9.54 | -7.35 | 102.23 | 85.43 | 3.69 | 0.995 |
| 324.10 | 100.00 | 9.55 | 3.97 | 5.58 | 9.55 | -7.43 | 102.51 | 85.65 | 3.70 | 0.995 |
| 324.20 | 100.00 | 9.56 | 3.99 | 5.57 | 9.56 | -7.50 | 102.80 | 85.87 | 3.71 | 0.995 |
| 324.30 | 100.00 | 9.57 | 4.02 | 5.55 | 9.57 | -7.58 | 103.10 | 86.09 | 3.72 | 0.995 |
| 324.40 | 100.00 | 9.58 | 4.04 | 5.54 | 9.58 | -7.66 | 103.39 | 86.31 | 3.72 | 0.995 |
| 324.50 | 100.00 | 9.59 | 4.07 | 5.52 | 9.59 | -7.74 | 103.69 | 86.52 | 3.73 | 0.995 |
| 324.60 | 100.00 | 9.59 | 4.10 | 5.49 | 9.59 | -7.82 | 104.00 | 86.73 | 3.74 | 0.995 |
| 324.70 | 100.00 | 9.60 | 4.12 | 5.48 | 9.60 | -7.90 | 104.30 | 86.94 | 3.75 | 0.995 |
| 324.80 | 100.00 | 9.61 | 4.15 | 5.46 | 9.61 | -7.99 | 104.61 | 87.15 | 3.76 | 0.995 |
| 324.90 | 100.00 | 9.62 | 4.18 | 5.44 | 9.62 | -8.07 | 104.92 | 87.36 | 3.76 | 0.995 |
| 325.00 | 100.00 | 9.63 | 4.20 | 5.43 | 9.63 | -8.15 | 105.24 | 87.56 | 3.77 | 0.995 |
| 325.10 | 100.00 | 9.64 | 4.23 | 5.41 | 9.64 | -8.24 | 105.56 | 87.76 | 3.78 | 0.995 |
| 325.20 | 100.00 | 9.65 | 4.25 | 5.40 | 9.65 | -8.32 | 105.88 | 87.96 | 3.79 | 0.995 |
| 325.30 | 100.00 | 9.65 | 4.28 | 5.37 | 9.65 | -8.41 | 106.20 | 88.16 | 3.79 | 0.995 |
| 325.40 | 100.00 | 9.66 | 4.31 | 5.35 | 9.66 | -8.50 | 106.53 | 88.35 | 3.80 | 0.995 |
| 325.50 | 100.00 | 9.67 | 4.33 | 5.34 | 9.67 | -8.58 | 106.86 | 88.55 | 3.81 | 0.995 |
| 325.60 | 100.00 | 9.68 | 4.36 | 5.32 | 9.68 | -8.67 | 107.20 | 88.74 | 3.82 | 0.995 |
| 325.70 | 100.00 | 9.69 | 4.38 | 5.31 | 9.69 | -8.76 | 107.54 | 88.92 | 3.82 | 0.995 |
| 325.80 | 100.00 | 9.70 | 4.41 | 5.29 | 9.70 | -8.85 | 107.88 | 89.11 | 3.83 | 0.995 |
| 325.90 | 100.00 | 9.71 | 4.44 | 5.27 | 9.71 | -8.95 | 108.23 | 89.29 | 3.84 | 0.995 |
| 326.00 | 100.00 | 9.71 | 4.46 | 5.25 | 9.71 | -9.04 | 108.58 | 89.47 | 3.85 | 0.995 |
| 326.10 | 100.00 | 9.72 | 4.49 | 5.23 | 9.72 | -9.13 | 108.93 | 89.64 | 3.85 | 0.995 |
| 326.20 | 100.00 | 9.73 | 4.51 | 5.22 | 9.73 | -9.23 | 109.28 | 89.81 | 3.86 | 0.995 |
| 326.30 | 100.00 | 9.74 | 4.54 | 5.20 | 9.74 | -9.32 | 109.64 | 89.98 | 3.87 | 0.995 |
| 326.40 | 100.00 | 9.75 | 4.56 | 5.19 | 9.75 | -9.42 | 110.01 | 90.15 | 3.87 | 0.995 |
| 326.50 | 100.00 | 9.75 | 4.59 | 5.16 | 9.75 | -9.51 | 110.37 | 90.31 | 3.88 | 0.995 |
| 326.60 | 100.00 | 9.76 | 4.62 | 5.14 | 9.76 | -9.61 | 110.74 | 90.47 | 3.89 | 0.995 |
| 326.70 | 100.00 | 9.77 | 4.64 | 5.13 | 9.77 | -9.71 | 111.12 | 90.63 | 3.89 | 0.995 |
| 326.80 | 100.00 | 9.78 | 4.67 | 5.11 | 9.78 | -9.80 | 111.49 | 90.78 | 3.90 | 0.995 |
| 326.90 | 100.00 | 9.79 | 4.69 | 5.10 | 9.79 | -9.90 | 111.87 | 90.93 | 3.91 | 0.995 |
| 327.00 | 100.00 | 9.79 | 4.72 | 5.07 | 9.79 | -10.00 | 112.26 | 91.07 | 3.91 | 0.995 |
| 327.10 | 100.00 | 9.80 | 4.74 | 5.06 | 9.80 | -10.10 | 112.65 | 91.21 | 3.92 | 0.995 |
| 327.20 | 100.00 | 9.81 | 4.77 | 5.04 | 9.81 | -10.20 | 113.04 | 91.35 | 3.92 | 0.995 |
| 327.30 | 100.00 | 9.82 | 4.79 | 5.03 | 9.82 | -10.30 | 113.43 | 91.48 | 3.93 | 0.995 |
| 327.40 | 100.00 | 9.83 | 4.82 | 5.01 | 9.83 | -10.40 | 113.83 | 91.61 | 3.94 | 0.995 |
| 327.50 | 100.00 | 9.83 | 4.84 | 4.99 | 9.83 | -10.50 | 114.23 | 91.73 | 3.94 | 0.995 |
| 327.60 | 100.00 | 9.84 | 4.87 | 4.97 | 9.84 | -10.60 | 114.64 | 91.85 | 3.95 | 0.995 |
| 327.70 | 100.00 | 9.85 | 4.89 | 4.96 | 9.85 | -10.70 | 115.05 | 91.96 | 3.95 | 0.995 |
| 327.80 | 100.00 | 9.86 | 4.92 | 4.94 | 9.86 | -10.80 | 115.46 | 92.07 | 3.96 | 0.995 |
| 327.90 | 100.00 | 9.86 | 4.94 | 4.92 | 9.86 | -10.90 | 115.88 | 92.18 | 3.96 | 0.995 |
| 328.00 | 100.00 | 9.87 | 4.96 | 4.91 | 9.87 | -11.00 | 116.29 | 92.28 | 3.97 | 0.995 |
| 328.10 | 100.00 | 9.88 | 4.99 | 4.89 | 9.88 | -11.10 | 116.72 | 92.37 | 3.97 | 0.995 |
| 328.20 | 100.00 | 9.89 | 5.01 | 4.88 | 9.89 | -11.20 | 117.14 | 92.46 | 3.98 | 0.995 |
| 328.30 | 100.00 | 9.89 | 5.04 | 4.85 | 9.89 | -11.30 | 117.57 | 92.55 | 3.98 | 0.995 |
| 328.40 | 100.00 | 9.90 | 5.06 | 4.84 | 9.90 | -11.39 | 118.01 | 92.62 | 3.99 | 0.995 |
| 328.50 | 100.00 | 9.91 | 5.08 | 4.83 | 9.91 | -11.49 | 118.44 | 92.70 | 3.99 | 0.995 |
| 328.60 | 100.00 | 9.91 | 5.11 | 4.80 | 9.91 | -11.58 | 118.88 | 92.76 | 4.00 | 0.995 |
| 328.70 | 100.00 | 9.92 | 5.13 | 4.79 | 9.92 | -11.68 | 119.33 | 92.82 | 4.00 | 0.995 |
| 328.80 | 100.00 | 9.93 | 5.16 | 4.77 | 9.93 | -11.77 | 119.77 | 92.88 | 4.00 | 0.995 |
| 328.90 | 100.00 | 9.94 | 5.18 | 4.76 | 9.94 | -11.86 | 120.22 | 92.93 | 4.01 | 0.995 |
| 329.00 | 100.00 | 9.94 | 5.20 | 4.74 | 9.94 | -11.94 | 120.67 | 92.97 | 4.01 | 0.995 |
| 329.10 | 100.00 | 9.95 | 5.23 | 4.72 | 9.95 | -12.03 | 121.13 | 93.00 | 4.01 | 0.995 |
| 329.20 | 100.00 | 9.96 | 5.25 | 4.71 | 9.96 | -12.11 | 121.59 | 93.03 | 4.02 | 0.995 |
| 329.30 | 100.00 | 9.96 | 5.27 | 4.69 | 9.96 | -12.19 | 122.05 | 93.05 | 4.02 | 0.995 |
| 329.40 | 100.00 | 9.97 | 5.29 | 4.68 | 9.97 | -12.27 | 122.52 | 93.07 | 4.02 | 0.995 |
| 329.50 | 100.00 | 9.98 | 5.32 | 4.66 | 9.98 | -12.34 | 122.98 | 93.08 | 4.03 | 0.995 |
| 329.60 | 100.00 | 9.98 | 5.34 | 4.64 | 9.98 | -12.41 | 123.45 | 93.08 | 4.03 | 0.995 |
| 329.70 | 100.00 | 9.99 | 5.36 | 4.63 | 9.99 | -12.47 | 123.93 | 93.07 | 4.03 | 0.995 |
| 329.80 | 100.00 | 10.00 | 5.38 | 4.62 | 10.00 | -12.53 | 124.40 | 93.05 | 4.03 | 0.995 |
| 329.90 | 100.00 | 10.00 | 5.41 | 4.59 | 10.00 | -12.59 | 124.88 | 93.03 | 4.04 | 0.995 |
| 330.00 | 100.00 | 10.01 | 5.43 | 4.58 | 10.01 | -12.64 | 125.36 | 93.00 | 4.04 | 0.995 |
| 330.10 | 100.00 | 10.01 | 5.45 | 4.56 | 10.01 | -12.69 | 125.84 | 92.96 | 4.04 | 0.995 |
| 330.20 | 100.00 | 10.02 | 5.47 | 4.55 | 10.02 | -12.73 | 126.33 | 92.92 | 4.04 | 0.995 |
| 330.30 | 100.00 | 10.03 | 5.49 | 4.54 | 10.03 | -12.76 | 126.82 | 92.86 | 4.04 | 0.995 |
| 330.40 | 100.00 | 10.03 | 5.51 | 4.52 | 10.03 | -12.79 | 127.31 | 92.80 | 4.04 | 0.995 |
| 330.50 | 100.00 | 10.04 | 5.54 | 4.50 | 10.04 | -12.82 | 127.80 | 92.72 | 4.05 | 0.995 |
| 330.60 | 100.00 | 10.04 | 5.56 | 4.48 | 10.04 | -12.84 | 128.29 | 92.64 | 4.05 | 0.995 |
| 330.70 | 100.00 | 10.05 | 5.58 | 4.47 | 10.05 | -12.85 | 128.78 | 92.55 | 4.05 | 0.995 |
| 330.80 | 100.00 | 10.06 | 5.60 | 4.46 | 10.06 | -12.86 | 129.28 | 92.45 | 4.05 | 0.995 |
| 330.90 | 100.00 | 10.06 | 5.62 | 4.44 | 10.06 | -12.85 | 129.78 | 92.34 | 4.05 | 0.995 |
| 331.00 | 100.00 | 10.07 | 5.64 | 4.43 | 10.07 | -12.85 | 130.28 | 92.22 | 4.05 | 0.995 |
| 331.10 | 100.00 | 10.07 | 5.66 | 4.41 | 10.07 | -12.83 | 130.78 | 92.09 | 4.05 | 0.995 |
| 331.20 | 100.00 | 10.08 | 5.68 | 4.40 | 10.08 | -12.81 | 131.28 | 91.95 | 4.05 | 0.995 |
| 331.30 | 100.00 | 10.08 | 5.70 | 4.38 | 10.08 | -12.79 | 131.78 | 91.80 | 4.05 | 0.995 |
| 331.40 | 100.00 | 10.09 | 5.72 | 4.37 | 10.09 | -12.75 | 132.28 | 91.64 | 4.05 | 0.995 |
| 331.50 | 100.00 | 10.10 | 5.74 | 4.36 | 10.10 | -12.72 | 132.79 | 91.47 | 4.05 | 0.995 |
| 331.60 | 100.00 | 10.10 | 5.76 | 4.34 | 10.10 | -12.67 | 133.29 | 91.29 | 4.04 | 0.995 |
| 331.70 | 100.00 | 10.11 | 5.78 | 4.33 | 10.11 | -12.62 | 133.79 | 91.10 | 4.04 | 0.995 |
| 331.80 | 100.00 | 10.11 | 5.80 | 4.31 | 10.11 | -12.56 | 134.29 | 90.90 | 4.04 | 0.995 |
| 331.90 | 100.00 | 10.12 | 5.82 | 4.30 | 10.12 | -12.50 | 134.80 | 90.68 | 4.04 | 0.995 |
| 332.00 | 100.00 | 10.12 | 5.84 | 4.28 | 10.12 | -12.43 | 135.30 | 90.46 | 4.04 | 0.995 |
| 332.10 | 100.00 | 10.13 | 5.86 | 4.27 | 10.13 | -12.35 | 135.80 | 90.22 | 4.04 | 0.995 |
| 332.20 | 100.00 | 10.13 | 5.87 | 4.26 | 10.13 | -12.27 | 136.30 | 89.97 | 4.03 | 0.995 |
| 332.30 | 100.00 | 10.14 | 5.89 | 4.25 | 10.14 | -12.19 | 136.80 | 89.71 | 4.03 | 0.995 |
| 332.40 | 100.00 | 10.14 | 5.91 | 4.23 | 10.14 | -12.10 | 137.29 | 89.44 | 4.03 | 0.995 |
| 332.50 | 100.00 | 10.14 | 5.93 | 4.21 | 10.14 | -12.00 | 137.79 | 89.15 | 4.02 | 0.995 |
| 332.60 | 100.00 | 10.15 | 5.95 | 4.20 | 10.15 | -11.91 | 138.28 | 88.86 | 4.02 | 0.995 |
| 332.70 | 100.00 | 10.15 | 5.96 | 4.19 | 10.15 | -11.80 | 138.77 | 88.55 | 4.02 | 0.995 |
| 332.80 | 100.00 | 10.16 | 5.98 | 4.18 | 10.16 | -11.70 | 139.26 | 88.23 | 4.01 | 0.995 |
| 332.90 | 100.00 | 10.16 | 6.00 | 4.16 | 10.16 | -11.59 | 139.74 | 87.89 | 4.01 | 0.995 |
| 333.00 | 100.00 | 10.17 | 6.02 | 4.15 | 10.17 | -11.53 | 140.22 | 87.54 | 4.00 | 0.995 |
| 333.10 | 100.00 | 10.17 | 6.03 | 4.14 | 10.17 | -11.60 | 140.70 | 87.18 | 4.00 | 0.995 |
| 333.20 | 100.00 | 10.17 | 6.05 | 4.12 | 10.17 | -11.66 | 141.17 | 86.81 | 3.99 | 0.995 |
| 333.30 | 100.00 | 10.18 | 6.07 | 4.11 | 10.18 | -11.72 | 141.64 | 86.43 | 3.99 | 0.995 |
| 333.40 | 100.00 | 10.18 | 6.08 | 4.10 | 10.18 | -11.77 | 142.10 | 86.03 | 3.98 | 0.995 |
| 333.50 | 100.00 | 10.19 | 6.10 | 4.09 | 10.19 | -11.82 | 142.56 | 85.61 | 3.98 | 0.995 |
| 333.60 | 100.00 | 10.19 | 6.11 | 4.08 | 10.19 | -11.86 | 143.02 | 85.19 | 3.97 | 0.995 |
| 333.70 | 100.00 | 10.19 | 6.13 | 4.06 | 10.19 | -11.90 | 143.46 | 84.75 | 3.97 | 0.995 |
| 333.80 | 100.00 | 10.20 | 6.14 | 4.06 | 10.20 | -11.94 | 143.91 | 84.30 | 3.96 | 0.995 |
| 333.90 | 100.00 | 10.20 | 6.16 | 4.04 | 10.20 | -11.97 | 144.34 | 83.83 | 3.95 | 0.995 |
| 334.00 | 100.00 | 10.20 | 6.18 | 4.02 | 10.20 | -11.99 | 144.77 | 83.35 | 3.95 | 0.995 |
| 334.10 | 100.00 | 10.21 | 6.19 | 4.02 | 10.21 | -12.01 | 145.19 | 82.86 | 3.94 | 0.995 |
| 334.20 | 100.00 | 10.21 | 6.20 | 4.01 | 10.21 | -12.02 | 145.60 | 82.35 | 3.93 | 0.995 |
| 334.30 | 100.00 | 10.21 | 6.22 | 3.99 | 10.21 | -12.03 | 146.01 | 81.83 | 3.93 | 0.995 |
| 334.40 | 100.00 | 10.22 | 6.23 | 3.99 | 10.22 | -12.03 | 146.40 | 81.30 | 3.92 | 0.995 |
| 334.50 | 100.00 | 10.22 | 6.25 | 3.97 | 10.22 | -12.03 | 146.79 | 80.75 | 3.91 | 0.995 |
| 334.60 | 100.00 | 10.22 | 6.26 | 3.96 | 10.22 | -12.02 | 147.17 | 80.19 | 3.90 | 0.995 |
| 334.70 | 100.00 | 10.23 | 6.27 | 3.96 | 10.23 | -12.00 | 147.54 | 79.61 | 3.89 | 0.995 |
| 334.80 | 100.00 | 10.23 | 6.29 | 3.94 | 10.23 | -11.98 | 147.89 | 79.03 | 3.88 | 0.995 |
| 334.90 | 100.00 | 10.23 | 6.30 | 3.93 | 10.23 | -11.95 | 148.24 | 78.43 | 3.87 | 0.995 |
| 335.00 | 100.00 | 10.24 | 6.31 | 3.93 | 10.24 | -11.92 | 148.57 | 77.81 | 3.86 | 0.995 |
| 335.10 | 100.00 | 10.24 | 6.33 | 3.91 | 10.24 | -11.88 | 148.90 | 77.18 | 3.85 | 0.995 |
| 335.20 | 100.00 | 10.24 | 6.34 | 3.90 | 10.24 | -11.83 | 149.21 | 76.54 | 3.84 | 0.995 |
| 335.30 | 100.00 | 10.24 | 6.35 | 3.89 | 10.24 | -11.78 | 149.51 | 75.89 | 3.83 | 0.995 |
| 335.40 | 100.00 | 10.25 | 6.36 | 3.89 | 10.25 | -11.73 | 149.79 | 75.22 | 3.82 | 0.995 |
| 335.50 | 100.00 | 10.25 | 6.37 | 3.88 | 10.25 | -11.67 | 150.06 | 74.55 | 3.81 | 0.995 |
| 335.60 | 100.00 | 10.25 | 6.39 | 3.86 | 10.25 | -11.60 | 150.32 | 73.86 | 3.80 | 0.995 |
| 335.70 | 100.00 | 10.25 | 6.40 | 3.85 | 10.25 | -11.53 | 150.56 | 73.15 | 3.79 | 0.995 |
| 335.80 | 100.00 | 10.25 | 6.41 | 3.84 | 10.25 | -11.46 | 150.79 | 72.44 | 3.78 | 0.995 |
| 335.90 | 100.00 | 10.26 | 6.42 | 3.84 | 10.26 | -11.38 | 151.00 | 71.71 | 3.77 | 0.995 |
| 336.00 | 100.00 | 10.26 | 6.43 | 3.83 | 10.26 | -11.30 | 151.20 | 70.97 | 3.75 | 0.995 |
| 336.10 | 100.00 | 10.26 | 6.44 | 3.82 | 10.26 | -11.21 | 151.38 | 70.23 | 3.74 | 0.995 |
| 336.20 | 100.00 | 10.26 | 6.45 | 3.81 | 10.26 | -11.12 | 151.54 | 69.47 | 3.73 | 0.995 |
| 336.30 | 100.00 | 10.26 | 6.46 | 3.80 | 10.26 | -11.03 | 151.69 | 68.70 | 3.72 | 0.995 |
| 336.40 | 100.00 | 10.27 | 6.47 | 3.80 | 10.27 | -10.93 | 151.82 | 67.92 | 3.70 | 0.995 |
| 336.50 | 100.00 | 10.27 | 6.48 | 3.79 | 10.27 | -10.83 | 151.93 | 67.13 | 3.69 | 0.995 |
| 336.60 | 100.00 | 10.27 | 6.49 | 3.78 | 10.27 | -10.73 | 152.02 | 66.33 | 3.68 | 0.995 |
| 336.70 | 100.00 | 10.27 | 6.50 | 3.77 | 10.27 | -10.62 | 152.09 | 65.52 | 3.66 | 0.995 |
| 336.80 | 100.00 | 10.27 | 6.51 | 3.76 | 10.27 | -10.52 | 152.14 | 64.71 | 3.65 | 0.995 |
| 336.90 | 100.00 | 10.27 | 6.51 | 3.76 | 10.27 | -10.41 | 152.18 | 63.88 | 3.63 | 0.995 |
| 337.00 | 100.00 | 10.27 | 6.52 | 3.75 | 10.27 | -10.30 | 152.19 | 63.05 | 3.62 | 0.995 |
| 337.10 | 100.00 | 10.27 | 6.53 | 3.74 | 10.27 | -10.19 | 152.18 | 62.21 | 3.60 | 0.995 |
| 337.20 | 100.00 | 10.28 | 6.54 | 3.74 | 10.28 | -10.07 | 152.15 | 61.37 | 3.59 | 0.995 |
| 337.30 | 100.00 | 10.28 | 6.55 | 3.73 | 10.28 | -9.96 | 152.10 | 60.52 | 3.57 | 0.995 |
| 337.40 | 100.00 | 10.28 | 6.55 | 3.73 | 10.28 | -9.84 | 152.03 | 59.66 | 3.56 | 0.995 |
| 337.50 | 100.00 | 10.28 | 6.56 | 3.72 | 10.28 | -9.72 | 151.93 | 58.80 | 3.54 | 0.995 |
| 337.60 | 100.00 | 10.28 | 6.57 | 3.71 | 10.28 | -9.61 | 151.82 | 57.94 | 3.52 | 0.995 |
| 337.70 | 100.00 | 10.28 | 6.57 | 3.71 | 10.28 | -9.49 | 151.68 | 57.07 | 3.51 | 0.995 |
| 337.80 | 100.00 | 10.28 | 6.58 | 3.70 | 10.28 | -9.37 | 151.51 | 56.20 | 3.49 | 0.995 |
| 337.90 | 100.00 | 10.28 | 6.59 | 3.69 | 10.28 | -9.25 | 151.33 | 55.32 | 3.47 | 0.995 |
| 338.00 | 100.00 | 10.28 | 6.59 | 3.69 | 10.28 | -9.13 | 151.12 | 54.45 | 3.46 | 0.995 |
| 338.10 | 100.00 | 10.28 | 6.60 | 3.68 | 10.28 | -9.01 | 150.88 | 53.57 | 3.44 | 0.995 |
| 338.20 | 100.00 | 10.28 | 6.60 | 3.68 | 10.28 | -8.89 | 150.63 | 52.69 | 3.42 | 0.995 |
| 338.30 | 100.00 | 10.28 | 6.61 | 3.67 | 10.28 | -8.77 | 150.35 | 51.81 | 3.40 | 0.995 |
| 338.40 | 100.00 | 10.28 | 6.61 | 3.67 | 10.28 | -8.65 | 150.04 | 50.93 | 3.38 | 0.995 |
| 338.50 | 100.00 | 10.28 | 6.62 | 3.66 | 10.28 | -8.53 | 149.72 | 50.06 | 3.37 | 0.995 |
| 338.60 | 100.00 | 10.28 | 6.62 | 3.66 | 10.28 | -8.42 | 149.36 | 49.18 | 3.35 | 0.995 |
| 338.70 | 100.00 | 10.28 | 6.62 | 3.66 | 10.28 | -8.30 | 148.99 | 48.31 | 3.33 | 0.995 |
| 338.80 | 100.00 | 10.28 | 6.63 | 3.65 | 10.28 | -8.18 | 148.58 | 47.45 | 3.31 | 0.995 |
| 338.90 | 100.00 | 10.28 | 6.63 | 3.65 | 10.28 | -8.06 | 148.16 | 46.58 | 3.29 | 0.995 |
| 339.00 | 100.00 | 10.28 | 6.63 | 3.65 | 10.28 | -7.94 | 147.71 | 45.73 | 3.27 | 0.995 |
| 339.10 | 100.00 | 10.28 | 6.64 | 3.64 | 10.28 | -7.83 | 147.24 | 44.87 | 3.25 | 0.995 |
| 339.20 | 100.00 | 10.28 | 6.64 | 3.64 | 10.28 | -7.71 | 146.74 | 44.03 | 3.23 | 0.995 |
| 339.30 | 100.00 | 10.28 | 6.64 | 3.64 | 10.28 | -7.60 | 146.22 | 43.19 | 3.21 | 0.995 |
| 339.40 | 100.00 | 10.28 | 6.64 | 3.64 | 10.28 | -7.48 | 145.67 | 42.36 | 3.19 | 0.995 |
| 339.50 | 100.00 | 10.28 | 6.65 | 3.63 | 10.28 | -7.40 | 145.10 | 41.53 | 3.17 | 0.995 |
| 339.60 | 100.00 | 10.28 | 6.65 | 3.63 | 10.28 | -7.38 | 144.51 | 40.72 | 3.15 | 0.995 |
| 339.70 | 100.00 | 10.28 | 6.65 | 3.63 | 10.28 | -7.36 | 143.89 | 39.92 | 3.13 | 0.995 |
| 339.80 | 100.00 | 10.28 | 6.65 | 3.63 | 10.28 | -7.33 | 143.25 | 39.12 | 3.11 | 0.995 |
| 339.90 | 100.00 | 10.28 | 6.65 | 3.63 | 10.28 | -7.31 | 142.59 | 38.34 | 3.08 | 0.995 |
| 340.00 | 100.00 | 10.28 | 6.65 | 3.63 | 10.28 | -7.28 | 141.90 | 37.57 | 3.06 | 0.995 |
| 340.10 | 100.00 | 10.28 | 6.65 | 3.63 | 10.28 | -7.25 | 141.20 | 36.81 | 3.04 | 0.995 |
| 340.20 | 100.00 | 10.28 | 6.65 | 3.63 | 10.28 | -7.22 | 140.47 | 36.07 | 3.02 | 0.995 |
| 340.30 | 100.00 | 10.28 | 6.65 | 3.63 | 10.28 | -7.19 | 139.72 | 35.34 | 3.00 | 0.995 |
| 340.40 | 100.00 | 10.28 | 6.65 | 3.63 | 10.28 | -7.16 | 138.95 | 34.62 | 2.98 | 0.995 |
| 340.50 | 100.00 | 10.28 | 6.65 | 3.63 | 10.28 | -7.12 | 138.16 | 33.92 | 2.95 | 0.995 |
| 340.60 | 100.00 | 10.28 | 6.65 | 3.63 | 10.28 | -7.09 | 137.34 | 33.24 | 2.93 | 0.995 |
| 340.70 | 100.00 | 10.27 | 6.65 | 3.62 | 10.27 | -7.05 | 136.51 | 32.57 | 2.91 | 0.995 |
| 340.80 | 100.00 | 10.27 | 6.64 | 3.63 | 10.27 | -7.02 | 135.66 | 31.91 | 2.89 | 0.995 |
| 340.90 | 100.00 | 10.27 | 6.64 | 3.63 | 10.27 | -6.98 | 134.79 | 31.28 | 2.86 | 0.995 |
| 341.00 | 100.00 | 10.27 | 6.64 | 3.63 | 10.27 | -6.94 | 133.90 | 30.66 | 2.84 | 0.995 |
| 341.10 | 100.00 | 10.27 | 6.64 | 3.63 | 10.27 | -6.90 | 132.99 | 30.06 | 2.82 | 0.995 |
| 341.20 | 100.00 | 10.27 | 6.64 | 3.63 | 10.27 | -6.86 | 132.07 | 29.48 | 2.79 | 0.995 |
| 341.30 | 100.00 | 10.27 | 6.63 | 3.64 | 10.27 | -6.82 | 131.13 | 28.92 | 2.77 | 0.995 |
| 341.40 | 100.00 | 10.27 | 6.63 | 3.64 | 10.27 | -6.77 | 130.17 | 28.37 | 2.75 | 0.995 |
| 341.50 | 100.00 | 10.27 | 6.63 | 3.64 | 10.27 | -6.73 | 129.20 | 27.85 | 2.72 | 0.995 |
| 341.60 | 100.00 | 10.26 | 6.62 | 3.64 | 10.26 | -6.69 | 128.22 | 27.35 | 2.70 | 0.995 |
| 341.70 | 100.00 | 10.26 | 6.62 | 3.64 | 10.26 | -6.64 | 127.21 | 26.87 | 2.68 | 0.995 |
| 341.80 | 100.00 | 10.26 | 6.61 | 3.65 | 10.26 | -6.60 | 126.20 | 26.40 | 2.65 | 0.995 |
| 341.90 | 100.00 | 10.26 | 6.61 | 3.65 | 10.26 | -6.55 | 125.17 | 25.96 | 2.63 | 0.995 |
| 342.00 | 100.00 | 10.26 | 6.60 | 3.66 | 10.26 | -6.51 | 124.13 | 25.54 | 2.61 | 0.995 |
| 342.10 | 100.00 | 10.26 | 6.60 | 3.66 | 10.26 | -6.46 | 123.08 | 25.15 | 2.58 | 0.995 |
| 342.20 | 100.00 | 10.26 | 6.59 | 3.67 | 10.26 | -6.41 | 122.01 | 24.77 | 2.56 | 0.995 |
| 342.30 | 100.00 | 10.26 | 6.59 | 3.67 | 10.26 | -6.36 | 120.94 | 24.42 | 2.54 | 0.995 |
| 342.40 | 100.00 | 10.25 | 6.58 | 3.67 | 10.25 | -6.31 | 119.86 | 24.09 | 2.51 | 0.995 |
| 342.50 | 100.00 | 10.25 | 6.58 | 3.67 | 10.25 | -6.27 | 118.76 | 23.78 | 2.49 | 0.995 |
| 342.60 | 100.00 | 10.25 | 6.57 | 3.68 | 10.25 | -6.22 | 117.66 | 23.49 | 2.47 | 0.995 |
| 342.70 | 100.00 | 10.25 | 6.56 | 3.69 | 10.25 | -6.17 | 116.55 | 23.23 | 2.44 | 0.995 |
| 342.80 | 100.00 | 10.25 | 6.56 | 3.69 | 10.25 | -6.12 | 115.43 | 22.99 | 2.42 | 0.995 |
| 342.90 | 100.00 | 10.25 | 6.55 | 3.70 | 10.25 | -6.07 | 114.30 | 22.77 | 2.40 | 0.995 |
| 343.00 | 100.00 | 10.25 | 6.54 | 3.71 | 10.25 | -6.02 | 113.17 | 22.57 | 2.37 | 0.995 |
| 343.10 | 100.00 | 10.25 | 6.53 | 3.72 | 10.25 | -5.97 | 112.04 | 22.40 | 2.35 | 0.995 |
| 343.20 | 100.00 | 10.24 | 6.53 | 3.71 | 10.24 | -5.92 | 110.89 | 22.25 | 2.33 | 0.995 |
| 343.30 | 100.00 | 10.24 | 6.52 | 3.72 | 10.24 | -5.87 | 109.75 | 22.12 | 2.31 | 0.995 |
| 343.40 | 100.00 | 10.24 | 6.51 | 3.73 | 10.24 | -5.82 | 108.60 | 22.01 | 2.28 | 0.995 |
| 343.50 | 100.00 | 10.24 | 6.50 | 3.74 | 10.24 | -5.76 | 107.44 | 21.93 | 2.26 | 0.995 |
| 343.60 | 100.00 | 10.24 | 6.49 | 3.75 | 10.24 | -5.71 | 106.29 | 21.87 | 2.24 | 0.995 |
| 343.70 | 100.00 | 10.24 | 6.48 | 3.76 | 10.24 | -5.66 | 105.13 | 21.83 | 2.22 | 0.995 |
| 343.80 | 100.00 | 10.24 | 6.47 | 3.77 | 10.24 | -5.61 | 103.97 | 21.82 | 2.20 | 0.995 |
| 343.90 | 100.00 | 10.24 | 6.46 | 3.78 | 10.24 | -5.56 | 102.81 | 21.82 | 2.18 | 0.995 |
| 344.00 | 100.00 | 10.24 | 6.45 | 3.79 | 10.24 | -5.51 | 101.65 | 21.85 | 2.15 | 0.995 |
| 344.10 | 100.00 | 10.23 | 6.44 | 3.79 | 10.23 | -5.46 | 100.49 | 21.90 | 2.13 | 0.995 |
| 344.20 | 100.00 | 10.23 | 6.43 | 3.80 | 10.23 | -5.41 | 99.32 | 21.97 | 2.11 | 0.995 |
| 344.30 | 100.00 | 10.23 | 6.42 | 3.81 | 10.23 | -5.36 | 98.16 | 22.06 | 2.09 | 0.995 |
| 344.40 | 100.00 | 10.23 | 6.41 | 3.82 | 10.23 | -5.31 | 97.01 | 22.18 | 2.08 | 0.995 |
| 344.50 | 100.00 | 10.23 | 6.40 | 3.83 | 10.23 | -5.26 | 95.85 | 22.31 | 2.06 | 0.995 |
| 344.60 | 100.00 | 10.23 | 6.39 | 3.84 | 10.23 | -5.21 | 94.69 | 22.47 | 2.04 | 0.995 |
| 344.70 | 100.00 | 10.23 | 6.37 | 3.86 | 10.23 | -5.16 | 93.54 | 22.64 | 2.02 | 0.995 |
| 344.80 | 100.00 | 10.23 | 6.36 | 3.87 | 10.23 | -5.11 | 92.39 | 22.84 | 2.00 | 0.995 |
| 344.90 | 100.00 | 10.23 | 6.35 | 3.88 | 10.23 | -5.06 | 91.25 | 23.05 | 1.99 | 0.995 |
| 345.00 | 100.00 | 10.23 | 6.34 | 3.89 | 10.23 | -5.01 | 90.11 | 23.29 | 1.97 | 0.995 |
| 345.10 | 100.00 | 10.23 | 6.33 | 3.90 | 10.23 | -4.97 | 88.97 | 23.54 | 1.95 | 0.995 |
| 345.20 | 100.00 | 10.23 | 6.31 | 3.92 | 10.23 | -4.92 | 87.84 | 23.81 | 1.94 | 0.995 |
| 345.30 | 100.00 | 10.23 | 6.30 | 3.93 | 10.23 | -4.87 | 86.71 | 24.10 | 1.93 | 0.995 |
| 345.40 | 100.00 | 10.23 | 6.29 | 3.94 | 10.23 | -4.82 | 85.59 | 24.41 | 1.91 | 0.995 |
| 345.50 | 100.00 | 10.23 | 6.27 | 3.96 | 10.23 | -4.78 | 84.47 | 24.74 | 1.90 | 0.995 |
| 345.60 | 100.00 | 10.23 | 6.26 | 3.97 | 10.23 | -4.73 | 83.36 | 25.09 | 1.89 | 0.995 |
| 345.70 | 100.00 | 10.23 | 6.24 | 3.99 | 10.23 | -4.68 | 82.26 | 25.45 | 1.88 | 0.995 |
| 345.80 | 100.00 | 10.23 | 6.23 | 4.00 | 10.23 | -4.64 | 81.16 | 25.83 | 1.87 | 0.995 |
| 345.90 | 100.00 | 10.23 | 6.22 | 4.01 | 10.23 | -4.59 | 80.07 | 26.23 | 1.86 | 0.995 |
| 346.00 | 100.00 | 10.23 | 6.20 | 4.03 | 10.23 | -4.55 | 78.99 | 26.64 | 1.85 | 0.995 |
| 346.10 | 100.00 | 10.23 | 6.19 | 4.04 | 10.23 | -4.50 | 77.91 | 27.08 | 1.85 | 0.995 |
| 346.20 | 100.00 | 10.23 | 6.17 | 4.06 | 10.23 | -4.46 | 76.84 | 27.52 | 1.84 | 0.995 |
| 346.30 | 100.00 | 10.23 | 6.16 | 4.07 | 10.23 | -4.42 | 75.78 | 27.98 | 1.84 | 0.995 |
| 346.40 | 100.00 | 10.23 | 6.14 | 4.09 | 10.23 | -4.37 | 74.73 | 28.46 | 1.84 | 0.995 |
| 346.50 | 100.00 | 10.23 | 6.12 | 4.11 | 10.23 | -4.33 | 73.68 | 28.95 | 1.83 | 0.995 |
| 346.60 | 100.00 | 10.23 | 6.11 | 4.12 | 10.23 | -4.29 | 72.65 | 29.46 | 1.84 | 0.995 |
| 346.70 | 100.00 | 10.23 | 6.09 | 4.14 | 10.23 | -4.25 | 71.62 | 29.98 | 1.84 | 0.995 |
| 346.80 | 100.00 | 10.23 | 6.08 | 4.15 | 10.23 | -4.20 | 70.60 | 30.51 | 1.84 | 0.995 |
| 346.90 | 100.00 | 10.23 | 6.06 | 4.17 | 10.23 | -4.16 | 69.59 | 31.06 | 1.85 | 0.995 |
| 347.00 | 100.00 | 10.23 | 6.04 | 4.19 | 10.23 | -4.12 | 68.59 | 31.62 | 1.85 | 0.995 |
| 347.10 | 100.00 | 10.24 | 6.03 | 4.21 | 10.24 | -4.08 | 67.59 | 32.20 | 1.86 | 0.995 |
| 347.20 | 100.00 | 10.24 | 6.01 | 4.23 | 10.24 | -4.05 | 66.61 | 32.79 | 1.87 | 0.995 |
| 347.30 | 100.00 | 10.24 | 5.99 | 4.25 | 10.24 | -4.01 | 65.63 | 33.39 | 1.88 | 0.995 |
| 347.40 | 100.00 | 10.24 | 5.97 | 4.27 | 10.24 | -3.97 | 64.67 | 34.00 | 1.90 | 0.995 |
| 347.50 | 100.00 | 10.24 | 5.96 | 4.28 | 10.24 | -3.93 | 63.71 | 34.63 | 1.91 | 0.995 |
| 347.60 | 100.00 | 10.25 | 5.94 | 4.31 | 10.25 | -3.90 | 62.76 | 35.26 | 1.93 | 0.995 |
| 347.70 | 100.00 | 10.25 | 5.92 | 4.33 | 10.25 | -3.86 | 61.82 | 35.91 | 1.95 | 0.995 |
| 347.80 | 100.00 | 10.25 | 5.90 | 4.35 | 10.25 | -3.82 | 60.89 | 36.57 | 1.97 | 0.995 |
| 347.90 | 100.00 | 10.25 | 5.89 | 4.36 | 10.25 | -3.79 | 59.98 | 37.24 | 1.99 | 0.995 |
| 348.00 | 100.00 | 10.26 | 5.87 | 4.39 | 10.26 | -3.75 | 59.07 | 37.92 | 2.02 | 0.995 |
| 348.10 | 100.00 | 10.26 | 5.85 | 4.41 | 10.26 | -3.72 | 58.16 | 38.61 | 2.05 | 0.995 |
| 348.20 | 100.00 | 10.26 | 5.83 | 4.43 | 10.26 | -3.69 | 57.27 | 39.31 | 2.08 | 0.995 |
| 348.30 | 100.00 | 10.27 | 5.82 | 4.45 | 10.27 | -3.66 | 56.39 | 40.02 | 2.11 | 0.995 |
| 348.40 | 100.00 | 10.27 | 5.80 | 4.47 | 10.27 | -3.62 | 55.52 | 40.73 | 2.14 | 0.995 |
| 348.50 | 100.00 | 10.27 | 5.78 | 4.49 | 10.27 | -3.59 | 54.66 | 41.46 | 2.18 | 0.995 |
| 348.60 | 100.00 | 10.28 | 5.76 | 4.52 | 10.28 | -3.58 | 53.81 | 42.20 | 2.22 | 0.995 |
| 348.70 | 100.00 | 10.28 | 5.74 | 4.54 | 10.28 | -3.59 | 52.97 | 42.95 | 2.26 | 0.995 |
| 348.80 | 100.00 | 10.29 | 5.72 | 4.57 | 10.29 | -3.60 | 52.13 | 43.70 | 2.30 | 0.995 |
| 348.90 | 100.00 | 10.29 | 5.71 | 4.58 | 10.29 | -3.60 | 51.31 | 44.46 | 2.34 | 0.995 |
| 349.00 | 100.00 | 10.30 | 5.69 | 4.61 | 10.30 | -3.61 | 50.50 | 45.23 | 2.39 | 0.995 |
| 349.10 | 100.00 | 10.30 | 5.67 | 4.63 | 10.30 | -3.62 | 49.69 | 46.01 | 2.44 | 0.995 |
| 349.20 | 100.00 | 10.31 | 5.65 | 4.66 | 10.31 | -3.62 | 48.90 | 46.80 | 2.50 | 0.995 |
| 349.30 | 100.00 | 10.31 | 5.63 | 4.68 | 10.31 | -3.63 | 48.11 | 47.59 | 2.55 | 0.995 |
| 349.40 | 100.00 | 10.32 | 5.62 | 4.70 | 10.32 | -3.64 | 47.34 | 48.39 | 2.61 | 0.995 |
| 349.50 | 100.00 | 10.33 | 5.60 | 4.73 | 10.33 | -3.64 | 46.57 | 49.20 | 2.67 | 0.995 |
| 349.60 | 100.00 | 10.33 | 5.58 | 4.75 | 10.33 | -3.65 | 45.81 | 50.02 | 2.73 | 0.995 |
| 349.70 | 100.00 | 10.34 | 5.56 | 4.78 | 10.34 | -3.68 | 45.07 | 50.84 | 2.80 | 0.994 |
| 349.80 | 100.00 | 10.35 | 5.55 | 4.80 | 10.35 | -3.73 | 44.33 | 51.67 | 2.87 | 0.994 |
| 349.90 | 100.00 | 10.35 | 5.53 | 4.82 | 10.35 | -3.79 | 43.60 | 52.50 | 2.94 | 0.994 |
| 350.00 | 100.00 | 10.36 | 5.51 | 4.85 | 10.36 | -3.85 | 42.88 | 53.34 | 3.02 | 0.994 |
| 350.10 | 100.00 | 10.37 | 5.49 | 4.88 | 10.37 | -3.91 | 42.17 | 54.19 | 3.10 | 0.994 |
| 350.20 | 100.00 | 10.38 | 5.48 | 4.90 | 10.38 | -3.97 | 41.47 | 55.04 | 3.18 | 0.994 |
| 350.30 | 100.00 | 10.38 | 5.46 | 4.92 | 10.38 | -4.03 | 40.78 | 55.90 | 3.27 | 0.994 |
| 350.40 | 100.00 | 10.39 | 5.45 | 4.94 | 10.39 | -4.10 | 40.09 | 56.76 | 3.36 | 0.994 |
| 350.50 | 100.00 | 10.40 | 5.43 | 4.97 | 10.40 | -4.16 | 39.42 | 57.63 | 3.45 | 0.994 |
| 350.60 | 100.00 | 10.41 | 5.42 | 4.99 | 10.41 | -4.23 | 38.75 | 58.51 | 3.55 | 0.994 |
| 350.70 | 100.00 | 10.42 | 5.40 | 5.02 | 10.42 | -4.29 | 38.09 | 59.39 | 3.65 | 0.994 |
| 350.80 | 100.00 | 10.43 | 5.39 | 5.04 | 10.43 | -4.36 | 37.45 | 60.27 | 3.76 | 0.994 |
| 350.90 | 100.00 | 10.44 | 5.37 | 5.07 | 10.44 | -4.43 | 36.81 | 61.16 | 3.87 | 0.994 |
| 351.00 | 100.00 | 10.45 | 5.36 | 5.09 | 10.45 | -4.50 | 36.18 | 62.06 | 3.98 | 0.994 |
| 351.10 | 100.00 | 10.46 | 5.35 | 5.11 | 10.46 | -4.57 | 35.55 | 62.95 | 4.10 | 0.994 |
| 351.20 | 100.00 | 10.47 | 5.33 | 5.14 | 10.47 | -4.64 | 34.94 | 63.86 | 4.23 | 0.994 |
| 351.30 | 100.00 | 10.48 | 5.32 | 5.16 | 10.48 | -4.72 | 34.33 | 64.77 | 4.36 | 0.994 |
| 351.40 | 100.00 | 10.49 | 5.31 | 5.18 | 10.49 | -4.79 | 33.74 | 65.68 | 4.49 | 0.994 |
| 351.50 | 100.00 | 10.51 | 5.30 | 5.21 | 10.51 | -4.87 | 33.15 | 66.59 | 4.63 | 0.994 |
| 351.60 | 100.00 | 10.52 | 5.29 | 5.23 | 10.52 | -4.94 | 32.57 | 67.51 | 4.78 | 0.994 |
| 351.70 | 100.00 | 10.53 | 5.28 | 5.25 | 10.53 | -5.02 | 32.00 | 68.44 | 4.93 | 0.994 |
| 351.80 | 100.00 | 10.54 | 5.27 | 5.27 | 10.54 | -5.10 | 31.43 | 69.37 | 5.08 | 0.994 |
| 351.90 | 100.00 | 10.56 | 5.27 | 5.29 | 10.56 | -5.18 | 30.88 | 70.30 | 5.25 | 0.994 |
| 352.00 | 100.00 | 10.57 | 5.26 | 5.31 | 10.57 | -5.27 | 30.33 | 71.23 | 5.42 | 0.994 |
| 352.10 | 100.00 | 10.59 | 5.26 | 5.33 | 10.59 | -5.35 | 29.79 | 72.17 | 5.59 | 0.994 |
| 352.20 | 100.00 | 10.60 | 5.25 | 5.35 | 10.60 | -5.44 | 29.25 | 73.12 | 5.78 | 0.994 |
| 352.30 | 100.00 | 10.62 | 5.25 | 5.37 | 10.62 | -5.53 | 28.73 | 74.06 | 5.97 | 0.994 |
| 352.40 | 100.00 | 10.63 | 5.25 | 5.38 | 10.63 | -5.62 | 28.21 | 75.01 | 6.16 | 0.994 |
| 352.50 | 100.00 | 10.65 | 5.25 | 5.40 | 10.65 | -5.71 | 27.70 | 75.96 | 6.37 | 0.994 |
| 352.60 | 100.00 | 10.66 | 5.25 | 5.41 | 10.66 | -5.81 | 27.20 | 76.92 | 6.58 | 0.994 |
| 352.70 | 100.00 | 10.68 | 5.25 | 5.43 | 10.68 | -5.90 | 26.71 | 77.88 | 6.80 | 0.994 |
| 352.80 | 100.00 | 10.69 | 5.25 | 5.44 | 10.69 | -6.00 | 26.22 | 78.84 | 7.03 | 0.994 |
| 352.90 | 100.00 | 10.71 | 5.25 | 5.46 | 10.71 | -6.10 | 25.74 | 79.81 | 7.27 | 0.994 |
| 353.00 | 100.00 | 10.73 | 5.26 | 5.47 | 10.73 | -6.20 | 25.27 | 80.78 | 7.52 | 0.994 |
| 353.10 | 100.00 | 10.75 | 5.27 | 5.48 | 10.75 | -6.30 | 24.80 | 81.75 | 7.77 | 0.994 |
| 353.20 | 100.00 | 10.77 | 5.28 | 5.49 | 10.77 | -6.41 | 24.34 | 82.72 | 8.04 | 0.994 |
| 353.30 | 100.00 | 10.78 | 5.29 | 5.49 | 10.78 | -6.52 | 23.89 | 83.70 | 8.31 | 0.994 |
| 353.40 | 100.00 | 10.80 | 5.30 | 5.50 | 10.80 | -6.62 | 23.45 | 84.68 | 8.60 | 0.994 |
| 353.50 | 100.00 | 10.82 | 5.31 | 5.51 | 10.82 | -6.74 | 23.01 | 85.66 | 8.90 | 0.994 |
| 353.60 | 100.00 | 10.84 | 5.33 | 5.51 | 10.84 | -6.85 | 22.58 | 86.65 | 9.21 | 0.994 |
| 353.70 | 100.00 | 10.86 | 5.34 | 5.52 | 10.86 | -6.96 | 22.16 | 87.64 | 9.53 | 0.994 |
| 353.80 | 100.00 | 10.88 | 5.36 | 5.52 | 10.88 | -7.08 | 21.75 | 88.63 | 9.86 | 0.994 |
| 353.90 | 100.00 | 10.90 | 5.38 | 5.52 | 10.90 | -7.20 | 21.34 | 89.62 | 10.20 | 0.994 |
| 354.00 | 100.00 | 10.92 | 5.41 | 5.51 | 10.92 | -7.32 | 20.94 | 90.62 | 10.56 | 0.994 |
| 354.10 | 100.00 | 10.95 | 5.43 | 5.52 | 10.95 | -7.44 | 20.54 | 91.61 | 10.93 | 0.994 |
| 354.20 | 100.00 | 10.97 | 5.46 | 5.51 | 10.97 | -7.56 | 20.15 | 92.62 | 11.31 | 0.994 |
| 354.30 | 100.00 | 10.99 | 5.49 | 5.50 | 10.99 | -7.69 | 19.77 | 93.62 | 11.71 | 0.994 |
| 354.40 | 100.00 | 11.01 | 5.52 | 5.49 | 11.01 | -7.81 | 19.40 | 94.63 | 12.12 | 0.994 |
| 354.50 | 100.00 | 11.04 | 5.55 | 5.49 | 11.04 | -7.94 | 19.03 | 95.63 | 12.54 | 0.994 |
| 354.60 | 100.00 | 11.06 | 5.59 | 5.47 | 11.06 | -8.06 | 18.66 | 96.64 | 12.98 | 0.994 |
| 354.70 | 100.00 | 11.08 | 5.63 | 5.45 | 11.08 | -8.19 | 18.31 | 97.66 | 13.44 | 0.994 |
| 354.80 | 100.00 | 11.11 | 5.67 | 5.44 | 11.11 | -8.32 | 17.96 | 98.67 | 13.91 | 0.994 |
| 354.90 | 100.00 | 11.13 | 5.71 | 5.42 | 11.13 | -8.44 | 17.62 | 99.69 | 14.40 | 0.994 |
| 355.00 | 100.00 | 11.16 | 5.76 | 5.40 | 11.16 | -8.56 | 17.28 | 100.71 | 14.91 | 0.994 |
| 355.10 | 100.00 | 11.18 | 5.81 | 5.37 | 11.18 | -8.68 | 16.95 | 101.73 | 15.44 | 0.994 |
| 355.20 | 100.00 | 11.21 | 5.86 | 5.35 | 11.21 | -8.80 | 16.63 | 102.76 | 15.98 | 0.994 |
| 355.30 | 100.00 | 11.23 | 5.91 | 5.32 | 11.23 | -8.92 | 16.31 | 103.78 | 16.54 | 0.994 |
| 355.40 | 100.00 | 11.26 | 5.97 | 5.29 | 11.26 | -9.03 | 16.00 | 104.81 | 17.12 | 0.994 |
| 355.50 | 100.00 | 11.28 | 6.02 | 5.26 | 11.28 | -9.13 | 15.69 | 105.84 | 17.72 | 0.994 |
| 355.60 | 100.00 | 11.31 | 6.08 | 5.23 | 11.31 | -9.23 | 15.39 | 106.88 | 18.34 | 0.994 |
| 355.70 | 100.00 | 11.33 | 6.15 | 5.18 | 11.33 | -9.32 | 15.10 | 107.91 | 18.98 | 0.994 |
| 355.80 | 100.00 | 11.36 | 6.21 | 5.15 | 11.36 | -9.40 | 14.82 | 108.95 | 19.64 | 0.994 |
| 355.90 | 100.00 | 11.39 | 6.28 | 5.11 | 11.39 | -9.47 | 14.54 | 109.99 | 20.33 | 0.994 |
| 356.00 | 100.00 | 11.41 | 6.35 | 5.06 | 11.41 | -9.53 | 14.26 | 111.03 | 21.03 | 0.994 |
| 356.10 | 100.00 | 11.44 | 6.42 | 5.02 | 11.44 | -9.57 | 13.99 | 112.08 | 21.76 | 0.994 |
| 356.20 | 100.00 | 11.47 | 6.50 | 4.97 | 11.47 | -9.60 | 13.73 | 113.12 | 22.51 | 0.994 |
| 356.30 | 100.00 | 11.49 | 6.57 | 4.92 | 11.49 | -9.62 | 13.47 | 114.17 | 23.29 | 0.994 |
| 356.40 | 100.00 | 11.52 | 6.65 | 4.87 | 11.52 | -9.67 | 13.22 | 115.22 | 24.08 | 0.994 |
| 356.50 | 100.00 | 11.54 | 6.74 | 4.80 | 11.54 | -10.01 | 12.98 | 116.28 | 24.90 | 0.994 |
| 356.60 | 100.00 | 11.57 | 6.82 | 4.75 | 11.57 | -10.36 | 12.74 | 117.33 | 25.75 | 0.994 |
| 356.70 | 100.00 | 11.60 | 6.90 | 4.70 | 11.60 | -10.73 | 12.51 | 118.39 | 26.62 | 0.993 |
| 356.80 | 100.00 | 11.62 | 6.99 | 4.63 | 11.62 | -11.11 | 12.28 | 119.45 | 27.51 | 0.993 |
| 356.90 | 100.00 | 11.65 | 7.08 | 4.57 | 11.65 | -11.51 | 12.06 | 120.51 | 28.43 | 0.993 |
| 357.00 | 100.00 | 11.67 | 7.17 | 4.50 | 11.67 | -11.92 | 11.85 | 121.57 | 29.37 | 0.993 |
| 357.10 | 100.00 | 11.70 | 7.26 | 4.44 | 11.70 | -12.35 | 11.64 | 122.64 | 30.34 | 0.993 |
| 357.20 | 100.00 | 11.72 | 7.36 | 4.36 | 11.72 | -12.78 | 11.44 | 123.70 | 31.33 | 0.993 |
| 357.30 | 100.00 | 11.74 | 7.45 | 4.29 | 11.74 | -13.21 | 11.24 | 124.77 | 32.35 | 0.993 |
| 357.40 | 100.00 | 11.77 | 7.55 | 4.22 | 11.77 | -13.63 | 11.05 | 125.84 | 33.39 | 0.993 |
| 357.50 | 100.00 | 11.79 | 7.65 | 4.14 | 11.79 | -14.03 | 10.86 | 126.92 | 34.45 | 0.993 |
| 357.60 | 100.00 | 11.81 | 7.75 | 4.06 | 11.81 | -14.39 | 10.68 | 127.99 | 35.53 | 0.993 |
| 357.70 | 100.00 | 11.83 | 7.85 | 3.98 | 11.83 | -14.69 | 10.51 | 129.07 | 36.64 | 0.993 |
| 357.80 | 100.00 | 11.85 | 7.95 | 3.90 | 11.85 | -14.91 | 10.34 | 130.15 | 37.77 | 0.993 |
| 357.90 | 100.00 | 11.87 | 8.05 | 3.82 | 11.87 | -15.02 | 10.18 | 131.23 | 38.92 | 0.993 |
| 358.00 | 100.00 | 11.89 | 8.15 | 3.74 | 11.89 | -15.01 | 10.02 | 132.32 | 40.09 | 0.993 |
| 358.10 | 100.00 | 11.90 | 8.25 | 3.65 | 11.90 | -14.87 | 9.88 | 133.40 | 41.28 | 0.993 |
| 358.20 | 100.00 | 11.92 | 8.36 | 3.56 | 11.92 | -14.62 | 9.73 | 134.49 | 42.48 | 0.993 |
| 358.30 | 100.00 | 11.93 | 8.46 | 3.47 | 11.93 | -14.25 | 9.59 | 135.58 | 43.70 | 0.993 |
| 358.40 | 100.00 | 11.95 | 8.56 | 3.39 | 11.95 | -13.93 | 9.46 | 136.67 | 44.94 | 0.993 |
| 358.50 | 100.00 | 11.96 | 8.66 | 3.30 | 11.96 | -15.00 | 9.33 | 137.76 | 46.19 | 0.993 |
| 358.60 | 100.00 | 11.97 | 8.76 | 3.21 | 11.97 | -16.24 | 9.21 | 138.86 | 47.45 | 0.993 |
| 358.70 | 100.00 | 11.98 | 8.86 | 3.12 | 11.98 | -17.69 | 9.10 | 139.96 | 48.71 | 0.993 |
| 358.80 | 100.00 | 11.98 | 8.97 | 3.01 | 11.98 | -19.40 | 8.99 | 141.06 | 49.99 | 0.993 |
| 358.90 | 100.00 | 11.99 | 9.06 | 2.93 | 11.99 | -21.43 | 8.89 | 142.16 | 51.26 | 0.993 |
| 359.00 | 100.00 | 11.99 | 9.16 | 2.83 | 11.99 | -23.74 | 8.79 | 143.26 | 52.54 | 0.994 |
| 359.10 | 100.00 | 11.99 | 9.26 | 2.73 | 11.99 | -25.80 | 8.70 | 144.37 | 53.82 | 0.994 |
| 359.20 | 100.00 | 11.99 | 9.36 | 2.63 | 11.99 | -26.06 | 8.61 | 145.48 | 55.09 | 0.994 |
| 359.30 | 100.00 | 11.99 | 9.45 | 2.54 | 11.99 | -24.18 | 8.54 | 146.59 | 56.36 | 0.994 |
| 359.40 | 100.00 | 11.99 | 9.54 | 2.45 | 11.99 | -21.73 | 8.46 | 147.70 | 57.62 | 0.994 |
| 359.50 | 100.00 | 11.98 | 9.63 | 2.35 | 11.98 | -19.50 | 8.40 | 148.81 | 58.86 | 0.994 |
| 359.60 | 100.00 | 11.97 | 9.72 | 2.25 | 11.97 | -17.61 | 8.33 | 149.93 | 60.09 | 0.994 |
| 359.70 | 100.00 | 11.96 | 9.81 | 2.15 | 11.96 | -15.99 | 8.28 | 151.04 | 61.30 | 0.994 |
| 359.80 | 100.00 | 11.95 | 9.89 | 2.06 | 11.95 | -14.59 | 8.23 | 152.16 | 62.49 | 0.994 |
| 359.90 | 100.00 | 11.93 | 9.98 | 1.95 | 11.93 | -13.37 | 8.19 | 153.29 | 63.65 | 0.994 |
| 360.00 | 100.00 | 11.91 | 10.05 | 1.86 | 11.91 | -12.29 | 8.15 | 154.41 | 64.79 | 0.994 |
| 360.10 | 100.00 | 11.89 | 10.13 | 1.76 | 11.89 | -11.63 | 8.12 | 155.54 | 65.89 | 0.994 |
| 360.20 | 100.00 | 11.86 | 10.21 | 1.65 | 11.86 | -11.25 | 8.10 | 156.66 | 66.96 | 0.994 |
| 360.30 | 100.00 | 11.84 | 10.28 | 1.56 | 11.84 | -10.86 | 8.08 | 157.79 | 67.99 | 0.994 |
| 360.40 | 100.00 | 11.81 | 10.35 | 1.46 | 11.81 | -10.45 | 8.07 | 158.92 | 68.98 | 0.994 |
| 360.50 | 100.00 | 11.78 | 10.42 | 1.36 | 11.78 | -10.04 | 8.06 | 160.06 | 69.93 | 0.994 |
| 360.60 | 100.00 | 11.74 | 10.48 | 1.26 | 11.74 | -9.64 | 8.06 | 161.19 | 70.83 | 0.994 |
| 360.70 | 100.00 | 11.70 | 10.54 | 1.16 | 11.70 | -9.34 | 8.07 | 162.33 | 71.68 | 0.994 |
| 360.80 | 100.00 | 11.66 | 10.60 | 1.06 | 11.66 | -9.00 | 8.08 | 163.47 | 72.48 | 0.994 |
| 360.90 | 100.00 | 11.62 | 10.65 | 0.97 | 11.62 | -8.62 | 8.10 | 164.61 | 73.24 | 0.994 |
| 361.00 | 100.00 | 11.57 | 10.70 | 0.87 | 11.57 | -8.22 | 8.12 | 165.75 | 73.94 | 0.995 |
| 361.10 | 100.00 | 11.53 | 10.75 | 0.78 | 11.53 | -7.81 | 8.16 | 166.89 | 74.58 | 0.995 |
| 361.20 | 100.00 | 11.47 | 10.80 | 0.67 | 11.47 | -7.39 | 8.19 | 168.04 | 75.17 | 0.995 |
| 361.30 | 100.00 | 11.42 | 10.84 | 0.58 | 11.42 | -6.98 | 8.24 | 169.19 | 75.70 | 0.995 |
| 361.40 | 100.00 | 11.36 | 10.88 | 0.48 | 11.36 | -6.65 | 8.29 | 170.34 | 76.17 | 0.995 |
| 361.50 | 100.00 | 11.30 | 10.91 | 0.39 | 11.30 | -6.34 | 8.35 | 171.49 | 76.59 | 0.995 |
| 361.60 | 100.00 | 11.24 | 10.95 | 0.29 | 11.24 | -6.03 | 8.41 | 172.64 | 76.95 | 0.995 |
| 361.70 | 100.00 | 11.18 | 10.98 | 0.20 | 11.18 | -5.74 | 8.48 | 173.80 | 77.25 | 0.995 |
| 361.80 | 100.00 | 11.11 | 11.00 | 0.11 | 11.11 | -5.82 | 8.56 | 174.95 | 77.49 | 0.995 |
| 361.90 | 100.00 | 11.05 | 11.03 | 0.02 | 11.05 | -5.95 | 8.65 | 176.11 | 77.68 | 0.995 |
| 362.00 | 100.00 | 10.97 | 11.05 | -0.08 | 10.97 | -6.07 | 8.74 | 177.27 | 77.81 | 0.995 |
| 362.10 | 100.00 | 10.90 | 11.07 | -0.17 | 10.90 | -6.17 | 8.84 | 178.43 | 77.89 | 0.995 |
| 362.20 | 100.00 | 10.83 | 11.08 | -0.25 | 10.83 | -6.46 | 8.94 | 179.60 | 77.91 | 0.995 |
| 362.30 | 100.00 | 10.75 | 11.10 | -0.35 | 10.75 | -6.75 | 9.05 | 180.76 | 77.88 | 0.996 |
| 362.40 | 100.00 | 10.67 | 11.11 | -0.44 | 10.67 | -7.01 | 9.17 | 181.93 | 77.80 | 0.996 |
| 362.50 | 100.00 | 10.59 | 11.12 | -0.53 | 10.59 | -7.26 | 9.30 | 183.10 | 77.68 | 0.996 |
| 362.60 | 100.00 | 10.51 | 11.12 | -0.61 | 10.51 | -7.47 | 9.43 | 184.26 | 77.50 | 0.996 |
| 362.70 | 100.00 | 10.43 | 11.12 | -0.69 | 10.43 | -7.65 | 9.57 | 185.44 | 77.29 | 0.996 |
| 362.80 | 100.00 | 10.34 | 11.12 | -0.78 | 10.34 | -7.79 | 9.71 | 186.61 | 77.03 | 0.996 |
| 362.90 | 100.00 | 10.25 | 11.12 | -0.87 | 10.25 | -7.89 | 9.87 | 187.78 | 76.73 | 0.996 |
| 363.00 | 100.00 | 10.17 | 11.12 | -0.95 | 10.17 | -7.95 | 10.03 | 188.96 | 76.39 | 0.996 |
| 363.10 | 100.00 | 10.08 | 11.11 | -1.03 | 10.08 | -7.97 | 10.19 | 190.13 | 76.02 | 0.996 |
| 363.20 | 100.00 | 9.99 | 11.10 | -1.11 | 9.99 | -7.94 | 10.37 | 191.31 | 75.61 | 0.996 |
| 363.30 | 100.00 | 9.90 | 11.09 | -1.19 | 9.90 | -7.88 | 10.55 | 192.49 | 75.17 | 0.996 |
| 363.40 | 100.00 | 9.80 | 11.08 | -1.28 | 9.80 | -7.78 | 10.74 | 193.67 | 74.70 | 0.996 |
| 363.50 | 100.00 | 9.71 | 11.07 | -1.36 | 9.71 | -7.64 | 10.94 | 194.85 | 74.21 | 0.996 |
| 363.60 | 100.00 | 9.62 | 11.05 | -1.43 | 9.62 | -7.49 | 11.14 | 196.03 | 73.69 | 0.996 |
| 363.70 | 100.00 | 9.52 | 11.04 | -1.52 | 9.52 | -7.30 | 11.35 | 197.22 | 73.14 | 0.997 |
| 363.80 | 100.00 | 9.43 | 11.02 | -1.59 | 9.43 | -7.10 | 11.57 | 198.40 | 72.58 | 0.997 |
| 363.90 | 100.00 | 9.33 | 11.00 | -1.67 | 9.33 | -6.89 | 11.80 | 199.59 | 72.00 | 0.997 |
| 364.00 | 100.00 | 9.24 | 10.98 | -1.74 | 9.24 | -6.66 | 12.03 | 200.77 | 71.40 | 0.997 |
| 364.10 | 100.00 | 9.14 | 10.96 | -1.82 | 9.14 | -6.43 | 12.27 | 201.96 | 70.78 | 0.997 |
| 364.20 | 100.00 | 9.05 | 10.93 | -1.88 | 9.05 | -6.19 | 12.52 | 203.15 | 70.15 | 0.997 |
| 364.30 | 100.00 | 8.95 | 10.91 | -1.96 | 8.95 | -5.95 | 12.78 | 204.34 | 69.51 | 0.997 |
| 364.40 | 100.00 | 8.85 | 10.88 | -2.03 | 8.85 | -5.70 | 13.04 | 205.53 | 68.86 | 0.997 |
| 364.50 | 100.00 | 8.76 | 10.86 | -2.10 | 8.76 | -5.47 | 13.31 | 206.72 | 68.21 | 0.997 |
| 364.60 | 100.00 | 8.66 | 10.83 | -2.17 | 8.66 | -5.23 | 13.59 | 207.91 | 67.54 | 0.997 |
| 364.70 | 100.00 | 8.56 | 10.80 | -2.24 | 8.56 | -4.99 | 13.88 | 209.10 | 66.87 | 0.997 |
| 364.80 | 100.00 | 8.47 | 10.77 | -2.30 | 8.47 | -4.77 | 14.18 | 210.30 | 66.19 | 0.997 |
| 364.90 | 100.00 | 8.37 | 10.74 | -2.37 | 8.37 | -4.66 | 14.48 | 211.49 | 65.51 | 0.997 |
| 365.00 | 100.00 | 8.28 | 10.71 | -2.43 | 8.28 | -4.56 | 14.79 | 212.68 | 64.83 | 0.997 |
| 365.10 | 100.00 | 8.18 | 10.68 | -2.50 | 8.18 | -4.46 | 15.11 | 213.88 | 64.14 | 0.997 |
| 365.20 | 100.00 | 8.09 | 10.64 | -2.55 | 8.09 | -4.36 | 15.44 | 215.07 | 63.46 | 0.997 |
| 365.30 | 100.00 | 7.99 | 10.61 | -2.62 | 7.99 | -4.27 | 15.77 | 216.27 | 62.78 | 0.997 |
| 365.40 | 100.00 | 7.90 | 10.58 | -2.68 | 7.90 | -4.18 | 16.12 | 217.46 | 62.09 | 0.997 |
| 365.50 | 100.00 | 7.80 | 10.54 | -2.74 | 7.80 | -4.09 | 16.47 | 218.66 | 61.41 | 0.997 |
| 365.60 | 100.00 | 7.71 | 10.51 | -2.80 | 7.71 | -4.00 | 16.83 | 219.86 | 60.74 | 0.997 |
| 365.70 | 100.00 | 7.62 | 10.48 | -2.86 | 7.62 | -3.91 | 17.20 | 221.05 | 60.06 | 0.997 |
| 365.80 | 100.00 | 7.52 | 10.44 | -2.92 | 7.52 | -3.83 | 17.57 | 222.25 | 59.39 | 0.997 |
| 365.90 | 100.00 | 7.43 | 10.41 | -2.98 | 7.43 | -3.75 | 17.96 | 223.44 | 58.73 | 0.997 |
| 366.00 | 100.00 | 7.34 | 10.37 | -3.03 | 7.34 | -3.67 | 18.35 | 224.64 | 58.07 | 0.997 |
| 366.10 | 100.00 | 7.25 | 10.33 | -3.08 | 7.25 | -3.59 | 18.75 | 225.83 | 57.41 | 0.998 |
| 366.20 | 100.00 | 7.16 | 10.30 | -3.14 | 7.16 | -3.52 | 19.16 | 227.03 | 56.76 | 0.998 |
| 366.30 | 100.00 | 7.07 | 10.26 | -3.19 | 7.07 | -3.45 | 19.58 | 228.22 | 56.12 | 0.998 |
| 366.40 | 100.00 | 6.99 | 10.22 | -3.23 | 6.99 | -3.38 | 20.01 | 229.42 | 55.49 | 0.998 |
| 366.50 | 100.00 | 6.90 | 10.19 | -3.29 | 6.90 | -3.31 | 20.45 | 230.61 | 54.86 | 0.998 |
| 366.60 | 100.00 | 6.81 | 10.15 | -3.34 | 6.81 | -3.24 | 20.89 | 231.80 | 54.24 | 0.998 |
| 366.70 | 100.00 | 6.72 | 10.11 | -3.39 | 6.72 | -3.18 | 21.34 | 232.99 | 53.62 | 0.998 |
| 366.80 | 100.00 | 6.64 | 10.08 | -3.44 | 6.64 | -3.12 | 21.80 | 234.18 | 53.01 | 0.998 |
| 366.90 | 100.00 | 6.56 | 10.04 | -3.48 | 6.56 | -3.06 | 22.28 | 235.38 | 52.41 | 0.998 |
| 367.00 | 100.00 | 6.47 | 10.00 | -3.53 | 6.47 | -3.00 | 22.75 | 236.56 | 51.82 | 0.998 |
| 367.10 | 100.00 | 6.39 | 9.96 | -3.57 | 6.39 | -2.95 | 23.24 | 237.75 | 51.24 | 0.998 |
| 367.20 | 100.00 | 6.31 | 9.93 | -3.62 | 6.31 | -2.89 | 23.74 | 238.94 | 50.66 | 0.998 |
| 367.30 | 100.00 | 6.23 | 9.89 | -3.66 | 6.23 | -2.84 | 24.24 | 240.13 | 50.09 | 0.998 |
| 367.40 | 100.00 | 6.14 | 9.85 | -3.71 | 6.14 | -2.79 | 24.76 | 241.31 | 49.53 | 0.998 |
| 367.50 | 100.00 | 6.07 | 9.82 | -3.75 | 6.07 | -2.74 | 25.28 | 242.49 | 48.98 | 0.998 |
| 367.60 | 100.00 | 5.99 | 9.78 | -3.79 | 5.99 | -2.69 | 25.81 | 243.68 | 48.44 | 0.998 |
| 367.70 | 100.00 | 5.91 | 9.74 | -3.83 | 5.91 | -2.64 | 26.36 | 244.86 | 47.90 | 0.998 |
| 367.80 | 100.00 | 5.83 | 9.70 | -3.87 | 5.83 | -2.60 | 26.91 | 246.03 | 47.37 | 0.998 |
| 367.90 | 100.00 | 5.75 | 9.67 | -3.92 | 5.75 | -2.56 | 27.46 | 247.21 | 46.85 | 0.998 |
| 368.00 | 100.00 | 5.68 | 9.63 | -3.95 | 5.68 | -2.51 | 28.03 | 248.39 | 46.34 | 0.998 |
| 368.10 | 100.00 | 5.60 | 9.59 | -3.99 | 5.60 | -2.47 | 28.61 | 249.56 | 45.84 | 0.998 |
| 368.20 | 100.00 | 5.53 | 9.56 | -4.03 | 5.53 | -2.43 | 29.20 | 250.73 | 45.34 | 0.998 |
| 368.30 | 100.00 | 5.46 | 9.52 | -4.06 | 5.46 | -2.40 | 29.79 | 251.90 | 44.85 | 0.998 |
| 368.40 | 100.00 | 5.39 | 9.48 | -4.09 | 5.39 | -2.36 | 30.39 | 253.06 | 44.37 | 0.998 |
| 368.50 | 100.00 | 5.31 | 9.45 | -4.14 | 5.31 | -2.32 | 31.01 | 254.23 | 43.90 | 0.998 |
| 368.60 | 100.00 | 5.24 | 9.41 | -4.17 | 5.24 | -2.29 | 31.63 | 255.39 | 43.43 | 0.998 |
| 368.70 | 100.00 | 5.17 | 9.37 | -4.20 | 5.17 | -2.26 | 32.26 | 256.55 | 42.97 | 0.998 |
| 368.80 | 100.00 | 5.10 | 9.34 | -4.24 | 5.10 | -2.25 | 32.90 | 257.70 | 42.52 | 0.998 |
| 368.90 | 100.00 | 5.04 | 9.30 | -4.26 | 5.04 | -2.24 | 33.55 | 258.86 | 42.08 | 0.998 |
| 369.00 | 100.00 | 4.97 | 9.26 | -4.29 | 4.97 | -2.24 | 34.21 | 260.01 | 41.65 | 0.998 |
| 369.10 | 100.00 | 4.90 | 9.23 | -4.33 | 4.90 | -2.24 | 34.88 | 261.16 | 41.22 | 0.998 |
| 369.20 | 100.00 | 4.84 | 9.19 | -4.35 | 4.84 | -2.23 | 35.55 | 262.30 | 40.79 | 0.998 |
| 369.30 | 100.00 | 4.77 | 9.16 | -4.39 | 4.77 | -2.23 | 36.24 | 263.44 | 40.38 | 0.998 |
| 369.40 | 100.00 | 4.71 | 9.12 | -4.41 | 4.71 | -2.22 | 36.94 | 264.58 | 39.97 | 0.998 |
| 369.50 | 100.00 | 4.64 | 9.09 | -4.45 | 4.64 | -2.21 | 37.64 | 265.71 | 39.57 | 0.998 |
| 369.60 | 100.00 | 4.58 | 9.05 | -4.47 | 4.58 | -2.21 | 38.35 | 266.84 | 39.18 | 0.998 |
| 369.70 | 100.00 | 4.52 | 9.02 | -4.50 | 4.52 | -2.20 | 39.08 | 267.97 | 38.79 | 0.998 |
| 369.80 | 100.00 | 4.45 | 8.98 | -4.53 | 4.45 | -2.20 | 39.81 | 269.09 | 38.41 | 0.998 |
| 369.90 | 100.00 | 4.39 | 8.95 | -4.56 | 4.45 | -2.19 | 40.55 | 270.21 | 38.03 | 0.998 |
| 370.00 | 100.00 | 4.33 | 8.91 | -4.58 | 4.45 | -2.19 | 41.30 | 271.33 | 37.66 | 0.998 |
| 370.10 | 100.00 | 4.27 | 8.88 | -4.61 | 4.46 | -2.18 | 42.06 | 272.44 | 37.30 | 0.998 |
| 370.20 | 100.00 | 4.22 | 8.85 | -4.63 | 4.47 | -2.18 | 42.82 | 273.54 | 36.94 | 0.998 |
| 370.30 | 100.00 | 4.16 | 8.81 | -4.65 | 4.47 | -2.17 | 43.60 | 274.65 | 36.59 | 0.998 |
| 370.40 | 100.00 | 4.10 | 8.78 | -4.68 | 4.48 | -2.16 | 44.39 | 275.74 | 36.25 | 0.998 |
| 370.50 | 100.00 | 4.04 | 8.74 | -4.70 | 4.49 | -2.16 | 45.18 | 276.84 | 35.91 | 0.998 |
| 370.60 | 100.00 | 3.99 | 8.71 | -4.72 | 4.49 | -2.15 | 45.98 | 277.93 | 35.57 | 0.998 |
| 370.70 | 100.00 | 3.93 | 8.68 | -4.75 | 4.50 | -2.15 | 46.80 | 279.01 | 35.25 | 0.998 |
| 370.80 | 100.00 | 3.88 | 8.65 | -4.77 | 4.50 | -2.14 | 47.62 | 280.09 | 34.92 | 0.998 |
| 370.90 | 100.00 | 3.82 | 8.61 | -4.79 | 4.51 | -2.14 | 48.45 | 281.16 | 34.61 | 0.998 |
| 371.00 | 100.00 | 3.77 | 8.58 | -4.81 | 4.52 | -2.13 | 49.29 | 282.23 | 34.29 | 0.998 |
| 371.10 | 100.00 | 3.71 | 8.55 | -4.84 | 4.52 | -2.13 | 50.13 | 283.29 | 33.99 | 0.998 |
| 371.20 | 100.00 | 3.66 | 8.52 | -4.86 | 4.53 | -2.12 | 50.99 | 284.35 | 33.68 | 0.998 |
| 371.30 | 100.00 | 3.61 | 8.48 | -4.87 | 4.53 | -2.12 | 51.86 | 285.40 | 33.39 | 0.998 |
| 371.40 | 100.00 | 3.56 | 8.45 | -4.89 | 4.54 | -2.11 | 52.73 | 286.45 | 33.09 | 0.998 |
| 371.50 | 100.00 | 3.51 | 8.42 | -4.91 | 4.54 | -2.11 | 53.61 | 287.49 | 32.81 | 0.998 |
| 371.60 | 100.00 | 3.46 | 8.39 | -4.93 | 4.54 | -2.10 | 54.50 | 288.52 | 32.52 | 0.998 |
| 371.70 | 100.00 | 3.41 | 8.36 | -4.95 | 4.55 | -2.10 | 55.40 | 289.55 | 32.25 | 0.998 |
| 371.80 | 100.00 | 3.36 | 8.33 | -4.97 | 4.55 | -2.09 | 56.31 | 290.57 | 31.97 | 0.998 |
| 371.90 | 100.00 | 3.31 | 8.30 | -4.99 | 4.56 | -2.09 | 57.23 | 291.59 | 31.70 | 0.998 |
| 372.00 | 100.00 | 3.26 | 8.26 | -5.00 | 4.56 | -2.08 | 58.15 | 292.60 | 31.44 | 0.998 |
| 372.10 | 100.00 | 3.21 | 8.23 | -5.02 | 4.56 | -2.08 | 59.08 | 293.60 | 31.18 | 0.998 |
| 372.20 | 100.00 | 3.17 | 8.20 | -5.03 | 4.57 | -2.08 | 60.02 | 294.59 | 30.92 | 0.998 |
| 372.30 | 100.00 | 3.12 | 8.17 | -5.05 | 4.57 | -2.07 | 60.97 | 295.58 | 30.67 | 0.998 |
| 372.40 | 100.00 | 3.07 | 8.14 | -5.07 | 4.58 | -2.07 | 61.93 | 296.56 | 30.42 | 0.998 |
| 372.50 | 100.00 | 3.03 | 8.11 | -5.08 | 4.58 | -2.06 | 62.89 | 297.54 | 30.17 | 0.998 |
| 372.60 | 100.00 | 2.98 | 8.08 | -5.10 | 4.58 | -2.06 | 63.87 | 298.51 | 29.93 | 0.998 |
| 372.70 | 100.00 | 2.94 | 8.05 | -5.11 | 4.58 | -2.06 | 64.85 | 299.47 | 29.69 | 0.998 |
| 372.80 | 100.00 | 2.89 | 8.02 | -5.13 | 4.59 | -2.06 | 65.84 | 300.42 | 29.46 | 0.998 |
| 372.90 | 100.00 | 2.85 | 8.00 | -5.15 | 4.59 | -2.05 | 66.83 | 301.37 | 29.23 | 0.998 |
| 373.00 | 100.00 | 2.81 | 7.97 | -5.16 | 4.59 | -2.05 | 67.84 | 302.30 | 29.00 | 0.998 |
| 373.10 | 100.00 | 2.76 | 7.94 | -5.18 | 4.60 | -2.05 | 68.85 | 303.24 | 28.78 | 0.998 |
| 373.20 | 100.00 | 2.72 | 7.91 | -5.19 | 4.60 | -2.05 | 69.87 | 304.16 | 28.56 | 0.998 |
| 373.30 | 100.00 | 2.68 | 7.88 | -5.20 | 4.60 | -2.04 | 70.89 | 305.07 | 28.34 | 0.998 |
| 373.40 | 100.00 | 2.64 | 7.85 | -5.21 | 4.61 | -2.04 | 71.93 | 305.98 | 28.13 | 0.998 |
| 373.50 | 100.00 | 2.59 | 7.82 | -5.23 | 4.61 | -2.04 | 72.97 | 306.88 | 27.92 | 0.998 |
| 373.60 | 100.00 | 2.55 | 7.80 | -5.25 | 4.61 | -2.04 | 74.01 | 307.77 | 27.72 | 0.998 |
| 373.70 | 100.00 | 2.51 | 7.77 | -5.26 | 4.61 | -2.04 | 75.07 | 308.65 | 27.51 | 0.998 |
| 373.80 | 100.00 | 2.47 | 7.74 | -5.27 | 4.62 | -2.03 | 76.13 | 309.52 | 27.31 | 0.998 |
| 373.90 | 100.00 | 2.43 | 7.71 | -5.28 | 4.62 | -2.03 | 77.20 | 310.39 | 27.12 | 0.998 |
| 374.00 | 100.00 | 2.39 | 7.69 | -5.30 | 4.62 | -2.03 | 78.27 | 311.25 | 26.92 | 0.998 |
| 374.10 | 100.00 | 2.35 | 7.66 | -5.31 | 4.62 | -2.03 | 79.35 | 312.10 | 26.73 | 0.998 |
| 374.20 | 100.00 | 2.31 | 7.63 | -5.32 | 4.62 | -2.03 | 80.44 | 312.93 | 26.54 | 0.998 |
| 374.30 | 100.00 | 2.28 | 7.60 | -5.32 | 4.63 | -2.03 | 81.53 | 313.76 | 26.36 | 0.998 |
| 374.40 | 100.00 | 2.24 | 7.58 | -5.34 | 4.63 | -2.03 | 82.63 | 314.58 | 26.17 | 0.998 |
| 374.50 | 100.00 | 2.20 | 7.55 | -5.35 | 4.63 | -2.03 | 83.74 | 315.40 | 25.99 | 0.998 |
| 374.60 | 100.00 | 2.16 | 7.52 | -5.36 | 4.63 | -2.03 | 84.85 | 316.20 | 25.81 | 0.998 |
| 374.70 | 100.00 | 2.13 | 7.50 | -5.37 | 4.63 | -2.03 | 85.97 | 316.99 | 25.64 | 0.998 |
| 374.80 | 100.00 | 2.09 | 7.47 | -5.38 | 4.64 | -2.03 | 87.09 | 317.78 | 25.47 | 0.998 |
| 374.90 | 100.00 | 2.05 | 7.45 | -5.40 | 4.64 | -2.03 | 88.22 | 318.55 | 25.30 | 0.998 |
| 375.00 | 100.00 | 2.02 | 7.42 | -5.40 | 4.64 | -2.03 | 89.35 | 319.32 | 25.13 | 0.998 |
| 375.10 | 100.00 | 1.98 | 7.39 | -5.41 | 4.64 | -2.03 | 90.49 | 320.07 | 24.96 | 0.998 |
| 375.20 | 100.00 | 1.94 | 7.37 | -5.43 | 4.64 | -2.03 | 91.64 | 320.82 | 24.80 | 0.998 |
| 375.30 | 100.00 | 1.91 | 7.34 | -5.43 | 4.64 | -2.03 | 92.79 | 321.56 | 24.64 | 0.998 |
| 375.40 | 100.00 | 1.87 | 7.32 | -5.45 | 4.65 | -2.03 | 93.94 | 322.28 | 24.48 | 0.998 |
| 375.50 | 100.00 | 1.84 | 7.29 | -5.45 | 4.65 | -2.03 | 95.10 | 323.00 | 24.33 | 0.998 |
| 375.60 | 100.00 | 1.80 | 7.27 | -5.47 | 4.65 | -2.03 | 96.27 | 323.70 | 24.17 | 0.998 |
| 375.70 | 100.00 | 1.77 | 7.24 | -5.47 | 4.65 | -2.03 | 97.44 | 324.40 | 24.02 | 0.998 |
| 375.80 | 100.00 | 1.74 | 7.22 | -5.48 | 4.65 | -2.04 | 98.61 | 325.09 | 23.87 | 0.998 |
| 375.90 | 100.00 | 1.70 | 7.19 | -5.49 | 4.65 | -2.04 | 99.79 | 325.76 | 23.72 | 0.998 |
| 376.00 | 100.00 | 1.67 | 7.17 | -5.50 | 4.65 | -2.04 | 100.97 | 326.43 | 23.58 | 0.998 |
| 376.10 | 100.00 | 1.63 | 7.14 | -5.51 | 4.66 | -2.04 | 102.16 | 327.09 | 23.44 | 0.998 |
| 376.20 | 100.00 | 1.60 | 7.12 | -5.52 | 4.66 | -2.04 | 103.34 | 327.73 | 23.29 | 0.998 |
| 376.30 | 100.00 | 1.57 | 7.10 | -5.53 | 4.66 | -2.04 | 104.54 | 328.37 | 23.16 | 0.998 |
| 376.40 | 100.00 | 1.54 | 7.07 | -5.53 | 4.66 | -2.05 | 105.74 | 329.00 | 23.02 | 0.998 |
| 376.50 | 100.00 | 1.50 | 7.05 | -5.55 | 4.66 | -2.05 | 106.94 | 329.61 | 22.88 | 0.998 |
| 376.60 | 100.00 | 1.47 | 7.02 | -5.55 | 4.66 | -2.05 | 108.14 | 330.22 | 22.75 | 0.998 |
| 376.70 | 100.00 | 1.44 | 7.00 | -5.56 | 4.66 | -2.05 | 109.34 | 330.81 | 22.62 | 0.998 |
| 376.80 | 100.00 | 1.41 | 6.98 | -5.57 | 4.66 | -2.05 | 110.56 | 331.39 | 22.49 | 0.998 |
| 376.90 | 100.00 | 1.38 | 6.95 | -5.57 | 4.67 | -2.06 | 111.77 | 331.96 | 22.36 | 0.998 |
| 377.00 | 100.00 | 1.34 | 6.93 | -5.59 | 4.67 | -2.06 | 112.98 | 332.53 | 22.23 | 0.998 |
| 377.10 | 100.00 | 1.31 | 6.91 | -5.60 | 4.67 | -2.06 | 114.20 | 333.08 | 22.11 | 0.998 |
| 377.20 | 100.00 | 1.28 | 6.89 | -5.61 | 4.67 | -2.07 | 115.42 | 333.62 | 21.98 | 0.998 |
| 377.30 | 100.00 | 1.25 | 6.86 | -5.61 | 4.67 | -2.07 | 116.64 | 334.15 | 21.86 | 0.998 |
| 377.40 | 100.00 | 1.22 | 6.84 | -5.62 | 4.67 | -2.07 | 117.87 | 334.67 | 21.74 | 0.998 |
| 377.50 | 100.00 | 1.19 | 6.82 | -5.63 | 4.67 | -2.07 | 119.09 | 335.18 | 21.62 | 0.998 |
| 377.60 | 100.00 | 1.16 | 6.80 | -5.64 | 4.68 | -2.08 | 120.32 | 335.68 | 21.51 | 0.998 |
| 377.70 | 100.00 | 1.13 | 6.77 | -5.64 | 4.68 | -2.08 | 121.55 | 336.17 | 21.39 | 0.998 |
| 377.80 | 100.00 | 1.10 | 6.75 | -5.65 | 4.68 | -2.08 | 122.78 | 336.64 | 21.28 | 0.998 |
| 377.90 | 100.00 | 1.07 | 6.73 | -5.66 | 4.69 | -2.09 | 124.01 | 337.11 | 21.16 | 0.998 |
| 378.00 | 100.00 | 1.04 | 6.71 | -5.67 | 4.69 | -2.09 | 125.25 | 337.57 | 21.05 | 0.998 |
| 378.10 | 100.00 | 1.01 | 6.69 | -5.68 | 4.69 | -2.10 | 126.48 | 338.01 | 20.94 | 0.998 |
| 378.20 | 100.00 | 0.98 | 6.66 | -5.68 | 4.70 | -2.10 | 127.72 | 338.45 | 20.84 | 0.998 |
| 378.30 | 100.00 | 0.95 | 6.64 | -5.69 | 4.70 | -2.10 | 128.95 | 338.87 | 20.73 | 0.998 |
| 378.40 | 100.00 | 0.93 | 6.62 | -5.69 | 4.70 | -2.11 | 130.19 | 339.29 | 20.62 | 0.998 |
| 378.50 | 100.00 | 0.90 | 6.60 | -5.70 | 4.70 | -2.11 | 131.43 | 339.69 | 20.52 | 0.998 |
| 378.60 | 100.00 | 0.87 | 6.58 | -5.71 | 4.71 | -2.12 | 132.66 | 340.08 | 20.42 | 0.998 |
| 378.70 | 100.00 | 0.84 | 6.56 | -5.72 | 4.71 | -2.12 | 133.90 | 340.46 | 20.32 | 0.998 |
| 378.80 | 100.00 | 0.81 | 6.54 | -5.73 | 4.71 | -2.13 | 135.14 | 340.83 | 20.22 | 0.998 |
| 378.90 | 100.00 | 0.78 | 6.52 | -5.74 | 4.72 | -2.13 | 136.38 | 341.19 | 20.12 | 0.998 |
| 379.00 | 100.00 | 0.76 | 6.50 | -5.74 | 4.72 | -2.13 | 137.61 | 341.54 | 20.02 | 0.998 |
| 379.10 | 100.00 | 0.73 | 6.48 | -5.75 | 4.72 | -2.14 | 138.85 | 341.88 | 19.92 | 0.998 |
| 379.20 | 100.00 | 0.70 | 6.46 | -5.76 | 4.72 | -2.14 | 140.09 | 342.21 | 19.83 | 0.998 |
| 379.30 | 100.00 | 0.67 | 6.44 | -5.77 | 4.73 | -2.15 | 141.32 | 342.53 | 19.73 | 0.998 |
| 379.40 | 100.00 | 0.65 | 6.42 | -5.77 | 4.73 | -2.15 | 142.56 | 342.84 | 19.64 | 0.998 |
| 379.50 | 100.00 | 0.62 | 6.40 | -5.78 | 4.73 | -2.16 | 143.79 | 343.14 | 19.55 | 0.998 |
| 379.60 | 100.00 | 0.59 | 6.38 | -5.79 | 4.74 | -2.16 | 145.02 | 343.42 | 19.46 | 0.998 |
| 379.70 | 100.00 | 0.56 | 6.36 | -5.80 | 4.74 | -2.17 | 146.25 | 343.70 | 19.37 | 0.998 |
| 379.80 | 100.00 | 0.54 | 6.34 | -5.80 | 4.74 | -2.17 | 147.48 | 343.97 | 19.28 | 0.998 |
| 379.90 | 100.00 | 0.51 | 6.32 | -5.81 | 4.74 | -2.18 | 148.71 | 344.22 | 19.19 | 0.998 |
| 380.00 | 100.00 | 0.48 | 6.30 | -5.82 | 4.75 | -2.18 | 149.94 | 344.47 | 19.11 | 0.998 |
| 380.10 | 100.00 | 0.46 | 6.28 | -5.82 | 4.75 | -2.19 | 151.16 | 344.71 | 19.02 | 0.998 |
| 380.20 | 100.00 | 0.43 | 6.26 | -5.83 | 4.75 | -2.20 | 152.38 | 344.94 | 18.94 | 0.998 |
| 380.30 | 100.00 | 0.40 | 6.24 | -5.84 | 4.75 | -2.20 | 153.60 | 345.15 | 18.86 | 0.998 |
| 380.40 | 100.00 | 0.38 | 6.22 | -5.84 | 4.76 | -2.21 | 154.82 | 345.36 | 18.77 | 0.998 |
| 380.50 | 100.00 | 0.35 | 6.20 | -5.85 | 4.76 | -2.21 | 156.04 | 345.56 | 18.69 | 0.998 |
| 380.60 | 100.00 | 0.33 | 6.18 | -5.85 | 4.76 | -2.22 | 157.25 | 345.74 | 18.61 | 0.998 |
| 380.70 | 100.00 | 0.30 | 6.16 | -5.86 | 4.76 | -2.22 | 158.46 | 345.92 | 18.53 | 0.998 |
| 380.80 | 100.00 | 0.27 | 6.14 | -5.87 | 4.76 | -2.23 | 159.67 | 346.09 | 18.46 | 0.998 |
| 380.90 | 100.00 | 0.25 | 6.13 | -5.88 | 4.77 | -2.24 | 160.88 | 346.25 | 18.38 | 0.998 |
| 381.00 | 100.00 | 0.22 | 6.11 | -5.89 | 4.77 | -2.24 | 162.08 | 346.40 | 18.30 | 0.998 |
| 381.10 | 100.00 | 0.20 | 6.09 | -5.89 | 4.77 | -2.25 | 163.28 | 346.54 | 18.23 | 0.998 |
| 381.20 | 100.00 | 0.17 | 6.07 | -5.90 | 4.77 | -2.25 | 164.47 | 346.67 | 18.15 | 0.998 |
| 381.30 | 100.00 | 0.15 | 6.05 | -5.90 | 4.78 | -2.26 | 165.67 | 346.79 | 18.08 | 0.998 |
| 381.40 | 100.00 | 0.12 | 6.03 | -5.91 | 4.78 | -2.27 | 166.86 | 346.90 | 18.01 | 0.998 |
| 381.50 | 100.00 | 0.10 | 6.02 | -5.92 | 4.78 | -2.27 | 168.04 | 347.01 | 17.93 | 0.998 |
| 381.60 | 100.00 | 0.07 | 6.00 | -5.93 | 4.78 | -2.28 | 169.22 | 347.10 | 17.86 | 0.998 |
| 381.70 | 100.00 | 0.05 | 5.98 | -5.93 | 4.78 | -2.29 | 170.40 | 347.19 | 17.79 | 0.998 |
| 381.80 | 100.00 | 0.02 | 5.96 | -5.94 | 4.79 | -2.29 | 171.58 | 347.26 | 17.72 | 0.998 |
| 381.90 | 100.00 | 0.00 | 5.95 | -5.95 | 4.79 | -2.30 | 172.75 | 347.33 | 17.65 | 0.998 |
| 382.00 | 100.00 | -0.03 | 5.93 | -5.96 | 4.79 | -2.31 | 173.91 | 347.39 | 17.59 | 0.998 |
| 382.10 | 100.00 | -0.05 | 5.91 | -5.96 | 4.79 | -2.31 | 175.08 | 347.44 | 17.52 | 0.998 |
| 382.20 | 100.00 | -0.08 | 5.89 | -5.97 | 4.79 | -2.32 | 176.23 | 347.48 | 17.45 | 0.998 |
| 382.30 | 100.00 | -0.10 | 5.88 | -5.98 | 4.80 | -2.33 | 177.39 | 347.52 | 17.39 | 0.998 |
| 382.40 | 100.00 | -0.13 | 5.86 | -5.99 | 4.80 | -2.33 | 178.54 | 347.54 | 17.32 | 0.998 |
| 382.50 | 100.00 | -0.15 | 5.84 | -5.99 | 4.80 | -2.34 | 179.68 | 347.56 | 17.26 | 0.998 |
| 382.60 | 100.00 | -0.18 | 5.83 | -6.01 | 4.80 | -2.35 | 180.82 | 347.57 | 17.20 | 0.998 |
| 382.70 | 100.00 | -0.20 | 5.81 | -6.01 | 4.80 | -2.35 | 181.96 | 347.57 | 17.13 | 0.998 |
| 382.80 | 100.00 | -0.22 | 5.79 | -6.01 | 4.81 | -2.36 | 183.09 | 347.57 | 17.07 | 0.998 |
| 382.90 | 100.00 | -0.25 | 5.78 | -6.03 | 4.81 | -2.37 | 184.21 | 347.55 | 17.01 | 0.998 |
| 383.00 | 100.00 | -0.27 | 5.76 | -6.03 | 4.81 | -2.38 | 185.33 | 347.53 | 16.95 | 0.998 |
| 383.10 | 100.00 | -0.30 | 5.74 | -6.04 | 4.81 | -2.38 | 186.45 | 347.50 | 16.89 | 0.998 |
| 383.20 | 100.00 | -0.32 | 5.73 | -6.05 | 4.81 | -2.39 | 187.56 | 347.47 | 16.83 | 0.998 |
| 383.30 | 100.00 | -0.34 | 5.71 | -6.05 | 4.81 | -2.40 | 188.67 | 347.42 | 16.77 | 0.998 |
| 383.40 | 100.00 | -0.37 | 5.69 | -6.06 | 4.82 | -2.41 | 189.77 | 347.37 | 16.72 | 0.998 |
| 383.50 | 100.00 | -0.39 | 5.68 | -6.07 | 4.82 | -2.41 | 190.86 | 347.31 | 16.66 | 0.998 |
| 383.60 | 100.00 | -0.42 | 5.66 | -6.08 | 4.82 | -2.42 | 191.95 | 347.25 | 16.60 | 0.998 |
| 383.70 | 100.00 | -0.44 | 5.65 | -6.09 | 4.82 | -2.43 | 193.04 | 347.18 | 16.55 | 0.998 |
| 383.80 | 100.00 | -0.46 | 5.63 | -6.09 | 4.82 | -2.44 | 194.12 | 347.10 | 16.49 | 0.998 |
| 383.90 | 100.00 | -0.49 | 5.62 | -6.11 | 4.82 | -2.44 | 195.19 | 347.01 | 16.44 | 0.998 |
| 384.00 | 100.00 | -0.51 | 5.60 | -6.11 | 4.83 | -2.45 | 196.26 | 346.92 | 16.38 | 0.998 |
| 384.10 | 100.00 | -0.53 | 5.58 | -6.11 | 4.83 | -2.46 | 197.32 | 346.82 | 16.33 | 0.998 |
| 384.20 | 100.00 | -0.56 | 5.57 | -6.13 | 4.83 | -2.47 | 198.38 | 346.72 | 16.28 | 0.998 |
| 384.30 | 100.00 | -0.58 | 5.55 | -6.13 | 4.83 | -2.48 | 199.43 | 346.61 | 16.23 | 0.998 |
| 384.40 | 100.00 | -0.60 | 5.54 | -6.14 | 4.83 | -2.48 | 200.48 | 346.49 | 16.17 | 0.998 |
| 384.50 | 100.00 | -0.63 | 5.52 | -6.15 | 4.83 | -2.49 | 201.51 | 346.37 | 16.12 | 0.998 |
| 384.60 | 100.00 | -0.65 | 5.51 | -6.16 | 4.84 | -2.50 | 202.55 | 346.24 | 16.07 | 0.998 |
| 384.70 | 100.00 | -0.67 | 5.49 | -6.16 | 4.84 | -2.51 | 203.58 | 346.10 | 16.02 | 0.998 |
| 384.80 | 100.00 | -0.70 | 5.48 | -6.18 | 4.84 | -2.52 | 204.60 | 345.96 | 15.97 | 0.998 |
| 384.90 | 100.00 | -0.72 | 5.46 | -6.18 | 4.84 | -2.52 | 205.61 | 345.82 | 15.93 | 0.998 |
| 385.00 | 100.00 | -0.74 | 5.45 | -6.19 | 4.84 | -2.53 | 206.62 | 345.66 | 15.88 | 0.998 |
| 385.10 | 100.00 | -0.77 | 5.43 | -6.20 | 4.84 | -2.54 | 207.63 | 345.51 | 15.83 | 0.998 |
| 385.20 | 100.00 | -0.79 | 5.42 | -6.21 | 4.84 | -2.55 | 208.62 | 345.34 | 15.78 | 0.998 |
| 385.30 | 100.00 | -0.81 | 5.41 | -6.22 | 4.85 | -2.56 | 209.62 | 345.18 | 15.74 | 0.998 |
| 385.40 | 100.00 | -0.83 | 5.39 | -6.22 | 4.85 | -2.57 | 210.60 | 345.00 | 15.69 | 0.998 |
| 385.50 | 100.00 | -0.86 | 5.38 | -6.24 | 4.85 | -2.57 | 211.58 | 344.83 | 15.64 | 0.998 |
| 385.60 | 100.00 | -0.88 | 5.36 | -6.24 | 4.85 | -2.58 | 212.55 | 344.64 | 15.60 | 0.998 |
| 385.70 | 100.00 | -0.90 | 5.35 | -6.25 | 4.85 | -2.59 | 213.52 | 344.46 | 15.55 | 0.998 |
| 385.80 | 100.00 | -0.92 | 5.33 | -6.25 | 4.85 | -2.60 | 214.48 | 344.27 | 15.51 | 0.998 |
| 385.90 | 100.00 | -0.95 | 5.32 | -6.27 | 4.85 | -2.61 | 215.44 | 344.07 | 15.47 | 0.998 |
| 386.00 | 100.00 | -0.97 | 5.31 | -6.28 | 4.85 | -2.62 | 216.38 | 343.87 | 15.42 | 0.998 |
| 386.10 | 100.00 | -0.99 | 5.29 | -6.28 | 4.86 | -2.63 | 217.32 | 343.66 | 15.38 | 0.998 |
| 386.20 | 100.00 | -1.02 | 5.28 | -6.30 | 4.86 | -2.63 | 218.26 | 343.46 | 15.34 | 0.998 |
| 386.30 | 100.00 | -1.04 | 5.27 | -6.31 | 4.86 | -2.64 | 219.19 | 343.24 | 15.30 | 0.998 |
| 386.40 | 100.00 | -1.06 | 5.25 | -6.31 | 4.86 | -2.65 | 220.11 | 343.02 | 15.26 | 0.998 |
| 386.50 | 100.00 | -1.08 | 5.24 | -6.32 | 4.86 | -2.66 | 221.03 | 342.80 | 15.21 | 0.998 |
| 386.60 | 100.00 | -1.11 | 5.23 | -6.34 | 4.86 | -2.67 | 221.94 | 342.58 | 15.17 | 0.998 |
| 386.70 | 100.00 | -1.13 | 5.21 | -6.34 | 4.86 | -2.68 | 222.84 | 342.35 | 15.13 | 0.998 |
| 386.80 | 100.00 | -1.15 | 5.20 | -6.35 | 4.86 | -2.69 | 223.74 | 342.12 | 15.09 | 0.998 |
| 386.90 | 100.00 | -1.17 | 5.19 | -6.36 | 4.87 | -2.70 | 224.63 | 341.88 | 15.06 | 0.998 |
| 387.00 | 100.00 | -1.19 | 5.17 | -6.36 | 4.87 | -2.71 | 225.52 | 341.64 | 15.02 | 0.998 |
| 387.10 | 100.00 | -1.22 | 5.16 | -6.38 | 4.87 | -2.72 | 226.40 | 341.40 | 14.98 | 0.998 |
| 387.20 | 100.00 | -1.24 | 5.15 | -6.39 | 4.87 | -2.72 | 227.27 | 341.15 | 14.94 | 0.998 |
| 387.30 | 100.00 | -1.26 | 5.13 | -6.39 | 4.87 | -2.73 | 228.14 | 340.90 | 14.90 | 0.998 |
| 387.40 | 100.00 | -1.28 | 5.12 | -6.40 | 4.87 | -2.74 | 229.00 | 340.65 | 14.87 | 0.998 |
| 387.50 | 100.00 | -1.31 | 5.11 | -6.42 | 4.87 | -2.75 | 229.85 | 340.40 | 14.83 | 0.998 |
| 387.60 | 100.00 | -1.33 | 5.10 | -6.43 | 4.87 | -2.76 | 230.70 | 340.14 | 14.79 | 0.998 |
| 387.70 | 100.00 | -1.35 | 5.08 | -6.43 | 4.87 | -2.77 | 231.54 | 339.88 | 14.76 | 0.998 |
| 387.80 | 100.00 | -1.37 | 5.07 | -6.44 | 4.87 | -2.79 | 232.37 | 339.62 | 14.72 | 0.998 |
| 387.90 | 100.00 | -1.39 | 5.06 | -6.45 | 4.88 | -2.80 | 233.20 | 339.35 | 14.69 | 0.998 |
| 388.00 | 100.00 | -1.42 | 5.05 | -6.47 | 4.88 | -2.81 | 234.02 | 339.08 | 14.65 | 0.998 |
| 388.10 | 100.00 | -1.44 | 5.03 | -6.47 | 4.88 | -2.83 | 234.84 | 338.81 | 14.62 | 0.998 |
| 388.20 | 100.00 | -1.46 | 5.02 | -6.48 | 4.88 | -2.84 | 235.65 | 338.54 | 14.58 | 0.998 |
| 388.30 | 100.00 | -1.48 | 5.01 | -6.49 | 4.88 | -2.85 | 236.45 | 338.26 | 14.55 | 0.998 |
| 388.40 | 100.00 | -1.50 | 5.00 | -6.50 | 4.88 | -2.87 | 237.25 | 337.99 | 14.52 | 0.998 |
| 388.50 | 100.00 | -1.53 | 4.99 | -6.52 | 4.88 | -2.88 | 238.04 | 337.71 | 14.48 | 0.998 |
| 388.60 | 100.00 | -1.55 | 4.97 | -6.52 | 4.88 | -2.89 | 238.83 | 337.43 | 14.45 | 0.998 |
| 388.70 | 100.00 | -1.57 | 4.96 | -6.53 | 4.88 | -2.90 | 239.61 | 337.14 | 14.42 | 0.998 |
| 388.80 | 100.00 | -1.59 | 4.95 | -6.54 | 4.88 | -2.92 | 240.38 | 336.86 | 14.39 | 0.998 |
| 388.90 | 100.00 | -1.61 | 4.94 | -6.55 | 4.89 | -2.93 | 241.15 | 336.57 | 14.36 | 0.998 |
| 389.00 | 100.00 | -1.64 | 4.93 | -6.57 | 4.89 | -2.94 | 241.91 | 336.28 | 14.32 | 0.998 |
| 389.10 | 100.00 | -1.66 | 4.92 | -6.58 | 4.89 | -2.96 | 242.66 | 335.99 | 14.29 | 0.998 |
| 389.20 | 100.00 | -1.68 | 4.90 | -6.58 | 4.89 | -2.97 | 243.41 | 335.70 | 14.26 | 0.998 |
| 389.30 | 100.00 | -1.70 | 4.89 | -6.59 | 4.89 | -2.98 | 244.15 | 335.41 | 14.23 | 0.998 |
| 389.40 | 100.00 | -1.72 | 4.88 | -6.60 | 4.89 | -3.00 | 244.89 | 335.12 | 14.20 | 0.998 |
| 389.50 | 100.00 | -1.75 | 4.87 | -6.62 | 4.89 | -3.01 | 245.62 | 334.82 | 14.17 | 0.998 |
| 389.60 | 100.00 | -1.77 | 4.86 | -6.63 | 4.89 | -3.02 | 246.35 | 334.52 | 14.14 | 0.998 |
| 389.70 | 100.00 | -1.79 | 4.85 | -6.64 | 4.89 | -3.04 | 247.07 | 334.23 | 14.12 | 0.998 |
| 389.80 | 100.00 | -1.81 | 4.84 | -6.65 | 4.89 | -3.05 | 247.78 | 333.93 | 14.09 | 0.998 |
| 389.90 | 100.00 | -1.83 | 4.83 | -6.66 | 4.89 | -3.06 | 248.49 | 333.63 | 14.06 | 0.998 |
| 390.00 | 100.00 | -1.85 | 4.81 | -6.66 | 4.89 | -3.08 | 249.19 | 333.33 | 14.03 | 0.998 |
| 390.10 | 100.00 | -1.88 | 4.80 | -6.68 | 4.89 | -3.09 | 249.89 | 333.03 | 14.00 | 0.998 |
| 390.20 | 100.00 | -1.90 | 4.79 | -6.69 | 4.90 | -3.10 | 250.58 | 332.73 | 13.98 | 0.998 |
| 390.30 | 100.00 | -1.92 | 4.78 | -6.70 | 4.90 | -3.12 | 251.26 | 332.43 | 13.95 | 0.998 |
| 390.40 | 100.00 | -1.94 | 4.77 | -6.71 | 4.90 | -3.13 | 251.94 | 332.12 | 13.92 | 0.998 |
| 390.50 | 100.00 | -1.96 | 4.76 | -6.72 | 4.90 | -3.14 | 252.62 | 331.82 | 13.90 | 0.998 |
| 390.60 | 100.00 | -1.98 | 4.75 | -6.73 | 4.90 | -3.16 | 253.29 | 331.52 | 13.87 | 0.998 |
| 390.70 | 100.00 | -2.01 | 4.74 | -6.75 | 4.90 | -3.17 | 253.95 | 331.21 | 13.84 | 0.998 |
| 390.80 | 100.00 | -2.03 | 4.73 | -6.76 | 4.90 | -3.18 | 254.61 | 330.91 | 13.82 | 0.998 |
| 390.90 | 100.00 | -2.05 | 4.72 | -6.77 | 4.90 | -3.20 | 255.26 | 330.60 | 13.79 | 0.998 |
| 391.00 | 100.00 | -2.07 | 4.71 | -6.78 | 4.90 | -3.21 | 255.91 | 330.30 | 13.77 | 0.998 |
| 391.10 | 100.00 | -2.09 | 4.70 | -6.79 | 4.90 | -3.22 | 256.55 | 329.99 | 13.74 | 0.998 |
| 391.20 | 100.00 | -2.11 | 4.69 | -6.80 | 4.90 | -3.24 | 257.19 | 329.68 | 13.72 | 0.998 |
| 391.30 | 100.00 | -2.14 | 4.68 | -6.82 | 4.90 | -3.25 | 257.82 | 329.38 | 13.69 | 0.998 |
| 391.40 | 100.00 | -2.16 | 4.67 | -6.83 | 4.90 | -3.26 | 258.44 | 329.07 | 13.67 | 0.998 |
| 391.50 | 100.00 | -2.18 | 4.66 | -6.84 | 4.90 | -3.28 | 259.06 | 328.77 | 13.65 | 0.998 |
| 391.60 | 100.00 | -2.20 | 4.65 | -6.85 | 4.90 | -3.29 | 259.68 | 328.46 | 13.62 | 0.998 |
| 391.70 | 100.00 | -2.22 | 4.64 | -6.86 | 4.90 | -3.30 | 260.29 | 328.16 | 13.60 | 0.998 |
| 391.80 | 100.00 | -2.24 | 4.63 | -6.87 | 4.91 | -3.32 | 260.90 | 327.85 | 13.58 | 0.998 |
| 391.90 | 100.00 | -2.27 | 4.62 | -6.89 | 4.91 | -3.33 | 261.50 | 327.55 | 13.55 | 0.998 |
| 392.00 | 100.00 | -2.29 | 4.61 | -6.90 | 4.91 | -3.35 | 262.09 | 327.24 | 13.53 | 0.998 |
| 392.10 | 100.00 | -2.31 | 4.60 | -6.91 | 4.91 | -3.36 | 262.68 | 326.94 | 13.51 | 0.998 |
| 392.20 | 100.00 | -2.33 | 4.59 | -6.92 | 4.91 | -3.37 | 263.27 | 326.64 | 13.49 | 0.998 |
| 392.30 | 100.00 | -2.35 | 4.58 | -6.93 | 4.91 | -3.39 | 263.85 | 326.33 | 13.46 | 0.998 |
| 392.40 | 100.00 | -2.37 | 4.57 | -6.94 | 4.91 | -3.40 | 264.42 | 326.03 | 13.44 | 0.998 |
| 392.50 | 100.00 | -2.40 | 4.56 | -6.96 | 4.91 | -3.41 | 265.00 | 325.73 | 13.42 | 0.998 |
| 392.60 | 100.00 | -2.42 | 4.55 | -6.97 | 4.91 | -3.43 | 265.56 | 325.42 | 13.40 | 0.998 |
| 392.70 | 100.00 | -2.44 | 4.54 | -6.98 | 4.91 | -3.44 | 266.12 | 325.12 | 13.38 | 0.998 |
| 392.80 | 100.00 | -2.46 | 4.54 | -7.00 | 4.91 | -3.46 | 266.68 | 324.82 | 13.36 | 0.998 |
| 392.90 | 100.00 | -2.48 | 4.53 | -7.01 | 4.92 | -3.47 | 267.24 | 324.52 | 13.34 | 0.998 |
| 393.00 | 100.00 | -2.50 | 4.52 | -7.02 | 4.92 | -3.48 | 267.78 | 324.22 | 13.32 | 0.998 |
| 393.10 | 100.00 | -2.52 | 4.51 | -7.03 | 4.92 | -3.50 | 268.33 | 323.93 | 13.30 | 0.998 |
| 393.20 | 100.00 | -2.55 | 4.50 | -7.05 | 4.92 | -3.51 | 268.87 | 323.63 | 13.28 | 0.998 |
| 393.30 | 100.00 | -2.57 | 4.49 | -7.06 | 4.92 | -3.53 | 269.40 | 323.33 | 13.26 | 0.998 |
| 393.40 | 100.00 | -2.59 | 4.48 | -7.07 | 4.93 | -3.54 | 269.93 | 323.04 | 13.24 | 0.998 |
| 393.50 | 100.00 | -2.61 | 4.47 | -7.08 | 4.93 | -3.55 | 270.46 | 322.74 | 13.22 | 0.998 |
| 393.60 | 100.00 | -2.63 | 4.47 | -7.10 | 4.93 | -3.57 | 270.98 | 322.45 | 13.20 | 0.998 |
| 393.70 | 100.00 | -2.65 | 4.46 | -7.11 | 4.93 | -3.58 | 271.50 | 322.16 | 13.18 | 0.998 |
| 393.80 | 100.00 | -2.68 | 4.45 | -7.13 | 4.93 | -3.60 | 272.01 | 321.86 | 13.17 | 0.998 |
| 393.90 | 100.00 | -2.70 | 4.44 | -7.14 | 4.94 | -3.61 | 272.52 | 321.57 | 13.15 | 0.998 |
| 394.00 | 100.00 | -2.72 | 4.43 | -7.15 | 4.94 | -3.62 | 273.03 | 321.28 | 13.13 | 0.998 |
| 394.10 | 100.00 | -2.74 | 4.42 | -7.16 | 4.94 | -3.64 | 273.53 | 320.99 | 13.11 | 0.998 |
| 394.20 | 100.00 | -2.76 | 4.41 | -7.17 | 4.94 | -3.65 | 274.03 | 320.71 | 13.09 | 0.998 |
| 394.30 | 100.00 | -2.78 | 4.41 | -7.19 | 4.94 | -3.67 | 274.52 | 320.42 | 13.08 | 0.998 |
| 394.40 | 100.00 | -2.81 | 4.40 | -7.21 | 4.95 | -3.68 | 275.01 | 320.14 | 13.06 | 0.998 |
| 394.50 | 100.00 | -2.83 | 4.39 | -7.22 | 4.95 | -3.69 | 275.50 | 319.85 | 13.04 | 0.998 |
| 394.60 | 100.00 | -2.85 | 4.38 | -7.23 | 4.95 | -3.71 | 275.98 | 319.57 | 13.02 | 0.998 |
| 394.70 | 100.00 | -2.87 | 4.37 | -7.24 | 4.95 | -3.72 | 276.46 | 319.29 | 13.01 | 0.998 |
| 394.80 | 100.00 | -2.89 | 4.37 | -7.26 | 4.95 | -3.74 | 276.93 | 319.01 | 12.99 | 0.998 |
| 394.90 | 100.00 | -2.91 | 4.36 | -7.27 | 4.95 | -3.75 | 277.40 | 318.73 | 12.98 | 0.998 |
| 395.00 | 100.00 | -2.93 | 4.35 | -7.28 | 4.96 | -3.77 | 277.87 | 318.45 | 12.96 | 0.998 |
| 395.10 | 100.00 | -2.96 | 4.34 | -7.30 | 4.96 | -3.78 | 278.33 | 318.18 | 12.94 | 0.998 |
| 395.20 | 100.00 | -2.98 | 4.33 | -7.31 | 4.96 | -3.79 | 278.79 | 317.90 | 12.93 | 0.998 |
| 395.30 | 100.00 | -3.00 | 4.33 | -7.33 | 4.96 | -3.81 | 279.25 | 317.63 | 12.91 | 0.998 |
| 395.40 | 100.00 | -3.02 | 4.32 | -7.34 | 4.96 | -3.82 | 279.70 | 317.36 | 12.90 | 0.998 |
| 395.50 | 100.00 | -3.04 | 4.31 | -7.35 | 4.97 | -3.84 | 280.15 | 317.09 | 12.88 | 0.998 |
| 395.60 | 100.00 | -3.06 | 4.30 | -7.36 | 4.97 | -3.85 | 280.60 | 316.82 | 12.87 | 0.998 |
| 395.70 | 100.00 | -3.09 | 4.30 | -7.39 | 4.97 | -3.87 | 281.04 | 316.55 | 12.85 | 0.998 |
| 395.80 | 100.00 | -3.11 | 4.29 | -7.40 | 4.97 | -3.88 | 281.48 | 316.29 | 12.84 | 0.998 |
| 395.90 | 100.00 | -3.13 | 4.28 | -7.41 | 4.97 | -3.90 | 281.92 | 316.02 | 12.82 | 0.998 |
| 396.00 | 100.00 | -3.15 | 4.27 | -7.42 | 4.97 | -3.91 | 282.35 | 315.76 | 12.81 | 0.998 |
| 396.10 | 100.00 | -3.17 | 4.27 | -7.44 | 4.98 | -3.92 | 282.78 | 315.50 | 12.79 | 0.998 |
| 396.20 | 100.00 | -3.19 | 4.26 | -7.45 | 4.98 | -3.94 | 283.20 | 315.24 | 12.78 | 0.998 |
| 396.30 | 100.00 | -3.22 | 4.25 | -7.47 | 4.98 | -3.95 | 283.63 | 314.98 | 12.77 | 0.998 |
| 396.40 | 100.00 | -3.24 | 4.25 | -7.49 | 4.98 | -3.97 | 284.05 | 314.73 | 12.75 | 0.998 |
| 396.50 | 100.00 | -3.26 | 4.24 | -7.50 | 4.98 | -3.98 | 284.46 | 314.47 | 12.74 | 0.998 |
| 396.60 | 100.00 | -3.28 | 4.23 | -7.51 | 4.98 | -4.00 | 284.88 | 314.22 | 12.73 | 0.998 |
| 396.70 | 100.00 | -3.30 | 4.22 | -7.52 | 4.99 | -4.01 | 285.29 | 313.97 | 12.71 | 0.998 |
| 396.80 | 100.00 | -3.33 | 4.22 | -7.55 | 4.99 | -4.03 | 285.70 | 313.72 | 12.70 | 0.998 |
| 396.90 | 100.00 | -3.35 | 4.21 | -7.56 | 4.99 | -4.04 | 286.10 | 313.47 | 12.69 | 0.998 |
| 397.00 | 100.00 | -3.37 | 4.20 | -7.57 | 4.99 | -4.06 | 286.51 | 313.23 | 12.67 | 0.998 |
| 397.10 | 100.00 | -3.39 | 4.20 | -7.59 | 4.99 | -4.07 | 286.91 | 312.98 | 12.66 | 0.998 |
| 397.20 | 100.00 | -3.41 | 4.19 | -7.60 | 4.99 | -4.09 | 287.30 | 312.74 | 12.65 | 0.998 |
| 397.30 | 100.00 | -3.43 | 4.18 | -7.61 | 4.99 | -4.10 | 287.70 | 312.50 | 12.64 | 0.998 |
| 397.40 | 100.00 | -3.46 | 4.18 | -7.64 | 5.00 | -4.12 | 288.09 | 312.26 | 12.63 | 0.998 |
| 397.50 | 100.00 | -3.48 | 4.17 | -7.65 | 5.00 | -4.13 | 288.48 | 312.02 | 12.61 | 0.998 |
| 397.60 | 100.00 | -3.50 | 4.16 | -7.66 | 5.00 | -4.14 | 288.86 | 311.79 | 12.60 | 0.998 |
| 397.70 | 100.00 | -3.52 | 4.16 | -7.68 | 5.00 | -4.16 | 289.25 | 311.55 | 12.59 | 0.998 |
| 397.80 | 100.00 | -3.54 | 4.15 | -7.69 | 5.00 | -4.17 | 289.63 | 311.32 | 12.58 | 0.998 |
| 397.90 | 100.00 | -3.57 | 4.14 | -7.71 | 5.00 | -4.19 | 290.01 | 311.09 | 12.57 | 0.998 |
| 398.00 | 100.00 | -3.59 | 4.14 | -7.73 | 5.00 | -4.20 | 290.38 | 310.86 | 12.56 | 0.998 |
| 398.10 | 100.00 | -3.61 | 4.13 | -7.74 | 5.01 | -4.22 | 290.76 | 310.64 | 12.54 | 0.998 |
| 398.20 | 100.00 | -3.63 | 4.13 | -7.76 | 5.01 | -4.23 | 291.13 | 310.41 | 12.53 | 0.998 |
| 398.30 | 100.00 | -3.65 | 4.12 | -7.77 | 5.01 | -4.25 | 291.50 | 310.19 | 12.52 | 0.998 |
| 398.40 | 100.00 | -3.67 | 4.11 | -7.78 | 5.01 | -4.26 | 291.87 | 309.97 | 12.51 | 0.998 |
| 398.50 | 100.00 | -3.70 | 4.11 | -7.81 | 5.01 | -4.28 | 292.23 | 309.75 | 12.50 | 0.998 |
| 398.60 | 100.00 | -3.72 | 4.10 | -7.82 | 5.01 | -4.29 | 292.59 | 309.53 | 12.49 | 0.998 |
| 398.70 | 100.00 | -3.74 | 4.09 | -7.83 | 5.01 | -4.31 | 292.95 | 309.32 | 12.48 | 0.998 |
| 398.80 | 100.00 | -3.76 | 4.09 | -7.85 | 5.02 | -4.32 | 293.31 | 309.10 | 12.47 | 0.998 |
| 398.90 | 100.00 | -3.78 | 4.08 | -7.86 | 5.02 | -4.34 | 293.67 | 308.89 | 12.46 | 0.998 |
| 399.00 | 100.00 | -3.81 | 4.08 | -7.89 | 5.02 | -4.36 | 294.02 | 308.68 | 12.45 | 0.998 |
| 399.10 | 100.00 | -3.83 | 4.07 | -7.90 | 5.02 | -4.37 | 294.37 | 308.47 | 12.44 | 0.998 |
| 399.20 | 100.00 | -3.85 | 4.07 | -7.92 | 5.02 | -4.39 | 294.72 | 308.27 | 12.43 | 0.998 |
| 399.30 | 100.00 | -3.87 | 4.06 | -7.93 | 5.02 | -4.40 | 295.07 | 308.06 | 12.42 | 0.998 |
| 399.40 | 100.00 | -3.89 | 4.05 | -7.94 | 5.02 | -4.42 | 295.42 | 307.86 | 12.41 | 0.998 |
| 399.50 | 100.00 | -3.92 | 4.05 | -7.97 | 5.03 | -4.43 | 295.76 | 307.66 | 12.40 | 0.998 |
| 399.60 | 100.00 | -3.94 | 4.04 | -7.98 | 5.03 | -4.45 | 296.10 | 307.46 | 12.40 | 0.998 |
| 399.70 | 100.00 | -3.96 | 4.04 | -8.00 | 5.03 | -4.46 | 296.44 | 307.26 | 12.39 | 0.998 |
| 399.80 | 100.00 | -3.98 | 4.03 | -8.01 | 5.03 | -4.48 | 296.78 | 307.07 | 12.38 | 0.998 |
| 399.90 | 100.00 | -4.01 | 4.03 | -8.04 | 5.03 | -4.49 | 297.11 | 306.88 | 12.37 | 0.998 |
| 400.00 | 100.00 | -4.03 | 4.02 | -8.05 | 5.03 | -4.51 | 297.45 | 306.69 | 12.36 | 0.998 |

**IMPORTANT NOTE:  AVG PWR GAIN _MUST_ BE IN THE RANGE 0.8-1.2 FOR A VALID NEC MODEL (IDEALLY = 1)!
VALUES OUTSIDE THIS RANGE INDICATE A FLAWED MODEL, USUALLY A RESULT OF INCORRECT SEGMENTATION.
SEE DISCUSSION, p.101, NEC 4.1 USER'S MANUAL, LLNL UCRL-MA-109338 Pt. I (Gerald J. Burke), 1992.